\pdfoutput=1

\documentclass[reqno]{gsm-l-odonnell}

\usepackage{aobf}
\newcommand{\rquers}{r}
\newcommand{\rejrate}{\lambda}
\newcommand{\pred}{\psi}
\newcommand{\bpred}{\bpsi}
\newcommand{\predset}{\Psi}         \newcommand{\Predset}{\predset}
\newcommand{\instance}{{\calP}}
\newcommand{\asgn}{F}

\makenomenclature
\makeindex

\begin{document}

\frontmatter

\begin{center}
    {\LARGE \textbf{ANALYSIS OF BOOLEAN FUNCTIONS}} \\

    \vspace*{.8in}

    {\LARGE Ryan O'Donnell}
\end{center}

\vspace*{1in}

\begin{center}
\includegraphics{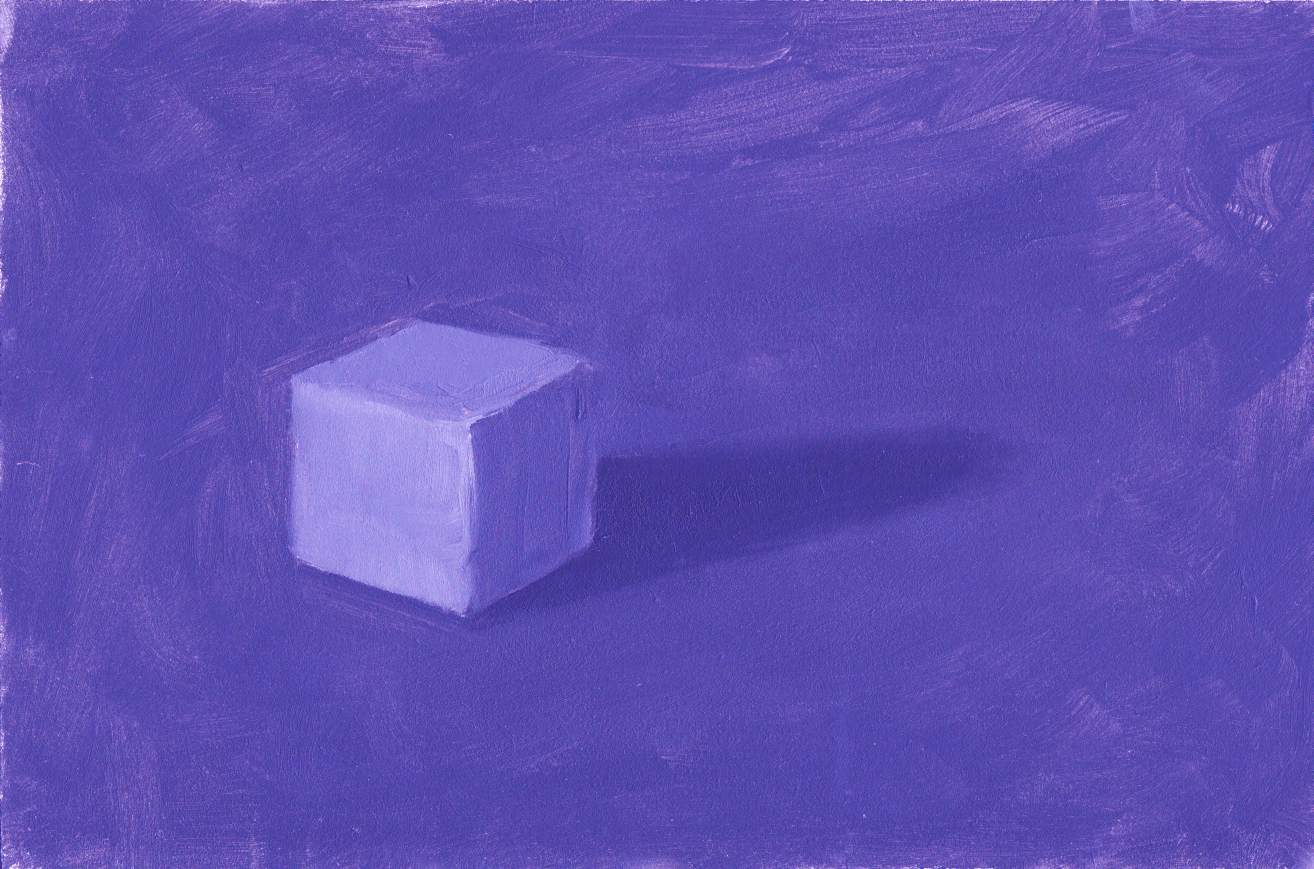}
\end{center}\par

\null\vfill

\noindent May 2021 \emph{arXiv Edition}.

\noindent Copyright \textcopyright\ Ryan O'Donnell, 2014, 2015, 2016, 2017, 2018, 2019, 2020, 2021.

\noindent Originally published April 2014 by Cambridge University Press.

\noindent Cover illustration by A.\ A.\ Williams.

\noindent {\tiny Single paper or electronic copies for noncommercial personal use may be made without explicit permission from the author or publisher. All other rights reserved.}

\cleardoublepage
\thispagestyle{empty}
\vspace*{16pc}
\begin{center}
  To Zeynep,\\[6pt]

  for her unending support and encouragement.
\end{center}
\cleardoublepage

\tableofcontents

\chapter*{Prefaces}

\noindent \textbf{\large Preface to the arXiv edition}

\medskip

The purpose of this May 2021 revision is to fix 100+ small typos and errors present in the first edition, and to make the result available on arXiv.
The numbering of all theorems, definitions, exercises, etc.\ is unchanged; there are only slight pagination differences.
Essentially no new mathematical content has been added, despite plenty of progress in the field; the book can be considered a ``snapshot'' of analysis of Boolean functions circa 2014.

\vspace{0.5in}

\noindent \textbf{\large Preface to the first edition}

\medskip

The subject of this textbook is the \emph{analysis of Boolean functions}.  Roughly speaking, this refers to studying Boolean functions $f \co \{0,1\}^n \to \{0,1\}$ via their Fourier expansion and other analytic means.  Boolean functions are perhaps the most basic object of study in theoretical computer science, and Fourier analysis has become an indispensable tool in the field.  The topic has also played a key role in several other areas of mathematics, from combinatorics, random graph theory, and statistical physics, to Gaussian geometry, metric/Banach spaces, and social choice theory.

The intent of this book is both to develop the foundations of the field and to give a wide (though far from exhaustive) overview of its applications.  Each chapter ends with a ``highlight'' showing the power of analysis of Boolean functions in different subject areas: property testing, social choice, cryptography, circuit complexity, learning theory, pseudorandomness, hardness of approximation, concrete complexity, and random graph theory.

The book can be used as a reference for working researchers or as the basis of a one-semester graduate-level course.  The author has twice taught such a course at Carnegie Mellon University, attended mainly by graduate students in computer science and mathematics but also by advanced undergraduates, postdocs, and researchers in adjacent fields.  In both years most of Chapters~1--5 and~7 were covered, along with parts of Chapters~6, 8, 9, and~11, and some additional material on additive combinatorics.  Nearly~$500$ exercises are provided at the ends of the book's chapters.

\vspace{0.5in}

\noindent \textbf{\large Acknowledgments}

\medskip

My foremost acknowledgment is to all of the people who have taught me analysis of Boolean functions, especially Guy Kindler and Elchanan Mossel.  I also learned a tremendous amount from my advisor Madhu Sudan, and my coauthors and colleagues Per Austrin, Eric Blais, Nader Bshouty, Ilias Diakonikolas, Irit Dinur, Uri Feige, Ehud Friedgut, Parikshit Gopalan, Venkat Guruswami, Johan H{\aa}stad, Gil Kalai, Daniel Kane, Subhash Khot, Adam Klivans, James Lee, Assaf Naor, Joe Neeman, Krzysztof Oleszkiewicz, Yuval Peres, Oded Regev, Mike Saks, Oded Schramm, Rocco Servedio, Amir Shpilka, Jeff Steif, Benny Sudakov, Li-Yang Tan, Avi Wigderson, Karl Wimmer, John Wright, Yi Wu, Yuan Zhou, and many others.  Ideas from all of them have strongly informed this book.

Many thanks to my PhD students who suffered from my inattention during the completion of this book: Eric Blais, Yuan Zhou, John Wright, and David Witmer.  I'd also like to thank the students who took my 2007 and 2012 courses on analysis of Boolean functions; special thanks to Deepak Bal, Carol Wang, and Patrick Xia for their very helpful course writing projects.

Thanks to my editor Lauren Cowles for her patience and encouragement, to the copyediting team of David Anderson and Rishi Gupta, and to Cambridge University Press for welcoming the free online publication of this book. Thanks also to Amanda Williams for the use of the cover image.

I'm very grateful to everyone who pointed out errors in earlier drafts of this work: Amirali Abdullah, Stefan Alders, anon, Arda Antikac{\i}o\u{g}lu, Albert Atserias, Per Austrin, Deepak Bal, Paul Beame, Tim Black, Ravi Boppana, Cl\'ement Canonne, Yongzhi Cao, Sankardeep Chakraborty, Bireswar Das, Andrew Drucker, Kirill Elagin, John Engbers, Diodato Ferraioli, Magnus Find, Michael Forbes, Malin Forsstr{\"o}m, Matt Franklin, David Gajser, David Garc{\'i}a Soriano, Dmitry Gavinsky, Daniele Gewurz, Mrinalkanti Ghosh, Sivakanth Gopi, Tom Gur, Zachary Hamaker, Sean Harrap, Prahladh Harsha, Justin Hilyard, Dmitry Itsykson, Hamidreza Jahanjou, Mitchell Johnston, Gautam Kamath, Shiva Kaul, Brian Kell, Ohad Klein, Pravesh Kothari, Chin Ho Lee, Euiwoong Lee, Holden Lee, Jerry Li, Noam Lifshitz, Chengyu Lin, Yang Liu, Tengyu Ma, Mladen Mik\v{s}a, Alexey Milovanov, Sidhanth Mohanty, Ashley Montanaro, Tobias M{\"u}ller, Aleksandar Nikolov, Krzysztof Oleszkiewicz, Pavithra KS, David Pritchard, Juspreet Sandhu, Swagato Sanyal, Pranav Senthilnathan, Igor Shinkar, Lior Silberman, Marla Slusky, Dmitry Sokolov, Aravind Srinivasan, Jeffrey Steif, Avishay Tal, Li-Yang Tan, Roei Tell, Suresh Venkatasubramanian, Marc Vinyals, Emanuele Viola, Poorvi Vora, Amos Waterland, Jake Wellens, Ryan Williams, Karl Wimmer, Chung Hoi Wong, Xi Wu, Yi Wu, Mingji Xia, Yuichi Yoshida, Shengyu Zhang, Yu Zhao, and Artsem Zhuk.  Special thanks in this group to Matt Franklin and Li-Yang Tan; extra-special thanks in this group to Noam Lifshitz and Jeff Steif.

I'm grateful to Denis Th\'{e}rien for inviting me to lecture at the Barbados Complexity Workshop, to Cynthia Dwork and the STOC~2008 PC for inviting me to give a tutorial, and to the Simons Foundation who arranged for me to co-organize a symposium together with Elchanan Mossel and Krzysztof Oleskiewicz, all on the topic of analysis of Boolean functions.  These opportunities greatly helped me to crystallize my thoughts on the topic.

I worked on this book while visiting the Institute for Advanced Study in 2010--2011 (supported by the Von Neumann Fellowship and in part by NSF grants DMS-0835373 and CCF-0832797); I'm very grateful to them for having me and for the wonderful working environment they provided.  The remainder of the work on this book was done at Carnegie Mellon; I'm of course very thankful to my colleagues there and to the Department of Computer Science.  ``Reasonable'' random variables were named after the department's ``Reasonable Person Principle''.  I was also supported in this book-writing endeavor by the National Science Foundation, specifically grants CCF-0747250 and CCF-1116594.  As usual: ``This material is based upon work supported by the National Science Foundation under grant numbers listed above. Any opinions, findings and conclusions or recommendations expressed in this material are those of the author and do not necessarily reflect the views of the National Science Foundation (NSF).''

Finally, I'd like to thank all of my colleagues, friends, and relatives who encouraged me to write and to finish the book, Zeynep most of all.

\bigskip
\bigskip
\bigskip

\begin{flushright}
 -- Ryan O'Donnell\\[-2pt]
 Pittsburgh\\[-2pt]
 October 2013
\end{flushright}

\renewcommand{\nomname}{List of Notation}

\renewcommand{\nompreamble}{

}

\printnomenclature[8em]

\mainmatter                      % Manual page number hack in chapter01.tex
\setcounter{page}{19}

\nomenclature[N]{$\N$}{$\{0, 1, 2, 3, \dots \}$}
\nomenclature[ln]{$\ln x$}{$\log_e x$}
\nomenclature[log]{$\log x$}{$\log_2 x$}

                                        \index{analysis of Boolean functions|(}%

\chapter{Boolean functions and the Fourier expansion}                       \label{chap:fourier-expansion}

In this chapter we describe the basics of analysis of Boolean functions.  We emphasize viewing the Fourier expansion of a Boolean function as its representation as a real multilinear polynomial.  The viewpoint based on harmonic analysis over $\F_2^n$ is mostly deferred to Chapter~\ref{chap:spectral-structure}.  We illustrate the use of basic Fourier formulas through the analysis of the Blum--Luby--Rubinfeld linearity test.

\section{On analysis of Boolean functions}                                  \label{sec:on-AOBF}	
This is a book about Boolean functions,
                                        \index{Boolean function}
\begin{equation*}
    f \co \zo^n \to \zo.
\end{equation*}
Here $f$ maps each length-$n$ binary vector, or \emph{string},
                                        \index{string}
into a single binary value, or \emph{bit}.
                                        \index{bit}
Boolean functions arise in many areas of computer science and mathematics.  Here are some examples:
\begin{itemize}
    \item In circuit design, a Boolean function may represent the desired behavior of a circuit with $n$ inputs and one output.
    \item In graph theory, one can identify $v$-vertex graphs $G$ with length-$\binom{v}{2}$ strings indicating which edges are present.  Then $f$ may represent a property of such graphs; e.g., $f(G) = 1$ if and only if $G$ is connected.
    \item In extremal combinatorics, a Boolean function $f$ can be identified with a ``set system''
                                        \index{set system}
          $\calF$ on $[n] = \{1, 2, \dots, n\}$,
                                        \nomenclature[n]{$[n]$}{$\{1, 2, 3, \dots, n\}$}
          where sets $X \subseteq [n]$ are identified with their $0$-$1$ indicators and $X \in \calF$ if and only if $f(X) = 1$.
    \item In coding theory, a Boolean function might be the indicator function for the set of messages in a binary error-correcting code of length~$n$.
    \item In learning theory, a Boolean function may represent a ``concept'' with $n$ binary attributes.
    \item In social choice theory, a Boolean function can be identified with a ``voting rule'' for an election with two candidates named $0$ and $1$.
\end{itemize}

We will be quite flexible about how bits
                                        \index{bit}
are represented. Sometimes we will use $\true$ and $\false$; sometimes we will use $-1$ and $1$, thought of as real numbers. Other times we will use $0$ and $1$, and these might be thought of as real numbers, as elements of the field $\F_2$
                                        \nomenclature[F2]{$\F_2$}{the finite field of size $2$}
of size~$2$, or just as symbols.  Most frequently we will use $-1$~and~$1$, so a Boolean function will look like
\begin{equation*}
    f \btb.
\end{equation*}
But we won't be dogmatic about the issue.

We refer to the domain of a Boolean function, $\{-1,1\}^n$, as the \emph{Hamming cube}
                                        \index{cube, Hamming}
                                        \index{Hamming cube|seeonly{cube, Hamming}}
(or hypercube,
                                        \index{hypercube|seeonly{cube, Hamming}}
$n$-cube,
                                        \index{n-cube@$n$-cube|seeonly{cube, Hamming}}
Boolean cube,
                                        \index{Boolean cube|seeonly{cube, Hamming}}
or discrete cube).
                                        \index{discrete cube|seeonly{cube, Hamming}}
The name ``Hamming cube'' emphasizes that we are often interested in the \emph{Hamming distance}
                                        \index{Hamming distance}
between strings $x, y \in \bitsn$, defined by
\begin{equation*}
    \hamdist(x,y) = \#\{i : x_i \neq y_i\}.
\end{equation*}%
                            \nomenclature[Delta]{$\hamdist(x,y)$}{the Hamming distance, $\#\{i : x_i \neq y_i\}$}%
Here we've used notation that will arise constantly: $x$ denotes a bit string, and $x_i$ denotes its $i$th coordinate.

Suppose you have a problem involving Boolean functions with the following two characteristics:
\begin{itemize}
  \item the Hamming distance is relevant;
  \item you are \emph{counting} strings, or the uniform probability distribution on $\{-1,1\}^n$ is involved.
\end{itemize}
These are the hallmarks of a problem for which \emph{analysis of Boolean functions}
                                    \index{analysis of Boolean functions}%
                                    \index{Fourier analysis of Boolean functions|seeonly{analysis of Boolean functions}}%
                                    \index{harmonic analysis of Boolean functions|seeonly{analysis of Boolean functions}}%
may help.  Roughly speaking, this means deriving information about Boolean functions by analyzing their \emph{Fourier expansion}.

\section{The ``Fourier expansion'': functions as multilinear polynomials}          \label{sec:fourier-expansion}

The \emph{Fourier expansion}
                                    \index{Fourier expansion|(}%
of a Boolean function $f \btb$ is simply its representation as a real, multilinear polynomial.  (\emph{Multilinear}
                                    \index{multilinear polynomial}%
means that no variable $x_i$ appears squared, cubed, etc.) For example, suppose $n = 2$ and $f = \maxt$, the
``maximum'' function on $2$ bits:
\iftex \begingroup\addtolength{\jot}{-.2em}\fi                
\begin{align*}
    \maxt(+1,+1) &= +1, \\
    \maxt(-1,+1) &= +1, \\
    \maxt(+1,-1) &= +1, \\
    \maxt(-1,-1) &= -1.
\end{align*}
\iftex \endgroup \fi
Then $\maxt$ can be expressed as a multilinear polynomial,
\begin{equation}                                    \label{eqn:min2-expansion}
    \maxt(x_1,x_2) = \tfrac12 + \tfrac12 x_1 + \tfrac12 x_2 - \tfrac12 x_1 x_2;
\end{equation}
this is the ``Fourier expansion'' of $\maxt$.  As another example, consider the \emph{majority function} on~$3$ bits, $\Maj_3 \co \bits^3 \to \bits$,
                                \index{majority}%
                                \nomenclature[majority]{$\maj_n$}{the majority function on $n$ bits}%
which outputs the $\pm 1$ bit occurring more frequently in its input. Then it's easy to verify the Fourier expansion
\begin{equation}                                    \label{eqn:maj3-expansion}
    \Maj_3(x_1,x_2,x_3) = \half x_1 + \half x_2 + \half x_3 - \half x_1x_2x_3.
\end{equation}
The functions $\maxt$ and $\maj_3$ will serve as running examples in this chapter.

Let's see how to obtain such multilinear polynomial representations in general.  Given an arbitrary Boolean function $f \btb$ there is a familiar method for finding a polynomial that interpolates the $2^n$ values that $f$ assigns to the points $\bits^n \subset \R^n$. For each point $a = (a_1, \dots, a_n) \in \bitsn$ the \emph{indicator polynomial}
                                \index{indicator polynomial}
\[
    1_{\{a\}}(x) = \left(\tfrac{1+a_1x_1}{2}\right)\left(\tfrac{1+a_2x_2}{2}\right) \cdots \left(\tfrac{1+a_nx_n}{2}\right)
\]
takes value $1$ when $x = a$ and value $0$ when $x \in \bitsn \setminus \{a\}$.  Thus $f$ has the polynomial representation
\[
    f(x) = \sum_{a \in \bits^n} f(a) 1_{\{a\}}(x).
\]
Illustrating with the $f = \maxt$ example again, we have
\iftex \begingroup\addtolength{\jot}{-.2em} \fi
\begin{align}
    \qquad\qquad\quad
    \maxt(x)\quad=\quad& \left(+1\right) \left(\tfrac{1+x_1}{2}\right)\left(\tfrac{1+x_2}{2}\right)\nonumber\\
             \quad+\quad& \left(+1\right) \left(\tfrac{1-x_1}{2}\right)\left(\tfrac{1+x_2}{2}\right) \label{eqn:min2-fourier-computation}\\
             \quad+\quad& \left(+1\right) \left(\tfrac{1+x_1}{2}\right)\left(\tfrac{1-x_2}{2}\right)\nonumber\\
             \quad+\quad& \left(-1\right) \left(\tfrac{1-x_1}{2}\right)\left(\tfrac{1-x_2}{2}\right)\nonumber
             \quad=\quad \tfrac12 + \tfrac12 x_1 + \tfrac12 x_2 - \tfrac12 x_1 x_2. \nonumber
\end{align}
\iftex \endgroup \fi
Let us make two remarks about this interpolation procedure.  First, it works equally well in the more general case of \emph{real-valued Boolean functions},
                                \index{Boolean function!real-valued}
$f \btR$.  Second, since the indicator polynomials are multilinear when expanded out, the interpolation always produces a multilinear polynomial.  Indeed, it makes sense that we can represent functions $f \btR$ with multilinear polynomials: since we only care about inputs $x$ where $x_i = \pm 1$, any factor of $x_i^2$ can be replaced by~$1$.

We have illustrated that every $f \btR$ can be represented by a real multilinear polynomial; as we will see in Section~\ref{sec:ortho-basis-parities}, this representation is unique.  The multilinear polynomial for~$f$ may have up to $2^n$ terms, corresponding to the subsets $S \subseteq [n]$.  We write the monomial corresponding to $S$ as
\[
\phantom{\text{(with $x^\emptyset = 1$ by convention)}} x^S = \prod_{i \in S} x_i \tag*{(with $x^\emptyset = 1$ by convention),}
\]
                        \nomenclature[xS]{$x^S$}{$\prod_{i \in S} x_i$, with the convention $x^\emptyset = 1$}%
and we use the following notation for its coefficient:
\begin{equation*}
    \wh{f}(S) = \text{coefficient on monomial $x^S$ in the multilinear representation of $f$}.
\end{equation*}%
                            \nomenclature[fS]{$\wh{f}(S)$}{the Fourier coefficient of $f$ on character $\chi_S$}%
                            \index{Fourier coefficient}%
This discussion is summarized by the \emph{Fourier expansion theorem}:
\begin{theorem}                                     \label{thm:unique-fourier-expansion}
    Every function $f \btR$ can be uniquely expressed as a multilinear polynomial,
    \begin{equation}                                \label{eqn:multilinear-expansion}
        f(x) = \sum_{S \subseteq [n]} \wh{f}(S)\,x^S.
    \end{equation}
    This expression is called the \emph{Fourier expansion} of~$f$, and the real number $\wh{f}(S)$ is called the \emph{Fourier coefficient of $f$ on $S$}.  Collectively, the coefficients are called the \emph{Fourier spectrum}
                                                    \index{Fourier spectrum}
    of~$f$.
\end{theorem}
As examples, from~\eqref{eqn:min2-expansion} and~\eqref{eqn:maj3-expansion} we obtain:
\[
    \wh{\maxt}(\emptyset) = \half, \quad \wh{\maxt}(\{1\}) = \half, \quad \wh{\maxt}(\{2\}) = \half, \quad \wh{\maxt}(\{1,2\}) = -\half;
\]
\[
    \wh{\maj_3}(\{1\}),\ \wh{\maj_3}(\{2\}),\ \wh{\maj_3}(\{3\}) = \half, \quad \wh{\maj_3}(\{1,2,3\}) = -\half, \quad \wh{\maj_3}(S) = 0 \text{ else.}
\]

We finish this section with some notation.  It is convenient to think of the monomial $x^S$ as a function on $x = (x_1, \dots, x_n) \in \R^n$; we write it
                \nomenclature[chiS]{$\chi_S(x)$}{when $x \in \R^n$, denotes $\prod_{i \in S} x_i$, where $S \subseteq [n]$; when $x \in \F_2^n$, denotes $(-1)^{\sum_{i \in S} x_i}$}
as
\[
    \chi_S(x) = \prod_{i \in S} x_i.
\]
Thus we sometimes write the Fourier expansion of $f \btR$ as
\[
    f(x) = \sumS \wh{f}(S)\,\chi_S(x).
\]
So far our notation makes sense only when representing the Hamming cube by $\bitsn \subseteq \R^n$.  The other frequent representation we will use for the cube is $\F_2^n$.  We can define the Fourier expansion for functions $f \ftR$ by ``encoding'' input bits $0, 1\in \F_2$ by the real numbers $-1,1 \in \R$.  We choose the encoding $\chi \co \F_2 \to \R$ defined by
\[
    \chi(0_{\F_2}) = +1, \quad \chi(1_{\F_2}) = -1.
\]%
                            \nomenclature[chi]{$\chi(b)$}{when $b \in \F_2^n$, denotes $(-1)^b \in \R$}%
This encoding is not so natural from the perspective of Boolean logic; e.g., it means the function $\max_2$ we have discussed represents logical~$\AND$.  But it's mathematically natural because for $b \in \F_2$ we have the formula $\chi(b) = (-1)^b$.  We now extend the $\chi_S$ notation:
\begin{definition}
    For $S \subseteq [n]$ we define $\chi_S \co \F_2^n \to \R$ by
    \[
        \chi_S(x) = \prod_{i \in S} \chi(x_i) = (-1)^{\sum_{i \in S} x_i},
    \]
    which satisfies
    \begin{equation}                                        \label{eqn:chi-character}
        \chi_S(x+y) = \chi_S(x)\chi_S(y).
    \end{equation}
\end{definition}
In this way, given any function $f \ftR$ it makes sense to write its Fourier expansion as
\begin{equation*}
    f(x) = \sum_{S \subseteq [n]} \wh{f}(S)\,\chi_S(x).
\end{equation*}

In fact, if we are really thinking of $\F_2^n$ the $n$-dimensional vector space over $\F_2$, it makes sense to identify subsets $S \subseteq [n]$ with vectors $\gamma \in \F_2^n$.  This will be discussed in
                                        \index{Fourier expansion|)}
Chapter~\ref{sec:dec-trees}.

\section{The orthonormal basis of parity functions}                          \label{sec:ortho-basis-parities}

For $x \in \bits^n$, the number $\chi_S(x) = \prod_{i \in S} x_i$ is in $\bits$.  Thus $\chi_S \btb$ is a Boolean function; it computes the logical \emph{parity},
                                        \index{parity}
or \emph{exclusive-or} (XOR),
                                        \index{exclusive-or|seeonly{parity}}%
                                        \index{XOR|seeonly{parity}}%
of the bits $(x_i)_{i \in S}$.  The parity functions play a special role in the analysis of Boolean functions: the Fourier expansion
\begin{equation}                                       \label{eqn:parity-basis-expanion}
    f = \sumS \wh{f}(S)\,\chi_S
\end{equation}
shows that any $f$ can be represented as a linear combination of parity functions (over the reals).

It's useful to explore this idea further from the perspective of linear algebra.  The set of all functions $f \btR$ forms a vector space~$V$, since we can add two functions (pointwise) and we can multiply a function by a real scalar.  The vector space $V$ is $2^n$-dimensional: if we like we can think of the functions in this vector space as vectors in $\R^{2^n}$, where we stack the $2^n$ values $f(x)$ into a tall column vector (in some fixed order). Here we illustrate the Fourier expansion~\eqref{eqn:min2-expansion} of the $\maxt$ function from this perspective:
\begin{equation}                                                    \label{eqn:min2-basis-epansion}
    \maxt   = \begin{bmatrix} +1 \\ +1 \\ +1 \\ -1 \end{bmatrix}
            \ =\ (1/2) \begin{bmatrix} +1 \\ +1 \\ +1 \\ +1 \end{bmatrix}
              \ +(1/2) \begin{bmatrix} +1 \\ -1 \\ +1 \\ -1 \end{bmatrix}
              \ +(1/2) \begin{bmatrix} +1 \\ +1 \\ -1 \\ -1 \end{bmatrix}
              \ +(-1/2) \begin{bmatrix} +1 \\ -1 \\ -1 \\ +1 \end{bmatrix}.
\end{equation}

More generally, the Fourier expansion~\eqref{eqn:parity-basis-expanion} shows that every function $f \btR$ in~$V$ is a linear combination of the parity functions; i.e., the parity functions are a \emph{spanning set} for~$V$.  Since the number of parity functions is $2^n = \dim V$, we can deduce that they are in fact a \emph{linearly independent basis} for $V$.  In particular this justifies the uniqueness of the Fourier expansion stated in Theorem~\ref{thm:unique-fourier-expansion}.

We can also introduce an inner product on pairs of function $f, g \btR$ in $V$.  The usual inner product on $\R^{2^n}$ would correspond to $\sum_{x \in \bits^n} f(x)g(x)$, but it's more convenient to scale this by a factor of $2^{-n}$, making it an average rather than a sum. In this way, a Boolean function $f \btb$ will have $\la f, f \ra = 1$, i.e., be a ``unit vector''.
\begin{definition}
    We define an inner product $\la \cdot , \cdot \ra$ on pairs of function $f, g \btR$ by
    \begin{equation}                                        \label{eqn:inner-product}
        \la f, g \ra = 2^{-n} \sum_{x \in \bits^n} f(x) g(x) = \Ex_{\bx \sim \bits^n} \left[f(\bx) g(\bx)\right].
    \end{equation}
    We also use the notation $\|f\|_2 = \sqrt{\la f, f \ra}$, and more generally,
    \[
        \|f\|_p = \Ex[|f(\bx)|^p]^{1/p}.
    \]
\end{definition}%
                                            \index{inner product}%
                                            \index{norm}%
                                            \nomenclature[fg]{$\la f, g \ra$}{$\E_{\bx}[f(\bx) g(\bx)]$}%
                                            \nomenclature{$\|f\|_p$}{$\E[\vert f \vert^p]^{1/p}$}%
                                            \nomenclature{$\|f(\bx,\by)\|_{p,\bx}$}{$\E_{\bx}[\vert f(\bx, \by) \vert^p]^{1/p}$, a function of~$\by$}%

Here we have introduced probabilistic notation that will be used heavily throughout the book:
\begin{notation}
     We write $\bx \sim \bits^n$
                    \nomenclature[x~2]{$\bx \sim \bits^n$}{the random variable $\bx$ is chosen uniformly from $\bits^n$}
     to denote that $\bx$ is a uniformly chosen random string from $\bits^n$. Equivalently, the $n$ coordinates $\bx_i$ are independently chosen to be $+1$ with probability $1/2$ and $-1$ with probability $1/2$.
                                            \index{uniform distribution}
    We always write random variables in \textbf{boldface}. Probabilities $\Pr$ and expectations $\E$ will always be with respect to a uniformly random $\bx \sim \bits^n$ unless otherwise specified.  Thus we might write the expectation in~\eqref{eqn:inner-product} as $\Ex_{\bx}[f(\bx) g(\bx)]$ or $\Ex[f(\bx) g(\bx)]$ or even $\E[f g]$.
\end{notation}%

Returning to the basis of parity functions for $V$, the crucial fact underlying all analysis of Boolean functions is that this is an \emph{orthonormal basis}.
\begin{theorem}                                     \label{thm:parities-orthormal}
    The $2^n$ parity functions $\chi_S \btb$ form an orthonormal basis for the vector space $V$ of functions $\bn \to \R$;
                                                \index{orthonormal}%
    i.e.,
    \begin{equation*}
        \la \chi_S, \chi_T \ra = \begin{cases}
          1 & \text{if $S = T$,}\\
          0 & \text{if $S \neq T$.}
        \end{cases}
    \end{equation*}
\end{theorem}
\noindent Recalling the definition $\la \chi_S, \chi_T \ra = \E[\chi_S(\bx) \chi_T(\bx)]$, Theorem~\ref{thm:parities-orthormal} follows immediately from two facts:
\begin{fact}  \label{fact:symdiff} For $x \in \bitsn$ it holds that $\chi_S(x) \chi_T(x) = \chi_{S \symdiff T}(x)$, where $S \symdiff T$ denotes symmetric difference.
                            \nomenclature{$\symdiff$}{symmetric difference of sets; i.e., $S \symdiff T = \{i : i \text{ is in exactly one of } S, T\}$}
\end{fact}
\begin{proof}
     $\displaystyle \chi_S(x) \chi_T(x) = \prod_{i \in S} x_i \prod_{i \in T} x_i = \prod_{i \in S \symdiff T} x_i \prod_{i \in S \cap T} x_i^2 = \prod_{i \in S \symdiff T} x_i = \chi_{S \symdiff T}(x).$
\end{proof}
\begin{fact} \label{fact:exp-character} \iftex $\displaystyle \fi \ifblog \[ \fi
        \E[\chi_S(\bx)] = \E\Bigl[\prod_{i \in S} \bx_i\Bigr] =  \begin{cases}
          1 & \text{if $S = \emptyset$,}\\
          0 & \text{if $S \neq \emptyset$.}
        \end{cases}
             \iftex $ \fi \ifblog \] \fi
\end{fact}
\begin{proof}
    If $S = \emptyset$ then $\E[\chi_S(\bx)] = \E[1] = 1$.  Otherwise,
    \[
    \E\Bigl[\prod_{i \in S} \bx_i\Bigr] = \prod_{i \in S} \E[\bx_i]
    \]
    because the random bits $\bx_1, \dots, \bx_n$ are independent. But each of the factors $\E[\bx_i]$ in the above (nonempty) product is $(1/2) (+1) + (1/2) (-1) = 0$.
\end{proof}

\section{Basic Fourier formulas}                          \label{sec:basic-fourier-formulas}

As we have seen, the Fourier expansion of $f \btR$ can be thought of as the representation of $f$ over the orthonormal basis of parity functions $(\chi_S)_{S \subseteq [n]}$.  In this basis, $f$ has $2^n$ ``coordinates'', and these are precisely the Fourier coefficients of~$f$.  The ``coordinate'' of $f$ in the $\chi_S$ ``direction'' is $\la f, \chi_S\ra$; i.e., we have the following formula for Fourier coefficients:
\begin{proposition}                                     \label{prop:fourier-coeff-formula}
    For $f \btR$ and $S \subseteq [n]$, the Fourier coefficient
                                                \index{Fourier coefficient!formula}
    of $f$ on $S$ is given by
    \begin{equation*}
        \wh{f}(S) = \la f, \chi_S \ra = \E_{\bx \sim \bits^n}[f(\bx) \chi_S(\bx)].
    \end{equation*}
\end{proposition}
\noindent We can verify this formula explicitly:
\begin{equation} \label{eqn:fourier-coeff-verification}
    \la f, \chi_S \ra = \left\la \sum_{T \subseteq [n]} \wh{f}(T)\,\chi_T, \chi_S \right\ra = \sum_{T \subseteq [n]} \wh{f}(T) \la \chi_T, \chi_S \ra = \wh{f}(S),
\end{equation}
where we used the Fourier expansion of $f$, the linearity of $\la \cdot, \cdot \ra$, and finally  Theorem~\ref{thm:parities-orthormal}.  This formula is the simplest way to calculate the Fourier coefficients of a given function; it can also be viewed as a streamlined version of the interpolation method illustrated in~\eqref{eqn:min2-fourier-computation}. Alternatively, this formula can be taken as the \emph{definition} of Fourier coefficients.

The orthonormal basis of parities also lets us measure the squared ``length'' ($2$-norm) of $f \btR$ efficiently: it's just the sum of the squares of $f$'s ``coordinates'' -- i.e., Fourier coefficients.  This simple but crucial fact is called \emph{Parseval's Theorem}.
\begin{named}{Parseval's Theorem}                                     \label{thm:parseval}
                                   \index{Parseval's Theorem}%
    For any $f \btR$,
    \[
        \la f, f \ra = \Ex_{\bx \sim \bn}[f(\bx)^2] = \sumS \wh{f}(S)^2.
    \]
    In particular, if $f \btb$ is Boolean-valued then
    \[
        \sumS \wh{f}(S)^2 = 1.
    \]
\end{named}

As examples we can recall the Fourier expansions of $\maxt$ and $\maj_3$:
\[
\maxt(x) = \tfrac12 + \tfrac12 x_1 + \tfrac12 x_2 - \tfrac12 x_1 x_2, \qquad \Maj_3(x) = \half x_1 + \half x_2 + \half x_3 - \half x_1x_2x_3.
\]
In both cases the sum of squares of Fourier coefficients is $4 \times (1/4) = 1$.

More generally, given two functions $f, g \btR$, we can compute $\la f, g \ra$ by taking the ``dot product'' of their coordinates in the orthonormal basis of parities.  The resulting formula is called \emph{Plancherel's Theorem}.
\begin{named}{Plancherel's Theorem}                                     \label{thm:plancherel}
                                                \index{Plancherel's Theorem}%
    For any $f, g \btR$,
    \[
        \la f, g \ra = \Ex_{\bx \sim \bn}[f(\bx)g(\bx)] = \sumS \wh{f}(S) \wh{g}(S).
    \]
\end{named}
We can verify this formula explicitly as we did in~\eqref{eqn:fourier-coeff-verification}:
\[
    \la f, g \ra = \Bigl\la \sum_{S \subseteq [n]} \wh{f}(S)\,\chi_S, \sum_{T \subseteq [n]} \wh{g}(T)\,\chi_T \Bigr\ra = \sum_{S, T \subseteq [n]} \wh{f}(S)\wh{g}(T) \la \chi_S, \chi_T \ra = \sum_{S\subseteq [n]} \wh{f}(S)\wh{g}(S).
\]

Now is a good time to remark that for Boolean-valued functions $f, g \btb$, the inner product $\la f, g \ra$ can be interpreted as a kind of ``correlation'' between $f$ and~$g$, measuring how similar they are.  Since $f(x)g(x) = 1$ if $f(x) = g(x)$ and $f(x)g(x) = -1$ if $f(x) \neq g(x)$, we have:
\begin{proposition}                                     \label{prop:ip-as-correlation}
    If $f, g \btb$,
    \[
        \la f, g \ra = \Pr[f(\bx) = g(\bx)] - \Pr[f(\bx) \neq g(\bx)] = 1 - 2\dist(f,g).
    \]
\end{proposition}
Here we are using the following definition:
\begin{definition}                                  \label{def:dist}
    Given $f, g \btb$, we define their \emph{(relative Hamming) distance}
                                                    \index{distance, relative Hamming}
    to be
    \[
       \dist(f,g) = \Pr_{\bx}[f(\bx) \neq g(\bx)],
    \]%
                        \nomenclature[dist]{$\dist(g,h)$}{the relative Hamming distance; i.e., the fraction of inputs on which $g$ and $h$ disagree}%
    the fraction of inputs on which they disagree.
\end{definition}

With a number of Fourier formulas now in hand we can begin to illustrate a basic theme in the analysis of Boolean functions: interesting combinatorial properties of a Boolean function $f$ can be ``read off'' from its Fourier coefficients.  Let's start by looking at one way to measure the ``bias'' of $f$:
\begin{definition}              \label{def:mean-bias}
    The \emph{mean}
                                                    \index{mean}
    of $f \btR$ is $\E[f]$.  When $f$ has mean $0$ we say that it is \emph{unbiased}, or \emph{balanced}.
                                                    \index{unbiased}%
                                                    \index{balanced|seeonly{unbiased}}%
    In the particular case that $f \btb$ is Boolean-valued, its mean is
    \[
        \E[f] = \Pr[f = 1] - \Pr[f = -1];
    \]
    thus $f$ is unbiased if and only if it takes value $1$ on exactly half of the points of the Hamming cube.
\end{definition}
\begin{fact}                                        \label{fact:empty-is-mean}
    If $f \btR$ then $\E[f] = \wh{f}(\emptyset)$.
\end{fact}
This formula holds simply because $\E[f] = \la f, 1 \ra = \wh{f}(\emptyset)$ (taking $S = \emptyset$ in Proposition~\ref{prop:fourier-coeff-formula}).  In particular, a Boolean function is unbiased if and only if its empty-set Fourier coefficient is $0$.

Next we obtain a formula for the \emph{variance}
                                                \index{variance}
of a real-valued Boolean function (thinking of $f(\bx)$ as a real-valued random variable):
\begin{proposition}                                     \label{prop:variance-formula}
    The \emph{variance} of $f \btR$ is
    \[
        \Var[f] = \la f - \E[f], f - \E[f] \ra = \E[f^2] - \E[f]^2 = \sum_{S \neq \emptyset} \wh{f}(S)^2.
    \]
\end{proposition}%
                        \nomenclature[Var]{$\Var[f]$}{the variance of $f$, $\Var[f] = \E[f^2] - \E[f]^2$}
\noindent The above Fourier formula follows immediately from Parseval's Theorem and Fact~\ref{fact:empty-is-mean}.  We also have:
\begin{fact}                                        \label{fact:Boolean-variance}
    For $f \btb$,
    \[
        \Var[f] = 1 - \E[f]^2 = 4 \Pr[f(\bx) = 1] \Pr[f(\bx) = -1] \in [0,1].
    \]
\end{fact}
In particular, a Boolean-valued function $f$ has variance~$1$ if it's unbiased and variance~$0$ if it's constant.  More generally, the variance of a Boolean-valued function is proportional to its ``distance from being constant''.
\begin{proposition}                                     \label{prop:variance-bias}
    Let $f \btb$.  Then $2 \eps \leq \Var[f] \leq 4\eps$, where
    \[
    \eps = \min\{\dist(f, 1), \dist(f, -1)\}.
    \]
\end{proposition}
\noindent The proof of Proposition~\ref{prop:variance-bias} is an exercise. See also Exercise~\ref{ex:Boolean-variance}.

By using Plancherel in place of Parseval, we get a generalization of Proposition~\ref{prop:variance-formula} for \emph{covariance}:
                                            \index{covariance}
\begin{proposition}                                     \label{prop:covariance-formula}
    The \emph{covariance} of $f, g \btR$ is
    \[
        \Cov[f,g] = \la f - \E[f], g - \E[g] \ra = \E[fg] - \E[f]\E[g] = \sum_{S \neq \emptyset} \wh{f}(S)\wh{g}(S).
    \]
\end{proposition}%
                    \nomenclature[Cov]{$\Cov[f,g]$}{the covariance of $f$ and $g$, $\Cov[f] = \E[fg] - \E[f]\E[g]$}%

We end this section by discussing the \emph{Fourier weight distribution} of Boolean functions.
\begin{definition}
     The \emph{(Fourier) weight}
                                            \index{Fourier weight}%
                                            \index{weight|seeonly{Fourier weight}}%
    of $f \btR$ on set $S$ is defined to be the squared Fourier coefficient, $\wh{f}(S)^2$.
\end{definition}
Although we lose some information about the Fourier coefficients when we square them, many Fourier formulas only depend on the weights of $f$.  For example, Proposition~\ref{prop:variance-formula} says that the variance of~$f$ equals its Fourier weight on nonempty sets.  Studying Fourier weights is particularly pleasant for Boolean-valued functions $f \btb$ since Parseval's Theorem says that they always have total weight~$1$.  In particular, they define a \emph{probability distribution} on subsets of~$[n]$.
\begin{definition}
     Given $f \btb$, the \emph{spectral sample} for $f$, denoted $\specsamp{f}$, is the probability distribution on subsets of $[n]$ in which the set $S$ has probability $\wh{f}(S)^2$.  We write $\bS \sim \specsamp{f}$ for a draw from this distribution.
\end{definition}
For example, the spectral sample for the $\maxt$ function is the uniform distribution on all four subsets of $[2]$; the spectral sample for $\maj_3$ is the uniform distribution on the four subsets of $[3]$ with odd cardinality.

Given a Boolean function it can be helpful to try to keep a mental picture of its weight distribution on the subsets of $[n]$, partially ordered by inclusion.  Figure~\ref{fig:maj-weight} is an example for the $\maj_3$ function, with the white circles indicating weight~$0$ and the shaded circles indicating weight~$1/4$.

\myfig{.5}{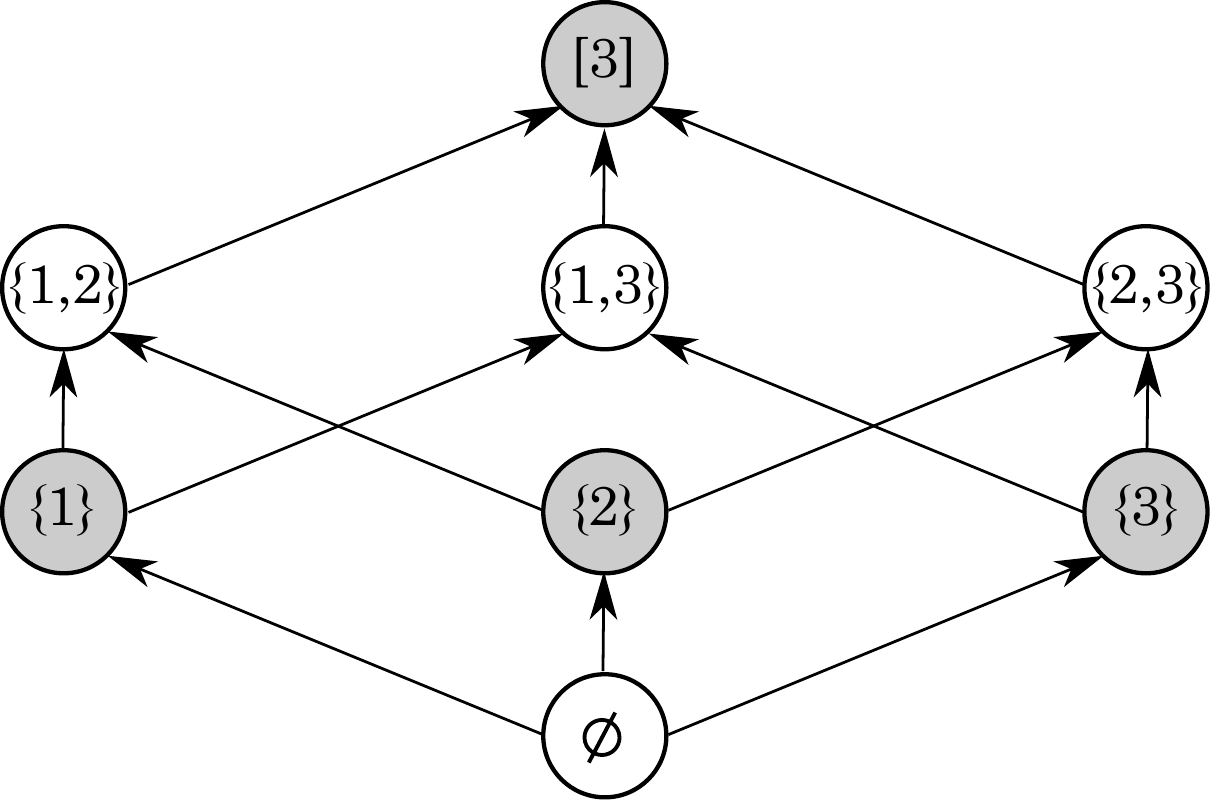}{Fourier weight distribution of the $\maj_3$ function}{fig:maj-weight}

Finally, as suggested by the diagram we often stratify the subsets $S \subseteq [n]$ according to their cardinality (also called ``height'' or ``level'').  Equivalently, this is the \emph{degree}
                                                    \index{degree}
of the associated monomial $x^S$.
\begin{definition}
    For $f \btR$ and $0 \leq k \leq n$, the \emph{(Fourier) weight of $f$ at degree~$k$} is
    \[
        \W{k}[f] = \sum_{\substack{S \subseteq [n] \\ |S| = k}} \wh{f}(S)^2.
    \]%
                               \nomenclature[Wk]{$\W{k}[f]$}{the Fourier weight of $f$ at degree~$k$}
    If $f \btb$ is Boolean-valued, an equivalent definition is
    \[
        \W{k}[f] = \Pr_{\bS \sim \specsamp{f}}[|S| = k].
    \]
    By Parseval's Theorem, $\W{k}[f] = \|f^{=k}\|_2^2$
                                \nomenclature[fk]{$f^{=k}$}{$\sum_{\vert S \vert =k} \wh{f}(S)\,\chi_S$}
    where
    \[
        f^{=k} = \sum_{|S| = k} \wh{f}(S)\,\chi_S
    \]
    is called the
                                            \index{degree~$k$ part}
    \emph{degree~$k$ part of $f$}.  We will also sometimes use notation
                                \nomenclature[Wk1]{$\W{> k}[f]$}{the Fourier weight of $f$ at degrees above~$k$}%
                                \nomenclature[fk1]{$f^{\leq k}$}{$\sum_{\vert S \vert \leq k} \wh{f}(S)\,\chi_S$}%
    like $\W{> k}[f] = \sum_{|S| > k} \wh{f}(S)^2$ and $f^{\leq k} = \sum_{|S| \leq k} \wh{f}(S)\,\chi_S$.
\end{definition}

\section{Probability densities and convolution}                          \label{sec:densities-and-convolution}

For variety's sake, in this section we write the Hamming cube as $\F_2^n$ rather than $\bn$.  In developing the Fourier expansion, we have generalized from \emph{Boolean-valued Boolean functions}
                                        \index{Boolean-valued function}
$f \co \F_2^n \to \bits$ to \emph{real-valued Boolean functions}
                                        \index{Boolean function!real-valued}
$f \co \F_2^n \to \R$. Boolean-valued functions arise more often in combinatorial problems, but there are important classes of real-valued Boolean functions.  One example is \emph{probability densities}.
\begin{definition}                          \label{def:prob-density}
     A \emph{(probability) density}
                                        \index{probability density}%
                                        \index{density function|seeonly{probability density}}%
    function on the Hamming cube $\F_2^n$ is any nonnegative function $\vphi \co \F_2^n \to \R^{\geq 0}$
                        \nomenclature[R]{$\R$}{the real numbers}%
                        \nomenclature[R0]{$\R^{\geq 0}$}{the nonnegative real numbers}%
    satisfying
     \[
        \Ex_{\bx \sim \F_2^n}[\vphi(\bx)] = 1.
     \]
     We write $\by \sim \vphi$
                        \nomenclature[xphi]{$\bx \sim \vphi$}{the random variable $\bx$ is chosen from the probability distribution with density $\vphi$}
     to denote that $\by$ is a random string drawn from the associated probability distribution, defined by
     \[
\phantom{ \quad \forall y \in \F_2^n.}  \Pr_{\by \sim \vphi}[\by = y] = \vphi(y) \frac{1}{2^n} \quad \forall y \in \F_2^n.
     \]
\end{definition}
Here you should think of $\vphi(y)$ as being the \emph{relative} density of~$y$ with respect to the uniform distribution on $\F_2^n$.
For example, we have:
\begin{fact}                                        \label{fact:density-expectation}
    If $\vphi$ is a density function and $g \co \F_2^n \to \R$, then
    \[
        \E_{\by \sim \vphi}[g(\by)] = \la \vphi, g \ra = \Ex_{\bx \sim \F_2^n}[\vphi(\bx) g(\bx)].
    \]
\end{fact}
The simplest example of a probability density is just the constant function~$1$, which corresponds to the uniform probability distribution on $\F_2^n$.  The most common case arises from the uniform distribution over some subset $A \subseteq \F_2^n$.
\begin{definition}
    If $A \subseteq \F_2^n$ we write $1_A \co \F_2^n \to \{0,1\}$
                            \nomenclature[1A]{$1_A$}{$0$-$1$ indicator function for $A$}
    for the $0$-$1$ \emph{indicator function}
                                                    \index{indicator function}
    of $A$; i.e.,
    \begin{equation*}
        1_A(x) = \begin{cases} 1 & \text{if $x \in A$,} \\ 0 & \text{if $x \notin A$.} \end{cases}
    \end{equation*}
    Assuming $A \neq \emptyset$ we write $\vphi_A$
                            \nomenclature[phiA]{$\vphi_A$}{the density function for the uniform probability distribution on $A$; i.e., $1_A/\E[1_A]$}
    for the density function associated to the uniform distribution
                                                    \index{uniform distribution on $A$}
    on~$A$; i.e.,
    \begin{equation*}
        \vphi_A = \tfrac{1}{\E[1_A]} 1_A.
    \end{equation*}
    We typically write $\by \sim A$
                            \nomenclature[x~0]{$\bx \sim A$}{the random variable $\bx$ is chosen uniformly from the set $A$}
    rather than $\by \sim \vphi_A$.
\end{definition}
A simple but useful example is when $A$ is the singleton set $A = \{0\}$. (Here~$0$ is denoting the vector $(0, 0, \dots, 0) \in \F_2^n$.) In this case the function $\vphi_{\{0\}}$ takes value $2^n$ on input $0 \in \F_2^n$ and is zero elsewhere on $\F_2^n$.  In Exercise~\ref{ex:compute-expansions} you will verify the Fourier expansion of $\vphi_{\{0\}}$:
\begin{fact}                                        \label{fact:expansion-0-dens}
    Every Fourier coefficient of $\vphi_{\{0\}}$ is $1$; i.e., its Fourier expansion is
    \[
        \vphi_{\{0\}}(y) = \sum_{S \subseteq [n]} \chi_S(y).
    \]
\end{fact}

                                \index{convolution|(}
We now introduce an operation on functions that interacts particularly nicely with density functions, namely, \emph{convolution}.
\begin{definition}
    Let $f, g \ftR$.  Their \emph{convolution} is the function $f \conv g \ftR$
                                    \nomenclature[*]{$\conv$}{the convolution operator}
    defined by
    \begin{equation*}
        (f \conv g)(x) = \Ex_{\by \sim \F_2^n}[f(\by) g(x - \by)] = \Ex_{\by \sim \F_2^n}[f(x-\by)g(\by)].
    \end{equation*}
    Since subtraction is equivalent to addition in $\F_2^n$ we may also write
    \begin{equation*}
        (f \conv g)(x) = \Ex_{\by \sim \F_2^n}[f(\by) g(x + \by)] = \Ex_{\by \sim \F_2^n}[f(x+\by)g(\by)].
    \end{equation*}
    If we were representing the Hamming cube by $\bits^n$ rather than $\F_2^n$ we would replace $x + \by$ with $x \circ \by$, where $\circ$
                                    \nomenclature{$\circ$}{entry-wise multiplication of vectors}
    denotes entry-wise multiplication.
\end{definition}

Exercise~\ref{ex:conv-ac} asks you to verify that convolution is associative and commutative:
\[
    f \conv (g \conv h) = (f \conv g) \conv h, \qquad f \conv g = g \conv f.
\]

Using Fact~\ref{fact:density-expectation} we can deduce the following two simple results:
\begin{proposition}                                        \label{prop:density-conv}
    If $\vphi$ is a density function on $\F_2^n$ and $g \ftR$ then
    \[
        \vphi \conv g(x) = \Ex_{\by \sim \vphi}[g(x - \by)] = \Ex_{\by \sim \vphi}[g(x + \by)].
    \]
    In particular, $\E_{\by \sim \vphi}[g(\by)] = \vphi \conv g(0)$.
\end{proposition}
\begin{proposition}                                        \label{prop:density-conv-density}
    If $g = \psi$ is itself a probability density function then so is $\vphi \conv \psi$; it represents the distribution on $\bx \in \F_2^n$ given by choosing $\by \sim \vphi$ and $\bz \sim \psi$ independently and setting $\bx = \by + \bz$.
\end{proposition}

The most important theorem about convolution is that it corresponds to multiplication of Fourier coefficients:
\begin{theorem}                                     \label{thm:convolution-theorem}
    Let $f, g \ftR$.  Then for all $S \subseteq [n]$,
    \begin{equation*}
        \wh{f \conv g}(S) = \wh{f}(S) \wh{g}(S).
    \end{equation*}
\end{theorem}
\begin{proof}
    We have
    \begin{align*}                                       \label{eqn:}
        \wh{f \conv g}(S)   &= \E_{\bx \sim \F_2^n}[ (f \conv g)(\bx) \chi_S(\bx)] \tag{the Fourier formula}\\
                            &= \E_{\bx \sim \F_2^n}\left[ \Ex_{\by \sim \F_2^n}[f(\by) g(\bx - \by)]  \chi_S(\bx)\right] \tag{by definition}\\
                            &= \E_{\substack{\by, \bz \sim \F_2^n \\ \text{independently}}}[f(\by) g(\bz)  \chi_S(\by + \bz)] \quad \ifblog \qquad\qquad\qquad\qquad \fi \tag{as $x - \by$ is uniform on $\F_2^n$ $\forall x$}\\
                            &= \E_{\by, \bz \sim \F_2^n}[f(\by)  \chi_S(\by) g(\bz)  \chi_S(\bz)] \tag{by identity \eqref{eqn:chi-character}}\\
                            &= \wh{f}(S) \wh{g}(S) \tag*{(Fourier formula, independence),}
    \end{align*}
    as
                                                \index{convolution|)}
    claimed.
\end{proof}

\section{Highlight: Almost linear functions and the BLR Test}                       \label{sec:BLR}

In linear algebra there are two equivalent definitions of what it means for a function to be linear:
                                                \index{linear (over $\F_2$)}
\begin{definition}                                                  \label{def:linear}
    A function $f \co \F_2^n \to \F_2$ is \emph{linear} if either of the following equivalent conditions hold:
    \begin{enumerate}
        \item \label{def:lin1} $f(x+y) = f(x) + f(y)$ for all $x, y \in \F_2^n$;
        \item \label{def:lin2} $f(x) = a \cdot x$ for some $a \in \F_2^n$; i.e., $f(x) = \sum_{i \in S} x_i$ for some $S \subseteq [n]$.
    \end{enumerate}
\end{definition}
\noindent Exercise~\ref{ex:linear-equivalence} asks you to verify that the conditions are indeed equivalent. If we encode the output of $f$ by $\pm 1 \in \R$ in the usual way then the ``linear'' functions $f \co \F_2^n \to \bits$ are precisely the $2^n$ parity functions $(\chi_S)_{S \subseteq [n]}$.

Let's think of what it might mean for a function $f \co \F_2^n \to \F_2$ to be \emph{approximately} linear.  Definition~\ref{def:linear} suggests two possibilities:
\begin{enumerate}
    \item[\hypertarget{lin1p}{($1'$)}] $f(x+y) = f(x) + f(y)$ for \emph{almost all} pairs $x, y \in \F_2^n$;
    \item[\hypertarget{lin2p}{($2'$)}] \label{def:lin2p} there is some $S \subseteq [n]$ such that $f(x) = \sum_{i \in S} x_i$ for \emph{almost all} $x \in \F_2^n$.
\end{enumerate}
Are these equivalent?  The proof of $\eqref{def:lin2} \implies \eqref{def:lin1}$ in Definition~\ref{def:linear} is ``robust'': it easily extends to show (\hyperlink{lin2p}{$2'$})~$\implies$~(\hyperlink{lin1p}{$1'$}) (see Exercise~\ref{ex:linear-equivalence}).  But the natural proof of $\eqref{def:lin1} \implies \eqref{def:lin2}$ in Definition~\ref{def:linear} does not have this robustness property.  The goal of this section is to show that (\hyperlink{lin1p}{$1'$})~$\implies$~(\hyperlink{lin2p}{$2'$}) nevertheless holds.

Motivation for this problem comes from an area of theoretical computer science called \emph{property testing},
                                            \index{testing}%
                                            \index{property testing|seeonly{testing}}%
which we will discuss in more detail in Chapter~\ref{chap:testing}.  Imagine that you have ``black-box'' access to a function $f \co \F_2^n \to \F_2$, meaning that the function $f$ is unknown to you but you can ``query'' its value on inputs $x \in \F_2^n$ of your choosing. The function $f$ is ``supposed'' to be a linear function, and you would like to try to verify this.

The only way you can be certain $f$ is indeed a linear function is to query its value on all $2^n$ inputs; unfortunately, this is very expensive.  The idea behind ``property testing'' is to try to verify that $f$ has a certain property -- in this case, linearity -- by querying its value on just a few random inputs.  In exchange for efficiency, we need to be willing to only approximately verify the property.
\begin{definition}              \label{def:eps-close}
    If $f$ and $g$ are Boolean-valued functions we say they are \emph{$\eps$-close} if $\dist(f,g) \leq \eps$; otherwise we say they are \emph{$\eps$-far}.
                                    \index{epsilon-close@$\eps$-close}
    If $\calP$ is a (nonempty) property of $n$-bit Boolean functions we define $\dist(f, \calP) = \min_{g \in \calP} \{\dist(f,g)\}$. We say that $f$ is $\eps$-close to $\calP$ if $\dist(f,\calP) \leq \eps$; i.e., $f$ is $\eps$-close to some $g$ satisfying $\calP$.
\end{definition}
In particular, in property testing we take property~(\hyperlink{lin2p}{$2'$}) above to be the notion of ``approximately linear'': we say $f$ is $\eps$-close to being linear if $\dist(f,g) \leq \eps$ for some truly linear $g(x) = \sum_{i \in S} x_i$.
                                    \index{testing!linearity}

In 1990 Blum, Luby, and Rubinfeld~\cite{BLR90} showed that indeed (\hyperlink{lin1p}{$1'$})~$\implies$~(\hyperlink{lin1p}{$2'$}) holds, giving the following ``test'' for the property of linearity that makes just~$3$ queries:

\begin{named}{BLR Test}
                                                \index{BLR (Blum--Luby--Rubinfeld) Test}
    Given query access to $f \co \F_2^n \to \F_2$:
    \begin{itemize}
        \item Choose $\bx \sim \F_2^n$ and $\by \sim \F_2^n$ independently.
        \item Query $f$ at $\bx$, $\by$, and $\bx + \by$.
        \item ``Accept'' if $f(\bx) + f(\by) = f(\bx + \by)$.
    \end{itemize}
\end{named}

We now show that if the BLR Test accepts $f$ with high probability then $f$ is close to being linear.  The proof works by directly relating the acceptance probability to the quantity $\sum_{S} \wh{f}(S)^3$; see equation~\eqref{eqn:BLR} below.
\begin{theorem}                                     \label{thm:blr-test}
    Suppose the BLR Test accepts $f \co \F_2^n \to \F_2$ with probability $1 - \eps$.  Then $f$ is $\eps$-close to being linear.
\end{theorem}
\begin{proof}
    In order to use the Fourier transform we encode $f$'s output by $\pm 1 \in \R$; thus the acceptance condition of the BLR Test becomes $f(\bx)f(\by) = f(\bx + \by)$.  Since
    \[
        \half + \half f(\bx)f(\by)f(\bx+\by) =  \begin{cases}
                                                    1 & \text{if $f(\bx)f(\by) = f(\bx + \by)$}, \\
                                                    0 & \text{if $f(\bx)f(\by) \neq f(\bx + \by)$,}
                                                \end{cases}
    \]
    we conclude
    \begin{align*}
        1 - \eps = \Pr[\text{BLR accepts $f$}]
                                         &= \Ex_{\bx, \by}[\half + \half f(\bx)f(\by)f(\bx+\by)] \\
                                         &= \half + \half \Ex_{\bx}[f(\bx) \cdot \Ex_{\by}[f(\by)f(\bx+\by)]]\\
                                         &= \half + \half \Ex_{\bx}[f(\bx) \cdot (f \conv f)(\bx)] \tag{by definition}\\
                                         &= \half + \half \sum_{S \subseteq [n]} \wh{f}(S) \wh{f \conv f}(S) \tag{Plancherel} \\
                                         &= \half + \half \sum_{S \subseteq [n]} \wh{f}(S)^3 \tag*{(Theorem~\ref{thm:convolution-theorem}).}
    \end{align*}
    We rearrange this equality and then continue:
    \begin{align}
        1 - 2\eps   &= \sum_{S \subseteq [n]} \wh{f}(S)^3 \label{eqn:BLR}\\
                    &\leq \max_{S \subseteq [n]} \{\wh{f}(S)\} \cdot \sum_{S \subseteq [n]} \wh{f}(S)^2 \nonumber\\
                    &= \max_{S \subseteq [n]} \{\wh{f}(S)\}  \tag*{(Parseval).}
    \end{align}
    But $\wh{f}(S) = \la f, \chi_S \ra = 1 - 2\dist(f, \chi_S)$ (Proposition~\ref{prop:ip-as-correlation}).
    Hence there exists some $S^* \subseteq [n]$ such that $1 - 2\eps \leq 1 - 2\dist(f,\chi_{S^*})$; i.e., $f$ is $\eps$-close to the linear function $\chi_{S^*}$.
\end{proof}
In fact, for small $\eps$ one can show that $f$ is more like $(\eps/3)$-close to linear, and this is sharp. See Exercise~\ref{ex:BLR}.

\medskip

The BLR Test shows that given black-box access to $f \co \F_2^n \to \{-1,1\}$, we can ``test'' whether $f$ is close to some linear function~$\chi_S$ using just~$3$ queries.  The test does not reveal \emph{which} linear function $\chi_S$ is close to (indeed, determining this takes at least~$n$ queries; see Exercise~\ref{ex:learning-parity}).  Nevertheless, we can still determine the value of $\chi_S(x)$ with high probability for \emph{every} $x \in \F_2^n$ of our choosing using just~$2$ queries.  This property is called \emph{local correctability}
                                        \index{locally correctable}
of linear functions.
\begin{proposition}                                     \label{prop:local-correcting-linearity}
    Suppose $f \co \F_2^n \to \{-1,1\}$ is $\eps$-close to the linear function~$\chi_S$.  Then for every $x \in \F_2^n$, the following algorithm outputs $\chi_S(x)$ with probability at least $1-2\eps$:
    \begin{itemize}
        \item Choose $\by \sim \F_2^n$.
        \item Query $f$ at $\by$ and $x+\by$.
        \item Output $f(\by)f(x+\by)$.
    \end{itemize}
\end{proposition}
We emphasize the order of quantifiers here: if we just output $f(x)$ then this will equal $\chi_S(x)$ for \emph{most} ~$x$; however, the above ``local correcting'' algorithm determines $\chi_S(x)$ (with high probability) for \emph{every}~$x$.
\begin{proof}
    Since $\by$ and $x+\by$ are both uniformly distributed on $\F_2^n$ (though not independently) we have $\Pr[f(\by) \neq \chi_S(\by)] \leq \eps$ and $\Pr[f(x+\by) \neq \chi_S(x+\by)] \leq \eps$ by assumption.  By the union bound, the probability of either event occurring is at most $2\eps$; when neither occurs,
    \[
        f(\by) f(x+\by) = \chi_S(\by) \chi_S(x + \by) = \chi_S(x)
    \]
    as desired.
\end{proof}

\section{Exercises and notes}
\begin{exercises}
    \item \label{ex:compute-expansions}
    Compute the Fourier expansions of the following functions:
    \begin{exercises}
        \item $\min_2 \co \{-1,1\}^2 \to \bits$, the minimum function on $2$ bits (also known as the logical $\OR$ function);
        \item $\min_3 \co \{-1,1\}^3 \to \bits$ and $\max_3 \co \{-1,1\}^3 \to \bits$;
        \item \label{ex:indic-transf} the indicator function
                                    \index{indicator function}
                $1_{\{a\}} \co \F_2^n \to \{0,1\}$, where $a \in \F_2^n$;
        \item \label{ex:single-point-density} the density function $\vphi_{\{a\}} \co \F_2^n \to \R^{\geq 0}$, where $a \in \F_2^n$;
        \item \label{ex:Ei-dist} the density function $\vphi_{\{a, a+e_i\}} \co \F_2^n \to \R^{\geq 0}$, where $a \in \F_2^n$ and $e_i = (0, \dots, 0, 1, 0, \dots, 0)$ with the $1$ in the $i$th coordinate;
        \item \label{ex:noise-dist} the density function corresponding to the product probability distribution on $\bn$ in which each coordinate has mean $\rho \in [-1,1]$;
        \item the \emph{inner product mod $2$ function},
                                    \index{inner product mod $2$ function}
                $\IP_{2n} \co \F_2^{2n} \to \{-1,1\}$ defined by $\IP_{2n}(x_1, \dots, x_n, y_1, \dots, y_n) = (-1)^{x \cdot y}$;
        \item \label{ex:equ} the \emph{equality function}
                                    \index{equality function}
            $\EQU_n \co \{-1,1\}^n \to \{0,1\}$, defined by $\EQU_n(x) = 1$ if and only if $x_1 = x_2 = \cdots = x_n$;
        \item \label{ex:NAE} the \emph{not-all-equal function}
                                    \index{not-all-equal (NAE) function}
            $\NAE_n \co \{-1,1\}^n \to \{0,1\}$, defined by $\NAE_n(x) = 1$ if and only if the bits $x_1, \dots, x_n$ are not all equal;
        \item \label{ex:SEL} the \emph{selection function},
                                    \index{selection function}
              $\SEL \co \bits^3 \to \{-1,1\}$, which outputs $x_2$ if $x_1 = -1$ and outputs $x_3$ if $x_1 = 1$;
        \item $\mathrm{mod}_3 \co \F_2^3 \to \{0,1\}$,
                                    \index{mod 3 function}
                which is $1$ if and only if the number of $1$'s in the input is divisible by~$3$;
        \item $\oxr \co \F_2^3 \to \{0,1\}$ defined by $\oxr(x_1,x_2,x_3) = x_1 \vee (x_2 \oplus x_3)$.  Here~$\vee$~denotes logical OR, $\oplus$ denotes logical XOR;
                                    \nomenclature{$\vee$}{logical OR}
                                    \nomenclature{$\wedge$}{logical AND}
                                    \nomenclature{$\neg$}{logical NOT}
                                    \nomenclature{$\oplus$}{logical XOR (exclusive-or)}
                                    \index{OXR function}
        \item \label{ex:sorted} the \emph{sortedness function}
                                    \index{sortedness function}%
                                    \index{Ambainis function|seeonly{sortedness function}}%
                $\SORT_4 \co \{-1,1\}^4 \to \{-1,1\}$, defined by $\SORT_4(x) = -1$ if and only if $x_1 \leq x_2 \leq x_3 \leq x_4$ or $x_1 \geq x_2 \geq x_3 \geq x_4$; 
        \item the \emph{hemi-icosahedron function} $\mathrm{HI} \co \bits^6 \to \bits$ (also known as the \emph{Kushilevitz function}), defined to be
                                    \index{hemi-icosahedron function}%
                                    \index{Kushilevitz function|seeonly{hemi-icosahedron function}}%
            the number of facets labeled $(+1, +1, +1)$ in Figure~\ref{fig:hemi-icosahedron}, minus the number of facets labeled $(-1,-1,-1)$, modulo~$3$.

            \myfig{.375}{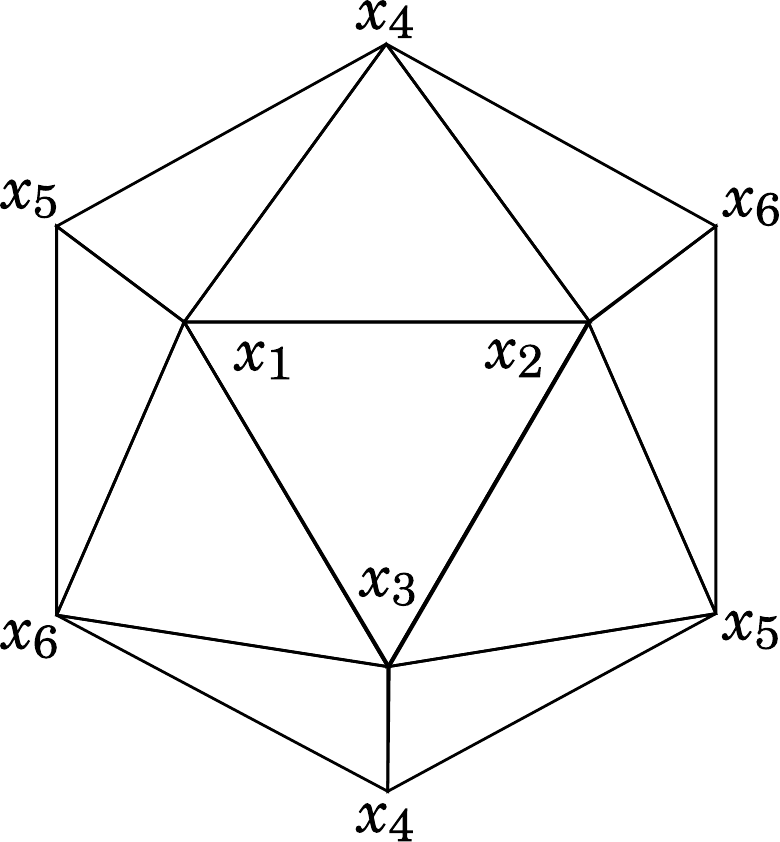}{The hemi-icosahedron}{fig:hemi-icosahedron}

            (Hint: First compute the real multilinear interpolation of the analogue $\mathrm{HI} \co \zo^6 \to \zo$.)
        \item the majority
                                    \index{majority}
                functions $\maj_5 \co \{-1,1\}^5 \to \bits$  and $\maj_7 \co \{-1,1\}^7 \to \bits$;
        \item the \emph{complete quadratic function} $\CQ_n \co \F_2^{n} \to \{-1,1\}$ defined by $\CQ_n(x) = \chi(\sum_{1 \leq i < j \leq n} x_ix_j)$.
                                    \index{complete quadratic function}
            (Hint: Determine $\CQ_n(x)$ as a function of the number of $1$'s in the input modulo~$4$.  You'll want to distinguish whether~$n$ is even or odd.)
    \end{exercises}
    \item How many Boolean functions $f \btb$ have exactly $1$ nonzero Fourier coefficient?
    \item \label{ex:odd-ones-full-sparsity} Let $f \ftzo$, $n > 1$, and suppose $\#\{x : f(x) = 1\}$ is odd.  Prove that all of $f$'s Fourier coefficients are nonzero.
    \item \label{ex:multilin-interp} Let $f \btR$ have Fourier expansion $f(x) = \sum_{S \subseteq [n]} \wh{f}(S)\,x^S$.  Let $\multilin{f} \co \R^n \to \R$ be the extension of $f$ which is also defined by $\multilin{f}(x) = \sum_{S \subseteq [n]} \wh{f}(S)\,x^S$.  Show that if $\mu = (\mu_1, \dots, \mu_n) \in [-1,1]^n$ then
        \[
            \multilin{f}(\mu) = \Ex_{\by}[f(\by)],
        \]
        where $\by$ is the random string in $\bits^n$ defined by having $\E[\by_i] = \mu_i$ independently for all $i \in [n]$.
    \item \label{ex:uniq-decoding} Prove that any $f \btb$ has at most one Fourier coefficient with magnitude exceeding $1/2$.  Is this also true for any $f \btR$ with $\|f\|_2 = 1$?
    \item Use Parseval's Theorem to prove uniqueness of the Fourier expansion.
    \item \label{ex:random-Fourier} Let $\boldf \btb$ be a random function (i.e., each $\boldf(x)$ is $\pm 1$ with
                                                \index{random function}
        probability $1/2$, independently for all $x \in \bn$).  Show that for each $S \subseteq [n]$, the random variable $\wh{\boldf}(S)$ has mean~$0$ and variance~$2^{-n}$.  (Hint: Parseval.)
    \item \label{ex:odd-even} The \emph{(Boolean) dual}
                                                \index{dual, Boolean}
            of $f \btR$ is the function $f^\booldual$
                                                \nomenclature[f+]{$f^\booldual$}{the Boolean dual defined by $f^\booldual(x) = -f(-x)$}
            defined by $f^\booldual(x) = -f(-x)$.  The function $f$ is said to be \emph{odd}
                                                \index{odd function}
            if it equals its dual; equivalently, if $f(-x) = -f(x)$ for all $x$.  The function~$f$ is said to be \emph{even}
                                                \index{even function}
        if $f(-x) = f(x)$ for all $x$. Given any function $f \btR$, its \emph{odd part} is the function $f^\odd \btR$ defined by $f^\odd(x) = (f(x) - f(-x))/2$,
                                        \nomenclature[fodd]{$f^\odd$}{the odd part of $f$, $(f(x) - f(-x))/2$}
        and its \emph{even part} is the function $f^\even \btR$
                                        \nomenclature[feven]{$f^\even$}{the even part of $f$, $(f(x) + f(-x))/2$}
        defined by $f^\even(x) = (f(x) + f(-x))/2$.
        \begin{exercises}
            \item Express $\wh{f^\booldual}(S)$ in terms of $\wh{f}(S)$.
            \item Verify that $f = f^\odd + f^\even$ and that $f$ is odd (respectively, even) if and only if $f = f^\odd$ (respectively, $f = f^\even$).
            \item Show that
            \begin{equation*}
                f^\odd = \sum_{\substack{S \subseteq [n] \\ |S|\ \odd}} \wh{f}(S)\,\chi_S, \qquad f^\even = \sum_{\substack{S \subseteq [n] \\ |S|\ \even}} \wh{f}(S)\,\chi_S.
            \end{equation*}
        \end{exercises}
    \item \label{ex:pm1tozo} In this problem we consider representing $\false, \true$ as $0, 1 \in \R$.
        \begin{exercises}
            \item Using the interpolation method from Section~\ref{sec:fourier-expansion}, show that every $f \co \{\false,\true\}^n \to \{\false,\true\}$ can be represented as a real multilinear polynomial
                \begin{equation} \label{eqn:zotzo-poly}
                    q(x) = \sum_{S \subseteq [n]} c_S \prod_{i \in S} x_i,
                \end{equation}
                ``over $\{0,1\}$'', meaning mapping $\{0,1\}^n \to \{0,1\}$.
                                                \index{0-1 multilinear representation}
            \item Show that this representation is unique.  (Hint: If $q$ as in~\eqref{eqn:zotzo-poly} has at least one nonzero coefficient, consider $q(a)$ where $a \in \{0,1\}^n$ is the indicator vector of a minimal~$S$ with $c_S \neq 0$.)
            \item \label{ex:zotzo-integer} Show that all coefficients $c_S$ in the representation~\eqref{eqn:zotzo-poly} will be integers in the range $[-2^n, 2^n]$.
            \item \label{ex:zotzo-halfs} Let $f \co \{\false, \true\}^n \to \{\false, \true\}$.  Let $p(x)$ be $f$'s multilinear representation when $\false, \true$ are $1, -1 \in \R$ (i.e., $p$ is the Fourier expansion of~$f$) and let $q(x)$ be $f$'s multilinear representation when $\false, \true$ are $0, 1 \in \R$.  Show that $q(x) = \half - \half p(1-2x_1, \dots, 1-2x_n)$.
        \end{exercises}
    \item \label{ex:degree} Let $f \btR$ be not identically $0$.  The \emph{(real) degree}
                                                \index{degree}
          of $f$, denoted $\deg(f)$,
                                                \nomenclature[deg]{$\deg(f)$}{the degree of $f$; the least $k$ such that $f$ is a real linear combination of $k$-juntas}
          is defined to be the degree of its multilinear (Fourier) expansion; i.e., $\max \{|S| : \wh{f}(S) \neq 0\}$.
          \begin{exercises}
              \item Show that $\deg(f) = \deg(a+bf)$ for any $a, b \in \R$ (assuming $b \neq 0$, $a+bf \neq 0$).
              \item \label{ex:degree-coord-free} Show that $\deg(f) \leq k$ if and only if $f$ is a real linear combination of functions $g_1, \dots, g_s$, each of which depends on at most $k$ input coordinates.
              \item Which functions in Exercise~\ref{ex:compute-expansions} have ``nontrivial'' degree? (Here $f \btR$ has ``nontrivial'' degree if $\deg(f) < n$.)
          \end{exercises}
    \item \label{ex:deg-dyadic}  Suppose that $f \btb$ has $\deg(f) = k \geq 1$.
           \begin{exercises}
                \item Show that $f$'s real multilinear representation over $\{0,1\}$ (see Exercise~\ref{ex:pm1tozo}), call it $q(x)$, also has $\deg(q) = k$.
                \item Using Exercise~\ref{ex:pm1tozo}\ref{ex:zotzo-integer},\ref{ex:zotzo-halfs}, deduce that $f$'s Fourier spectrum is ``$2^{1-k}$-granular'', meaning each $\wh{f}(S)$ is an integer multiple of
                                            \index{granularity, Fourier spectrum}
                    $2^{1-k}$.
                                                    \index{Fourier norm!$1$-}
                \item Show that $\sum_{S \subseteq [n]} |\wh{f}(S)| \leq 2^{k-1}$.
           \end{exercises}
    \item \label{ex:fast-walsh} A \emph{Hadamard Matrix}
                                                \index{Hadamard Matrix}%
        is any $N \times N$ real matrix with $\pm 1$ entries and orthogonal rows.  Particular examples are the \emph{Walsh--Hadamard Matrices}
                                                \index{Walsh--Hadmard Matrix}%
        $H_N$, inductively defined for $N = 2^n$ as follows: $H_1 = \begin{bmatrix} 1 \end{bmatrix}$, $H_{2^{n+1}} = \begin{bmatrix} H_{2^n} & H_{2^n} \\ H_{2^n} & -H_{2^n} \end{bmatrix}$.
        \begin{exercises}
            \item \label{ex:hadamard-a} Let's index the rows and columns of $H_{2^n}$ by the integers $\{0, 1, 2, \dots, 2^n-1\}$ rather than $[2^n]$. Further, let's identify such an integer $i$ with its binary expansion $(i_0, i_1, \dots, i_{n-1}) \in \F_2^n$, where $i_0$ is the least significant bit and $i_{n-1}$ the most.  For example, if $n = 3$, we identify the index $i = 6$ with $(0, 1, 1)$.  Now show that the $(\gamma, x)$ entry of $H_{2^n}$ is $(-1)^{\gamma \cdot x}$.
            \item \label{ex:hadamard-b} Show that if $f \ftR$ is represented as a column vector in $\R^{2^n}$ (according to the  indexing scheme from part~\ref{ex:hadamard-a}) then $2^{-n} H_{2^n} f = \wh{f}$. Here we think of $\wh{f}$ as also being a  function $\F_2^n \to \R$, identifying subsets $S \subseteq \{0, 1, \dots, n-1\}$ with their indicator vectors.
            \item  Show how to compute $H_{2^n} f$ using just $n 2^n$ additions and subtractions (rather than $2^{2n}$ additions and subtractions as the usual matrix-vector multiplication algorithm would require). This computation is called the \emph{Fast Walsh--Hadamard Transform}
                                                \index{Fast Walsh--Hadamard Transform}
                and is the method of choice for computing the Fourier expansion of a generic function $f \ftR$ when $n$ is large.
            \item Show that taking the Fourier transform is essentially an ``involution'': $\wh{\wh{f}} = 2^{-n} f$ (using the notations from part~\ref{ex:hadamard-b}).
        \end{exercises}
    \item \label{ex:norm-monotonicity}
            Let $f \btR$ and let $0 < p \leq q < \infty$. Show that $\|f\|_p \leq \|f\|_q$.  (Hint: Use Jensen's inequality with the convex function $t \mapsto t^{q/p}$.)  Extend the inequality to the case $q = \infty$, where $\|f\|_\infty$ is defined to be $\max_{x \in \bits^n}\{|f(x)|\}$.
    \item Compute the mean and variance of each function from Exercise~\ref{ex:compute-expansions}.
    \item \label{ex:restrict-mean} Let $f \btR$.  Let $K \subseteq [n]$ and let $z \in \bits^K$.  Suppose $g \co \bits^{[n] \setminus K} \to \R$ is the subfunction of $f$ gotten by restricting the $K$-coordinates to be~$z$. Show that $\E[g] = \sum_{T \subseteq K} \wh{f}(T)\,z^T$.
    \item \label{ex:variance-bias}
            If $f \btb$, show that $\Var[f] = 4 \cdot \dist(f, 1) \cdot \dist(f, -1)$. Deduce Proposition~\ref{prop:variance-bias}.
    \item \label{ex:Boolean-variance}  Extend Fact~\ref{fact:Boolean-variance} by proving the following: If $\bF$ is a $\bits$-valued random variable with mean~$\mu$ then
        \[
            \Var[\bF] = \E[(\bF - \mu)^2] = \half\E[(\bF - \bF')^2] = 2\Pr[\bF \neq \bF'] = \E[|\bF - \mu|],
        \]
        where $\bF'$ is an independent copy of~$\bF$. (The first two equalities do not require $\bF$ to be $\bits$-valued.)
    \item For any $f \btR$, show that
                \[
                    \la f^{=k}, f^{=\ell} \ra = \begin{cases}\W{k}[f] & \text{if $k = \ell$,} \\ 0 & \text{if $k \neq \ell$.} \end{cases}
                \]
    \item \label{ex:weight-1-dictator} Let $f \btb$.
          \begin{exercises}
              \item \label{ex:weight-deg-1} Suppose $\W{1}[f] = 1$.  Show that $f(x) = \pm \chi_S$ for some $|S| = 1$.
              \item Suppose $\W{\leq 1}[f] = 1$.  Show that $f$ depends on at most $1$ input coordinate. 
              \item Suppose $\W{\leq 2}[f] = 1$.  Must $f$ depend on at most $2$ input coordinates?  At most $3$ input coordinates?  What if we assume $\W{2}[f] = 1$?
          \end{exercises}
    \item \label{ex:fkn-helper}  Let $f \btR$ satisfy $f = f^{=1}$.  Show that $\Var[f^2] = 2\sum_{i \neq j} \wh{f}(i)^2\wh{f}(j)^2$.
    \item \label{ex:sparsity1} Prove that there are no functions $f \btb$ with exactly~$2$ nonzero Fourier coefficients.  What about exactly~$3$ nonzero Fourier coefficients?
    \item \label{ex:density-convolution}
            Verify Propositions~\ref{prop:density-conv} and~\ref{prop:density-conv-density}.
    \item \label{ex:prob-distances} In this exercise you will prove some basic facts about ``distances'' between probability distributions. 
        Let $\vphi$ and $\psi$ be probability densities on~$\F_2^n$.
        \begin{exercises}
        \item Show that the \emph{total variation distance} between $\vphi$ and $\psi$, defined by
                                            \index{total variation distance}%
              \[
                   \dtv{\vphi}{\psi} = \max_{A \subseteq \F_2^n} \Bigl\{\Bigl|\Pr_{\by \sim \vphi}[\by \in A] - \Pr_{\by \sim \psi}[\by \in A]\Bigr|\Bigr\},
              \]
              is equal to $\half \|\vphi - \psi\|_1$.
                                            \nomenclature[dTV]{$\dtv{\vphi}{\psi}$}{total variation distance between the distributions with densities $\vphi$, $\psi$}
        \item Show that the \emph{collision probability}
                                            \index{collision probability}%
               of $\vphi$, defined to be
               \[
                    \Pr_{\substack{\by, \by' \sim \vphi \\ \text{independently}}}[\by = \by'],
               \]
               is equal to $\|\vphi\|_2^2/2^n$.
        \item The \emph{$\chi^2$-distance of $\vphi$ from~$\psi$} is defined by
                                            \index{chi-squared distance}%
              \[
                \dchi{\vphi}{\psi} = \E_{\by \sim \psi}\Bigl[\Bigl(\frac{\vphi(\by)}{\psi(\by)} - 1\Bigr)^2\Bigr],
              \]
              assuming $\psi$ has full support.  Show that the $\chi^2$-distance of $\vphi$ from uniform is equal to $\Var[\vphi]$.
                                            \nomenclature[dchi]{$\dchi{\vphi}{1}$}{chi-squared distance of the distribution with density $\vphi$ from the uniform distribution}
        \item Show that the total variation distance of $\vphi$ from uniform is at most $\half \sqrt{\Var[\vphi]}$.
        \end{exercises}
    \item Let $A \subseteq \bits^n$ have ``volume'' $\delta$, meaning $\E[1_A] = \delta$.  Suppose $\vphi$ is a probability density \emph{supported} on $A$, meaning $\vphi(x) = 0$ when $x \notin A$.  Show that $\|\vphi\|_2^2 \geq 1/\delta$ with equality if $\vphi = \vphi_A$, the uniform density on~$A$. 
    \item \label{ex:conv-ac}
            Show directly from the definition that the convolution operator is associative and commutative.
    \item \label{ex:linear-equivalence}
            Verify that $\eqref{def:lin1} \iff \eqref{def:lin2}$ in Definition~\ref{def:linear}.
    \item \label{ex:learning-parity} Suppose an algorithm is given query access to a linear function $f \co \F_2^n \to \F_2$ and its task is to determine \emph{which} linear function $f$ is.  Show that querying $f$ on $n$ inputs is necessary and sufficient.
    \item \label{ex:BLR}
        \begin{exercises}
            \item Generalize Exercise~\ref{ex:uniq-decoding} as follows: Let $f \co \F_2^n \to \bits$ and suppose that $\dist(f, \chi_{S^*}) = \delta$.  Show that $|\wh{f}(S)| \leq 2\delta$ for all $S \neq S^*$. (Hint: Use the union bound.) 
            \item Deduce that the BLR Test rejects $f$ with probability at least $3\delta - 10\delta^2 + 8\delta^3$.
            \item Show that this lower bound cannot be improved to $c \delta - O(\delta^2)$ for any $c > 3$.
        \end{exercises}
    \item \label{ex:BLR4}
        \begin{exercises}
            \item We call $f \co \F_2^n \to \F_2$ an \emph{affine} function
                                                            \index{affine function}
                  if $f(x) = a \cdot x + b$ for some $a \in \F_2^n$, $b \in \F_2$.  Show that $f$ is affine if and only if $f(x) + f(y) + f(z) = f(x+y+z)$ for all $x, y, z, \in \F_2^n$
            \item \label{ex:gowers2} Let $f \ftR$.  Suppose we choose $\bx, \by, \bz \sim \F_2^n$ independently and uniformly.  Show that $\E[f(\bx)f(\by)f(\bz)f(\bx+\by+\bz)] = \sum_{S} \wh{f}(S)^4$.
                                                        \index{Fourier norm!$4$-}
            \item Give a $4$-query test for a function $f \co \F_2^n \to \F_2$ with the following property: if the test accepts with probability $1-\eps$ then $f$ is $\eps$-close to being affine.  All four query inputs should have the uniform distribution on~$\F_2^n$ (but of course need not be independent).
            \item Give an alternate $4$-query test for being affine in which three of the query inputs are uniformly distributed and the fourth is not random.  (Hint: Show that $f$ is affine if and only if $f(x) + f(y) + f(0) = f(x+y)$ for all $x, y \in \F_2^n$.)
        \end{exercises}
    \item \label{ex:golomb}
        Permutations $\pi \in S_n$
                                    \nomenclature[Sn]{$S_n$}{the symmetric group on $[n]$}
        act on strings $x \in \bitsn$ in the natural way: $(x^\pi)_i$ = $x_{\pi(i)}$. They also act on functions $f \btR$ via $f^\pi(x) = f(x^\pi)$ for all $x \in \bits^n$.  We say that functions $g,\ h \btb$ are \emph{{(permutation-)isomorphic}}
                                                \index{isomorphic}
        if $g = h^\pi$ for some $\pi \in S_n$.  We call $\Aut(f) = \{\pi \in S_n : f^\pi = f\}$ the \emph{{(permutation-)automorphism} group} of~$f$.
                                                \index{automorphism group}%
                                                \nomenclature[Autf]{$\Aut(f)$}{the group of automorphisms of Boolean function $f$}%

        \smallskip

        \begin{exercises}
            \item Show that $\wh{f^{\pi}}(S) = \wh{f}(\pi^{-1}(S))$ for all $S \subseteq [n]$. \label{ex:golomb-symm}
        \end{exercises}

        \smallskip

        For future reference, when we write $(\wh{f}(S))_{|S|=k}$, we mean the sequence of degree-$k$ Fourier coefficients of $f$, listed in lexicographic order of the $k$-sets $S$.

        Given complete truth tables of some $g$ and $h$ we might wish to determine whether they are isomorphic.  One way to do this would be to define a \emph{canonical form} $\text{can}(f) \btb$ for each $f \btb$, meaning that: (i)~$\text{can}(f)$ is isomorphic to $f$; (ii)~if $g$ is isomorphic to $h$ then $\text{can}(g) = \text{can}(h)$.  Then we can determine whether $g$ is isomorphic to $h$ by checking whether $\text{can}(g) = \text{can}(h)$.  Here is one possible way to define a canonical form for $f$:

        \begin{enumerate}[leftmargin=2.5\parindent,label=\arabic*.]
            \item Set $P_0 = S_n$.

            \item For each $k = 1, 2, 3, \dots, n$,

            \begin{enumerate}[leftmargin=1.3em,align=left]
            \item[3.] Define $P_k$ to be the set of all $\pi \in P_{k-1}$ that make the sequence $(\wh{f^\pi}(S))_{|S|=k}$ maximal in lexicographic order on $\R^{\binom{n}{k}}$.
            \end{enumerate}

            \item[4.] Let $\text{can}(f) = f^{\pi}$ for (any) $\pi \in P_n$.
        \end{enumerate}

        \begin{exercises}[resume*]
            \item Show that this is well-defined, meaning that $\text{can}(f)$ is the same function for any choice of $\pi \in P_n$.
            \item Show that $\text{can}(f)$ is indeed a canonical form; i.e., it satisfies~(i) and~(ii) above.
            \item Show that if $\wh{f}(\{1\}), \dots, \wh{f}(\{n\})$ are distinct numbers then $\text{can}(f)$ can be computed in $\wt{O}(2^n)$ time.
            \item We could more generally consider $g, h \btb$ to be isomorphic if  $g(x) = h(\pm x_{\pi(1)}, \dots, \pm x_{\pi(n)})$ for some permutation $\pi$ on $[n]$ and some choice of signs.  Extend the results of this exercise to handle this definition.
        \end{exercises}
\end{exercises}

\subsection*{Notes.}
The Fourier expansion for real-valued Boolean functions dates back to Walsh~\cite{Wal23}
                                            \index{Walsh functions}
who introduced a complete orthonormal basis for $L^2([0,1])$ consisting of $\pm 1$-valued functions, constant on dyadic intervals. Using the ordering introduced by Paley~\cite{Pal32}, the $n$th Walsh basis function $w_n \co [0,1] \to \{-1,1\}$ is defined by $w_n(x) = \prod_{i=0}^{\infty}r_i(x)^{n_i}$, where $n = \sum_{i=0}^{\infty} n_i 2^i$ and $r_i(x)$ (the ``$i$th Rademacher function
                                            \index{Rademacher functions}
at $x$'') is defined to be $(-1)^{x_i}$, with $x = \sum_{i=0}^\infty x_i 2^{-(i+1)}$ for non-dyadic $x \in [0,1]$.  Walsh's interest was in comparing and contrasting the properties of this basis with the usual basis of trigonometric polynomials and also Haar's basis~\cite{Haa10}.

The first major study of the Walsh functions came in the remarkable paper of Paley~\cite{Pal32}, which included strong results on the $L^p$-norms of truncations of Walsh series.  Sadly, Paley died in an avalanche one year later (at age~26) while skiing near Banff.  The next major development in the study of Walsh series was conceptual, with Vilenkin~\cite{Vil47} and Fine~\cite{Fin49} independently suggesting the more natural viewpoint of the Walsh functions as characters of the discrete group $\Z_2^n$.  There was significant subsequent work in the 1950s and 1960s, but it's somewhat unnatural from our point of view because it relies fundamentally on ordering the Rademacher and Walsh functions according to binary expansions. Bonami~\cite{Bon68} and Kiener~\cite{Kie69} seem to have been the first authors to take our viewpoint, treating bits $x_1, x_2, x_3, \dots$ symmetrically and ordering Fourier characters $\chi_S$ according to~$|S|$ rather than~$\max(S)$.  Bonami also obtained the first \emph{hypercontractivity}
                                            \index{hypercontractivity}%
result for the Boolean cube. This proved to be a crucial tool for analysis of Boolean functions; see Chapter~\ref{chap:hypercontractivity}.  For an early survey on Walsh series, see Balashov and Rubinshtein~\cite{BR73}.

Turning to Boolean functions and computer science, the idea of using Boolean logic to study ``switching functions'' (as engineers originally called Boolean functions) dates to the late 1930s and is usually credited to Nakashima~\cite{Nak35}, Shannon~\cite{Sha37}, and Shestakov~\cite{She38}.  Muller~\cite{Mul54} seems to be the first  to have used Fourier coefficients in the study of Boolean functions; he mentions computing them while classifying all functions $f \co \{0,1\}^4 \to \{0,1\}$ up to certain equivalences.  The first publication devoted to Boolean Fourier coefficients was by Ninomiya~\cite{Nin58}, who expanded on Muller's use of Fourier coefficients for the classification of Boolean functions up to various isomorphisms. Golomb~\cite{Gol59} independently pursued the same project (his work is the content of Exercise~\ref{ex:golomb}); he was also the first to recognize the connection to Walsh series.  The use of ``Fourier--Walsh analysis'' in the study of Boolean functions quickly became well known in the early 1960s. Several symposia on applications of Walsh functions took place in the early 1970s, with Lechner's 1971 monograph~\cite{Lec71} and Karpovsky's 1976 book~\cite{Kar76} becoming the standard references. However, the use of Boolean analysis in theoretical computer science seemed to wane until 1988, when the outstanding work of Kahn, Kalai, and Linial~\cite{KKL88} ushered in a new area of sophistication. 

The original analysis by Blum, Luby, and Rubinfeld~\cite{BLR90} for their linearity test was combinatorial; our proof of Theorem~\ref{thm:blr-test} is the elegant analytic one due to  Bellare, Coppersmith, \Hastad, Kiwi, and Sudan~\cite{BCH+96}.  In fact, the essence of this analysis appears already in the 1953 work of Roth~\cite{Rot53} (in the context of the cyclic group $\Z_N$ rather than $\F_2^n$).  The work of Bellare et~al.\ also gives additional analysis improving the results of Theorem~\ref{thm:blr-test} and Exercise~\ref{ex:BLR}.  See also the work of Kaufman, Litsyn, and Xie~\cite{KLX10} for further slight improvement.

In Exercise~\ref{ex:compute-expansions}, the sortedness function was introduced by Ambainis~\cite{Amb03,LLS06}; the hemi-icosahedron function was introduced by Kushilevitz~\cite{NW95}.  The fast algorithm for computing the Fourier transform mentioned in Exercise~\ref{ex:fast-walsh} is due to Lechner~\cite{Lec63}.

\chapter{Basic concepts and social choice}                          \label{chap:concepts}

In this chapter we introduce a number of important basic concepts including influences and noise stability.  Many of these concepts are nicely motivated using the language of \emph{social choice}.  The chapter is concluded with Kalai's Fourier-based proof of Arrow's Theorem.

\section{Social choice functions}                                   \label{sec:social-choice}

In this section we describe some rudiments of the mathematics of \emph{social choice},
                                            \index{social choice}
a topic studied by economists, political scientists, mathematicians, and computer scientists.  The fundamental question in this area is how best to \emph{aggregate} the opinions of many agents.  Examples where this problem arises include citizens voting in an election, committees deciding on alternatives, and independent computational agents making collective decisions. Social choice theory also provides very appealing interpretations for a number of important functions and concepts in the analysis of Boolean functions.

A Boolean function $f \btb$ can be thought of as a \emph{voting rule}
                                            \index{voting rule|seeonly{social choice function}}
or \emph{social choice function}
                                            \index{social choice function}
for an election with~$2$ candidates and~$n$ voters; it maps the votes of the voters to the winner of the election.  Perhaps the most familiar voting rule is the majority function:
\begin{definition}
    For $n$ odd, \emph{the} \emph{majority}
                                            \index{majority}
    function $\Maj_n \btb$ is defined by $\Maj_n(x) = \sgn(x_1 + x_2 + \cdots + x_n)$.  (Occasionally, for $n$ even we say that $f$ is \emph{a} majority function if $f(x)$ equals the sign of $x_1 + \cdots + x_n$ whenever this number is nonzero.)
\end{definition}
The Boolean AND and OR functions correspond to voting rules in which a certain candidate is always elected unless all voters are unanimously opposed. Recalling our somewhat nonintuitive convention that~$-1$ represents $\true$ and~$+1$ represents $\false$:
\begin{definition}
    The function $\AND_n \btb$
                                            \index{AND function}
                                            \nomenclature[AND]{$\AND_n$}{the logical AND function on $n$ bits: $\False$ unless all inputs are $\True$}
    is defined by $\AND_n(x) = +1$ unless $x = (-1, -1, \dots, -1)$.  The function $\OR_n \btb$
                                            \index{OR function}
                                            \nomenclature[OR]{$\OR_n$}{the logical OR function on $n$ bits: $\True$ unless all inputs are $\False$}
    is defined by $\OR_n(x) = -1$ unless $x = (+1, +1, \dots, +1)$.
\end{definition}

Another voting rule commonly encountered in practice:
\begin{definition}
    The $i$th \emph{dictator}
                                            \index{dictator}
    function $\chi_i \btb$ is defined by $\chi_i(x) = x_i$.
\end{definition}
Here we are simplifying notation for the singleton monomial from $\chi_{\{i\}}$ to $\chi_i$.  Even though they are extremely simple functions, the dictators play a very important role in analysis of Boolean functions; to highlight this we prefer the colorful terminology ``dictator functions'' to the more mathematically staid ``projection functions''.  Generalizing:
\begin{definition}
    A function $f \btb$ is called a \emph{$k$-junta}
                                            \index{junta}
    for $k \in \N$ if it depends on at most $k$ of its input coordinates; i.e., $f(x) = g(x_{i_1}, \dots, x_{i_k})$ for some $g \co \bits^k \to \bits$ and $i_1, \dots, i_k \in [n]$.   Informally, we say that $f$ is a ``junta'' if it depends on only a ``constant'' number of coordinates.
\end{definition}
\noindent For example, the number of functions $f \btb$ which are $1$-juntas is precisely $2n+2$: the $n$ dictators, the $n$ negated-dictators, and the $2$ constant functions $\pm 1$.

The European Union's Council of Ministers adopts decisions based on a weighted majority voting rule:
\begin{definition}
    A function $f \btb$ is called a \emph{weighted majority}
                                            \index{weighted majority|seeonly{linear threshold function}}%
    or \emph{(linear) threshold function}
                                            \index{linear threshold function}%
                                            \index{halfspace|seeonly{linear threshold function}}%
                                            \index{LTF|seeonly{linear threshold function}}%
                                            \index{threshold function|seeonly{linear threshold function}}%
    if it is expressible as $f(x) = \sgn(a_0 + a_1 x_1 + \cdots + a_n x_n)$ for some $a_0, a_1, \dots, a_n \in \R$.
\end{definition}
\noindent Exercise~\ref{ex:all-thresholds} has you verify that majority, AND, OR, dictators, and constants are all linear threshold functions.

The leader of the United States (and many other countries) is elected via a kind of ``two-level majority''.  We make a natural definition along these lines:
\begin{definition}                      \label{def:rec-maj}
    The \emph{depth-$d$ recursive majority of $n$}
                                            \index{recursive majority}%
    function, denoted $\Maj_n^{\otimes d}$,%
                                            \nomenclature[f x g]{$f \otimes g$}{if $f \co \bits^m \to \bits$ and $g \btb$, denotes the function $h \co \bits^{mn} \to \bits$ defined by $h(x^{(1)}, \dots, x^{(m)}) = f(g(x^{(1)}), \dots, g(x^{(m)}))$}%
                                            \nomenclature[f x gd]{$f^{\otimes d}$}{if $f \btb$, then $f^{\otimes d} \co \bits^{n^d} \to \bits$ is defined inductively by $f^{\otimes 1} = f$, $f^{\otimes(d+1)} = f \otimes f^{\otimes d}$}
    is the Boolean function of $n^d$ bits defined inductively as follows: $\Maj_n^{\otimes 1} = \Maj_n$, and $\Maj_n^{\otimes(d+1)}(x^{(1)}, \dots, x^{(n)}) = \Maj_n(\Maj_n^{\otimes d}(x^{(1)}), \dots, \Maj_n^{\otimes d}(x^{(n)}))$ for $x^{(i)} \in \bits^{n^d}$.
\end{definition}

In our last example of a $2$-candidate voting rule, the voters are divided into ``tribes'' of equal size and the outcome is $\true$ if and only if at least one tribe is unanimously in favor of $\true$.  This rule is only somewhat plausible in practice, but it plays a very important role in the analysis of Boolean functions:
\begin{definition}
    The \emph{tribes}
                                            \index{tribes function}
    function of width $w$ and size $s$, $\tribes_{w,s} \co \bits^{sw} \to \bits$, is defined by $\tribes_{w,s}(x^{(1)}, \dots, x^{(s)}) = \OR_s(\AND_w(x^{(1)}), \dots, \AND_w(x^{(s)}))$, where $x^{(i)} \in \bits^w$.
\end{definition}

Here are some natural properties of $2$-candidate social choice functions which may be considered desirable:
\begin{definition}                          \label{def:Boolean-social-choice-properties}
    We say that a function $f \btb$ is:
    \begin{itemize}
        \item \emph{monotone}
                                        \index{monotone function}
                if $f(x) \leq f(y)$ whenever $x \leq y$ coordinate-wise;
        \item \emph{odd}
                                        \index{odd function}
                if $f(-x) = -f(x)$;
        \item \emph{unanimous} if $f(1, \dots, 1) = 1$ and $f(-1, \dots, -1) = -1$;
        \item \emph{symmetric}
                                        \index{symmetric function}
                if $f(x^\pi) = f(x)$ for all permutations $\pi \in S_n$ (using the notation from Exercise~\ref{ex:golomb}); i.e., $f(x)$ only depends on the number of $1$'s in $x$.
    \end{itemize}
    The definitions of monotone, odd, and symmetric are also natural for $f \btR$.
\end{definition}
\begin{examples}
The majority function (for $n$ odd) has all four properties in Definition~\ref{def:Boolean-social-choice-properties}; indeed, \emph{May's Theorem}
                                        \index{May's Theorem}
(Exercise~\ref{ex:may-theorem}) states that it is the only monotone, odd, symmetric function. The dictator functions have the first three properties above, as do recursive majority functions.  The AND and OR functions are monotone, unanimous, and symmetric, but not odd.  The tribes functions are monotone and unanimous; although they are not symmetric they have an important weaker property:
\end{examples}
\begin{definition}                  \label{def:transitive-symmetric}
    A function $f \btb$ is \emph{transitive-symmetric}
                                        \index{transitive-symmetric function}%
    if for all $i, i' \in [n]$ there exists a permutation $\pi \in S_n$ taking $i$ to $i'$ such $f(x^\pi) = f(x)$ for all $x \in \bn$.
\end{definition}
\noindent Intuitively, a function is transitive-symmetric if any two coordinates $i, j \in [n]$ are ``equivalent''.

One more natural desirable property of a $2$-candidate voting rule is that it be \emph{unbiased} as defined in Chapter~\ref{sec:basic-fourier-formulas}, i.e., ``equally likely'' to elect~$\pm 1$.  Of course, this presupposes the uniform probability distribution on votes.
\begin{definition}
    The \emph{impartial culture assumption}
                                        \index{impartial culture assumption}%
    is that the $n$ voters' preferences are independent and uniformly random.
\end{definition}
Although this assumption might seem somewhat unrealistic, it gives a good basis for comparing voting rules in the absence of other information.  One might also consider it as a model for the votes of just the ``undecided'' or ``party-independent'' voters.
		
\section{Influences and derivatives}                                            \label{sec:influences}
                                            \index{influence|(}
Given a voting rule $f \btb$ it's natural to try to measure the ``influence'' or ``power'' of the $i$th voter.  One can define this to be the ``probability that the $i$th vote affects the outcome''.
\begin{definition}
    We say that coordinate $i \in [n]$ is \emph{pivotal} for $f \btb$
                                            \index{pivotal}%
    on input $x$ if $f(x) \neq f(x^{\oplus i})$.  Here we have used the
                                            \nomenclature[xoi]{$x^{\oplus i}$}{$(x_1, \dots, x_{i-1}, -x_i, x_{i+1}, \dots, x_n)$}
    notation $x^{\oplus i}$ for the string $(x_1, \dots, x_{i-1}, -x_i, x_{i+1}, \dots, x_n)$.
\end{definition}
\begin{definition}                                      \label{def:Boolean-influence}
    The \emph{influence} of coordinate $i$ on $f \btb$ is defined to be the probability that $i$ is pivotal for a random input:
                                    \nomenclature[Infi]{$\Inf_i[f]$}{the influence of coordinate $i$ on $f$}
    \[
        \Inf_i[f] = \Pr_{\bx \sim \bn}[f(\bx) \neq f(\bx^{\oplus i})].
    \]
\end{definition}
Influences can be equivalently defined in terms of ``geometry'' of the Hamming cube:
\begin{fact}                                        \label{fact:influence-equals-bdry}
    For $f \btb$, the influence $\Inf_i[f]$ equals the fraction of \emph{dimension-$i$ edges} in the Hamming cube which are \emph{boundary}
                                            \index{edge boundary}
    edges.  Here $(x,y)$ is a dimension-$i$ edge if $y = x^{\oplus i}$; it is a boundary edge if $f(x) \neq f(y)$.
\end{fact}

\myfig{.5}{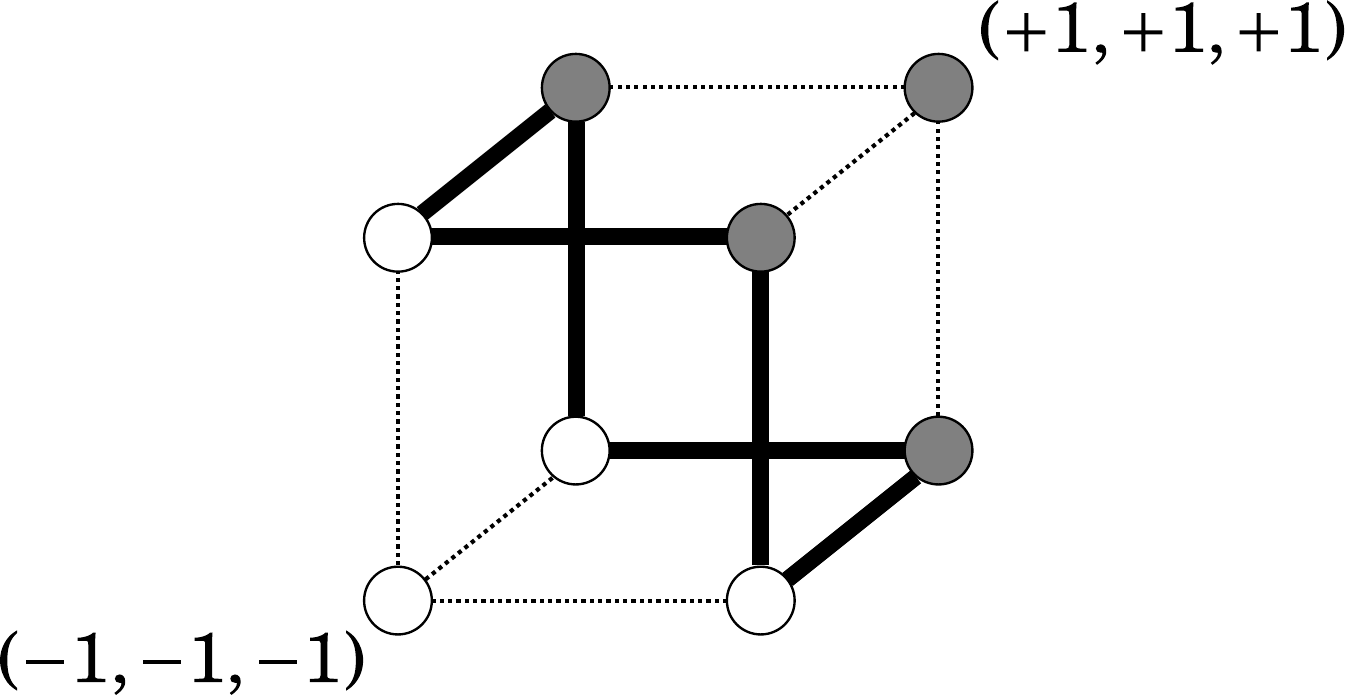}{Boundary edges of the $\maj_3$ function}{fig:maj3-tinf}

\begin{examples} \label{egs:influences} For the $i$th dictator function $\chi_i$ we have that coordinate~$i$ is pivotal for every input $x$; hence $\Inf_i[\chi_i] = 1$.  On the other hand, if $j \neq i$ then coordinate~$j$ is never pivotal; hence $\Inf_j[\chi_i] = 0$ for $j \neq i$.  Note that the same two statements are true about the negated-dictator functions.  For the constant functions $\pm 1$, all influences are~$0$. For the $\OR_n$ function, coordinate~$1$ is pivotal for exactly two inputs, $(-1, 1, 1, \dots, 1)$ and $(1, 1, 1, \dots, 1)$; hence $\Inf_1[\OR_n] = 2^{1-n}$. Similarly, $\Inf_i[\OR_n] = \Inf_i[\AND_n] = 2^{1-n}$ for all $i \in [n]$.  The $\maj_3$ is depicted in Figure~\ref{fig:maj3-tinf}; the points where it's $+1$ are colored gray and the points where it's $-1$ are colored white.  Its boundary edges are highlighted in black; there are $2$ of them in each of the $3$~dimensions. Since there are $4$ total edges in each dimension, we conclude $\Inf_i[\Maj_3] = 2/4 = 1/2$ for all $i \in [3]$.  For majority in higher dimensions, $\Inf_i[\Maj_n]$ equals the probability that among $n-1$ random bits, exactly half of them are~$1$.  This is roughly $\frac{\sqrt{2/\pi}}{\sqrt{n}}$ for large~$n$; see Exercise~\ref{ex:maj-influences} or Chapter~\ref{sec:majority}.
\end{examples}

Influences can also be defined more ``analytically'' by introducing the \emph{derivative} operators.
\begin{definition}
    The \emph{$i$th (discrete) derivative operator}
                                            \index{discrete derivative|seeonly{derivative operator}}
                                            \index{derivative operator}
    $\D_i$
                                            \nomenclature[D]{$\D_i$}{the $i$th discrete derivative: $\D_i f(x) = \frac{f(x^{(i \mapsto 1)}) - f(x^{(i \mapsto -1)})}{2}$}
    maps the function $f \btR$ to the function $\D_i f \btR$ defined by
    \[
        \D_i f (x) = \frac{f(x^{(i\mapsto 1)}) - f(x^{(i \mapsto -1)})}{2}.
    \]
    Here we have used the
                                            \nomenclature[xib]{$x^{(i \mapsto b)}$}{the string $(x_1, \dots, x_{i-1}, b, x_{i+1}, \dots, x_n)$}
    notation $x^{(i \mapsto b)} = (x_1, \dots, x_{i-1}, b, x_{i+1}, \dots, x_n)$.
    Notice that $\D_if(x)$ does not actually depend on $x_i$.  The operator $\D_i$ is a linear operator: i.e., $\D_i(f+g) = \D_i f + \D_i g$.
\end{definition}

If $f \btb$ is Boolean-valued then
\begin{equation}
    \D_if(x) =  \begin{cases}
                    0 & \text{if coordinate $i$ is not pivotal for $x$,} \\
                    \pm 1 & \text{if coordinate $i$ is pivotal for $x$.}
                \end{cases} \label{eqn:derivative-of-Boolean}
\end{equation}
Thus $\D_if(x)^2$ is the $0$-$1$ indicator for whether $i$ is pivotal for $x$ and we conclude that $\Inf_i[f] = \E[\D_if(\bx)^2]$.  We take this formula as a \emph{definition} for the influences of real-valued Boolean functions.
\begin{definition}
    We generalize Definition~\ref{def:Boolean-influence} to functions $f \btR$ by defining the influence of coordinate $i$ on $f$ to be
    \[
        \Inf_i[f] = \E_{\bx \sim \bn}[\D_if(\bx)^2] = \|\D_i f\|_2^2.
    \]
\end{definition}
\begin{definition} \label{def:relevant}
    We say that coordinate $i \in [n]$ is \emph{relevant}
                                            \index{relevant coordinate}
    for $f \btR$ if and only if $\Inf_i[f] > 0$; i.e., $f(x^{(i\mapsto 1)}) \neq f(x^{(i \mapsto -1)})$ for at least one $x \in \bn$.
\end{definition}

The discrete derivative operators are quite analogous to the usual partial derivatives.  For example, $f \btR$ is monotone if and only if $\D_i f(x) \geq 0$ for all $i$ and $x$.  Further, $\D_i$ acts like formal differentiation on Fourier expansions:
\begin{proposition}                                     \label{prop:Di-action}
    Let $f \btR$ have the multilinear expansion $f(x) = \sum_{S \subseteq [n]} \wh{f}(S)\,x^S$.  Then
                                            \nomenclature{$\ni$}{$S \ni i$ is equivalent to $i \in S$}
    \begin{equation}                                                \label{eqn:deriv-formula}
        \D_i f(x) = \sum_{\substack{S \subseteq [n] \\ S \ni i} } \wh{f}(S)\,x^{S \setminus \{i\}}.
    \end{equation}
\end{proposition}
\begin{proof}
    Since $\D_i$ is a linear operator, the claim follows immediately from the observation that
    \[
        \D_i x^S = \begin{cases} x^{S \setminus \{i\}} & \text{if $i \in S$,} \\ 0 & \text{if $i \not \in S$.} \end{cases} \qedhere
    \]
\end{proof}

By applying Parseval's Theorem to the Fourier expansion~\eqref{eqn:deriv-formula}, we obtain a Fourier formula for influences:
\begin{theorem}                                     \label{thm:influences-formula}
    For $f \btR$ and $i \in [n]$,
    \[
        \Inf_i[f] = \sum_{S \ni i} \wh{f}(S)^2.
    \]
\end{theorem}
In other words, the influence of coordinate $i$ on $f$ equals the sum of $f$'s Fourier weights on sets containing~$i$. This is another good example of being able to ``read off'' an interesting combinatorial property of a Boolean function from its Fourier expansion.  In the special case that $f \btb$ is monotone there is a much simpler way to read off its influences: they are the degree-$1$ Fourier coefficients.  In what follows, we write $\wh{f}(i)$
                                                \nomenclature[fhati]{$\wh{f}(i)$}{shorthand for $\wh{f}(\{i\})$ when $i \in \N$}
in place of $\wh{f}(\{i\})$.
\begin{proposition}                                     \label{prop:monotone-influences}
    If $f \btb$ is monotone, then $\Inf_i[f] = \wh{f}(i)$.
\end{proposition}
\begin{proof}
    By monotonicity, the $\pm 1$ in~\eqref{eqn:derivative-of-Boolean} is always~$1$; i.e., $\D_if(x)$ is the $0$-$1$ indicator that $i$ is pivotal for $x$.  Hence $\Inf_i[f] = \E[\D_i f] = \wh{\D_if}(\emptyset) = \wh{f}(i)$, where the third equality used Proposition~\ref{prop:Di-action}.
\end{proof}
                                                \index{influence|)}
This formula allows us a neat proof that for any $2$-candidate voting rule that is monotone and transitive-symmetric, all of the voters have \emph{small influence}:
\begin{proposition} Let $f \btb$ be transitive-symmetric and monotone.  Then $\Inf_i[f] \leq 1/\sqrt{n}$ for all $i \in [n]$. \label{prop:mono-transitive-implies-small-influences}
\end{proposition}
\begin{proof}
    Transitive-symmetry of $f$ implies that $\wh{f}(i) = \wh{f}(i')$ for all $i, i' \in [n]$ (using Exercise~\ref{ex:golomb}\ref{ex:golomb-symm}); thus by monotonicity, $\Inf_i[f] = \wh{f}(i) = \wh{f}(1)$ for all $i \in [n]$.  But by Parseval, $1 = \sum_S \wh{f}(S)^2 \geq \sum_{i=1}^n \wh{f}(i)^2 = n \wh{f}(1)^2$; hence $\wh{f}(1) \leq 1/\sqrt{n}$.
\end{proof}
\noindent This bound is slightly improved in Proposition~\ref{prop:weight-1-same} and Exercise~\ref{ex:weight-1-same}.

The derivative operators are very convenient for functions defined on $\bitsn$. However they are less natural if we think of the Hamming cube as $\{\True, \False\}^n$; for the more general domains we'll look at in Chapter~\ref{chap:generalized-domains} they don't even make sense.  We end this section by introducing some useful definitions that will generalize better later.
\begin{definition}                              \label{def:expectation-operator}
    The \emph{$i$th expectation operator}
                                        \index{expectation operator}
    $\uE_i$
                                        \nomenclature[Ei]{$\uE_i$}{the $i$th expectation operator: $\uE_if(x) = \Ex_{\bx_i}[f(x_1, \dots, x_{i-1}, \bx_i, x_{i+1}, \dots, x_n))]$}
    is the linear operator on functions $f \btR$ defined by
    \[
        \uE_i f (x) = \E_{\bx_i}[f(x_1, \dots, x_{i-1}, \bx_i, x_{i+1}, \dots, x_n)].
    \]
\end{definition}
Whereas $\D_i f$ isolates the part of $f$ depending on the $i$th coordinate, $\uE_i f$ isolates the part \emph{not} depending on the $i$th coordinate.  Exercise~\ref{ex:expectation-facts} asks you to verify the following:
\begin{proposition} For $f \btR$,                                    \label{prop:expectation-facts}
    \begin{itemize}
        \item $\displaystyle \uE_i f (x) = \frac{f(x^{(i \mapsto 1)}) + f(x^{(i \mapsto -1)})}{2}$,
        \item $\displaystyle \uE_i f (x) = \sum_{S \not \ni i} \wh{f}(S)\,x^{S}$,
        \item $\displaystyle f(x) = x_i \D_i f(x) + \uE_i f(x)$.
    \end{itemize}
\end{proposition}
Note that in the decomposition $f = x_i \D_i f + \uE_i f$, neither $\D_i f$ nor $\uE_i f$ depends on~$x_i$.  This decomposition is very useful for proving facts about Boolean functions by induction on~$n$.

Finally, we will also define an operator very similar to $\D_i$ called the \emph{$i$th Laplacian}:%
                                        \index{Laplacian operator!$i$th coordinate}%
\begin{definition}                      \label{def:coordinate-laplacianop}
    The \emph{$i$th coordinate Laplacian operator} $\Lap_i$
                                        \nomenclature[Li]{$\Lap_i$}{the $i$th coordinate  Laplacian operator: $\Lap_i f = f - \uE_i f$}
    is defined by
    \[
        \Lap_i f = f - \uE_i f.
    \]
    Notational warning: Elsewhere you might see the negated definition, $\uE_i f - f$.
\end{definition}
Exercise~\ref{ex:dir-laplacian-facts} asks you to verify the following:
\begin{proposition} For $f \btR$,                                    \label{prop:dir-laplacian-facts}
    \begin{itemize}
        \item $\displaystyle \Lap_i f (x) = \frac{f(x)- f(x^{\oplus i})}{2}$,
        \item $\displaystyle \Lap_i f (x) = x_i \D_i f(x) = \sum_{S \ni i} \wh{f}(S)\,x^{S}$,
        \item $\displaystyle \la f, \Lap_i f \ra = \la \Lap_i f, \Lap_i f \ra = \Inf_i[f]$.
    \end{itemize}
\end{proposition}

\section{Total influence}		                                     \label{sec:total-influence}
A very important quantity in the analysis of a Boolean function is the sum
                                            \index{total influence|(}
of its influences.
\begin{definition}                                          \label{def:total-influence}
    The \emph{total influence}
    of $f \btR$ is defined to be
    \[
        \Tinf[f] = \sum_{i=1}^n \Inf_i[f].
    \]
\end{definition}
For Boolean-valued functions $f \btb$ the total influence has several additional interpretations.  First, it is often referred to as the \emph{average sensitivity}
                                        \index{average sensitivity|seeonly{total influence}}
of $f$ because of the following proposition:
\begin{proposition}                                     \label{prop:total-inf-equals-average-sens}
    For $f \btb$
    \[
        \Tinf[f] = \Ex_{\bx}[\sens_f(\bx)],
    \]
    where $\sens_f(x)$
                                        \nomenclature[sens]{$\sens_f(x)$}{the number of pivotal coordinates for $f$ at $x$}%
    is the \emph{sensitivity}
                                        \index{sensitivity}
    of $f$ at $x$, defined to be the number of pivotal coordinates for $f$ on input $x$.
\end{proposition}
\begin{proof}
\begin{multline*}
    \Tinf[f] = \sum_{i=1}^n \Inf_i[f] = \sum_{i=1}^n \Pr_{\bx}[f(\bx) \neq f(\bx^{\oplus i})] \\ = \sum_{i=1}^n \E_{\bx}[\bone_{f(\bx) \neq f(\bx^{\oplus i})}] = \E_{\bx}\left[\sum_{i=1}^n \bone_{f(\bx) \neq f(\bx^{\oplus i})}\right] = \E_{\bx}[\sens_f(\bx)]. \qedhere
\end{multline*}
\end{proof}
                                        \nomenclature[1B]{$\bone_B$}{$0$-$1$ indicator random variable for event $B$}
The total influence of $f \btb$ is also closely related to the size of its \emph{edge boundary};
                                        \index{edge boundary}
from Fact~\ref{fact:influence-equals-bdry} we deduce:
\begin{fact}                                        \label{fact:total-influence-equals-bdry}
    The fraction of edges in the Hamming cube $\bn$ which are boundary edges for $f \btb$ is equal to $\frac{1}{n} \Tinf[f]$.
\end{fact}
\begin{examples} (Recall Example~\ref{egs:influences}.) For Boolean-valued functions $f \btb$ the total influence ranges between $0$ and~$n$.  It is minimized by the constant functions $\pm 1$ which have total influence~$0$. It is maximized by the parity function $\chi_{[n]}$ and its negation which have total influence~$n$; every coordinate is pivotal on every input for these functions.  The dictator functions (and their negations) have total influence~$1$.  The total influence of $\OR_n$ and $\AND_n$ is very small: $n2^{1-n}$.  On the other hand, the total influence of $\Maj_n$ is fairly large: roughly $\sqrt{2/\pi}\sqrt{n}$ for large~$n$.
\end{examples}

By virtue of Proposition~\ref{prop:monotone-influences} we have another interpretation for the total influence of \emph{monotone} functions:
\begin{proposition}                                     \label{prop:monotone-total-influence}
    If $f \btb$ is monotone, then
    \[
        \Tinf[f] = \sum_{i=1}^n \wh{f}(i).
    \]
\end{proposition}
This sum of the degree-$1$ Fourier coefficients has a natural interpretation in social choice:
\begin{proposition}                                     \label{prop:sum-deg-1-social-choice}
    Let $f \btb$ be a voting rule for a $2$-candidate election.  Given votes $\bx = (\bx_1, \dots, \bx_n)$, let $\bw$ be the number of votes that agree with the outcome of the election, $f(\bx)$.  Then
    \[
        \E[\bw] = \frac{n}{2} + \frac12 \sum_{i=1}^n \wh{f}(i).
    \]
\end{proposition}
\begin{proof}
    By the formula for Fourier coefficients,
    \begin{equation}                                \label{eqn:deg-1-sum}
        \sum_{i=1}^n \wh{f}(i) = \sum_{i=1}^n \E_{\bx}[f(\bx) \bx_i] = \Ex_{\bx}[f(\bx)(\bx_1 + \bx_2 + \cdots + \bx_n)].
    \end{equation}
    Now $\bx_1 + \cdots + \bx_n$ equals the difference between the number of votes for candidate~$1$ and the number of votes for candidate~$-1$.  Hence $f(\bx)(\bx_1 + \cdots + \bx_n)$ equals the difference between the number of votes for the winner and the number of votes for the loser; i.e., $\bw - (n-\bw) = 2\bw - n$.  The result  follows.
\end{proof}
Rousseau~\cite{Rou62} suggested that the ideal voting rule is one which maximizes the number of votes that agree with the outcome. Here we show that the majority rule has this property (at least when $n$ is odd):
\begin{theorem}                                     \label{thm:maj-maximizes-deg-1-sum}
    The unique maximizers of $\sum_{i=1}^n \wh{f}(i)$ among all $f \btb$ are the majority functions.
                                      \index{total influence!monotone functions}
    In particular, $\Tinf[f] \leq \Tinf[\Maj_n] = \sqrt{2/\pi}\sqrt{n} + O(n^{-1/2})$ for
                                        \index{majority!total influence}
    all monotone~$f$.
\end{theorem}
\begin{proof}
    From~\eqref{eqn:deg-1-sum},
    \[
        \sum_{i=1}^n \wh{f}(i) = \Ex_{\bx}[f(\bx)(\bx_1 + \bx_2 + \cdots + \bx_n)] \leq \Ex_{\bx}[|\bx_1 + \bx_2 + \cdots + \bx_n|],
    \]
    since $f(\bx) \in \{-1,1\}$ always.  Equality holds if and only if $f(x) = \sgn(x_1 + \cdots + x_n)$ whenever $x_1 + \cdots + x_n \neq 0$. The second statement of the theorem follows from                         Proposition~\ref{prop:monotone-total-influence} and Exercise~\ref{ex:maj-influences}.
\end{proof}

Let's now take a look at more analytic expressions for the total influence.  By definition, if $f \btR$, then
\begin{equation} \label{eqn:tinf-gradient}
    \Tinf[f] = \sum_{i=1}^n \Inf_i[f] = \sum_{i=1}^n \Ex_{\bx}[\D_i f(\bx)^2] = \Ex_{\bx}\left[\sum_{i=1}^n \D_i f(\bx)^2\right].
\end{equation}
This motivates the following definition:
\begin{definition}                              \label{def:discrete-grad}
    The \emph{(discrete) gradient operator}
                                            \index{discrete gradient|seeonly{gradient operator}}%
                                            \index{gradient operator}%
    $\grad$
                                            \nomenclature{$\grad$}{the gradient: $\grad f(x) = (\D_1 f(x), \dots, \D_n f(x))$}
    maps the function $f \btR$ to the function $\grad f \co \bits^n \to \R^n$ defined by
    \[
        \grad f(x) = (\D_1 f(x), \D_2 f(x), \dots, \D_n f(x)).
    \]
\end{definition}
Note that for $f \btb$ we have $\|\grad f(x)\|_2^2 = \sens_f(x)$, where $\| \cdot \|_2$ is the usual Euclidean norm in $\R^n$.  In general, from~\eqref{eqn:tinf-gradient} we deduce:
\begin{proposition}                                     \label{prop:tinf-gradient-formula}
    For $f \btR$,
    \[
        \Tinf[f] = \Ex_{\bx}[\|\grad f(\bx)\|_2^2].
    \]
\end{proposition}
An alternative analytic definition involves introducing the
                                        \index{Laplacian operator}%
\emph{Laplacian}:
\begin{definition}                      \label{def:laplacianop}
    The \emph{Laplacian operator} $\Lap$
                                        \nomenclature[Lf]{$\Lap f$}{the Laplacian operator applied to the Boolean function~$f$, defined by $\Lap f = \sum_{i=1}^n \Lap_i f$ (or, the Ornstein--Uhlenbeck operator if $f$ is a function on Gaussian space)}
    is the linear operator on functions $f \btR$ defined by $\Lap = \sum_{i=1}^n \Lap_i$.
\end{definition}
Exercise~\ref{ex:laplacian-facts} asks you to verify the following:
\begin{proposition} For $f \btR$,                                    \label{prop:laplacian-facts}
    \begin{itemize}
        \item $\displaystyle \Lap f (x) = (n/2)\bigl(f(x) - \avg_{i \in [n]} \{f(x^{\oplus i})\}\bigr)$,
        \item $\displaystyle \Lap f (x) = f(x) \cdot \sens_f(x)$ \quad if $f \btb$,
        \item $\displaystyle \Lap f = \sum_{S \subseteq [n]} |S| \wh{f}(S)\,\chi_S$,
        \item $\displaystyle \la f, \Lap f \ra = \Tinf[f]$.
    \end{itemize}
\end{proposition}

We can obtain a Fourier formula for the total influence of a function using Theorem~\ref{thm:influences-formula}; when we sum that theorem over all $i \in [n]$ the Fourier weight $\wh{f}(S)^2$ is counted exactly $|S|$ times.  Hence:
\begin{theorem}                                     \label{thm:total-influence-formula}
    For $f \btR$,
    \begin{equation}                                \label{eqn:total-influence-formula}
        \Tinf[f] = \sum_{S \subseteq [n]} |S| \wh{f}(S)^2 = \sum_{k=0}^n k \cdot \W{k}[f].
    \end{equation}
    For $f \btb$ we can express this using the spectral sample:
    \[
        \Tinf[f] = \Ex_{\bS \sim \specsamp{f}}[|\bS|].
    \]
\end{theorem}
Thus the total influence of $f \btb$ also measures the average ``height'' or degree of its Fourier weights.

Finally, from Proposition~\ref{prop:variance-formula} we have $\Var[f] = \sum_{k > 0} \W{k}[f]$; comparing this with~\eqref{eqn:total-influence-formula} we immediately deduce a simple but important fact called the \emph{Poincar\'{e} Inequality}.
\begin{named}{Poincar\'{e} Inequality}
                                                \index{Poincar\'{e} Inequality}%
    For any $f \btR$, $\Var[f] \leq \Tinf[f]$.
\end{named}
Equality holds in the Poincar\'{e} Inequality if and only if all of $f$'s Fourier weight is at degrees $0$ and~$1$; i.e., $\W{\leq 1}[f] = \E[f^2]$.  For Boolean-valued $f \btb$, Exercise~\ref{ex:weight-1-dictator} tells us this can only occur if $f = \pm 1$ or $f = \pm \chi_i$ for some~$i$.

For Boolean-valued $f \btR$, the Poincar\'{e} Inequality can be viewed as an (edge-)isoperimetric inequality,
                                            \index{isoperimetric inequality!Hamming cube}%
or \emph{(edge-)expansion bound},
                                            \index{expansion}%
for the Hamming cube.  If we think of $f$ as the indicator function for a set $A \subseteq \bitsn$ of ``measure'' $\alpha = |A|/2^n$, then $\Var[f] = 4\alpha(1-\alpha)$ (Fact~\ref{fact:Boolean-variance}) whereas $\Tinf[f]$ is $n$ times the (fractional) size of $A$'s edge boundary.  In particular, the Poincar\'{e} Inequality says that subsets $A \subseteq \bitsn$ of measure $\alpha = 1/2$ must have edge boundary at least as large as those of the dictator sets.

For $\alpha \notin \{0, 1/2, 1\}$ the Poincar\'{e} Inequality is not sharp as an edge-isoperimetric inequality for the Hamming cube; for small $\alpha$ even the asymptotic dependence is not optimal.  Precisely optimal edge-isoperimetric results (and also vertex-isoperimetric results) are known for the Hamming cube. The following simplified theorem is optimal for $\alpha$ of the form $2^{-i}$:
\begin{theorem}                                     \label{thm:edge-iso}
    For $f \btb$ with $\alpha = \min\{\Pr[f = 1], \Pr[f=-1]\}$,
    \[
        2\alpha \log(1/\alpha) \leq \Tinf[f].
    \]
\end{theorem}
This result illustrates an important recurring concept in the analysis of Boolean functions: The Hamming cube is a ``small-set expander''.
                                            \index{expansion!small-set}%
Roughly speaking, this is the idea that ``small'' subsets $A \subseteq \bn$ have unusually large ``boundary size''.
                                            \index{total influence|)}

\section{Noise stability}                                                       \label{sec:noise-stab}

Suppose $f \btb$ is a voting rule for a $2$-candidate election.  Making the impartial culture assumption, the~$n$ voters independently and uniformly randomly choose their votes $\bx = (\bx_1, \dots, \bx_n)$.  Now imagine that when each voter goes to the ballot box there is some chance that their vote is \emph{misrecorded}.  Specifically, say that each vote is correctly recorded with probability $\rho \in [0,1]$ and is garbled -- i.e., changed to a random bit -- with probability $1-\rho$.  Writing $\by = (\by_1, \dots, \by_n)$ for the votes that are finally recorded, we may ask about the probability that $f(\bx) = f(\by)$, i.e., whether the misrecorded votes affected the outcome of the election.  This has to do with
                                            \index{noise stability|(}
the \emph{noise stability} of~$f$.

\begin{definition}                                      \label{def:noise}
    Let $\rho \in [0,1]$.  For fixed $x \in \bits^n$ we write $\by \sim N_\rho(x)$
                                            \nomenclature[Nrhox]{$N_\rho(x)$}{when $x \in \bn$, denotes the probability distribution generating a string $\rho$-correlated to $x$}
    to denote that the random string $\by$ is drawn as follows: for each $i \in [n]$ independently,
    \[
        \by_i = \begin{cases} x_i & \text{with probability $\rho$,}\\
                            \text{uniformly random} & \text{with probability $1-\rho$.}
                 \end{cases}
    \]
    We extend the notation to all $\rho \in [-1,1]$ as follows:
    \[
        \by_i = \begin{cases} x_i & \text{with probability $\half + \half \rho$,}\\
                            -x_i & \text{with probability $\half - \half \rho$.}
                 \end{cases}
    \]
    We say that $\by$ is
                                            \index{correlated strings}%
                                            \index{rho-correlated strings@$\rho$-correlated strings|seeonly{correlated strings}}
    \emph{$\rho$-correlated} to $x$.
\end{definition}
\begin{definition}                      \label{def:rho-correlated pair}
      If $\bx \sim \bitsn$ is drawn uniformly at random and then $\by \sim N_\rho(\bx)$, we say that $(\bx, \by)$ is a \emph{$\rho$-correlated pair} of random strings.  This definition is symmetric in $\bx$ and $\by$; it is equivalent to saying that independently for each $i \in [n]$, the pair of random bits $(\bx_i, \by_i)$ satisfies $\E[\bx_i] = \E[\by_i] = 0$ and $\E[\bx_i \by_i] = \rho$.
\end{definition}

With these definitions in hand we can now define the important concept of noise stability, which measures the correlation between $f(\bx)$ and $f(\by)$ when $(\bx,\by)$ is a $\rho$-correlated pair.
\begin{definition}
    For $f \btR$ and $\rho \in [-1,1]$, the \emph{noise stability of $f$ at $\rho$}
                                            \nomenclature[Stabrho]{$\Stab_\rho[f]$}{the noise stability of $f$ at $\rho$: $\E[f(\bx)f(\by)]$ where $\bx,\by$ are a $\rho$-correlated pair}
    is
    \[
        \Stab_\rho[f] = \Es{(\bx, \by) \\ \text{ $\rho$-correlated}}[f(\bx) f(\by)].
    \]
    If $f \btb$ we have
    \begin{align*}
        \Stab_\rho[f] &= \Pr_{\substack{(\bx, \by) \\ \text{ $\rho$-correlated}}}[f(\bx) = f(\by)] \quad- \Pr_{\substack{(\bx, \by) \\ \text{ $\rho$-correlated}}}[f(\bx) \neq f(\by)] \\
        &= 2\Pr_{\substack{(\bx, \by) \\ \text{ $\rho$-correlated}}}[f(\bx) = f(\by)] - 1.
    \end{align*}
\end{definition}
In the voting scenario described above, the probability that the misrecording of votes doesn't affect the election outcome is $\half + \half \Stab_\rho[f]$.

When $\rho$ is close to~$1$ (i.e., the ``noise'' is small) it's sometimes more natural to ask about the probability that reversing a small fraction of the votes reverses the outcome of the election.
\begin{definition}
    For $f \btb$ and $\delta \in [0,1]$ we write $\NS_\delta[f]$ for
                                            \nomenclature[NSf]{$\NS_\delta[f]$}{the noise sensitivity of $f$ at $\delta$; i.e., $\frac12 - \frac12 \Stab_{1-2\delta}[f]$}
    \emph{noise sensitivity of $f$ at $\delta$},
                                            \index{noise sensitivity}
    defined to be the probability that $f(\bx) \neq f(\by)$ when $\bx \sim \bn$ is uniformly random and $\by$ is formed from $\bx$ by reversing each bit independently with probability~$\delta$.  In other words,
    \[
        \NS_\delta[f] = \frac12 - \frac12 \Stab_{1-2\delta}[f].
    \]
\end{definition}

\begin{examples}  The constant functions $\pm 1$ have noise stability~$1$ for every~$\rho$.  The dictator functions $\chi_i$ satisfy $\Stab_\rho[\chi_i] = \rho$ for all $\rho$ (equivalently, $\NS_\delta[\chi_i] = \delta$ for all $\delta$).  More generally,
\[
\Stab_\rho[\chi_S] = \Es{(\bx, \by) \\ \text{ $\rho$-correlated}}[\bx^S \by^S] = \E\left[\prod_{i \in S} (\bx_i \by_i)\right] = \prod_{i \in S} \E[\bx_i \by_i] = \prod_{i \in S} \rho = \rho^{|S|},
\]
where we used the fact that the bit pairs $(\bx_i,\by_i)$ are independent across~$i$ to convert the expectation of a product to a product of an expectation.
\end{examples}
There is no convenient expression for the noise stability of the majority function $\Stab_\rho[\Maj_n]$.  However, for a fixed noise rate, the noise stability/sensitivity tends to a nice limit as $n \to \infty$:
\begin{theorem}                                     \label{thm:maj-stab}
    For
                                                \index{majority!noise stability}%
    any $\rho \in [-1,1]$,
    \[
        \lim_{\substack{n \to \infty \\ n \text{ odd}}} \Stab_\rho[\Maj_n]  = \tfrac{2}{\pi} \arcsin \rho = 1 - \tfrac{2}{\pi} \arccos \rho.
    \]
    Equivalently, for $\delta \in [0,1]$,
    \[
        \lim_{\substack{n \to \infty \\ n \text{ odd}}} \NS_\delta[\Maj_n] = \tfrac{1}{\pi}\arccos(1-2\delta).
    \]
    Using $\cos(z) = 1 - \tfrac12 z^2 + O(z^4)$, hence $\arccos(1- 2\delta) = 2\sqrt{\delta} + O(\delta^{3/2})$, we deduce
    \[
        \lim_{\substack{n \to \infty \\ n \text{ odd}}} \NS_\delta[\Maj_n] = \tfrac{2}{\pi} \sqrt{\delta} + O(\delta^{3/2}).
    \]
\end{theorem}
\myfig{.5}{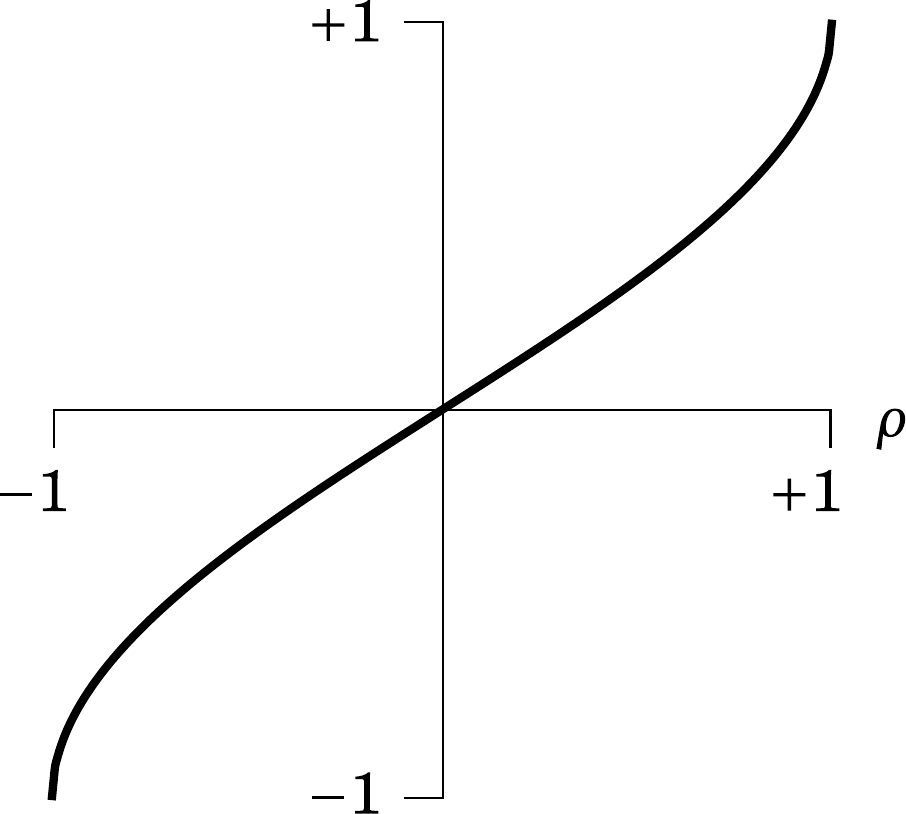}{Plot of $\tfrac{2}{\pi} \arcsin \rho$ as a function of $\rho$}{fig:maj-stab-plot}
We prove Theorem~\ref{thm:maj-stab} in Chapter~\ref{sec:majority}.

There is a simple Fourier formula for the noise stability of a Boolean function; it's one of the most powerful links between the combinatorics of Boolean functions and their Fourier spectra.  To determine it, we begin by introducing the most important operator in analysis of Boolean functions: the
                                        \index{noise operator}%
                                        \index{T$_\rho$|seeonly{noise operator}}%
\emph{noise operator}, denoted $\T_\rho$
                                        \nomenclature[Trho]{$\T_\rho$}{the noise operator: $\T_\rho f(x) = \Ex_{\by \sim N_\rho(x)}[f(\by)]$}
for historical reasons.
\begin{definition}
    For $\rho \in [-1,1]$, the \emph{noise operator with parameter $\rho$} is the linear operator $\T_\rho$ on functions $f \btR$ defined by
    \[
        \T_\rho f(x) = \Ex_{\by \sim N_\rho(x)}[f(\by)].
    \]
\end{definition}
\begin{proposition}                                     \label{prop:T-rho-formula}
    For $f \btR$, the Fourier expansion of $\T_\rho f$ is given by
    \[
        \T_\rho f = \sum_{S \subseteq [n]} \rho^{|S|} \wh{f}(S)\,\chi_S = \sum_{k=0}^n \rho^k f^{=k}.
    \]
\end{proposition}
\begin{proof}
    Since $\T_\rho$ is a linear operator, it suffices to verify that $\T_\rho \chi_S = \rho^{|S|} \chi_S$:
    \[
        \T_\rho \chi_S(x) = \Ex_{\by \sim N_\rho(x)}[\by^S] = \prod_{i \in S} \Ex_{\by \sim N_\rho(x)}[\by_i] = \prod_{i \in S} (\rho x_i) = \rho^{|S|} \chi_S(x).
    \]
    Here we used the fact that for $\by \sim N_\rho(x)$ the bits $\by_i$ are independent and satisfy $\E[\by_i] = \rho x_i$.
\end{proof}
\noindent Exercise~\ref{ex:Trho-plug-in} gives an alternate way of looking at this proof.  Yet another proof using probability densities and convolution is outlined in Exercise~\ref{ex:conv-Trho}.

The connection between $\T_\rho$ and noise stability is that
\[
\Stab_\rho[f] = \Es{\bx \sim \bn \\ \by \sim N_\rho(\bx)}[f(\bx) f(\by)] = \Ex_{\bx} \left[f(\bx) \Ex_{\by \sim N_\rho(\bx)}[f(y)]\right];
\]
hence:
\begin{fact}                                        \label{fact:stab-ip}
    $\Stab_\rho[f] = \la f, \T_\rho f \ra$.
\end{fact}
From Plancherel's Theorem and Proposition~\ref{prop:T-rho-formula} we deduce the Fourier formula for noise stability:
\begin{theorem}                                     \label{thm:stab-formula}
    For $f \btR$,
    \[
        \Stab_\rho[f] = \sum_{S \subseteq [n]} \rho^{|S|}\wh{f}(S)^2 = \sum_{k=0}^n \rho^k \cdot \W{k}[f].
    \]
    Hence for $f \btb$ we have
    \begin{gather}
        \Stab_\rho[f] = \Ex_{\bS \sim \specsamp{f}}[\rho^{|\bS|}], \\
        \NS_\delta[f] = \half \sum_{k=0}^n (1 - (1-2\delta)^{k}) \cdot \W{k}[f].
    \end{gather}
\end{theorem}
Thus the noise stability of $f$ at $\rho$ is equal to the sum of its Fourier weights, attenuated by a factor which decreases exponentially with degree.  A simple but important corollary is that dictators (and their negations) maximize noise stability:
\begin{proposition}                                     \label{prop:dicts-max-noise-stab}
    Let $\rho \in (0,1)$.  If $f \btb$ is unbiased, then $\Stab_\rho[f] \leq \rho$, with equality if and only if $f = \pm \chi_i$ for some $i \in [n]$.
\end{proposition}
\begin{proof}
    For unbiased $f$ we have $\W{0}[f] = 0$ and hence $\Stab_\rho[f] = \sum_{k \geq 1} \rho^k \W{k}[f]$.  Since $\rho^k < \rho$ for all $k > 1$, noise stability is maximized if all of $f$'s Fourier weight is on degree~$1$.  This occurs if and only if $f = \pm \chi_i$, by Exercise~\ref{ex:weight-1-dictator}\ref{ex:weight-deg-1}.
\end{proof}

For a fixed function $f$, it's often interesting to see how $\Stab_\rho[f]$ varies as a function of $\rho$.  From Theorem~\ref{thm:stab-formula} we see that $\Stab_\rho[f]$ is a (univariate) \emph{polynomial} with nonnegative coefficients; in particular, it's an increasing function of $\rho$ on $[0,1]$.  The derivatives of this polynomial at $0$ and $1$ have nice interpretations, as can be immediately deduced from Theorem~\ref{thm:stab-formula}:
\begin{proposition}                                     \label{prop:tinf-is-stab-deriv}
    For $f \btR$,
    \begin{align*}
        \frac{d}{d \rho} \Stab_\rho[f]\Bigm|_{\rho = 0} &= \W{1}[f], \\
        \frac{d}{d \rho} \Stab_\rho[f]\Bigm|_{\rho = 1} &= \Tinf[f].
    \end{align*}
\end{proposition}
\noindent For $f \btb$ we have that $\NS_\delta[f]$ is an increasing function of $\delta$ on $[0, 1/2]$, and the second identity is equivalent to
\[
    \frac{d}{d \delta} \NS_\delta[f]\Bigm|_{\delta = 0} = \Tinf[f].
\]
                                            \index{noise stability|)}

We conclude this section by introducing a version of influences that also incorporates noise.
\begin{definition}                          \label{def:stable-influence}
    For $f \btR$, $\rho \in [0,1]$ and $i \in [n]$, the \emph{$\rho$-stable influence}
                                        \index{stable influence}%
                                        \index{influence!$\rho$-stable|seeonly{stable influence}}%
                                        \index{noisy influence|seeonly{stable influence}}%
                                        \index{attenuated influence|seeonly{stable influence}}%
    of $i$ on $f$ is
    \[
        \Inf_i^{(\rho)}[f] = \Stab_\rho[\D_i f] = \sum_{S \ni i} \rho^{|S|-1} \wh{f}(S)^2,
    \]
    with
                                        \nomenclature[Infirho]{$\Inf_i^{(\rho)}[f]$}{the $\rho$-stable influence, $\Stab_\rho[\D_i f]$}
    $0^0$ interpreted as $1$.  We also define $\Tinf^{(\rho)}[f] = \sum_{i=1}^n \Inf_i^{(\rho)}[f]$.
\end{definition}
Exercise~\ref{ex:stable-tinf} asks you to verify the following:
\begin{fact} \label{fact:stable-tinf}
    $\Tinf^{(\rho)}[f] = \frac{d}{d \rho} \Stab_\rho[f] = \sum_{k=1}^n k \rho^{k-1}  \cdot \W{k}[f]$.
\end{fact}

The $\rho$-stable influence $\Inf_i^{(\rho)}[f]$ increases from $\wh{f}(i)^2$ up to $\Inf_i[f]$ as $\rho$ increases from~$0$ to~$1$.  For $0 < \rho < 1$ there isn't an especially natural combinatorial interpretation for $\Inf_i^{(\rho)}[f]$ beyond $\Stab_\rho[\D_i f]$; however, we will see later that the stable influences are technically very useful.  One reason for this is that every function $f \btb$ has at most ``constantly'' many ``stably-influential'' coordinates:
\begin{proposition}                                     \label{prop:few-stable-influences}
    Suppose $f \btR$ has $\Var[f] \leq 1$.  Given $0 < \delta, \eps \leq 1$, let $J = \{ i \in [n] : \Inf_i^{(1-\delta)}[f] \geq \eps\}$.  Then $|J| \leq \frac{1}{\delta \eps}$.
\end{proposition}
\begin{proof}
    Certainly $|J| \leq \Tinf^{(1-\delta)}[f]/\eps$ so it remains to verify $\Tinf^{(1-\delta)}[f] \leq 1/\delta$. Comparing Fact~\ref{fact:stable-tinf} with $\Var[f] = \sum_{k \neq 0} \W{k}[f]$ term by term, it suffices to show that $(1-\delta)^{k-1} k \leq 1/\delta$ for all $k \geq 1$. This is the easy Exercise~\ref{ex:geom-bound}.
\end{proof}
It's good to think of the set $J$ in this proposition as the ``notable''
                                                    \index{notable coordinates}
coordinates for function~$f$.  Had we used the usual influences in place of stable influences, we would not have been guaranteed a bounded number of ``notable'' coordinates (since, e.g., the parity function $\chi_{[n]}$ has all~$n$ of its influences equal to~$1$).

\section{Highlight: Arrow's Theorem}                                            \label{sec:arrow}

When there are just $2$ candidates, the majority function possesses all of the mathematical properties that seem desirable in a voting rule (e.g., May's Theorem and Theorem~\ref{thm:maj-maximizes-deg-1-sum}).  Unfortunately, as soon as there are $3$ (or more) candidates the problem of social choice becomes much more difficult.  For example, suppose we have candidates $a$, $b$, and $c$, and each of $n$ voters has a ranking of them.  How should we aggregate these preferences to produce a winning candidate?

                                            \index{Condorcet Paradox|(}
In his 1785 \emph{Essay on the Application of Analysis to the Probability of Majority Decisions}~\cite{dC85}, Condorcet suggested using the voters' preferences to conduct the three possible pairwise elections, $a$~vs.~$b$, $b$~vs.~$c$, and $c$~vs.~$a$.  This calls for the use of a $2$-candidate voting rule $f \btb$; Condorcet suggested $f = \Maj_n$ but we might consider any such rule.  Thus a ``$3$-candidate Condorcet election'' using $f$ is conducted as follows:

\iftex
\begin{center}
\begin{tabular}{c|ccccc|c}
\hline
 & \multicolumn{5}{c|}{Voters' Preferences}  & \\
 & $\#1$ & $\#2$ & $\#3$ & $\cdots$ & & \text{Societal Aggregation} \\
\hline
$a$\ {\tiny ($+1$)} \text{ vs. } $b$\ {\tiny ($-1$)} & $+1$ & $+1$ & $-1$ & $\cdots$ & $= x$ & $f(x)$ \\
$b$\ {\tiny ($+1$)} \text{ vs. } $c$\ {\tiny ($-1$)} & $+1$ & $-1$ & $+1$ & $\cdots$ & $= y$ & $f(y)$ \\
$c$\ {\tiny ($+1$)} \text{ vs. } $a$\ {\tiny ($-1$)} & $-1$ & $-1$ & $+1$ & $\cdots$ & $= z$ & $f(z)$ \\
\hline
\end{tabular}
\end{center}
\fi \ifblog HAND-CODE THIS \fi

\medskip

In the above example, voter~$\#1$ ranked the candidates $a > b > c$, voter~$\#2$ ranked them $a > c > b$, voter~$\#3$ ranked them $b > c > a$, etc. Note that the $i$th voter has one of $3! = 6$ possible rankings, and these translate into a triple of bits $(x_i, y_i, z_i)$ from the following set:
\[
\Bigl\{(+1, +1, -1), (+1, -1, -1), (-1,+1,-1), (-1,+1,+1), (+1,-1,+1), (-1, -1, +1)\Bigr\}.
\]
These are precisely the triples satisfying the \emph{not-all-equal}
                                    \index{not-all-equal (NAE) function}
predicate $\NAE_3$ (see Exercise~\ref{ex:compute-expansions}\ref{ex:NAE}).

In the example above, if $n = 3$ and $f = \Maj_3$ then the societal outcome would be $(+1,+1,-1)$, meaning that society elects $a$~over~$b$, $b$~over~$c$, and $a$~over~$c$.  In this case it is only natural to declare $a$ the overall winner.
\begin{definition}
    In an election employing Condorcet's method with voting rule $f \btb$, we say that a candidate is a \emph{Condorcet winner}
    if it wins all of the pairwise elections in which it participates.
\end{definition}
Unfortunately, as Condorcet himself noted, there may not \emph{be} a Condorcet winner.  In the example above, if voter~$\#2$'s ranking was instead $c > a > b$ (corresponding to $(+1, -1, +1)$), we would obtain the ``paradoxical'' outcome $(+1, +1, +1)$: society prefers $a$~over~$b$, $b$~over~$c$, and $c$~over~$a$!  This lack of a Condorcet winner is termed \emph{Condorcet's Paradox}; it occurs when the outcome $(f(x), f(y), f(z))$ is one of the two ``all-equal'' triples $\{(-1, -1, -1), (+1,+1,+1)\}$.
                                            \index{Condorcet Paradox|)}

One might wonder if the Condorcet Paradox can be avoided by using a voting rule $f \btb$ other than majority.  However, in 1950 Arrow~\cite{Arr50} famously showed that the only means of avoidance is an unappealing one:
\begin{named}{Arrow's Theorem}
    Suppose $f \btb$ is a unanimous voting rule used in a $3$-candidate Condorcet election.
                                            \index{Arrow's Theorem}
    If there is \emph{always} a Condorcet winner, then $f$ must be a dictatorship.
\end{named}
\noindent (In fact, Arrow's Theorem is slightly stronger than this; see Exercise~\ref{ex:full-arrow}.)

In 2002 Kalai gave a new proof of Arrow's Theorem; it takes its cue from the title of Condorcet's work and computes the \emph{probability} of a Condorcet winner.  This is done under the ``impartial culture assumption'' for $3$-candidate elections: each voter independently chooses one of the $6$ possible rankings uniformly at random.
\begin{theorem}                                     \label{thm:prob-condorcet}
    Consider a $3$-candidate Condorcet election using $f \btb$.  Under the impartial culture assumption, the probability of a Condorcet winner is precisely $\tfrac34 - \tfrac34 \Stab_{-1/3}[f]$.
\end{theorem}
\begin{proof}
    Let $\bx, \by, \bz \in \bits^n$ be the votes for the elections $a$~vs.~$b$, $b$~vs.~$c$, and $c$~vs.~$a$, respectively.  Under impartial culture, the bit triples $(\bx_i, \by_i, \bz_i)$ are independent and each is drawn uniformly from the $6$ triples satisfying the not-all-equal predicate $\NAE_3 \co \bits^3 \to \{0,1\}$.  There is a Condorcet winner if and only if $\NAE_3(f(\bx),f(\by),f(\bz)) = 1$.  Hence
    \begin{equation} \label{eqn:prob-cond}
        \Pr[\exists \text{ Condorcet winner}] = \E[\NAE_3(f(\bx),f(\by),f(\bz))].
    \end{equation}
    The multilinear (Fourier) expansion of $\NAE_3$ is
    \[
        \NAE_3(w_1, w_2, w_3) = \tfrac34 - \tfrac14w_1 w_2 -\tfrac14 w_1 w_3 - \tfrac14 w_2w_3;
    \]
    thus
    \[
        \eqref{eqn:prob-cond} = \tfrac34 - \tfrac14\E[f(\bx)f(\by)] - \tfrac14\E[f(\bx)f(\bz)]- \tfrac14\E[f(\by)f(\bz)].
    \]
    In the joint distribution of $\bx, \by$ the $n$ bit pairs $(\bx_i, \by_i)$ are independent.  Further, by inspection we see that $\E[\bx_i] = \E[\by_i] = 0$ and that $\E[\bx_i \by_i] = (2/6)(+1) + (4/6)(-1) = -1/3$.  Hence $\E[f(\bx)f(\by)]$ is precisely $\Stab_{-1/3}[f]$.  Similarly we have $\E[f(\bx)f(\bz)] = \E[f(\by)f(\bz)] = \Stab_{-1/3}[f]$ and the proof is complete.
\end{proof}

Arrow's Theorem is now an easy corollary:

\begin{proof}[Proof of Arrow's Theorem] By assumption, the probability of a Condorcet winner is~$1$; hence
\[
    1 =  \tfrac34 - \tfrac34 \Stab_{-1/3}[f] =  \frac34 - \frac34 \sum_{k=0}^n(-1/3)^{k} \W{k}[f].
\]
Since $(-1/3)^k \geq -1/3$ for all $k$, the equality above can only occur if all of $f$'s Fourier weight is on degree~$1$; i.e., $\W{1}[f] = 1$.  By Exercise~\ref{ex:weight-1-dictator}\ref{ex:weight-deg-1} this implies that $f$ is either a dictator or a negated-dictator.  Since $f$ is unanimous, it must in fact be a dictator.
\end{proof}

An advantage of Kalai's analytic proof of Arrow's Theorem is that we can deduce several more interesting results about the probability of a Condorcet winner.  For example, combining Theorem~\ref{thm:prob-condorcet} with Theorem~\ref{thm:maj-stab} we get \emph{Guilbaud's Formula}:
                                        \index{Guilbaud's Formula}
\begin{named}{Guilbaud's Formula}
    In a $3$-candidate Condorcet election using $\maj_n$, the probability of a Condorcet winner tends to
    \[
        \tfrac{3}{2\pi}\arccos(-1/3) \approx 91.2\%.
    \]
    as $n \to \infty$.
\end{named}
This is already a fairly high probability.  Unfortunately, if we want to improve on it while still using a reasonably fair election scheme, we can only set our hopes higher by a sliver:
\begin{theorem}                                     \label{thm:91.9}
    In a $3$-candidate Condorcet election using an $f \btb$ with all $\wh{f}(i)$ equal, the probability of a Condorcet winner is at most $\frac79 + \frac{4}{9\pi} + o_n(1) \approx 91.9\%$.
\end{theorem}
The condition in Theorem~\ref{thm:91.9} seems like it would be satisfied by most reasonably fair voting rules $f \btb$ (e.g., it is satisfied if $f$ is transitive-symmetric or is monotone with all influences equal).  In fact, we will show that Theorem~\ref{thm:91.9}'s hypothesis can be relaxed in Chapter~\ref{sec:weight-level-1}; we will further show in Chapter~\ref{sec:MIST} that $\frac79 + \frac{4}{9\pi}$ can be improved to the tight value $\tfrac{3}{2\pi} \arccos(-1/3)$ of majority.  To return to Theorem~\ref{thm:91.9}, it is an immediate consequence of the following two results, the first being Exercise~\ref{ex:weight-1-same} and the second being an easy corollary of Theorem~\ref{thm:prob-condorcet}.
\begin{proposition}                                     \label{prop:weight-1-same}
    Suppose $f \btb$ has all $\wh{f}(i)$ equal.
                                                \index{Fourier weight!degree-1}%
    Then $\W{1}[f] \leq 2/\pi + o_n(1)$.
\end{proposition}
\begin{corollary}                                     \label{cor:condorcet-weight-1}
    In a $3$-candidate Condorcet election using $f \btb$, the probability of a Condorcet winner is at most $\frac79 + \frac29 \W{1}[f]$.
\end{corollary}
\begin{proof}
    From Theorem~\ref{thm:prob-condorcet}, the probability is
    \begin{align*}
        \tfrac34 - \tfrac34 \Stab_{-1/3}[f] &= \tfrac34 - \tfrac34(\W{0}[f] - \tfrac13 \W{1}[f] + \tfrac19 \W{2}[f] - \tfrac{1}{27} \W{3}[f] + \cdots)\\
        &\leq \tfrac34 + \tfrac14 \W{1}[f] + \tfrac{1}{36} \W{3}[f] + \tfrac{1}{324} \W{5}[f] + \cdots\\
        &\leq \tfrac34 + \tfrac14 \W{1}[f] + \tfrac{1}{36} (\W{3}[f] + \W{5}[f] + \cdots)\\
        &\leq \tfrac34 + \tfrac14 \W{1}[f] + \tfrac{1}{36} (1 - \W{1}[f]) \quad=\quad \tfrac79 + \tfrac29\W{1}[f]. \qedhere
    \end{align*}
\end{proof}

Finally, using Corollary~\ref{cor:condorcet-weight-1} we can prove a ``robust'' version of Arrow's Theorem, showing that a Condorcet election is \emph{almost} paradox-free only if it is \emph{almost} a dictatorship (possibly negated).
\begin{corollary}                                     \label{cor:almost-arrow}
    Suppose that in a $3$-candidate Condorcet election using $f \btb$, the probability of a Condorcet winner is $1 - \epsilon$.  Then $f$ is $O(\eps)$-close to $\pm \chi_i$ for some $i \in [n]$.
\end{corollary}
\begin{proof}
    From Corollary~\ref{cor:condorcet-weight-1} we obtain that $\W{1}[f] \geq 1 - \tfrac92 \epsilon$.  The conclusion
    now follows from the FKN~Theorem.
\end{proof}
\begin{named}{Friedgut--Kalai--Naor (FKN) Theorem}  Suppose $f \btb$ has $\W{1}[f] \geq 1-\delta$.
                                                    \index{FKN Theorem}%
Then $f$ is $O(\delta)$-close to $\pm \chi_i$ for some $i \in [n]$.
\end{named}
We will see the proof of the FKN~Theorem in Chapter~\ref{sec:4th-moment}. We'll also show in Chapter~\ref{sec:weight-level-1} that the $O(\delta)$ closeness can be improved to~$\delta/4 + O(\delta^2 \log(2/\delta))$.

\section{Exercises and notes}
\begin{exercises}
    \item For each function in Exercise~\ref{ex:compute-expansions}, determine if it is odd, transitive-symmetric, and/or symmetric.
    \item \label{ex:all-thresholds}  Show that the $n$-bit functions majority, AND, OR, $\pm \chi_i$, and $\pm 1$ are all linear threshold functions.
    \item \label{ex:may-theorem} Prove \emph{May's Theorem}:
        \begin{exercises}
            \item Show that $f \btb$ is symmetric and monotone if and only if it can be expressed as a weighted majority with $a_1 = a_2 = \cdots = a_n = 1$.
            \item Suppose $f \btb$ is symmetric, monotone, and odd.  Show that $n$ must be odd, and that $f = \Maj_n$.
        \end{exercises}
    \item Subset $A \subseteq \bn$ is called a \emph{Hamming ball}
                                                    \index{Hamming ball}
          if $A = \{x : \hamdist(x,z) < r\}$ for some $z \in \bn$ and real~$r$.  Show that $f \btb$ is the indicator of a Hamming ball if and only if it's expressible as a linear threshold function $f(x) = \sgn(a_0 + a_1 x_1 + \cdots + a_n x_n)$ with $|a_1| = |a_2| = \cdots = |a_n|$.
    \item \label{ex:deg-1-vs-inf} Let $f \btb$ and $i \in [n]$. We say that $f$ is \emph{unate in the $i$th direction}
                                                \index{unate}
     if either $f(x^{(i \mapsto -1)}) \leq f(x^{(i \mapsto 1)})$ for all $x$ (\emph{monotone in the $i$th direction}) or $f(x^{(i \mapsto -1)}) \geq f(x^{(i\mapsto 1)})$ for all $x$ (\emph{antimonotone in the $i$th direction}).  We say that $f$ is \emph{unate} if it is unate in all~$n$ directions.
        \begin{exercises}
            \item \label{ex:deg-1-vs-inf-a} Show that $|\wh{f}(i)| \leq \Inf_i[f]$ with equality if and only if $f$ is unate in the $i$th direction.
            \item Show that the second statement of Theorem~\ref{thm:maj-maximizes-deg-1-sum} holds even for all unate~$f$.
        \end{exercises}
    \item \label{ex:ltf-unate} Show that linear threshold functions are unate.
    \item For each function $f$ in Exercise~\ref{ex:compute-expansions}, compute $\Inf_1[f]$.
    \item \label{ex:inf-at-most-var} Let $f \btb$.  Without using Fourier formulas, show that $\Inf_i[f] \leq \Var[f]$ for each $i \in [n]$.  (Hint: Show  $\Inf_i[f] \leq 2\min\{\Pr[f = -1], \Pr[f = 1]\}$.)
    \item \label{ex:banzhaf} Let $f \co \zo^6 \to \bits$ be given by the weighted majority $f(x) = \sgn(-58 + 31x_1 + 31x_2 + 28x_3 + 21x_4 + 2x_5 + 2x_6)$.  Compute $\Inf_i[f]$ for all $i \in [6]$.
    \item \label{ex:one-way-inf}  Given $b \in \bits$, say that coordinate $i$ is \emph{$b$-pivotal}
                                                    \index{pivotal}%
            for $f \btb$ on input~$x$ if $f(x) = b$ and $f(x^{\oplus i}) \neq b$.  Show that $\Pr_{\bx}[i \text{ is $b$-pivotal on } \bx] = \half \Inf_i[f]$.  Deduce that $\Tinf[f] = 2 \E_{\bx}[\# \text{ $b$-pivotal coordinates on } \bx]$.
    \item \label{ex:nonzero-coeff-relevance}  Let $f \btb$ and suppose $\wh{f}(S) \neq 0$.  Show that each coordinate $i \in S$ is relevant for~$f$.
    \item \label{ex:random-influence} Let $\boldf \btb$ be a random function (as in Exercise~\ref{ex:random-Fourier}).  Compute
                                                \index{random function}
        $\E[\Inf_1[\boldf]]$ and $\E[\Tinf[\boldf]]$.
    \item \label{ex:basic-tribes} Let $w \in \N$, $n = w 2^w$, and write $f$ for
                                                \index{tribes function}
            $\tribes_{w,2^w} \btb$.
          \begin{exercises}
               \item Compute $\E[f]$ and $\Var[f]$, and estimate them asymptotically in terms of~$n$.
               \item Describe the function $\D_1 f$.
               \item Compute $\Inf_1[f]$ and $\Tinf[f]$ and estimate them asymptotically.
          \end{exercises}
    \item \label{ex:abs-decreases-influences} Let $f \btR$, and write $g = |f|$.  Show that $|\D_i g| \leq |\D_i f|$ pointwise.  Deduce that $\Inf_i[g] \leq \Inf_i[f]$ and $\Tinf[g] \leq \Tinf[f]$.
    \item \label{ex:expectation-facts} Prove Proposition~\ref{prop:expectation-facts}.
    \item \label{ex:dir-laplacian-facts} Prove Proposition~\ref{prop:dir-laplacian-facts}.
    \item \label{ex:laplacian-facts} Prove Proposition~\ref{prop:laplacian-facts}.
    \item \label{ex:laplacian-as-deriv} Let $f \btR$.  Show that
            \[
                \Lap f(x) = \frac{d}{d\rho} \T_\rho f(x)\Bigr|_{\rho = 1} = -\frac{d}{dt} \T_{e^{-t}} f(x)\Bigr|_{t = 0}.
             \]
    \item \label{ex:xixj}  Suppose $f, g \btR$ have the property that $f$ does not depend on the $i$th coordinate and $g$ does not depend on the $j$th coordinate ($i \neq j$).  Show that $\E[\bx_i \bx_j f(\bx) g(\bx)] = \E[\D_j f(\bx) \D_i g(\bx)]$.
    \item For $f \btb$ we have that $\E[\sens_f(\bx)] = \Ex_{\bS \sim \specsamp{f}}[|\bS|]$.  Show that also $\E[\sens_f(\bx)^2] = \Ex[|\bS|^2]$. (Hint: Use Proposition~\ref{prop:laplacian-facts}.)  Is it true that $\E[\sens_f(\bx)^3] = \Ex[|\bS|^3]$?
    \item \label{ex:inf-var}  Let $f \btR$ and $i \in [n]$.
            \begin{exercises}
                \item Define $\uVar_i f \btR$
                                        \nomenclature[Vari]{$\uVar_i$}{the operator defined by $\uVar_if(x) = \Var_{\bx_i}[f(x_1, \dots, x_{i-1}, \bx_i, x_{i+1}, \dots, x_n))]$}
            by 
            \[
                \uVar_i f(x) = \Var_{\bx_i}[f(x_1, \dots, x_{i-1}, \bx_i, x_{i+1}, \dots, x_n)].
            \]
            Show that $\Inf_i[f] = \E_{\bx}[\uVar_i f(\bx)]$.
                \item Show that
                    \[
                        \Inf_i[f] = \half \E_{\substack{\bx_i, \bx_i' \sim \bits \\ \text{independent}}}\left[\left\|\restr{f}{}{\bx_i} - \restr{f}{}{\bx_i'}\right\|_2^2\right],
                    \]
                    where $\restr{f}{}{b}$ denotes the function of $n-1$ variables gotten by fixing the $i$th input of~$f$ to bit~$b$.
            \end{exercises}
    \item \label{ex:maj-influences}
        \begin{exercises}
            \item Show that $\Inf_i[\Maj_n] = \binom{n-1}{\frac{n-1}{2}}2^{1-n}$ for all $i \in [n]$.
            \item Show that $\Inf_1[\Maj_n]$ is a decreasing function of (odd) $n$.
            \item Use Stirling's Formula
                                            \index{Stirling's Formula}
                  $m! =(m/e)^m(\sqrt{2\pi m} + O(m^{-1/2}))$ to deduce that $\Inf_1[\Maj_n] = \frac{\sqrt{2/\pi}}{\sqrt{n}} + O(n^{-3/2})$. (Here the $O(\cdot)$ terms are nonnegative.)
            \item Deduce that $2/\pi \leq \W{1}[\Maj_n] \leq 2/\pi + O(n^{-1})$.
            \item Deduce that $\sqrt{2/\pi}\sqrt{n} \leq \Tinf[\Maj_n] \leq \sqrt{2/\pi}\sqrt{n} + O(n^{-1/2})$.
            \item Suppose $n$ is even and $f \btb$ is a majority function.  Show that $\Tinf[f] = \Tinf[\Maj_{n-1}] = \sqrt{2/\pi}\sqrt{n} + O(n^{-1/2})$.
        \end{exercises}
    \item \label{ex:simple-mono-tinf} Using only Cauchy--Schwarz and Parseval, give a very simple proof of the following weakening of Theorem~\ref{thm:maj-maximizes-deg-1-sum}: If $f \btb$ is monotone then $\Tinf[f] \leq \sqrt{n}$.  Extend also to the case of $f$~unate (see Exercise~\ref{ex:deg-1-vs-inf}).
    \item \label{ex:weight-1-same} Prove Proposition~\ref{prop:weight-1-same} with $O(n^{-1})$ in place of $o_n(1)$.  (Hint: Show $\wh{f}(i) \leq \frac{\sqrt{2/\pi}}{\sqrt{n}} + O(n^{-3/2})$ using Theorem~\ref{thm:maj-maximizes-deg-1-sum}.)
    \item \label{ex:Trho-plug-in} Deduce $\T_\rho f(x) = \sum_{S} \rho^{|S|} \wh{f}(S)\,x^S$ using Exercise~\ref{ex:multilin-interp}.
    \item For each function $f$ in Exercise~\ref{ex:compute-expansions}, compute $\Tinf[f]$.
    \item Which functions $f \btb$ with $\#\{x : f(x) = 1\} = 3$ maximize~$\Tinf[f]$?
    \item \label{ex:even-poincare} Suppose $f \btR$ is an even function (recall Exercise~\ref{ex:odd-even}). Show the improved Poincar\'{e} Inequality $\Var[f] \leq \half \Tinf[f]$.
    \item \label{ex:very-weak-kkl} Let $f \btb$ be unbiased, $\E[f] = 0$, and let
                                            \nomenclature[MaxInf]{$\MaxInf[f]$}{$\max_i \{\Inf_i[f]\}$}%
        $\MaxInf[f]$ denote $\max_{i \in [n]}\{\Inf_i[f]\}$.
        \begin{exercises}
            \item Use the Poincar\'{e} Inequality to show $\MaxInf[f] \geq 1/n$.
            \item Prove that $\Tinf[f] \geq 2 - n \MaxInf[f]^2$.  (Hint: Prove $\Tinf[f] \geq \W{1}[f] + 2(1-\W{1}[f])$ and use Exercise~\ref{ex:deg-1-vs-inf}.)  Deduce that $\MaxInf[f] \geq \frac{2}{n} - \frac{4}{n^2}$.
        \end{exercises}
    \item \label{ex:conv-Trho} Use Exercises~\ref{ex:compute-expansions}\ref{ex:Ei-dist},\ref{ex:noise-dist} to deduce the formulas $\uE_i f = \sum_{S \not \ni i} \wh{f}(S)\, \chi_S$ and $\T_\rho f = \sum_S \rho^{|S|} \wh{f}(S)\,\chi_S$.
    \item \label{ex:positivity-preserving} Show that $\T_\rho$ is \emph{positivity-preserving} for $\rho \in [-1,1]$; i.e., $f \geq 0 \implies \T_\rho f \geq 0$.
        Show that $\T_\rho$ is \emph{positivity-improving} for $\rho \in (-1,1)$; i.e., $f \geq 0, f \neq 0 \implies \T_\rho f > 0$.
    \item \label{ex:semigroup} Show that $\T_\rho$ satisfies the
                                                \index{semigroup property}%
            \emph{semigroup property}: $\T_{\rho_1} \T_{\rho_2} = \T_{\rho_1 \rho_2}$.
    \item \label{ex:T-contracts} For $\rho \in [-1,1]$, show that $\T_\rho$ is a \emph{contraction on $L^p(\bn)$} for all $p \geq 1$; i.e., $\|\T_\rho f\|_p \leq \|f\|_p$ for all $f \btR$.
    \item \label{ex:T-and-abs} Show that $|\T_\rho f| \leq T_\rho |f|$ pointwise for any $f \btR$. Further show that for $-1 < \rho < 1$, equality occurs if and only if $f$ is everywhere nonnegative or everywhere nonpositive.
    \item \label{ex:individual-T} For $i \in [n]$ and $\rho \in \R$, let $\T^i_\rho$ be the operator on functions $f \btR$ defined by
        \[
            \T^i_\rho f = \rho f + (1-\rho) \uE_i f = \uE_i f + \rho \Lap_i f.
        \]
        \begin{exercises}
            \item Show that for $\rho \in [-1,1]$ we have
                \[
                    \T^i_\rho f(x) = \E_{\by_i \sim N_\rho(x_i)}[f(x_1, \dots, x_{i-1}, \by_i, x_{i+1}, \dots, x_n)].
                \]
            \item Show that $\T^i_{\rho_1} \T^i_{\rho_2} = \T^i_{\rho_1 \rho_2}$ (cf.~Exercise~\ref{ex:semigroup}) and that any two operators $\T^i_{\rho}$ and $\T^j_{\rho'}$ commute.
            \item For $(\rho_1, \dots, \rho_n) \in \R^n$ we define $\T_{(\rho_1, \dots, \rho_n)} = \T^1_{\rho_1} \T^2_{\rho_2} \cdots \T^n_{\rho_n}$.  Show that $\T_{(\rho, \dots, \rho)}$ is simply $\T_\rho$ and that $\T_{(1, \dots, 1, \rho, 1, \dots, 1)}$ (with the $\rho$ in the $i$th position) is $\T^i_\rho$.
            \item For $\rho_1, \dots, \rho_n \in [-1,1]$, show that $\T_{(\rho_1, \dots, \rho_n)}$ is a contraction on $L^p(\bn)$ for all $p \geq 1$ (cf.~Exercise~\ref{ex:T-contracts}).
        \end{exercises}
    \item Show that $\Stab_{-\rho}[f] = -\Stab_{\rho}[f]$ if $f$ is odd and $\Stab_{-\rho}[f] = \Stab_{\rho}[f]$ if $f$ is even.
    \item For each function $f$ in Exercise~\ref{ex:compute-expansions}, compute $\Stab_\rho[f]$.
    \item Compute $\Stab_\rho[\tribes_{w,s}]$.
    \item Suppose $f \btb$ has $\min(\Pr[f = 1], \Pr[f = -1]) = \alpha$.  Show that $\NS_\delta[f] \leq 2\alpha$ for all $\delta \in [0,1]$.
    \item \label{ex:stable-tinf} Verify Fact~\ref{fact:stable-tinf}.
    \item Fix $f \btR$. Show that $\Stab_\rho[f]$ is a convex function of $\rho$ on $[0,1]$.
    \item \label{ex:tinf-bounds-ns} Let $f \btb$.  Show that $\NS_\delta[f] \leq \delta \Tinf[f]$ for all $\delta \in [0,1]$.
    \item \label{ex:ns-tails} \begin{exercises}
        \item \label{ex:average-influence}  Define the \emph{average influence}
                                                \index{influence!average}
            of $f \btR$ to be $\Ainf[f] = \frac{1}{n} \Tinf[f]$.  Now for $f \btb$, show
            \[
              \AInf[f] = \Pr_{\substack{\bx \sim \bits^n \\ \bi \sim [n]}}[f(\bx) \neq f(\bx^{\oplus \bi})] \quad \text{and} \quad
            \tfrac{1-e^{-2}}{2} \AInf[f] \leq \NS_{1/n}[f] \leq \AInf[f].
            \]
        \item Given $f \btb$ and integer $k \geq 2$, define
            \[
                A_k = \frac{1}{k}(\W{\geq 1}[f] + \W{\geq 2}[f] + \cdots + \W{\geq k}[f]),
            \] the ``average of the first~$k$ tail weights''.  Generalizing the second statement in part~\ref{ex:average-influence}, show that $\tfrac{1-e^{-2}}{2} A_k \leq \NS_{1/k}[f] \leq A_k$.
        \end{exercises}
    \item \label{ex:ns-union-bound}  Suppose $f_1, \dots, f_s \btb$ satisfy $\NS_\delta[f_i] \leq \eps_i$.  Let $g \co \bits^s \to \bits$ and define $h \btb$ by $h = g(f_1, \dots, f_s)$.  Show that $\NS_\delta[h] \leq \sum_{i=1}^s \eps_i$.
    \item \label{ex:geom-bound} Complete the proof of Proposition~\ref{prop:few-stable-influences} by showing that $(1-\delta)^{k-1} k \leq 1/\delta$ for all $0 < \delta \leq 1$ and $k \in \N^+$.
                                                \nomenclature[N+]{$\N^+$}{$\{1, 2, 3, \dots \}$}
        (Hint: Compare both sides with $1 + (1-\delta) + (1-\delta)^2 + \cdots + (1-\delta)^{k-1}$.)
    \item \label{ex:stab-lipschitz} Fixing $f \btR$, show the following Lipschitz bound for $\Stab_\rho[f]$ when $0 \leq \rho-\eps \leq \rho < 1$:
            \[
                \bigl|\Stab_{\rho}[f] - \Stab_{\rho-\eps}[f]\bigr| \leq \eps \cdot \frac{1}{1-\rho} \cdot \Var[f].
            \]
        (Hint: Use the Mean Value Theorem and Exercise~\ref{ex:geom-bound}.)
    \item \label{ex:transitive-regular}
                                                \index{automorphism group}%
                                                \index{transitive-symmetric function}%
        Let $f \btb$ be a transitive-symmetric function; in the notation of Exercise~\ref{ex:golomb}, this means the group $\Aut(f)$ acts transitively on~$[n]$.  Show that $\Pr_{\bpi \sim \Aut(f)}[\bpi(i) = j] = 1/n$ for all $i, j \in [n]$.
    \item \label{ex:convex-functionals} Suppose that $\mathop{\bf F}$ is a functional on functions $f \btR$ expressible as $\mathop{\bf F\/}[f] = \sum_{S} c_S \wh{f}(S)^2$ where $c_S \geq 0$ for all $S \subseteq [n]$.  (Examples include $\Var$, $\W{k}$, $\Inf_i$, $\Tinf$, $\Inf^{(1-\delta)}_i$, and $\Stab_\rho$ for $\rho \geq 0$.)  Show that $\mathop{\bf F}$ is convex, meaning $\mathop{\bf F\/}[\lambda f + (1-\lambda) g] \leq \lambda \mathop{\bf F\/}[f] + (1-\lambda) \mathop{\bf F\/}[g]$ for all $f$, $g$, and $\lambda \in [0,1]$.
    \item \label{ex:FKN-0-and-1}  Extend the FKN~Theorem as follows: Suppose $f \btb$ has $\W{\leq 1}[f] \geq 1-\delta$.  Show that $f$ is $O(\delta)$-close to a $1$-junta.  (Hint: Consider $g(x_0, x) = x_0 f(x_0 x)$.)
    \item Compute the precise probability of a Condorcet winner (under impartial culture) in a $3$-candidate, $3$-voter election using $f = \Maj_3$.
    \item \label{ex:full-arrow}
        \begin{exercises}
            \item Arrow's Theorem for $3$ candidates is slightly more general than what we stated: it allows for three \emph{different} unanimous functions $f, g, h \btb$ to be used in the three pairwise elections.  But show that if using $f$, $g$, $h$ always gives rise to a Condorcet winner then $f = g = h$.  (Hint: First show $g(x) = -f(-x)$ for all $x$ by using the fact that $x$, $y = -x$, and $z = (f(x), \dots, f(x))$ is always a valid possibility for the votes.)
            \item Extend Arrow's Theorem to the case of Condorcet elections with more than~$3$ candidates.
        \end{exercises}
    \item \label{ex:polarization} The \emph{polarizations} of~$f \btR$ (also known as
                                                \index{polarization}%
                                                \index{compression|seeonly{polarization}}%
                                                \index{shifting|seeonly{polarization}}%
        compressions, downshifts, or two-point rearrangements) are defined as follows.  For $i \in [n]$, the $i$-polarization of $f$ is the function $f^{\sigma_i} \btR$ defined by
        \[
            f^{\sigma_i}(x) = \begin{cases}
                                  \max\{f(x^{(i \mapsto +1)}), f(x^{(i \mapsto -1)})\} & \text{if $x_i = +1$,} \\
                                  \mathrlap{\min}\phantom{\max}\{f(x^{(i \mapsto +1)}), f(x^{(i \mapsto -1)})\} & \text{if $x_i = -1$.}
                              \end{cases}
        \]
        \begin{exercises}
            \item Show that $\E[f^{\sigma_i}] = \E[f]$ and $\|f^{\sigma_i}\|_p = \|f\|_p$ for all~$p$.
            \item Show that $\Inf_{j}[f^{\sigma_i}] \leq \Inf_{j}[f]$ for all $j \in [n]$.
            \item Show that $\Stab_\rho[f^{\sigma_i}] \geq \Stab_\rho[f]$ for all $0 \leq \rho \leq 1$.
            \item Show that $f^{\sigma_i}$ is monotone in the~$i$th direction (recall Exercise~\ref{ex:deg-1-vs-inf}).  Further, show that if $f$ is monotone in the $j$th direction for some $j \in [n]$ then $f^{\sigma_i}$ is still monotone in the $j$th direction.
            \item Let $f^* = f^{\sigma_1 \sigma_2 \cdots \sigma_n}$.  Show that $f^*$ is monotone, $\E[f^*] = \E[f]$, $\Inf_j[f^*] \leq \Inf_j[f]$ for all $j \in [n]$, and $\Stab_\rho[f^*] \geq \Stab_\rho[f]$ for all $0 \leq \rho \leq 1$.
        \end{exercises}
    \item \label{ex:enflo} The Hamming distance $\hamdist(x,y) = \#\{i : x_i \neq y_i\}$ on the discrete cube $\bits^n$ is an example of an \emph{$\ell_1$ metric space}.  For $D \geq 1$, we say that the discrete cube can be \emph{embedded into $\ell_2$ with distortion~$D$} if there is a mapping $F \co \bn \to \R^m$ for some $m \in \N$ such that:
            \begin{align*}
                \|F(x) - F(y)\|_2 &\geq \hamdist(x,y) \text{ for all $x$, $y$;} \tag{``no contraction''}\\
                \|F(x) - F(y)\|_2 &\leq D \cdot \hamdist(x,y) \text{ for all $x$, $y$.} \tag{``expansion at most $D$''}
            \end{align*}
        In this exercise you will show that the least distortion possible is $D = \sqrt{n}$.
        \begin{exercises}
             \item Recalling the definition of $f^\odd$ from Exercise~\ref{ex:odd-even}, show that for any $f \btR$ we have  $\|f^\odd\|_2^2 \leq \Tinf[f]$  and hence
                 \[
                    \Ex_{\bx}[(f(\bx) - f(-\bx))^2] \leq \sum_{i = 1}^n \Ex_{\bx}\Bigl[\bigl(f(\bx) - f(\bx^{\oplus i})\bigr)^2\Bigr].
                 \]
             \item Suppose $F \co \bn \to \R^m$, and write $F(x) = (f_1(x), f_2(x), \dots, f_m(x))$ for functions $f_i \co \bn \to \R$.  By summing the above inequality over $i \in [m]$, show that any $F$ with no contraction must have expansion at least $\sqrt{n}$.
             \item Show that there is an embedding $F$ achieving distortion $\sqrt{n}$.
        \end{exercises}
    \item Give a Fourier-free proof of the Poincar\'{e} Inequality by induction on~$n$.
    \item \label{ex:lo-kk} Let $V$ be a vector space with norm $\| \cdot \|$ and fix $w_1, \dots, w_n \in V$. Define $g \btR$ by $g(x) = \|\sum_{i=1}^n x_i w_i\|$.
        \begin{exercises}
            \item Show that $\Lap g \leq g$ pointwise.  (Hint: Triangle inequality.)
            \item Deduce $2\Var[g] \leq \E[g^2]$ and thus the following
                                                \index{Khintchine(--Kahane) Inequality}%
            \emph{Khintchine--Kahane Inequality}:
            \[
                \E_{\bx}\left[\left\|\littlesum_{i=1}^n \bx_i w_i\right\|\right] \geq \frac{1}{\sqrt{2}} \cdot \E_{\bx}\left[\left\|\littlesum_{i=1}^n \bx_i w_i\right\|^2\right]^{1/2}.
            \]
            (Hint: Exercise~\ref{ex:even-poincare}.)
            \item Show that the constant $\frac{1}{\sqrt{2}}$ above is optimal, even if $V = \R$.
        \end{exercises}
    \item \label{ex:nicd} In the \emph{correlation distillation} problem,
                                                    \index{correlation distillation}
          a \emph{source} chooses $\bx \sim \bits^n$ uniformly at random and broadcasts it to $q$ \emph{parties}.  We assume that the transmissions suffer from some kind of noise, and therefore the players receive imperfect copies $\by^{(1)}, \dots, \by^{(q)}$ of $\bx$.  The parties are not allowed to communicate, and despite having imperfectly correlated information they wish to agree on a single random bit. In other words, the $j$th party will output a bit $f_j(\by^{(j)}) \in \bits$, and the goal is to find functions $f_1, \dots, f_q$ that maximize the probability that $f_1(\by^{(1)}) = f_2(\by^{(2)}) = \cdots = f_q(\by^{(q)})$. To avoid trivial deterministic solutions, we insist that $\E[f_j(\by^{(j)})]$ be $0$ for all $j \in [q]$.
          \begin{exercises}
                \item \label{ex:nicd2} Suppose $q = 2$, $\rho \in (0,1)$, and $\by^{(j)} \sim N_\rho(\bx)$ independently for each $j$. Show that the optimal solution is $f_1 = f_2 = \pm \chi_i$ for some $i \in [n]$.  (Hint: You'll need Cauchy--Schwarz.)
                \item \label{ex:nicd3} Show the same result for $q = 3$.
                \item \label{ex:nicd-ow} Let $q = 2$ and $\rho \in (\frac12, 1)$.  Suppose that $\by^{(1)} = \bx$ exactly, but $\by^{(2)} \in \{-1, 0, 1\}^n$ has \emph{erasures}: it's formed from $\bx$ by setting $\by^{(2)}_i = \bx_i$ with probability~$\rho$ and $\by^{(2)}_i = 0$ with probability $1-\rho$, independently for all $i \in [n]$.  Show that the optimal success probability is $\half + \half \rho$ and there is an optimal solution in which $f_1 = \pm \chi_i$ for any $i \in [n]$.  (Hint: Eliminate the source, and introduce a fictitious party $1'$\dots)
                \item \label{ex:nicd-ow2} Consider the previous scenario but with $\rho \in (0,\frac12)$.  Show that if $n$ is sufficiently large, then the optimal solution does \emph{not} have $f_1 = \pm \chi_i$.
          \end{exercises}
    \item \label{ex:weak-kkl}
        \begin{exercises}
            \item Let $g \co \bn \to \R^{\geq 0}$ have $\E[g] = \delta$.  Show that for any $\rho \in [0,1]$,
                    \[
                        \rho \sum_{j=1}^n |\wh{g}(j)| \leq \delta + \sum_{k = 2}^n \rho^k \|g^{=k}\|_\infty.
                    \]
                  (Hint: Exercise~\ref{ex:positivity-preserving}.)
            \item Assume further that $g \co \bn \to \{0,1\}$.  Show that $\|g^{=k}\|_\infty \leq \sqrt{\delta} \sqrt{\binom{n}{k}}$.  (Hint: First bound $\|g^{=k}\|_2^2$.)  Deduce $\rho \sum_{j=1}^n |\wh{g}(j)| \leq \delta + 2\rho^2 \sqrt{\delta} n$,  assuming $\rho \leq \frac{1}{2\sqrt{n}}$.
            \item Show that $\sum_{j=1}^n |\wh{g}(j)| \leq 2\sqrt{2} \delta^{3/4} \sqrt{n}$ (assuming $\delta \leq 1/4$). Deduce $\W{1}[g] \leq 2\sqrt{2} \cdot \delta^{7/4} \sqrt{n}$. (Hint: show $|\wh{g}(j)| \leq \delta$ for all~$j$.)
            \item Suppose $f \btb$ is monotone and $\MaxInf[f] \leq \delta$.  Show $\W{2}[f] \leq \sqrt{2} \cdot \delta^{3/4} \cdot \Tinf[f] \cdot \sqrt{n}$.
            \item Suppose further that $f$ is unbiased. Show that $\MaxInf[f] \leq o(n^{-2/3})$ implies $\Tinf[f] \geq 3 - o(1)$; conclude $\MaxInf[f] \geq \frac{3}{n} - o(1/n)$.  (Hint: Extend Exercise~\ref{ex:very-weak-kkl}.)  Use Exercise~\ref{ex:polarization} to remove the assumption that~$f$ is monotone for these statements.
        \end{exercises}
    \item \label{ex:generalized-range} Let $V$ be a vector space (over~$\R$) with norm $\| \cdot \|_V$.  If $f \co \bn \to V$ we can define its Fourier coefficients $\wh{f}(S) \in V$ by the usual formula $\wh{f}(S) = \Ex_{\bx \in \bits^n}[f(\bx) \bx^S]$.  We may also define $\|f\|_p = \E_{\bx \in \bits^n}[\|f(\bx)\|_V^p]^{1/p}$.  Finally, if the norm $\| \cdot \|_V$ arises from an inner product $\la \cdot, \cdot \ra_V$ on $V$ we can define an inner product on functions $f, g \co \bn \to V$ by $\la f, g \ra = \Ex_{\bx \in \bits^n}[\la f(\bx), g(\bx)\ra_V]$.   The material developed so far in this book has used $V = \R$ with $\la \cdot, \cdot \ra_V$ being multiplication.  Explore the extent to which this material extends to the more general setting.
\end{exercises}

\subsection*{Notes.}

The mathematical study of social choice began in earnest in the late 1940s; see Riker~\cite{Rik61} for an early survey or the compilation~\cite{BGR09} for some modern results.  Arrow's Theorem was the field's first major result; Arrow proved it in 1950~\cite{Arr50} under the extra assumption of  monotonicity (and with a minor error~\cite{Bla57}), with the refined version appearing in 1963~\cite{Arr63}.  He was awarded the Nobel Prize for this work in 1972.  May's Theorem is from 1952~\cite{May52}. Guilbaud's Formula is also from 1952~\cite{Gui52}, though Guilbaud only stated it in a footnote and wrote that it is computed ``by the usual means in combinatorial analysis''.  The first published proof appears to be due to Garman and Kamien~\cite{GK68}; they also introduced the impartial culture assumption.  The term ``junta'' appears to have been introduced by Parnas, Ron, and Samorodnitsky~\cite{PRS01}.

The notion of influence $\Inf_i[f]$ was originally introduced by the geneticist Penrose~\cite{Pen46}, who observed that $\Inf_i[\Maj_n] \sim \frac{\sqrt{2/\pi}}{\sqrt{n}}$.  It was rediscovered by the lawyer Banzhaf in 1965~\cite{Ban65}; he sued the Nassau County (NY) Board after proving that the voting system it used (the one in Exercise~\ref{ex:banzhaf}) gave some towns zero influence.  Influence is sometimes referred to as the Banzhaf, Penrose--Banzhaf, or Banzhaf--Coleman index (Coleman being another rediscoverer~\cite{Col71}).  Influences were first studied in the computer science literature by Ben-Or and Linial~\cite{BL85}; they introduced also introduced ``tribes''
                                                \index{tribes function}
as an example of a function with constant variance yet small influences.  The Fourier formulas for influence may have first appeared in the work of Chor and Ger\'eb-Graus~\cite{CGG87}.

Total influence of Boolean functions has long been studied in combinatorics, since it is equivalent to edge-boundary size for subsets of the Hamming cube.  For example, the edge-isoperimetric inequality was first proved by Harper in 1964~\cite{Har64}.  In the context of Boolean functions, Karpovsky \cite{Kar76} proposed $\Tinf[f]$ as a measure of the computational complexity of~$f$, and Hurst, Miller, and Muzio~\cite{HMM82} gave the Fourier formula $\sum_S |S|\wh{f}(S)^2$.  The terminology ``Poincar\'{e} Inequality'' comes from the theory of functional inequalities and Markov chains; the inequality is equivalent to the \emph{spectral gap} for the discrete cube graph.

The noise stability of Boolean functions was first studied explicitly by Benjamini, Kalai, and Schramm in 1999~\cite{BKS99}, though it plays an important role in the earlier work of H{\aa}stad~\cite{Has97}.  See O'Donnell~\cite{O'D03} for a survey.  The noise operator was introduced by Bonami~\cite{Bon70} and independently by Beckner~\cite{Bec75}, who used the notation $\T_\rho$ which was standardized by Kahn, Kalai, and Linial~\cite{KKL88}.  For nonnegative noise rates it's often natural to use the alternate parameterization $\T_{e^{-t}}$ for $t \in [0,\infty]$.

The Fourier approach to Arrow's Theorem is due to Kalai~\cite{Kal02}; he also proved Theorem~\ref{thm:91.9} and Corollary~\ref{cor:almost-arrow}.  The FKN~Theorem is due to Friedgut, Kalai, and Naor~\cite{FKN02}; the observation from Exercise~\ref{ex:FKN-0-and-1} is due to Kindler.

The polarizations from Exercise~\ref{ex:polarization} originate in Kleitman~\cite{Kle66}. Exercise~\ref{ex:enflo} is a theorem of Enflo from 1970~\cite{Enf70}.  Exercise~\ref{ex:lo-kk} is a theorem of Lata{\l}a and Oleszkiewicz~\cite{LO94}.  In Exercise~\ref{ex:nicd}, part~\ref{ex:nicd3} is due to Mossel and O'Donnell~\cite{MO05}; part~\ref{ex:nicd-ow} was conjectured by Yang~\cite{Yan04} and proved by O'Donnell and Wright~\cite{OW12}.  Exercise~\ref{ex:weak-kkl} is a polishing of the 1987 work by Chor and Ger\'{e}b-Graus~\cite{CGG87,CGG88}, a precursor of the KKL~Theorem. The weaker Exercise~\ref{ex:very-weak-kkl} is also due to them and Noga Alon independently.

\chapter{Spectral structure and learning}                       \label{chap:spectral-structure}

One reasonable way to assess the ``complexity'' of a Boolean function is in terms how complex its Fourier spectrum is. For example, functions with sufficiently simple Fourier spectra can be efficiently \emph{learned} from examples.  This chapter will be concerned with understanding the location, magnitude, and structure of a Boolean function's Fourier spectrum.
	
\section{Low-degree spectral concentration}                    \label{sec:low-degree}

One way a Boolean function's Fourier spectrum can be ``simple'' is for it to be mostly concentrated at small degree.  \begin{definition}                                          \label{def:conc-degree}
    We say that the Fourier spectrum of $f \btR$ is \emph{$\eps$-concentrated on degree up to~$k$}
                                                \index{concentration, spectral}%
                                                \index{spectral concentration|seeonly{concentration, spectral}}%
    if
    \[
        \W{>k}[f] = \sum_{\substack{S \subseteq [n] \\|S| > k}} \wh{f}(S)^2 \leq \eps.
    \]
    For $f \btb$ we can express this condition using the spectral sample: $\Pr_{\bS \sim \specsamp{f}}[|\bS| > k] \leq \eps$.
\end{definition}
It's possible to show such a concentration result combinatorially by showing that a function has small total influence:
\begin{proposition}                                     \label{prop:tinf-concentration}
    For any $f \btR$ and $\eps > 0$, the Fourier spectrum of $f$ is $\eps$-concentrated on degree up to $\Tinf[f]/\eps$.
\end{proposition}
\begin{proof}
    This follows immediately from Theorem~\ref{thm:total-influence-formula}, $\Tinf[f] = \sum_{k=0}^n k \cdot \W{k}[f]$. For $f \btb$, this is Markov's inequality applied to the cardinality of the spectral sample.
\end{proof}
For example, in Exercise~\ref{ex:basic-tribes} you showed that $\TInf[\tribes_{w,2^w}] \leq O(\log n)$, where $n = w2^w$; thus this function's spectrum is $.01$-concentrated on degree up to $O(\log n)$, a rather low level.  Proving this by explicitly calculating Fourier coefficients would be quite painful.

Another means of showing low-degree spectral concentration is through noise stability/sensitivity:
\begin{proposition}                                     \label{prop:ns-concentration}
    For any $f \btb$ and $\delta \in (0, 1/2]$, the Fourier spectrum of $f$ is $\eps$-concentrated on degree up to $1/\delta$ for
    \[
       \eps = \tfrac{2}{1-e^{-2}}\NS_\delta[f] \leq 3 \NS_\delta[f].
    \]
\end{proposition}
\begin{proof}
    Using the Fourier formula from Theorem~\ref{thm:stab-formula},
    \begin{align*}
        2\NS_\delta[f] &= \E_{\bS \sim \specsamp{f}}[1 - (1-2\delta)^{|\bS|}] \\
        &\geq (1-(1-2\delta)^{1/\delta}) \cdot \Pr_{\bS \sim \specsamp{f}}[|\bS| \geq 1/\delta] \\
        &\geq (1-e^{-2}) \cdot \Pr_{\bS \sim \specsamp{f}}[|\bS| \geq 1/\delta],
    \end{align*}
    where the first inequality used that $1-(1-2\delta)^k$ is a nonnegative nondecreasing function of $k$.  The claim follows.
\end{proof}
As an example, Theorem~\ref{thm:maj-stab} tells us that for $\delta > 0$ sufficiently small and~$n$ sufficiently large (as a function of~$\delta$), $\NS_\delta[\Maj_n] \leq \sqrt{\delta}$.  Hence the Fourier spectrum of $\Maj_n$ is $3\sqrt{\delta}$-concentrated on degree up to $1/\delta$; equivalently, it is $\eps$-concentrated on degree up to $9/\eps^2$.  (We will give sharp constants for majority's spectral concentration in Chapter~\ref{sec:maj-coefficients}.)  This example also shows there is no simple converse to Proposition~\ref{prop:tinf-concentration}; although $\Maj_n$ has its spectrum $.01$-concentrated on degree up to~$O(1)$, its total influence is $\Theta(\sqrt{n})$.

Finally, suppose a function $f \btb$ has its Fourier spectrum \emph{$0$-concentrated} up to degree~$k$; in other words, $f$ has real degree $\deg(f) \leq k$.  In this case $f$ must be somewhat simple; indeed, if~$k$ is a constant, then~$f$ is a junta:
\begin{theorem}                                     \label{thm:nisan-szegedy}
    Suppose $f \btb$ has $\deg(f) \leq k$.  Then $f$ is a $k2^{k-1}$-junta.
\end{theorem}
The bound $k2^{k-1}$ cannot be significantly improved; see Exercise~\ref{ex:read-once-dt}.  The key to proving Theorem~\ref{thm:nisan-szegedy} is the following lemma, the proof of which is outlined in Exercise~\ref{ex:schwartz-zippel-like}:
\begin{lemma}                                       \label{lem:schwartz-zippel-like}
    Suppose $\deg(f) \leq k$, where $f \btR$ is not identically~$0$.  Then $\Pr[f(\bx) \neq 0] \geq 2^{-k}$.
\end{lemma}
Since $\deg(\D_i f) \leq k-1$ when $\deg(f) \leq k$ (by the ``differentiation'' formula) and since $\Inf_i[f] = \Pr[\D_i f(\bx) \neq 0$] for Boolean-valued~$f$, we immediately infer:
\begin{proposition}                                     \label{prop:degree-influences}
    If $f \btb$ has $\deg(f) \leq k$ then $\Inf_i[f]$ is either~$0$ or at least $2^{1-k}$ for all $i \in [n]$.
\end{proposition}
We can now give the proof of Theorem~\ref{thm:nisan-szegedy}. From Proposition~\ref{prop:degree-influences} the number of coordinates which have nonzero influence on~$f$ is at most $\Tinf[f]/2^{1-k}$, and this in turn is at most $k2^{k-1}$ by the following fact:
\begin{fact}                                        \label{fact:tinf-at-most-degree}
    For $f \btb$, $\Tinf[f] \leq \deg(f)$.
\end{fact}
\noindent Fact~\ref{fact:tinf-at-most-degree} is immediate from the Fourier formula for total influence. 

We remark that the FKN Theorem (stated in Chapter~\ref{sec:arrow}) is a ``robust'' version of Theorem~\ref{thm:nisan-szegedy} for $k=1$.  
In Chapter~\ref{sec:KKL} we will see Friedgut's Junta Theorem, a related robust result showing that if $\Tinf[f] \leq k$ then~$f$ is $\eps$-close to a $2^{O(k/\eps)}$-junta.

\section{Subspaces and decision trees}              \label{sec:dec-trees}
In this section we will treat the domain of a Boolean function as $\F_2^n$, an $n$-dimensional vector space over the field~ $\F_2$.  As mentioned in Chapter~\ref{sec:fourier-expansion}, it can be natural to index the Fourier characters $\chi_S \ftb$ not by subsets $S \subseteq [n]$ but by their $0$-$1$ indicator vectors $\gamma \in \F_2^n$; thus
\[
    \chi_\gamma(x) = (-1)^{\gamma \cdot x},
\]
with the dot product $\gamma \cdot x$ being carried out in $\F_2^n$.  For example, in this notation we'd write $\chi_0$ for the constantly~$1$ function and $\chi_{e_i}$ for the $i$th dictator.  Fact~\ref{fact:symdiff} now becomes
\begin{equation}        \label{eqn:dual-char}
    \chi_\beta \chi_\gamma = \chi_{\beta + \gamma} \quad \forall \beta, \gamma.
\end{equation}
Thus the characters form a group under multiplication, which is isomorphic to the group $\F_2^n$ under addition.  To distinguish this group from the input domain we write it as $\dualF$; we also tend to identify the character with its index.
                                        \nomenclature[F2nhat]{$\dualF$}{the group (vector space) indexing the Fourier characters of functions $f \ftR$}%
                                        \nomenclature[Zmnhat]{$\wh{\Z_m^n}$}{the group indexing the Fourier characters of functions $f \co \Z_m^n \to \C$}%
Thus the Fourier expansion of $f \ftR$ can be written as
\[
    f(x) = \sum_{\gamma \in \dualF} \wh{f}(\gamma) \chi_\gamma(x).
\]

The Fourier transform of $f$ can be thought of as a function $\wh{f} \co \dualF \to \R$.  We can measure its complexity with various norms.
\begin{definition}
    The \emph{Fourier (or spectral) $p$-norm}
                                            \index{Fourier norm}%
                                            \index{spectral norm|seeonly{Fourier norm}}%
                                            \nomenclature{$\snorm{f}_p$}{$(\sum_{\gamma \in \dualF} \vert \wh{f}(\gamma) \vert^p)^{1/p}$}%
    of $f \btR$ is
    \[
        \snorm{f}_p = \left(\sum_{\gamma \in \dualF} |\wh{f}(\gamma)|^p\right)^{1/p}.
    \]
\end{definition}
Note that we use the ``counting measure'' on $\dualF$, and hence we have a nice rephrasing of Parseval's Theorem: $\|f\|_2 = \snorm{f}_2$.  We make two more definitions relating to the simplicity of $\wh{f}$:
\begin{definition}
    The \emph{Fourier (or spectral) sparsity} of
                                            \index{Fourier sparsity}%
                                            \index{spectral sparsity|seeonly{Fourier sparsity}}%
    $f \btR$
                                            \nomenclature[suppf]{$\supp(f)$}{if $f$ is a function, denotes the set of inputs where $f$ is nonzero}%
                                            \nomenclature[sparsity]{$\sparsity{f}$}{$\Pr_{\bx}[f(\bx) \neq 0]$}%
                                            \nomenclature[sparsitys]{$\ssparsity{f}$}{$\vert \supp(\wh{f}) \vert$}%
    is
    \[
        \ssparsity{f} = |\supp(\wh{f})| = \#\bigl\{\gamma \in \dualF : \wh{f}(\gamma) \neq 0\bigr\}.
    \]
\end{definition}
\begin{definition}
    We say that $\wh{f}$ is \emph{$\eps$-granular} if
                                            \index{granularity, Fourier spectrum}
    $\wh{f}(\gamma)$ is an integer multiple of~$\eps$ for all $\gamma \in \dualF$.
\end{definition}

To gain some practice with this notation, let's look at the Fourier transforms of some indicator functions $1_A \co \F_2^n \to \{0,1\}$ and probability density functions $\vphi_A$, where $A \subseteq \F_2^n$.  First, suppose $A \leq \F_2^n$ is
                                            \index{subspaces}
a \emph{subspace}.  Then one way to characterize $A$ is by its \emph{perpendicular subspace}
                                            \index{perpendicular subspace}%
                                            \index{orthogonal complement|seeonly{perpendicular subspace}}%
                                            \nomenclature[Aperp]{$A^\perp$}{$\{\gamma : \gamma \cdot x = 0 \text{ for all }x \in A\}$}%
$A^\perp$:
\[
    A^\perp = \{\gamma \in \dualF : \gamma \cdot x = 0 \text{ for all } x\in A\}.
\]
It holds that $\dim A^\perp = n - \dim A$ (this is called the \emph{codimension} of~$A$) and that $A = (A^\perp)^\perp$.
\begin{proposition}                                     \label{prop:subspace-transf}
    If $A \leq \F_2^n$ has $\codim A = \dim A^\perp = k$,
                                            \index{codimension}%
                                            \nomenclature[codim]{$\codim H$}{for a subspace $H \leq \F^n$, denotes $n - \dim H$}%
    then
    \[
        1_A = \sum_{\gamma \in A^\perp} 2^{-k} \chi_\gamma, \qquad \vphi_A = \sum_{\gamma \in A^\perp} \chi_\gamma.
    \]
\end{proposition}
\begin{proof}
Let $\gamma_1, \dots, \gamma_k$ form a basis of $A^\perp$.  Since $A = (A^\perp)^\perp$ it follows that $x \in A$ if and only if $\chi_{\gamma_i}(x) = 1$ for all $i \in [k]$.  We therefore have
\[
1_A(x) = \prod_{i=1}^k \Bigl(\half + \half \chi_{\gamma_i}(x)\Bigr) = 2^{-k}\sum_{\gamma \in \spn\{\gamma_1, \dots, \gamma_k\}} \chi_\gamma(x)
\]
as claimed, where the last equality used~\eqref{eqn:dual-char}.  The Fourier expansion of $\vphi_A$ follows because $\E[1_A] = 2^{-k}$.
\end{proof}
More generally, suppose $A$ is \emph{affine subspace} (or \emph{coset})
                                            \index{affine subspace}%
                                            \index{coset|seeonly{affine subspace}}%
of $\F_2^n$; i.e., $A = H + a$ for some $H \leq \F_2^n$ and $a \in \F_2^n$, or equivalently
\[
    A = \{x \in \F_2^n : \gamma \cdot x = \gamma \cdot a \text{ for all } \gamma \in H^\perp\}.
\]
Then it is easy (Exercise~\ref{ex:coset-transf}) to extend Proposition~\ref{prop:subspace-transf} to:
\begin{proposition}                                     \label{prop:coset-transf}
    If $A = H+a$ is an affine subspace of codimension~$k$, then
    \[
        \wh{1_A}(\gamma) =  \begin{cases}
                                \chi_{\gamma}(a) 2^{-k} & \text{if $\gamma \in H^\perp$} \\
                                0 & \text{else;}
                            \end{cases}
    \]
    hence $\vphi_A = \sum_{\gamma \in H^\perp} \chi_{\gamma}(a) \chi_\gamma$.  We have $\ssparsity{1_A} = 2^k$, $\wh{1_A}$ is $2^{-k}$-granular,
                                                \index{granularity, Fourier spectrum}
    $\snorm{1_A}_\infty = 2^{-k}$, and $\snorm{1_A}_1 = 1$.
\end{proposition}

In computer science terminology, any $f \ftzo$ that is a conjunction of parity conditions is the indicator of an affine subspace (or the zero function).  In the simple case that the parity conditions are all of the form ``$x_i = a_i$'', the function is a logical AND of \emph{literals}, and we call the affine subspace a
                                                    \index{subcube}%
\emph{subcube}.

Another class of Boolean functions with simple Fourier spectra are the ones computable by simple \emph{decision trees}:
\begin{definition}
    A \emph{decision tree}
                                                    \index{decision tree}
    $T$ is a representation of a Boolean function $f \ftR$.  It consists of a rooted binary tree in which the internal nodes are labeled by coordinates $i \in [n]$, the outgoing edges of each internal node are labeled~$0$ and~$1$, and the leaves are labeled by real numbers.  We insist that no coordinate $i \in [n]$ appears more than once on any root-to-leaf path.

    On input $x \in \F_2^n$, the tree $T$ constructs a \emph{computation path} from the root node to a leaf.  Specifically, when the computation path reaches an internal node labeled by coordinate $i \in [n]$ we say that $T$ \emph{queries} $x_i$; the computation path then follows the outgoing edge labeled by $x_i$.  The output of $T$ (and hence~$f$) on input~$x$ is the label of the leaf reached by the computation path.  We often identify a tree with the function it computes.
\end{definition}
For decision trees, a picture is worth a thousand words; see Figure~\ref{fig:sort3-dt}.

\myfig{.75}{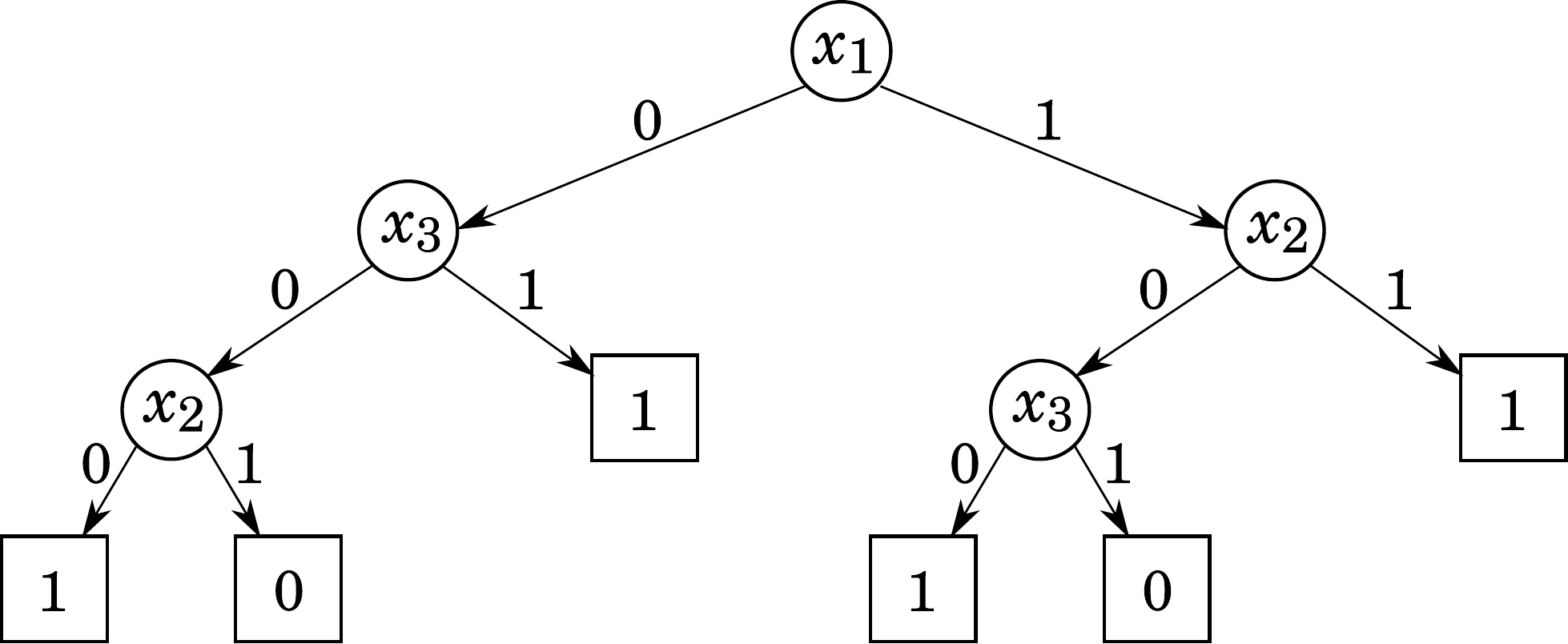}{Decision tree computing $\SORT_3$}{fig:sort3-dt}

(It's traditional to write $x_i$ rather than $i$ for the internal node labels.) For example, the computation path of the above tree on input $x = (0,1,0) \in \F_2^3$ starts at the root, queries $x_1$, proceeds left, queries $x_3$, proceeds left, queries $x_2$, proceeds right, and reaches a leaf labeled~$0$.  In fact, this tree computes the function $\SORT_3$ defined by $\SORT_3(x) = 1$ if and only if $x_1 \leq x_2 \leq x_3$ or $x_1 \geq x_2 \geq x_3$.

\begin{definition}
    The \emph{size} $s$ of a decision tree $T$ is
                                                \index{decision tree!size}%
                                                \index{decision tree!depth}%
    the total number of leaves. The \emph{depth}~$k$ of $T$ is the maximum length of any root-to-leaf path. For decision trees over $\F_2^n$ we have $k \leq n$ and $s \leq 2^k$.  Given $f \ftR$ we write $\DT(f)$
                                                \nomenclature[DT]{$\DT(f)$}{least possible depth of a decision tree computing $f$}
    (respectively, $\DTsize(f)$)
                                                \nomenclature[DTsize]{$\DTsize(f)$}{least possible size of a decision tree computing $f$}
    for the least depth (respectively, size) of a decision tree computing~$f$. (Note that these are not necessarily achieved by the same tree.)
\end{definition}
The example decision tree above has size~$6$ and depth~$3$.

Let $T$ be a decision tree computing $f \co \F_2^n \to \R$ and let $P$ be one of its root-to-leaf paths.  The set of inputs $x$ that follow computation path~$P$ in~$T$ is precisely a subcube of $\F_2^n$, call it $C_P$.  The function $f$ is constant on~$C_P$; we will call its value there~$f(P)$.  Further, since every input $x$ follows a unique path in $T$, the subcubes $\{C_P : P \text{ a path in $T$}\}$ form a \emph{partition} of $\F_2^n$.  These observations yield the following ``spectral simplicity'' results for decision trees:
\begin{fact}                                     \label{fact:dt-spectrum}
    Let $f \ftR$ be computed by a decision tree~$T$. Then
    \[
        f = \sum_{\text{paths $P$ of $T$}} f(P)\cdot 1_{C_P}.
    \]
\end{fact}
\begin{proposition}                                     \label{prop:dt-spectrum}
    Let $f \ftR$ be computed by a decision tree~$T$ of size $s$ and depth $k$.
                                                \index{decision tree!Fourier spectrum}%
    Then:
    \begin{itemize}
        \item $\deg(f) \leq k$;
        \item $\ssparsity{f} \leq s 2^k \leq 4^k$;
        \item $\snorm{f}_1 \leq \|f\|_\infty \cdot s \leq \|f\|_\infty \cdot 2^k$;
        \item $\wh{f}$ is $2^{-k}$-granular
                                                    \index{granularity, Fourier spectrum}
        assuming $f \co \F_2^n \to \Z$.
    \end{itemize}
\end{proposition}
\begin{proposition}                                     \label{prop:dt-trunc-spectrum}
    Let $f \ftb$ be computable by a decision tree of size~$s$ and let $\eps \in (0,1]$.  Then the spectrum of $f$ is $\eps$-concentrated on degree up to $\log(s/\eps)$.
\end{proposition}
You are asked to prove these propositions in Exercises~\ref{ex:dt-spectrum} and~\ref{ex:dt-trunc-spectrum}.
Similar spectral simplicity results hold for some generalizations of the decision tree representation (``subcube partitions'', ``parity decision trees''); see Exercise~\ref{ex:coset-partitions}.

\section{Restrictions}                                 \label{sec:restrictions}

A common operation on Boolean functions $f \btR$ is \emph{restriction}
                                                \index{restriction|(}
to subcubes.  Suppose $[n]$ is partitioned into two sets, $J$ and
                                                \nomenclature[Jbar]{$\ol{J}$}{if $J \subseteq [n]$, denotes $[n] \setminus J$}
$\ol{J} = [n] \setminus J$.  If the inputs bits in $\ol{J}$ are fixed to constants, the result is a function $\bits^J \to \R$.  For example, if we take the function $\Maj_5 \co \bits^5 \to \bits$ and restrict the $4$th and $5$th coordinates to be~$1$ and~$-1$ respectively, we obtain the function $\Maj_3 \co \bits^3 \to \bits$.  If we further restrict the $3$rd coordinate to be $-1$, we obtain the two-bit function which is~$1$ if and only if both input bits are~$1$.

We introduce following notation:
\begin{definition} \label{def:restrict-subcube}
    Let $f \btR$ and let $(J, \ol{J})$ be a partition of $[n]$.  Let $z \in \bits^{\ol{J}}$.  Then we write $\restr{f}{J}{z} \co \bits^J \to \R$ (pronounced ``the restriction of $f$ to $J$ using $z$'') for
                                                \nomenclature[fJz]{$\restr{f}{J}{z}$}{if $f \co \Omega^n \to \R$, $J \subseteq [n]$, and $z \in \Omega^{\ol{J}}$, denotes the restriction of~$f$ given by fixing the coordinates in $\ol{J}$ to $z$}%
                                                \nomenclature[fJJz]{$\restr{f}{}{z}$}{if $f \co \Omega^n \to \R$, $J \subseteq [n]$, and $z \in \Omega^{\ol{J}}$, denotes the restriction of $f$ given by fixing the coordinates in $\ol{J}$ to $z$}%
    the subfunction of $f$ given by fixing the coordinates in $\ol{J}$ to the bit values~$z$.  When the partition $(J, \ol{J})$ is understood we may write simply $\restr{f}{}{z}$.  If $y \in \bits^J$ and $z \in \bits^{\ol{J}}$ we will sometimes write $(y,z)$
                                                \nomenclature[yz]{$(y,z)$}{if $J \subseteq [n]$, $y \in \bits^J$, $z \in \bits^{\ol{J}}$, denotes the natural composite string in $\bits^n$}
    for the composite string in $\bits^n$, even though $y$ and $z$ are not literally concatenated; with this notation, $\restr{f}{J}{z}(y) = f(y,z)$.
\end{definition}

Let's examine how restrictions affect the Fourier transform by considering an example.
\begin{example} \label{ex:restrict} Let $f \co \bits^4 \to \bits$ be the function defined by
\begin{equation} \label{eqn:restr-eg1}
f(x) = 1 \quad \iff \quad x_3 = x_4 = -1 \text{\ \ or\ \ } x_1 \geq x_2 \geq x_3 \geq x_4 \text{\ \ or\ \ } x_1 \leq x_2 \leq x_3 \leq x_4.
\end{equation}
You can check that $f$ has the Fourier expansion
\begin{align}
    f(x) = &+\tfrac18-\tfrac18x_1+\tfrac18x_2-\tfrac18x_3-\tfrac18x_4 \nonumber\\
    &+\tfrac38x_1x_2+\tfrac18x_1x_3-\tfrac38x_1x_4+\tfrac38x_2x_3-\tfrac18x_2x_4+\tfrac58x_3x_4     \label{eqn:restr-eg2}    \\ &+\tfrac18x_1x_2x_3+\tfrac18x_1x_2x_4-\tfrac18x_1x_3x_4+\tfrac18x_2x_3x_4-\tfrac18x_1x_2x_3x_4. \nonumber
\end{align}
Consider the restriction $x_3 = 1$, $x_4 = -1$, and let $f' = \restr{f}{\{1,2\}}{(1,-1)}$ be the restricted function of~$x_1$ and~$x_2$. From the original definition~\eqref{eqn:restr-eg1} of~$f$ we see that $f'(x_1,x_2)$ is~$1$ if and only if $x_1 = x_2 = 1$. This is the $\mint$ function of~$x_1$ and~$x_2$, which we know has Fourier expansion
\begin{equation} \label{eqn:restr-eg3}
    f'(x_1,x_2) = \mint(x_1,x_2) = -\tfrac12 + \tfrac12 x_1 + \tfrac12 x_2 + \tfrac12x_1x_2.
\end{equation}
We can of course obtain this expansion simply by plugging $x_3 = 1, x_4 = -1$ into~\eqref{eqn:restr-eg2}. Now suppose we only wanted to know the coefficient on~$x_1$ in the Fourier expansion of~$f'$.  We can find it as follows: Consider all monomials in~\eqref{eqn:restr-eg2} that contain $x_1$ and possibly also $x_3$, $x_4$; substitute $x_3 = 1$, $x_4 = -1$ into the associated terms; and sum the results.  The relevant terms in~\eqref{eqn:restr-eg2} are $-\tfrac18x_1$, $+\tfrac18x_1x_3$, $-\tfrac38x_1x_4$, $-\tfrac18x_1x_3x_4$, and substituting in $x_3 = 1, x_4 = -1$ gives us $-\tfrac18 +\tfrac18 +\tfrac38 +\tfrac18 = \tfrac12$, as expected from~\eqref{eqn:restr-eg3}.
\end{example}

Now we work out these ideas more generally.  In the setting of Definition~\ref{def:restrict-subcube} the restricted function $\restr{f}{J}{z}$ has $\bits^J$ as its domain.  Thus its Fourier coefficients are indexed by subsets of~$J$.  Let's introduce notation for the Fourier coefficients of a restricted function:
\begin{definition} \label{def:frestr-subcube}
    Let $f \btR$ and let $(J, \ol{J})$ be a partition of $[n]$.  Let $S \subseteq J$.  Then we write
                                                \nomenclature[FSJf]{$\frestr{S}{\ol{J}}{f}(z)$}{for $S \subseteq J \subseteq [n]$, denotes $\wh{\restr{f}{J}{z}}(S)$}%
    $\frestr{S}{\ol{J}}{f} \co \bits^{\ol{J}} \to \R$ for the
                                                    \index{restriction!Fourier}
    function $\wh{\restr{f}{J}{\bullet}}(S)$; i.e.,
    \[
        \frestr{S}{\ol{J}}{f}(z) = \wh{\restr{f}{J}{z}}(S).
    \]
    When the partition $(J, \ol{J})$ is understood we may write simply $\frestr{S}{}{f}$.
\end{definition}

In Example~\ref{ex:restrict} we considered $\ol{J} = \{3,4\}$, $S = \{1\}$, and $z = (1,-1)$.  See Figure~\ref{fig:restriction} for an illustration of a typical restriction scenario.

\myfig{.75}{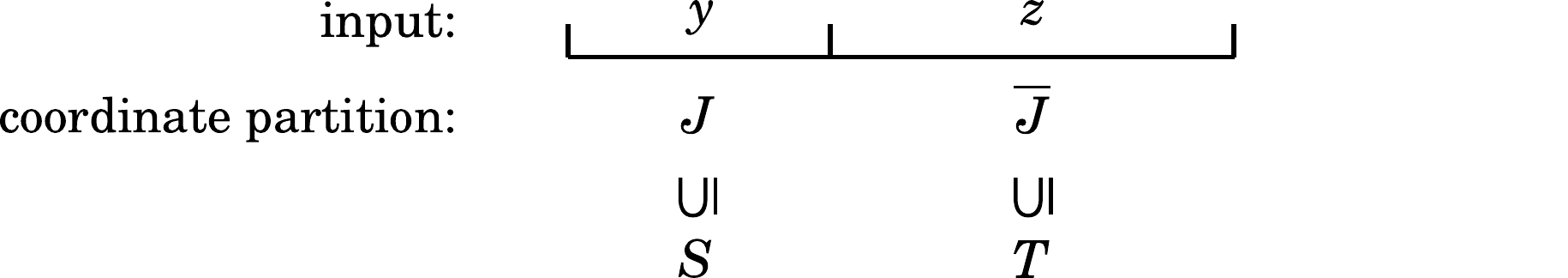}{Notation for a typical restriction scenario.  Note that $J$ and $\ol{J}$ need not be literally contiguous.}{fig:restriction}

In general, for a fixed partition $(J, \ol{J})$ of~$[n]$ and a fixed  $S \subseteq J$, we may wish to know what $\wh{\restr{f}{J}{z}}(S)$ is as a function of~$z \in \bits^{\ol{J}}$.  This is precisely asking for the Fourier transform of~$\frestr{S}{\ol{J}}{f}$.  Since the function $\frestr{S}{\ol{J}}{f}$ has domain $\bits^{\ol{J}}$, its Fourier transform has coefficients indexed by subsets of~$\ol{J}$.  The formula for this Fourier transform generalizes the computation we used at the end of Example~\ref{ex:restrict}:
\begin{proposition}                                     \label{prop:frestr-subcube-formula}
    In the setting of Definition~\ref{def:frestr-subcube} we have the Fourier expansion
    \[
        \frestr{S}{\ol{J}}{f}(z) = \sum_{T \subseteq \ol{J}} \wh{f}(S \cup T) z^T;
    \]
    i.e.,
    \[
        \wh{\frestr{S}{\ol{J}}{f}}(T) = \wh{f}(S \cup T).
    \]
\end{proposition}
\begin{proof}
    (The $S = \emptyset$  case here is Exercise~\ref{ex:restrict-mean}.)     Every $U \subseteq [n]$ indexing $f$'s Fourier coefficients can be written as a disjoint union $U = S \cup T$, where $S \subseteq J$ and $T \subseteq \ol{J}$. We can also decompose any $x \in \bits^n$ into two substrings $y \in \bits^J$ and $z \in \bits^{\ol{J}}$.  We have $x^U = y^Sz^T$ and so
    \[
        f(x) = \sum_{U \subseteq [n]} \wh{f}(U)\,x^U = \sum_{\substack{S \subseteq J \\ T \subseteq \ol{J}}} \wh{f}(S \cup T)\,y^Sz^T
        = \sum_{S \subseteq J\vphantom{\ol{J}}} \Bigl( \sum_{T \subseteq \overline{J}} \wh{f}(S \cup T)\,z^T\Bigr) y^S.
    \]
    Thus when $z$ is fixed, the resulting function of $y$ indeed has $\sum_{T \subseteq \overline{J}} \wh{f}(S \cup T)\,z^T$ as its Fourier coefficient on the monomial~$y^S$.
\end{proof}

\begin{corollary}                   \label{cor:expected-restrict-coeffs}
    Let $f \btR$, let $(J, \ol{J})$ be a partition of $[n]$, and fix $S \subseteq J$.  Suppose $\bz \sim \bits^{\ol{J}}$ is chosen uniformly at random.  Then
    \begin{align*}
        \E_{\bz}[\wh{\restr{f}{J}{\bz}}(S)] &= \wh{f}(S),\\
        \E_{\bz}[\wh{\restr{f}{J}{\bz}}(S)^2] &= \sum_{T \subseteq \ol{J}} \wh{f}(S \cup T)^2.
    \end{align*}
\end{corollary}
\begin{proof}
    The first statement is immediate from Proposition~\ref{prop:frestr-subcube-formula}, taking $T = \emptyset$ and unraveling the definition.  As for the second statement,
    \begin{align*}
        \E_{\bz}[\wh{\restr{f}{J}{\bz}}(S)^2] &= \E_{\bz}[\frestr{S}{\ol{J}}{f}(\bz)^2]     \tag{by definition} \\
                                        &= \sum_{T \subseteq \ol{J}} \wh{\frestr{S}{\ol{J}}{f}}(T)^2 \tag{Parseval}\\
                                        &= \sum_{T \subseteq \ol{J}} \wh{f}(S \cup T)^2 \tag*{(Proposition~\ref{prop:frestr-subcube-formula}) \qedhere}
    \end{align*}
\end{proof}
                                                \index{restriction|)}%
                                                \index{restriction!to subspaces|(}
We move on to discussing a more general kind of restriction; namely, restricting a function $f \ftR$ to an affine subspace $H + z$.  This generalizes  restriction to subcubes as we've seen so far, by considering $H = \spn\{e_i : i \in J\}$ for a given subset $J \subseteq [n]$.  For restrictions to a subspace $H \leq \F_2^n$ we have a natural definition:
\begin{definition} \label{def:restrict-subspace}
    If $f \ftR$ and $H \leq \F_2^n$ is a subspace,
                                            \nomenclature[fH]{$f_H$}{if $f \ftR$, $H \leq \F_2^n$, denotes the restriction of $f$ to $H$}%
    we write $f_H \co H \to \R$ for the restriction of~$f$ to~$H$.
\end{definition}
For restrictions to \emph{affine} subspaces, we run into difficulties if we try to extend our notation for restrictions to subcubes.  Unlike in the subcube case of $H = \spn\{e_i : i \in J\}$, we don't in general have a canonical isomorphism between~$H$ and a coset~$H+z$. Thus it's not natural to introduce notation such as $\restr{f}{H}{z} \co H \to \R$ for the function $h \mapsto f(h+z)$, because such a definition depends on the choice of representative for~$H+z$.  As an example consider $H = \{(0,0), (1,1)\} \leq \F_2^2$, a $1$-dimensional subspace (which satisfies $H^\perp = H$).  Here the nontrivial coset is $H + (1,0) = H + (0,1) = \{(1,0), (0,1)\}$, which has no canonical representative.

To get around this difficulty we can view restriction to a coset $H+z$ as consisting of two steps: first, translation of the domain by a fixed representative~$z$, and then restriction to the subspace~$H$.  Let's introduce some notation for the first operation:
\begin{definition} \label{def:translate-domain}
    Let $f \ftR$ and let $z \in \F_2^n$.  We define the function
                                                \nomenclature[f+z]{$f^{+z}$}{if $f \ftR$, $z \in \F_2^n$, denotes the function $f^{+z}(x) = f(x+z)$}%
    $f^{+z} \ftR$ by $f^{+z}(x) = f(x+z)$.
\end{definition}
By substituting $x = x+z$ into the Fourier expansion of~$f$, we deduce:
\begin{fact} \label{fact:translation-expansion} The Fourier coefficients of $f^{+z}$ are given by $\wh{f^{+z}}(\gamma) = (-1)^{\gamma \cdot z} \wh{f}(\gamma)$; i.e.,
    \[
        f^{+z}(x) = \sum_{\gamma \in \dualF} \chi_{\gamma}(z)\wh{f}(\gamma)\,\chi_\gamma(x).
    \]
\end{fact}
\noindent (This fact also follows by noting that $f^{+z} = \vphi_{\{z\}} \conv f$; see Exercise~\ref{ex:conv-trans}.)

We can now give notation for the restriction of a function to an affine subspace:
\begin{definition}
    Let $f \ftR$, $z \in \F_2^n$, $H \leq \F_2^n$.  We write $f^{+z}_H \co H \to \R$ for the function $(f^{+z})_H$;
                                    \nomenclature[f+zH]{$f^{+z}_H$}{denotes $(f^{+z})_H$}%
    namely, the restriction of~$f$ to coset $H+z$ with the representative~$z$ made explicit.
\end{definition}

Finally, we would like to consider Fourier coefficients of restricted functions $f^{+z}_H$.  These can be indexed by the cosets of $H^\perp$ in~$\dualF$.  However, we again have a notational difficulty since the only coset with a canonical representative  is $H^\perp$ itself, with representative~$0$.  There is no need to introduce extra notation for $\wh{f^{+z}_H}(0)$, the average value of $f$ on coset $H+z$, since it is just
\[
    \Ex_{\bh \sim H}[f(\bh+z)] = \la \vphi_H, f^{+z} \ra.
\]
Applying Plancherel on the right-hand side, as well as Proposition~\ref{prop:subspace-transf} and Fact~\ref{fact:translation-expansion}, we deduce the following classical
                            \index{Poisson summation formula}
fact:
\begin{named}{Poisson Summation Formula}
    Let $f \ftR$, $H \leq \F_2^n$, $z \in \F_2^n$.  Then
    \[
        \Ex_{\bh \sim H}[f(\bh + z)] = \sum_{\gamma \in H^\perp} \chi_\gamma(z) \wh{f}(\gamma).
    \]
\end{named}
                                                \index{restriction!to subspaces|)}

\section{Learning theory}                                          \label{sec:learning}

\emph{Computational learning theory}
                                                \index{learning theory|(}
is an area of algorithms research devoted to the following task: Given a source of ``examples'' $(x, f(x))$ from an unknown function~$f$, compute a ``hypothesis'' function~$h$ that is good at predicting $f(y)$ on future inputs~$y$.  In this book we will focus on just one possible formulation of the task:
\begin{definition}
    In the model of \emph{PAC (``Probably Approximately Correct'') learning under the uniform distribution on $\bits^n$},
                                                    \index{PAC learning|seeonly{learning theory}}
    a learning problem is identified with a \emph{concept class}~$\calC$, which is just a collection of functions~$f \btb$.  A \emph{learning algorithm}~$A$ for $\calC$ is a randomized algorithm which has limited access to an unknown \emph{target function} $f \in \calC$. The two access models, in increasing order of strength, are:
    \begin{itemize}
        \item \emph{random examples}, meaning $A$ can draw pairs $(\bx, f(\bx))$ where $\bx \in \bits^n$ is uniformly random;
        \item \emph{queries}, meaning $A$ can request the value $f(x)$ for any $x \in \bits^n$ of its choice.
    \end{itemize}
    In addition, $A$ is given as input an \emph{accuracy parameter}~$\eps \in [0,1/2]$.  The output of~$A$ is required to be (the circuit representation of) a \emph{hypothesis} function $h \btb$.  We say that $A$ \emph{learns $\calC$ with error~$\eps$} if for any $f \in \calC$, with high probability $A$ outputs an $h$ which is $\eps$-close to $f$: i.e., satisfies $\dist(f,h) \leq \eps$.
\end{definition}
In the above definition, the phrase ``with high probability'' can be fixed to mean, say, ``except with probability at most~$1/10$''.  (As is common with randomized algorithms, the choice of constant $1/10$ is unimportant; see Exercise~\ref{ex:learning-delta}.)

For us, the main desideratum of a learning algorithm is efficient \emph{running time}.  One can easily learn \emph{any} function $f$ to error~$0$ in time $\wt{O}(2^n)$ (see Exercise~\ref{ex:trivial-learning}); however, this is not very efficient.  If the concept class $\calC$ contains very complex functions, then such exponential running time is necessary; however, if $\calC$ contains only relatively ``simple'' functions, then more efficient learning may be possible.  For example, the results of Section~\ref{sec:goldreich-levin} show that the concept class
\[
    \calC = \{f \ftb \mid \DTsize(f) \leq s\}
\]
can be learned with queries to error~$\eps$ by an algorithm whose running time is $\poly(s, n, 1/\eps)$.

A common way of trying to learn an unknown target $f \btb$ is by discovering ``most of'' its Fourier spectrum.  To formalize this, let's generalize Definition~\ref{def:conc-degree}:
\begin{definition}                                          \label{def:concentration}
    Let $\calF$ be a collection of subsets $S \subseteq [n]$.  We say that the Fourier spectrum of $f \btR$ is \emph{$\eps$-concentrated on $\calF$}
                                                \index{concentration, spectral}%
    if
    \[
        \sum_{\substack{S \subseteq [n] \\ S \notin \calF}} \wh{f}(S)^2 \leq \eps.
    \]
    For $f \btb$ we can express this condition using the spectral sample: $\Pr_{\bS \sim \specsamp{f}}[\bS \notin \calF] \leq \eps$.
\end{definition}

Most functions don't have their Fourier spectrum concentrated on a small collection (see Exercise~\ref{ex:no-conc}).  But for those that do, we may hope to discover ``most of'' their Fourier coefficients.  The main result of this section is a kind of ``meta-algorithm'' for learning an unknown target~$f$.  It reduces the problem of learning $f$ to the problem of identifying a collection of characters on which $f$'s Fourier spectrum is concentrated.
\begin{theorem}                                     \label{thm:meta-learning}
    Assume learning algorithm $A$ has (at least) random example access to target $f \btb$.  Suppose that $A$ can -- somehow -- identify a collection $\calF$ of subsets on which $f$'s Fourier spectrum is $\eps/2$-concentrated.  Then using $\poly(|\calF|, n, 1/\eps)$ additional time, $A$ can with high probability output a hypothesis~$h$ that is $\eps$-close to $f$.
\end{theorem}

The idea of the theorem is that~$A$ will estimate all of $f$'s Fourier coefficients in~$\calF$, obtaining a good approximation to $f$'s Fourier expansion.  Then $A$'s hypothesis will be the \emph{sign} of this approximate Fourier expansion.

The first tool we need to prove Theorem~\ref{thm:meta-learning} is the ability to accurately estimate any fixed Fourier coefficient:
\begin{proposition}                                     \label{prop:learn-one-coeff}
    Given access to random examples from $f \btb$, there is a randomized algorithm which takes as input $S \subseteq [n]$, $0 < \delta, \eps \leq 1/2$, and outputs an estimate $\wt{f}(S)$ for $\wh{f}(S)$ that satisfies
    \[
        |\wt{f}(S) - \wh{f}(S)| \leq \eps
    \]
    except with probability at most $\delta$.  The running time is $\poly(n, 1/\eps) \cdot \log(1/\delta)$.
\end{proposition}
\begin{proof}
    We have $\wh{f}(S) = \Ex_{\bx}[f(\bx)\chi_S(\bx)]$.  Given random examples $(\bx, f(\bx))$, the algorithm can compute $f(\bx) \chi_S(\bx) \in \bits$ and therefore empirically estimate $\Ex_{\bx}[f(\bx)\chi_S(\bx)]$.  A standard application of the Chernoff bound implies that $O(\log(1/\delta)/\eps^2)$ examples are sufficient to obtain an estimate within~$\pm \eps$ with probability at least $1-\delta$.
\end{proof}

The second observation we need to prove Theorem~\ref{thm:meta-learning} is the following:
\begin{proposition}                                     \label{prop:sign-conc}
    Suppose that $f \btb$ and $g \btR$ satisfy $\|f - g\|_2^2 \leq \eps$.  Let $h \btb$ be defined by $h(x) = \sgn(g(x))$, with
                                        \nomenclature[sgn]{$\sgn(t)$}{$+1$ if $t \geq 0$, $-1$ if $t < 0$}
    $\sgn(0)$ chosen arbitrarily from $\bits$.  Then $\dist(f,h) \leq \eps$.
\end{proposition}
\begin{proof}
    Since $|f(x) - g(x)|^2 \geq 1$ whenever $f(x) \neq \sgn(g(x))$, we conclude
    \[
        \dist(f,h) = \Pr_{\bx}[f(\bx) \neq h(\bx)] = \Ex_{\bx}[\bone_{f(\bx) \neq \sgn(g(\bx))}] \leq \Ex_{\bx}[|f(\bx) - g(\bx)|^2] = \|f - g\|_2^2. \qedhere
    \]
\end{proof}
\noindent (See Exercise~\ref{ex:avrim-trick} for an improvement to this argument.)

We can now prove Theorem~\ref{thm:meta-learning}:
\begin{proof}[Proof of Theorem~\ref{thm:meta-learning}]
    For each $S \in \calF$ the algorithm uses Proposition~\ref{prop:learn-one-coeff} to produce an estimate $\wt{f}(S)$ for $\wh{f}(S)$ which satisfies $|\wt{f}(S) - \wh{f}(S)| \leq \sqrt{\eps}/(2\sqrt{|\calF|})$ except with probability at most $1/(10 |\calF|)$.  Overall this requires $\poly(|\calF|, n, 1/\eps)$ time, and by the union bound, except with probability at most $1/10$ all $|\calF|$ estimates have the desired accuracy.  Finally, $A$ forms the real-valued function $g = \sum_{S \in \calF} \wt{f}(S) \chi_S$ and outputs hypothesis $h = \sgn(g)$.  By Proposition~\ref{prop:sign-conc}, it suffices to show that $\|f - g\|_2^2 \leq \eps$.  And indeed,
    \begin{align*}
        \|f - g\|_2^2 &= \sum_{S \subseteq [n]} \wh{f - g}(S)^2 \tag{Parseval} \\
        &= \sum_{S \in \calF} (\wh{f}(S) - \wt{f}(S))^2 + \sum_{S \notin \calF} \wh{f}(S)^2 \\
        &\leq \sum_{S \in \calF} \left(\frac{\sqrt{\eps}}{2\sqrt{|\calF|}}\right)^2 + \eps/2 \tag{estimates, concentration assumption} \\
        &= \eps/4 + \eps/2 \quad\leq\quad \eps,
    \end{align*}
    as desired.
\end{proof}

As we described, Theorem~\ref{thm:meta-learning} reduces the algorithmic task of learning~$f$ to the algorithmic task of identifying a collection~$\calF$ on which $f$'s Fourier spectrum is concentrated.  In Section~\ref{sec:goldreich-levin} we will describe the Goldreich--Levin algorithm, a sophisticated way to find such an~$\calF$ assuming query access to~$f$.  For now, though, we observe that for several interesting concept classes we don't need to do any algorithmic searching for~$\calF$; we can just take $\calF$ to be all sets of small cardinality. This works whenever all functions in $\calC$ have low-degree spectral concentration.

\begin{named}{The ``Low-Degree Algorithm''} Let $k \geq 1$
                                                \index{Low-Degree Algorithm}
and let $\calC$ be a concept class for which every function $f \btb$ in $\calC$ is $\eps/2$-concentrated up to degree~$k$. Then $\calC$ can be learned from random examples only with error~$\eps$ in time $\poly(n^k, 1/\eps)$.
\end{named}
\begin{proof}
    Apply Theorem~\ref{thm:meta-learning} with $\calF = \{S \subseteq [n] : |S| \leq k\}$.  We have $|\calF| = \sum_{j=0}^k \binom{n}{j} \leq O(n^k)$.
\end{proof}

The Low-Degree Algorithm reduces the \emph{algorithmic} problem of learning~$\calC$ from random examples to the \emph{analytic} task of showing low-degree spectral concentration for the functions in~$\calC$.  Using the results of Section~\ref{sec:low-degree} we can quickly obtain some learning-theoretic results.  For example:
\begin{corollary}                                       \label{cor:learn-tinf}
    For $t \geq 1$, let $\calC = \{f \btb \mid \Tinf[f] \leq t\}$.    Then $\calC$ is learnable from random examples with error~$\eps$ in time $n^{O(t/\eps)}$.
\end{corollary}
\begin{proof}
    Use the Low-Degree Algorithm with $k = 2t/\eps$; the result follows from Proposition~\ref{prop:tinf-concentration}.
\end{proof}
\begin{corollary}                                       \label{cor:learn-mono}
    Let $\calC = \{f \btb \mid f \text{ is monotone}\}$.
                                                    \index{monotone function!learning}%
    Then $\calC$ is learnable from random examples with error~$\eps$ in time $n^{O(\sqrt{n}/\eps)}$.
\end{corollary}
\begin{proof}
    Follows from the previous corollary and Theorem~\ref{thm:maj-maximizes-deg-1-sum}.
\end{proof}
You might be concerned that a running time such as $n^{O(\sqrt{n})}$ does not seem very efficient.  Still, it's much better than the trivial running time of $\wt{O}(2^n)$.  Further, as we will see in the next section, learning algorithms are sometimes used in attacks on cryptographic schemes, and in this context even subexponential-time algorithms are considered dangerous.  

Continuing with applications of the Low-Degree Algorithm:
\begin{corollary}                                       \label{cor:learn-ns}
    For $\delta \in (0,1/2]$, let $\calC = \{f \btb \mid \NS_\delta[f] \leq \eps/6\}$.    Then $\calC$ is learnable from random examples with error~$\eps$ in time $\poly(n^{1/\delta}, 1/\eps)$.
\end{corollary}
\begin{proof}
    Follows from Proposition~\ref{prop:ns-concentration}.
\end{proof}
\begin{corollary}                                       \label{cor:learn-dt}
    Let $\calC = \{f \btb \mid \DTsize(f) \leq s\}$. Then $\calC$ is learnable from random examples with error~$\eps$ in time $n^{O(\log(s/\eps))}$.
\end{corollary}
\begin{proof}
    Follows from Proposition~\ref{prop:dt-trunc-spectrum}.
\end{proof}

With a slight extra twist one can also \emph{exactly} learn the class of degree-$k$ functions in time~$\poly(n^k)$; see Exercise~\ref{ex:learn-low-degree}:
\begin{theorem}                                       \label{thm:learn-low-degree}
    Let $k \geq 1$ and let $\calC = \{f \btb \mid \deg(f) \leq k\}$ (e.g.,~$\calC$ contains all depth-$k$ decision trees).
                                                    \index{decision tree!learning}%
    Then $\calC$ is learnable from random examples with error~$0$ in time $n^{k} \cdot \poly(n, 2^k)$.
\end{theorem}
                                                \index{learning theory|)}

\section{Highlight: the Goldreich--Levin Algorithm}                 \label{sec:goldreich-levin}
                                                \index{Kushilevitz--Mansour Algorithm|seeonly{Goldreich--Levin Algorithm}}

We close this chapter by briefly describing a topic which is in some sense the ``opposite'' of learning theory: \emph{cryptography}.
                                                \index{cryptography}
At the highest level, cryptography is concerned with constructing functions which are computationally easy to compute but computationally difficult to invert.  Intuitively, think about the task of encrypting secret messages: You would like a scheme where it's easy to take any message~$x$ and produce an encrypted version $e(x)$, but where it's hard for an adversary to compute $x$ given $e(x)$.  Indeed, even with examples $e(x^{(1)})$, \dots, $e(x^{(m)})$ of several encryptions, it should be hard for an adversary to learn anything about the encrypted messages, or to predict (``forge'') the encryption of future messages.

                                                \index{Goldreich--Levin Algorithm|(}
A basic task in cryptography is building stronger cryptographic functions from weaker ones.  Often the first example in ``Cryptography~$101$'' is the \emph{Goldreich--Levin Theorem}, which is used to build a ``pseudorandom generator'' from a ``one-way permutation''.  We sketch the meaning of these terms and the analysis of the construction in Exercise~\ref{ex:GL}; for now, suffice it to say that the key to the analysis of Goldreich and Levin's construction is a \emph{learning algorithm}.  Specifically, the Goldreich--Levin learning algorithm solves the following problem: Given \emph{query} access to a target function $f \co \F_2^n \to \F_2$, find all of the linear functions (in the sense of Chapter~\ref{sec:BLR}) with which~$f$ is at least slightly correlated.  Equivalently, find all of the noticeably large Fourier coefficients of~$f$.
\begin{named}{Goldreich--Levin Theorem}
    Given query access to a target $f \btb$ as well as input $0 < \tau \leq 1$, there is a $\poly(n, 1/\tau)$-time algorithm that with high probability outputs a list $L = \{U_1, \dots, U_\ell\}$ of subsets of $[n]$ such that:
    \begin{itemize}
        \item $|\wh{f}(U)| \geq \tau\ \ \implies\ \ U \in L$;
        \item $U \in L\ \ \implies\ \ |\wh{f}(U)| \geq \tau/2$.
    \end{itemize}
    (By Parseval's Theorem, the second guarantee implies that $|L| \leq 4/\tau^2$.)
\end{named}

Although the Goldreich--Levin Theorem was originally developed for cryptography, it was soon put to use for learning theory.  Recall that the ``meta-algorithm'' of Theorem~\ref{thm:meta-learning} reduces learning an unknown target $f \btb$ to identifying a collection~$\calF$ of sets on which $f$'s Fourier spectrum is $\eps/2$-concentrated.  Using the Goldreich--Levin Algorithm, a learner with query access to~$f$ can ``collect up'' its largest Fourier coefficients until only~$\eps/2$ Fourier weight remains unfound.  This strategy straightforwardly yields the following result (see Exercise~\ref{ex:deduce-KM}):
\begin{theorem}                                     \label{thm:KM}
    Let $\calC$ be a concept class such that every $f \btb$ in $\calC$ has its Fourier spectrum $\eps/4$-concentrated on a collection of at most~$M$ sets.  Then $\calC$ can be learned using queries with error~$\eps$ in time $\poly(M, n, 1/\eps)$.
\end{theorem}
The algorithm of Theorem~\ref{thm:KM} is often called the \emph{Kushilevitz--Mansour Algorithm}.  Much like the Low-Degree Algorithm, it reduces the computational problem of learning $\calC$ (using queries) to the analytic problem of proving that the functions in $\calC$ have concentrated Fourier spectra.  The advantage of the Kushilevitz--Mansour Algorithm is that it works so long as the Fourier spectrum of~$f$ is concentrated on \emph{some} small collection of sets; the Low-Degree Algorithm requires that the concentration specifically be on the low-degree characters.  The disadvantage of the Kushilevitz--Mansour Algorithm is that it requires query access to~$f$, rather than just random examples. An example concept class for which the Kushilevitz--Mansour Algorithm works well is the set of all~$f$ for which $\snorm{f}_1$ is not too large:
\begin{theorem}                                     \label{thm:learn-low-l1}
                                                    \index{Fourier norm!$1$-}
    Let $\calC = \{f \btb \mid \snorm{f}_1 \leq s\}$ (e.g., $\calC$ contains any~$f$ computable by a decision tree of size at most~$s$). Then $\calC$ is learnable from queries with error~$\eps$ in time $\poly(n, s, 1/\eps)$.
\end{theorem}
This is proved in Exercise~\ref{ex:learn-low-l1}.

Let's now return to the Goldreich--Levin Algorithm itself, which seeks the Fourier coefficients $\wh{f}(U)$ with magnitude at least~$\tau$. Given any candidate~$U \subseteq [n]$, Proposition~\ref{prop:learn-one-coeff} lets us easily distinguish whether the associated coefficient is large, $|\wh{f}(U)| \geq \tau$, or small, $|\wh{f}(U)| \leq \tau/2$.  The trouble is that there are $2^n$ potential candidates.  The Goldreich--Levin Algorithm overcomes this difficulty using a divide-and-conquer strategy that measures the Fourier weight of $f$ on various collections of sets.  
Let's make a definition:
\begin{definition}
    Let $f \btR$ and $S \subseteq J \subseteq [n]$.  We write
    \[
        \W{S\mid\overline{J}}[f] = \sum_{T \subseteq \ol{J}} \wh{f}(S \cup T)^2
    \]
    for the Fourier weight of $f$ on sets whose restriction to $J$ is $S$.
\end{definition}
The crucial tool for the Goldreich--Levin Algorithm is Corollary~\ref{cor:expected-restrict-coeffs}, which says that \begin{equation} \label{eqn:GL-key}
    \W{S\mid\overline{J}}[f] = \Ex_{\bz \sim \bits^{\ol{J}}}[\wh{\restr{f}{J}{\bz}}(S)^2].
\end{equation}
This identity lets a learning algorithm with query access to~$f$ efficiently estimate any $\W{S\mid\overline{J}}[f]$ of its choosing. Intuitively, query access to $f$ allows query access to $\restr{f}{J}{z}$ for any $z \in \bits^{\ol{J}}$; with this one can estimate any $\wh{\restr{f}{J}{z}}(S)$ and hence~\eqref{eqn:GL-key}.  More precisely:
\begin{proposition}                                       \label{prop:GL-substep}
    For any $S \subseteq J \subseteq [n]$ an algorithm with query access to $f \btb$ can compute an estimate of $\W{S\mid\ol{J}}[f]$ that is accurate to within $\pm \eps$ (except with probability at most~$\delta$) in time $\poly(n, 1/\eps) \cdot \log(1/\delta)$.
\end{proposition}
\begin{proof}
    From~\eqref{eqn:GL-key},
    \begin{align*}
        \W{S\mid\overline{J}}[f] = \Ex_{\bz \sim \bits^{\ol{J}}}[\wh{\restr{f}{J}{\bz}}(S)^2]  &=  \Ex_{\bz \sim \bits^{\ol{J}}}\left[\Ex_{\by \sim \bits^J\vphantom{\bits^{\ol{J}}}}[f(\by,\bz)\chi_S(\by)]^2\right] \\
        &= \Ex_{\bz \sim \bits^{\ol{J}}}\ \Ex_{\by, \by' \sim \bits^J\vphantom{\bits^{\ol{J}}}}[f(\by, \bz)\chi_S(\by) \cdot f(\by', \bz) \chi_S(\by')],
    \end{align*}
    where $\by$, $\by'$ are independent.  As in Proposition~\ref{prop:learn-one-coeff}, $f(\by, \bz)\chi_S(\by) \cdot f(\by', \bz) \chi_S(\by')$ is a $\pm 1$-valued random variable that the algorithm can sample from using queries to~$f$.  A Chernoff bound implies that $O(\log(1/\delta)/\eps^2)$ samples are sufficient to estimate its mean with accuracy $\eps$ and confidence $1-\delta$.
\end{proof}

We're now ready to prove the Goldreich--Levin Theorem.
\begin{proof}[Proof of the Goldreich--Levin Theorem] We begin with an overview of how the algorithm works.  Initially, all $2^n$ sets $U$ are (implicitly) put in a single ``bucket''.  The algorithm then repeats the following loop:
\begin{itemize}
    \item Select any bucket $\calB$ containing $2^m$ sets, $m \geq 1$.
    \item Split it into two buckets $\calB_1$, $\calB_2$ of $2^{m-1}$ sets each.
    \item ``Weigh'' each $\calB_i$, $i = 1, 2$; i.e., estimate $\sum_{U \in \calB_i} \wh{f}(U)^2$.
    \item Discard $\calB_1$ or $\calB_2$ if its weight estimate is at most $\tau^2/2$.
\end{itemize}
The algorithm stops once all buckets contain just~$1$ set; it then outputs the list of these sets.

We now fill in the details.  First we argue the correctness of the algorithm, assuming all weight estimates are accurate (this assumption is removed later).  On one hand, any set $U$ with $|\wh{f}(U)| \geq \tau$ will never be discarded, since it always contributes weight at least $\tau^2 \geq \tau^2/2$ to the bucket it's in.  On the other hand, no set $U$ with $|\wh{f}(U)| \leq \tau/2$ can end up in a singleton bucket because such a bucket, when created, would have weight only $\tau^2/4 \leq \tau^2/2$ and thus be discarded.  Notice that this correctness proof does not rely on the weight estimates being exact; it suffices for them to be accurate to within~$\pm \tau^2/4$.

The next detail concerns running time.  Note that any ``active'' (undiscarded) bucket has weight at least $\tau^2/4$, even assuming the weight estimates are only accurate to within $\pm \tau^2/4$. Therefore Parseval tells us there can only ever be at most~$4/\tau^2$ active buckets.  Since a bucket can be split only~$n$ times, it follows that the algorithm repeats its main loop at most $4n/\tau^2$ times.  Thus as long as the buckets can be maintained and accurately weighed in $\poly(n, 1/\tau)$ time, the overall running time will be $\poly(n, 1/\tau)$ as claimed.

Finally, we describe the bucketing system.  The buckets are indexed (and thus maintained implicitly) by an integer $0 \leq k \leq n$ and a subset $S \subseteq [k]$.  The bucket $\calB_{k,S}$ is defined by
\[
    \calB_{k,S} = \Bigl\{S \cup T : T \subseteq \{k+1, k+2, \dots, n\}\Bigr\}.
\]
Note that $|\calB_{k,S}| = 2^{n-k}$.  The initial bucket is $\calB_{0, \emptyset}$. The algorithm always splits a bucket $\calB_{k,S}$ into the two buckets $\calB_{k+1,S}$ and $\calB_{k+1,S \cup \{k+1\}}$.  The final singleton buckets are of the form $\calB_{n,S} = \{S\}$.  Finally, the weight of bucket~$\calB_{k,S}$ is precisely $\W{S\mid\{k+1, \dots, n\}}[f]$. Thus it can be estimated to accuracy $\pm \tau^2/4$ with confidence $1 - \delta$ in time $\poly(n, 1/\tau)\cdot \log(1/\delta)$ using Proposition~\ref{prop:GL-substep}.  Since the main loop is executed at most $4n/\tau^2$ times, the algorithm overall needs to make at most $8n/\tau^2$ weighings; by setting $\delta = \tau^2/(80n)$ we ensure that \emph{all} weighings are accurate with high probability (at least $9/10$).  The overall running time is therefore indeed $\poly(n, 1/\tau)$.
\end{proof}
                             \index{Goldreich--Levin Algorithm|)}

\section{Exercises and notes}
\begin{exercises}
    \item \label{ex:lintransf-fourier} Let $M \co \F_2^n \to \F_2^n$ be an invertible linear transformation.  Given $f \ftR$, let $f \circ M \ftR$ be defined by $f \circ M(x) = f(Mx)$.  Show that $\wh{f \circ M}(\gamma) = \wh{f}(M^{-\top}\gamma)$.  What if $M$ is an invertible \emph{affine} transformation? What if $M$ is not invertible?
    \item Show that $\frac{2}{1-e^{-2}}$ is smallest constant (not depending on~$\delta$ or~$n$) that can be taken in Proposition~\ref{prop:ns-concentration}.
    \item \label{ex:ns-concentration} Generalize Proposition~\ref{prop:ns-concentration} by showing that any $f \btR$ is $\eps$-concentrated on degree up to $1/\delta$ for $\eps = (\E[f^2] - \Stab_{1-\delta}[f])/(1-1/e)$.
    \item \label{ex:schwartz-zippel-like}  Prove Lemma~\ref{lem:schwartz-zippel-like} by induction on~$n$.  (Hint: If one of the subfunctions $f(x_1, \dots, x_n, \pm 1)$ is identically~$0$, show that the other has degree at most~$k-1$.)
    \item Verify for all $p \in [1, \infty]$ that $\snorm{\cdot}_p$ is a norm on the vector space of functions $f \ftR$.
    \item \label{ex:algebra-norm} Show that $\snorm{fg}_1 \leq \snorm{f}_1 \snorm{g}_1$ for all $f, g \ftR$.
    \item \label{ex:restrict-norm} Let $f \btR$ and let $J \subseteq [n]$, $z \in \bits^{\ol{J}}$.
        \begin{exercises}
            \item Show that restriction reduces spectral $1$-norm: $\snorm{\restr{f}{J}{z}}_1 \leq \snorm{f}_1$.
            \item Show that it also reduces Fourier sparsity: $\ssparsity{\restr{f}{J}{z}} \leq \ssparsity{f}$.
        \end{exercises}
    \item Let $f \btR$ and let $0 < p \leq q \leq \infty$.  Show that $\snorm{f}_p \geq \snorm{f}_q$.  (Cf.~Exercise~\ref{ex:norm-monotonicity}.)
    \item \label{ex:hausdorff-young} Let $f \btR$.  Show that $\snorm{f}_\infty \leq \|f\|_1$ and $\|f\|_\infty \leq \snorm{f}_1$.  (These are easy special cases of the
                                                \index{Hausdorff--Young Inequality}
        \emph{Hausdorff--Young Inequality}.)
    \item \label{ex:monotone-max-fourier01} Suppose $f \btb$ is monotone.  Show that $|\wh{f}(S)| \leq \wh{f}(i)$ whenever $i \in S \subseteq [n]$.  Deduce that $\snorm{f}_\infty = \max_{S} \{|\wh{f}(S)|\}$ is achieved by an~$S$ of cardinality~$0$ or~$1$.  (Hint: Apply the previous exercise to $f$'s derivatives.)
    \item \label{ex:coset-transf} Prove Proposition~\ref{prop:coset-transf}.
    \item Verify Parseval's Theorem for the Fourier expansion of subspaces given in Proposition~\ref{prop:subspace-transf}.
    \item \label{ex:3/2} Let $f \ftzo$ be the indicator of $A \subseteq \F_2^n$. We know that $\snorm{f}_1 = 1$ if $A$ is an affine subspace. So assume that~$A$ is \emph{not} an affine subspace.
                                                            \index{Fourier norm!$1$-}
            \begin{exercises}
                \item Show that there exists an affine subspace $B$ of dimension~$2$ on which~$f$ takes the value~$1$ exactly~$3$ times.
                \item Let $b$ be the point in $B$ where $f$ is~$0$ and let $\psi = \vphi_B - (1/2) \vphi_{b}$.  Show that $\snorm{\psi}_\infty = 1/2$.
                \item Show that $\la \psi, f \ra = 3/4$ and deduce $\snorm{f}_1 \geq 3/2$.
            \end{exercises}
    \item \label{ex:max-1-norm}
                                                            \index{Fourier norm!$1$-}
            Suppose $f \btR$ satisfies $\E[f^2] \leq 1$.  Show that $\snorm{f}_1 \leq 2^{n/2}$, and show that for any even~$n$ the upper bound can be achieved by a function $f \btb$.
    \item \label{ex:uncertainty} Given $f \ftR$, define its \emph{(fractional) sparsity}
                                                \index{sparsity (fractional)}
          to be 
          \[
            \sparsity{f} = |\supp(f)|/2^n = \Pr_{\bx \in \F_2^n}[f(\bx) \neq 0].
          \]
          In this exercise you will prove the \emph{uncertainty principle}:
                                                \index{uncertainty principle}%
          If $f$ is nonzero, then $\sparsity{f} \cdot \ssparsity{f} \geq 1$.
          \begin{exercises}
              \item Show that we may assume $\|f\|_1 = 1$.
              \item Suppose $\calF = \{ \gamma : \wh{f}(\gamma) \neq 0\}$.  Show that $\snorm{f}^2_2 \leq |\calF|$.
              \item Suppose $\calG = \{ x : f(x) \neq 0\}$.  Show that $\|f\|_2^2 \geq 2^n/|\calG|$, and deduce the uncertainty principle.
              \item Identify all cases of equality.
          \end{exercises}
    \item \label{ex:l1-conc} Let $f \btR$ and let $\eps > 0$.  Show that $f$ is
                                                        \index{Fourier norm!$1$-}
        $\eps$-concentrated on a collection $\calF \subseteq 2^{[n]}$ with $|\calF| \leq \snorm{f}_1^2/\eps$.
    \item \label{ex:close-conc} Suppose the Fourier spectrum of $f \btR$ is $\eps_1$-concentrated on $\calF$ and that $g \btR$ satisfies $\|f - g\|_2^2 \leq \eps_2$.  Show that the Fourier spectrum of $g$ is~$2(\eps_1+\eps_2)$-concentrated on~$\calF$.
    \item Show that every function $f \ftR$ is computed by a decision tree with depth at most~$n$ and size at most~$2^n$.
    \item Let $f \ftR$ be computable by a decision tree of size~$s$ and depth~$k$  Show that $-f$ and the Boolean dual $f^\booldual$ are also computable by decision trees of size~$s$ and depth~$k$.
    \item For each function in Exercise~\ref{ex:compute-expansions} with $4$ or fewer inputs, give a decision tree computing it.  Try primarily to use the least possible depth, and secondarily to use the least possible size.
    \item \label{ex:dt-spectrum}  Prove Proposition~\ref{prop:dt-spectrum}.
    \item \label{ex:dt-trunc-spectrum}  Let $f \ftb$ be computed by a decision tree~$T$ of size~$s$ and let $\eps \in (0,1]$.  Suppose each path in~$T$ is truncated (if necessary) so that its length does not exceed~$\log(s/\eps)$; new leaves with labels $-1$ and $1$ may be created in an arbitrary way as necessary. Show that the resulting decisions tree $T'$ computes a function that is $\eps$-close to $f$.  Deduce Proposition~\ref{prop:dt-trunc-spectrum}.
    \item \label{ex:decision-list} A \emph{decision list}
                                                \index{decision list}
          is a decision tree in which every internal node has an outgoing edge to at least one leaf.  Show that any function computable by a decision list is a linear threshold function.
    \item \label{ex:read-once-dt} A \emph{read-once} decision tree
                                                \index{decision tree!read-once}
          is one in which every internal node queries a distinct variable.  Bearing this in mind, show that the bound~$k2^{k-1}$ in Theorem~\ref{thm:nisan-szegedy} cannot be reduced below~$2^k - 1$.
    \item Suppose that $f$ is computed by a read-once decision tree in which every root-to-leaf path has length~$k$ and every internal node at the deepest level has one child (leaf) labeled~$-1$ and one child labeled~$1$.  Compute the influence of each coordinate on $f$, and compute $\Tinf[f]$.
    \item \label{ex:coset-partitions} The following are generalizations of decision trees:

            \emph{Subcube partition}:
                                                \index{subcube partition}
            This is defined by a collection $C_1, \dots, C_s$ of subcubes that form a partition of $\F_2^n$, along with values $b_1, \dots, b_s \in \R$.  It computes the function $f \ftR$ which has value $b_i$ on all inputs in $C_i$.  The subcube partition's size is $s$ and its ``codimension''~$k$ (analogous to depth) is the maximum codimension of the cubes $C_i$.

            \emph{Parity decision tree}:
                                                \index{parity decision tree}
            This is similar to a decision tree except that the internal nodes are labeled by vectors $\gamma \in \F_2^n$.  At such a node the computation path on input~$x$ follows the edge labeled $\gamma \cdot x$.  We insist that for each root-to-leaf path, the vectors appearing in its internal nodes are linearly independent.  Size $s$ and depth~$k$ are defined as with normal decision trees.

            \emph{Affine subspace partition}: This is similar to a subcube partition except the subcubes $C_i$ may be arbitrary affine subspaces.

        \begin{exercises}
            \item Show that subcube partition size/codimension and parity decision tree size/depth generalize normal decision tree size/depth, and are generalized by affine subspace partition size/codimension.
            \item Show that Proposition~\ref{prop:dt-spectrum} holds also for the generalizations, except that the statement about degree need not hold for parity decision trees and affine subspace partitions.
            \item Show that the class of functions with affine subspace partition size at most~$s$ is learnable from queries with error~$\eps$ in time $\poly(n, s, 1/\eps)$.
        \end{exercises}
    \item \label{ex:DT-vs-deg} Define $\EQU_3 \co \bits^3 \to \{-1,1\}$ by $\EQU_3(x) = -1$ if and only if $x_1 = x_2 = x_3$.
        \begin{exercises}
            \item Show that $\deg(\EQU_3) = 2$.
            \item Show that $\DT(\EQU_3) = 3$.
            \item Show that $\EQU_3$ is computable by a parity decision tree of codimension~$2$.
            \item For $d \in \N$, define $f \bits^{3^d} \to \bits$ by $f = \EQU_3^{\otimes d}$ (using the notation from Definition~\ref{def:rec-maj}).  Show that $\deg(f) = 2^d$ but $\DT(f) = 3^d$.
        \end{exercises}
    \item \label{ex:orthog1} Let $f \btR$ and $J \subseteq [n]$.  Define $f^{\subseteq J} \btR$ by
                                                        \index{projection onto coordinates}%
            $f(x) = \Ex_{\by \sim \bits^{\ol{J}}}[f(x_J, \by)]$, where $x_J \in \bits^J$ is the projection of $x$ to coordinates~$J$.  Verify the Fourier expansion
            \[
                f^{\subseteq J} = \sum_{S \subseteq J} \wh{f}(S)\,\chi_S.
            \]
    \item Let $\vphi \co \F_2^n \to \R^{\geq 0}$ be a probability density function corresponding to probability distribution $\phi$ on $\F_2^n$.  Let $J \subseteq [n]$.
         \begin{exercises}
            \item Consider the marginal probability distribution of $\phi$ on coordinates~$J$.  What is its probability density function (a function $\F_2^J \to \R^{\geq 0}$) in terms of $\vphi$?
            \item Consider the probability distribution of $\phi$ conditioned on a substring $z \in \F_2^{\ol{J}}$.  Assuming it's well defined, what is its probability density function in terms of $\vphi$?
         \end{exercises}
    \item \label{ex:weak-learn1} Suppose $f \btR$ is computable by a decision tree that has a leaf at depth~$k$ labeled~$b$.  Show that $\snorm{f}_\infty \geq |b|/2^k$.  (Hint: You may find Exercise~\ref{ex:orthog1} helpful.)
    \item \label{ex:conv-trans} Prove Fact~\ref{fact:translation-expansion} by using Theorem~\ref{thm:convolution-theorem} and Exercise~\ref{ex:compute-expansions}\ref{ex:single-point-density}.
    \item \label{ex:sparsity2} \begin{exercises}
              \item \label{ex:sparsity2a} Suppose $f \ftR$ has $\ssparsity{f} < 2^n$.  Show that for any $\gamma \in \supp(\wh{f})$ there exists nonzero $\beta \in \dualF$ such that $f_{\beta^\perp}$ has $\wh{f}(\gamma)$ as a Fourier coefficient.            
              \item Prove by induction on $n$ that if $f \ftb$ has $\ssparsity{f} = s > 1$ then $\wh{f}$ is $2^{1-\lfloor \log s \rfloor}$-granular.
                                                              \index{granularity, Fourier spectrum}
                  (Hint: Distinguish the cases $s = 2^n$ and $s < 2^n$.  In the latter case use part~\ref{ex:sparsity2a}.)
              \item Prove that there are no functions $f \btb$
                                                          \index{Fourier sparsity}%
                    with $\ssparsity{f} \in \{2, 3, 5, 6, 7, 9\}$.
          \end{exercises}
    \item \label{ex:trivial-learning} Show that one can learn \emph{any} target $f \btb$ with error~$0$ from random examples only in time $\wt{O}(2^n)$.
    \item \label{ex:avrim-trick} Improve Proposition~\ref{prop:sign-conc} as follows.  Suppose $f \btb$ and $g \btR$ satisfy $\|f - g\|_1 \leq \eps$.  Pick $\btheta \in [-1,1]$ uniformly at random and define $\bh \btb$ by $\bh(x) = \sgn(g(x) - \btheta)$.  Show that $\E[\dist(f,\bh)] \leq \eps/2$.
    \item \label{ex:no-conc}
        \begin{exercises}
            \item For $n$ even, find a function $f \btb$ such that $f$ is not $1/2$-concentrated on
                                                \nomenclature[2A]{$2^A$}{the set of all subsets of~$A$}
                    any $\calF \subseteq 2^{[n]}$ with $|\calF| < 2^{n-1}$.  (Hint: Exercise~\ref{ex:compute-expansions}.)
            \item Let $\boldf \btb$ be a random function as in Exercise~\ref{ex:random-Fourier}.  Show that with
                                                        \index{random function}
            probability at least~$1/2$, $\boldf$ is not $1/4$-concentrated on degree up to $\lfloor n/2 \rfloor$.
        \end{exercises}
    \item \label{ex:learn-low-degree} Prove Theorem~\ref{thm:learn-low-degree}. (Hint: In light of Exercise~\ref{ex:deg-dyadic} you may round off certain estimates with confidence.)
    \item \label{ex:learn-sparse}  Show that each of the following classes $\calC$ (ordered by inclusion) can be learned exactly (i.e., with error~$0$) using queries in time $\poly(n, 2^k)$:
            \begin{exercises}
              \item \label{ex:learn-juntas} $\calC = \{f \btb \mid f \text{ is a $k$-junta}\}$. (Hint: Estimate influences.)%
                                              \index{junta!learning}%
              \item \label{ex:learn-dt-queries} $\calC = \{f \btb \mid \DT(f) \leq k\}$.
                            \index{decision tree!learning}%
              \item \label{ex:learn-sparse-queries} $\calC = \{f \btb \mid \ssparsity{f} \leq 2^{O(k)}\}$. (Hint: Exercise~\ref{ex:sparsity2}.)
          \end{exercises}
    \item \label{ex:learn-low-l1} Prove Theorem~\ref{thm:learn-low-l1}.  (Hint: Exercise~\ref{ex:l1-conc}.)
    \item \label{ex:deduce-KM}  Deduce Theorem~\ref{thm:KM} from the Goldreich--Levin Algorithm.
    \item \label{ex:learning-delta} Suppose $A$ learns $\calC$ from random examples with error $\eps/2$ in time~$T$ --  with probability at least $9/10$.
          \begin{exercises}
              \item After producing hypothesis $h$ on target~$f \btb$, show that $A$ can ``check'' whether $h$ is a good hypothesis in time $\poly(n, T, 1/\eps) \cdot \log(1/\delta)$.  Specifically, except with probability at most~$\delta$, $A$ should output `YES' if $\dist(f, h) \leq \eps/2$ and `NO' if $\dist(f, h) > \eps$. (Hint: Time~$\poly(T)$ may be required for $A$ to evaluate $h(x)$.)
              \item Show that for any $\delta \in (0, 1/2]$, there is a learning algorithm that learns $\calC$ with error $\eps$ in time $\poly(n, T, \eps)\cdot \log(1/\delta)$ -- with probability at least $1-\delta$.
          \end{exercises}
    \item \begin{exercises}
        \item Our description of the Low-Degree Algorithm with degree~$k$ and error~$\eps$ involved using a new batch of random examples to estimate each low-degree Fourier coefficient.  Show that one can instead simply draw a single batch $\calE$ of $\poly(n^k, 1/\eps)$ examples and use $\calE$ to estimate each of the low-degree
                                        \index{Low-Degree Algorithm}
            coefficients.
        \item Show that when using the above form of the Low-Degree Algorithm, the final hypothesis $h \btb$ is of the form
            \[
                h(y) = \sgn\left(\sum_{(x, f(x)) \in \mathcal{E}} w(\hamdist(y,x)) \cdot f(x) \right),
            \]
            for some function $w \co \{0, 1, \dots, n\} \to \R$. In other words, the hypothesis on a given~$y$ is equal to a weighted vote over all examples seen, where an example's weight depends only on its Hamming distance to~$y$.  Simplify your expression for $w$ as much as you can.
        \end{exercises}
    \item Extend the Goldreich--Levin Algorithm so that it works also for functions $f \co \bn \to [-1,1]$.  (The learning model for targets $f \co \bn \to [-1,1]$ assumes that $f(x)$ is always a rational number expressible by $\poly(n)$ bits.)
    \item \label{ex:learn-f2-parity}
          \begin{exercises}
              \item Assume $\gamma, \gamma' \in \dualF$ are distinct.  Show that $\Pr_{\bx}[\gamma \cdot \bx = \gamma' \cdot \bx] = 1/2$.
              \item Fix $\gamma \in \dualF$ and suppose $\bx^{(1)}, \dots, \bx^{(m)} \sim \F_2^n$ are drawn uniformly and independently.  Show that if $m = Cn$ for $C$ a sufficiently large constant then with high probability, the only $\gamma' \in \dualF$ satisfying $\gamma' \cdot \bx^{(i)} = \gamma \cdot \bx^{(i)}$ for all $i \in [m]$ is $\gamma' = \gamma$.
              \item Essentially improve on Exercise~\ref{ex:learning-parity} by showing that the concept class of all linear functions $\F_2^n \to \F_2$ can be learned from random examples only, with error~$0$, in time~$\poly(n)$.  (Remark: If $\omega \in \R$ is such that $n \times n$ matrix multiplication can be done in $O(n^\omega)$ time, then the learning algorithm also requires only $O(n^\omega)$ time.)
          \end{exercises}
    \item Let $\tau \geq 1/2 + \eps$ for some constant $\eps > 0$.  Give an algorithm simpler than Goldreich and Levin's that solves the following problem with high probability: Given query access to $f \btb$, in time $\poly(n, 1/\eps)$ find the unique $U \subseteq [n]$ such that $|\wh{f}(U)| \geq \tau$, assuming it exists.  (Hint: Use Proposition~\ref{prop:local-correcting-linearity} and Exercise~\ref{ex:learning-parity}.)
    \item \label{ex:GL} Informally: a ``one-way permutation'' is a bijective function $f \co \F_2^n \to \F_2^n$ that
                                                \index{cryptography}
        is easy to compute on all inputs but hard to invert on more than a negligible fraction of inputs; a ``pseudorandom generator'' is a function $g \co \F_2^k \to \F_2^m$ for $m > k$ whose output on a random input ``looks unpredictable'' to any efficient algorithm.  Goldreich and Levin proposed the following construction of the latter from the former:  for $k = 2n$, $m = 2n+1$, define
        \[
            g(r, s) = (r, f(s), r \cdot s),
        \]
        where $r, s \in \F_2^n$.  When $g$'s input $(\br, \bs)$ is uniformly random, then so is the first $2n$ bits of its output (using the fact that $f$ is a bijection).  The key to the analysis is showing that the final bit, $\br \cdot \bs$, is highly unpredictable to efficient algorithms even \emph{given} the first $2n$ bits $(\br, f(\bs))$.  This is proved by contradiction.
        \begin{exercises}
            \item Suppose that an adversary has a deterministic, efficient algorithm~$A$ good at predicting the bit $\br \cdot \bs$:
                \[
                    \Pr_{\br, \bs \sim \F_2^n}[A(\br, f(\bs)) = \br \cdot \bs] \geq \frac12 + \gamma.
                \]
                Show there exists $B \subseteq \F_2^n$ with $|B|/2^n \geq \frac12 \gamma$ such that
                \[
                    \Pr_{\br \sim \F_2^n}[A(\br, f(s)) = \br \cdot s] \geq \frac12 + \frac12 \gamma
                \]
                for all $s \in B$.
            \item Switching to $\pm 1$ notation in the output, deduce $\wh{\restr{A}{}{f(s)}}(s) \geq \gamma$ for all $s \in B$.
            \item Show that the adversary can efficiently compute $s$ given~$f(s)$ (with high probability) for any $s \in B$.  If $\gamma$ is nonnegligible, this contradicts the assumption that $f$ is ``one-way''. (Hint: Use the Goldreich--Levin Algorithm.)
            \item Deduce the same conclusion even if $A$ is a randomized algorithm.
        \end{exercises}
\end{exercises}

\subsection*{Notes.}                                        \label{sec:spectral-structure-notes}

The fact that the Fourier characters $\chi_{\gamma} \co \F_2^n \to \{-1,1\}$ form a group isomorphic to $\F_2^n$ is not a coincidence; the analogous result holds for any finite abelian group and is a special case of the theory of Pontryagin duality in harmonic analysis.  We will see further examples of this in Chapter~\ref{chap:generalized-domains}.

Regarding spectral structure, Karpovsky~\cite{Kar76} proposed $\ssparsity{f}$ as a measure of complexity for the function~$f$.  Brandman's thesis~\cite{Bra87} (see also~\cite{BOH90}) is an early work connecting decision tree and subcube partition complexity to Fourier analysis. The notation introduced for restrictions in Section~\ref{sec:restrictions} is not standard; unfortunately there is no standard notation.  The uncertainty principle from Exercise~\ref{ex:uncertainty} dates back to Matolcsi and Sz\"{u}cs~\cite{MS73}. The result of Exercise~\ref{ex:3/2} is due to Green and Sanders~\cite{GS08}, with inspiration from Saeki~\cite{Sae68}.  The main result of Green and Sanders is the sophisticated theorem that any $f \ftzo$ with $\snorm{f}_1 \leq s$
                                                    \index{Fourier norm!$1$-}
can be expressed as $\sum_{i = 1}^L \pm 1_{H_i}$, where $L \leq 2^{2^{\poly(s)}}$ and each $H_i \leq \F_2^n$.

Theorem~\ref{thm:nisan-szegedy} is due to Nisan and Szegedy~\cite{NS94}.  That work also showed a nontrivial kind of converse to the first statement in Proposition~\ref{prop:dt-spectrum}: Any $f \btb$ is computable by a decision tree of depth at most $\poly(\deg(f))$. The best upper bound currently known is $\deg(f)^3$ due to Midrij\=anis~\cite{Mid04}. Nisan and Szegedy also gave the example in Exercise~\ref{ex:DT-vs-deg} showing the dependence cannot be linear.

The field of computational learning theory was introduced by Valiant in 1984~\cite{Val84}; for a good survey with focus on learning under the uniform distribution, see the thesis by Jackson~\cite{Jac95}.  Linial, Mansour, and Nisan~\cite{LMN93} pioneered the Fourier approach to learning, developing the Low-Degree Algorithm. We present their strong results on constant-depth circuits in Chapter~\ref{chap:circuits}.  The noise sensitivity approach to the Low-Degree Algorithm is from Klivans, O'Donnell, and Servedio~\cite{KOS04}.  Corollary~\ref{cor:learn-mono} is due to Bshouty and Tamon~\cite{BT96} who also gave certain matching lower bounds. Goldreich and Levin's work dates from 1989~\cite{GL89}. Besides its applications to cryptography and learning, it is important in coding theory and complexity as a \emph{local list-decoding algorithm} for the Hadamard code. The Kushilevitz--Mansour algorithm is from their 1993 paper~\cite{KM93}; they also are responsible for the results of Exercise~\ref{ex:learn-sparse}\ref{ex:learn-dt-queries} and~\ref{ex:learn-low-l1}. The results of
Exercise~\ref{ex:sparsity2} and~\ref{ex:learn-sparse}\ref{ex:learn-sparse-queries} are from Gopalan et~al.~\cite{GOS+11}.

\chapter{DNF formulas and small-depth circuits}                     \label{chap:circuits}

In this chapter we investigate Boolean functions representable by small DNF formulas and constant-depth circuits; these are significant generalizations of decision trees.  Besides being natural from a computational point of view, these representation classes are close to the limit of what complexity theorists can ``understand'' (e.g., prove explicit lower bounds for).  One reason for this is that functions in these classes have strong Fourier concentration properties.

\section{DNF formulas}                                              \label{sec:DNFs}

One of the commonest ways of representing a Boolean function $f \zotzo$ is by a
                                            \index{DNF}
DNF formula:
\begin{definition}
    A \emph{DNF (disjunctive normal form) formula} over Boolean variables $x_1, \dots, x_n$ is defined to be a logical~OR of \emph{terms},
                                            \index{term (DNF)}
    each of which is a logical~AND of \emph{literals}.
                                            \index{literal}
    A \emph{literal} is either a variable~$x_i$ or its logical negation~$\ol{x}_i$.  We insist that no term contains both a variable and its negation.  The number of literals in a term is called its \emph{width}.  We often identify a DNF formula with the Boolean function $f \zotzo$ it computes.
\end{definition}
\begin{example} \label{eg:sort3-dnf} Recall the function $\Sort_3$, defined by $\Sort_3(x_1, x_2, x_3) = 1$ if and only if $x_1 \leq x_2 \leq x_3$ or $x_1 \geq x_2 \geq x_3$.  We can represent it by a DNF formula as follows:
\[
    \Sort_3(x_1, x_2, x_3)\ =\ (x_1 \wedge x_2)\ \vee\ (\ol{x}_2 \wedge \ol{x}_3)\ \vee\ (\ol{x}_1 \wedge x_3) \ \vee\ (x_1 \wedge \ol{x}_3).
\]
The DNF representation says that the bits are sorted if either the first two bits are~$1$, or the last two bits are~$0$, or the first bit is~$0$ and the last bit is~$1$, or the first bit is~$1$ and the last bit is~$0$.
\end{example}

The complexity of a DNF formula is measured by its size and width:
\begin{definition}
    The \emph{size}
                                            \index{DNF!size}
    of a DNF formula is its number of terms.  The \emph{width} is the maximum width of its terms.
                                            \index{DNF!width}
    Given $f \btb$ we write $\DNFsize(f)$
                                                \nomenclature[DNFsize]{$\DNFsize(f)$}{least possible size of a DNF formula computing $f$}
    (respectively, $\DNFwidth(f)$)
                                                \nomenclature[DNFwidth]{$\DNFwidth(f)$}{least possible width of a DNF formula computing $f$}
    for the least size (respectively, width) of a DNF formula computing~$f$.

\end{definition}
The DNF formula for $\Sort_3$ from Example~\ref{eg:sort3-dnf} has size~$3$ and width~$2$.  Every function $f \zotzo$ can be computed by a DNF of size at most~$2^n$ and width at most~$n$ (Exercise~\ref{ex:triv-DNF}).

There is also a ``dual'' notion to DNF formulas:
\begin{definition}
    A \emph{CNF (conjunctive normal form) formulas}
                                            \index{CNF}
    is a logical~AND of \emph{clauses}, each of which is a logical~OR of literals.  Size and width are defined as for DNFs.
\end{definition}
Some functions can be represented much more compactly by CNFs than DNFs (see Exercise~\ref{ex:tribes-CNF}). On the other hand, if we take a CNF computing~$f$ and switch its ANDs and ORs, the result is a DNF computing the dual function~$f^\booldual$ (see Exercises~\ref{ex:odd-even} and~\ref{ex:CNF-DNF-duality}).  Since~$f$ and~$f^\dagger$ have essentially the same Fourier expansion, there isn't much difference between CNFs and DNFs when it comes to Fourier analysis.  We will therefore focus mainly on DNFs.

DNFs and CNFs are more powerful than decision trees for representing Boolean-valued functions, as the following proposition shows:
\begin{proposition}                                     \label{prop:DNF-DT}
    Let $f \zotzo$ be computable by a decision tree~$T$ of size~$s$ and depth~$k$.  Then $f$ is computable by a DNF (and also a CNF) of size at most~$s$ and width at most~$k$.
\end{proposition}
\begin{proof}
    Take each path in~$T$ from the root to a leaf labeled~$1$ and form the logical AND of the literals describing the path.  These are the terms of the required DNF.  (For the CNF clauses, take paths to label~$0$ and negate all literals describing the path.)
\end{proof}
\begin{example}
    If we perform this conversion on the decision tree computing $\Sort_3$ in Figure~\ref{fig:sort3-dt} we get the DNF
    \[
        (\ol{x}_1 \wedge \ol{x}_3 \wedge \ol{x}_2)\ \vee\ (\ol{x}_1 \wedge x_3)\ \vee\ (x_1 \wedge \ol{x}_2 \wedge \ol{x}_3)\ \vee\ (x_2 \wedge x_3).
    \]
    This has size~$4$ (indeed at most the decision tree size~$6$) and width~$3$ (indeed at most the decision tree depth~$3$).  It is not as simple as the equivalent DNF from Example~\ref{eg:sort3-dnf}, though; DNF representation is not unique.
\end{example}

The class of functions computable by small DNFs is intensively studied in learning theory.  This is one reason why the problem of analyzing spectral concentration for DNFs is important.  Let's begin with the simplest method for this: understanding low-degree concentration via total influence.  We will switch to~$\pm 1$ notation.
\begin{proposition}                                     \label{prop:DNF-tinf}
    Suppose that $f \btb$ has $\DNFwidth(f) \leq w$.
                                        \index{total influence!DNF formulas}
    Then $\Tinf[f] \leq 2w$.
\end{proposition}
\begin{proof}
    We use Exercise~\ref{ex:one-way-inf}, which states that
    \[
        \Tinf[f] = 2\Ex_{\bx \sim \bn}[\# \text{ $(-1)$-pivotal coordinates for $f$ on $\bx$}],
    \]
    where coordinate~$i$ is ``$(-1)$-pivotal'' on input~$x$ if $f(x) = -1$ (logical $\true$) but $f(x^{\oplus i}) = 1$ (logical $\false$). It thus suffices to show that on \emph{every} input~$x$ there are at most~$w$ coordinates which are $(-1)$-pivotal.  To have any $(-1)$-pivotal coordinates at all on~$x$ we must have $f(x) = -1$ ($\true$); this means that at least one term~$T$ in $f$'s width-$w$ DNF representation must be made $\true$ by~$x$.  But now if $i$ is a $(-1)$-pivotal coordinate then either $x_i$ or $\ol{x}_i$ must appear in~$T$; otherwise, $T$ would still be made true by~$x^{\oplus i}$.  Thus the number of $(-1)$-pivotal coordinates on~$x$ is at most the number of literals in~$T$, which is at most~$w$.
\end{proof}
Since $\Tinf[f^\booldual] = \Tinf[f]$ the proposition is also true for CNFs of width at most~$w$. The proposition is very close to being tight: The parity function $\chi_{[w]} \btb$ has $\Tinf[\chi_{[w]}] = w$ and $\DNFwidth(\chi_{[w]}) \leq w$ (the latter being true for all $w$-juntas). In fact, the proposition can be improved to give the tight upper bound~$w$ (Exercise~\ref{ex:amano}).

Using Proposition~\ref{prop:tinf-concentration} we deduce:
\begin{corollary}                                       \label{cor:DNF-low-deg-conc}
    Let $f \btb$ have $\DNFwidth(f) \leq w$.  Then for $\eps > 0$, the Fourier spectrum of $f$ is $\eps$-concentrated on degree up to $2w/\eps$.
\end{corollary}
The dependence here on $w$ is of the correct order (by the example of the parity $\chi_{[w]}$ again), but the dependence on~$\eps$ can be significantly improved as we will see in Section~\ref{sec:switching-lemma}.

There's usually more interest in DNF \emph{size} than in DNF width; for example, learning theorists are often interested in the class of $n$-variable DNFs of size $\poly(n)$.  The following fact (similar to Exercise~\ref{ex:dt-trunc-spectrum}) helps relate the two, suggesting $O(\log n)$ as an analogous width bound:
\begin{proposition}                                     \label{prop:DNF-size-to-width}
    Let $f \btb$ be computable by a DNF (or CNF) of size~$s$ and let $\eps \in (0, 1]$.  Then~$f$ is~$\eps$-close to a function $g$ computable by a DNF of width~$\log(s/\eps)$.
\end{proposition}
\begin{proof}
    Take the DNF computing $f$ and delete all terms with more than~$\log(s/\eps)$ literals; let $g$ be the function computed by the resulting DNF.  For any deleted term~$T$, the probability a random input $\bx \sim \bn$ makes~$T$ true is at most $2^{-\log(s/\eps)} = \eps/s$.  Taking a union bound over the (at most $s$) such terms shows that $\Pr[g(\bx) \neq f(\bx)] \leq \eps$.   (A similar proof works for CNFs.)
\end{proof}
By combining Proposition~\ref{prop:DNF-size-to-width} and Corollary~\ref{cor:DNF-low-deg-conc} we can deduce (using Exercise~\ref{ex:close-conc}) that DNFs of size~$s$ have Fourier spectra $\eps$-concentrated up to degree $O(\log(s/\eps)/\eps)$. Again, the dependence on $\eps$ will be improved in Section~\ref{sec:switching-lemma}.  We will also later show in Section~\ref{sec:random-restrictions} that size-$s$ DNFs have total influence at most $O(\log s)$, something we cannot deduce immediately from Proposition~\ref{prop:DNF-tinf}.

In light of the Kushilevitz--Mansour learning algorithm it would also be nice to show that $\poly(n)$-size DNFs have their Fourier spectra concentrated on small collections (not necessarily low-degree).  In Section~\ref{sec:switching-lemma} we will show they are $\eps$-concentrated on collections of size $n^{O(\log \log n)}$ for any constant $\eps>0$.  It has been conjectured that this can be improved to $\poly(n)$:
\begin{named}{Mansour's Conjecture}  Let $f \btb$ be computable by a
                                                \index{Mansour's Conjecture}%
                                                \index{DNF!Fourier spectrum}%
DNF of size $s > 1$ and let $\eps \in (0,1/2]$.  Strong conjecture: $f$'s Fourier spectrum is $\eps$-concentrated on a collection $\calF$ with $|\calF| \leq s^{O(\log(1/\eps))}$.  Weaker conjecture: if $s \leq \poly(n)$ and $\eps > 0$ is any fixed constant, then we have the bound $|\calF| \leq \poly(n)$.
\end{named}

\section{Tribes}                                                    \label{sec:tribes}
                                                \index{tribes function|(}
In this section we study the \emph{tribes} DNF formulas, which serve as an important examples and counterexamples in analysis of Boolean functions.  Perhaps the most notable feature of the tribes function is that (for a suitable choice of parameters) it is essentially unbiased and yet all of its influences are quite tiny.

Recall from Chapter~\ref{sec:social-choice} that the function $\tribes_{w,s} \co \bits^{sw} \to \bits$ is defined by its width-$w$, size-$s$ DNF representation:
\begin{multline*}
    \tribes_{w,s}(x_1, \dots, x_w, \dots, x_{(s-1)w+1}, \dots, x_{sw}) \\
    = (x_1 \wedge \cdots \wedge x_w)\ \vee\ \cdots\ \vee\ (x_{(s-1)w+1} \wedge \cdots \wedge x_{sw}).
\end{multline*}
(We are using the notation where~$-1$ represents logical $\true$ and~$1$ represents logical $\false$.)
As is computed in Exercise~\ref{ex:basic-tribes} we have:
\begin{fact}                                        \label{fact:tribes-mean}
    $\Pr_\bx[\tribes_{w,s}(\bx) = -1] = 1 - (1-2^{-w})^s$.
\end{fact}
The most interesting setting of parameters makes this probability as close to~$1/2$ as possible (a slightly different choice than the one in Exercise~\ref{ex:basic-tribes}):
\begin{definition}  \label{def:tribes-careful}
    For~$w \in \N^+$, let $s = s_w$ be the largest integer such that ${1 - (1-2^{-w})^s} \leq 1/2$.  Then for $n = n_w = sw$ we define $\tribes_n \btb$ to be $\tribes_{w,s}$.  Note this is only defined only for certain~$n$: $1$, $4$, $15$, $40$, \dots
\end{definition}
Here $s \approx \ln(2) 2^w$, hence $n \approx \ln(2) w 2^w$ and therefore $w \approx \log n - \log \ln n$ and $s \approx n/\log n$.  A slightly more careful accounting (Exercise~\ref{ex:tribes-careful}) yields:
\begin{proposition}                                     \label{prop:tribes-facts}
    For the $\tribes_n$ function as in Definition~\ref{def:tribes-careful}:
    \begin{itemize}
        \item $s = \ln(2) 2^w - \Theta_w(1)$;
        \item $n = \ln(2) w 2^w - \Theta(w)$, thus $n_{w+1} = (2 + o(1)) n_w$;
        \item $w = \log n - \log \ln n + o_n(1)$, and $2^w = \frac{n}{\ln n} (1+o_n(1))$;
        \item $\Pr[\tribes_{n}(\bx) = -1] = 1/2 - O\left(\frac{\log n}{n}\right)$.
    \end{itemize}
\end{proposition}

Thus with this setting of parameters $\tribes_n$ is essentially unbiased.  Regarding its influences:
\begin{proposition}                                     \label{prop:tribes-influences}
    $\Inf_i[\Tribes_n] = \frac{\ln n}{n}(1\pm o(1))$ for each $i \in [n]$ and hence $\Tinf[\Tribes_n] = (\ln n)(1\pm o(1))$.
\end{proposition}
\begin{proof}
    Thinking of $\Tribes_n = \Tribes_{w,s}$ as a voting rule, voter~$i$ is pivotal if and only if: (a)~all other voters in $i$'s ``tribe'' vote $-1$ ($\true$); (b)~all other tribes produce the outcome~$1$ ($\false$).  The probability of this is indeed
    \[
        2^{-(w-1)} \cdot (1-2^{-w})^{s-1} = \tfrac{2}{2^w - 1} \cdot \Pr[\tribes_n = 1] = \tfrac{\ln n}{n}(1\pm o(1)),
    \]
    where we used Fact~\ref{fact:tribes-mean} and then  Proposition~\ref{prop:tribes-facts}.
\end{proof}
Thus if we are interested in (essentially) unbiased voting rules in which every voter has small influence, $\Tribes_n$ is a much stronger example than $\Maj_n$ where each voter has influence $\Theta(1/\sqrt{n})$.  You may wonder if the maximum influence can be even \emph{smaller} than $\Theta\bigl(\frac{\ln n}{n}\bigr)$ for unbiased voting rules.  Certainly it can't be smaller than $\frac{1}{n}$, since the Poincar\'{e} Inequality says that $\Tinf[f] \geq 1$ for unbiased~$f$.  In fact the famous KKL Theorem shows that the $\Tribes_n$ example is tight up to constants:
\begin{named}{Kahn--Kalai--Linial (KKL) Theorem}  For
                                                    \index{Kahn--Kalai--Linial Theorem|seeonly{KKL Theorem}}%
                                                    \index{KKL Theorem}%
any $f \btb$,
\[
    \MaxInf[f] = \max_{i \in [n]} \{\Inf_i[f] \} \geq \Var[f] \cdot \Omega\Bigl(\frac{\log n}{n}\Bigr).
\]
\end{named}
\noindent We prove the KKL~Theorem in Chapter~\ref{chap:hypercontractivity}.

We conclude this section by recording a formula for the Fourier coefficients of $\Tribes_{w,s}$.  The proof is Exercise~\ref{ex:tribes-coeffs}.
\begin{proposition}                                     \label{prop:tribes-coeffs}
    Suppose we index the Fourier coefficients of the function $\tribes_{w,s} \bits^{sw} \to \bits$ by sets $T = (T_1, \dots, T_s) \subseteq [sw]$, where $T_i$ is the intersection of $T$ with the $i$th ``tribe''.  Then
    \[
        \wh{\Tribes_{w,s}}(T) = \begin{cases}
                                    2(1-2^{-w})^s - 1 & \text{if $T = \emptyset$,} \\
                                    2(-1)^{k + |T|}2^{-kw} (1-2^{-w})^{s-k} & \text{if $k  = \#\{i : T_i \neq \emptyset\} > 0$.}
                                \end{cases}
    \]
\end{proposition}
                                                \index{tribes function|)}

\section{Random restrictions}                                               \label{sec:random-restrictions}
                                                \index{restriction!random|(}
In this section we describe the method of applying \emph{random restrictions}.  This is a very ``Fourier-friendly'' way of simplifying a Boolean function.  As motivation, let's consider the problem of bounding total influence for size-$s$ DNFs.  One plan  is to use the results from Section~\ref{sec:DNFs}:  size-$s$ DNFs are $.01$-close to width-$O(\log s)$ DNFs, which in turn have total influence $O(\log s)$.  This suggests that size-$s$ DNFs themselves have total influence $O(\log s)$. To prove this though we'll need to reverse the steps of the plan; instead of truncating DNFs to a fixed width and arguing that a random input is unlikely to notice, we'll first pick a random (partial) input and argue that this is likely to make the width small.

Let's formalize the notion of a random partial input, or restriction:
\begin{definition}                                  \label{def:delta-random-subset}
    For $\delta \in [0,1]$, we say that $\bJ$ is a \emph{$\delta$-random subset}
                                                \index{random subset}
    of $N$ if it is formed by including each element of~$N$ independently with probability~$\delta$. We define a \emph{$\delta$-random restriction on $\bn$} to be a pair $(\bJ \mid \bz)$, where first $\bJ$ is chosen to be a $\delta$-random subset of $[n]$ and then $\bz \sim \bits^{\ol{\bJ}}$ is chosen uniformly at random.  We say that coordinate~$i \in [n]$ is \emph{free} if $i \in \bJ$ and is \emph{fixed} if $i \notin \bJ$.  An equivalent definition is that each coordinate $i$ is (independently) free with probability $\delta$ and fixed to $\pm 1$ with probability $(1-\delta)/2$ each.
\end{definition}
Given $f \btR$ and a random restriction $(\bJ \mid \bz)$, we can form the restricted function $\restr{f}{\bJ}{\bz} \co \bits^{\bJ} \to \R$ as usual.  However, it's inconvenient that the domain of this function depends on the random restriction.  Thus when dealing with random restriction we usually invoke the following convention:
\begin{definition}
    Given $f \btR$, $I \subseteq [n]$, and $z \in \bits^{\ol{I}}$, we may identify the restricted function $\restr{f}{I}{z} \co \bits^I \to \R$ with its extension $\restr{f}{I}{z} \btR$ in which the input coordinates $\bits^{\ol{I}}$ are ignored.
\end{definition}
As mentioned, random restrictions interact nicely with Fourier expansions:
\begin{proposition}                                     \label{prop:rand-restrict-expansion}
    Fix $f \btR$ and $S \subseteq [n]$.  Then if $(\bJ \mid \bz)$ is a $\delta$-random restriction on $\bits^n$,
    \[
        \E[\wh{\restr{f}{\bJ}{\bz}}(S)] = \Pr[S \subseteq \bJ] \cdot \wh{f}(S) = \delta^{|S|} \wh{f}(S),
    \]
    and
    \[
        \E[\wh{\restr{f}{\bJ}{\bz}}(S)^2] = \sum_{U \subseteq [n]} \Pr[U \cap \bJ = S]\cdot\wh{f}(U)^2 = \sum_{U \supseteq S} \delta^{|S|} (1-\delta)^{|U \setminus S|} \wh{f}(U)^2,
    \]
    where we are treating $\restr{f}{\bJ}{\bz}$ as a function $\bits^n \to \R$.
\end{proposition}
\begin{proof}
    Suppose first that $J \subseteq [n]$ is fixed.  When we think of restricted functions $\restr{f}{J}{\bz}$ as having domain $\bits^n$, Corollary~\ref{cor:expected-restrict-coeffs} may be stated as saying that for any $S \subseteq [n]$,
    \begin{align*}
        \E_{\bz \sim \bits^{\ol{J}}}[\wh{\restr{f}{J}{\bz}}(S)] &= \wh{f}(S) \cdot \bone_{S \subseteq J},\\
        \E_{\bz \sim \bits^{\ol{J}}}[\wh{\restr{f}{J}{\bz}}(S)^2] &= \sum_{U \subseteq [n]} \wh{f}(U)^2 \cdot \bone_{U \cap J = S}.
    \end{align*}
    The proposition now follows by taking the expectation over~$\bJ$.
\end{proof}
\begin{corollary}                                       \label{cor:rand-restrict-tinf}
    Fix $f \btR$ and $i \in [n]$.  If $(\bJ \mid \bz)$ is a $\delta$-random restriction, then $\E[\Inf_i[\restr{f}{\bJ}{\bz}]] = \delta \Inf_i[f]$.  Hence also $\E[\Tinf[\restr{f}{\bJ}{\bz}]] = \delta \Tinf[f]$.
\end{corollary}
\begin{proof}
    We have
    \begin{multline*}
        \E[\Inf_i[\restr{f}{\bJ}{\bz}]] = \E\left[\sum_{S \ni i} \wh{\restr{f}{\bJ}{\bz}}(S)^2\right]     = \sum_{S \ni i} \sum_{U \subseteq [n]} \Pr[U \cap \bJ = S] \wh{f}(U)^2 \\
        = \sum_{U \subseteq [n]} \Pr[U \cap \bJ \ni i] \wh{f}(U)^2 = \sum_{U \ni i} \delta \wh{f}(U)^2 = \delta \Inf_i[f],
    \end{multline*}
    where the second equality used Proposition~\ref{prop:rand-restrict-expansion}.
\end{proof}
\noindent (Proving Corollary~\ref{cor:rand-restrict-tinf} via Proposition~\ref{prop:rand-restrict-expansion} is a bit more elaborate than necessary; see Exercise~\ref{ex:rand-restrict-tinf}.)

Corollary~\ref{cor:rand-restrict-tinf} lets us bound the total influence of a function~$f$ by bounding the (expected) total influence of a random restriction of~$f$.  This is useful if $f$ is computable by a DNF formula of small size, since a random restriction is very likely to make this DNF have small width.  This is a consequence of the following lemma:
\begin{lemma}                                     \label{lem:rand-restr-dnf}
    Let $T$ be a DNF term over $\bn$ and fix $w \in \N^+$.  Let $(\bJ \mid \bz)$ be a $(1/2)$-random restriction on $\bn$.  Then $\Pr[\text{width}(\restr{T}{\bJ}{\bz}) \geq w] \leq (3/4)^w$.
\end{lemma}
\begin{proof}
    We may assume the initial width of $T$ is at least~$w$, as otherwise its restriction under $(\bJ \mid \bz)$ cannot have width at least~$w$.  Now if any literal appearing in~$T$ is fixed to $\false$ by the random restriction, the restricted term $\restr{T}{\bJ}{\bz}$ will be constantly $\false$ and thus have width~$0 < w$.  Each literal is fixed to $\false$ with probability $1/4$; hence the probability no literal in~$T$ is fixed to $\false$ is at most $(3/4)^w$.
\end{proof}

We can now bound the total influence of small DNF formulas.
\begin{theorem}                                     \label{thm:dnf-tinf}
    Let $f \btb$ be computable by a DNF of size~$s$.
                                                \index{total influence!DNF formulas}
    Then $\Tinf[f] \leq O(\log s)$.
\end{theorem}
\begin{proof}
    Let $(\bJ \mid \bz)$ be a $(1/2)$-random restriction on $\bn$ and write $\bw = \DNFwidth(\restr{f}{\bJ}{\bz})$.  By a union bound and Lemma~\ref{lem:rand-restr-dnf} we have that $\Pr[\bw \geq w] \leq s (3/4)^w$.  Hence
    \begin{multline*}
        \E[\bw] = \sum_{w = 1}^\infty \Pr[\bw \geq w] \leq 3 \log s + \sum_{w > 3 \log s} s (3/4)^w  \\
        \leq 3 \log s + 4s (3/4)^{3 \log s} \leq 3\log s + 4/s^{0.2} = O(\log s).
    \end{multline*}
    From Proposition~\ref{prop:DNF-tinf} we obtain $\E[\Tinf[\restr{f}{\bJ}{\bz}]] \leq 2 \cdot O(\log s) = O(\log s)$.  And so from Corollary~\ref{cor:rand-restrict-tinf} we conclude $\Tinf[f] = 2\E[\Tinf[\restr{f}{\bJ}{\bz}]] \leq O(\log s)$.
\end{proof}
                                                \index{restriction!random|)}

\section{H{\aa}stad's Switching Lemma and the spectrum of DNFs}               \label{sec:switching-lemma}
Let's further investigate how random restrictions can simplify DNF formulas.  Suppose $f$ is computable by a DNF formula of width~$w$, and we apply to it a $\delta$-random restriction with $\delta \ll 1/w$.  For each term~$T$ in the DNF, one of three things may happen to it under the random restriction.  First and by far most likely, one of its literals may be fixed to $\false$, allowing us to delete it.  If this doesn't happen, the second possibility is that all of $T$'s literals are made $\true$, in which case the whole DNF reduces to the constantly $\true$ function.  With $\delta \ll 1/w$, this is in turn much more likely than the third possibility, which is that at least one of $T$'s literals is left free, but all the fixed literals are made $\true$.  Only in this third case is $T$ not trivialized by the random restriction.

This reasoning might suggest that $f$ is likely to become a constant function under the random restriction.  Indeed, this is true, as the following theorem shows:
\begin{named}{Baby Switching Lemma}  Let $f \btb$ be computable by
                                            \index{Switching Lemma!Baby}
a DNF or CNF of width at most~$w$ and let $(\bJ \mid \bz)$ be a $\delta$-random restriction.  Then
\[
\Pr[\restr{f}{\bJ}{\bz} \text{ is not a constant function}] \leq 5 \delta w.
\]
\end{named}
This is in fact the $k=1$ case of the following much more powerful theorem:
\begin{named}{\Hastad's Switching Lemma}  Let $f \btb$ be
                                            \index{Switching Lemma!H{\aa}stad's}
computable by a DNF or CNF of width at most~$w$ and let $(\bJ \mid \bz)$ be a $\delta$-random restriction.  Then for any $k \in \N$,
\[
\Pr[\DT(\restr{f}{\bJ}{\bz}) \geq k] \leq (5 \delta w)^k.
\]
\end{named}
What is remarkable about this result is that it has no dependence on the \emph{size} of the DNF, or on~$n$. In words, \Hastad's Switching Lemma says that when $\delta \ll 1/w$, it's exponentially unlikely (in~$k$) that applying a $\delta$-random restriction to a width-$w$ DNF does not convert (``switch'') it to a decision tree of depth less than~$k$.  The result is called a ``lemma'' for historical reasons; in fact, its proof requires some work.  You are asked to prove the Baby Switching Lemma in Exercise~\ref{ex:baby-switch}; for \Hastad's Switching Lemma, consult \Hastad's original proof~\cite{Has87} or the alternate proof of Razborov~\cite{Raz93,Bea94}.

Since we have strong results about the Fourier spectra of decision trees (Proposition~\ref{prop:dt-spectrum}), and since we know random restrictions interact nicely with Fourier coefficients (Proposition~\ref{prop:rand-restrict-expansion}), \Hastad's Switching Lemma allows us to prove some strong results about Fourier concentration of narrow DNF formulas.
                                            \index{DNF!Fourier spectrum|(}
We start with an intermediate result which will be of use:
\begin{lemma}                                       \label{lem:rand-restr-dt}
    Let $f \btb$ and let $(\bJ \mid \bz)$ be a $\delta$-random restriction, $\delta > 0$.  Fix $k \in \N^+$ and write $\eps = \Pr[\DT(\restr{f}{\bJ}{\bz}) \geq k]$.  Then the Fourier spectrum of $f$ is $3\eps$-concentrated on degree up to $3k/\delta$.
\end{lemma}
\begin{proof}
    The key observation is that $\DT(\restr{f}{\bJ}{\bz}) < k$ implies $\deg(\restr{f}{\bJ}{\bz}) < k$ (Proposition~\ref{prop:dt-spectrum}), in which case the Fourier weight of $\restr{f}{\bJ}{\bz}$ at degree~$k$ and above is~$0$.  Since this weight at most~$1$ in all cases we conclude
    \[
        \Ex_{(\bJ \mid \bz)}\Bigl[\sum_{\substack{S \subseteq [n] \\ |S| \geq k}} \wh{\restr{f}{\bJ}{\bz}}(S)^2 \Bigr] \leq \eps.
    \]
    Using Proposition~\ref{prop:rand-restrict-expansion} we have
    \[
        \Ex_{(\bJ \mid \bz)}\Bigl[\sum_{\substack{S \subseteq [n] \\ |S| \geq k}} \wh{\restr{f}{\bJ}{\bz}}(S)^2 \Bigr] = \sum_{\substack{S \subseteq [n] \\ |S| \geq k}} \Ex_{(\bJ \mid \bz)}[\wh{\restr{f}{\bJ}{\bz}}(S)^2] = \sum_{U \subseteq [n]} \Pr_{(\bJ \mid \bz)}[|U \cap \bJ| \geq k] \cdot \wh{f}(U)^2.
    \]
    The distribution of random variable $|U \cap \bJ|$ is $\text{Binomial}(|U|,\delta)$.  When $|U| \geq 3k/\delta$ this random variable has mean at least~$3k$, and a Chernoff bound shows $\Pr[|U \cap \bJ| < k] \leq \exp(-\frac23 k) \leq 2/3$.  Thus
    \[
        \eps \geq \sum_{U \subseteq [n]} \Pr_{(\bJ \mid \bz)}[|U \cap \bJ| \geq k] \cdot \wh{f}(U)^2\geq \sum_{|U| \geq 3k/\delta} (1-2/3) \cdot \wh{f}(U)^2
    \]
    and hence $\sum_{|U| \geq 3k/\delta} \wh{f}(U)^2 \leq 3\eps$ as claimed.
\end{proof}

We can now improve the dependence on $\eps$ in Corollary~\ref{cor:DNF-low-deg-conc}'s low-degree spectral concentration for DNFs:
\begin{theorem}                                     \label{thm:dnf-low-degree}
    Suppose $f \btb$ is computable by a DNF of width~$w$.  Then $f$'s Fourier spectrum
    is $\eps$-concentrated on degree    up to~$O(w \log(1/\eps))$.
\end{theorem}
\begin{proof}
    This follows immediately from \Hastad's Switching Lemma together with Lemma~\ref{lem:rand-restr-dt}, taking $\delta = \frac{1}{10w}$ and $k = C \log(1/\eps)$ for a sufficiently large constant~$C$.
\end{proof}

In Lemma~\ref{lem:rand-restr-dt}, instead of using the fact that depth-$k$ decision trees have no Fourier weight above degree~$k$, we could have used the fact that their Fourier $1$-norm is at most~$2^k$.  As you are asked to show in Exercise~\ref{ex:rand-restr-dt2}, this would yield:
\begin{lemma}                                       \label{lem:rand-restr-dt2}
    Let $f \btb$ and let $(\bJ \mid \bz)$ be a $\delta$-random restriction.  Then
    \[
        \sum_{U \subseteq [n]} \delta^{|U|} \cdot |\wh{f}(U)| \leq \Ex_{(\bJ \mid \bz)}[2^{\DT(\restr{f}{\bJ}{\bz})}].
    \]
\end{lemma}
We can combine this with the Switching Lemma to deduce that width-$w$ DNFs have small Fourier $1$-norm at low degree:
\begin{theorem}                                     \label{thm:dnf-low-l1}
    Suppose $f \btb$ is computable by a DNF of width~$w$.  Then for any $k$,
    \[
        \sum_{|U| \leq k} |\wh{f}(U)| \leq 2 \cdot (20w)^k.
    \]
\end{theorem}
\begin{proof}
    Apply \Hastad's Switching Lemma to $f$ with $\delta = \frac{1}{20 w}$ to deduce
    \[
        \Ex_{(\bJ \mid \bz)}[2^{\DT(\restr{f}{\bJ}{\bz})}] \leq \sum_{d=0}^\infty \bigl(\tfrac{5}{20}\bigr)^d \cdot 2^d = 2.
    \]
    Thus from Lemma~\ref{lem:rand-restr-dt2} we get
    \[
        2 \geq \sum_{U \subseteq [n]} \bigl(\tfrac{1}{20 w}\bigr)^{|U|} \cdot |\wh{f}(U)| \geq \bigl(\tfrac{1}{20 w}\bigr)^{k} \cdot \sum_{|U| \leq k} |\wh{f}(U)|,
    \]
    as needed.
\end{proof}
Our two theorems about the Fourier structure of DNF are \emph{almost} enough to prove Mansour's Conjecture:
\begin{theorem}                                     \label{thm:almost-Mansour}
    Let $f \btb$ be computable by a DNF of width~$w \geq 2$.  Then for any $\eps \in (0,1/2]$, the Fourier spectrum of $f$ is $\eps$-concentrated on a collection $\calF$ with $|\calF| \leq w^{O(w \log(1/\eps))}$.
\end{theorem}
\begin{proof}
    Let $k = C w \log(4/\eps)$ and let $g = f^{\leq k}$.  If $C$ is a large enough constant, then Theorem~\ref{thm:dnf-low-degree} tells us that $\|f - g\|_2^2 \leq \eps/4$.  Furthermore, Theorem~\ref{thm:dnf-low-l1} gives $\snorm{g}_1 \leq w^{O(w \log(1/\eps))}$.  By Exercise~\ref{ex:l1-conc}, $g$ is $(\eps/4)$-concentrated  on some collection $\calF$ with $|\calF| \leq 4\snorm{g}_1^2/\eps \leq w^{O(w \log(1/\eps))}$.  And so by Exercise~\ref{ex:close-conc}, $f$ is $\eps$-concentrated on this same collection.
\end{proof}
For the interesting case of DNFs of width $O(\log n)$ and constant $\eps$, we get concentration on a collection of cardinality $O(\log n)^{O(\log n)} = n^{O(\log \log n)}$, nearly polynomial.  Using Proposition~\ref{prop:DNF-size-to-width} (and Exercise~\ref{ex:close-conc}) we get the same deduction for DNFs of size $\poly(n)$; more generally, for size~$s$ we have $\eps$-concentration on a collection of cardinality at most $(s/\eps)^{O(\log \log(s/\eps) \log(1/\eps))}$.
                                            \index{DNF!Fourier spectrum|)}

\section{Highlight: LMN's work on constant-depth circuits}                          \label{sec:LMN}
                                            \index{AC0@$\AC^0$|seeonly{constant-depth circuits}}%
                                            \index{circuits|see{constant-depth circuits}}%
                                            \index{constant-depth circuits|(}
Having derived strong results about the Fourier spectrum of small DNFs and CNFs, we will now extend to the case of \emph{constant-depth circuits}.  We begin by describing how H{\aa}stad applied his Switching Lemma to constant-depth circuits.  We then describe some Fourier-theoretic consequences coming from a very early (1989) work in analysis of Boolean functions by Linial, Mansour, and Nisan (LMN).

To define constant-depth circuits it is best to start with a picture.  Here is an example of a depth-$3$ circuit:
\myfig{.75}{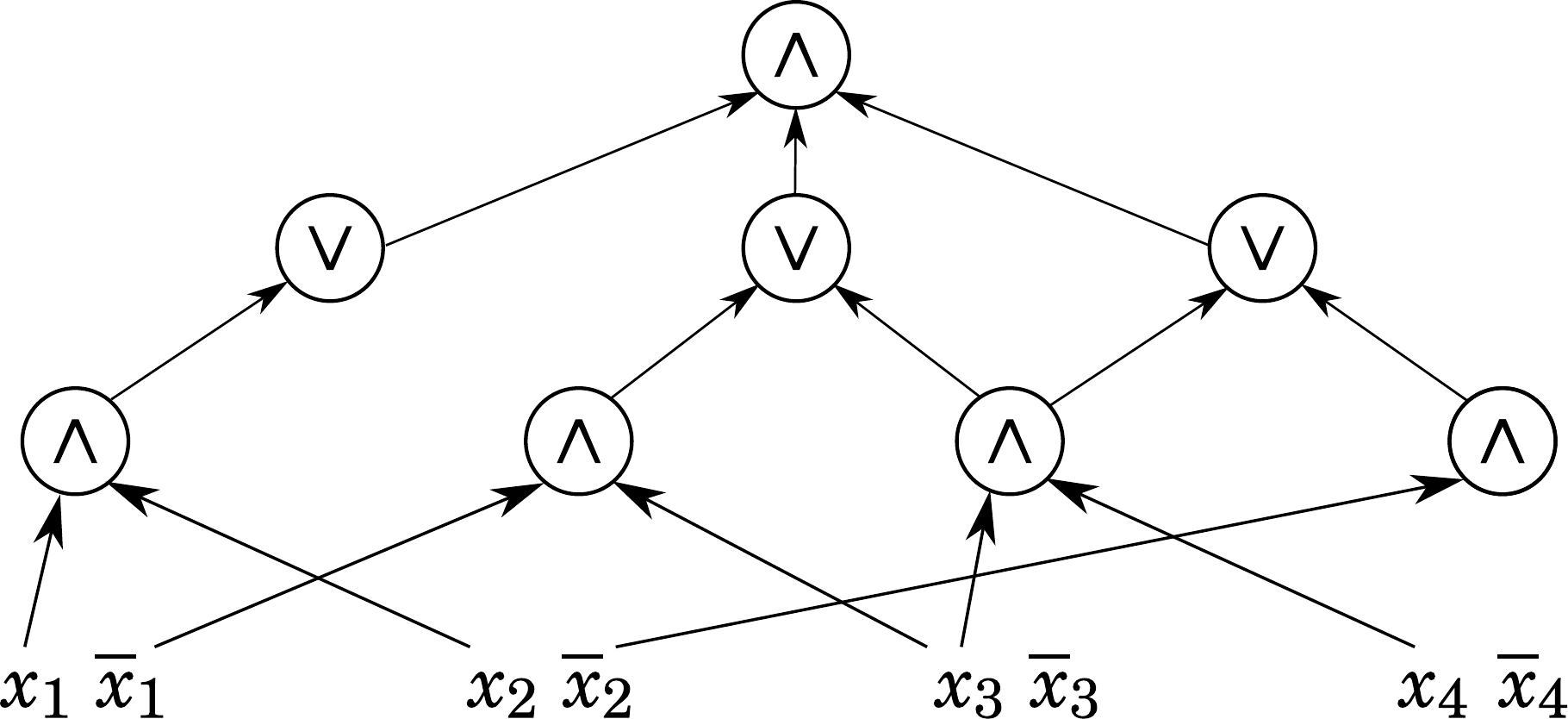}{Example of a depth-$3$ circuit, with the layer~$0$ nodes at the bottom and the layer~$3$ node at the top}{fig:circuit-example}
\noindent This circuit computes the function
\[
    x_1x_2\ \wedge\ (\overline{x}_1 x_3\ \vee\ x_3  x_4)\ \wedge\ (x_3 x_4\ \vee\ \overline{x}_2),
\]
where we suppressed the $\wedge$ in concatenated literals.  To be precise:
\begin{definition}  \label{def:constant-depth-circuit}
    For an integer $d \geq 2$, we define a \emph{depth-$d$ circuit} over Boolean variables $x_1, \dots, x_n$ as follows: It is a directed acyclic graph in which the nodes (``gates'') are arranged in $d+1$ layers, with all arcs (``wires'') going from layer~$j-1$ to layer~$j$ for some $j \in [d]$. There are exactly $2n$ nodes in layer~$0$ (the ``inputs'') and exactly~$1$ node in layer~$d$ (the ``output'').  The nodes in layer~$0$ are labeled by the $2n$ literals.  The nodes in layers $1$, $3$, $5$, etc.\ have the same label, either $\wedge$  or $\vee$, and the nodes in layers $2$, $4$, $6$, etc.\ have the other label.  Each node ``computes'' a function $\bn \to \bits$: the literals compute themselves and the $\wedge$ (respectively, $\vee$) nodes compute the logical AND (respectively, OR) of the functions computed by their incoming nodes.  The circuit itself is said to compute the function computed by its output node.
\end{definition}
In particular, DNFs and CNFs are depth-$2$ circuits.  We extend the definitions of size and width appropriately:
\begin{definition}
    The \emph{size} of a depth-$d$ circuit is defined to be the number of nodes in layers~$1$ through~$d-1$.   Its \emph{width} is the maximum in-degree of any node at layer~$1$.  (As with DNFs and CNFs, we insist that no node at layer~$1$ is connected to a variable or its negation more than once.)
\end{definition}
The layering we assume in our definition of depth-$d$ circuits can be achieved with a factor-$2d$ size overhead for any ``unbounded fan-in AND/OR/NOT circuit''.  We will not discuss any other type of Boolean circuit in this section.

We now show that \Hastad's Switching Lemma can be usefully applied not just to DNFs and CNFs but more generally to constant-depth circuits:
                                            \index{Switching Lemma!H{\aa}stad's|(}
\begin{lemma}                                       \label{lem:switch-circuits}
    Let $f \btb$ be computable by a depth-$d$ circuit  of size~$s$ and width~$w$, and let $\eps \in (0,1]$.  Set
    \[
        \delta = \frac{1}{10w} \left(\frac{1}{10\ell}\right)^{d-2}, \quad \text{where } \ell = \log(2s/\eps).
    \]
    Then if $(\bJ \mid \bz)$ is a $\delta$-random restriction, $\Pr[\DT(\restr{f}{\bJ}{\bz}) \geq \log(2/\eps)] \leq \eps$.
\end{lemma}
\begin{proof}
    The $d = 2$ case is immediate from \Hastad's Switching Lemma, so we assume $d \geq 3$.

    The first important observation is that random restrictions ``compose''. That is, making a $\delta_1$-random restriction followed by a $\delta_2$-random restriction to the free coordinates is equivalent to making a $\delta_1\delta_2$-random restriction.  Thus we can think of $(\bJ \mid \bz)$ as being produced as follows:
    \begin{enumerate}
        \item make a $\frac{1}{10w}$-random restriction;
        \item make $d-3$ subsequent $\frac{1}{10\ell}$-random restrictions;
        \item make a final $\frac{1}{10\ell}$-random restriction.
    \end{enumerate}

    Without loss of generality, assume the nodes at layer~$2$ of the circuit are labeled~$\vee$.  Thus any node~$g$ at layer~$2$ computes a DNF of width at most~$w$.  By \Hastad's Switching Lemma, after the initial $\frac{1}{10w}$-random restriction $g$~can be replaced by a decision tree of depth at most~$\ell$ except with probability at most~$2^{-\ell}$.  In particular, it can be replaced by a CNF of width at most~$\ell$, using Proposition~\ref{prop:DNF-DT}.  If we write $s_2$ for the number of nodes at layer~$2$, a union bound lets us conclude:
    \begin{equation} \label{eqn:bad-switch}
        \Pr_{\substack{\frac{1}{10w}\text{-random} \\\text{restriction}}}[\text{not all nodes at layer $2$ replaceable by width-$\ell$ CNFs}] \leq s_2 \cdot 2^{-\ell}.
    \end{equation}

    We now come to the second important observation: If all nodes at layer~$2$ can be switched to width-$\ell$ CNFs, then layers~$2$ and~$3$ can be ``compressed'', producing a depth-$(d-1)$ circuit of width at most~$\ell$. More precisely, we can form an equivalent circuit by shortening all length-$2$ paths from layer~$1$ to layer~$3$ into single arcs, and then deleting the nodes at layer~$2$.  We give an illustration of this in Figure~\ref{fig:switching-example}:
    \myfig{.5}{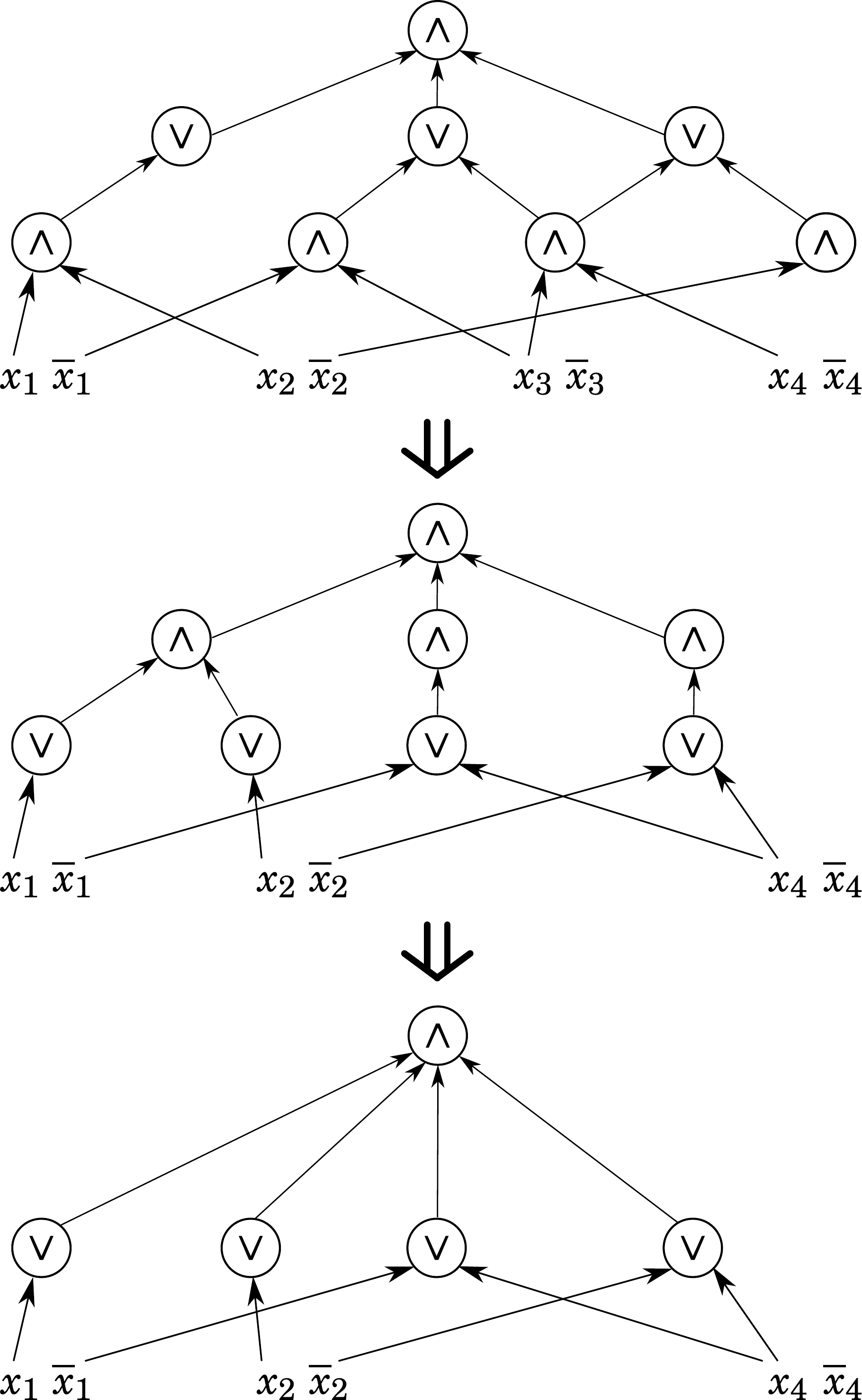}{At top is the initial circuit.  Under the restriction fixing $x_3 = \true$, all three DNFs at layer~$2$ may be replaced by CNFs of width at most~$2$.  Finally, the nodes at layers~$2$ and~$3$ may be compressed.}{fig:switching-example}
    Assuming the event in~\eqref{eqn:bad-switch} does not occur, the initial $\frac{1}{10w}$-random restriction reduces the circuit to having depth-$(d-1)$ and width at most~$\ell$.  The number of $\wedge$-nodes at the new layer~$2$ is at most~$s_3$, the number of nodes at layer~$3$ in the \emph{original} circuit.

    Next we make a $\frac{1}{10\ell}$-random restriction.  As before, by \Hastad's Switching Lemma this reduces all width-$\ell$ CNFs at the new layer~$2$ to depth-$\ell$ decision trees (hence width-$\ell$ DNFs), except with probability at most~$s_3 \cdot 2^{-\ell}$.  We may then compress layers and reduce depth again.

    Proceeding for all $\frac{1}{10\ell}$-random restrictions except the final one, a union bound gives
    \begin{multline*}
        \smash{\Pr_{\substack{\frac{1}{10w}\left(\frac{1}{10\ell}\right)^{d-3}\text{-random} \\\text{restriction}}}[\text{circuit does not reduce to depth $2$ and width $\ell$}]} \\ \leq s_2 \cdot 2^{-\ell} + s_3 \cdot 2^{-\ell} + \cdots + s_{d-1} \cdot 2^{-\ell} \leq s \cdot 2^{-\ell} = \eps/2.
    \end{multline*}
    Assuming the event above does not occur, \Hastad's Switching Lemma tells us that the final $\frac{1}{10\ell}$-random restriction reduces the circuit to a decision tree of depth less than $\log(2/\eps)$ except with probability at most~$\eps/2$.  This completes the proof.
\end{proof}
                                            \index{Switching Lemma!H{\aa}stad's|)}
We may now obtain the main theorem of Linial, Mansour, and Nisan:
                                            \index{constant-depth circuits!spectrum}
\begin{named}{LMN Theorem}  Let $f \btb$ be computable by a depth-$d$ circuit of size~$s > 1$ and let $\eps \in (0, 1/2]$.
                                            \index{LMN Theorem}
Then $f$'s Fourier spectrum is $\eps$-concentrated up to degree $O(\log(s/\eps))^{d-1} \cdot \log(1/\eps)$.
\end{named}
\begin{proof}
    If the circuit for $f$ also had width at most~$w$, we would be able to deduce $3\eps$-concentration up to degree $30w \cdot (10 \log(2s/\eps))^{d-2} \cdot \log(2/\eps)$ by combining Lemma~\ref{lem:switch-circuits} with Lemma~\ref{lem:rand-restr-dt}.  But if we simply delete all layer-$1$ nodes of width at least~$\log(s/\eps)$, the resulting circuit computes a function which is $\eps$-close to~$f$, as in the proof of Proposition~\ref{prop:DNF-size-to-width}.  Thus (using Exercise~\ref{ex:close-conc}) $f$'s spectrum is $O(\eps)$-concentrated up to degree $O(\log(2s/\eps))^{d-1} \cdot \log(2/\eps)$, and the result follows by adjusting constants.
\end{proof}
\begin{remark} \Hastad~\cite{Has01a} has slightly sharpened the degree in the LMN Theorem to $O(\log(s/\eps))^{d-2} \cdot \log(s) \cdot \log(1/\eps)$.
\end{remark}
In Exercise~\ref{ex:boppana} you are asked to use a simpler version of this proof, along the lines of Theorem~\ref{thm:dnf-tinf}, to show the following:
\begin{theorem}                                     \label{thm:boppana}
    Let $f \btb$ be computable by a depth-$d$ circuit of size~$s$.  Then $\Tinf[f] \leq O(\log s)^{d-1}$.
\end{theorem}

These rather strong Fourier concentration results for constant-depth circuits have several applications.  By introducing the Low-Degree Algorithm for learning, Linial--Mansour--Nisan gave as their main application:
                                            \index{constant-depth circuits!learning}
\begin{theorem}                                     \label{thm:learn-ac0}
    Let $\calC$ be the class of functions $f \btb$ computable depth-$d$ $\poly(n)$-size circuits.  Then $\calC$ can be learned from random examples with error any $\eps = 1/\poly(n)$ in time $n^{O(\log n)^d}$.
\end{theorem}
In complexity theory the class of poly-size, constant-depth circuits is referred to as $\AC^0$.  Thus the above theorem may be summarized as ``$\AC^0$ is learnable in quasipolynomial time''.  In fact, under a strong enough assumption about the intractability of factoring certain integers, it is known that quasipolynomial time is \emph{required} to learn $\AC^0$ circuits, even with query access~\cite{Kha93}.

The original motivation of the line of work leading to \Hastad's Switching Lemma was to show that the parity function $\chi_{[n]}$ cannot be computed in $\AC^0$.  \Hastad even showed that $\AC^0$ cannot even approximately compute
                                            \index{parity}
parity. We can derive this result from the LMN~Theorem:
\begin{corollary}                                       \label{cor:approx-parity}
    Fix any constant $\eps_0 > 0$.  Suppose $C$ is a depth-$d$ circuit over $\bits^n$ with $\Pr_{\bx}[C(\bx) = \chi_{[n]}(x)] \geq 1/2 + \eps_0$.  Then the size of $C$ is at least $2^{\Omega(n^{1/(d-1)})}$.
\end{corollary}
\begin{proof}
    The hypothesis on $C$ implies $\wh{C}([n]) \geq 2\eps_0$.  The result then follows by taking $\eps = 2\eps_0^2$ in the LMN~Theorem.
\end{proof}
This corollary is close to being tight, since the parity $\chi_{[n]}$ \emph{can} be computed by a depth-$d$ circuit of size $n2^{n^{1/(d-1)}}$ for any $d \geq 2$; see Exercise~\ref{ex:compute-parity}.  The simpler result Theorem~\ref{thm:boppana} is often handier for showing that certain functions can't be computed by $\AC^0$ circuits.  For example, we know that $\Tinf[\Maj_n] = \Theta(\sqrt{n})$; hence any constant-depth circuit computing $\Maj_n$ must have size at least~$2^{n^{\Omega(1)}}$.

Finally, Linial, Mansour, and Nisan gave an application to cryptography.
                                                \index{cryptography}
Informally, a function $f \co \bits^m \times \bits^n \to \bits$ is said to be a ``pseudorandom function generator with seed length~$m$'' if, for any efficient algorithm~$A$,
\[
    \left|\Pr_{\bs \sim \bits^m}[A(f(\bs, \cdot)) = \text{``accept''}] - \Pr_{\bg \sim \bits^{\bn}}[A(\bg) = \text{``accept''}]\right| \leq 1/n^{\omega(1)}.
\]
Here the notation $A(h)$ means that $A$ has query access to target function~$h$, and $\bg \sim \bits^{\bn}$ means that~$\bg$ is a uniformly random $n$-bit function.  In other words, for almost all ``seeds''~$\bs$ the function $f(\bs, \cdot) \btb$ is nearly indistinguishable (to efficient algorithms) from a truly random function.  Theorem~\ref{thm:boppana} shows that pseudorandom function generators cannot be computed by~$\AC^0$ circuits.  To see this, consider the algorithm $A(h)$ which chooses $\bx \sim \bn$ and $\bi \in [n]$ uniformly at random, queries $h(\bx)$ and $h(\bx^{\oplus \bi})$, and accepts if these values are unequal.  If $h$ is a uniformly random function, $A(h)$ will accept with probability~$1/2$.  In general, $A(h)$ accepts with probability $\Tinf[h]/n$.  Thus Theorem~\ref{thm:boppana} implies that if $h$ is computable in $\AC^0$ then $A(h)$ accepts with probability at most $\polylog(n)/n \ll 1/2$.
                                    \index{constant-depth circuits|)}
		
\section{Exercises and notes}
\begin{exercises}
    \item \label{ex:triv-DNF}  Show that every function $f \zotzo$ can be represented by a DNF formula of size at most~$2^n$ and width at most~$n$.
    \item \label{ex:CNF-DNF-duality}  Suppose we have a certain CNF computing $f \zotzo$.  Switch ANDs with ORs in the CNF.  Show that the result is a DNF computing the Boolean dual $f^\booldual \zotzo$.
    \item A DNF formula is said to be \emph{monotone}
                                                \index{monotone!DNF}
          if its terms contain only unnegated variables. Show that monotone DNFs compute monotone functions and that any monotone function can be computed by a monotone DNF, but that a nonmonotone DNF may compute a monotone function.
    \item \label{ex:start-Jackson} Let $f \btb$ be computable by a DNF of size~$s$.
            \begin{exercises}
                \item Show there exists $S \subseteq [n]$ with $|S| \leq \log(s) + O(1)$ and $|\wh{f}(S)| \geq \Omega(1/s)$.  (Hint: Use Proposition~\ref{prop:DNF-size-to-width} and Exercise~\ref{ex:weak-learn1}.)
                \item Let $\calC$ be the concept class of functions $\btb$ computable by DNF formulas of size at most~$s$.  Show that $\calC$ is learnable using queries with error $\half - \Omega(1/s)$ in time $\poly(n, s)$.  (Such a result, with error bounded away from $\half$, is called \emph{weak learning}.)
            \end{exercises}
    \item \label{ex:tribes-careful} Verify Proposition~\ref{prop:tribes-facts}.
    \item \label{ex:tribes-coeffs} Verify Proposition~\ref{prop:tribes-coeffs}.
    \item For each $n$ that is an input length for $\Tribes_n$, show that there exists a function $f \btb$ that is truly unbiased ($\E[f] = 0$) and has $\Inf_i[f] \leq O\bigl(\frac{\log n}{n}\bigr)$ for all $i \in [n]$.
    \item \label{ex:read-once-dnf-l1} Suppose $f \btb$ is computed by a \emph{read-once} DNF
                                        \index{DNF!read-once}
          (meaning no variable is involved in more than one term) in which all terms have width exactly~$w$. Compute
                                         \index{DNF!Fourier spectrum|)}
        $\snorm{f}_1$ exactly.  Deduce that $\snorm{\Tribes_n}_1 = 2^{\frac{n}{\log n}(1\pm o(1))}$ and that there are
                                        \index{tribes function}
        $n$-variable width-$2$ DNFs with Fourier $1$-norm $\Omega(\sqrt{3/2}^n)$.
    \item \label{ex:rand-restrict-tinf}  Give a direct (Fourier-free) proof of Corollary~\ref{cor:rand-restrict-tinf}. (Hint: Condition on whether $i \in \bJ$.)
    \item Tighten the constant factor on $\log s$ in Theorem~\ref{thm:dnf-tinf} as much as you can (avenues of improvement include the argument in Lemma~\ref{lem:rand-restr-dnf}, the choice of~$\delta$, and Exercise~\ref{ex:amano}).
    \item \label{ex:rand-restr-dt2} Prove Lemma~\ref{lem:rand-restr-dt2}.
    \item \label{ex:compute-parity}
        \begin{exercises}
            \item \label{ex:compute-parity-1} Show that the parity
                                                    \index{parity}
            function $\chi_{[n]} \btb$ can be computed by a DNF (or a CNF) of size~$2^{n-1}$.
            \item Show that the bound $2^{n-1}$ above is exactly tight. (Hint: Show that every term must have width exactly~$n$.)
            \item Show that there is a depth-$3$ circuit of size $O(n^{1/2}) \cdot 2^{n^{1/2}}$ computing~$\chi_{[n]}$.  (Hint: Break up the input into $n^{1/2}$ blocks of size $n^{1/2}$ and use~\ref{ex:compute-parity-1} twice.  How can you compress the result from depth~$4$ to depth~$3$?)
            \item More generally, show there is a depth-$d$ circuit of size $O(n^{1-1/(d-1)}) \cdot 2^{n^{1/(d-1)}}$ computing~$\chi_{[n]}$.
        \end{exercises}
    \item \label{ex:general-circuit}  In this exercise we define the most standard class of Boolean circuits.  A
                                                \index{circuits (De Morgan)}
        \emph{(De Morgan) circuit}~$C$ over Boolean variables $x_1, \dots, x_n$ is a directed acyclic graph in which each node (``gate'') is labeled with either an~$x_i$ or with $\wedge$, $\vee$, or~$\neg$ (logical NOT).  Each~$x_i$ is used as label exactly once; the associated nodes are called ``input'' gates and must have in-degree~$0$.  Each~$\wedge$ and~$\vee$ node must have in-degree~$2$, and each~$\neg$ node must have in-degree~$1$.  Each node ``computes'' a Boolean function of the inputs as in Definition~\ref{def:constant-depth-circuit}.  Finally, one node of~$C$ is designated as the ``output'' gate, and~$C$ itself is said to compute the function computed by the output node.
        For this type of circuit we define its $\emph{size}$, denoted $\size(C)$, to be the number of nodes.

        Show that each of the following $n$-input functions can be computed by De~Morgan circuits of size~$O(n)$:
        \begin{exercises}
            \item The logical AND function.
            \item The parity function.
                                                    \index{parity}
            \item The complete quadratic function from Exercise~\ref{ex:compute-expansions}.
                                                    \index{complete quadratic function}
        \end{exercises}
    \item \label{ex:tribes-CNF} Show that computing $\Tribes_{w,s}$ by a CNF formula requires size at least~$w^s$.
    \item \label{ex:tribes-not-a-junta} Show that there is a universal constant $\eps_0 > 0$ such that the following holds: Every $\frac34n$-junta~$g \btb$ is $\eps_0$-far from $\tribes_n$ (assuming~$n > 1$). (Hint: Letting~$J$ denote the coordinates on which~$g$ depends, show that if~$J$ has non-full intersection with at least~$\frac14$ of the tribes/terms then when $\bx \sim \bits^J$, there is a constant chance that $\Var[\restr{f}{}{\bx}] \geq \Omega(1)$.)
    \item \label{ex:KKL-transitive} Using the KKL~Theorem, show that if $f \btb$ is a transitive-symmetric function with $\Var[f] \geq \Omega(1)$, then $\Tinf[f] \geq \Omega(\log n)$.
    \item \label{ex:amano} Let $f \co \{\true, \false\}^n \to \{\true, \false\}$ be computable by a CNF~$C$ of width~$w$ over variables $x_1, \dots, x_n$.
                                                    \index{total influence!DNF formulas}
        In this exercise you will show that $\Tinf[f] \leq w$.

        Consider the following algorithm~$\calA$, which takes as input a permutation $\pi \in S_n$ and a ``seed'' $r \in \{\true,\false\}^n$, and which ``tries'' to output a string~$z$ satisfying~$C$:

        $\calA(\pi,r):$

        \quad For $i = \pi(1), \pi(2), \dots, \pi(n)$:

        \quad \quad  If $C$ contains the clause $(x_{i})$ \emph{and} the clause $(\ol{x}_{i})$, abort.

        \quad \quad  Else if $C$ contains just the clause $(x_{i})$, set $z_{i} = \true$.

        \quad \quad  Else if $C$ contains just the clause $(\ol{x}_{i})$, set $z_{i} = \false$.

        \quad \quad  Else set $z_{i} = r_{i}$ and say coordinate $i$ was ``unforced''.

        \quad \quad  Syntactically simplify~$C$ under the restriction $x_{i} = z_i$.

        \quad Output $z$.

        \noindent We write $F_j(\pi,r)$ for the $0$-$1$ indicator that coordinate~$j$ was \emph{forced} in the execution of~$\calA(\pi,r)$.

        \begin{exercises}
            \item Show that if $\calA(\pi,r)$ does not abort, then its output~$z$ satisfies~$C$.
            \item Fix any $y$ satisfying~$C$ and write $p(y) = \Pr_{\bpi,\br}[\calA(\bpi,\br) = y]$, where $\bpi$ and $\br$ are uniformly random.  Show that $p(y) = \E_{\bpi}[\prod_{j=1}^n (1/2)^{1-F_j(\bpi,y)}]$.
            \item Deduce $2^n p(y) \geq 2\sum_{j=1}^n \E_{\bpi}[F_j(\bpi,y)]$.
            \item Suppose further that $y^{\oplus j}$ does \emph{not} satisfy~$C$.  Show $\E_{\bpi}[F_j(\bpi,y)] \geq 1/w$.
            \item Deduce $\Tinf[f] \leq w$.
        \end{exercises}
    \item \label{ex:talagrand-random-dnf} Given Boolean variables $x_1, \dots, x_n$, a ``random monotone term of width~$w \in \N^+$'' is defined to be the logical AND of $x_{\bi_{1}}, \dots, x_{\bi_{w}}$, where $\bi_1$, \dots, $\bi_w$ are chosen independently and uniformly at random from~$[n]$.  (If the $\bi_j$'s are not all distinct then the resulting term will in fact have width strictly less than~$w$.)  A ``random monotone DNF of width~$w$ and size~$s$'' is defined to be the logical~OR of~$s$ independent random monotone terms.  For this exercise we assume~$n$ is a sufficiently large perfect square, and we let $\bvphi$ be a random monotone DNF of width $\sqrt{n}$ and size $2^{\sqrt{n}}$.
        \begin{exercises}
            \item Fix an input $x \in \bits^n$ and define $u = (\sum_{i=1}^n x_i)/\sqrt{n} \in [-\sqrt{n}, \sqrt{n}]$.  Let~$\bV_j$ be the event that the $j$th term of $\bvphi$ is made~$1$ (logical $\false$) by~$x$. Compute $\Pr[\bV_j]$ and $\Pr[\bvphi(x) = 1]$, and show that the latter is at least $10^{-9}$ assuming $|u| \leq 2$.
            \item Let $\bU_j$ be the event that the $j$th term of $\bvphi$ has exactly one~$1$ on input~$x$.  Show that $\Pr[\bU_j \mid \bV_j] \geq \Omega(w 2^{-w})$ assuming $|u| \leq 2$.
            \item Suppose we condition on $\bvphi(x) = 1$; i.e., $\cap_{j} \bV_j$.  Argue that the events~$\bU_j$ are independent.  Further, argue that for the $\bU_j$'s that do occur, the indices of their uniquely-$1$ variables are independent and uniformly random among the $1$'s of~$x$.
            \item Show that $\Pr[\sens_{\bvphi}(x) \geq c \sqrt{n} \mid \bvphi(x) = 1] \geq 1- 10^{-10}$ for $c > 0$ a sufficiently small constant.
            \item Show that $\Pr_{\bx}[|(\sum_{i=1}^n \bx_i)/\sqrt{n}| \leq 2] \geq \Omega(1)$.
            \item Deduce that there exists a monotone function $f \btb$ with the property that $\Pr_{\bx}[\sens_f(\bx) \geq c' \sqrt{n}] \geq c'$ for some universal constant $c' > 0$.
            \item Both $\Maj_n$ and the function~$f$ from the previous exercise have average sensitivity~$\Theta(\sqrt{n})$.  Contrast the ``way'' in which this occurs for the two functions.
        \end{exercises}
    \item \label{ex:baby-switch} In this exercise you will prove the Baby Switching
                                            \index{Switching Lemma!Baby}%
            Lemma with constant~$3$ in place of~$5$.  Let $\phi = T_1 \vee T_2 \vee \cdots \vee T_s$ be a DNF of width~$w \geq 1$ over variables $x_1, \dots, x_n$.  We may assume $\delta \leq 1/3$, else the theorem is trivial.
          \begin{exercises}
              \item \label{ex:baby1} Suppose $R = (J \mid z)$ is a ``bad'' restriction, meaning that $\restr{\phi}{J}{z}$ is not a constant function. Let $i$ be minimal such that $\restr{(T_i)}{J}{z}$ is neither constantly $\true$ or $\false$, and let $j$ be minimal such that $x_j$ or $\overline{x}_j$ appears in this restricted term.  Show there is a unique restriction $R' = (J \setminus \{j\} \mid z')$ extending~$R$ that doesn't falsify $T_i$.
              \item Suppose we enumerate all bad restrictions $R$, and for each we write the associated~ $R'$ as in~\ref{ex:baby1}. Show that no restriction is written more than~$w$ times.
              \item If $(\bJ \mid \bz)$ is a $\delta$-random restriction and $R$ and $R'$ are as in~\ref{ex:baby1}, show that $\Pr[(\bJ \mid \bz) = R] = \frac{2\delta}{1-\delta} \Pr[(\bJ \mid \bz) = R']$.
              \item Complete the proof by showing $\Pr[(\bJ \mid \bz) \text{ is bad}] \leq 3\delta w$.
          \end{exercises}
    \item \label{ex:boppana}  In this exercise you will prove Theorem~\ref{thm:boppana}. Say that a ``$(d, w, s')$-circuit'' is a depth-$d$ circuit with width at most~$w$ and with at most $s'$ nodes at layers~$2$ through~$d$ (i.e., excluding layers~$0$ and~$1$).
        \begin{exercises}
            \item Show by induction on $d \geq 2$ that any $f \btb$ computable by a $(d,w,s')$-circuit satisfies $\Tinf[f] \leq w O(\log s')^{d-2}$.
            \item Deduce Theorem~\ref{thm:boppana}.
        \end{exercises}
\end{exercises}

\subsection*{Notes.}

Mansour's Conjecture dates from 1994~\cite{Man94}.  Even the weaker version would imply that the Kushilevitz--Mansour algorithm learns the class of $\poly(n)$-size DNF with any constant error, using queries, in time $\poly(n)$.  In fact, this learning result was subsequently obtained in a celebrated work of Jackson~\cite{Jac97}, using a different method (which begins with Exercise~\ref{ex:start-Jackson}).  Nevertheless, the Mansour Conjecture remains important for learning theory since Gopalan, Kalai, and Klivans~\cite{GKK08} have shown that it implies the same learning result in the more challenging and realistic model of ``agnostic learning''.  Theorems~\ref{thm:dnf-low-l1} and~\ref{thm:almost-Mansour} are also due to Mansour~\cite{Man95}.

The method of random restrictions dates back to Subbotovskaya~\cite{Sub61}.  \Hastad's Switching Lemma~\cite{Has87} and his Lemma~\ref{lem:switch-circuits} are the culmination of a line of work due to Furst, Saxe, and Sipser~\cite{FSS84}, Ajtai~\cite{Ajt83}, and Yao~\cite{Yao85}.  Linial, Mansour, and Nisan~\cite{LMN89,LMN93} proved Lemma~\ref{lem:rand-restr-dt}, which allowed them to deduce the LMN~Theorem and its consequences.
An additional cryptographic application of the LMN~Theorem is found in Goldmann and Russell~\cite{GR00a}.  The strongest lower bound currently known for approximately computing parity in~$\AC^0$ is due to Impagliazzo, Matthews, and Paturi~\cite{IMP12} and independently to H{\aa}stad~\cite{Has12a}.

Theorem~\ref{thm:dnf-tinf} and its generalization Theorem~\ref{thm:boppana} are from the work of Boppana~\cite{Bop97}; Linial, Mansour, and Nisan had given the weaker bound $O(\log s)^d$. Exercise~\ref{ex:amano} is due to Amano~\cite{Ama11}, and Exercise~\ref{ex:talagrand-random-dnf} is due to Talagrand~\cite{Tal96}.

\chapter{Majority and threshold functions}                              \label{chap:thresholds}

This chapter is devoted to linear threshold functions, their generalization to higher degrees, and their exemplar the majority function.  The study of LTFs leads naturally to the introduction of the Central Limit Theorem and Gaussian random variables -- important tools in analysis of Boolean functions.  We will first use these tools to analyze the Fourier spectrum of the $\maj_n$ function, which in some sense ``converges'' as $n \to \infty$.  We'll then extend to analyzing the degree-$1$ Fourier weight, noise stability, and total influence of general linear threshold functions.

\section{Linear threshold functions and polynomial threshold functions} \label{sec:ltfs-ptfs}

                    \index{linear threshold function|(}
Recall from Chapter~\ref{sec:social-choice} that a linear threshold function (abbreviated~LTF) is a Boolean-valued function $f \btb$ that can be represented as
\begin{equation} \label{eqn:generic-LTF}
    f(x) = \sgn(a_0 + a_1 x_1 + \cdots + a_n x_n)
\end{equation}
for some constants $a_0, a_1, \dots, a_n \in \R$.  (For definiteness we'll take $\sgn(0) = 1$.  If we're using the representation $f \btzo$, then $f$ is an LTF if it can be represented as $f(x) = 1_{\{a_0 + a_1 x_1 + \cdots + a_n x_n > 0\}}$.) Examples include majority, AND, OR, dictators, and decision lists (Exercise~\ref{ex:decision-list}).  Besides representing ``weighted majority'' voting schemes, LTFs play an important role in learning theory and in circuit complexity.

There is also a geometric perspective on LTFs.  Writing $\ell(x) = a_0 + a_1 x_1 + \cdots + a_n x_n$, we can think of $\ell$ as an affine function $\R^n \to \R$. Then $\sgn(\ell(x))$ is the $\pm 1$-indicator of a \emph{halfspace} in $\R^n$.  A Boolean LTF is thus the restriction of such a halfspace-indicator to the discrete cube $\bn \subset \R^n$.  Equivalently, a function $f \btb$ is an LTF if and only if it has a ``linear separator''; i.e., a hyperplane in $\R^n$ that separates the points $f$ labels~$1$ from the points $f$ labels~$-1$.

An LTF $f \btb$ can have several different representations as in~\eqref{eqn:generic-LTF} -- in fact it always has infinitely many.  This is clear from the geometric viewpoint; any small enough perturbation to a linear separator will not change the way it partitions the discrete cube.  Because we can make these perturbations, we may ensure that $a_0 + a_1 x_1 + \cdots + a_n x_n \neq 0$ for every $x \in \bn$.  We'll usually insist that LTF representations have this property so that the nuisance of $\sgn(0)$ doesn't arise.  We also observe that we can scale all of the coefficients in an LTF representation by the same positive constant without changing the LTF. These observations can be used to show it's always possible to take the $a_i$'s to be integers (Exercise~\ref{ex:LTF-integer}).  However, we will most often scale so that $\sum_{i=1}^n a_i^2 = 1$; this is convenient when using the Central Limit Theorem.
                    \index{linear threshold function|)}

The most elegant result connecting LTFs and Fourier expansions is
                                            \index{Chow's Theorem}
Chow's Theorem, which says that a Boolean LTF is completely determined by its degree-$0$ and degree-$1$ Fourier coefficients.  In fact, it's determined not just within the class of LTFs but within the class of all Boolean functions:
\begin{theorem}                                     \label{thm:chow}
    Let $f \btb$ be an LTF and let $g \btb$ be any function.  If $\wh{g}(S) = \wh{f}(S)$ for all $|S| \leq 1$, then $g = f$.
\end{theorem}
\begin{proof}
    Let $f(x) = \sgn(\ell(x))$, where $\ell \btR$ has degree at most~$1$ and is never~$0$ on $\bn$.  For any $x \in \bn$ we have $f(x)\ell(x) = |\ell(x)| \geq g(x)\ell(x)$, with equality if and only if $f(x) = g(x)$ (here we use $\ell(x) \neq 0$).  Using this observation along with Plancherel's Theorem (twice) we have
    \[
        \sum_{|S| \leq 1} \wh{f}(S) \wh{\ell}(S) = \E[f(\bx)\ell(\bx)] \geq \E[g(\bx) \ell(\bx)] = \sum_{|S| \leq 1} \wh{g}(S) \wh{\ell}(S).
    \]
    But by assumption, the left-hand and right-hand sides above are equal.  Thus the inequality must be an equality for every value of~$\bx$; i.e., $f(x) = g(x)$~$\forall x$.
\end{proof}
\noindent In light of Chow's Theorem, the $n+1$ numbers $\wh{g}(\emptyset)$, $\wh{g}(\{1\}), \dots, \wh{g}(\{n\})$ are sometimes called the \emph{Chow parameters}
                                            \index{Chow parameters}
of the Boolean function~$g$.

As we will show in Section~\ref{sec:peres}, linear threshold functions are very noise-stable; hence they have a lot of
                                \index{linear threshold function!Fourier weight|(}
their Fourier weight at low degrees.  Here is a simple result along these lines:
\begin{theorem}                                     \label{thm:gotsman-linial1}
    Let $f \btb$ be an LTF.
                                            \index{Gotsman--Linial Theorem}%
    Then $\W{\leq 1}[f] \geq 1/2$.
\end{theorem}
\begin{proof}
    Writing $f(x) = \sgn(\ell(x))$ we have
    \[
        \|\ell\|_1 = \E[|\ell(\bx)|] = \la f, \ell \ra = \la f^{\leq 1}, \ell \ra \leq \|f^{\leq 1}\|_2 \|\ell\|_2 = \sqrt{\W{\leq 1}[f]} \cdot \|\ell\|_2,
    \]
    where the third equality follows from Plancherel and the inequality is Cauchy--Schwarz.  Assume first that $\ell(x) = a_1 x_1 + \cdots + a_n x_n$ (i.e., $\ell(x)$ has no constant term).  The Khintchine--Kahane Inequality
                                                \index{Khintchine(--Kahane) Inequality}%
    (Exercise~\ref{ex:lo-kk}) states that $\|\ell\|_1 \geq \frac{1}{\sqrt{2}} \|\ell\|_2$, and hence we deduce
    \[
        \tfrac{1}{\sqrt{2}} \|\ell\|_2 \leq \sqrt{\W{\leq 1}[f]} \cdot \|\ell\|_2.
    \]
    The conclusion $\W{\leq 1}[f] \geq 1/2$ follows immediately (since $\|\ell\|_2$ cannot be~$0$).  The case when $\ell(x)$ has a constant term is handled in~Exercise~\ref{ex:finish-gotsman-linial1}.
\end{proof}
From Exercise~\ref{ex:maj-influences} we know that $\W{\leq 1}[\Maj_n] = \W{1}[\Maj_n] \geq 2/\pi$ for all~$n$; it is reasonable to conjecture that majority is extremal for Theorem~\ref{thm:gotsman-linial1}.  This is an open problem.
\begin{conjecture}  \label{conj:ltf-weight-1} Let $f \btb$ be an LTF.  Then $\W{\leq 1}[f] \geq 2/\pi$.
\end{conjecture}
                                \index{linear threshold function!Fourier weight|)}

A natural generalization of linear threshold functions is \emph{polynomial threshold functions}:
\begin{definition} A function $f \btb$ is called a \emph{polynomial threshold function (PTF)} of degree at most~$k$ if it is
            \index{polynomial threshold function|(}%
            \index{PTF|seeonly{polynomial threshold function}}%
expressible as $f(x) = \sgn(p(x))$ for some real polynomial $p \btR$ of degree at most~$k$.
\end{definition}
\begin{example} \label{eg:ptf}
    Let $f \co \bits^4 \to \bits$ be the $4$-bit equality function, which is~$1$ if and only if all input bits are equal.  Then $f$ is a degree-$2$ PTF because it has the representation $f(x) = \sgn(-3 + x_1x_2 + x_1x_3 + x_1x_4 + x_2x_3 + x_2x_4 + x_3x_4)$.
\end{example}
\emph{Every} Boolean function $f \btb$ is a PTF of degree at most~$n$, since we can take the sign of its Fourier expansion.  Thus we are usually interested in the case when the degree $k$ is ``small'', say, $k = O_n(1)$.  Low-degree PTFs arise frequently in learning theory, for example, as hypotheses in the Low-Degree Algorithm and many other practical learning algorithms.  Indeed, any function with low noise sensitivity is close to being a low-degree PTF; by combining Propositions~\ref{prop:ns-concentration} and~\ref{prop:sign-conc} we immediately obtain:
\begin{proposition}                                     \label{prop:noise-stable-implies-PTF}
    Let $f\btb$ and let $\delta \in (0, 1/2]$.  Then $f$ is $(3\NS_\delta[f])$-close to a PTF of degree~$1/\delta$.
\end{proposition}
\noindent For a kind of converse to this proposition, see Section~\ref{sec:peres}.

PTFs also  arise in circuit complexity, wherein a PTF representation
\[
    f(x) = \sgn\left(\sum_{i=1}^s {a_i} x^{T_i}\right)
\]
is thought of as a ``threshold-of-parities circuit'':
                                                    \index{threshold-of-parities circuit}
i.e., a depth-$2$ circuit with $s$ ``parity gates''~$x^{T_i}$ at layer~$1$ and a single ``(linear) threshold gate'' at layer~$2$. From this point of view, the size of the circuit corresponds to the \emph{sparsity} of the PTF representation:
\begin{definition}
    We say a PTF representation $f(x) = \sgn(p(x))$ has \emph{sparsity} at most~$s$ if $p(x)$ is a multilinear
                        \index{polynomial threshold function!sparsity}
    polynomial with at most~$s$ terms.
\end{definition}
\noindent For example, the PTF representation of the $4$-bit equality function from Example~\ref{eg:ptf} has sparsity~$7$.
                            \index{polynomial threshold function|)}%

Let's extend the two theorems about LTFs we proved above to the case of PTFs.  The generalization of Chow's Theorem is straightforward; its proof is left as Exercise~\ref{ex:ptf-chow}:
\begin{theorem}                                     \label{thm:ptf-chow}
    Let $f \btb$ be a PTF of degree at most~$k$ and let $g \btb$ be any function.
                                                        \index{Chow's Theorem!for polynomial threshold functions}
    If $\wh{g}(S) = \wh{f}(S)$ for all $|S| \leq k$, then $g = f$.
\end{theorem}
We also have the following extension of Theorem~\ref{thm:gotsman-linial1}:
\begin{theorem}                                     \label{thm:gotsman-linial2}
    Let $f \btb$ be a degree-$k$ PTF.
                                            \index{Gotsman--Linial Theorem}%
    Then
                                    \index{polynomial threshold function!Fourier spectrum|(}
    $\W{\leq k}[f] \geq e^{-2k}$.
\end{theorem}
\begin{proof}
    Writing $f(x) = \sgn(p(x))$ for $p$ of degree~$k$, we again have
    \[
        \|p\|_1 = \E[|p(\bx)|] = \la f, p \ra = \la f^{\leq k}, p \ra \leq \|f^{\leq k}\|_2 \|p\|_2 = \sqrt{\W{\leq k}[f]} \cdot \|p\|_2.
    \]
    To complete the proof we need the fact that $\|p\|_2 \leq e^k \|p\|_1$ for any degree-$k$ polynomial $p \btR$.  We will prove this much later in Theorem~\ref{thm:p-2-bonami} of Chapter~\ref{chap:hypercontractivity} on
                                                    \index{hypercontractivity}%
    hypercontractivity.
\end{proof}
\noindent The $e^{-2k}$ in this theorem cannot be improved beyond $2^{1-k}$; see Exercise~\ref{ex:no-beat-gl}.

We close this section by discussing PTF sparsity.  We begin with a (simpler) variant of Theorem~\ref{thm:gotsman-linial2}, which is useful for proving PTF sparsity lower bounds:
\begin{theorem}                                     \label{thm:bruck}
    Let $f \btb$ be expressible as a PTF over the collection of monomials $\calF \subseteq 2^{[n]}$; i.e., $f(x) = \sgn(p(x))$ for some polynomial $p(x) = \sum_{S \in \calF} \wh{p}(S) x^S$.  Then $\sum_{S \in \calF} |\wh{f}(S)| \geq 1$.
\end{theorem}
\begin{proof}
    Define $g \btR$ by $g(x) = \sum_{S \in \calF} \wh{f}(S)\,x^S$.  Since $\snorm{p}_\infty \leq \|p\|_1$ (Exercise~\ref{ex:hausdorff-young}) we have
    \[
       \snorm{p}_\infty \leq \|p\|_1 = \E[f(\bx)p(\bx)] = \sum_{S \subseteq [n]} \wh{f}(S) \wh{p}(S) = \sum_{S \in \calF} \wh{g}(S) \wh{p}(S) \leq \snorm{g}_1 \snorm{p}_\infty,
    \]
    and hence $\snorm{g}_1 \geq 1$ as claimed.
\end{proof}
                                    \index{polynomial linear threshold function!Fourier spectrum|)}
We can use this result to show that the ``inner product mod $2$ function'' (see Exercise~\ref{ex:compute-expansions}) requires huge
                                                    \index{threshold-of-parities circuit}
threshold-of-parities circuits:
\begin{corollary}                                       \label{cor:ip-no-sparse-PTF}
    Any PTF representation of the
                                                    \index{inner product mod $2$ function}
    inner product mod $2$ function $\IP_{2n} \co \F_2^{2n} \to \{-1,1\}$ has sparsity at least~$2^n$.
\end{corollary}
\begin{proof}
    This follows immediately from Theorem~\ref{thm:bruck} and the fact that $|\wh{\IP}_{2n}(S)| = 2^{-n}$ for all $S \subseteq [2n]$ (Exercise~\ref{ex:compute-expansions}).
\end{proof}

We can also show that any function $f \btb$ with small Fourier $1$-norm $\snorm{f}_1$ has a sparse PTF representation.  In fact a stronger result holds: such a function can be additively \emph{approximated} by a sparse
                                                    \index{approximating polynomial}
polynomial:
\begin{theorem}                                     \label{thm:bruck-smolensky}
    Let $f \btR$ be nonzero, let $\delta > 0$, and let $s \geq 4n\snorm{f}_1^2/\delta^2$ be an integer.  Then there is a multilinear polynomial $q \btR$ of sparsity at most~$s$ such that $\|f - q\|_\infty < \delta$.
\end{theorem}
\begin{proof}
    The proof is by the probabilistic method.  Let $\bT \subseteq [n]$ be randomly chosen according to the distribution ${\Pr[\bT = T]} = \frac{|\wh{f}(T)|}{\snorm{f}_1}$.  Let $\bT_1, \dots, \bT_s$ be independent draws from this distribution and define the multilinear polynomial
    \[
        \bp(x) = \sum_{i=1}^s \sgn(\wh{f}(\bT_i))\,x^{\bT_i}.
    \]
    When $x \in \bn$ is fixed, each monomial $\sgn(\wh{f}(\bT_i))\,x^{\bT_i}$ becomes a $\pm 1$-valued random variable with expectation
    \[
        \sum_{T \subseteq [n]} \tfrac{|\wh{f}(T)|}{\snorm{f}_1}\cdot \sgn(\wh{f}(T))\,x^{T} = \tfrac{1}{\snorm{f}_1} \sum_{T \subseteq [n]} \wh{f}(T)\,x^T = \tfrac{f(x)}{\snorm{f}_1}.
    \]
    Thus by a Chernoff bound, for any $\eps > 0$,
    \[
        \Pr_{\bT_1, \dots, \bT_s}\left[\Bigl|\bp(x) - \tfrac{f(x)}{\snorm{f}_1} s\Bigr| \geq \eps s\right] \leq 2\exp(-\eps^2 s/2).
    \]
    Selecting $\eps = \delta/\snorm{f}_1$ and using $s \geq 4n\snorm{f}_1^2/\delta^2$, the above probability is at most $2\exp(-2n) < 2^{-n}$.  Taking a union bound over all $2^n$ choices of $x \in \bn$, we conclude that there exists some $p(x) = \sum_{i=1}^s \sgn(\wh{f}(T_i))\,x^{T_i}$ such that for all $x \in \bn$,
    \[
         \Bigl|p(x) - \tfrac{f(x)}{\snorm{f}_1} s\Bigr| < \eps s = \tfrac{\delta}{\snorm{f}_1} s \quad\implies\quad           \Bigl|\tfrac{\snorm{f}_1}{s} \cdot p(x) - f(x) \Bigr| < \delta.
    \]
    Thus we may take $q = \frac{\snorm{f}_1}{s} \cdot p$.
\end{proof}
\begin{corollary}                                     \label{cor:bruck-smolensky}
    Let $f \btb$.  Then $f$ is expressible as a PTF of sparsity at most $s = \lceil 4n\snorm{f}_1^2\rceil$.  Indeed, $f$
                                                \index{polynomial threshold function!sparsity}%
     can be represented as a majority of~$s$ parities or negated-parities.
\end{corollary}
\begin{proof}
    Apply the previous theorem with $\delta = 1$;  we then have $f(x) = \sgn(q(x))$.  Since this is also equivalent to $\sgn(p(x))$, the terms $\sgn(\wh{f}(T_i))\,x^{T_i}$ are the required parities/negated-parities.
\end{proof}
Though functions computable by small DNFs need not have small Fourier $1$-norm, it is a further easy corollary that they can be computed by sparse PTFs: see Exercise~\ref{ex:krause-pudlak}.  We also remark that there is no good converse to Corollary~\ref{cor:bruck-smolensky}: the $\maj_n$ function has a PTF (indeed, an LTF) of sparsity~$n$ but has exponentially large Fourier $1$-norm (Exercise~\ref{ex:L1-maj}).

\section{Majority, and the Central Limit Theorem}                  \label{sec:majority}
Majority is one of the more important functions in Boolean analysis, and its study motivates the introduction of one of the more important tools: the Central Limit Theorem~(CLT).  In this section we will show how the CLT can be used to estimate the total influence and the noise stability of $\Maj_n$.  Though we already determined $\Tinf[\Maj_n] \sim \sqrt{2/\pi}\sqrt{n}$ in Exercise~\ref{ex:maj-influences} using binomial coefficients and Stirling's Formula, computations using the~CLT are more flexible and extend to other linear threshold functions.

We begin with a reminder about the CLT.
										 \index{CLT|seeonly{Central Limit Theorem}}%
                                        \index{Central Limit Theorem}%
Suppose $\bX_1, \dots, \bX_n$ are independent random variables and $\bS = \bX_1 + \cdots + \bX_n$.  Roughly speaking, the CLT says that so long as no $\bX_i$ is too dominant in terms of variance, the distribution of~$\bS$ is close to that of a Gaussian random variable with the same mean and variance.   Recall:
\begin{notation}
    We write $\bZ \sim \normal(0,1)$ denote that~$\bZ$ is a standard Gaussian random variable.  We use the notation
    \[
        \vphi(z) = \tfrac{1}{\sqrt{2\pi}}e^{-z^2/2}, \quad \Phi(t) = \int_{-\infty}^t \phi(z)\,dz, \quad \olPhi(t) = \Phi(-t) = \int_{t}^\infty \phi(z)\,dz
    \]
    for the pdf, cdf, and complementary cdf of this random variable.
                                         \index{Gaussian random variable}%
                                         \index{normal random variable|seeonly{Gaussian random variable}}%
                                         \nomenclature[Nz(0,1)]{$\normal(0,1)$}{the standard Gaussian distribution}%
                                         \nomenclature[Nz(0,1)d]{$\normal(0,1)^d$}{the distribution of $d$ independent standard Gaussians; i.e., $\normal(0, I_{d \times d})$}%
                                         \nomenclature[Nz(mu,Sigma)]{$\normal(\mu,\Sigma)$}{for $\mu \in \R^d$ and $\Sigma \in \R^{d \times d}$ positive semidefinite, the $d$-variate Gaussian distribution with mean $\mu$ and covariance matrix~$\Sigma$}%
                                        \nomenclature[phi]{$\phi$}{the standard Gaussian pdf, $\phi(z) = \tfrac{1}{\sqrt{2\pi}}e^{-z^2/2}$}%
                                        \nomenclature[Phi1]{$\Phi$}{the standard Gaussian cdf, $\Phi(t) = \int_{-\infty}^t \phi(z)\,dz$}%
                                        \nomenclature[Phi2]{$\olPhi$}{the standard Gaussian complementary cdf, $\olPhi(t) = \int_{t}^\infty \phi(z)\,dz$}%
    More generally, if $\mu \in \R^d$ and $\Sigma \in \R^{d \times d}$ is a positive semidefinite matrix, we write $\bZ \sim \normal(\mu, \Sigma)$ to denote that $\bZ$ is a $d$-dimensional random vector with mean~$\mu$ and covariance matrix~$\Sigma$.
\end{notation}
We give a precise statement of the CLT below in the form of the \emph{Berry--Esseen Theorem}.  The CLT also extends to the \emph{multidimensional} case (sums of independent random vectors); we give a precise statement in Exercise~\ref{ex:ltf-stab-error}.  In Chapter~\ref{chap:invariance} we will show one way to prove such CLTs.

											 \index{majority!total influence|(}
Let's see how we can use the CLT to obtain the estimate $\Tinf[\Maj_n] \sim \sqrt{2/\pi}\sqrt{n}$.  Recall the proof of Theorem~\ref{thm:maj-maximizes-deg-1-sum}, which shows that $\Maj_n$ maximizes $\sum_{i=1}^n \wh{f}(i)$ among all $f \btb$.  In it we saw that
\begin{equation} \label{eqn:sum-deg-1-maj}
     \Tinf[\Maj_n] = \sum_{i=1}^n \wh{\Maj_n}(i) =  \Ex_{\bx}[\Maj_n(\bx)(\littlesum_i \bx_i)] = \Ex_{\bx}[|\littlesum_i \bx_i|].
\end{equation}
When using the CLT, it's convenient to define majority (equivalently) as
\[
	\Maj_n(x) = \sgn\Bigl(\littlesum_{i=1}^n \tfrac{1}{\sqrt{n}} x_i\Bigr).
\]
This motivates writing~\eqref{eqn:sum-deg-1-maj} as
\begin{equation} \label{eqn:sum-deg-1-maj2}
     \Tinf[\Maj_n] = \sqrt{n} \cdot \Ex_{\bx \sim \bn}[|\littlesum_i \tfrac{1}{\sqrt{n}}\bx_i|].
\end{equation}
If we introduce $\bS = \sum_{i=1}^n \tfrac{1}{\sqrt{n}} \bx_i$, then $\bS$ has mean~$0$ and variance $\sum_i (1/\sqrt{n})^2 = 1$.  Thus the CLT tells us that the distribution of~$\bS$ is close (for large~$n$) to that of a standard Gaussian, $\bZ \sim \normal(0,1)$.
                             \index{Gaussian random variable}%
                             \index{normal random variable|seeonly{Gaussian random variable}}%
So as $n \to \infty$ we have
\begin{equation} \label{eqn:abs-first-moment-gaussian}
\Ex_{\bx}[|\bS|] \sim \Ex_{\bZ \sim \normal(0,1)}[|\bZ|] = 2\int_{0}^\infty z \cdot \tfrac{1}{\sqrt{2\pi}} e^{-z^2/2}\,dz = -\sqrt{2/\pi} e^{-z^2/2}\Bigm|_{0}^\infty = \sqrt{2/\pi},
\end{equation}
which when combined with~\eqref{eqn:sum-deg-1-maj2} gives us the estimate $\Tinf[\Maj_n] \sim \sqrt{2/\pi} \sqrt{n}$.
											 \index{majority!total influence|)}

To make this kind of estimate more precise we state the Berry--Esseen Theorem, which is a strong version
                                            \index{Berry--Esseen Theorem}%
                                            \index{Central Limit Theorem}%
of the CLT giving explicit error bounds rather than just limiting statements.
\begin{named}{Berry--Esseen (Central Limit) Theorem}
    Let $\bX_1, \dots, \bX_n$ be independent random variables with $\E[\bX_i] = 0$ and $\Var[\bX_i] = \sigma_i^2$, and assume $\sum_{i=1}^n \sigma_i^2 = 1$.  Let $\bS = \sum_{i=1}^n \bX_i$ and let $\bZ \sim \normal(0,1)$ be a standard Gaussian. Then for all $u \in \R$,
    \[
        |\Pr[\bS \leq u] - \Pr[\bZ \leq u]| \leq c \gamma,
    \]
    where
    \[
        \gamma = \sum_{i=1}^n \|\bX_i\|_3^3
    \]
    and $c$ is a universal constant.  (For definiteness, $c = .56$ is acceptable.)
\end{named}
\begin{remark}  \label{rem:simpler-BE}
    If all of the $\bX_i$'s satisfy $|\bX_i| \leq \eps$ with probability~$1$, then we can use the bound
    \[
        \gamma = \sum_{i=1}^n \E[|\bX_i|^3] \leq \eps \cdot \sum_{i=1}^n \E[|\bX_i|^2] = \eps \cdot \sum_{i=1}^n \sigma_i^2 = \eps.
    \]
\end{remark}
\noindent See Exercises~\ref{ex:BE-interval} and~\ref{ex:no-restrict} for some additional observations.

Our most frequent use of the Berry--Esseen Theorem will be in analyzing random sums
\[
    \bS = \sum_{i=1}^n a_i \bx_i,
\]
where $\bx \sim \bn$ and the constants $a_i \in \R$ are normalized so that $\sum_i a_i^2 = 1$. For majority, all of the $a_i$'s were equal to $\tfrac{1}{\sqrt{n}}$. But from Remark~\ref{rem:simpler-BE} we see that $\bS$ is close in distribution to a standard Gaussian so long as each $|a_i|$ is small.  For example, in Exercise~\ref{ex:l1-be} you are asked to show the following:
\begin{theorem}                                     \label{thm:l1-be}
    Let $a_1, \dots, a_n \in \R$ satisfy $\sum_i a_i^2 = 1$ and $|a_i| \leq \eps$ for all~$i$.  Then
    \[
        \left|\Ex_{\bx \sim \bn}[|\littlesum_i a_i \bx_i|] - \sqrt{2/\pi}\right| \leq C \eps,
    \]
    where $C$ is a universal constant.
\end{theorem}
Theorem~\ref{thm:l1-be} justifies~\eqref{eqn:abs-first-moment-gaussian} with an error bound of~$O(1/\sqrt{n})$, yielding the more precise estimate $\Tinf[\Maj_n] = \sqrt{2/\pi}\sqrt{n} \pm O(1)$ (cf.~Exercise~\ref{ex:maj-influences}, which gives an even better error bound).

                                                \index{majority!noise stability|(}
Now let's turn to the noise stability of majority.  Theorem~\ref{thm:maj-stab} stated the  formula
\begin{equation} \label{eqn:re-maj-stab}
	\lim_{n \to \infty} \Stab_\rho[\Maj_n] = \tfrac{2}{\pi} \arcsin \rho = 1 - \tfrac{2}{\pi} \arccos \rho.
\end{equation}
Let's now spend some time justifying this using the multidimensional CLT. (For complete details, see Exercise~\ref{ex:ltf-stab-error}.) By definition,
\begin{equation} \label{eqn:stab-maj-justify1}
	\Stab_\rho[\Maj_n] =  \Ex_{\substack{(\bx, \by) \\ \mathclap{\text{ $\rho$-correlated}}}}[\Maj_n(\bx) \cdot \Maj_n(\by)] = \Es{(\bx, \by) \\ \mathclap{\text{ $\rho$-correlated}}}[\sgn(\littlesum_i \tfrac{1}{\sqrt{n}} \bx_i) \cdot \sgn(\littlesum_i \tfrac{1}{\sqrt{n}} \by_i)].
\end{equation}
For each $i \in [n]$ let's stack $\tfrac{1}{\sqrt{n}} \bx_i$ and $\tfrac{1}{\sqrt{n}} \by_i$ into a $2$-dimensional vector and then write
\begin{equation} \label{eqn:Svec-cov}
    \vec{\bS} = \sum_{i=1}^n \begin{bmatrix}
                           \tfrac{1}{\sqrt{n}}\bx_i \\
                           \tfrac{1}{\sqrt{n}}\by_i
                       \end{bmatrix} \in \R^2.
\end{equation}
We are summing $n$ independent random vectors, so the multidimensional CLT tells us that the distribution of $\vec{\bS}$ is close to that of
											 \index{Central Limit Theorem!multidimensional}
a $2$-dimensional Gaussian $\vec{\bZ}$ with the same mean and covariance matrix, namely (see Exercise~\ref{ex:Svec-cov})
\[
    \vec{\bZ} \sim \normal\left(\begin{bmatrix} 0 \\ 0 \end{bmatrix}, \begin{bmatrix} 1 & \rho \\ \rho & 1 \end{bmatrix}\right).
\]
Continuing from~\eqref{eqn:stab-maj-justify1},
\begin{align*}
    \Stab_\rho[\Maj_n] &= \E[\sgn(\vec{\bS}_1)\cdot \sgn(\vec{\bS}_2)] \\
    &= \Pr[\sgn(\vec{\bS}_1) = \sgn(\vec{\bS}_2)] - \Pr[\sgn(\vec{\bS}_1) \neq \sgn(\vec{\bS}_2)] \\
    &= 2\Pr[\sgn(\vec{\bS}_1) = \sgn(\vec{\bS}_2)] - 1 = 4\Pr[\vec{\bS} \in Q_{--}] - 1,
\end{align*}
where $Q_{--}$ denotes the lower-left quadrant of $\R^2$ and the last step uses the symmetry $\Pr[\vec{\bS} \in Q_{++}] = \Pr[\vec{\bS} \in Q_{--}]$.  Since $Q_{--}$ is convex, the $2$-dimensional CLT lets us deduce
\[
    \lim_{n \to \infty} \Pr[\vec{\bS} \in Q_{--}] = \Pr[\vec{\bZ} \in Q_{--}].
\]
So to justify the noise stability formula~\eqref{eqn:re-maj-stab} for majority, it remains to verify
\[
4\Pr[\vec{\bZ} \in Q_{--}] - 1 = 1 - \tfrac{2}{\pi} \arccos \rho \quad\iff\quad \Pr[\vec{\bZ} \in Q_{--}] =  \frac12 - \frac{1}{2} \frac{\arccos \rho}{\pi}.
\]
And this in turn is a $19$th-century identity known as
                                                    \index{Sheppard's Formula}%
                                                    \index{Gaussian quadrant probability}%
\emph{Sheppard's Formula}:
\begin{named}{Sheppard's Formula}  Let $\bz_1$, $\bz_2$ be standard Gaussian random variables with correlation $\E[\bz_1 \bz_2] = \rho \in [-1,1]$.  Then
\[
    \Pr[\bz_1 \leq 0, \bz_2 \leq 0] = \frac12 - \frac{1}{2}\frac{\arccos \rho}{\pi}.
\]
\end{named}
\noindent Proving Sheppard's Formula is a nice exercise using the rotational symmetry of a pair of independent standard Gaussians; we defer the proof till Example~\ref{eg:sheppard} in Chapter~\ref{sec:gaussian}.  This completes the justification of formula~\eqref{eqn:re-maj-stab} for the limiting noise stability of majority.

You may have noticed that once we applied the $2$-dimensional CLT to~\eqref{eqn:stab-maj-justify1}, the remainder of the derivation had nothing to do with majority.  In fact, the same analysis works for \emph{any} linear threshold function $\sgn(a_1 x_1 + \cdots + a_n x_n)$, the only difference being the ``error term'' arising from the CLT.  As in Theorem~\ref{thm:l1-be}, this error is small so long as no coefficient~$a_i$ is too dominant:
\begin{theorem}                                     \label{thm:ltf-stab-error}
    Let $f \btb$ be an unbiased LTF, $f(x) = \sgn(a_1 x_1 + \cdots + a_n x_n)$ with $\sum_i a_i^2 = 1$ and $|a_i| \leq \eps$ for all~$i$.  Then
                                            \index{linear threshold function!noise stability}
    for any $\rho \in (-1,1)$,
    \[
        \Bigl|\Stab_\rho[f] - \tfrac{2}{\pi} \arcsin \rho\Bigr| \leq O\Bigl(\tfrac{\eps}{\sqrt{1-\rho^2}}\Bigr).
    \]
\end{theorem}
You are asked to prove Theorem~\ref{thm:ltf-stab-error} in Exercise~\ref{ex:ltf-stab-error}.  In the particular case of $\Maj_n$ where $a_i = \tfrac{1}{\sqrt{n}}$ for all~$i$ we can make a slightly stronger claim (see Exercise~\ref{ex:maj-stab-error2}):
\begin{theorem}                                     \label{thm:maj-stab-precise}
    For any $\rho \in [0,1)$, $\Stab_\rho[\Maj_n]$ is a decreasing function of~$n$, with
    \[
        \tfrac{2}{\pi} \arcsin \rho \leq \Stab_\rho[\Maj_n] \leq \tfrac{2}{\pi} \arcsin \rho + O\Bigl(\tfrac{1}{\sqrt{1-\rho^2}\sqrt{n}}\Bigr).
    \]
\end{theorem}

We end this section by mentioning another way in which the majority function is extremal: among all unbiased functions with small influences, it has (essentially) the largest noise stability.
\begin{named}{Majority Is Stablest Theorem}
                                    \index{Majority Is Stablest Theorem}%
Fix $\rho \in (0,1)$.  Then for any $f \btI$ with $\E[f] = 0$ and $\MaxInf[f] \leq \tau$,
\[
\Stab_\rho[f] \leq \tfrac{2}{\pi} \arcsin \rho + o_\tau(1) = 1 - \tfrac{2}{\pi} \arccos \rho + o_\tau(1).
\]
\end{named}
\noindent For sufficiently small~$\rho$, we'll prove this in Section~\ref{sec:weight-level-1}.   The proof of the full Majority Is Stablest Theorem will have to wait until Chapter~\ref{chap:invariance}.
									 \index{majority!noise stability|)}

\section{The Fourier coefficients of Majority}                         \label{sec:maj-coefficients}
                                        \index{majority!Fourier weight|(}%
In this section we will analyze the Fourier coefficients of $\maj_n$.  In fact, we give an explicit formula for them in Theorem~\ref{thm:maj-coeffs} below.  But most of the time this formula is not too useful; instead, it's better to understand the Fourier coefficients of $\maj_n$ asymptotically as $n \to \infty$.

Let's begin with a few basic observations.  First, $\maj_n$ is a symmetric function and hence $\wh{\maj_n}(S)$ only depends on~$|S|$ (Exercise~\ref{ex:golomb}).  Second, $\maj_n$ is an odd function and hence $\wh{\maj_n}(S) = 0$ whenever~$|S|$ is even (Exercise~\ref{ex:odd-even}).  It remains to determine the Fourier coefficients $\wh{\Maj_n}(S)$ for $|S|$ odd. By symmetry, $\wh{\Maj_n}(S)^2 = \W{k}[\Maj_n]/\binom{n}{k}$ for all $|S| = k$, so if we are content to know the magnitudes of $\Maj_n$'s Fourier coefficients, it suffices to determine the quantities~$\W{k}(\Maj_n)$.

In fact, for each $k \in \N$ the quantity $\W{k}(\Maj_n)$ converges to a fixed constant as $n \to \infty$.  We can deduce this using our analysis of the noise stability of majority.  From the previous section we know that for all $|\rho| \leq 1$,
\begin{equation} \label{eqn:maj-stab-series}
    \lim_{n \to \infty} \Stab_\rho[\Maj_n] = \tfrac{2}{\pi} \arcsin \rho = \tfrac{2}{\pi}\Bigl(\rho + \tfrac16 \rho^3 + \tfrac{3}{40} \rho^5 + \tfrac{5}{112} \rho^7 + \cdots \Bigr),
\end{equation}
where we have used the power series for $\arcsin$,
\begin{equation} \label{eqn:arcsin}
    \arcsin z = \sum_{k \text{ odd}} \ \frac{2}{k2^k} \binom{k-1}{\frac{k-1}{2}} \cdot z^k,
\end{equation}
valid for $|\rho| \leq 1$ (see Exercise~\ref{ex:arcsin}). Comparing~\eqref{eqn:maj-stab-series} with the formula
\[
    \Stab_\rho[\Maj_n] = \sum_{k \geq 0} \W{k}[\Maj_n] \cdot \rho^k
\]
suggests the following:  For each fixed $k \in \N$,
\begin{equation} \label{eqn:maj-one-function}
    \lim_{n \to \infty} \W{k}[\Maj_n] = [\rho^k] (\tfrac{2}{\pi} \arcsin \rho) = \begin{cases}
                                \frac{4}{\pi k2^k} \binom{k-1}{\frac{k-1}{2}} & \text{\textnormal{if $k$ odd,}} \\
                                0 & \text{\textnormal{if $k$ even.}}
                                                                                 \end{cases}
\end{equation}
\noindent (Here  $[z^k]F(z)$ denotes the coefficient
                \nomenclature[{[zk]}]{$[z^k]F(z)$}{coefficient on $z^k$ in the power series $F(z)$}
on $z^k$ in power series $F(z)$.) Indeed, we prove this identity below in Theorem~\ref{thm:maj-asymptotic-weight}. The noise stability method that suggests it can also be made formal (Exercise~\ref{ex:maj-stab-series}).

Identity~\eqref{eqn:maj-one-function} is one way to formulate precisely the statement that the ``Fourier spectrum of $\maj_n$ converges''. Introducing notation such as ``$\W{k}(\Maj)$'' for the quantity in~\eqref{eqn:maj-one-function}, we have the further asymptotics
\begin{equation} \label{eqn:maj-asympt-asympt}
\begin{aligned}
\text{for $k$ odd,} \qquad    \W{k}(\Maj) &\sim \left(\tfrac{2}{\pi}\right)^{3/2} k^{-3/2},\\
    \W{>k}(\Maj) &\sim \left(\tfrac{2}{\pi}\right)^{3/2} k^{-1/2} \qquad \text{as $k \to \infty$}.
\end{aligned}
\end{equation}
(See Exercise~\ref{ex:maj-asympt-asympt}.)  The estimates~\eqref{eqn:maj-asympt-asympt}, together with the precise value $\W{1}(\Maj) = \tfrac{2}{\pi}$, are usually all you need to know about the Fourier coefficients of majority.

Nevertheless, let's now compute the Fourier coefficients of $\Maj_n$ exactly.
\begin{theorem}                                     \label{thm:maj-coeffs}
    If $|S|$ is even, then $\wh{\Maj_n}(S) = 0$.
                                            \index{majority!Fourier coefficients}%
    If $|S| = k$ is odd,
    \[
        \wh{\Maj_n}(S) = %\begin{cases}
                            (-1)^{\frac{k-1}{2}}
                            \frac{\binom{\frac{n-1}{2}}{\frac{k-1}{2}}}{\binom{n-1}{k-1}}  \cdot \tfrac{2}{2^n} {\textstyle \binom{n-1}{\frac{n-1}{2}}}.
    \]
\end{theorem}
\begin{proof}
The first statement holds because $\Maj_n$ is an odd function; henceforth we assume $|S| = k$ is odd.  The trick will be to compute the Fourier expansion of majority's \emph{derivative} $\D_n \Maj_n = \Half_{n-1} \co \bits^{n-1} \to \{0,1\}$, the $0$-$1$ indicator of the set of $(n-1)$-bit strings with exactly half of their coordinates equal to~$-1$.  By the derivative formula and the fact that $\Maj_n$ is symmetric, $\wh{\Maj_n}(S) = \wh{\Half_{n-1}}(T)$ for any $T \subseteq [n-1]$ with $|T| = k-1$.  So  writing $n-1 = 2m$ and $k-1 = 2j$, it suffices to show
\begin{equation} \label{eqn:half-formula}
    \wh{\Half_{2m}}([2j]) = (-1)^{j} \frac{\binom{m}{j}}{\binom{2m}{2j}}\cdot \tfrac{1}{2^{2m}}{\textstyle \binom{2m}{m}}.
\end{equation}

By the probabilistic definition of $\T_\rho$, for any $\rho \in [-1,1]$ we have
\[
    \T_\rho \Half_{2m}(1, 1, \dots, 1) = \E_{\bx \sim N_\rho((1, 1, \dots, 1))}[\Half_{2m}(\bx)] =
    \Pr[\bx \text{ has $m$ $1$'s and $m$ $-1$'s}],
\]
where each coordinate of $\bx$ is $1$ with probability $\half + \half \rho$. Thus
\begin{equation} \label{eqn:half1}
    \T_\rho \Half_{2m}(1, 1, \dots, 1) = {\textstyle \binom{2m}{m}}(\half + \half \rho)^{m} (\half - \half \rho)^{m} = \tfrac{1}{2^{2m}} {\textstyle \binom{2m}{m}} (1-\rho^2)^m.
\end{equation}
On the other hand, by the Fourier formula for $\T_\rho$ and the fact that $\Half_{2m}$ is symmetric we have
\begin{equation} \label{eqn:half2}
    \T_\rho \Half_{2m}(1, 1, \dots, 1) = \sum_{U \subseteq [2m]} \wh{\Half_{2m}}(U) \rho^{|U|} = \sum_{i=0}^{2m} {\textstyle \binom{2m}{i}} \wh{\Half_{2m}}([i]) \rho^{i}.
\end{equation}
Since we have equality $\eqref{eqn:half1} = \eqref{eqn:half2}$ between two degree-$2m$ polynomials of~$\rho$ on all of $[-1,1]$, we can equate coefficients. In particular, for $i = 2j$ we have
\[
    {\textstyle \binom{2m}{2j}} \wh{\Half_{2m}}([2j]) = \tfrac{1}{2^{2m}} {\textstyle \binom{2m}{m}} \cdot [\rho^{2j}](1-\rho^2)^m = \tfrac{1}{2^{2m}} {\textstyle \binom{2m}{m}} \cdot (-1)^j {\textstyle \binom{m}{j}},
\]
confirming~\eqref{eqn:half-formula}.
\end{proof}

You are asked to prove the following corollaries in Exercises~\ref{ex:maj-coeffs-symm},~\ref{ex:maj-weight-decreasing}:
\begin{corollary}                                       \label{cor:maj-coeffs-symm}
    $\wh{\Maj_n}(S) = (-1)^{\frac{n-1}{2}} \wh{\Maj_n}(T)$ whenever $|S| + |T| = n+1$. Hence also $\W{n-k+1}[\Maj_n] = \frac{k}{n-k+1} \W{k}[\Maj_n]$.
\end{corollary}
\begin{corollary}                                       \label{cor:maj-weight-decreasing}
    For any odd $k$, $\W{k}[\Maj_n]$ is a strictly decreasing function of~$n$ (for $n \geq k$ odd).
\end{corollary}
We can now prove the identity~\eqref{eqn:maj-one-function}:
\begin{theorem} \label{thm:maj-asymptotic-weight} For each fixed odd $k$,
    \[
    \W{k}[\Maj_n] \searrow [\rho^k] (\tfrac{2}{\pi} \arcsin \rho) = \tfrac{4}{\pi k2^k} {\textstyle \binom{k-1}{\frac{k-1}{2}}}
    \]
    as $n \geq k$ tends to $\infty$ (through the odd numbers).  Further, we have the error bound
    \begin{equation} \label{eqn:maj-weight-err}
        [\rho^k] (\tfrac{2}{\pi} \arcsin \rho) \leq \W{k}[\Maj_n] \leq (1+2k/n) \cdot [\rho^k] (\tfrac{2}{\pi} \arcsin \rho)
    \end{equation}
    for all $k < n/2$.  (For $k > n/2$ you can use Corollary~\ref{cor:maj-coeffs-symm}.)
\end{theorem}
\begin{proof}
    Corollary~\ref{cor:maj-weight-decreasing} tells us that $\W{k}[\Maj_n]$ is decreasing in $n$; hence we only need to justify~\eqref{eqn:maj-weight-err}.  Using the formula from Theorem~\ref{thm:maj-coeffs} we have
    \[
        \frac{\W{k}[\Maj_n]}{[\rho^k] (\tfrac{2}{\pi} \arcsin \rho)} = \frac{\binom{n}{k}\tfrac{4}{2^{2n}}\binom{n-1}{\frac{n-1}{2}}^2\left.\binom{\frac{n-1}{2}}{\frac{k-1}{2}}^2\middle/\binom{n-1}{k-1}^2\right.}{\frac{4}{\pi k2^k} \binom{k-1}{\frac{k-1}{2}}} = \tfrac{\pi}{2} n \cdot 2^{k-n}{\textstyle \binom{n-k}{\frac{n-k}{2}}} \cdot 2^{1-n}{\textstyle \binom{n-1}{\frac{n-1}{2}}},
    \]
    where the second identity is verified by expanding all binomial coefficients to factorials.
    By Stirling's approximation we have $2^{-m}\binom{m}{m/2} \nearrow \sqrt{\frac{2}{\pi m}}$, meaning that the ratio of the left side to the right side increases to~$1$ as $m \to \infty$.  Thus
    \[
        \frac{\W{k}[\Maj_n]}{[\rho^k] (\tfrac{2}{\pi} \arcsin \rho)} \nearrow \frac{n}{\sqrt{n-k}\sqrt{n-1}} = (1-\tfrac{k+1}{n} +\tfrac{k}{n^2})^{-1/2},
    \]
    and the right-hand side is at most $1+2k/n$ for $1 \leq k \leq n/2$ by Exercise~\ref{ex:maj-annoying}.
\end{proof}
Finally, we can deduce the asymptotics~\eqref{eqn:maj-asympt-asympt} from this theorem (see Exercise~\ref{ex:maj-asympt-asympt}):
\begin{corollary}                                       \label{cor:maj-asympt-asympt}
    Let $k \in \N$ be odd and assume $n = n(k) \geq 2k^2$.  Then
    \begin{align*}
        \W{k}(\Maj_n) &= \left(\tfrac{2}{\pi}\right)^{3/2} k^{-3/2} \cdot (1\pm O(1/k)),  \\
        \W{>k}(\Maj_n) &= \left(\tfrac{2}{\pi}\right)^{3/2} k^{-1/2} \cdot (1\pm O(1/k)),
    \end{align*}
    and hence the Fourier spectrum of $\Maj_n$ is $\eps$-concentrated on degree up to $\frac{8}{\pi^3} \eps^{-2} + O_{\eps}(1)$.
\end{corollary}
                                        \index{majority!Fourier weight|)}

\section{Degree-$1$ weight}                                                    \label{sec:weight-level-1}
                                                \index{level-1 Fourier weight|seeonly{Fourier weight, degree-1}}%
                                                \index{degree-1 Fourier weight|seeonly{Fourier weight, degree-1}}%
                                                \index{Fourier weight!degree-1|(}%
In this section we prove two theorems about the degree-$1$ Fourier weight of Boolean functions:
\[
    \W{1}[f] = \sum_{i=1}^n \wh{f}(i)^2.
\]
This important quantity can be given a combinatorial interpretation thanks to the noise stability formula $\Stab_\rho[f] = \sum_{k\geq 0}\rho^k \cdot \W{k}[f]$:
\[
    \text{For $f \btR$,} \quad \W{1}[f] = \frac{d}{d \rho} \Stab_\rho[f]\Bigm|_{\rho = 0}.
\]
Thinking of $\|f\|_2$ as constant and $\rho \to 0$, the noise stability formula implies
\[
    \Stab_\rho[f] = \E[f]^2 + \W{1}[f]\rho \pm O(\rho^2),
\]
or equivalently,
\[
\Cov_{\substack{(\bx, \by) \\ \mathclap{\text{ $\rho$-correlated}}}}[f(\bx),f(\by)] = \W{1}[f] \rho \pm O(\rho^2).
\]
In other words, for $f \btb$ the degree-$1$ weight quantifies the extent to which $\Pr[f(\bx) = f(\by)]$ increases when $\bx$ and $\by$ go from being uncorrelated to being slightly correlated.

There is an additional viewpoint if we think of $f$ as the indicator of a subset $A \subseteq \bn$ and its noise sensitivity $\NS_{\delta}[f]$ as a notion of $A$'s ``surface area'', or ``noisy boundary size''.  For nearly maximal noise rates -- i.e., $\delta = \half - \half \rho$ where $\rho$ is small -- we have that $A$'s noisy boundary size is ``small'' if and only if $\W{1}[f]$ is ``large'' (vis-\`a-vis $A$'s measure).
                                                \index{Fourier weight!degree-1|)}%

Two examples suggest themselves when thinking of subsets of the Hamming cube with small ``boundary'': subcubes and Hamming balls.
\begin{proposition}  \label{prop:cube-weight1}
                                                \index{subcube!degree-1 weight}%
Let $f \ftzo$ be the indicator of a subcube of codimension~$k \geq 1$ (e.g., the $\AND_k$ function).  Then $\E[f] = 2^{-k}$, $\W{1}[f] = k 2^{-2k}$.
\end{proposition}
\begin{proposition}  \label{prop:ball-weight1} Fix $t \in \R$.
                                                \index{Hamming ball!degree-1 weight}
Consider the sequence of LTFs $f_n \co \bits^n \to \{0,1\}$ defined by $f_n(x) = 1$ if and only if $\sum_{i=1}^n \tfrac{1}{\sqrt{n}} x_i > t$.  (That is, $f_n$ is the indicator of the Hamming ball $\{x : \hamdist(x, (1, \dots, 1)) < \frac{n}{2} - \frac{t}{2}\sqrt{n}\}$.) Then
\[
    \lim_{n \to \infty} \E[f_n] = \olPhi(t), \qquad \lim_{n \to \infty} \W{1}[f_n] = \phi(t)^2.
\]
\end{proposition}
You are asked to verify these facts in Exercises~\ref{ex:cube-weight1},~\ref{ex:ball-weight1}. Regarding Proposition~\ref{prop:ball-weight1}, it's natural for $\phi(t)$ to arise since $\W{1}[f_n]$ is related to the influences of $f_n$, and coordinates are  influential for $f_n$ if and only if $\sum_{i=1}^n \tfrac{1}{\sqrt{n}} x_i {\approx t}$.  If we write $\alpha = \lim_{n \to \infty} \E[f_n]$ then this proposition can be thought of as saying that $\W{1}[f_n] \to \GIso(\alpha)^2$, where~$\GIso$ is defined as follows:
\begin{definition}                              \label{def:giso}
    The \emph{Gaussian isoperimetric function}
                                                \nomenclature[U]{$\Giso$}{the Gaussian isoperimetric function, $\Giso = \phi \circ \Phi^{-1}$}%
                                                \index{Gaussian isoperimetric function}%
    $\Giso \co [0,1] \to [0,\frac{1}{\sqrt{2\pi}}]$ is defined by $\Giso = \phi \circ \Phi^{-1}$.  This function is symmetric about~$1/2$; i.e., $\Giso = \phi \circ \olPhi{}^{-1}$.
\end{definition}
The name of this function will be explained when we study the Gaussian Isoperimetric Inequality in Chapter~\ref{sec:gaussian-surface-area}.  For now we'll just use the following fact:
\begin{proposition}                                     \label{prop:giso-asympt}
    For $\alpha \to 0^+$, $\Giso(\alpha) \sim  \alpha \sqrt{2 \ln(1/\alpha)}$.
\end{proposition}
\begin{proof}
    Write $\alpha = \olPhi(t)$, where $t \to \infty$.  We use the well-known fact that $\olPhi(t) \sim \phi(t)/t$. Thus
    \begin{gather*}
        \alpha \sim \tfrac{1}{\sqrt{2\pi} t}\exp(-t^2/2) \qRq t \sim \sqrt{2\ln (1/\alpha)}, \\
        \phi(t) \sim \olPhi(t) \cdot t \qRq \Giso(\alpha) \sim \alpha \cdot t \sim \alpha \sqrt{2\ln(1/\alpha)}.\qedhere
    \end{gather*}
\end{proof}

Given Propositions~\ref{prop:cube-weight1} and~\ref{prop:ball-weight1}, let's consider the degree-$1$ Fourier weight of subcubes and Hamming balls asymptotically as their ``volume'' $\alpha = \E[f]$ tends to~$0$. For the subcubes we have $\W{1}[f] = \alpha^2 \log(1/\alpha)$.  For the Hamming balls we have $\W{1}[f_n] \to \GIso(\alpha)^2 \sim 2\alpha^2 \ln(1/\alpha)$.  So in both cases we have an upper bound of $O(\alpha^2 \log(1/\alpha))$.

You should think of this upper bound $O(\alpha^2 \log(1/\alpha))$ as being unusually small.  The obvious a priori upper bound, given that $f \btzo$ has $\E[f] = \alpha$, is
\[
    \W{1}[f] \leq \Var[f] = \alpha(1-\alpha) \sim \alpha.
\]
Yet subcubes and Hamming balls have degree-$1$ weight which is almost quadratically smaller.  In fact the first theorem we will show in this section is the following:
\begin{named}{Level-1 Inequality}
                                                \index{Level-1 Inequality}
Let $f \btzo$ have mean $\E[f] = \alpha \leq 1/2$.  Then
\[
\W{1}[f] \leq O(\alpha^2\log(1/\alpha)).
\]
(For the case $\alpha \geq 1/2$, replace $f$ by $1-f$.)
\end{named}
Thus \emph{all} small subsets of $\bits^n$ have unusually small $\W{1}[f]$; or equivalently (in some sense), unusually large ``noisy boundary''.  This is another key illustration of the idea that the Hamming cube is a ``small-set expander''.
                                            \index{expansion!small-set}
\begin{remark} \label{rem:improved-level-1} The bound in the Level-1 Inequality has a sharp form, $\W{1}[f] \leq 2\alpha^2\ln(1/\alpha)$.  Thus Hamming balls are in fact the ``asymptotic maximizers'' of $\W{1}[f]$ among sets of small volume~$\alpha$. Also, the inequality holds more generally for $f \btI$ with $\alpha = \E[|f|]$.
\end{remark}
\begin{remark} The name ``Level-1 Inequality'' is not completely standard; e.g., in additive combinatorics the result would be called
                                            \index{Chang's Inequality|seeonly{Level-1 Inequality}}
\emph{Chang's Inequality}.  We use this name because we will also generalize to ``Level-$k$ Inequalities'' in Chapter~\ref{sec:hypercon-apps}.
\end{remark}

So far we considered maximizing degree-$1$ weight among subsets of the Hamming cube of a fixed small volume,~$\alpha$.  The second theorem in this section is concerned with what happens when there is no volume constraint.  In this case, maximizing examples tend to have volume $\alpha = 1/2$; switching the notation to $f \btb$, this corresponds to $f$ being unbiased ($\E[f] = 0$).  The unbiased Hamming ball is $\Maj_n$, which we know has $\W{1}[\Maj_n] \to \tfrac{2}{\pi}$.  This is quite large.  But unbiased subcubes are just the dictators $\chi_i$ and their negations; these have $\W{1}[\pm \chi_i] = 1$ which is obviously maximal.

Thus the question of which $f \btb$ maximizes $\W{1}[f]$ has a trivial answer.  But this answer is arguably unsatisfactory, since dictators (and their negations) are not ``really'' functions of~$n$ bits.  Indeed, when we studied social choice in Chapter~\ref{chap:concepts} we were motivated to rule out functions~$f$ having a coordinate with unfairly large influence.  And in fact Proposition~\ref{prop:weight-1-same} showed that if all $\wh{f}(i)$ are equal (and hence small) then $\W{1}[f] \leq \frac{2}{\pi} + o_n(1)$.  The second theorem of this section significantly generalizes Proposition~\ref{prop:weight-1-same}:
\begin{named}{The ${\frac{\mathbf 2}{\bpi}}$ Theorem}
    Let $f \btb$ satisfy $|\wh{f}(i)| \leq \eps$ for all $i \in [n]$.
                                                    \index{2pi Theorem@$\frac{2}{\pi}$ Theorem}%
    Then
    \begin{equation} \label{eqn:regular-W1-bound}
        \W{1}[f] \leq \tfrac{2}{\pi} +O(\eps).
    \end{equation}
    Further, if $\W{1}[f] \geq \tfrac{2}{\pi} - \eps$, then $f$ is $O(\sqrt{\eps})$-close to the LTF $\sgn(f^{=1})$.
\end{named}
Functions $f$ with $|\wh{f}(i)| \leq \eps$ for all $i \in [n]$ are called \emph{$(\eps,1)$-regular}; see Chapter~\ref{sec:pseudorandomness-notions}.  So the $\frac{2}{\pi}$~Theorem says (roughly speaking) that within the class of $(\eps,1)$-regular functions, the maximal degree-$1$ weight is $\frac{2}{\pi}$, and any function achieving this is an unbiased LTF.  Further, from Theorem~\ref{thm:ltf-stab-error} we know that \emph{all} unbiased LTFs which are $(\eps,1)$-regular achieve this.
\begin{remark}
                                                    \index{Majority Is Stablest Theorem}%
Since we have $\Stab_\rho[f] \approx \W{1}[f] \rho$ and $\frac{2}{\pi} \arcsin \rho \approx \frac{2}{\pi} \rho$ when $\rho$ is small, the $\frac{2}{\pi}$~Theorem gives the Majority Is Stablest Theorem in the limit $\rho \to 0^+$.
\end{remark}

Let's now discuss how we'll prove our two theorems about degree-$1$ weight.  Let $f \btzo$ and $\alpha = \E[f]$; we think of $\alpha$ as small for the Level-1 Inequality and $\alpha = 1/2$ for the $\frac{2}{\pi}$~Theorem. By Plancherel, $\W{1}[f] = \E[f(\bx) L(\bx)]$, where
\[
    L(x) = f^{=1}(x) = \wh{f}(1)x_1 + \cdots + \wh{f}(n) x_n.
\]
To upper-bound $\E[f(\bx) L(\bx)]$, consider that as $\bx$ varies the real number $L(\bx)$ may be rather large or small, but $f(\bx)$ is always~$0$ or~$1$.   Given that $f(\bx)$ is~$1$ on only a $\alpha$ fraction of $\bx$'s, the ``worst case'' for $\E[f(\bx) L(\bx)]$ would be if $f(x)$ were~$1$ precisely on the $\alpha$ fraction of $x$'s where $L(x)$ is largest.  In other words,
\begin{equation} \label{eqn:talagrand1}
    \W{1}[f] = \E[f(\bx) L(\bx)] \leq \E[\bone_{\{L(\bx) \geq t\}} \cdot L(\bx)],
\end{equation}
where $t$ is chosen so that
\begin{equation} \label{eqn:what-is-t}
    \Pr[L(\bx) \geq t] \approx \alpha.
\end{equation}
But now we can analyze~\eqref{eqn:talagrand1} quite effectively using tools such as Hoeffding's bound and the CLT, since $L(\bx)$ is just a linear combination of independent~$\pm 1$ random bits. In particular $L(\bx)$ has mean~$0$ and standard deviation $\sigma = \sqrt{\W{1}[f]}$ so by the CLT it acts like the Gaussian $\bZ \sim \normal(0,\sigma^2)$, at least if we assume all $|\wh{f}(i)|$ are small.  If we are thinking of $\alpha = 1/2$, then $t = 0$ and we get
\[
    \sigma^2 = \W{1}[f] \leq \E[\bone_{\{L(\bx) \geq 0\}} \cdot L(\bx)] \approx \E[\bone_{\{\bZ \geq 0\}} \cdot \bZ] = \tfrac{1}{\sqrt{2\pi}} \sigma;
\]
This implies $\sigma^2 \lessapprox \tfrac{1}{2\pi}$, as claimed in the $\frac{2}{\pi}$~Theorem (after adjusting $f$'s range to~$\bits$).  If we are instead thinking of $\alpha$ as small then~\eqref{eqn:what-is-t} suggest taking $t \sim \sigma \sqrt{2\ln(1/\alpha)}$ so that $\Pr[\bZ \geq t] \approx \alpha$.  Then a calculation akin to the one in Proposition~\ref{prop:giso-asympt} implies
\[
    \W{1}[f] \leq \E[\bone_{\{L(\bx) \geq t\}} \cdot L(\bx)] \approx \alpha \cdot \sigma \sqrt{2\ln(1/\alpha)},
\]
from which the Level-1 Inequality follows.  In fact, we don't even need all $|\wh{f}(i)|$ small for this latter analysis; for large~$t$ it's possible to upper-bound~\eqref{eqn:talagrand1} using only Hoeffding's bound:
\begin{lemma}  \label{lem:linear-tail} Let $\ell(x) = a_1 x_1 + \cdots + a_n x_n$, where $\sum_{i} a_i^2 = 1$.  Then for any $s \geq 1$,
\[
    \E[\bone_{\{|\ell(\bx)| > s\}} \cdot |\ell(\bx)|] \leq (2s+2)\exp(-\tfrac{s^2}{2}).
\]
\end{lemma}
\begin{proof}
    We have
    \begin{align*}
        \E[\bone_{\{|\ell(\bx)| > s\}} \cdot |\ell(\bx)|] &= s\Pr[|\ell(\bx)| > s] + \int_{s}^\infty \Pr[|\ell(\bx)| > u]\,du \\
        &\leq 2s\exp(-\tfrac{s^2}{2}) + \int_{s}^\infty 2\exp(-\tfrac{u^2}{2})\,du,
    \end{align*}
    using Hoeffding's bound.  But for $s \geq 1$,
    \[
        \int_{s}^\infty 2\exp(-\tfrac{u^2}{2})\,du \leq \int_{s}^\infty u \cdot 2\exp(-\tfrac{u^2}{2})\,du = 2\exp(-\tfrac{s^2}{2}). \qedhere
    \]
\end{proof}

We now give formal proofs of the two theorems, commenting that rather than~$L(x)$ it's more convenient to work with \[
    \ell(x) = \tfrac{1}{\sigma} f^{=1}(x) = \tfrac{\wh{f}(1)}{\sigma}x_1 + \cdots + \tfrac{\wh{f}(n)}{\sigma}x_n.
\]

\begin{proof}[Proof of the Level-1 Inequality]
    Following Remark~\ref{rem:improved-level-1} we let $f \btI$ and $\alpha = \E[|f|]$.  We may assume $\sigma = \sqrt{\W{1}[f]} > 0$.  Writing $\ell = \frac{1}{\sigma} f^{=1}$ we have
    $
        \la f, \ell \ra = \tfrac{1}{\sigma} \la f, f^{=1} \ra = \tfrac{1}{\sigma} \W{1}[f] = \sigma
    $ and
    hence
    \[
        \sigma = \la f, \ell \ra = \E[\bone_{\{|\ell(\bx)| \leq s\}} \cdot f(\bx) \ell(\bx)] + \E[\bone_{\{|\ell(\bx)| > s\}} \cdot f(\bx) \ell(\bx)]
    \]
    holds for any $s \geq 1$.  The first expectation above is at most $\E[s |f(\bx)|] = \alpha s$, and the second is at most $(2+2s)\exp(-s^2/2) \leq 4s\exp(-s^2/2)$ by Lemma~\ref{lem:linear-tail}. Hence
    \[
        \sigma \leq \alpha s + 4s \exp(-s^2/2).
    \]
    The optimal choice of $s$ is $s = (\sqrt{2} + o_\alpha(1)) \sqrt{\ln(1/\alpha)}$, yielding
    \[
        \sigma \leq (\sqrt{2} + o(1)) \alpha\sqrt{\ln(1/\alpha)}.
    \]
    Squaring this establishes the claim $\sigma^2 \leq (2+o_\alpha(1)) \alpha^2\ln(1/\alpha)$.
\end{proof}

\begin{proof}[Proof of the $\frac{2}{\pi}$~Theorem]
    We may assume $\sigma = \sqrt{\W{1}[f]} \geq 1/2$: for the theorem's first statement this is because otherwise there is nothing to prove; for the theorem's second statement this is because we may assume $\eps$ sufficiently small.

    We start by proving~\eqref{eqn:regular-W1-bound}.   Let $\ell = \frac{1}{\sigma} f^{=1}$, so $\|\ell\|_2 = 1$ and $|\wh{\ell}(i)| \leq 2\eps$ for all $i \in [n]$.  We have
    \begin{equation} \label{eqn:ltf-test1}
        \sigma = \la f, \ell \ra \leq \E[|\ell|] \leq \sqrt{\tfrac{2}{\pi}} + C \eps
    \end{equation}
    for some constant~$C$, where we used Theorem~\ref{thm:l1-be}.  Squaring this proves~\eqref{eqn:regular-W1-bound}.  We observe that~\eqref{eqn:regular-W1-bound} therefore holds even for $f \btI$.

    Now suppose we also have $\W{1}[f] \geq \tfrac{2}{\pi} - \eps$; i.e.,
    \[
        \sigma \geq \sqrt{\tfrac{2}{\pi} - \eps} \geq \sqrt{\tfrac{2}{\pi}} - 2\eps.
    \]
    Thus the first inequality in~\eqref{eqn:ltf-test1} must be close to tight; specifically,
    \begin{equation} \label{eqn:ltf-test-contra}
        (C+2)\eps \geq \E[|\ell|] - \la f, \ell\ra = \E[(\sgn(\ell(\bx)) - f(\bx)) \cdot \ell(\bx)].
    \end{equation}
    By the Berry--Esseen Theorem (and Remark~\ref{rem:simpler-BE}, Exercise~\ref{ex:BE-interval}),
    \[
        \Pr[|\ell| \leq K\sqrt{\eps}] \leq \Pr[|\normal(0,1)| \leq K\sqrt{\eps}] + .56 \cdot 2\eps \leq \tfrac{1}{\sqrt{2\pi}} \cdot 2K\sqrt{\eps} + 1.12 \eps \leq 2K\sqrt{\eps}
    \]
    for any constant $K \geq 1$.  We therefore have the implication
    \begin{align*}
        \Pr[f \neq \sgn(\ell)] \geq 3K\sqrt{\eps} &\implies \Pr[f(\bx) \neq \sgn(\ell(\bx))\ \wedge\ |\ell(\bx)| > K\sqrt{\eps}] \geq K \sqrt{\eps} \\ &\implies \E[(\sgn(\ell(\bx)) - f(\bx)) \cdot \ell(\bx)] \geq K\sqrt{\eps} \cdot 2(K\sqrt{\eps}) = 2K^2 \eps.
    \end{align*}
    This contradicts~\eqref{eqn:ltf-test-contra} for $K = \sqrt{C+2}$, say.  Thus $\Pr[f \neq \sgn(\ell)] \leq 3\sqrt{C+2}\sqrt{\eps}$, completing the proof.
\end{proof}

For an interpolation between these two theorems, see Exercise~\ref{ex:refined-level-1}.

We conclude this section with an application of the Level-1 Inequality.  First, a quick corollary which we leave for Exercise~\ref{ex:level-1-quick}:
\begin{corollary} \label{cor:level-1-quick}
    Let $f \btb$ have $|\E[f]| \geq 1 - \delta \geq 0$.  Then $\W{1}[f] \leq 4\delta^2 \log(2/\delta)$.
\end{corollary}
In Chapter~\ref{sec:arrow} we stated the FKN~Theorem, which says that if $f \btb$ has $\W{1}[f] \geq 1 - \delta$ then it must be $O(\delta)$-close to a dictator or negated-dictator.
                                                    \index{FKN Theorem}
The following theorem shows that once the FKN~Theorem is proved, it can be strengthened to give an essentially optimal (Exercise~\ref{ex:fkn-optimality}) closeness bound:
\begin{theorem}                                     \label{thm:FKN-improvement}
    Suppose the FKN Theorem holds with closeness bound $C\delta$, where $C \geq 1$ is a universal constant.  Then in fact it holds with bound $\delta/4 + \eta$, where $\eta = 16C^2 \delta^2 \max(\log(1/C\delta),1)$.
\end{theorem}
\begin{proof}
    Suppose $f \btb$ has $\W{1}[f] \geq 1 - \delta \geq 0$.  By assumption $f$ is $C\delta$-close to $\pm \chi_i$ for some $i \in [n]$, say $i = n$. Thus we have
    \[
        |\wh{f}(n)| \geq 1 - 2C\delta
    \]
    and our task is to show that in fact $|\wh{f}(n)| \geq 1 - \delta/2 - 2\eta$. We may assume $\delta \leq \frac{1}{10C}$ as otherwise $1 - \delta/2 - 2\eta < 0$ (Exercise~\ref{ex:FKN-improvement}) and there is nothing to prove.  By employing the trick from Exercise~\ref{ex:FKN-0-and-1} we may also assume $\E[f] = 0$.

    Consider the restriction of $f$ given by fixing coordinate~$n$ to $b \in \bits$; i.e., $\restr{f}{[n-1]}{b}$.  For both choices of $b$ we have $|\E[\restr{f}{[n-1]}{b}]| \geq 1 - 2C\delta$ and so Corollary~\ref{cor:level-1-quick} implies $\W{1}[\restr{f}{[n-1]}{b}] \leq 16C^2\delta^2 \log(1/C\delta)$.  Thus
    \[
        16C^2\delta^2 \log(1/C\delta) \geq \E_{\bb}[\W{1}[\restr{f}{[n-1]}{\bb}]] = \sum_{j < n} (\wh{f}(\{j\})^2 + \wh{f}(\{j, n\})^2) \geq \sum_{j < n} \wh{f}(j)^2,
    \]
    by Corollary~\ref{cor:expected-restrict-coeffs}.  It follows that
    \[
        \wh{f}(n)^2 = \W{1}[f] - \sum_{j<n}\wh{f}(j)^2 \geq 1- \delta - 16C^2\delta^2 \log(1/C\delta),
    \]
    and the proof is completed by the fact that
    \[
        1- \delta - 16C^2\delta^2 \log(1/C\delta) \geq (1-\delta/2 - 2\eta)^2
    \]
    when $\delta \leq \frac{1}{10C}$ (Exercise~\ref{ex:FKN-improvement}).
\end{proof}

\section{Highlight: Peres's Theorem and uniform noise stability}                               \label{sec:peres}
                                                \index{linear threshold function!noise stability|(}
Theorem~\ref{thm:ltf-stab-error} says that if $f$ is an unbiased linear threshold function $f(x) = \sgn(a_1 x_1 + \cdots + a_n x_n)$ in which all $a_i$'s are ``small'', then the noise stability $\Stab_\rho[f]$ is at least (roughly) $\frac{2}{\pi} \arcsin \rho$.  Rephrasing in terms of noise sensitivity, this means $\NS_\delta[f]$ is at most (roughly) $\tfrac{2}{\pi} \sqrt{\delta} + O(\delta^{3/2})$ (see the statement of Theorem~\ref{thm:maj-stab}).  On the other hand, if some $a_i$ were particularly \emph{large} then~$f$ would be pushed in the direction of the dictator function $\chi_i$, which has $\NS_\delta[\chi_i] = \delta \ll \sqrt{\delta}$.  This observation suggests that \emph{all} unbiased LTFs $f$ should have $\NS_\delta[f] \leq O(\sqrt{\delta})$.  The unbiasedness assumption also seems inessential, since biasing a function should tend to decrease its noise sensitivity.

Indeed, the idea here is correct, as was shown by Peres in 1999:
\begin{named}{Peres's Theorem}
                                                \index{Peres's Theorem}
Let $f \btb$ be any linear threshold function.  Then $\NS_\delta[f] \leq O(\sqrt{\delta})$.
\end{named}

Pleasantly, the proof is quite simple and uses no heavy tools like the Central Limit Theorem.  Before getting to it, let's make some remarks.  First, Peres's Theorem shows that the class of all linear threshold functions is what's called \emph{uniformly noise-stable}.
                                                \index{uniformly noise-stable}%
                                                \index{noise stability!uniform|seeonly{uniformly noise-stable}}%
\begin{definition}
    Let $\calB$ be a class of Boolean-valued functions.  We say that $\calB$ is \emph{uniformly noise-stable} if there exists $\eps \co [0,1/2] \to [0,1]$ with $\eps(\delta) \to 0$ as $\delta \to 0^+$ such that $\NS_\delta[f] \leq \eps(\delta)$ holds for all $f \in \calB$.
\end{definition}
This definition is only interesting for infinite classes~$\calB$. (Any class containing functions of only finitely many input lengths is vacuously uniformly noise-stable; see Exercise~\ref{ex:vac-noise-stable}.)  By Proposition~\ref{prop:noise-stable-implies-PTF} we see that functions in a uniformly noise-stable class have ``almost all of their Fourier weight at constant degree''; i.e., for all $\eps > 0$ there is a $k \in \N$ such that $\W{>k}[f] \leq \eps$ for all $f \in \calB$.  In particular, from Corollary~\ref{cor:learn-ns} we get that if $\calB$ is a uniformly noise-stable class then its restriction to $n$-input functions is learnable from random examples to any constant error in $\poly(n)$ time.
                                                \index{learning theory}%

Let's make these observations more concrete in the context of linear threshold functions.  Peres's Theorem immediately gives that LTFs have their Fourier
                                                \index{linear threshold function!learning}%
spectrum $\eps$-concentrated up to degree $O(1/\eps^2)$ (Proposition~\ref{prop:ns-concentration}) and hence the class of LTFs is learnable from random examples with error~$\eps$ in time~$n^{O(1/\eps^2)}$ (Corollary~\ref{cor:learn-ns}).  The latter result is not too impressive since it's been long known that LTFs are learnable in time $\poly(n, 1/\eps)$ using linear programming. However, the noise sensitivity approach is much more flexible.  Consider the concept class
\[
    \calC = \{h = g(f_1, \dots, f_s) \mid \text{$f_1, \dots, f_s \btb$ are LTFs}\}.
\]
For each $h \btb$ in $\calC$, Peres's Theorem and a union bound (Exercise~\ref{ex:ns-union-bound}) imply that $\NS_\delta[h] \leq O(s \sqrt{\delta})$.  Thus from Corollary~\ref{cor:learn-ns} we get that the class $\calC$ is learnable in time~$n^{O(s^2/\eps^2)}$.  This is the only known way of showing even that an AND of two LTFs is learnable with error~$.01$ in time $\poly(n)$.

\medskip

The trick for proving Peres's Theorem is to employ a fairly general technique for bounding noise sensitivity using \emph{average sensitivity} (total influence):
\begin{theorem}                                     \label{thm:ns-to-as}
                                                \index{noise sensitivity!vs. total influence}%
    Let $\delta \in (0,1/2]$ and let $A \co \N^+ \to \R$.  Let $\calB$ be a class of Boolean-valued functions closed under negation and identification of input variables.  Suppose that each $f \in \calB$ with domain $\bn$ has $\Tinf[f] \leq A(n)$.  Then each $f \in \calB$ has $\NS_\delta[f] \leq \frac{1}{m}A(m)$, where $m = \lfloor 1/\delta \rfloor$.
\end{theorem}
\begin{proof}
    Fix any $f \btb$ from $\calB$.   Since noise sensitivity is an increasing function of the noise parameter (see the discussion surrounding Proposition~\ref{prop:tinf-is-stab-deriv}) we may replace~$\delta$ by~$1/m$.  Thus our task is to upper-bound $\NS_{1/m}[f] = \Pr[f(\bx) \neq f(\by)]$ where $\bx \sim \bn$ is uniformly random and $\by \in \bn$ is formed from $\bx$ by negating each bit independently with probability~$1/m$.  The rough idea of the proof is that this is equivalent to randomly partitioning $\bx$'s bits into $m$ parts and then negating a randomly chosen part.

    More precisely, let $z \in \bits^n$ and let $\pi \co [n] \to [m]$ be a partition of $[n]$ into $m$ parts.  Define
    \[
        g_{z, \pi} \co \bits^m \to \bits, \quad g_{z, \pi}(w) = f(z \circ w^{\pi}),
    \]
    where $\circ$ denotes entry-wise multiplication and $w^{\pi} = (w_{\pi(1)}, \dots, w_{\pi(n)}) \in \bits^n$. Since $g_{z, \pi}$ is derived from~$f$ by negating and identifying input variables it follows that $g_{z,\pi} \in \calB$.  So by assumption $g_{z,\pi}$ has total influence $\Tinf[g_{z,\pi}] \leq A(m)$ and hence \emph{average} influence
                                                \index{influence!average}
    $\AInf[g_{z,\pi}] \leq \frac{1}{m} A(m)$ (see Exercise~\ref{ex:ns-tails}\ref{ex:average-influence}).

    Now suppose $\bz \sim \bits^n$ and $\bpi \co [n] \to [m]$ are chosen uniformly at random.  We certainly have
    \[
        \E_{\bz, \bpi}[\AInf[g_{\bz,\bpi}]] \leq \tfrac{1}{m}A(m).
    \]
    To complete the proof we will show that the left-hand side above is precisely $\NS_{1/m}[f]$.  Recall that in the experiment for average influence $\AInf[g]$ we choose $\bw \sim \bits^m$ and $\bj \sim [m]$ uniformly at random and check if $g(\bw) \neq g(\bw^{\oplus \bj})$.  Thus
    \[
        \E_{\bz, \bpi}[\AInf[g_{\bz,\bpi}]] = \Pr_{\bz, \bpi, \bw, \bj}[g_{\bz, \bpi}(\bw) \neq g_{\bz, \bpi}(\bw^{\oplus \bj})]
        = \Pr_{\bw, \bpi, \bj, \bz}[f(\bz \circ \bw^{\bpi}) \neq f(\bz \circ (\bw^{\oplus \bj})^{\bpi})].
    \]
    It is not hard to see that the joint distribution of $\bz \circ \bw^{\bpi}$, $\bz \circ (\bw^{\oplus \bj})^{\bpi}$ is the same as that of $\bx$, $\by$.  To be precise, define $\bJ = \bpi^{-1}(\bj)$, distributed as a random subset of~$[n]$ in which each coordinate is included with probability~$1/m$, and define $\blambda \in \bn$ by $\blambda_i = -1$ if and only if $i \in \bJ$.  Then
    \[
        \Pr_{\bw, \bpi, \bj, \bz}[f(\bz \circ \bw^{\bpi}) \neq f(\bz \circ (\bw^{\oplus \bj})^{\bpi})] = \Pr_{\bw, \bpi, \bj, \bz}[f(\bz \circ \bw^{\bpi}) \neq f(\bz \circ \bw^{\bpi} \circ \blambda)].
    \]
    But for every outcome of $\bw$, $\bpi$, $\bj$ (and hence $\bJ$, $\blambda$), we may replace $\bz$ with $\bz \circ \bw^{\bpi}$ since they have the same distribution, namely uniform on $\bn$.  Then the above becomes
    \[
        \Pr_{\bw, \bpi, \bj, \bz}[f(\bz) \neq f(\bz \circ \blambda)] = \NS_{1/m}[f],
    \]
    as claimed.
\end{proof}

Peres's Theorem is now a simple corollary of Theorem~\ref{thm:ns-to-as}.

\begin{proof}[Proof of Peres's Theorem]
    Let $\calB$ be the class of all linear threshold functions.  This class is indeed closed under negating and identifying variables.  Since each linear threshold function on~$m$ bits is \emph{unate} (i.e., monotone up to negation of some
                                                        \index{unate}
    input coordinates, see Exercises~\ref{ex:deg-1-vs-inf},~\ref{ex:ltf-unate}), its total influence is at most~$\sqrt{m}$ (see Exercise~\ref{ex:simple-mono-tinf}).  Applying Theorem~\ref{thm:ns-to-as} we get that for any LTF $f$ and any $\delta \in (0,1/2]$,
    \begin{align*}
        \NS_\delta[f] \leq \tfrac{1}{m}\sqrt{m} &= 1/\sqrt{m} \qquad \text{(for $m = \lfloor 1/\delta \rfloor$)} \\
        &\leq O(\sqrt{\delta}). \qedhere
    \end{align*}
\end{proof}

\begin{remark} \label{rem:peres} Our proof of Peres's Theorem attains the upper bound $\sqrt{1/\lfloor 1/\delta \rfloor}$.  This is at most $\sqrt{3/2} \sqrt{\delta}$ for all $\delta \in (0,1/2]$ and it's also $\sqrt{\delta} + O(\delta^{3/2})$ for small~$\delta$.  To further improve the constant we can use Theorem~\ref{thm:maj-maximizes-deg-1-sum} in place of Exercise~\ref{ex:simple-mono-tinf}; it implies that all unate $m$-bit functions have total influence at most $\sqrt{2/\pi}\sqrt{m} + O(m^{-1/2})$.  This lets us obtain the bound $\NS_\delta[f] \leq \sqrt{2/\pi} \sqrt{\delta} + O(\delta^{3/2})$ for all LTF~$f$.
\end{remark}

Recall from Theorem~\ref{thm:maj-stab} that $\NS_\delta[\maj_n] \sim \tfrac{2}{\pi} \sqrt{\delta}$ for large~$n$.  Thus the constant $\sqrt{2/\pi}$ in the bound from Remark~\ref{rem:peres} is fairly close to optimal.  It seems quite likely that majority's $\tfrac{2}{\pi}$ is the correct constant here.  There is still slack in Peres's proof because the random functions $g_{\bz, \bpi}$ arising in Theorem~\ref{thm:ns-to-as} are unlikely to be majorities, even if~$f$~is.  The most elegant possible result in this direction would be to prove the following conjecture of Benjamini, Kalai, and Schramm:
\begin{named}{Majority Is Least Stable Conjecture}
                                                    \index{Majority Is Least Stable Conjecture}
Let $f \btb$ be a linear threshold function, $n$~odd.  Then for all $\rho \in [0,1]$, $\Stab_\rho[f] \geq \Stab_\rho[\maj_n]$.
\end{named}
\noindent (This is a precise statement about majority's noise stability within the class of LTFs; the Majority Is Stablest Theorem refers to its noise stability within the class of small-influence functions.) However, Sivakanth Gopi and others found a counterexample to the above conjecture, already for $n=5$.  A plausible replacement would be to conjecture that $\Stab_\rho[f] \geq \tfrac{2}{\pi} \arcsin \rho$ for all linear threshold functions~$f$.

                                                \index{linear threshold function!noise stability|)}

\medskip

                                                \index{polynomial threshold function!noise stability|(}
A challenging problem in this area is to extend Peres's Theorem to \emph{polynomial threshold functions}.  Let
\[
    \calP_{n,k} = \{f \btb \mid f \text{ is a PTF of degree at most $k$}\}, \quad \calP_k = \bigcup_n \calP_{n,k}.
\]
Peres's Theorem shows that the class $\calP_1$ (i.e., LTFs) is uniformly noise-stable.  Is the same true of~$\calP_2$?  What about~$\calP_{100}$?  More quantitatively, what upper bound can we prove on $\NS_\delta[f]$ for $f \in \calP_k$?  Since $\calP_k$ is closed under negating and identifying variables, a natural approach to bounding the noise sensitivity of PTFs is to again use Theorem~\ref{thm:ns-to-as}.  For example, if we could show that $\Tinf[f] = o(n)$ for all $f \in \calP_k$ we could conclude that $\NS_\delta[f] = o_\delta(1)$ for all $f \in \calP_k$; i.e., that $\calP_k$ is uniformly noise-stable. (In fact, the total influence approach to bounding noise sensitivity is not just sufficient but is also necessary;
                                                \index{polynomial threshold function!noise stability|)}%
                                                \index{polynomial threshold function!total influence|(}%
see Exercise~\ref{ex:gl-equiv}.)  More ambitiously, if we could show that $\Tinf[f] \leq O_k(1) \sqrt{n}$ for all $f \in \calP_{n,k}$ then it would follow that $\NS_\delta[f] \leq O_k(1) \sqrt{\delta}$ for all $f \in \calP_k$, strictly generalizing Peres's Theorem.  In fact, a conjecture of Gotsman and Linial dating back to 1990 proposes an even more refined bound:
\begin{named}{Gotsman--Linial Conjecture}  Let $f \in \calP_{n,k}$.
                                                \index{Gotsman--Linial Conjecture}
Then $\Tinf[f] \leq O_k(1) \sqrt{n}$.  More strongly, $\Tinf[f] \leq O(k) \sqrt{n}$.  Most strongly, the $f \in \calP_{n,k}$ of maximal total influence is the symmetric one $f(x) = \sgn(p(x_1 + \cdots + x_n))$, where $p$ is a degree-$k$ univariate polynomial which alternates sign on the $k+1$ values of $x_1 + \cdots + x_n$ closest to~$0$.
\end{named}
The strongest form of the Gotsman--Linial Conjecture  is true when $k = 1$, by Theorem~\ref{thm:maj-maximizes-deg-1-sum}. However, even for $k = 2$ there was no progress on the conjecture for close to 20~years. At that point two independent works~\cite{DHK+10,HKM10a} showed that every $f \in \calP_{n,k}$ satisfies both $\TInf[f] \leq O(n^{1-1/2^{k}})$ and $\Tinf[f] \leq 2^{O(k)} n^{1 - 1/O(k)}$.  The former (essentially weaker) bound has the advantage of an elementary proof; see Exercise~\ref{ex:gl-weak}.  It  also suffices to show that~$\calP_k$, the class of degree-$k$ PTFs, is indeed uniformly noise-stable. This gives a nice kind of converse to Proposition~\ref{prop:noise-stable-implies-PTF}, which showed that every function in a uniformly noise-stable class is close to being a constant-degree PTF.

The latest progress on the Gotsman--Linial Conjecture is the following theorem of Kane~\cite{Kan12a}, which comes quite close to proving it:
\begin{theorem}                                 \label{thm:kane}
    Every $f \in \calP_{n,k}$ satisfies $\Tinf[f] \leq \sqrt{n} \cdot (2^k \log n)^{O(k \log k)}$.  It follows (via Theorem~\ref{thm:ns-to-as}) that for a \emph{fixed} $k \in \N^+$, every $f \in \calP_k$ satisfies $\NS_\delta[f] \leq \sqrt{\delta} \cdot \polylog(1/\delta)$.
\end{theorem}
                          \index{polynomial threshold function!total influence|)}

\section{Exercises and notes}

\begin{exercises}
    \item \label{ex:LTF-integer}
        \begin{exercises}
            \item Suppose $f \btb$ is an LTF.  Show that it can be expressed as $f(x) = \sgn(a_0 + a_1 x_1 + \cdots a_n x_n)$ where the $a_i$'s are integers.  (Hint: First obtain rational $a_i$'s by a perturbation.)
            \item Show also that a degree-$d$ PTF has a representation in which all of the degree-$d$ polynomial's coefficients are integers.
        \end{exercises}
    \item Let $f(x) = \sgn(a_0 + a_1 x_1 + \cdots a_n x_n)$ be an LTF.
        \begin{exercises}
        \item Show that if $a_0 = 0$, then $\E[f] = 0$.  (Hint: Show that~$f$ is in fact an odd function.)
        \item Show that if $a_0 \geq 0$, then $\E[f] \geq 0$. Show that the converse need not hold.
        \item Suppose $g \btb$ is an LTF with $\E[g] = 0$.  Show that $g$ can be represented as $g(x) = \sgn(c_1 x_1 + \cdots + c_n x_n)$.
        \end{exercises}
    \item Suppose $f(x) = \sgn(a_0 + a_1 x_1 + \cdots a_n x_n)$ is an LTF with $|a_1| \geq |a_2| \geq \cdots \geq |a_n|$.  Show that $\Inf_1[f] \geq \Inf_2[f] \geq \cdots \geq \Inf_n[f]$.  (Hint: Why does it suffice to prove this for $n=2$?)
    \item \label{ex:chow-counting} \begin{exercises}
        \item Show that the number of functions $f \btb$ that are LTFs is at most $2^{n^2 + O(n)}$.  (Hint: Chow's Theorem.)
        \item More generally, show that the number of functions $f \btb$ that are degree-$k$ PTFs is at most $2^{n^{k+1} + O(n)}$.
        \end{exercises}
    \item \label{ex:finish-gotsman-linial1}
        \begin{exercises}
            \item Suppose $\ell \btR$ is defined by $\ell(x) = a_0 + a_1 x_1 + \cdots + a_n x_n$.  Define $\wt{\ell} \co \bits^{n+1} \to \R$ by $\wt{\ell}(x_0, \dots, x_n) = a_0 x_0 + a_1 x_1 + \cdots a_n x_n$.  Show that $\|\wt{\ell}\|_1 = \|\ell\|_1$ and $\|\wt{\ell}\|^2_2 = \|\ell\|_2^2$.
            \item Complete the proof of Theorem~\ref{thm:gotsman-linial1}.
        \end{exercises}
    \item \label{ex:LTF-KKL} Let $f \btb$ be an unbiased linear threshold function.  Show that $\Inf_i[f] \geq \frac{1}{\sqrt{2 n}}$ for some $i \in [n]$, improving the KKL Theorem for LTFs.
    \item \label{ex:titsworth} Consider the following ``correlation distillation'' problem (cf.~Exercise~\ref{ex:nicd}).
                \index{correlation distillation}
          For each $i \in [n]$ there is a number $\rho_i \in [-1,1]$ and an independent sequence of pairs of $\rho_i$-correlated bits, $(\ba_i^{(1)}, \bb_i^{(1)})$, $(\ba_i^{(2)}, \bb_i^{(2)})$, $(\ba_i^{(3)}, \bb_i^{(3)})$, etc.  Party~$A$ on Earth has access to the stream of $n$-bit strings $\ba^{(1)}$, $\ba^{(2)}$, $\ba^{(3)}$,~\dots and a party~$B$ on Venus has access to the stream $\bb^{(1)}$, $\bb^{(2)}$, $\bb^{(3)}$,~\dots.  Neither party knows the numbers $\rho_1, \dots, \rho_n$. The goal is for~$B$ to estimate these correlations.  To assist in this,~$A$ can send a small number of bits to~$B$.  A~reasonable strategy is for~$A$ to send $f(\ba^{(1)})$, $f(\ba^{(2)})$, $f(\ba^{(3)})$, \dots to~$B$, where $f \btb$ is some Boolean function.  Using this information~$B$ can try to estimate $\E[f(\ba) \bb_i]$ for each~$i$.
          \begin{exercises}
            \item Show that $\E[f(\ba) \bb_i] = \wh{f}(i) \rho_i$.
            \item \label{ex:titsworthb} This motivates choosing an $f$ for which all $\wh{f}(i)$ are large.  If we also insist all $\wh{f}(i)$ be equal, show that majority functions~$f$ maximize this common value.
          \end{exercises}
    \item \label{ex:random-Fourier2}  For $n \geq 2$, let $\boldf \btb$ be a randomly chosen function (as in
                                                \index{random function}
        Exercise~\ref{ex:random-Fourier}).  Show that $\snorm{\boldf}_\infty \leq 2\sqrt{n} 2^{-n/2}$ except with probability at most~$2^{-n}$.
    \item \label{ex:ptf-chow} Prove Theorem~\ref{thm:ptf-chow}.
    \item \label{ex:parity-ptf}
          \begin{exercises}
            \item Give as simple a proof as you can that the parity function $\chi_{[n]} \btb$ is not a PTF of degree~$n-1$.
            \item Show that if $f \btb$ is not $\pm \chi_{[n]}$, then it \emph{is} a PTF of degree~$n-1$.  (Hint: Consider $f^{\leq n-1}$.)
          \end{exercises}
    \item \label{ex:no-beat-gl} For each $k \in \N^+$, show that there is a degree-$k$ PTF~$f$ with $\W{\leq k}[f] < 2^{1-k}$.
    \item \label{ex:bruck-separation} In this exercise you will show that threshold-of-parities circuits can be effectively
                                                \index{threshold-of-parities circuit}
                simulated by threshold-of-threshold circuits, but not the converse.
            \begin{exercises}
                 \item Let $f \btb$ be a symmetric function.  Show that $f$ is computable as the \emph{sum} of at most~$2n$ LTFs, plus a constant.
                 \item Deduce that if $f \btb$ is computable by a size-$s$ threshold-of-parities circuit, then it is also computable by a size-$2ns$ threshold-of-thresholds circuit.
                 \item Show that the complete quadratic function $\CQ_n \co \F_2^{n} \to \{-1,1\}$ (see Exercise~\ref{ex:compute-expansions})
                                                     \index{complete quadratic function}
                     is computable by a size-$2n$ threshold-of-thresholds circuit.
                 \item Assume $n$ even.  Show that any threshold-of-parities circuit for $\CQ_n$ requires size~$2^{n/2}$.
            \end{exercises}
    \item \label{ex:krause-pudlak}  Let $f \btb$ be computable by a DNF of size~$s$. Show that $f$ has a PTF representation of sparsity~$O(n s^3)$.  (Hint: Approximate the ANDs using Theorem~\ref{thm:bruck-smolensky}.)  Can you improve this bound to $O(n s^2)$?
    \item \label{ex:bruck-separation2}  In contrast to the previous exercise, show that there is a function $f \btb$ computable by a depth-$3$ $\AC^0$~circuit (see Chapter~\ref{sec:LMN})
                                              \index{constant-depth circuits}%
        but requiring
                                                \index{threshold-of-parities circuit}
        threshold-of-parities circuits of size at least~$n^{\log n}$.  (Hint: Involve the inner product mod~$2$ function and Exercise~\ref{ex:compute-parity}.)
    \item \label{ex:PTF-extremal1}  Let $\calF$ be a nonempty collection of subsets $S \subseteq [n]$.  For each $a \in \bn$, write $1_{\{a\}} \btzo$ for the indicator of~$\{a\}$, write $1_{\{a\}}^\calF \btR$ for $\sum_{S \in \calF} \wh{1_{\{a\}}}(S)\,\chi_S$, and write $\psi_a  = \frac{2^n}{|\calF|} \cdot 1_{\{a\}}^\calF$.
        \begin{exercises}
            \item Show that $\psi_a(a) = 1$ and $\E[\psi_a^2] = \tfrac{1}{|\calF|}$.  Show also that for all $x \in \bn$, $\psi_a(x) = \psi_x(a)$ and $\sum_{a : a \neq x} \psi_a(x)^2 = \frac{2^n}{|\calF|} - 1$.
            \item Fix $0 < \eps < 1$ and suppose $|\calF| \geq (1 - \tfrac{\eps^2}{6n}) 2^n$.  Let $\boldf \btb$
                                            \index{random function}
            be a random function as in Exercise~\ref{ex:random-Fourier}.  Show that for each $x \in \bn$, except with probability at most $4^{-n}$ we have $|\sum_{a : a \neq x} \boldf(a) \psi_a(x)| < \eps$.
            \item Deduce that for all but a $2^{-n}$ fraction of functions $f \btb$, there a multilinear polynomial $q \btR$ supported on the monomials $\{\chi_S : S \in \calF\}$ such that $\|f - q\|_\infty < \eps$.
                                    \index{polynomial threshold function!sparsity}%
                                    \index{polynomial threshold function!degree}%
                                    \index{approximating polynomial}%
            \item Deduce that all but a $2^{-n}$ fraction of functions $f \btb$ have PTF representation of degree at most $n/2 + O(\sqrt{n \log n})$.
        \end{exercises}
    \item \label{ex:BE-interval}
        \begin{exercises}
            \item Show that in the Berry--Esseen Theorem we can also conclude
                \[
                    |\Pr[\bS < u] - \Pr[\bZ < u]| \leq c \gamma.
                \]
                (Hint: You'll need that $\lim_{\delta \to 0^+} \Pr[\bZ \leq u - \delta] = \Pr[\bZ \leq u]$.)
            \item Deduce that if $I \subseteq \R$ is any interval, we can also conclude
				\[
					|\Pr[\bS \in I] - \Pr[\bZ \in I]| \leq 2c \gamma.
				\]			
        \end{exercises}
    \item \label{ex:no-restrict} Show that the assumptions $\E[\bX_i] = 0$ and $\sum_{i=1}^n \Var[\bX_i] = 1$ in the Berry--Esseen Theorem are not restrictive, as follows.  Let $\bX_1, \dots, \bX_n$ be independent random variables with finite means and variances.  Let $\bS = \sum_{i=1}^n \bX_i$ and let $\bZ \sim \normal(\mu,\sigma^2)$, where $\mu = \sum_{i=1}^n \E[\bX_i]$ and $\sigma^2 = \sum_{i=1}^n \Var[\bX_i]$.  Assuming $\sigma^2 > 0$, show that for all $u \in \R$,
        \[
            |\Pr[\bS \leq u] - \Pr[\bZ \leq u]| \leq c \eps / \sigma^3,
        \]
        where
        \[
            \eps = \sum_{i=1}^n \|\bX_i - \E[\bX_i]\|_3^3.
        \]
    \item  \label{ex:arcsin}
            \begin{exercises}
                \item Use the generalized Binomial Theorem to compute the power series for $(1-z^2)^{-1/2}$, valid for $|z| < 1$.
                \item Integrate to obtain the power series for $\arcsin z$ given in~\eqref{eqn:arcsin}, valid for $|z| < 1$.
                \item Confirm that equality holds also for $z = \pm 1$.
            \end{exercises}
	\item \label{ex:Svec-cov} Verify that the random vector $\vec{\bS}$ defined in~\eqref{eqn:Svec-cov} has $\E[\vec{\bS}_1] = \E[\vec{\bS}_2] = 0$, $\E[\vec{\bS}_1^2] = \E[\vec{\bS}_2^2] = 1$, $\E[\vec{\bS}_1\vec{\bS}_2] = \rho$; i.e., $\E[\vec{\bS}] = \begin{bmatrix} 0 \\ 0 \end{bmatrix}$ and $\Cov[\vec{\bS}] = \begin{bmatrix} 1 & \rho \\ \rho & 1 \end{bmatrix}$.
    \item \label{ex:maj-coeffs-symm} Prove Corollary~\ref{cor:maj-coeffs-symm}.
    \item \label{ex:maj-coeffs-decr} Fix $n$ odd.  Using Theorem~\ref{thm:maj-coeffs} show that $|\wh{\Maj_n}(S)|$ is a decreasing function of $|S|$ for odd $1 \leq |S| \leq \frac{n-1}{2}$. Deduce (using also Corollary~\ref{cor:maj-coeffs-symm}) that $\snorm{\Maj_n}_\infty = \Maj_n(\{1\}) \sim \frac{\sqrt{2/\pi}}{\sqrt{n}}$.
    \item \label{ex:maj-weight-decreasing} Prove Corollary~\ref{cor:maj-weight-decreasing}.
	\item \label{ex:maj-stab-error2}  %
							\index{majority!noise stability}%
            Prove Theorem~\ref{thm:maj-stab-precise}.  (Hint: Corollary~\ref{cor:maj-weight-decreasing}.)
    \item \label{ex:maj-annoying} Complete the proof of Theorem~\ref{thm:maj-asymptotic-weight} by showing that $(1-\tfrac{k+1}{n} +\tfrac{k}{n^2})^{-1/2} \leq 1+2k/n$ for all $1 \leq k \leq n/2$.
    \item \label{ex:maj-stab-series} Using just the facts that $\Stab_\rho[\Maj_n] \to \frac{2}{\pi} \arcsin \rho$ for all $\rho \in [-1,1]$ and that $\Stab_\rho[\Maj_n] = \sum_{k \geq 0} \W{k}[\Maj_n] \rho^k$, deduce that $\lim_{n \to \infty} \W{k}[\Maj_n] \to [\rho^k](\frac{2}{\pi} \arcsin \rho)$ for all $k \in \N$.  (Hint:
        By induction on~$k$, always taking $\rho$ ``small enough''.)
    \item \label{ex:L1-maj}              \begin{exercises}
            \item For $0 \leq j \leq m$ integers, show that $\snorm{\Maj_{2m+1}^{=2j+1}}_1 = \binom{m}{j} \frac{1}{2j+1} \cdot \frac{2m+1}{2^{2m}}\binom{2m}{m}$.
            \item Deduce  $\snorm{\Maj_{2m+1}}_1 = \E\left[\frac{1}{2\bX+1}\right] \cdot \frac{2m+1}{2^{m}}\binom{2m}{m}$, where $\bX \sim \Binomial(m,1/2)$.
            \item Deduce  $\snorm{\Maj_n}_1 \sim \frac{2}{\sqrt{\pi}} \frac{1}{\sqrt{n}} 2^{n/2}$.
            \end{exercises}
    \item \label{ex:maj-asympt-asympt}
        \begin{exercises}
            \item Show that for each odd $k \in \N$,
                    \[
                        \left(\tfrac{2}{\pi}\right)^{3/2} k^{-3/2} \leq [\rho^k] (\tfrac{2}{\pi} \arcsin \rho) \leq \left(\tfrac{2}{\pi}\right)^{3/2} k^{-3/2}(1+O(1/k)).
                    \]
                  (Hint: Stirling's approximation.)
            \item Prove Corollary~\ref{cor:maj-asympt-asympt}.  (Hint: For the second statement you'll need to approximate the sum $\sum_{\text{odd } j > k} \left(\tfrac{2}{\pi}\right)^{3/2} j^{-3/2}$ by an integral.)
        \end{exercises}
    \item \label{ex:kravchuk}  For integer $0 \leq j \leq n$, define $\Krav_j \btR$ by $\Krav_j(x) = \sum_{|S| = j} x^S$.  Since $\Krav_j$ is symmetric, the value $\Krav_j(x)$ depends only on the number~$z$ of $-1$'s in~$x$;  or equivalently, on $\sum_{i=1}^n x_i$.  Thus we may define $\krav_j \co \{0, 1, \dots, n\} \to \R$ by $\krav_j(z) = \Krav_j(x)$ for any $x$ with $\sum_i x_i = n - 2z$.
        \begin{exercises}
            \item Show that $\krav_j(z)$ can be expressed as a degree-$j$ polynomial in~$z$.  It is called the \emph{Kravchuk (\emph{or} Krawtchouk) polynomial} of degree~$j$.
                                                    \index{Kravchuk polynomials}%
                                                    \index{Krawtchouk polynomials|seeonly{Kravchuk polynomials}}%
                (The dependence on~$n$ is usually implicit.)
            \item Show that $\sum_{j=0}^n \Krav_j(x) = 2^n \cdot 1_{(1, \dots, 1)}(x)$.
            \item Show for $\rho \in [-1,1]$ that $\sum_{j=0}^n \Krav_j(x)\rho^j  = 2^n\Pr[\by = (1, \dots, 1)]$, where~$\by = N_\rho(x)$.
            \item Deduce the generating function identity $\krav_j(z) = [\rho^j]((1-\rho)^z(1+\rho)^{n-z})$.
        \end{exercises}
    \item \label{ex:cube-weight1} Prove Proposition~\ref{prop:cube-weight1}.
    \item \label{ex:ball-weight1} Prove Proposition~\ref{prop:ball-weight1} using the Central Limit Theorem.  (Hint for $\W{1}[f_n]$: use symmetry to show it equals the square of $\E[f_n(\bx)\littlesum \frac{1}{\sqrt{n}} \bx_i]$.)
    \item \label{ex:l1-be} Consider the setting of Theorem~\ref{thm:l1-be}.  Let $\bS = \sum_i a_i \bx_i$ where $\bx \sim \bn$, and let $\bZ \sim \normal(0,1)$.
        \begin{exercises}
            \item Show that $\Pr[|\bS| \geq t]$, $\Pr[|\bZ| \geq t] \leq 2\exp(-t^2/2)$ for all $t \geq 0$.
            \item Recalling $\E[|\bY|] = \int_0^\infty \Pr[|\bY| \geq t]\,dt$ for any random variable~$\bY$, use the Berry--Esseen Theorem (and Remark~\ref{rem:simpler-BE}, Exercise~\ref{ex:BE-interval}) to show
                \[
                    \Bigl|\E[|\bS|] - \E[|\bZ|]\Bigr| \leq O(\eps T + \exp(-T^2/2))
                \]
                for any $T \geq 1$.
            \item Deduce $|\Ex[|\bS|] - \sqrt{2/\pi}| \leq O(\eps\sqrt{\log(1/\eps)})$.
            \item Improve $O(\eps\sqrt{\log(1/\eps)})$  to the bound~$O(\eps)$ stated in Theorem~\ref{thm:l1-be} by using the \emph{nonuniform Berry--Esseen Theorem},
                                             \index{Berry--Esseen Theorem!nonuniform}
                which states that the bound~$c\gamma$ in the Berry--Esseen Theorem can be improved to~$C\gamma\cdot\frac{1}{1+|u|^3}$ for some constant~$C$.
        \end{exercises}
    \item \label{ex:ball-stab} Consider the sequence of LTFs defined in Proposition~\ref{prop:ball-weight1}. Show that
        \[
            \lim_{n \to \infty} \Stab_\rho[f_n] = \GaussQuad_\rho(\mu).
        \]
        Here $\mu = \olPhi(t)$ and $\GaussQuad_\rho(\mu)$ is the \emph{Gaussian quadrant probability}
                                            \index{Gaussian quadrant probability}%
                                            \nomenclature[Lambdarhomu]{$\GaussQuad_\rho(\alpha)$}{denotes $\GaussQuad_\rho(\alpha,\alpha)$}%
        defined by $\GaussQuad_\rho(\mu) = \Pr[\bz_1 > t, \bz_2 > t]$, where $\bz_1, \bz_2$ are standard Gaussians with correlation $\E[\bz_1\bz_2] = \rho$.  Verify also that $\GaussQuad_\rho(\alpha) = \Pr[\bz_1 \leq t, \bz_2 \leq t]$ where $\alpha = \Phi(t)$.
	\item \label{ex:ltf-stab-error}  %
							\index{Central Limit Theorem!multidimensional}%
							\index{majority!noise stability}%
							\index{linear threshold function!noise stability}%
							\index{Berry--Esseen Theorem!multidimensional}%
            In this exercise you will complete the justification of Theorem~\ref{thm:ltf-stab-error} using the following multidimensional Berry-Esseen Theorem:
            \begin{theorem}             \label{thm:multidim-BE}
                Let $\bX_1, \dots, \bX_n$ be independent $\R^d$-valued random vectors, each having mean zero.  Write $\bS = \sum_{i=1}^n \bX_i$ and assume $\Sigma = \Cov[\bS]$ is invertible. Let $\bZ \sim \normal(0,\Sigma)$ be a $d$-dimensional Gaussian with the same mean and covariance matrix as~$\bS$. Then for all convex sets $U \subseteq \R^d$,
                \[
                    |\Pr[\bS \in U] - \Pr[\bZ \in U]| \leq C d^{1/4} \gamma,
                \]
                where $C$ is a universal constant, $\gamma = \sum_{i=1}^n \E[\|\Sigma^{-1/2} \bX_i\|_2^3]$, and $\|\cdot\|_2$ denotes the Euclidean norm on~$\R^d$.
            \end{theorem}
            \begin{exercises}
                \item Let $\Sigma = \begin{bmatrix} 1 & \rho \\ \rho & 1 \end{bmatrix}$ where $\rho \in (-1,1)$.  Show that
                    \[
                        \Sigma^{-1} = \begin{bmatrix} 1 & -\rho \\ 0 & 1 \end{bmatrix}\begin{bmatrix} 1 & 0 \\ 0 & (1 -\rho^2)^{-1} \end{bmatrix} \begin{bmatrix} 1 & 0 \\ -\rho & 1 \end{bmatrix}.
                    \]
                \item Compute $y^\top \Sigma^{-1} y$ for $y = \begin{bmatrix} \pm a \\ \pm a \end{bmatrix} \in \R^2$.
                \item Complete the proof of Theorem~\ref{thm:ltf-stab-error}.
            \end{exercises}
    \item \label{ex:vac-noise-stable}  Let $\calB$ be a class of Boolean-valued functions, all of input length at most~$n$.  Show that $\NS_\delta[f] \leq n\delta$ for all $f \in \calB$ and hence~$\calB$ is uniformly noise-stable (in a sense, vacuously).  (Hint: Exercise~\ref{ex:tinf-bounds-ns}.)
    \item \label{ex:erics-fkn} Give a simple proof of the following fact, which is a robust form of the edge-isoperimetric inequality (for volume~$1/2$) and a weak form of the FKN~Theorem: If $f \btb$ has $\E[f] = 0$ and $\Tinf[f] \leq 1 + \delta$, then~$f$ is $O(\delta)$-close to $\pm \chi_i$ for some $i \in [n]$.  In fact, you should be able to achieve $\delta$-closeness (which can be further improved using Theorem~\ref{thm:FKN-improvement}).
                                                    \index{isoperimetric inequality!Hamming cube}%
        (Hint: Upper- and lower-bound $\sum_i \wh{f}(i)^2 \leq (\max_{i} |\wh{f}(i)|)(\sum_i |\wh{f}(i)|)$ using Proposition~\ref{prop:tinf-concentration} and Exercise~\ref{ex:deg-1-vs-inf}\ref{ex:deg-1-vs-inf-a}.)
    \item \label{ex:fkn-optimality} Show that Theorem~\ref{thm:FKN-improvement} is essentially optimal by exhibiting functions $f \btb$ with $\wh{f}(1) = 1-\delta/2$ and $\W{1}[f] \geq 1 - \delta + \Omega(\delta^2 \log(1/\delta))$, for a sequence of~$\delta$ tending to~$0$.
    \item \label{ex:level-1-quick} Prove Corollary~\ref{cor:level-1-quick}.
    \item \label{ex:FKN-improvement}  Fill in the details of the proof of Theorem~\ref{thm:FKN-improvement}.
    \item \label{ex:ns-deriv-LTFs} Show that if $f \btb$ is an LTF, then $\frac{d}{d\delta} \NS_\delta[f] \leq O(1/\sqrt{\delta})$.  (Hint: The only fact needed about LTFs is the corollary of Peres's Theorem that  $\W{\geq k}[f] \leq O(1/\sqrt{k})$ for all~$k$.)
    \item \label{ex:gl-equiv}  As discussed in Section~\ref{sec:peres}, Theorem~\ref{thm:ns-to-as} implies that an upper bound on the total influence of degree-$k$ PTFs is sufficient to derive an upper bound on their noise sensitivity.  This exercise asks you to show necessity as well.  More precisely, suppose $\NS_\delta[f] \leq \eps(\delta)$ for all $f \in \calP_k$.  Show that $\Tinf[f] \leq O(\eps(1/n) \cdot n)$ for all $f \in \calP_{n,k}$.  Deduce that $\calP_{k}$ is uniformly noise-stable if and only if $\Tinf[f] = o(n)$ for all $f \in \calP_{n,k}$ and that $\NS_{\delta}[f] \leq O(k \sqrt{\delta})$ for all $f \in \calP_{k}$ if and only if $\Tinf[f] \leq O(k \sqrt{n})$ for all $f \in \calP_{n,k}$.
                                                \index{polynomial threshold function!noise stability}%
                                                \index{polynomial threshold function!total influence}%
        (Hint: Exercise~\ref{ex:ns-tails}\ref{ex:average-influence}.)
    \item Estimate carefully the asymptotics of $\Tinf[f]$, where $f \in \mathrm{PTF}_{n,k}$ is as in the strongest form of the Gotsman--Linial Conjecture.
    \item \label{ex:rocco-problem} Let $A \subseteq \bn$ have cardinality $\alpha 2^n$, $\alpha \leq 1/2$.  Thinking of $\bn \subset \R^n$, let $\mu_A \in \R^n$ be the center of mass of $A$.  Show that $\mu_A$ is close to the origin in Euclidean distance: $\|\mu_A\|_2 \leq O(\sqrt{\log(1/\alpha)})$.
    \item \label{ex:giso-diffequ} Show that the Gaussian isoperimetric function
                                                    \index{Gaussian isoperimetric function}%
            satisfies $\GIso'' = -1/\Giso$ on $(0,1)$. Deduce that $\GIso$ is concave.
    \item \label{ex:refined-level-1} Fix $\alpha \in (0, 1/2)$.  Let $f \btI$ satisfy $\E[|f|] \leq \alpha$ and $|\wh{f}(i)| \leq \eps$ for all $i \in [n]$.  Show that
                                                        \index{Fourier weight!degree-1}%
        $\W{1}[f] \leq \Giso(\alpha)^2 + C \eps$, where $\Giso$ is the Gaussian isoperimetric function and where the constant~$C$ may depend on~$\alpha$.  (Hint: You will need the nonuniform Berry--Esseen Theorem from Exercise~\ref{ex:l1-be}.)
    \item \label{ex:gl-weak} In this exercise you will show by induction on~$k$ that
                                                \index{polynomial threshold function!noise stability}%
                                                \index{polynomial threshold function!total influence}%
            $\Inf[f] \leq 2 n^{1-1/2^k}$ for all degree-$k$ PTFs $f \btb$.  The $k = 0$ case is trivial.  So for $k > 0$, suppose $f = \sgn(p)$ where $p \btR$ is a degree-$k$ polynomial that is never~$0$.
        \begin{exercises}
            \item Show for $i \in [n]$ that $\E[f(\bx) \bx_i \sgn(\D_i p(\bx))] = \Inf_i[f]$.  (Hint: First use the decomposition $f = x_i \D_i f + \uE_i f$  to reach $\E[\D_i f \cdot \sgn(\D_i p)]$; then show that $\D_if = \sgn(\D_i p)$ whenever $\D_i f \neq 0$.)
            \item Conclude that $\Tinf[f] \leq \E[|\sum_i \bx_i \sgn(\D_i p(\bx))|]$.  Remark: When $k = 2$ and thus each $\sgn(\D_i p)$ is an LTF, it is conjectured that this bound is still~$O(\sqrt{n})$.
            \item Apply Cauchy--Schwarz and deduce
                \[
                    \Tinf[f] \leq \sqrt{n + \sum_{i \neq j} \E[\bx_i \bx_j \sgn(\D_i p(\bx)) \sgn(\D_j p(\bx))]}.
                \]
            \item Use Exercise~\ref{ex:xixj} and the AM-GM inequality to obtain 
                \[
                    \Tinf[f] \leq \sqrt{n + \sum_i \Tinf[\sgn(\D_i p)]}.
                \]
            \item Complete the induction.
            \item Finally, deduce that the class of degree-$k$ PTFs is uniformly noise-stable, specifically, that every degree-$k$ PTF $f$ satisfies $\NS_\delta[f] \leq 3\delta^{1/2^k}$ for all $\delta \in (0,1/2]$.  (Hint: Theorem~\ref{thm:ns-to-as}.)
        \end{exercises}
\end{exercises}

\subsection*{Notes.}

Chow's Theorem was proved by independently by Chow~\cite{Cho61} and by Tannenbaum~\cite{Tan61} in 1961; see also Elgot~\cite{Elg61}.  The generalization to PTFs (Theorem~\ref{thm:ptf-chow}) is due to Bruck~\cite{Bru90}, as is Theorem~\ref{thm:bruck} and Exercise~\ref{ex:bruck-separation}.  Theorems~\ref{thm:gotsman-linial1} and~\ref{thm:gotsman-linial2} are from Gotsman and Linial~\cite{GL94} and may be called the Gotsman--Linial Theorems; this work also contains the Gotsman--Linial Conjecture and Exercise~\ref{ex:no-beat-gl}. Conjecture~\ref{conj:ltf-weight-1} should be considered folklore.  Corollary~\ref{cor:bruck-smolensky} was proved by Bruck and Smolensky~\cite{BS92}; they also essentially proved Theorem~\ref{thm:bruck-smolensky} (but see~\cite{SB91}). Exercise~\ref{ex:krause-pudlak} is usually credited to Krause and Pudl\'{a}k~\cite{KP97}. The upper bound in Exercise~\ref{ex:chow-counting} is asymptotically sharp~\cite{Zue89}. Exercise~\ref{ex:PTF-extremal1} is from O'Donnell and Servedio~\cite{OS08a}.

Theorem~\ref{thm:maj-maximizes-deg-1-sum} and Proposition~\ref{prop:weight-1-same}, discussed in Section~\ref{sec:majority}, were essentially proved by Titsworth in 1962~\cite{Tit62}; see also~\cite{Tit63}. More precisely, Titsworth solved a version of the problem from Exercise~\ref{ex:titsworth}.  His motivation was in fact the construction of ``interplanetary ranging systems'' for measuring deep space distances, e.g., the distance from Earth to Venus.  The connection between ranging systems and Boolean functions was suggested by his advisor, Solomon Golomb.  Titsworth~\cite{Tit62} was also the first to compute the Fourier expansion of $\maj_n$.  His approach involved generating functions and contour integration.  Other approaches have used special properties of binomial coefficients~\cite{Bra87} or of Kravchuk polynomials~\cite{Kal02}.  The asymptotics of $\W{k}[\Maj_n]$ described in Section~\ref{sec:maj-coefficients} may have first appeared in Kalai~\cite{Kal02}, with the error bounds being from O'Donnell~\cite{O'D03}.  Kravchuk polynomials were introduced by Kravchuk~\cite{Kra29}.

The Berry--Esseen Theorem is due independently to Berry~\cite{Ber41} and Esseen~\cite{Ess42}.  Shevtsova~\cite{She13} has the record for the smallest known constant~$B$ that works therein: roughly~$.5514$.  The nonuniform version described in Exercise~\ref{ex:l1-be} is due to Bikelis~\cite{Bik66}. The multidimensional version Theorem~\ref{thm:multidim-BE} stated in Exercise~\ref{ex:ltf-stab-error} is due to Bentkus~\cite{Ben04}.  Sheppard proved his formula in 1899~\cite{She99}.  The results of Theorem~\ref{thm:maj-stab-precise} may have appeared first in O'Donnell~\cite{O'D04,O'D03}.

The Level-1 Inequality should probably be considered folklore; it was perhaps first published in Talagrand~\cite{Tal96} and we have followed his proof.  The first half of the $\frac{2}{\pi}$~Theorem is from Khot et~al.~\cite{KKMO07}; the second half is from Matulef et~al.~\cite{MORS10}.  Theorem~\ref{thm:FKN-improvement}, which improves the FKN Theorem to achieve ``closeness''~$\delta/4$, was independently obtained by Jendrej, Oleszkiewicz, and Wojtaszczyk~\cite{JOW12}, as was Exercise~\ref{ex:fkn-optimality} showing optimality of this closeness.  The closeness achieved in the original proof of the FKN~Theorem~\cite{FKN02} was~$\delta/2$; that proof (like ours) relies on having a separate proof of closeness~$O(\delta)$.  Kindler and Safra~\cite{KS02,Kin02} gave a self-contained proof of the $\delta/2$ bound relying only on the Hoeffding bound.  The content of Exercise~\ref{ex:erics-fkn} was communicated to the author by Eric Blais.  The result of Exercise~\ref{ex:refined-level-1} is from~\cite{KKMO07}; Exercise~\ref{ex:rocco-problem} was suggested by Rocco Servedio.

Peres's Theorem was published in 2004~\cite{Per04} but was mentioned as early as 1999 by Benjamini, Kalai, and Schramm~\cite{BKS99}.  The work~\cite{BKS99} introduced the definition of uniform noise stability and showed that the class of all LTFs satisfies it; however, their upper bound on the noise sensitivity of LTFs was~$O(\delta^{1/4})$, worse than Peres's.  The proof of Peres's Theorem that we presented is a simplification due to Parikshit Gopalan and incorporates an idea of Diakonikolas et~al.~\cite{DHK+10,HKM10a}.  Regarding the total influence of PTFs, the work of Kane~\cite{Kan12a} shows that every degree-$k$ PTF on~$n$ variables has $\Tinf[f] \leq \poly(k) n^{1 - 1/O(k)}$, which is better than Theorem~\ref{thm:kane} for certain superconstant values of~$k$.  Exercise~\ref{ex:ns-deriv-LTFs} was suggested by Nitin Saurabh.

\chapter{Pseudorandomness and \texorpdfstring{$\F_2$}{F2}-polynomials}              \label{chap:pseudorandomness}
In this chapter we discuss various notions of pseudorandomness for Boolean functions; by this we mean properties of a fixed Boolean function that are in some way characteristic of randomly chosen functions.  We will see some deterministic constructions of pseudorandom probability density functions with small support; these have algorithmic application in the field of derandomization.  Finally, several of the results in the chapter will involve interplay between the representation of $f \zotzo$ as a polynomial over the reals and its representation as a polynomial over~$\F_2$.
	
\section{Notions of pseudorandomness}                      \label{sec:pseudorandomness-notions}
The most obvious spectral property of a truly random function $\boldf \btb$ is that all of its Fourier coefficients are very small (as we saw in Exercise~\ref{ex:random-Fourier2}).  Let's switch notation to $\boldf \btzo$; in this case $\boldf(\emptyset)$ will not be very small but rather very close to~$1/2$.  Generalizing:
\begin{proposition}                                     \label{prop:random-Fourier}
                                            \index{random function}
    Let $n > 1$ and let $\boldf \btzo$ be a $p$-biased random function; i.e., each $\boldf(x)$ is~$1$ with probability~$p$ and~$0$ with probability $1-p$, independently for all $x \in \bn$.  Then except with probability at most~$2^{-n}$, all of the following hold:
    \[
        |\wh{\boldf}(\emptyset) - p| \leq 2\sqrt{n} 2^{-n/2}, \qquad \forall S \neq \emptyset \quad |\wh{\boldf}(S)| \leq 2\sqrt{n} 2^{-n/2}.
    \]
\end{proposition}
\begin{proof}
    We have $\wh{\boldf}(S) = \sum_{x} \frac{1}{2^n} x^S \boldf(x)$, where the random variables $\boldf(x)$ are independent.  If $S = \emptyset$, then the coefficients $\frac{1}{2^n} x^S$ sum to~$1$ and the mean of~$\wh{\boldf}(S)$ is~$p$; otherwise the coefficients sum to~$0$ and the mean of~$\wh{\boldf}(S)$ is~$0$.  Either way we may apply the Hoeffding bound to conclude that
    \[
        \Pr[|\wh{\boldf}(S) - \E[\wh{\boldf}(S)]| \geq t] \leq 2\exp(-t^2\cdot 2^{n-1})
    \]
    for any~$t > 0$.  Selecting $t = 2\sqrt{n} 2^{-n/2}$, the above bound is $2\exp(-2n) \leq 4^{-n}$.  The result follows by taking a union bound over all $S \subseteq [n]$.
\end{proof}

This proposition motivates the following basic notion of ``pseudorandomness'':
\begin{definition}
    A function $f \btR$
                                                    \index{epsilon-regular@$\eps$-regular}%
                                                    \index{regular|seeonly{$\eps$-regular}}%
                                                    \index{epsilon-uniform@$\eps$-uniform|seeonly{$\eps$-regular}}%
    is \emph{$\eps$-regular} (sometimes called \emph{$\eps$-uniform}) if $|\wh{f}(S)| \leq \eps$ for all $S \neq \emptyset$.
\end{definition}
\begin{remark}
    By Exercise~\ref{ex:hausdorff-young}, every function~$f$ is $\eps$-regular for $\eps = \|f\|_1$.  We are often concerned with $f \btI$, in which case we focus on $\eps \leq 1$.
\end{remark}
\begin{examples}                        \label{ex:regularity}
    Proposition~\ref{prop:random-Fourier} states that a random $p$-biased function is \linebreak $(2\sqrt{n} 2^{-n/2})$-regular with very high probability.  A function is $0$-regular if and only if it is constant (even though you might not think of a constant function as very ``random'').  If $A \subseteq \F_2^n$ is an affine subspace of codimension~$k$ then $1_A$ is $2^{-k}$-regular (Proposition~\ref{prop:coset-transf}).  For $n$ even the inner
                                        \index{inner product mod $2$ function}
    product mod~$2$ function and the complete quadratic function, $\IP_{n}, \CQ_n \co \F_2^{n} \to \{0,1\}$,  are
                                            \index{complete quadratic function}
    $2^{-n/2-1}$-regular (Exercise~\ref{ex:compute-expansions}).  On the other hand, the parity functions $\chi_S \btb$ are not $\eps$-regular for any $\eps < 1$ (except for $S = \emptyset$). By Exercise~\ref{ex:maj-coeffs-decr}, $\Maj_n$ is $\tfrac{1}{\sqrt{n}}$-regular.
\end{examples}

The notion of regularity can be particularly useful for probability density functions; in this case it is traditional to use an alternate name:
\begin{definition}
    If $\vphi \co \F_2^n \to \R^{\geq 0}$ is a probability density which is $\eps$-regular, we call it an \emph{$\eps$-biased density}.
                                            \index{probability density!$\eps$-biased}%
                                            \index{epsilon-biased set@$\eps$-biased set|seeonly{probability density, $\eps$-biased density}}%
    Equivalently, $\vphi$ is an $\eps$-biased density if and only if $|\Ex_{\bx \sim \vphi}[\chi_\gamma(\bx)]| \leq \eps$ for all $\gamma \in \dualF \setminus \{0\}$; thus one can think of ``$\eps$-biased'' as meaning ``at most $\eps$-biased on subspaces''.   Note that the marginal of such a distribution on any set of coordinates $J \subseteq [n]$ is also $\eps$-biased. If $\vphi$ is $\vphi_A = 1_A/\E[1_A]$ for some $A \subseteq \F_2^n$ we call $A$ an \emph{$\eps$-biased set}.
\end{definition}
\begin{examples}  For $\vphi$ a probability density we have $\|\vphi\|_1 = \E[\vphi] = 1$, so every density is $1$-biased.  The density corresponding to the uniform distribution on~$\F_2^n$, namely $\vphi \equiv 1$, is the only $0$-biased density. Densities corresponding to the uniform distribution on smaller affine subspaces are ``maximally biased'': if $A \subseteq \F_2^n$ is an affine subspace of dimension less than~$n$, then $\vphi_A$ is not $\eps$-biased for any $\eps < 1$ (Proposition~\ref{prop:coset-transf} again).  If $E = \{(0, \dots, 0), (1, \dots, 1)\}$, then $E$ is a $1/2$-biased set (an easy computation, see also Exercise~\ref{ex:compute-expansions}\ref{ex:equ}).
\end{examples}

There is a ``combinatorial'' property of functions~$f$ that is roughly equivalent to $\eps$-regularity.  Recall from Exercise~\ref{ex:BLR4} that $\snorm{f}_4^4$ has an equivalent non-Fourier formula:
                                        \index{Fourier norm!$4$-}
$\E_{\bx,\by,\bz}[f(\bx)f(\by)f(\bz)f(\bx+\by+\bz)]$.  We show (roughly speaking) that $f$ is regular if and only if this expectation is not much bigger than $\E[f]^4 = \E_{\bx,\by,\bz,\bw}[f(\bx)f(\by)f(\bz)f(\bw)]$:
\begin{proposition}                                     \label{prop:regularity-norm}
    Let $f \ftR$.  Then
    \begin{enumerate}
        \item If $f$ is $\eps$-regular, then $\snorm{f}_4^4 - \E[f]^4 \leq \eps^2 \cdot \Var[f]$.
        \item If $f$ is not $\eps$-regular, then $\snorm{f}_4^4 - \E[f]^4 \geq \eps^4$.
    \end{enumerate}
\end{proposition}
\begin{proof}
    If $f$ is $\eps$-regular, then
    \[
        \snorm{f}_4^4 - \E[f]^4 = \sum_{S \neq \emptyset} \wh{f}(S)^4 \leq \max_{S \neq \emptyset} \{\wh{f}(S)^2\} \cdot \sum_{S \neq \emptyset} \wh{f}(S)^2 \leq \eps^2 \cdot \Var[f].
    \]
    On the other hand, if $f$ is not $\eps$-regular, then $|\wh{f}(T)| \geq \eps$ for some $T \neq \emptyset$; hence $\snorm{f}_4^4$ is at least $\wh{f}(\emptyset)^4 + \wh{f}(T)^4 \geq \E[f]^4 + \eps^4$.
\end{proof}

The condition of $\eps$-regularity -- that \emph{all} non-empty-set coefficients are small -- is quite strong.  As we saw when investigating the $\frac{2}{\pi}$~Theorem in Chapter~\ref{sec:weight-level-1} it's also interesting to consider~$f$ that merely have $|\wh{f}(i)| \leq \eps$ for all $i \in [n]$; for monotone~$f$ this is the same as saying $\Inf_i[f] \leq \eps$ for~$i$.  This suggests two weaker possible notions of pseudorandomness: having all low-degree Fourier coefficients small, and having all influences small.  We will consider both possibilities, starting with the second.

Now a randomly chosen $\boldf \btb$ will \emph{not} have all of its influences small; in fact as we saw in Exercise~\ref{ex:random-influence}, each $\Inf_i[\boldf]$ is $1/2$ in expectation.  However, for any $\delta > 0$ it will have all of its \emph{$(1-\delta)$-stable} influences exponentially small (recall Definition~\ref{def:stable-influence}).
                                        \index{stable influence}%
In Exercise~\ref{ex:rand-stable-influences} you will show:
\begin{fact}                                        \label{fact:rand-stable-influences}
    Fix $\delta \in [0,1]$ and let $\boldf \btb$ be a randomly chosen function. Then for any $i \in [n]$,
    \[
        \E[\Inf_i^{(1-\delta)}[f]] = \frac{(1-\delta/2)^n}{2-\delta}.
    \]
\end{fact}
This motivates a very important notion of pseudorandomness in the analysis of Boolean functions: having all stable-influences small.  Recalling the discussion surrounding Proposition~\ref{prop:few-stable-influences}, we can also describe this as having no ``notable'' coordinates.
\begin{definition}                                          \label{def:small-influences}%
                            \index{epsilon delta-small stable influences@$(\eps,\delta)$-small stable influences}%
                            \index{notable coordinates}%
                            \index{small stable influences|seeonly{$(\eps,\delta)$-small stable influences}}%
    We say that $f \btR$ has \emph{$(\eps,\delta)$-small stable influences}, or \emph{no $(\eps,\delta)$-notable coordinates}, if $\Inf_{i}^{(1-\delta)}[f] \leq \eps$ for each $i \in [n]$.  This condition gets stronger as~$\eps$ and~$\delta$ decrease: when $\delta = 0$, meaning $\Inf_{i}[f] \leq \eps$ for all~$i$, we simply say~$f$ has \emph{$\eps$-small~influences}.
\end{definition}
\begin{examples} \label{eg:small-stable-influences}
    Besides random functions, important examples of Boolean-valued functions with no notable coordinates are constants, majority, and large parities.  Constant functions are the ultimate in this regard: they have $(0,0)$-small stable influences.  (Indeed, constant functions are the only ones with $0$-small influences.)  The $\maj_n$ function has $\frac{1}{\sqrt{n}}$-small influences.  To see the distinction between influences and stable influences, consider the parity functions~$\chi_S$. Any parity function~$\chi_S$ (with $S \neq \emptyset$) has at least one coordinate with maximal influence,~$1$. But if $|S|$ is ``large'' then all of its \emph{stable} influences will be small: We have $\Inf^{(1-\delta)}_i[\chi_S]$ equal to $(1-\delta)^{|S|-1}$ when $i \in S$ and equal to~$0$ otherwise; i.e., $\chi_S$ has $((1-\delta)^{|S|-1}, \delta)$-small stable influences.  In particular, $\chi_S$ has $(\eps, \delta)$-small stable influences whenever $|S| \geq \frac{\ln(e/\epsilon)}{\delta}$.

    The prototypical example of a function $f \btb$ that does \emph{not} have small stable influences is an unbiased $k$-junta.  Such a function has $\Var[f] = 1$ and hence from Fact~\ref{fact:stable-tinf} the sum of its $(1-\delta)$-stable influences is at least $(1-\delta)^{k-1}$.  Thus $\Inf^{(1-\delta)}_i[f] \geq (1-\delta)^{k-1}/k$ for at least one~$i$; hence~$f$ does \emph{not} have $((1-\delta)^k/k, \delta)$-small stable influences for any $\delta \in (0,1)$.  A somewhat different example is the function $f(x) = x_0 \maj_n(x_1, \dots, x_n)$, which has $\Inf_0^{(1-\delta)}[f] \geq 1 - \sqrt{\delta}$; see Exercise~\ref{ex:pseudorandomness-compare}\ref{ex:x0Maj}.
\end{examples}

Let's return to considering the interesting condition that $|\wh{f}(i)| \leq \eps$ for all $i \in [n]$.  We will call this condition \emph{$(\eps,1)$-regularity}.  It is equivalent to saying that $f^{\leq 1}$ is $\eps$-regular, or that $f$ has at most~$\eps$ ``correlation'' with every dictator: $|\la f, \pm \chi_i \ra| \leq \eps$ for all~$i$.  Our third notion of pseudorandomness extends this condition to higher degrees:
\begin{definition}
    A function $f \btR$
                                                \index{epsilon k-regular@$(\eps,k)$-regular}
    is \emph{$(\eps,k)$-regular} if $|\wh{f}(S)| \leq \eps$ for all $0 < |S| \leq k$; equivalently, if $f^{\leq k}$ is $\eps$-regular.  For $k = n$ (or $k = \infty$), this condition coincides with $\eps$-regularity.  When $\vphi \co \F_2^n \to \R^{\geq 0}$ is an $(\eps,k)$-regular probability density, it is more usual to call $\vphi$ (and the associated probability distribution) \emph{$(\eps,k)$-wise independent}.
                                                \index{epsilon k-wise independent@$(\eps,k)$-wise independent}%
                                                \index{almost $k$-wise independent|seeonly{$(\eps,k)$-wise independent}}%
\end{definition}
Below we give two alternate characterizations of $(\eps, k)$-regularity; however, they are fairly ``rough'' in the sense that they have exponential losses on~$k$.  This can be acceptable if~$k$ is thought of as a constant.  The first characterization is that~$f$ is $(\eps, k)$-regular if and only if fixing~$k$ input coordinates changes~$f$'s mean by at most~$O(\eps)$.  The second characterization is the condition that $f$ has $O(\eps)$ covariance with every $k$-junta.
\begin{proposition}                                     \label{prop:eps-delta-reg-character}
    Let $f \btR$ and let $\eps \geq 0$, $k \in \N$.
    \begin{enumerate}
        \item If $f$ is $(\eps,k)$-regular then any restriction of at most~$k$ coordinates changes~$f$'s mean by at most $2^{k} \eps$.
        \item If $f$ is not $(\eps,k)$-regular then some restriction to at most~$k$ coordinates changes~$f$'s mean by more than~$\eps$.
    \end{enumerate}
                                                    \index{mean}
\end{proposition}
\begin{proposition}                                     \label{prop:eps-delta-reg-cov}
    Let $f \btR$ and let $\eps \geq 0$, $k \in \N$.
    \begin{enumerate}
        \item \label{eqn:eps-delta-reg-cov} If $f$ is $(\eps,k)$-regular, then $\Cov[f,h] \leq \snorm{h}_1 \eps$ for any $h \btR$ with $\deg(h) \leq k$.  In particular, $\Cov[f,h] \leq 2^{k/2} \eps$ for any $k$-junta $h \btb$.
        \item If $f$ is not $(\eps,k)$-regular, then $\Cov[f,h] > \eps$ for some $k$-junta $h \btb$.
    \end{enumerate}
\end{proposition}
\noindent We will prove Proposition~\ref{prop:eps-delta-reg-character}, leaving the proof of Proposition~\ref{prop:eps-delta-reg-cov} to the exercises.
\begin{proof}[Proof of Proposition~\ref{prop:eps-delta-reg-character}]
    For the first statement, suppose $f$ is $(\eps,k)$-regular and let $J \subseteq [n]$, $z \in \bits^{J}$,  where $|J| \leq k$.  Then the statement holds because
    \[
        \E[\restr{f}{\ol{J}}{z}] = \wh{f}(\emptyset) + \sum_{\emptyset \neq T \subseteq J} \wh{f}(T)\,z^T
    \]
    (Exercise~\ref{ex:restrict-mean}) and each of the at most~$2^{k}$ terms $|\wh{f}(T)\,z^T| = |\wh{f}(T)|$ is at most~$\eps$.

    For the second statement, suppose that $|\wh{f}(J)| > \eps$, where $0 < |J| \leq k$.  Then a given restriction $z \in \bits^J$ changes $f$'s mean by
    \[
        h(z) = \sum_{\emptyset \neq T  \subseteq J} \wh{f}(T)\,z^T.
    \]
    We need to show that $\|h\|_\infty > \eps$, and this follows from
    \[
        \|h\|_\infty = \|h \chi_J\|_\infty \geq |\E[h\chi_J]| = |\wh{h}(J)| = |\wh{f}(J)| > \eps.\qedhere
    \]
\end{proof}

Taking $\eps = 0$ in the above two propositions we obtain:
\begin{corollary}                                       \label{cor:xiao-massey}
    For $f \btR$, the following are equivalent:
    \begin{enumerate}
        \item $f$ is $(0, k)$-regular.
        \item \label{item:xm} Every restriction of at most~$k$ coordinates leaves $f$'s mean unchanged.
        \item $\Cov[f,h] = 0$ for every $k$-junta $h \btb$.
    \end{enumerate}
    If $f$ is a probability density, condition~(3) is equivalent to $\E_{\bx \sim f}[h(\bx)] = \E[h]$ for every $k$-junta $h \btb$.
\end{corollary}
For such functions, additional terminology is used:
\begin{definition}
    If $f \btb$ is $(0,k)$-regular, it is also called \emph{$k$th-order correlation immune}.
                                                \index{correlation immune}
    If $f$ is in addition unbiased, then it is called \emph{$k$-resilient}.
                                                \index{resilient}
    Finally, if $\vphi \co \F_2^n \to \R^{\geq 0}$ is a $(0,k)$-regular probability density, then we call $\vphi$ (and the associated probability distribution) \emph{$k$-wise independent}.
                                                \index{k-wise independent@$k$-wise independent}%
\end{definition}
\begin{examples}  \label{ex:corr-immune} Any parity function $\chi_S \btb$ with $|S| = k+1$ is $k$-resilient.  More generally, so is  $\chi_S \cdot g$ for any $g \btb$ that does not depend on the coordinates in~$S$.  For a good example of a correlation immune function that is not resilient, consider $h \co \bits^{3m} \to \bits$ defined by $h = \chi_{\{1, \dots, 2m\}} \wedge \chi_{\{m+1, \dots, 3m\}}$.  This $h$ is not unbiased, being $\True$ on only a $1/4$-fraction of inputs. However, its bias does not change unless at least $2m$ input bits are fixed; hence $h$ is $(2m-1)$th-order correlation immune.
\end{examples}

We conclude this section with Figure~\ref{fig:pseudorandomness-compare}, indicating how our various notions of pseudorandomness compare:
\myfig{.75}{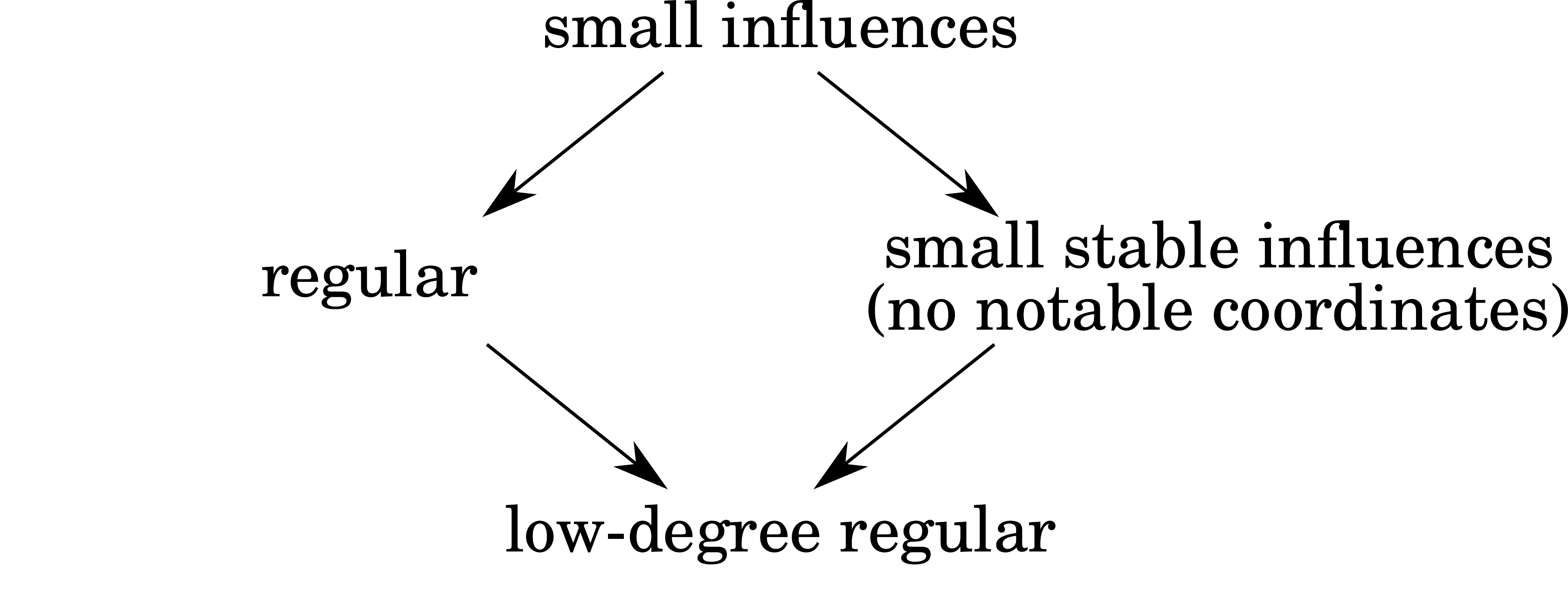}{Comparing notions of pseudorandomness: arrows go from stronger notions to (strictly) weaker ones}{fig:pseudorandomness-compare}
\noindent For precise quantitative statements, counterexamples showing that no other relationships are possible, and explanations for why these notions essentially coincide for monotone functions, see Exercise~\ref{ex:pseudorandomness-compare}.

\section{\texorpdfstring{$\F_2$}{F2}-polynomials}
                               \index{algebraic normal form|seeonly{$\F_2$-polynomial representation}}
                               \index{F2-polynomial representation@$\F_2$-polynomial representation|(}
We began our study of Boolean functions in Chapter~\ref{sec:fourier-expansion} by considering their polynomial representations over the real field.  In this section we take a brief look at their polynomial representations over the field~$\F_2$, with $\False$, $\true$ being represented by $0, 1 \in \F_2$ as usual.  Note that in the field $\F_2$, the arithmetic operations~$+$ and~$\cdot$ correspond to logical~XOR and logical~AND, respectively.

\begin{example}
    Consider the logical parity (XOR)
                                            \index{parity}
    function on $n$ bits, $\chi_{[n]}$.  To represent it over the reals (as we have done so far) we encode $\false, \true$ by $\pm 1 \in \R$; then $\chi_{[n]} \btb$ has the polynomial representation $\chi_{[n]}(x) = x_1 x_2 \cdots x_n$.  Suppose instead we encode $\false, \true$ by $0, 1 \in \F_2$; then $\chi_{[n]} \co \F_2^n \to \F_2$ has the polynomial representation $\chi_{[n]}(x) = x_1 + x_2 + \cdots + x_n$.  Notice this polynomial has degree~$1$, whereas the representation over the reals has degree~$n$.
\end{example}

In general, let $f \co \F_2^n \to \F_2$ be any Boolean function.  Just as in Chapter~\ref{sec:fourier-expansion} we can find a (multilinear) polynomial representation for it by interpolation.  The indicator function $1_{\{a\}} \co \F_2^n \to \F_2$ for $a \in \F_2^n$ can be written as
\begin{equation} \label{eqn:interp-f2-a}
    1_{\{a\}}(x) = \prod_{i : a_i = 1} x_i \prod_{i : a_i = 0}(1-x_i),
\end{equation}
a degree-$n$ multilinear polynomial.  (We could have written $1+x_i$ rather than $1-x_i$ since these are the same in~$\F_2$.) Hence $f$ has the multilinear polynomial expression
\begin{equation} \label{eqn:interp-f2-b}
    f(x) = \sum_{a \in \F_2^n} f(a)1_{\{a\}}(x).
\end{equation}
After simplification, this may be put in the form
\begin{equation} \label{eqn:f2-poly-rep}
    f(x) = \sum_{S \subseteq [n]} c_S x^S,
\end{equation}
where $x^S = \prod_{i \in S} x_i$ as usual, and each coefficient $c_S$ is in~$\F_2$.  We call~\eqref{eqn:f2-poly-rep} the \emph{$\F_2$-polynomial representation of~$f$}.  As an example, if $f = \chi_{[3]}$ is the parity function on~$3$ bits, its interpolation is
\begin{align}
    \chi_{[3]}(x) &= (1-x_1)(1-x_2)x_3 + (1-x_1)x_2(1-x_3) + x_1(1-x_2)(1-x_3) + x_1x_2x_3 \nonumber\\
    &= x_1 + x_2 + x_3 -2(x_1x_2 + x_1 x_3 + x_2 x_3)  + 4x_1x_2x_3  \label{eqn:chi3Z}\\
    &= x_1 + x_2 + x_3 \nonumber
\end{align}
as expected.  We also have uniqueness of the $\F_2$-polynomial representation; the quickest way to see this is to note that there are $2^{2^n}$ functions $\F_2^n \to \F_2$ and also $2^{2^n}$ possible choices for the coefficients~$c_S$. Summarizing:
\begin{proposition}                                     \label{prop:f2-poly-rep}
    Every $f \co \F_2^n \to \F_2$ has a unique $\F_2$-polynomial representation as in~\eqref{eqn:f2-poly-rep}.
\end{proposition}
\begin{examples}
    The logical AND function $\AND_n \co \F_2^n \to \F_2$ has the simple expansion $\AND_n(x) = x_1 x_2 \cdots x_n$.  The inner product mod $2$ function
                                    \index{inner product mod $2$ function}
    has the degree-$2$ expansion $\IP_{2n}(x_1, \dots, x_n, y_1, \dots, y_n) = x_1y_1 + x_2y_2 + \cdots + x_ny_n$.
\end{examples}
Since the $\F_2$-polynomial representation is unique we may define $\F_2$-degree:
\begin{definition}
    The \emph{$\F_2$-degree}
                               \index{F2-degree@$\F_2$-degree}
    of a Boolean function $f \co \{\false, \true\}^n \to \{\false, \true\}$, denoted $\deg_{\F_2}(f)$,
                                        \nomenclature[degF2]{$\deg_{\F_2}(f)$}{for Boolean-valued $f$, the degree of its $\F_2$-polynomial representation}
    is the degree of its $\F_2$-polynomial representation.  We reserve the notation $\deg(f)$ for the
                                \index{degree}%
    degree of $f$'s Fourier expansion.
\end{definition}
We can also give a formula for the coefficients of the $\F_2$-polynomial representation:
\begin{proposition}                                     \label{prop:zo-coeffs}
    Suppose $f \co \F_2^n \to \F_2$ has $\F_2$-polynomial representation $f(x) = \sum_{S \subseteq [n]} c_S x^S$.  Then $c_S = \sum_{\supp(x) \subseteq S} f(x)$.
\end{proposition}
\begin{corollary}                                       \label{cor:zo-top}
    Let $f \co \{\false, \true\}^n \to \{\false, \true\}$. Then $\deg_{\F_2}(f) = n$ if and only if $f(x) = \true$ for an odd number of inputs~$x$.
\end{corollary}
\noindent The proof of Proposition~\ref{prop:zo-coeffs} is left for Exercise~\ref{ex:mobius}; Corollary~\ref{cor:zo-top} is just the case~$S = [n]$. You can also directly see that $c_{[n]} = \sum_x f(x)$ by observing what happens with the monomial $x_1x_2 \cdots x_n$ in the interpolation~\eqref{eqn:interp-f2-a}, \eqref{eqn:interp-f2-b}.

Given a generic Boolean function $f \co \{\false, \true\}^n \to \{\false, \true\}$ it's natural to ask about the relationship between its Fourier expansion (i.e., polynomial representation over~$\R$) and its $\F_2$-polynomial representation.  In fact you can easily derive the $\F_2$-representation from the $\R$-representation.  Suppose $p(x)$ is the Fourier expansion of $f$; i.e., $f$'s $\R$-multilinear representation when we interpret $\false$, $\true$ as $\pm 1 \in \R$.  From Exercise~\ref{ex:pm1tozo}, $q(x) = \half - \half p(1-2x_1, \dots, 1-2x_n)$ is the unique $\R$-multilinear representation for $f$ when we interpret $\false$, $\true$ as $0, 1\in \R$.  But we can also obtain $q(x)$ by carrying out the interpolation in~\eqref{eqn:interp-f2-a},~\eqref{eqn:interp-f2-b} over~$\Z$.  Thus the $\F_2$ representation of~$f$ is obtained simply by reducing $q(x)$'s (integer) coefficients modulo~$2$.

We saw an example of this derivation above with $\chi_{[3]}$.  The $\pm 1$-representation is $x_1x_2x_3$.  The representation over $\{0,1\} \in \Z \subseteq \R$ is $\half - \half (1-2x_1)(1-2x_2)(1-2x_3)$, which when expanded equals~\eqref{eqn:chi3Z} and has integer coefficients.  Finally, we obtain the $\F_2$ representation $x_1+x_2+x_3$ by reducing the coefficients of~\eqref{eqn:chi3Z} modulo~$2$.

One thing to note about this transformation from Fourier expansion to $\F_2$-representation is that it can only decrease degree.  As noted in Exercise~\ref{ex:deg-dyadic}, the first step, forming $q(x) = \half - \half p(1-2x_1, \dots, 1-2x_n)$, does not change the degree at all (except if $p(x) \equiv 1$, $q(x) \equiv 0$).  And the second step, reducing $q$'s coefficients modulo~$2$, cannot increase the degree.  We conclude:
\begin{proposition}                                     \label{prop:R-F2-degree}
                               \index{F2-degree@$\F_2$-degree}%
                               \index{degree}%
    Let $f \btb$.  Then $\deg_{\F_2}(f) \leq \deg(f)$.
\end{proposition}
                               \index{F2-polynomial representation@$\F_2$-polynomial representation|)}
Here is an interesting consequence of this proposition.
                                            \index{Siegenthaler's Theorem|(}
Suppose that $f \btb$ is $k$-resilient; i.e., $\wh{f}(S) = 0$ for all $|S| \leq k < n$.  Let $g = \chi_{[n]} \cdot f$; thus $\wh{g}(S) = \wh{f}([n] \setminus S)$ and hence $\deg(g) \leq n-k-1$.  From Proposition~\ref{prop:R-F2-degree} we deduce $\deg_{\F_2}(g) \leq n-k-1$.  But if we interpret $f, g \co \F_2^n \to \F_2$, then $g = x_1 + \cdots + x_n + f$ and hence $\deg_{\F_2}(g) = \deg_{\F_2}(f)$ (unless $f$ is parity or its negation).  Thus:
\begin{proposition}                                     \label{prop:weak-siegenthaler}
    Let $f \btb$ be $k$-resilient, $k < n-1$.  Then $\deg_{\F_2}(f) \leq n-k-1$.
\end{proposition}
This proposition was shown by Siegenthaler, a cryptographer who was studying stream ciphers; his motivation is discussed further in the notes in Section~\ref{sec:pseudorandomness-notes}. More generally, Siegenthaler proved the following result (the proof does not require Fourier analysis):
\begin{named}{Siegenthaler's Theorem}  Proposition~\ref{prop:weak-siegenthaler} holds.  Further, if $f$ is merely $k$th-order correlation immune, then we still have $\deg_{\F_2}(f) \leq n-k$ (for $k < n$).
\end{named}
\begin{proof}
    Pick any monomial $x^J$ of maximal degree $d = \deg_{\F_2}(f)$ in $f$'s $\F_2$-polynomial representation; we may assume $d > 1$ else we are done.  Make an arbitrary restriction to the $n-d$ coordinates outside of~$J$, forming function~$g \co \F_2^J \to \F_2$.  The monomial $x^J$ still appears in $g$'s $\F_2$-polynomial representation; thus by Corollary~\ref{cor:zo-top}, $g$ is~$1$ for an odd number of inputs.

    Let us first show Proposition~\ref{prop:weak-siegenthaler}.  Assuming $f$ is $k$-resilient, it is unbiased.  But $g$ is~$1$ for an odd number of inputs so it cannot be unbiased (since $2^{d-1}$ is even for $d > 1$).  Thus the restriction changed $f$'s bias, and we must have $n-d > k$, hence $d \leq n-k-1$.

    Suppose now $f$ is merely $k$th-order correlation immune.  Pick an arbitrary input coordinate for~$g$ and suppose  its two possible restrictions give subfunctions~$g_0$ and~$g_1$. Since $g$ has an odd number of~$1$'s, one of $g_0$ has an odd number of~$1$'s and the other has an even number.  In particular, $g_0$ and $g_1$ have different biases. One of these biases must differ from $f$'s.  Thus $n-d+1 > k$, hence $d \leq n-k$.
\end{proof}
                                            \index{Siegenthaler's Theorem|)}

We end this section by mentioning another bound related to correlation immunity:
\begin{theorem}                                     \label{thm:correlation-immune-2/3}
    Suppose $f \btb$ is $k$th-order correlation immune but \emph{not} $k$-resilient (i.e., $\E[f] \neq 0$).  Then $k +1 \leq \frac{2}{3}n$.
\end{theorem}
The proof of this theorem (left to Exercise~\ref{ex:correlation-immune-2/3}) uses the Fourier expansion rather than the $\F_2$-representation.  The bounds in both Siegenthaler's Theorem and Theorem~\ref{thm:correlation-immune-2/3} can be sharp in many cases; see Exercise~\ref{ex:sieg-sharp}.

\section{Constructions of various pseudorandom functions}               \label{sec:constructing-pr-functions}

In this section we give some constructions of Boolean functions with strong pseudorandomness properties.  We begin by discussing
                                                \index{bent functions|(}
\emph{bent} functions:
\begin{definition}
    A function $f \ftb$ (with $n$ even) is called \emph{bent} if $|\wh{f}(\gamma)| = 2^{-n/2}$ for all $\gamma \in \dualF$.
\end{definition}
Bent functions are $2^{-n/2}$-regular.  If the definition of $\eps$-regularity were changed so that even $|\wh{f}(0)|$ needed to be at most~$\eps$, then bent functions would be the most regular possible functions.  This is because $\sum_\gamma \wh{f}(\gamma)^2 = 1$ for any $f \ftb$ and hence at least one $|\wh{f}(\gamma)|$ must be at least $2^{-n/2}$.  In particular, bent functions are those that are maximally distant from the class of affine functions, $\{\pm \chi_\gamma : \gamma \in \dualF\}$.

We have encountered some bent functions already.  The canonical example is the inner product mod $2$ function,
                                    \index{inner product mod $2$ function}
$\IP_{n}(x) = \chi(x_1 x_{n/2+1} + x_2 x_{n/2+2} + \cdots + x_{n/2} x_n)$.  (Recall the notation $\chi(b) = (-1)^b$.) For $n = 2$ this is just the $\AND_2$ function $\half + \half x_1 + \half x_2 - \half x_1x_2$, which is bent by inspection.  For general~$n$, the bentness is a consequence of the following fact (proved in Exercise~\ref{ex:oplus-bent}):
\begin{proposition}                                     \label{prop:oplus-bent}
    Let $f \ftb$ and $g \co \F_2^{n'} \to \bits$ be bent.  Then $f \oplus g \co \F_2^{n+n'} \to \bits$ defined
                                        \nomenclature[f + g]{$f \oplus g$}{if $f \co \bits^m \to \bits$ and $g \btb$, denotes the function $h \co \bits^{m+n} \to \bits$ defined by $h(x,y) = f(x)g(y)$}%
    by $(f \oplus g)(x, x') = f(x) g(x')$ is also bent.
\end{proposition}
Another example of a bent function is the complete quadratic function
                                    \index{complete quadratic function}
$\CQ_n(x) = \chi(\sum_{1 \leq i < j \leq n} x_ix_j)$ from Exercise~\ref{ex:compute-expansions}.  Actually, in some sense it is the ``same'' example, as we now explain.
\begin{proposition}                                     \label{prop:bent-to-bent}
    Let $f \ftb$ be bent.  Then $\pm \chi_\gamma \cdot f$ is bent for any $\gamma \in \dualF$, as is $f \circ M$ for any invertible linear transformation $M \co \F_2^n \to \F_2^n$.
\end{proposition}
\begin{proof}
    Multiplying by $-1$ does not change bentness, and both $\chi_\gamma \cdot f$ and $f \circ M$ have the same Fourier coefficients as~$f$ up to a permutation (see Exercise~\ref{ex:lintransf-fourier}).
\end{proof}
We claim that $\CQ_n$ arises from $f = \IP_n$ as in Proposition~\ref{prop:bent-to-bent}.  In the case $n = 4$, this is because $\littlesum_{1 \leq i < j \leq 4} x_ix_j = (x_1 + x_3)(x_2+x_3) + (x_1+x_2+x_3)x_4 + x_3$ over~$\F_2$; thus
\[
    \CQ_4(x) = \IP_4(Mx) \cdot \chi_{(0,0,1,0)}(x), \quad \text{where } M = \begin{bmatrix} 1 & 0 & 1 & 0 \\ 1 & 1 & 1 & 0 \\ 0 & 1 & 1 & 0 \\ 0 & 0 & 0 & 1 \end{bmatrix} \text{ is invertible}.
\]
The general case is left to Exercise~\ref{ex:cq-to-ip}.  In fact, every bent~$f$ with $\deg_{\F_2}(f) \leq 2$ arises by applying Proposition~\ref{prop:bent-to-bent} to the inner product mod~$2$ function; see Exercise~\ref{ex:dickson}. There are other large families of bent functions; however, the problem of classifying all bent functions is open and seems difficult.  We content ourselves by describing one more family:
\begin{proposition}                                     \label{prop:rothaus-bent}
    Let $f \co \F_2^{2n} \to \bits$ be defined by $f(x,y) = \IP_{2n}(x,y)g(y)$ where $g \btb$ is arbitrary.  Then~$f$ is bent.
\end{proposition}
\begin{proof}
    We will think of $y \in \dualF$, so $\IP_{2n}(x,y) = \chi_y(x)$.  We'll also write a generic $\gamma \in \wh{\F_2^{2n}}$ as $(\gamma_1, \gamma_2)$.  Then indeed
    \begin{multline*}
        \wh{f}(\gamma) = \Ex_{\bx, \by}[\chi_{\by}(\bx) g(\by) \chi_{(\gamma_1, \gamma_2)}(\bx, \by)]  = \Ex_{\by} \left[g(\by) \chi_{\gamma_2}(\by) \Ex_{\bx}[\chi_{\by+\gamma_1}(\bx)]\right] \\= \Ex_{\by} [g(\by) \chi_{\gamma_2}(\by) \bone_{\{\by+\gamma_1 = 0\}}] = 2^{-n} g(\gamma_1)\chi_{\gamma_2}(\gamma_1) = \pm 2^{-n}. \qedhere
    \end{multline*}
\end{proof}
                                                \index{bent functions|)}

                                            \index{probability density!$\eps$-biased|(}
We next discuss explicit constructions of small $\eps$-biased sets, which are of considerable use in the field of algorithmic derandomization. The most basic step in a randomized algorithm is drawing a string $\bx \sim \F_2^n$ from the uniform distribution; however, this has the ``cost'' of generating~$n$ independent, random bits.  But sometimes it's not necessary that~$\bx$ precisely have the uniform distribution; it may suffice that $\bx$ be drawn from an $\eps$-biased density. If we can deterministically find an $\eps$-biased (multi-)set~$A$ of cardinality, say, $2^\ell$, then we can generate $\bx \sim \vphi_A$ using just~$\ell$ independent random bits.  We will see some example derandomizations of this nature in Section~\ref{sec:testing-learning-apps}; for now we discuss constructions.

Fix $\ell \in \N^+$ and recall that there exists a finite field $\F_{2^\ell}$ with exactly $2^\ell$ elements.
                                        \index{F2\ell@$\F_{2^\ell}$ (finite field)}%
 It is easy to find an explicit representation for $\F_{2^\ell}$ -- a complete addition and multiplication table, say -- in time~$2^{O(\ell)}$. (In fact, one can compute within $\F_{2^\ell}$ even in deterministic $\poly(\ell)$ time.) The field elements $x \in \F_{2^\ell}$ are naturally encoded by distinct~$\ell$-bit vectors; we will write $\enc \co \F_{2^\ell} \to \F_2^\ell$ for this encoding.  The encoding is linear; i.e., it satisfies $\enc(0) = (0, \dots, 0)$ and $\enc(x+y) = \enc(x) + \enc(y)$ for all $x, y \in \F_{2^\ell}$.

\begin{theorem}                                     \label{thm:aghp}
    There is a deterministic algorithm that, given $n \geq 1$ and $0 < \eps \leq 1/2$,
    runs in $\poly(n/\eps)$ time and outputs a multiset $A \subseteq \F_2^n$ of cardinality at most $16(n/\eps)^2$ with the property that $\vphi_A$ is an $\eps$-biased density.
\end{theorem}
\begin{proof}
    It suffices to obtain cardinality $(n/\eps)^2$ under the assumption that $\eps = 2^{-t}$ and $n = 2^{\ell - t}$ are integer powers of~$2$. We will describe a probability density~$\vphi$ on $\F_2^n$ by giving a procedure for drawing a string $\by \sim \vphi$ which uses~$2\ell$ independent random bits. $A$ will be the multiset of $2^{2\ell} = (n/\eps)^2$ possible outcomes for~$\by$. It will be clear that~$A$ can be generated in deterministic polynomial time. The goal will be to show that $\vphi$ is $2^{-t}$-biased.

    To draw $\by \sim \vphi$, first choose $\br, \bs \sim \F_{2^\ell}$ independently and uniformly.  This uses~$2\ell$ independent random bits. Then define the $i$th coordinate of $\by$ by
    \[
        \by_i = \la \enc(\br^i), \enc(\bs) \ra,     \quad i \in [n],
    \]
    where the inner product $\la \cdot, \cdot \ra$ takes place in $\F_2^\ell$. Fixing $\gamma \in \dualF \setminus \{0\}$, we need to argue that $|\E[\chi_\gamma(\by)]| \leq 2^{-t}$.  Now over $\F_2^\ell$,
    \begin{align*}
        \la \gamma, \by \ra = \sum_{i=1}^n \gamma_i \la \enc(\br^i), \enc(\bs) \ra = \Bigl\la \sum_{i=1}^n \gamma_i \enc(\br^i), \enc(\bs) \Bigr\ra = \la \enc(\littlesum_{i=1}^n \gamma_i \br^i), \enc(\bs) \ra,
    \end{align*}
    where the last step used linearity of~$\enc$.  Thus
    \begin{equation} \label{eqn:aghp}
        \E[\chi_\gamma(\by)] = \E[(-1)^{\la \gamma, \by \ra}] = \E_{\br}\left[\Ex_{\bs}[(-1)^{\la \enc(p_\gamma(\br)), \enc(\bs)\ra}]\right],
    \end{equation}
    where $p_\gamma \co \F_{2^\ell} \to \F_{2^\ell}$ is the polynomial $a \mapsto \gamma_1 a + \gamma_2 a^2 + \cdots + \gamma_n a^n$.  This polynomial is of degree at most~$n$, and is nonzero since $\gamma \neq 0$.  Hence it has at most~$n$ roots (zeroes) over the field $\F_{2^\ell}$.  Whenever~$\br$ is one of these roots, $\enc(p_\gamma(\br)) = 0$ and the inner expectation in~\eqref{eqn:aghp} is~$1$.  But whenever $\br$ is not a root of $p_\gamma$ we have $\enc(p_\gamma(\br)) \neq 0$ and so the inner expectation is~$0$.  (We are using Fact~\ref{fact:exp-character} here.)  We deduce that
    \[
        0 \leq \E[\chi_\gamma(\by)] \leq \Pr[\br \text{ is a root of } p_\gamma] \leq \frac{n}{2^\ell} = 2^{-t},
    \]
    which is stronger than what we need.
\end{proof}
The bound of $O(n/\eps)^2$ in this theorem is fairly close to being optimally small; see Exercise~\ref{ex:rand-eps-biased} and the notes for this chapter.
                                            \index{probability density!$\eps$-biased|)}

                                                \index{$k$-wise independent|(}
Another useful tool in derandomization is that of $k$-wise independent distributions.  Sometimes a randomized algorithm using~$n$ independent random bits will still work assuming only that every subset of~$k$ of the bits is independent.  Thus as with $\eps$-biased sets, it's worthwhile to come up with deterministic constructions of small sets~$A \subset \F_2^n$ such that the density function $\vphi_A$ is $k$-wise independent (i.e., $(0, k)$-regular).  The best known examples have the additional pleasant feature that~$A$ is a linear subspace of $\F_2^n$; in this case, $k$-wise independence is easy to characterize:
\begin{proposition}                                     \label{prop:linear-sub-k-wise}
    Let $H$ be an $m \times n$ matrix over $\F_2$ and let $A \leq \F_2^n$ be the span of $H$'s rows.  Then $\vphi_A$ is $k$-wise independent if and only if any sum of at most $k$ columns of $H$ is nonzero in~$\F_2^m$.  (We exclude the ``empty'' sum.)
\end{proposition}
\begin{proof}
    Since $\vphi_A = \sum_{\gamma \in A^\perp} \chi_\gamma$ (Proposition~\ref{prop:subspace-transf}), $\vphi_A$ is $k$-wise independent if and only if $|\gamma| > k$ for every $\gamma \in A^\perp \setminus \{0\}$.
                                    \nomenclature{$|\gamma|$}{for $\gamma \in \F_2^n$, denotes the Hamming weight, $\#\{i \in [n] : \gamma_i = 1\}$}
    But $\gamma \in A^\perp$ if and only if $H \gamma = 0$.
\end{proof}
Here is a simple construction of such a matrix with $m \sim k \log n$:
\begin{theorem}                                     \label{thm:dual-bch-simple}
    Let $k, \ell \in \N^+$ and assume $n = 2^{\ell} \geq k$.  Then for $m = (k-1)\ell +1$, there is a matrix $H \in \F_2^{m \times n}$ such that any sum of at most $k$ columns of $H$ is nonzero in $\F_2^m$.
\end{theorem}
\begin{proof}
    Write $\alpha_1, \dots, \alpha_n$ for the elements of the finite field $\F_n$, and consider the following matrix $H' \in \F_{n}^{k \times n}$:
    \[
        H' = \begin{bmatrix} 1 & 1 & \cdots & 1 \\
                            \alpha_1 & \alpha_2 & \cdots & \alpha_n \\
                            \alpha_1^2 & \alpha_2^2 & \cdots & \alpha_n^2 \\
                            \vdots & \vdots & \ddots & \vdots \\
                            \alpha_1^{k-1} & \alpha_2^{k-1} & \cdots & \alpha_n^{k-1}
            \end{bmatrix}.
    \]
    Any submatrix of $H'$ formed by choosing $k$ columns is a Vandermonde matrix and is therefore nonsingular. Hence any subset of $k$ columns of $H'$ is linearly independent in $\F_{n}^k$.  In particular, any sum of at most~$k$ columns of~$H'$ is nonzero in $\F_n^k$.  Now form $H \in \F_2^{m \times n}$ from~$H'$ by replacing each entry $\alpha_j^i$ ($i > 0$) with $\enc(\alpha_j^i)$, thought of as a column vector in $\F_2^\ell$.  Since $\enc$ is a linear map we may conclude that any sum of at most~$k$ columns of~$H$ is nonzero in~$\F_2^{m}$.
\end{proof}
\begin{corollary}                                       \label{cor:k-wise-indep-simple}
    There is a deterministic algorithm that, given integers $1 \leq k \leq n$, runs in $\poly(n^k)$ time and outputs a subspace $A \leq \F_2^n$ of cardinality at most $2^kn^{k-1}$ such that $\vphi_A$ is $k$-wise independent.
\end{corollary}
\begin{proof}
    It suffices to assume $n =2^\ell$ is a power of~$2$ and then obtain cardinality $2n^{k-1} = 2^{(k-1)\ell+1}$.  In this case, the algorithm constructs~$H$ as in Theorem~\ref{thm:dual-bch-simple} and takes $A$ to be the span of its rows.  The fact that $\vphi_A$ is $k$-wise independent is immediate from Proposition~\ref{prop:linear-sub-k-wise}.
\end{proof}
For constant~$k$ this upper bound of $O(n^{k-1})$ is close to optimal. It can be improved to~$O(n^{\lfloor k/2 \rfloor})$, but there is a lower bound of $\Omega(n^{\lfloor k/2 \rfloor})$ for constant~$k$; see Exercises~\ref{ex:better-k-wise},~\ref{ex:k-wise-lower}.
                                                \index{$k$-wise independent|)}

                                                \index{epsilon k-wise independent@$(\eps,k)$-wise independent|(}
We conclude this section by noting that taking an $\eps$-biased density within a $k$-wise independent subspace yields an $(\eps,k)$-wise independent density:
\begin{lemma}                                       \label{lem:eps-k-wise}
    Suppose $H \in \F_2^{m \times n}$ is such that any sum of at most $k$ columns of~$H$ is nonzero in $\F_2^m$.  Let~$\vphi$ be an $\eps$-biased density on $\F_2^m$.  Consider drawing $\by \sim \vphi$ and setting $\bz = \by^\top H \in \F_2^n$.  Then the density of $\bz$ is $(\eps,k)$-wise independent.
\end{lemma}
\begin{proof}
    Suppose $\gamma \in \dualF$ has $0 < |\gamma| \leq k$.  Then $H\gamma$ is nonzero by assumption and hence
    $
        |\E[\chi_\gamma(\bz)]| = |\E_{\by \sim \vphi}[(-1)^{\by^\top H \gamma}]| \leq \eps
    $
    since $\vphi$ is $\eps$-biased.
\end{proof}

As a consequence, combining the constructions of Theorem~\ref{thm:aghp} and Theorem~\ref{thm:dual-bch-simple} gives an $(\eps,k)$-wise independent distribution that can be sampled from using only $O(\log k + \log \log(n) + \log(1/\eps))$ independent random bits:
\begin{theorem}                                     \label{thm:eps-k-wise}
    There is a deterministic algorithm that, given integers $1 \leq k \leq n$ and also $0 < \eps \leq 1/2$, runs in time $\poly(n/\eps)$ and outputs a multiset $A \subseteq \F_2^n$ of cardinality $O(k \log(n)/\eps)^2$ (a power of~$2$) such that $\vphi_A$ is $(\eps,k)$-wise independent.
\end{theorem}
                                                \index{epsilon k-wise independent@$(\eps,k)$-wise independent|)}

\section{Applications in learning and testing}             \label{sec:testing-learning-apps}

In this section we describe some applications of our study of pseudorandomness.

                                              \index{junta!learning|(}
We begin with a notorious open problem from learning theory, that of learning juntas.  Let $\calC = \{f \co \F_2^n \to \F_2 \mid f \text{ is a } k\text{-junta}\}$; we will always assume that $k \leq O(\log n)$.  In the query access model, it is quite easy to learn $\calC$ exactly (i.e., with error~$0$) in $\poly(n)$ time (Exercise~\ref{ex:learn-sparse}\ref{ex:learn-juntas}).  However, in the model of random examples, it's not obvious how to learn $\calC$ more efficiently than in the $n^{k} \cdot \poly(n)$ time required by the Low-Degree Algorithm (see Theorem~\ref{thm:learn-low-degree}). Unfortunately, this is superpolynomial as soon as $k > \omega(1)$.  The state of affairs is the same in the case of depth-$k$ decision trees (a superclass of~$\calC$), and is similar in the case of $\poly(n)$-size DNFs and CNFs.  Thus if we wish to learn, say, $\poly(n)$-size decision trees or DNFs from random examples only, a necessary prerequisite is doing the same for $O(\log n)$-juntas.

Whether or not $\omega(1)$-juntas can be learned from random examples in polynomial time is a longstanding open problem.  Here we will show a modest improvement on the $n^{k}$-time algorithm:
\begin{theorem}                                     \label{thm:mos}
    For $k \leq O(\log n)$, the class $\calC = \{f \co \F_2^n \to \F_2 \mid f \text{ is a } k\text{-junta}\}$ can be exactly learned from random examples in time $n^{(3/4)k} \cdot \poly(n)$.
\end{theorem}
\noindent (The $3/4$ in this theorem can in fact be replaced by $\omega/(\omega + 1)$, where $\omega$ is any number such that $n \times n$ matrices can be multiplied in time $O(n^{\omega})$.)

The first observation we will use to prove Theorem~\ref{thm:mos} is that to learn $k$-juntas, it suffices to be able to identify a single coordinate that is relevant (see Definition~\ref{def:relevant}).  The proof of this is fairly simple and is left for Exercise~\ref{ex:learn-one-rel}:
\begin{lemma}                                       \label{lem:learn-one-rel}
    Theorem~\ref{thm:mos} follows from the existence of a learning algorithm that, given random examples from a nonconstant $k$-junta $f \co \F_2^n \to \F_2$, finds at least one relevant coordinate for~$f$ (with probability at least $1-\delta$) in time $n^{(3/4)k} \cdot \poly(n) \cdot \log(1/\delta)$.
\end{lemma}

Assume then that we have random example access to a (nonconstant) $k$-junta $f \co \F_2^n \to \F_2$.  As in the Low-Degree Algorithm we will estimate the Fourier coefficients~$\wh{f}(S)$ for all $1 \leq |S| \leq d$, where $d \leq k$ is a parameter to be chosen later.  Using Proposition~\ref{prop:learn-one-coeff} we can ensure that all estimates are accurate to within~$(1/3)2^{-k}$, except with probability most~$\delta/2$, in time $n^d \cdot \poly(n) \cdot \log(1/\delta)$.  (Recall that $2^k \leq \poly(n)$.)  Since $f$ is a $k$-junta, all of its Fourier coefficients are either~$0$ or at least $2^{-k}$ in magnitude; hence we can exactly identify the sets $S$ for which $\wh{f}(S) \neq 0$.  For any such $S$, \emph{all} of the coordinates $i \in S$ are relevant for~$f$ (Exercise~\ref{ex:nonzero-coeff-relevance}).  So unless $\wh{f}(S) = 0$ for all $1 \leq |S| \leq d$, we can find a relevant coordinate for~$f$ in time $n^{d} \cdot \poly(n) \cdot \log(1/\delta)$ (except with probability at most~$\delta/2$).

To complete the proof of Theorem~\ref{thm:mos} it remains to handle the case that $\wh{f}(S) = 0$ for all $1 \leq |S| \leq d$; i.e., $f$ is $d$th-order correlation immune.  In this case, by Siegenthaler's Theorem
                                            \index{Siegenthaler's Theorem}
we know that $\deg_{\F_2}(f) \leq k-d$. (Note that $d < k$ since $f$ is not constant.) But there is a learning algorithm running in time $O(n)^{3\ell} \cdot \log(1/\delta)$ that exactly learns any $\F_2$-polynomial of degree at most~$\ell$ (except with probability at most~$\delta/2$).  Roughly speaking, the algorithm draws $O(n)^\ell$ random examples and then solves an $\F_2$-linear system to determine the coefficients of the unknown polynomial; see Exercise~\ref{ex:learn-f2-poly} for details.  Thus in time $n^{3(k-d)} \cdot \poly(n) \cdot \log(1/\delta)$ this algorithm will exactly determine~$f$, and in particular find a relevant coordinate.

By choosing $d = \left\lceil \frac34 k \right\rceil$ we balance the running time of the two algorithms.  Regardless of whether~$f$ is $d$th-order correlation immune, at least one of the two algorithms will find a relevant coordinate for~$f$ (except with probability at most $\delta/2 + \delta/2 = \delta$) in time $n^{(3/4)k} \cdot \poly(n) \cdot \log(1/\delta)$.  This completes the proof of Theorem~\ref{thm:mos}.
                                                    \index{junta!learning|)}

\medskip

                                                \index{derandomization|(}
                                                \index{learning theory|(}
Our next application of pseudorandomness involves using $\eps$-biased distributions to give a \emph{deterministic} version of the Goldreich--Levin Algorithm
                                    \index{Fourier norm!$1$-|(}%
(and hence the Kushilevitz--Mansour learning algorithm) for functions~$f$ with small~$\snorm{f}_1$. We begin with a basic lemma showing that you can get a good estimate for the mean of such functions using an  $\eps$-biased
                                    \index{probability density,$\eps$-biased}%
distribution:
\begin{lemma}                                       \label{lem:eps-biased-mean}
    If $f \btR$ and $\vphi \btR$ is an $\eps$-biased density, then
    \[
        \left|\E_{\bx \sim \vphi}[f(\bx)] - \E[f]\right| \leq \snorm{f}_1 \eps.
    \]
\end{lemma}
\noindent This lemma follows from Proposition~\ref{prop:eps-delta-reg-cov}.\eqref{eqn:eps-delta-reg-cov}, but we provide a separate proof:
\begin{proof}
    By Plancherel,
    \[
        \E_{\bx \sim \vphi}[f(\bx)] = \la \vphi, f\ra = \wh{f}(\emptyset) + \sum_{S \neq \emptyset} \wh{\vphi}(S) \wh{f}(S),
    \]
    and the difference of this from $\E[f] = \wh{f}(\emptyset)$ is, in absolute value, at most
    \[
        \sum_{S \neq \emptyset} |\wh{\vphi}(S)| \cdot |\wh{f}(S)| \leq \eps \cdot \sum_{S \neq \emptyset} |\wh{f}(S)| \leq \snorm{f}_1  \eps. \qedhere
    \]
\end{proof}
Since $\snorm{f^2}_1 \leq \snorm{f}_1^2$ (Exercise~\ref{ex:algebra-norm}), we also have the following immediate corollary:
\begin{corollary}                                       \label{cor:eps-biased-mean}
    If $f \btR$ and $\vphi \btR$ is an $\eps$-biased density, then
    \[
        \left|\E_{\bx \sim \vphi}[f(\bx)^2] - \E[f^2]\right| \leq \snorm{f}_1^2 \eps.
    \]
\end{corollary}

                                    \index{Goldreich--Levin Algorithm|(}%
We can use the first lemma to get a deterministic version of Proposition~\ref{prop:learn-one-coeff}, the learning algorithm that estimates a specified Fourier coefficient.
\begin{proposition}                                     \label{prop:learn-one-coeff-determ}
    There is a \emph{deterministic} algorithm that, given query access to a function $f \btR$ as well as $U \subseteq [n]$, $0 < \eps \leq 1/2$, and $s \geq 1$, outputs an estimate $\wt{f}(U)$ for $\wh{f}(U)$ satisfying
    \[
        |\wt{f}(U) - \wh{f}(U)| \leq \eps,
    \]
    provided $\snorm{f}_1 \leq s$.  The running time is $\poly(n, s, 1/\eps)$.
\end{proposition}
\begin{proof}
    It suffices to handle the case $U = \emptyset$ because for general~$U$, the algorithm can simulate query access to $f \cdot \chi_U$ with $\poly(n)$ overhead, and $\wh{f \cdot \chi_U}(\emptyset) = \wh{f}(U)$.   The algorithm will use Theorem~\ref{thm:aghp} to construct an $(\eps/s)$-biased density~$\vphi$ that is uniform over a (multi-)set of cardinality $O(n^2 s^2/\eps^2)$.  By enumerating over this set and using queries to~$f$, it can deterministically output the estimate $\wt{f}(\emptyset) = \E_{\bx \sim \vphi}[f(\bx)]$
    in time $\poly(n, s, 1/\eps)$.  The error bound now follows from Lemma~\ref{lem:eps-biased-mean}.
\end{proof}
The other key ingredient needed for the Goldreich--Levin Algorithm was Proposition~\ref{prop:GL-substep}, which let us estimate
\begin{equation} \label{eqn:gl-key2}
    \W{S\mid\overline{J}}[f] = \sum_{T \subseteq \ol{J}} \wh{f}(S \cup T)^2 = \Ex_{\bz \sim \bits^{\ol{J}}}[\wh{\restr{f}{J}{\bz}}(S)^2]
\end{equation}
for any $S \subseteq J \subseteq [n]$.
Observe that for any $z \in \bits^{\ol{J}}$ we can use Proposition~\ref{prop:learn-one-coeff-determ} to deterministically estimate $\wh{\restr{f}{J}{z}}(S)$ to accuracy $\pm \eps$. The reason is that we can simulate query access to the restricted function $\wh{\restr{f}{J}{z}}$, the $(\eps/s)$-biased density~$\vphi$ remains $(\eps/s)$-biased on $\bits^{J}$, and most importantly $\snorm{\restr{f}{J}{z}}_1 \leq \snorm{f}_1 \leq s$ by Exercise~\ref{ex:restrict-norm}.  It is not much more difficult to deterministically estimate~\eqref{eqn:gl-key2}:
\begin{proposition}                                     \label{prop:gl1-determ}
    There is a \emph{deterministic} algorithm that, given query access to a function $f \btb$ as well as $S \subseteq J \subseteq [n]$, $0 < \eps \leq 1/2$, and $s \geq 1$, outputs an estimate~$\beta$ for $\W{S\mid\ol{J}}[f]$ that satisfies
    \[
        |\W{S\mid\ol{J}}[f] - \beta| \leq \eps,
    \]
    provided $\snorm{f}_1 \leq s$.  The running time is $\poly(n, s, 1/\eps)$.
\end{proposition}
\begin{proof}
    Recall the notation $\frestr{S}{\ol{J}}{f}$ from Definition~\ref{def:frestr-subcube}; by~\eqref{eqn:gl-key2}, the algorithm's task is to estimate $\Ex_{\bz \sim \bits^{\ol{J}}}[(\frestr{S}{\ol{J}}{f})^2(\bz)]$.  If $\vphi \co \bits^{\ol{J}} \to \R^{\geq 0}$ is an $\tfrac{\eps}{4s^2}$-biased density, Corollary~\ref{cor:eps-biased-mean} tells us that
    \begin{equation} \label{eqn:gl-det-est}
        \Bigl| \Ex_{\bz \sim \vphi}[(\frestr{S}{\ol{J}}{f})^2(\bz)] - \Ex_{\bz \sim \bits^{\ol{J}}}[(\frestr{S}{\ol{J}}{f})^2(\bz)] \Bigr| \leq \snorm{\frestr{S}{\ol{J}}{f}}_1^2 \cdot \tfrac{\eps}{4s^2}\leq \snorm{f}_1^2 \cdot \tfrac{\eps}{4s^2} \leq \tfrac{\eps}{4},
    \end{equation}
    where the second inequality is immediate from  Proposition~\ref{prop:frestr-subcube-formula}.
    We now show the algorithm can approximately compute $\Ex_{\bz \sim \vphi}[(\frestr{S}{\ol{J}}{f})^2(\bz)]$.  For each $z \in \bits^{\ol{J}}$, the algorithm can use $\vphi$ to deterministically estimate $(\frestr{S}{\ol{J}}{f})(z) = \wh{\restr{f}{J}{z}}(S)$ to within $\pm s \cdot \tfrac{\eps}{4s^2} \leq \tfrac{\eps}{4}$ in $\poly(n,s, 1/\eps)$ time, just as was described in the text following~\eqref{eqn:gl-key2}.  Since $|\wh{\restr{f}{J}{z}}(S)| \leq 1$, the square of this estimate is within, say, $\tfrac{3\eps}{4}$ of $(\frestr{S}{\ol{J}}{f})^2(z)$.  Hence by enumerating over the support of~$\vphi$, the algorithm can in deterministic $\poly(n,s, 1/\eps)$ time estimate $\Ex_{\bz \sim \vphi}[(\frestr{S}{\ol{J}}{f})^2(\bz)]$ to within~$\pm \tfrac{3\eps}{4}$, which by~\eqref{eqn:gl-det-est} gives an estimate to within $\pm \eps$ of the desired quantity $\Ex_{\bz \sim \bits^{\ol{J}}}[(\frestr{S}{\ol{J}}{f})^2(\bz)]$.
\end{proof}

Propositions~\ref{prop:learn-one-coeff-determ} and~\ref{prop:gl1-determ} are the only two ingredients needed for a derandomization of the Goldreich--Levin Algorithm.  We can therefore state a derandomized version of its corollary Theorem~\ref{thm:learn-low-l1} on learning functions with small Fourier $1$-norm:
\begin{theorem}                                     \label{thm:learn-low-l1-derand}
    Let $\calC = \{f \btb \mid \snorm{f}_1 \leq s\}$.   Then $\calC$ is \emph{deterministically} learnable from queries with error~$\eps$ in time $\poly(n, s, 1/\eps)$.
\end{theorem}
Since any $f \btb$ with $\ssparsity{f} \leq s$ also has $\snorm{f}_1 \leq s$, we may also deduce from Exercise~\ref{ex:learn-sparse}\ref{ex:learn-sparse-queries}:
\begin{theorem}                                     \label{thm:learn-sparse-derand}
    Let $\calC = \{f \btb \mid \ssparsity{f} \leq 2^{O(k)}\}$.   Then $\calC$ is \emph{deterministically} learnable exactly ($0$~error) from queries in time $\poly(n, 2^k)$.
\end{theorem}
\noindent Example functions that fall into the  concept classes of these theorems are decision trees of size at most~$s$,
                                    \index{decision tree!learning}%
and decision trees of depth at most~$k$, respectively.
                                    \index{Goldreich--Levin Algorithm|)}%
                                                \index{learning theory|)}%
                                        \index{Fourier norm!$1$-|)}%

\medskip

We conclude this section by discussing a derandomized version of the Blum--Luby--Rubinfeld linearity test
                                    \index{BLR (Blum--Luby--Rubinfeld) Test!derandomized}%
from Chapter~\ref{sec:BLR}:

\begin{named}{Derandomized BLR Test}
    Given query access to $f \co \F_2^n \to \F_2$:
    \begin{enumerate}
        \item Choose $\bx \sim \F_2^n$ and $\by \sim \vphi$, where $\vphi$ is an $\eps$-biased density.
        \item Query $f$ at $\bx$, $\by$, and $\bx + \by$.
        \item ``Accept'' if $f(\bx) + f(\by) = f(\bx + \by)$.
    \end{enumerate}
\end{named}
Whereas the original BLR Test required exactly $2n$ independent random bits, the above derandomized version needs only~$n + O(\log(n/\eps))$.  This is very close to minimum possible; a test using only, say, $.99n$ random bits would only be able to inspect a $2^{-.01 n}$ fraction of~$f$'s values.

If $f$ is $\F_2$-linear then it is still accepted by the Derandomized BLR Test with probability~$1$.  As for the approximate converse, we'll have to make a slight concession: We'll show that any function accepted with probability close to~$1$ must be close to an \emph{affine} function,  i.e., satisfy $\deg_{\F_2}(f) \leq 1$.  This concession is necessary: the function $f \co \F_2^n \to \F_2$ might be~$1$ everywhere except on the (tiny) support of~$\vphi$.  In that case the acceptance criterion $f(\bx) + f(\by) = f(\bx + \by)$ will almost always be $1 + 0 = 1$; yet $f$ is very far from every linear function.  It is, however, very close to the affine function~$1$.

\begin{theorem} \label{thm:BLR-derand} Suppose the Derandomized BLR Test accepts $f \co \F_2^n \to \F_2$ with probability $\half + \half \theta$.  Then $f$ has correlation at least $\sqrt{\theta^2 - \eps}$ with some affine $g \co \F_2^n \to \F_2$; i.e., $\dist(f,g) \leq \half - \half \sqrt{\theta^2 - \eps}$.
\end{theorem}
\begin{remark}  The bound in this theorem works well both when $\theta$ is close to~$0$ and when $\theta$ is close to~$1$; e.g., for $\theta  = 1-2\delta$ we get that if $f$ is accepted with probability $1-\delta$, then $f$ is nearly $\delta$-close to an affine function, provided~$\eps \ll \delta$.
\end{remark}
\begin{proof}
    As in the analysis of the BLR Test (Theorem~\ref{thm:blr-test}) we encode $f$'s outputs by $\pm 1 \in \R$.  Using the first few lines of that analysis we see that our hypothesis is equivalent to
    \[
        \theta \leq \Es{\bx \sim \F_2^n \\ \by \sim \vphi}[f(\bx)f(\by)f(\bx+\by)] = \Ex_{\by \sim \vphi}[f(\by) \cdot (f \conv f)(\by)].
    \]
    By Cauchy--Schwarz,
    \[
        \Ex_{\by \sim \vphi}[f(\by) \cdot (f \conv f)(\by)] \leq \sqrt{\Ex_{\by \sim \vphi}[f(\by)^2]}\sqrt{\Ex_{\by \sim \vphi}[(f \conv f)^2(\by)]} = \sqrt{\Ex_{\by \sim \vphi}[(f \conv f)^2(\by)]},
    \]
    and hence
    \[
        \theta^2 \leq \Ex_{\by \sim \vphi}[(f \conv f)^2(\by)] \leq \E[(f \conv f)^2] + \snorm{f \conv f}_1 \eps = \sum_{\gamma \in \dualF} \wh{f}(\gamma)^4 + \eps,
    \]
    where the inequality is Corollary~\ref{cor:eps-biased-mean} and we used $\wh{f \conv f}(\gamma) = \wh{f}(\gamma)^2$.  The conclusion of the proof is as in the original analysis
                                        \index{Fourier norm!$4$-}
    (cf.~Proposition~\ref{prop:regularity-norm}, Exercise~\ref{ex:BLR4}):
    \[
        \theta^2 - \eps \leq \sum_{\gamma \in \dualF} \wh{f}(\gamma)^4 \leq \max_{\gamma \in \dualF} \{\wh{f}(\gamma)^2\} \cdot \sum_{\gamma \in \dualF} \wh{f}(\gamma)^2 = \max_{\gamma \in \dualF} \{\wh{f}(\gamma)^2\},
    \]
    and hence there exists $\gamma^*$ such that $|\wh{f}(\gamma^*)| \geq \sqrt{\theta^2-\eps}$.
\end{proof}
                                                \index{derandomization|)}

\section{Highlight: Fooling \texorpdfstring{$\F_2$}{F2}-polynomials}        \label{sec:viola}

Recall that a density $\vphi$ is said to be $\eps$-biased if its correlation with every $\F_2$-linear function~$f$ is at most~$\eps$ in magnitude.  In the lingo of pseudorandomness, one says that $\vphi$ \emph{fools}
the class of $\F_2$-linear functions:
\begin{definition}
    Let $\vphi \co \F_2^n \to \R^{\geq 0}$ be a density function and let $\calC$ be a class of functions $\F_2^n \to \R$.  We say that $\vphi$ \emph{$\eps$-fools}
                                            \index{fooling}%
                                            \index{epsilon-fools@$\eps$-fools|seeonly{fooling}}%
    $\calC$ if
    \[
        \Bigl|\E_{\by \sim \vphi}[f(\by)] - \E_{\bx \sim \F_2^n}[f(\bx)]\Bigr| \leq \eps
    \]
    for all $f \in \calC$.
\end{definition}

Theorem~\ref{thm:aghp} implies that using just $O(\log(n/\eps))$ independent random bits, one can generate a density that $\eps$-fools the class of $f \co \F_2^n \to \bits$ with $\deg_{\F_2}(f) \leq 1$.  A natural problem in the field of derandomization is: How many independent random bits are needed to generate a density which $\eps$-fools all functions of $\F_2$-degree at most~$d$?
                                           \index{F2-degree@$\F_2$-degree}
A naive hope might be that $\eps$-biased densities automatically fool functions of $\F_2$-degree $d > 1$.  The next example shows that this hope fails badly, even for $d = 2$:
\begin{example}
    Recall the inner product mod~$2$ function, $\IP_{n} \co \F_2^n \to \{0,1\}$, which has $\F_2$-degree~$2$.
                                        \index{inner product mod $2$ function}
    Let $\vphi \co \F_2^n \to \R^{\geq 0}$ be the density of the uniform distribution on the support of~$\IP_n$.  Now $\IP_n$ is an extremely regular function (see Example~\ref{ex:regularity}), and indeed $\vphi$ is a roughly $2^{-n/2}$-biased density (see Exercise~\ref{ex:bias-of-ip2}).  But $\vphi$ is very bad at fooling at least one function of $\F_2$-degree~$2$, namely $\IP_n$ itself:
    \[
        \Ex_{\bx \sim \F_2^n}[\IP_n(\bx)] \approx 1/2, \qquad  \Ex_{\by \sim \vphi}[\IP_n(\by)] = 1.
    \]
\end{example}

The problem of using few random bits to fool $n$-bit, $\F_2$-degree-$d$ functions was first taken up by Luby, Veli\v{c}kovi\'{c}, and Wigderson~\cite{LVW93}.  They showed how to generate a fooling distribution using $\exp(O(\sqrt{d \log(n/d) + \log(1/\eps)}))$ independent random bits.  There was no improvement on this for $14$ years, at which point Bogdanov and Viola~\cite{BV07} achieved $O(\log(n/\eps))$ random bits for $d = 2$ and $O(\log n) + \exp(\poly(1/\eps))$ random bits for $d = 3$.  In general, they suggested that $\F_2$-degree-$d$ functions might be fooled by the sum of~$d$ independent draws from a small-bias distribution.  Soon thereafter Lovett~\cite{Lov08} showed that a sum of~$2^d$ independent draws from a small-bias distribution suffices, implying that $\F_2$-degree-$d$ functions can be fooled using just $2^{O(d)} \cdot \log(n/\eps)$ random bits.  More precisely, if $\vphi$ is any $\eps$-biased density on $\F_2^n$, Lovett showed that
\[
    \Bigl|\E_{\by^{(1)}, \dots, \by^{(2^d)} \sim \vphi}[f(\by^{(1)} + \cdots + \by^{(2^d)})] - \E_{\bx \sim \F_2^n}[f(\bx)]\Bigr| \leq O(\eps^{1/4^d}).
\]
In other words, the $2^d$-fold convolution $\vphi^{\conv 2^d}$
                                            \nomenclature[f*n]{$f^{\conv n}$}{the $n$-fold convolution, $f \conv f \conv \cdots \conv f$}%
density fools functions of $\F_2$-degree~$d$.

The current state of the art for this problem is Viola's Theorem~\cite{Vio09}, which shows that the original idea of Bogdanov and Viola~\cite{BV07} works:  Summing~$d$ independent draws from an $\eps$-biased distribution fools $\F_2$-degree-$d$ polynomials.
\begin{named}{Viola's Theorem}
                                            \index{Viola's Theorem}
    Let $\vphi$ be any $\eps$-biased density on $\F_2^n$, $0 \leq \eps \leq 1$. Let $d \in \N^+$ and define $\eps_d = 9\eps^{1/2^{d-1}}$.  Then the class of all $f \co \F_2^n \to \bits$ with $\deg_{\F_2}(f) \leq d$ is $\eps_d$-fooled by     the $d$-fold convolution $\vphi^{\conv d}$;
                                            \nomenclature[f*n]{$f^{\conv n}$}{the $n$-fold convolution, $f \conv f \conv \cdots \conv f$}%
    i.e.,
    \[
        \Bigl|\E_{\by^{(1)}, \dots, \by^{(d)} \sim \vphi}[f(\by^{(1)} + \cdots + \by^{(d)})] - \E_{\bx \sim \F_2^n}[f(\bx)]\Bigr| \leq 9\eps^{1/2^{d-1}}.
    \]
\end{named}
In light of Theorem~\ref{thm:aghp}, Viola's Theorem implies that one can $\eps$-fool $n$-bit functions of $\F_2$-degree~$d$ using only $O(d\log n) + O(d 2^d \log(1/\eps))$ independent random bits.

The proof of Viola's Theorem is an induction on~$d$.  To reduce the case of degree~$d+1$ to degree~$d$, Viola makes use of a simple concept: directional derivatives.
\begin{definition}
    Let $f \co \F_2^n \to \F_2$ and let $y \in \F_2^n$.  The \emph{directional derivative}
                                            \index{directional derivative}%
                                            \nomenclature[Deltatf]{$\dD_y f$}{for $f \co \F_2^n \to \F_2$, the function $\F_2^n \to \F_2$ defined by $\dD_y f(x) = f(x+y) - f(x)$}%
    $\dD_y f \co \F_2^n \to \F_2$ is defined by
    \[
        \dD_y f(x) = f(x+y) - f(x).
    \]
    Over $\F_2$ we may equivalently write $\dD_y f(x) = f(x+y) + f(x)$.
\end{definition}
As expected, taking a derivative reduces degree by~$1$:
\begin{fact}                                        \label{fact:dir-deriv}
    For any $f \co \F_2^n \to \F_2$ and $y \in \F_2^n$ we have $\deg_{\F_2}(\dD_y f) \leq \deg_{\F_2}(f) -1$.
\end{fact}
In fact, we'll prove a slightly stronger statement:
\begin{proposition}                                     \label{prop:dir-deriv}
    Let $f \co \F_2^n \to \F_2$ have $\deg_{\F_2}(f) = d$ and fix $y, y' \in \F_2^n$.  Define $g \co\F_2^n \to \F_2$ by $g(x) = f(x+y) - f(x+y')$. Then $\deg_{\F_2}(g) \leq d-1$.
\end{proposition}
\begin{proof}
    In passing from the $\F_2$-polynomial representation of $f(x)$ to that of~$g(x)$, each monomial $x^S$ of maximal degree~$d$ is replaced by $(x+y)^S - (x+y')^S$.  Upon expansion the monomials $x^S$ cancel, leaving a polynomial of degree at most $d-1$.
\end{proof}

We are now ready to give the proof of Viola's Theorem.

\begin{proof}[Proof of Viola's Theorem]
    The proof is by induction on~$d$.  The $d = 1$ case is immediate (even without the factor of~$9$) because $\vphi$ is $\eps$-biased.  Assume that the theorem holds for general $d \geq 1$ and let $f \co \F_2^n \to \bits$ have $\deg_{\F_2}(f) \leq d+1$.  We split into two cases, depending on whether the bias of~$f$ is large or small.

    \paragraph{Case 1: $\E[f]^2 > \eps_d$.}  In this case,
    \begin{align*}
        &\quad \sqrt{\eps_d} \cdot \Bigl|\E_{\bz \sim \vphi^{\conv(d+1)}}[f(\bz)] - \E_{\bx \sim \F_2^n}[f(\bx)]\Bigr| \\
        &< |\E[f]| \cdot \Bigl|\E_{\bz \sim \vphi^{\conv(d+1)}}[f(\bz)] - \E_{\bx \sim \F_2^n}[f(\bx)]\Bigr| \\
        &= \Bigl|\E_{\bx' \sim \F_2^n, \bz \sim \vphi^{\conv(d+1)}}[f(\bx')f(\bz)] - \E_{\bx', \bx \sim \F_2^n}[f(\bx')f(\bx)]\Bigr| \\
        &= \Bigl|\E_{\by \sim \F_2^n,  \bz \sim \vphi^{\conv(d+1)}}[f(\bz + \by)f(\bz)] - \E_{\by, \bx \sim \F_2^n}[f(\bx + \by)f(\bx)]\Bigr| \\
        &= \Bigl|\E_{\by \sim \F_2^n,  \bz \sim \vphi^{\conv(d+1)}}[\dD_{\by}f(\bz)] - \E_{\by, \bx \sim \F_2^n}[\dD_{\by}f(\bx)]\Bigr| \\
        &\leq \E_{\by \sim \F_2^n}\Bigl[\Bigl|\E_{\bz \sim \vphi^{\conv(d+1)}}[\dD_{\by}f(\bz)] - \E_{\bx \sim \F_2^n}[\dD_{\by}f(\bx)\Bigr|\Bigr].
    \end{align*}
    For each outcome $\by=y$ the directional derivative $\dD_{y} f$ has $\F_2$-degree at most~$d$ (Fact~\ref{fact:dir-deriv}).  By induction we know that $\vphi^{\conv d}$ $\eps_d$-fools any such polynomial, and it follows from Exercise~\ref{ex:conv-still-fools} that $\vphi^{\conv (d+1)}$ does too.  Thus each quantity in the expectation over~$\by$ is at most $\eps_d$, and we conclude
    \[
        \Bigl|\E_{\bz \sim \vphi^{\conv(d+1)}}[f(\bz)] - \E_{\bx \sim \F_2^n}[f(\bx)]\Bigr|  \leq \frac{\eps_d}{\sqrt{\eps_d}} = \sqrt{\eps_d} = \tfrac13 \eps_{d+1} \leq \eps_{d+1}.
    \]

    \paragraph{Case 2: $\E[f]^2 \leq \eps_d$.}  In this case we want to show that $\E_{\bw \sim \vphi^{\conv(d+1)}}[f(\bw)]^2$ is nearly as small.  By Cauchy--Schwarz,
    \begin{multline*}
        \E_{\bw \sim \vphi^{\conv(d+1)}}[f(\bw)]^2 = \E_{\bz \sim \vphi^{\conv d}}\Bigl[\E_{\by \sim \vphi}[f(\bz+\by)]\Bigr]^2 \leq \E_{\bz \sim \vphi^{\conv d}}\Bigl[\E_{\by \sim \vphi}[f(\bz+\by)]^2\Bigr] \\= \E_{\bz \sim \vphi^{\conv d}}\Bigl[\E_{\by, \by' \sim \vphi}[f(\bz+\by)f(\bz+\by')]\Bigr] = \E_{\by, \by' \sim \vphi}\Bigl[\E_{\bz \sim \vphi^{\conv d}}[f(\bz+\by)f(\bz+\by')]\Bigr].
    \end{multline*}
    For each outcome of $\by=y$, $\by'=y'$, the function $f(\bz+y)f(\bz+y')$ is of $\F_2$-degree at most~$d$ in the variables~$\bz$, by Proposition~\ref{prop:dir-deriv}. Hence by induction we have
    \begin{align*}
        \E_{\by, \by' \sim \vphi}\Bigl[\E_{\bz \sim \vphi^{\conv d}}[f(\bz+\by)f(\bz+\by')]\Bigr] & \leq \E_{\by, \by' \sim \vphi}\Bigl[\E_{\bx \sim \F_2^n}[f(\bx+\by)f(\bx+\by')]\Bigr] + \eps_d\\
         &= \Ex_{\bx \sim \F_2^n}[(\vphi \conv f)(\bx)^2] +\eps_d\\
         &= \sum_{\gamma \in \dualF} \wh{\vphi}(\gamma)^2 \wh{f}(\gamma)^2 +\eps_d\\
         &\leq \wh{f}(0)^2 + \eps^2 \sum_{\gamma \neq 0} \wh{f}(\gamma)^2 + \eps_d\\
         &\leq 2\eps_d + \eps^2,
    \end{align*}
    where the last step used the hypothesis of Case~$2$.  We have thus shown
    \[
             \E_{\bw \sim \vphi^{\conv(d+1)}}[f(\bw)]^2 \leq 2\eps_d + \eps^2 \leq 3\eps_d \leq 4 \eps_d,
    \]
    and hence $|\E[f(\bw)]| \leq 2 \sqrt{\eps_d}$.  Since we are in Case~2, $|\E[f]| \leq \sqrt{\eps_d}$, and so
    \[
        \Bigl| \E_{\bw \sim \vphi^{\conv(d+1)}}[f(\bw)] - \E[f] \Bigr| \leq 3 \sqrt{\eps_d} = \eps_{d+1},
    \]
    as needed.
\end{proof}

We end this section by discussing the tightness of parameters in Viola's Theorem.  First, if we ignore the error parameter, then the result is sharp: a counting argument (see~\cite{BV07}) shows that the $d$-fold convolution of $\eps$-biased densities cannot in general fool functions of $\F_2$-degree~$d+1$.  More explicitly, for any $d \in \N^+$, $\ell \geq 2d+1$, Lovett and Tzur~\cite{LT09} gave an explicit $\frac{\ell}{2^n}$-biased density on $\F_2^{(\ell+1)n}$
and an explicit function $f \co \F_2^{(\ell+1)n} \to \bits$ of degree~$d+1$ for which
\[
    \Bigl|\E_{\bw \sim \vphi^{\conv d}}[f(\bw)] - \E[f]\Bigr| \geq 1-\frac{2d}{2^n}.
\]
Regarding the error parameter in Viola's Theorem, it is not known whether the quantity $\eps^{1/2^{d-1}}$ can be improved, even in the case~$d = 2$.  However, obtaining even a modest improvement to $\eps^{1/1.99^d}$ (for $d$ as large as $\log n$) would constitute a major advance since it would imply progress on the notorious problem of ``correlation bounds for polynomials''; see Viola~\cite{Vio09a}.

\section{Exercises and notes} \label{sec:pseudorandomness-notes}

\begin{exercises}
    \item Let $\boldf$ be chosen as in Proposition~\ref{prop:random-Fourier}.  Compute $\Var[\wh{\boldf}(S)]$ for each
                                                \index{random function}
        $S \subseteq [n]$.
    \item \label{ex:rand-stable-influences} Prove Fact~\ref{fact:rand-stable-influences}.
    \item Show that any nonconstant $k$-junta has $\Inf_i^{(1-\delta)}[f] \geq (1/2 - \delta/2)^{k-1}/k$ for at least one coordinate~$i$.
    \item Let $\vphi \co \F_2^n \to \R^{\geq 0}$ be an $\eps$-biased density.  For each $d \in \N^+$ show that the $d$-fold convolution $\vphi^{\conv d}$ is an $\eps^d$-biased density.
    \item \label{ex:pseudorandomness-compare} \begin{exercises}
            \item Show that if $f \btR$ has $\eps$-small influences, then it is $\sqrt{\eps}$-regular.
            \item Show that for all even $n$ there exists $f \btb$ that is $2^{-n/2}$-regular but does not have $\eps$-small influences for any $\eps < 1/2$.
            \item Show that there is a function $f \btb$ with $((1-\delta)^{n-1}, \delta)$-small stable influences that is not $\eps$-regular for any $\eps < 1$.
            \item \label{ex:x0Maj} Verify that the function $f(x) = x_0 \maj_n(x_1, \dots, x_n)$ from Example~\ref{eg:small-stable-influences} satisfies $\Inf_0^{(1-\delta)}[f] = \Stab_{1-\delta}[\Maj_n]$ for $\delta \in (0,1)$, and thus does not have $(\eps, \delta)$-small stable influences unless $\eps \geq 1 - \sqrt{\delta}$.
            \item Show that the function $f \co \bits^{n+1} \to \bits$ from part~\ref{ex:x0Maj} is $\frac{1}{\sqrt{n}}$-regular.
            \item Suppose $f \btR$ has $(\eps,\delta)$-small stable influences.  Show that $f$ is $(\eta,k)$-regular for $\eta = \sqrt{\eps/(1-\delta)^{k-1}}$.
            \item Show that $f$ has $(\eps,1)$-small stable influences if and only if $f$ is $(\sqrt{\eps}, 1)$-regular.
            \item Let $f \btb$ be monotone.  Show that if $f$ is $(\eps,1)$-regular then $f$ is $\eps$-regular and has $\eps$-small influences.
          \end{exercises}
    \item \begin{exercises}
            \item Let $f \btR$.  Let $(J,\ol{J})$ be a partition of $[n]$ and let $z \in \bits^{\ol{J}}$.  For $\bz \sim \bits^{\ol{J}}$ uniformly random, give a formula for $\Var_{\bz}[\E[\restr{f}{J}{\bz}]]$ in terms of $f$'s Fourier coefficients.  (Hint: Direct application of Corollary~\ref{cor:expected-restrict-coeffs}.)
            \item Using the above formula and the probabilistic method, give an alternate proof of the second statement of Proposition~\ref{prop:eps-delta-reg-character}.
          \end{exercises}
    \item \label{ex:bias-of-ip2} Let $\vphi \co \F_2^n \to \R^{\geq 0}$ be the density corresponding to the uniform distribution on the support of $\IP_n \co \F_2^n \to \{0,1\}$.  Show that $\vphi$ is $\eps$-biased for $\eps = 2^{-n/2}/(1-2^{-n/2})$, but not for smaller~$\eps$.
    \item Prove Proposition~\ref{prop:eps-delta-reg-cov}.
    \item Compute the $\F_2$-polynomial representation of the equality function
                                        \index{equality function}
            $\EQU_n \co \{0,1\}^n \to \{0,1\}$, defined by $\EQU_n(x) = 1$ if and only if $x_1 = x_2 = \cdots = x_n$.
    \item \label{ex:mobius} \begin{exercises}
            \item Let $f \co \{0,1\}^n \to \R$ and let $q(x) = \sum_{S \subseteq [n]} c_S x^S$ be the (unique) multilinear polynomial representation of~$f$ over~$\R$.  Show that 
                \[
                    c_S = \sum_{R \subseteq S} (-1)^{|S| - |R|} f(R),
                \]where we identify $R \subseteq [n]$ with its $0$-$1$ indicator string.  This formula is sometimes called \emph{M\"obius inversion}.
                                                \index{M\"obius inversion}
            \item Prove Proposition~\ref{prop:zo-coeffs}.
          \end{exercises}
    \item \label{ex:schwartz-zippel-like-f2}  (Cf.\ Lemma~\ref{lem:schwartz-zippel-like}.)  Let $f \co \F_2^n \to \F_2$ be nonzero and suppose $\deg_{\F_2}(f) \leq k$.  Show that $\Pr[f(\bx) \neq 0] \geq 2^{-k}$.  (Hint: As in the similar Exercise~\ref{ex:schwartz-zippel-like}, use induction on~$n$.)
    \item \label{ex:granularity-vs-deg} Let $f \btzo$.
            \begin{exercises}
                \item \label{ex:bernasconi-codenotti} Show that $\deg_{\F_2}(f) \leq \log(\ssparsity{f})$.  (Hint: You will need Exercise~\ref{ex:restrict-norm}, Corollary~\ref{cor:zo-top}, and Exercise~\ref{ex:odd-ones-full-sparsity}.)
                \item \label{ex:granularity-bc} Suppose $\wh{f}$ is $2^{-k}$-granular.
                                                            \index{granularity, Fourier spectrum}
                    Show that $\deg_{\F_2}(f) \leq k$.  (This is a stronger result than part~\ref{ex:bernasconi-codenotti}, by Exercise~\ref{ex:sparsity2}.)
            \end{exercises}
    \item \label{ex:ms-bent} Let $f \btb$ be bent, $n > 2$.  Show that $\deg_{\F_2}(f) \leq n/2$.  (Note that the upper bound $n/2+1$ follows from Exercise~\ref{ex:granularity-vs-deg}\ref{ex:granularity-bc}.)
    \item \label{ex:correlation-immune-2/3} In this exercise you will prove Theorem~\ref{thm:correlation-immune-2/3}.
        \begin{exercises}
            \item Suppose $p(x) = c_0 + c_S x^S + r(x)$ is a real multilinear polynomial over $x_1, \dots, x_n$ with $c_0, c_S \neq 0$, $|S| > \frac23n$, and $|T| > \frac23n$ for all monomials $x^T$ appearing in~$r(x)$. Show that after expansion and multilinear reduction (meaning $x_i^2 \mapsto 1$), $p(x)^2$ contains the term~$2c_0c_S x^S$.
            \item Deduce Theorem~\ref{thm:correlation-immune-2/3}.
        \end{exercises}
    \item \label{ex:sieg-sharp}  In this exercise you will explore the sharpness of Siegenthaler's Theorem and Theorem~\ref{thm:correlation-immune-2/3}.
        \begin{exercises}
            \item For all $n$ and $k < n-1$, find an $f \zotzo$ that is $k$-resilient and has $\deg_{\F_2}(f) = n-k-1$.
            \item For all $n \geq 3$, find an $f \zotzo$ that is $1$st-order correlation immune and has $\deg_{\F_2}(f) = n-1$.
            \item For all $n$ divisible by~$3$, find a biased $f \zotzo$ that is $(\frac23n - 1)$th-order correlation immune.
        \end{exercises}
    \item \label{ex:oplus-bent} Prove Proposition~\ref{prop:oplus-bent}.
    \item Bent functions come in pairs: Show that if $f \ftb$ is bent, then $2^{n/2} \wh{f}$ is also a bent function (with domain~$\dualF$).
    \item \label{ex:maiorana-mcfarland} Extend Proposition~\ref{prop:rothaus-bent} to show that if $\pi$ is any permutation on~$\F_2^n$, then $f(x,y) = \IP_{2n}(x, \pi(y)) g(y)$ is bent.
    \item \label{ex:dickson} \emph{Dickson's Theorem}
                                    \index{Dickson's Theorem}
            says the following: Any polynomial $p \co \F_2^n \to \F_2$ of degree at most~$2$ can be expressed as
            \begin{equation} \label{eqn:dickson}
                p(x) = \ell_0(x) + \sum_{j = 1}^k \ell_j(x) \ell'_j(x),
            \end{equation}
            where $\ell_0$ is an affine function and $\ell_1, \ell_1', \dots, \ell_k, \ell_k'$ are linearly independent linear functions. Here $k$ depends only on~$p$ and is called the ``rank'' of $p$. Show that for $n$ even, $g \co \F_2^n \to \bits$ defined by $g(x) = \chi(p(x))$ is bent if and only if $k = n/2$, if and only if $g$ arises from $\IP_n$ as in Proposition~\ref{prop:bent-to-bent}.
    \item \label{ex:cq-to-ip} Without appealing to Dickson's Theorem, prove that the complete quadratic $x \mapsto \sum_{1 \leq i < j \leq n} x_i x_j$ can be expressed as in~\eqref{eqn:dickson}, with $k = \lfloor n/2 \rfloor$.  (Hint: Induction on~$n$, with different steps depending on the parity of~$n$.)
    \item \label{ex:mod3}  Define $\mathrm{mod}_3 \co \{-1,1\}^n \to \{0,1\}$ by
                                        \index{mod 3 function}
        $\mathrm{mod}_3(x) = 1$ if and only if $\sum_{j=1}^n x_i$ is divisible by~$3$. Derive the Fourier expansion
        \[
            \mathrm{mod}_3(x) = \tfrac13 + \tfrac23 (-1/2)^n \sum_{\substack{S \subseteq [n] \\ |S| \text{ even}}} (-1)^{(|S| \text{ mod } 4)/2} \sqrt{3}^{|S|} x^S
        \]
        and conclude that $\mathrm{mod}_3$ is $\frac23(\frac{\sqrt{3}}{2})^n$-regular.  (Hint: Consider $\prod_{j=1}^n (-\half +  \frac{\sqrt{-3}}{2})x_j$.)
    \item In Theorem~\ref{thm:aghp}, show that given $\br, \bs$ any fixed bit $\by_i$ can be obtained in deterministic~$\poly(\ell)$ time.
    \item \label{ex:better-aghp} \begin{exercises}
            \item Slightly modify the construction in Theorem~\ref{thm:aghp} to obtain a $(2^{-t} - 2^{-\ell})$-biased density.  (Hint: Arrange for $p_\gamma$ to have degree at most~$n-1$.)
            \item Since $\F_{2^\ell}$ is a dimension-$\ell$ vector space over~$\F_2$, it has some basis $v_1, \dots, v_\ell$.  Suppose we modify the construction in Theorem~\ref{thm:aghp} so that $\vphi$ is a density on~$\F_2^{n\ell}$, with $\by_{ij} = \la \enc(v_j \br^i), \enc(\bs) \ra$ for $i \in [n], j \in [\ell]$. Show that $\vphi$ remains $2^{-t}$-biased.
          \end{exercises}
    \item \label{ex:rand-eps-biased} Fix $\eps \in (0,1)$ and $n \in \N$.  Let $A \subseteq \F_2^n$ be a randomly chosen multiset in which $\lceil C n/\eps^2 \rceil$ elements are included, independently and uniformly.  Show that if $C$ is a large enough constant, then~$A$ is $\eps$-biased except with probability at most $2^{-n}$.
    \item \label{ex:freivalds} Consider the problem of computing the matrix multiplication $C = AB$, where $A, B \in \F_2^{n \times n}$.  There is an algorithm~\cite{LeG14} for solving this problem in time $O(n^{\omega})$, where $\omega < 2.373$; however, the algorithm is very complicated.  Suppose you are given $A$, $B$, and the outcome~$C'$ of running this algorithm; you want to test that indeed $C' = AB$.
        \begin{exercises}
            \item \label{ex:frei} Give an algorithm using $n$ random bits and time $O(n^2)$ with the following property: If $C' = AB$, then the algorithm ``accepts'' with probability~$1$; if $C' \neq AB$, then the algorithm ``accepts'' with probability at most~$1/2$. (Hint: Compute $C'\bx$ and $AB\bx$ for a random $\bx \in \F_2^n$.)
            \item \label{ex:frei-nn} Show how to reduce the number of random bits used to $O(\log n)$ at the expense of making the false acceptance probability~$2/3$, while keeping the running time~$O(n^2)$.  (You may use the fact that in Theorem~\ref{thm:aghp}, the time required to compute~$\by$ given~$\br$ and~$\bs$ is~$n \cdot \polylog(\ell)$.)
        \end{exercises}
    \item \label{ex:2-wise} Simplify the exposition and analysis of Theorem~\ref{thm:dual-bch-simple} and Corollary~\ref{cor:k-wise-indep-simple} in the case of $k=2$, and show that you can take $m$ to be one less (i.e., $m = \ell$).
    \item \label{ex:better-k-wise}  Consider the matrix $H' \in \F_n^{k \times n}$ constructed in Theorem~\ref{thm:dual-bch-simple}, and suppose we delete all rows corresponding to even (nonzero) powers of the~$\alpha_j$'s.  Show that~$H'$ retains the property that any sum of at most~$k$ columns of~$H'$ is nonzero in~$\F_n^k$.  (Hint: Prove and use that $(\sum_j \beta_j)^2 = \sum_j \beta_j^2$ for any sequence of $\beta_j \in \F_n$.)  Deduce that the cardinality of~$A$ in Corollary~\ref{cor:k-wise-indep-simple} can be decreased to $2(2n)^{\lfloor k/2 \rfloor}$.
    \item \label{ex:k-wise-lower}
        Let $A \subseteq \bn$ be a multiset and suppose that the probability density~$\phi_A$ is $k$-wise independent.  In this exercise you will prove the lower bound $|A| \geq \Omega(n^{\lfloor k/2 \rfloor})$ (for~$k$ constant).
        \begin{exercises}
            \item Suppose $\calF \subseteq 2^{[n]}$ is a collection of subsets of~$[n]$ such that $|S \cup T| \leq k$ for all $S,T \in \calF$.  For each $S \in \calF$ define $\chi_S^A \in \bits^{|A|} \subseteq \R^{|A|}$ to be the real vector with entries indexed by~$A$ whose $a$th entry is $a^{S} = \prod_{i \in S} a_i$.  Show that the set of vectors $\{\frac{1}{\sqrt{|A|}} \chi_S^A : S \in \calF\}$ is orthonormal and hence $|A| \geq |\calF|$.
            \item Show that we can find $\calF$ satisfying $|\calF| \geq \sum_{j = 0}^{k/2} \binom{n}{j}$ if~$k$ is even and $|\calF| \geq \sum_{j = 0}^{(k-1)/2} \binom{n}{j} + \binom{n-1}{(k-1)/2}$ if~$k$ is odd.
        \end{exercises}
    \item \label{ex:conv-still-fools} Let $\calC$ be a class of functions $\F_2^n \to \R$ that is closed under translation; i.e., $f^{+z} \in \calC$ whenever $f \in \calC$ and $z \in \F_2^n$ (recall Definition~\ref{def:translate-domain}).  An example is the class of functions of $\F_2$-degree at most~$d$.  Show that if $\psi$ is a density that $\eps$-fools $\calC$, then $\psi \conv \vphi$ also $\eps$-fools $\calC$ for any density $\vphi$.
    \item \label{ex:learn-f2-poly}
                                   \index{F2-polynomial representation@$\F_2$-polynomial representation!learning}
            Fix an integer $\ell \geq 1$.  In this exercise you will generalize Exercise~\ref{ex:learn-f2-parity} by showing how to exactly learn $\F_2$-polynomials of degree at most~$\ell$.
            \begin{exercises}
              \item Fix $p \co \F_2^n \to \F_2$ with $\deg_{\F_2}(p) \leq \ell$ and suppose that $\bx^{(1)}, \dots, \bx^{(m)} \sim \F_2^n$ are drawn uniformly and independently from~$\F_2^n$.  Assume that $m \geq C \cdot 2^\ell (n^\ell + \log(1/\delta))$ for $0 < \delta \leq 1/2$ and~$C$ a sufficiently large constant.  Show that except with probability at most~$\delta$, the only $q \co \F_2^n \to \F_2$ with $\deg_{\F_2}(q) \leq \ell$ that satisfies $q(\bx^{(i)}) = p(\bx^{(i)})$ for all $i \in [m]$ is $q = p$.  (Hint: Exercise~\ref{ex:schwartz-zippel-like-f2} with $q-p$.)
              \item  Show that the concept class of all polynomials $\F_2^n \to \F_2$ of degree at most~$\ell$ can be learned from random examples only, with error~$0$, in time
                  $O(n)^{3\ell}$. (Remark: As in Exercise~\ref{ex:learn-f2-parity}, since the key step is solving a linear system, the learning algorithm can also be done in $O(n)^{\omega \ell}$ time, assuming matrix multiplication can be done in $O(n^\omega)$ time.)
              \item Extend this learning algorithm so that in running time $O(n)^{3\ell}\cdot \log(1/\delta)$ it achieves success probability at least $1 - \delta$.  (Hint: Similar to Exercise~\ref{ex:learning-delta}.)
        \end{exercises}
    \item \label{ex:learn-one-rel} In this exercise you will
                                                \index{junta!learning}
                prove Lemma~\ref{lem:learn-one-rel}.
            \begin{exercises}
                \item Give a $\poly(n, 2^k)\cdot \log(1/\delta)$-time learning algorithm that, given random examples from a $k$-junta $\F_2^n \to \F_2$, determines (except with probability at most~$\delta$) if $f$ is a constant function, and if so, which one.
                \item Given access to random examples from a $k$-junta $f \co \F_2^n \to \F_2$, let $P \subseteq [n]$ be a set of relevant coordinates for~$f$ and let $z \in \F_2^P$.  Show how to obtain~$M$ independent random examples from the $(k-|P|)$-junta~$\restr{f}{\overline{P}}{z}$ in time $\poly(n, 2^k)\cdot M \cdot \log(1/\delta)$ (except with probability at most~$\delta$).
                \item Complete the proof of Lemma~\ref{lem:learn-one-rel}.  (Hint: Build a depth-$k$ decision tree for~$f$.)
            \end{exercises}
    \item \label{ex:improved-bsvw} \begin{exercises}
            \item Improve the bound in Lemma~\ref{lem:eps-biased-mean} to $\snorm{f}_1 \eps - |\wh{f}(\emptyset)|\eps$ and the bound in Corollary~\ref{cor:eps-biased-mean} to $\snorm{f}_1^2 \eps - \|f\|_2^2\eps$.
            \item Improve the bound in Theorem~\ref{thm:BLR-derand} to $\sqrt{\theta^2 - \eps}/\sqrt{1-\eps}$.
          \end{exercises}
    \item  \label{ex:BLR-derand}
        Improve on Theorem~\ref{thm:BLR-derand} by a factor of roughly~$2$ in the case of acceptance probability near~$1$.  Specifically, show that if $f$ passes the Derandomized BLR Test with probability $1-\delta$, then there exists $\gamma^* \in \dualF$ with $|\wh{f}(\gamma^*)| \geq \sqrt{1-2\delta -\eps}/\sqrt{1-\epsilon}$.
    \item \label{ex:gowers-norm}    Fix an integer $k \in \N^+$.  Let $(f_s)_{s \in \zo^k}$ be a collection of functions indexed by length-$k$ binary sequences, each $f_s \co \F_2^n \to \R$.  Define the \emph{$k$th Gowers ``inner product''}
                                                        \index{Gowers norm}%
        $\la (f_s)_s \ra_{U^k} \in \R$ by
        \[
              \la (f_s)_s \ra_{U^k} = \E_{\bx, \by_1, \dots, \by_k}\left[\prod_{s \in \zo^k} f_s(\bx + \littlesum_{i : s_i = 1} \by_i)\right],
        \]
        where the $k+1$ random vectors $\bx_, \by_1, \dots, \by_k$ are independent and uniformly distributed on~$\F_2^n$.  Define the \emph{$k$th Gowers norm} of a function $f \co \F_2^n \to \R$ by
        \[
            \|f\|_{U^k} = \la (f, f, \dots, f) \ra_{U^k}^{1/2^{k}},
        \]
        where $(f, f, \dots, f)$ denotes that all $2^k$ functions in the collection equal~$f$.  (You will later verify that $\la (f, f, \dots, f) \ra_{U^k}$ is always nonnegative.)
        \begin{exercises}
            \item Check that $\la f_0, f_1 \ra_{U^1} = \E[f_0]\E[f_1]$ and therefore $\|f\|_{U^1}^2 = \E[f]^2$.
            \item Check that
                \[
                    \la f_{00}, f_{10}, f_{01}, f_{11} \ra_{U^2} = \sum_{\gamma \in \dualF} \wh{f_{00}}(\gamma)\wh{f_{10}}(\gamma)\wh{f_{01}}(\gamma)\wh{f_{11}}(\gamma)
                \]
                and therefore $\|f\|_{U^2}^4 = \snorm{f}_4^4$.
                                                 \index{Fourier norm!$4$-}%
                (Cf.\ Exercise~\ref{ex:BLR4}\ref{ex:gowers2}.)
            \item Show that
                \begin{equation} \label{eqn:gowers-inter}
                    \la (f_s)_s \ra_{U^k} = \E_{\by_1, \dots, \by_{k-1}}\left[\E_{\bx} \left[\prod_{s : s_k = 0} f_s(\bx + \littlesum_{i : s_i = 1} \by_i)\right]\cdot \E_{\bx'} \left[\prod_{s : s_k = 1} f_{s}(\bx' + \littlesum_{i : s_i = 1} \by_i)\right]\right],
                \end{equation}
                where $\bx'$ is independent of $\bx, \by_1, \dots, \by_{k-1}$ and uniformly distributed.
            \item Show that $\la (f, f, \dots, f) \ra_{U^k}$ is always nonnegative, as promised.
            \item Using~\eqref{eqn:gowers-inter} and Cauchy--Schwarz, show that
                \[
                    \la (f_s)_s \ra_{U^k} \leq \sqrt{\la (f_{(s_1, \dots, s_{k-1}, 0)})_S \ra_{U^k}} \sqrt{\la (f_{(s_1, \dots, s_{k-1}, 1)})_s \ra_{U^k}}.
                \]
            \item Show that
                    \begin{equation} \label{eqn:gowers-triangle}
                        \la (f_s)_s \ra_{U^k} \leq \prod_{s \in \zo^k} \|f_s\|_{U^k}.
                    \end{equation}
            \item Fixing $f \co \F_2^n \to \R$, show that $\|f\|_{U^k} \leq \|f\|_{U^{k+1}}$.  (Hint: Consider \linebreak $(f_s)_{s \in \zo^{k+1}}$ defined by $f_s = f$ if $s_{k+1}=0$ and $f_s = 1$ if $s_{k+1} = 1$.)
            \item Show that $\|\cdot\|_{U^k}$ satisfies the triangle inequality and is therefore a seminorm. (Hint: First show that
                \[
                    \|f_0 + f_1\|_{U^k}^{2^k} = \sum_{S \subseteq \zo^k} \la (f_{\bone[s \in S]})_{s \in \zo^k} \ra_{U^k}
                \]
                and then use~\eqref{eqn:gowers-triangle}.)
            \item Show that $\|\cdot\|_{U^k}$ is in fact a norm for all $k \geq 2$; i.e., $\|f\|_{U^k} = 0 \implies f = 0$.
        \end{exercises}
\end{exercises}

\subsection*{Notes.}

                               \index{F2-polynomial representation@$\F_2$-polynomial representation}
The $\F_2$-polynomial representation of a Boolean function $f$ is often called its algebraic normal form.  It seems to have first been explicitly introduced by Zhegalkin in 1927~\cite{Zhe27}.

For functions $f \co \Z_n \to \R$, the idea of $\eps$-regularity as a pseudorandomness notion dates back to Chung and Graham~\cite{CG92}, as does the equivalent combinatorial condition Proposition~\ref{prop:regularity-norm}.  (In the context of quasirandom graphs, the ideas date further back to Thomason~\cite{Tho87a} and to Chung, Graham, and Wilson~\cite{CGW89}.)  The idea of treating functions with small (stable) influences as being ``generic'' has its origins in the work of Kahn, Kalai, and Linial~\cite{KKL88}.  The notion was brought to the fore in work on hardness of approximation -- implicitly, by \Hastad~\cite{Has96,Has99}, and later more explicitly by Khot, Kindler, Mossel, and O'Donnell~\cite{KKMO07}.

                               \index{probability density, $\eps$-biased}%
The notion of $\eps$-biased sets (and also $(\eps,k)$-wise independent distributions) was introduced by Naor and Naor~\cite{NN93} (see also the independent work of Peralta~\cite{Per90}).  The construction in Theorem~\ref{thm:aghp} is due to Alon, Goldreich, \Hastad, and Peralta~\cite{AGHP92} (as is Exercise~\ref{ex:better-aghp}).  As noted by Naor and Naor~\cite{NN93}, $\eps$-biased sets are closely related to error-correcting codes over~$\F_2$; indeed, they are equivalent to linear error-correcting in which all pairs of codewords have relative distance in $[\half - \half \eps, \half + \half \eps]$.  In particular, the construction in Theorem~\ref{thm:aghp} is the concatenation of the well-known Reed--Solomon and Hadamard codes (see, e.g., MacWilliams and Sloane~\cite{MS77} for definitions).  The nonconstructive upper bound in Exercise~\ref{ex:rand-eps-biased} is essentially the Gilbert--Varshamov bound and is close to known lower bound of $\Omega(\frac{n}{\eps^2 \log(1/\eps)})$ (assuming $\eps \geq 2^{-\Omega(n)}$), which follows from the work of McEliece, Rodemich, Rumsey, and Welch~\cite{MRRW77} (see~\cite{MS77}).  Additionally, constructive upper bounds of $O(\frac{n}{\eps^3})$ and $O(\frac{n^{5/4}}{\eps^{5/2}})$ are known using tools from coding theory; see the work of Ben-Aroya and Ta-Shma~\cite{BT09} and Matthews and Peachey~\cite{MP11}.

                                                \index{correlation immune}
                                                \index{resilient}
The probabilistic notion of correlation immunity -- i.e., condition~\eqref{item:xm} of Corollary~\ref{cor:xiao-massey} -- was first introduced by Siegenthaler~\cite{Sie84}; we further discuss his work below. Independently and shortly thereafter, Chor, Friedman, Goldreich, \Hastad, Rudich, and Smolensky~\cite{CFG+85} introduced the definition of resilience and also connected it to $(0,k)$-regularity of the Fourier spectrum; i.e., they proved Corollary~\ref{cor:xiao-massey}. (In the cryptography literature, Corollary~\ref{cor:xiao-massey} is called the Xiao--Massey Theorem~\cite{XM88}.) The work~\cite{CFG+85} also essentially contains Theorem~\ref{thm:correlation-immune-2/3} and the relevant function from Example~\ref{ex:corr-immune}; cf.~the work of Mossel et~al.~\cite{MOS04}.

                                                \index{$k$-wise independent}
The problem of constructing explicit $k$-wise distributions of small support arose in different guises in different areas -- in the study of orthogonal arrays (in statistics), error-correcting codes, and algorithmic derandomization. Alon, Babai, and Itai~\cite{ABI85} gave the construction in Theorem~\ref{thm:dual-bch-simple} -- in fact, the stronger one from Exercise~\ref{ex:better-k-wise} -- based on the analysis of dual BCH codes in MacWilliams and Sloane~\cite{MS77}.  The lower bound from Exercise~\ref{ex:k-wise-lower} is essentially due to Rao~\cite{Rao47}; see also independent proofs~\cite{CFG+85,ABI85}.

Siegenthaler's Theorem dates from 1984~\cite{Sie84}.
                                                    \index{Siegenthaler's Theorem}
His motivation was the study of cryptographic stream ciphers in cryptography.  In this application, a short random sequence of bits (``secret key'') is transformed via some scheme into a very long sequence of pseudorandom bits (``keystream''), which can then be used as a one-time pad for encryption.  A basic component of most schemes is a linear feedback shift register (LFSR), which can efficiently generate long, fairly statistically-uniform sequences.  However, due to its $\F_2$-linearity, it suffers from some simple cryptanalytic attacks.  An early idea for combating this is to take~$n$ independent LFSR streams and combine them via some function~$f \co \F_2^n \to \F_2$.  Effective attacks are possible in such a scheme if~$f$ is correlated with any of its input bits -- or indeed (as Siegenthaler pointed out) any input pair, triple, etc.  This led Siegenthaler to define the probabilistic notion of correlation-immunity.  Although $\chi_{[n]}$ is the maximally correlation-immune function, it is not suitable as a LFSR combining function precisely because of its $\F_2$-linearity; the same is true of any function of low $\F_2$-degree.  Siegenthaler precisely captured this tradeoff between correlation-immunity and $\F_2$-degree in his theorem.

Bent functions were named and first studied by Rothaus
                                                \index{bent functions}%
around~1966; he didn't publish the notion until 1976, however~\cite{Rot76}, at which point there were already several works on subject, see, e.g.,~\cite{Dil72}.  Bent functions have application in cryptography and coding theory; see, e.g., Carlet's survey~\cite{Car10}. The basic constructions presented in Section~\ref{sec:constructing-pr-functions} are due to Rothaus; the class of bent functions described in Exercise~\ref{ex:maiorana-mcfarland} is called the Maiorana--McFarland family.  Dickson's Theorem is from a 1901 publication~\cite[Theorem~199]{Dic01}; see also MacWilliams and Sloane~\cite[Theorem~15.4]{MS77}.

                                              \index{junta!learning}
Theorem~\ref{thm:mos} is from Mossel et~al.~\cite{MOS04}; there is an improved algorithm for learning $k$-juntas that runs in time roughly $n^{.6024 k} \poly(n)$, due to Gregory Valiant~\cite{Val12}.  Avrim Blum offers a prize of \$1,000 for solving the case of $k = \log \log n$ in $\poly(n)$ time~\cite{Blu03}.  Theorem~\ref{thm:learn-low-l1-derand} is due to Kushilevitz and Mansour~\cite{KM93}. The Derandomized BLR Test and Theorem~\ref{thm:BLR-derand} (and
                                    \index{BLR (Blum--Luby--Rubinfeld) Test!derandomized}%
Exercise~\ref{ex:improved-bsvw}) are due to Ben-Sasson, Sudan, Vadhan, and Wigderson~\cite{BSVW03}.

The result of Exercise~\ref{ex:schwartz-zippel-like-f2} is due to Muller~\cite[Theorem~6]{Mul54a}; deriving Exercise~\ref{ex:learn-f2-poly} from it and from Blumer et~al.~\cite{BEHW87} is folklore. The result of Exercise~\ref{ex:granularity-vs-deg}\ref{ex:bernasconi-codenotti} is due to Bernasconi and Codenotti~\cite{BC99}; Exercise~\ref{ex:ms-bent} is from MacWilliams and Sloane~\cite{MS77}.  In Exercise~\ref{ex:freivalds}, part~\ref{ex:frei} is due to Freivalds~\cite{Fre79} and part~\ref{ex:frei-nn} to Naor and Naor~\cite{NN93}. The Gowers norm and results of Exercise~\ref{ex:gowers-norm} are from Gowers~\cite{Gow01}. Our proof of the second statement in Proposition~\ref{prop:eps-delta-reg-character} was suggested by Noam Lifshitz.

\chapter{Property testing, PCPPs, and CSPs}                                        \label{chap:testing}

In this chapter we study several closely intertwined topics: property testing, probabilistically checkable proofs of proximity (PCPPs), and constraint satisfaction problems (CSPs).  All of our work will be centered around the task of testing whether an unknown Boolean function is a dictator.  We begin by extending the BLR~Test to give a $3$-query property testing algorithm for the class of dictator functions.  This in turn allows us to give a $3$-query testing algorithm for \emph{any} property, so long as the right ``proof'' is provided.  We then introduce CSPs, which are in fact identical to string testing algorithms.  Finally, we explain how dictator tests can be translated into computational complexity results for CSPs, and we sketch the proofs of some of H{\aa}stad's optimal inapproximability results.

\section{Dictator testing}                              \label{sec:dictator-testing}

                                            \index{testing|(}
In Chapter~\ref{sec:BLR} we described the BLR property testing algorithm: Given query access to an unknown function $f \zotzo$, this algorithm queries~$f$ on a few random inputs and approximately determines whether~$f$ has the property of being linear over~$\F_2$.  The field of \emph{property testing} for Boolean functions is concerned with coming up with similar algorithms for other properties.  In general, a ``property'' can be any collection~$\calC$ of $n$-bit Boolean functions; it's the same as the notion of ``concept class'' from learning theory. Indeed, before running an algorithm to try to learn an unknown $f \in \calC$, one might first run a property testing algorithm to try to verify that indeed $f \in \calC$.

Let's encapsulate the key aspects of the BLR linearity test with some definitions:
\begin{definition}                      \label{def:function-tester}
    An \emph{$\rquers$-query function testing algorithm} for Boolean functions $f \zotzo$
    is a randomized algorithm that:
    \begin{itemize}
        \item chooses $\rquers$ (or fewer) strings $\bx^{(1)}, \dots, \bx^{(\rquers)} \in \zo^n$ according to some probability distribution;
        \item queries $f(\bx^{(1)}), \dots, f(\bx^{(\rquers)})$;
        \item based on the outcomes, decides (deterministically) whether to ``accept''~$f$.
    \end{itemize}
\end{definition}
\begin{definition}                      \label{def:local-function-tester}
    Let $\calC$ be a ``property'' of $n$-bit Boolean functions, i.e., a collection of functions $\zo^n \to \zo$.  We say a function testing algorithm is a \emph{local tester for $\calC$} (with \emph{rejection rate}~$\rejrate > 0$)
                                            \index{property testing!local tester}
    if it satisfies the following:
    \begin{itemize}
        \item If $f \in \calC$, then the tester  accepts with probability~$1$.
        \item For all $0 \leq \eps \leq 1$, if $\dist(f,\calC) > \eps$ (in the sense of Definition~\ref{def:eps-close}), then the tester rejects~$f$ with probability greater than $\rejrate \cdot \eps$.
        \item[] Equivalently, if the tester accepts~$f$ with probability at least $1 - \rejrate \cdot \eps$, then $f$ is $\eps$-close to $\calC$; i.e., $\exists g \in \calC$ such that $\dist(f,g) \leq \eps$.
    \end{itemize}
\end{definition}
By taking $\eps = 0$ in the above definition you see that any local tester gives a characterization of $\calC$: a function is in $\calC$ if and only if it is accepted by the tester with probability~$1$.  But a local tester furthermore gives a ``robust'' characterization: Any function accepted with probability \emph{close} to~$1$ must be \emph{close} to satisfying~$\calC$.
\begin{example}
    By Theorem~\ref{thm:blr-test}, the BLR~Test is a $3$-query local tester
                                                    \index{BLR (Blum--Luby--Rubinfeld) Test}
    for the property $\calC = \{f \co \F_2^n \to \F_2 \mid f \text{ is linear}\}$ (with rejection rate~$1$).
\end{example}
\begin{remark}                                  \label{rem:testing-families}
    To be pedantic, the BLR linearity test is actually a family of local testers, one for each value of~$n$.  This is a common scenario: We will usually be interested in testing natural \emph{families} of properties $(\calC_n)_{n \in \N^+}$, where $\calC_n$ contains functions $\zo^n \to \zo$.  In this case we need to describe a family of testers, one for each~$n$.  Generally, these testers will ``act the same'' for all values of~$n$ and will have the property that the rejection rate $\rejrate > 0$ is a universal constant independent of~$n$.
\end{remark}

There are a number of standard variations of Definition~\ref{def:local-function-tester} that one could consider.  One variation is to allow for an \emph{adaptive} testing algorithm, meaning that the algorithm can decide how to generate $\bx^{(t)}$ based on the query outcomes $f(\bx^{(1)}), \dots, f(\bx^{(t-1)})$.  However, in this book we will only consider nonadaptive testing.  Another variation is to relax the requirement that $\eps$-far functions be rejected with probability $\Omega(\eps)$; one could allow for smaller rates such as $\Omega(\eps^2)$, or $\Omega(\eps/\log n)$.  For simplicity, we will stick with the strict demand that the rejection probability be linear in~$\eps$. Finally, the most common definition of property testing allows the number of queries to be a function $\rquers(\eps)$ of~$\eps$ but requires that any function $\eps$-far from~$\calC$ be rejected with probability at least~$1/2$.  This is easier to achieve than satisfying Definition~\ref{def:local-function-tester}; see Exercise~\ref{ex:local-test-to-standard}.
                                            \index{testing|)}

So far we have seen that the property of being linear over~$\F_2$ is locally testable.  We'll now spend some time discussing local testability of an even simpler property, the property of being a \emph{dictator}.
                                    \index{dictator testing|seeonly{testing, dictatorship}}%
                                    \index{testing!dictatorship}%
In other words, we'll consider the property
\[
    \calD = \{f \zotzo \mid f(x) = x_i \text{ for some $i \in [n]$}\}.
\]
As we will see, dictatorship is in some ways the most important property to be able to test.

We begin with a reminder: Even though $\calD$ is a subclass of the linear functions and we have a local tester for linearity, this doesn't mean we automatically have a local tester for dictatorship.  (This is in contrast to learning theory, where a learning algorithm for a concept class automatically works for any subclass.)  The reason is that the non-dictator linear functions -- i.e.,~$\chi_S$ for $|S| \neq 1$ -- are at distance $\half$ from~$\calD$ but are accepted by any linearity test with probability~$1$.

Still, we could use a linearity test as a first component of a test for dictatorship; this essentially reduces the problem to testing if an unknown \emph{linear} function is a dictator. Historically, the first local testers for dictatorship~\cite{BGS95,PRS01} worked this way; after testing linearity, they chose $\bx, \by \sim \{0,1\}^n$ uniformly and independently, set $\bz = \bx \wedge \by$ (the bitwise logical AND), and tested whether $f(\bz) = f(\bx) \wedge f(\by)$.  The idea is that the only parity functions that satisfy this ``AND test'' with probability~$1$ are the dictators (and the constant~$0$).  The analysis of the test takes a bit of work; see Exercise~\ref{ex:BGS-test} for details.

Here we will describe a simpler dictatorship test.  Recall  we have already seen an important result that characterizes dictatorship: Arrow's Theorem, from Chapter~\ref{sec:arrow}.
                                        \index{Arrow's Theorem}
Furthermore the robust version of Arrow's Theorem (Corollary~\ref{cor:almost-arrow}) involves evaluating a $3$-candidate Condorcet election under the impartial culture assumption, and this is the same as querying the election rule~$f$ on~$3$ correlated random inputs.  This suggests a dictatorship testing component we call the ``NAE Test'':
\begin{named}{NAE Test}
                                                \index{NAE Test}
    Given query access to $f \btb$:
    \begin{itemize}
        \item Choose $\bx, \by, \bz \in \bn$ by letting each triple $(\bx_i, \by_i, \bz_i)$ be drawn independently and uniformly at random from among the $6$ triples satisfying the not-all-equal predicate $\NAE_3 \co \bits^3 \to \{0,1\}$.
        \item Query $f$ at $\bx$, $\by$, $\bz$.
        \item Accept if $\NAE_3(f(\bx), f(\by), f(\bz))$ is satisfied.
    \end{itemize}
\end{named}
The NAE Test by itself is \emph{almost} a $3$-query local tester for the property of being a dictator.  Certainly if~$f$ is a dictator then the NAE Test accepts with probability~$1$.  Furthermore, in Chapter~\ref{sec:arrow} we proved:
\begin{theorem}[Restatement of Corollary~\ref{cor:almost-arrow}]             \label{thm:NAE-test}
    If the NAE Test accepts $f$ with probability $1-\eps$, then $\W{1}[f] \geq 1 - \frac92\eps$, and hence $f$ is $O(\eps)$-close to $\pm \chi_i$ for some $i \in [n]$ by the FKN~Theorem.
\end{theorem}
There are two slightly unsatisfactory aspects to this theorem.  First, it gives a local tester only for the property of being a dictator or a negated-dictator.  Second, though the deduction $\W{1}[f] \geq 1 - \frac92\eps$ requires only simple Fourier analysis, the conclusion that~$f$ is close to a (negated-)dictator relies on the non-trivial FKN~Theorem.  Fortunately we can fix both issues simply by adding in the
                                                    \index{BLR (Blum--Luby--Rubinfeld) Test}
BLR~Test:
\begin{theorem}                                     \label{thm:blr+nae1}
    Given query access to $f \btb$, perform both the BLR~Test and the NAE~Test.  This is a $6$-query local tester for the property of being a dictator (with rejection rate~$.1$).
\end{theorem}
\begin{proof}
    The first condition in Definition~\ref{def:local-function-tester} is easy to check: If $f \btb$ is a dictator, then both tests accept~$f$ with probability~$1$.  To check the second condition, fix $0 \leq \eps \leq 1$ and assume the overall test accepts~$f$ with probability at least $1 - .1 \eps$.  Our goal is to show that~$f$ is $\eps$-close to some dictator.

    Since the overall test accepts with probability at least $1-.1 \eps$, both the BLR and the NAE tests must individually accept~$f$ with probability at least $1-.1\eps$.  By the analysis of the NAE~Test we deduce that $\W{1}[f] \geq 1-\frac92 \cdot .1\eps = 1-.45\eps$.  By the analysis of the BLR~Test (Theorem~\ref{thm:blr-test}) we deduce that $f$ is $.1\eps$-close to some parity function; i.e., $\wh{f}(S^*) \geq 1-.2\eps$ for some $S^* \subseteq [n]$.  Now if $|S^*| \neq 1$ we would have
    \[
        1 = \sum_{k=0}^n \W{k}[f] \geq (1 - .45\eps) + (1-.2\eps)^2 \geq 2 - .85 \eps > 1,
    \]
    a contradiction.  Thus we must have $|S^*| = 1$ and hence $f$ is $.1\eps$-close to the dictator $\chi_{S^*}$, stronger than what we need.
\end{proof}

As you can see, we haven't been particularly careful about obtaining the largest possible rejection rate.  Instead, we will be more interested in using as few queries as possible (while maintaining some positive constant rejection rate). Indeed we now show a small trick which lets us reduce our $6$-query local tester for dictatorship down to a $3$-query one. This is best possible since dictatorship can't be locally tested with $2$~queries (see Exercise~\ref{ex:no-2-query}).
\begin{named}{BLR+NAE Test}
    Given query
                                               \index{BLR+NAE Test}
    access to $f \btb$:
    \begin{itemize}
        \item With probability~$1/2$, perform the BLR~Test on~$f$.
        \item With probability~$1/2$, perform the NAE~Test on~$f$.
    \end{itemize}
\end{named}
\begin{theorem}                                     \label{thm:blr+nae2}
    The BLR+NAE Test is a $3$-query local tester for the property of being a dictator (with rejection rate~$.05$).
\end{theorem}
\begin{proof}
    The only observation we need to make is that if the BLR+NAE~Test accepts with probability $1-.05\eps$ then both the BLR and the NAE~tests individually must accept~$f$ with probability at least $1-.1\eps$.  The result then follows from the analysis of Theorem~\ref{thm:blr+nae1}.
\end{proof}
\begin{remark}          \label{rem:max-queries}
    In general, this trick lets us take the \emph{maximum} of the query complexities when we combine tests, rather than the sum (at the expense of worsening the rejection rate). Suppose we wish to combine~$t = O(1)$ different testing algorithms, where the $i$th tester uses~$\rquers_i$ queries. We make an overall test that performs each subtest with probability $1/t$.  This gives a $\max(\rquers_1, \dots, \rquers_t)$-query testing algorithm with the following guarantee: If the overall test accepts~$f$ with probability $1-\frac{\rejrate}{t} \eps $ then \emph{every} subtest must accept~$f$ with probability at least $1-\rejrate\eps$.
\end{remark}

We can now explain one reason why dictatorship is a particularly important property to be able to test locally.  Given the BLR Test for linear functions it still took us a little thought to find a local test for the subclass~$\calD$ of dictators.  But given our dictatorship test, it's easy to give a $3$-query local tester for \emph{any} subclass of~$\calD$.  (On a related note, Exercise~\ref{ex:test-any-subspace} asks you to give a $3$-query local tester for any affine subspace of the linear functions.)
\begin{theorem}                             \label{thm:test-any-dict-subclass}
    Let $\calS$ be any subclass of $n$-bit dictators; i.e., let $S \subseteq [n]$ and let
    \[
        \calS = \{\chi_i \zotzo \mid i \in S\}.
    \]
    Then there is a $3$-query local tester for $\calS$ (with rejection rate~$.01$).
\end{theorem}
\begin{proof}
    Let $1_S \in \zo^n$ denote the indicator string for the subset~$S$.  Given access to $f \zotzo$, the test is as follows:
    \begin{itemize}
        \item With probability $1/2$, perform the BLR+NAE Test on $f$.
        \item With probability $1/2$, apply the local correcting routine of Proposition~\ref{prop:local-correcting-linearity} to~$f$ on string~$1_S$; accept if and only if the output value is~$1$.
    \end{itemize}
    This test always makes either~$2$ or~$3$ queries, and whenever $f \in \calS$ it accepts with probability~$1$.  Now let $0 \leq \eps \leq 1$ and suppose the test accepts~$f$ with probability at least $1 - \rejrate\eps$, where $\rejrate = .01$.  Our goal will be to show that $f$ is $\eps$-close to a dictator $\chi_i$ with $i \in S$.

    Since the overall test accepts~$f$ with probability at least $1-\rejrate \eps$, the BLR+NAE Test must accept~$f$ with probability at least $1-2\rejrate\eps$.  By Theorem~\ref{thm:blr+nae2} we may deduce that~$f$ is $40\rejrate\eps$-close to some dictator~$\chi_i$.  Our goal is to show that $i \in S$; this will complete the proof because $40\rejrate\eps \leq \eps$ (by our choice of $\rejrate = .01$).

    So suppose by way of contradiction that $i \not \in S$; i.e., $\chi_i(1_S) = 0$.  Since~$f$ is $40\rejrate\eps$-close to the parity function~$\chi_i$, Proposition~\ref{prop:local-correcting-linearity} tells us that
    \[
        \Pr[\text{locally correcting~$f$ on input~$1_S$ produces the output~$\chi_i(1_S) = 0$}] \geq 1 - 80\rejrate\eps.
    \]
    On the other hand, since the overall test accepts~$f$ with probability at least $1-\rejrate \eps$, the second subtest must accept~$f$ with probability at least $1-2\rejrate\eps$. This means
    \[
        \Pr[\text{locally correcting~$f$ on input~$1_S$ produces the output~$0$}] \leq 2\rejrate\eps.
    \]
    But this is a contradiction, since $2\rejrate\eps < 1-80\rejrate\eps$ for all $0 \leq \eps \leq 1$ (by our choice of~$\rejrate=.01$).  Hence $i \in S$ as desired.
\end{proof}
	
\section{Probabilistically Checkable Proofs of Proximity}                                         \label{sec:pcpps}

In the previous section we saw that every subproperty of the dictatorship property has a $3$-query local tester.  In this section we will show that \emph{any property whatsoever} has a $3$-query local tester -- if an appropriate ``proof'' is provided.

To make sense of this statement let's first generalize the setting in which we study property testing. Definitions~\ref{def:function-tester} and~\ref{def:local-function-tester} are concerned with testing a Boolean function $f \zotzo$ by querying its values on various inputs.  If we think of $f$'s truth table as a Boolean string of length~$N = 2^n$, then a testing algorithm simply queries various coordinates of this string.  It makes sense to generalize to the notion of testing properties of $N$-bit \emph{strings}, for any length~$N$.  Here a property~$\calC$ will just be a collection $\calC \subseteq \zo^N$ of strings, and we'll be concerned with relative Hamming distance $\dist(w, w') = \frac{1}{N}\hamdist(w,w')$ between strings.  For simplicity, we'll begin to write~$n$ instead of~$N$.
\begin{definition}                          \label{def:string-tester}
    An \emph{$\rquers$-query string testing algorithm} for strings $w \in \zo^n$ is a randomized algorithm that:
    \begin{itemize}
        \item chooses $\rquers$ (or fewer) indices $\bi_1, \dots, \bi_\rquers \in [n]$ according to some probability distribution;
        \item queries $w_{\bi_1}, \dots, w_{\bi_r}$;
        \item based on the outcomes, decides (deterministically) whether to ``accept''~$w$.
    \end{itemize}
    We may also generalize this definition to testing strings $w \in \Omega^n$ over finite alphabets $\Omega$ of cardinality larger than~$2$.
\end{definition}
\begin{definition}                      \label{def:local-string-tester}
    Let $\calC \subseteq \zo^n$ be a ``property'' of $n$-bit Boolean strings.  We say a string testing algorithm is a \emph{local tester for $\calC$} (with \emph{rejection rate~$\rejrate > 0$})
                                            \index{property testing!local tester}
    if it satisfies the following:
    \begin{itemize}
        \item If $w \in \calC$, then the tester  accepts with probability~$1$.
        \item For all $0 \leq \eps \leq 1$, if $\dist(w, \calC) > \eps$, then the tester rejects~$w$ with probability greater than $\rejrate \cdot \eps$.
        \item[] Equivalently, if the tester accepts~$w$ with probability at least $1 - \rejrate \cdot \eps$, then~$w$ is $\eps$-close to $\calC$; i.e., $\exists w' \in \calC$ such that $\dist(w,w') \leq \eps$.
    \end{itemize}
\end{definition}
\begin{examples}            \label{eg:string-local-testers}
    Let $\calZ = \{(0, 0, \dots, 0)\} \subseteq \zo^n$ be the property of being the all-zeroes string.  Then the following is a $1$-query local tester for $\calZ$ (with rejection rate~$1$): Pick a uniformly random index~$\bi$ and accept if $w_{\bi} = 0$.

    Let $\calE = \{(0, 0, \dots, 0), (1, 1, \dots, 1)\} \subseteq \zo^n$ be the property of having all coordinates equal.  Then the following is a $2$-query local tester for $\calE$: Pick two independent and uniformly random indices~$\bi$ and~$\bj$ and accept if $w_{\bi} = w_{\bj}$.  In Exercise~\ref{ex:equality-string-test} you are asked to show that if $\dist(w,\calE) = \eps$, then this tester rejects~$w$ with probability $\half - \half(1-2\eps)^2 \geq \eps$.

    Let $\calO = \{w \in \F_2^n : \text{$w$ has an odd number of $1$'s}\}$.  This property does \emph{not} have a local tester making few queries.  In fact, in Exercise~\ref{ex:odd-1s-untestable} you are asked to show that any local tester for~$\calO$ must make the maximum number of queries,~$n$.
\end{examples}

As the last example shows, not every property has a local tester making a small number of queries; indeed, most properties of $n$-bit strings do not.  This is rather too bad: Imagine that for any large~$n$ and any complicated property $\calC \subseteq \zo^n$ there were an $O(1)$-query local tester.  Then if anyone supplied you with a string~$w$ claiming it satisfied~$\calC$, you wouldn't have to laboriously check this yourself, nor would you have to trust the supplier; you could simply spot-check~$w$ in a constant number of coordinates and become convinced that~$w$ is (close to being) in~$\calC$.

                                                    \index{PCPP|(}
But what if, in addition to $w \in \zo^n$, you could require the supplier to give you some additional side information $\Pi \in \zo^\ell$ about~$w$ so as to assist you in testing that $w \in \calC$?   One can think of $\Pi$ as a kind of ``proof'' that $w$ satisfies~$\calC$.  In this case it's possible that you can spot-check $w$ and~$\Pi$ together in a constant number of coordinates and become convinced that~$w$ is (close to being) in~$\calC$ -- all without having to ``trust'' the supplier of the string $w$ and the purported proof~$\Pi$.  These ideas lead to the notion of
                                                    \index{probabilistically checkable proof of proximity|seeonly{PCPP}}%
                                                    \index{assignment tester|seeonly{PCPP}}%
                                                    \index{assisted proof|seeonly{PCPP}}%
                                                    \index{locally testable proof|seeonly{PCPP}}%
\emph{probabilistically checkable proofs of proximity}~(PCPPs).
\begin{definition}                          \label{def:pcpp-system}
    Let $\calC \subseteq \zo^n$ be a property of $n$-bit Boolean strings and let $\ell \in \N$.  We say that $\calC$ has an \emph{$\rquers$-query, length-$\ell$ probabilistically checkable proof of proximity (PCPP) system} (with rejection rate~$\rejrate > 0$) when the following holds:  There exists an $\rquers$-query testing algorithm~$T$ for $(n+\ell)$-bit strings, thought of as pairs $w \in \zo^n$ and $\Pi \in \zo^\ell$, such that:
    \begin{itemize}
        \item (``Completeness.'') If $w \in \calC$, then there exists a ``proof'' $\Pi \in \zo^\ell$ such that~$T$ accepts with probability~$1$.
        \item (``Soundness.'')  For all $0 \leq \eps \leq 1$, if $\dist(w, \calC) > \eps$, then for \emph{every} ``proof'' $\Pi \in \zo^\ell$ the tester~$T$ rejects with probability greater than $\rejrate \cdot \eps$.
        \item[] Equivalently, if there exists $\Pi \in \zo^\ell$ that causes~$T$ to accept with probability at least $1 - \rejrate \cdot \eps$, then~$w$ must be $\eps$-close to $\calC$.
    \end{itemize}
    PCPP systems are also known as assisted testers, locally testable proofs, or assignment testers.
\end{definition}
\begin{remark}  A word on the three parameters:  We are usually interested in fixing the number of queries $\rquers$ to a very small universal constant (such as~$3$) while trying to keep the proof length $\ell = \ell(n)$ relatively small (e.g., $\poly(n)$ is a good goal).  We are usually not very concerned with the rejection rate $\rejrate$ so long as it's a positive universal constant (independent of~$n$).
\end{remark}

\begin{example} \label{eg:pcpp-linearity}
    In Example~\ref{eg:string-local-testers} we stated that $\calO = \{w \in \F_2^n : w_1 + \cdots + w_n = 1\}$ has no local tester making fewer than~$n$ queries.  But it's easy to give a $3$-query PCPP system for $\calO$ with proof length~$n-1$ (and rejection rate~$1$).  The idea is to require the proof string $\Pi$ to contain the partial sums of~$w$:
    \[
        \Pi_{j} = \sum_{i=1}^{j+1} w_i \pmod{2}.
    \]
    The tester will perform one of the following checks, uniformly at random:
    \begin{align*}
        \Pi_1 &= w_1 + w_2 \\
        \Pi_2 &= \Pi_1 + w_3 \\
        \Pi_3 &= \Pi_2 + w_4 \\
        &\cdots \\
        \Pi_{n-1} &= \Pi_{n-2} + w_n \\
        \Pi_{n-1} &= 1
    \end{align*}
    Evidently the tester always makes at most $3$ queries.  Further, in the ``completeness'' case $w \in \calO$, if $\Pi$ is a correct list of partial sums then the tester will accept with probability~$1$.  It remains to analyze the ``soundness'' case, $w \not \in \calO$.  Here we are significantly aided by the fact that $\dist(w,\calO)$ must be exactly~$1/n$ (since every string is at Hamming distance either~$0$ or~$1$ from~$\calO$). Thus to confirm  the claimed rejection rate of~$1$, we only need to observe that if $w \not \in \calO$ then at least one of the tester's~$n$ checks must fail.
\end{example}

This example generalizes to give a very efficient PCPP system for testing that $w$ satisfies any fixed $\F_2$-linear equation.  What about testing that $w$ satisfies a fixed system of $\F_2$-linear equations?
This interesting question is explored in Exercise~\ref{ex:pcpp-code}, which serves as a good warmup for our next result.

We now extend Theorem~\ref{thm:test-any-dict-subclass} to show the rather remarkable fact that \emph{any} property of~$n$-bit strings has a $3$-query PCPP system.  (The proof length, however, is enormous.)
\begin{theorem}                                     \label{thm:pcpp-for-any-ppty}
    Let $\calC \subseteq \zo^n$ be any class of strings.  Then there is a $3$-query, length-$2^{2^n}$ PCPP system for $\calC$ (with rejection rate~$.001$).
\end{theorem}
\begin{proof}
    Let $N = 2^n$ and fix an arbitrary bijection $\iota \co \zo^n \to [N]$.  The tester will interpret the string $w \in \zo^n$ to be tested as an index $\iota(w) \in [N]$ and will interpret the $2^N$-length proof $\Pi$ as a function $\Pi \co \zo^N \to \zo$.  The idea is for the tester to require that $\Pi$ be the dictator function corresponding to index~$\iota(w)$; i.e., $\chi_{\iota(w)} \co \zo^N \to \zo$.

    Now under the identification $\iota$, we can think of the string property $\calC$ as a subclass of all $N$-bit dictators, namely
    \[
        \calC' = \{\chi_{\iota(w')} \co \zo^N \to \zo \mid w' \in \calC\}.
    \]
    In particular, $\calC'$ is a property of $N$-bit functions.  We can now state the twofold goal of the tester:
    \begin{enumerate}
        \item check that $\Pi \in \calC'$;
        \item given that $\Pi$ is indeed some dictator $\chi_{\iota(w')} \co \zo^N \to \zo$ with $w' \in \calC$, check that $w' = w$.
    \end{enumerate}
    To accomplish the latter the tester would like to check $w_{\bj} = w'_{\bj}$ for a random $\bj \in [n]$.  The tester can query any $w_{j}$ directly but accessing~$w'_{j}$ requires a little thought. The trick is to prepare the string
    \[
        X^{(j)} \in \zo^N \text{ defined by } X^{(j)}_{\iota(y)} = y_j.
    \]
    and then to locally correct $\Pi$ on $X^{(j)}$ (using Proposition~\ref{prop:local-correcting-linearity}).

    Thus the tester is defined as follows:
    \begin{enumerate}
        \item \label{item:pcpp1} With probability $1/2$, locally test the function property $\calC'$ using Theorem~\ref{thm:test-any-dict-subclass}.
        \item \label{item:pcpp2} With probability $1/2$, pick $\bj \sim [n]$ uniformly at random; locally correct $\Pi$ on the string~$X^{(\bj)}$ and accept if the outcome equals $w_{\bj}$.
    \end{enumerate}
    Note that the tester makes~$3$ queries in both of the subtests.

    Verifying ``completeness'' of this PCPP system is easy: if $w \in \calC$ and $\Pi$ is indeed the (truth table of) $\chi_{\iota(w)} \co \zo^N \to \zo$ then the test will accept with probability~$1$.  It remains to verify the ``soundness'' condition.  Fix $w \in \zo^n$, $\Pi \co \zo^N \to \zo$, and $0 \leq \eps \leq 1$ and suppose that the tester accepts $(w, \Pi)$ with  probability at least $1-\rejrate \eps$, where $\rejrate = .001$.  Our goal is to show that $w$ is $\eps$-close to some string $w' \in \calC$.

    Since the overall test accepts with probability at least $1-\rejrate \eps$, subtest~\eqref{item:pcpp1} above accepts with probability at least $1-2\rejrate \eps$.  Thus by Theorem~\ref{thm:test-any-dict-subclass}, $\Pi$ must be $200\rejrate \eps$-close to some dictator $\chi_{\iota(w')}$ with $w' \in \calC$.  Since dictators are parity functions, Proposition~\ref{prop:local-correcting-linearity} tells us that
    \begin{equation}    \label{eqn:pcpp-correcting}
        \forall j,\ \Pr[\text{locally correcting $\Pi$ on $X^{(j)}$ produces $\chi_{\iota(w')}(X^{(j)}) = w'_j$}] \geq 1 - 400\rejrate \eps \geq 1/2,
    \end{equation}
    where we used $400 \rejrate \eps < 400 \rejrate \leq 1/2$ by the choice $\rejrate = .001$.

    On the other hand, since the overall test accepts with probability at least $1-\rejrate \eps$, subtest~\eqref{item:pcpp2} above rejects with probability at most $2\rejrate \eps$.  This means
    \[
        \Ex_{\bj \sim [n]}\left[\Pr[\text{locally correcting $\Pi$ on  $X^{(\bj)}$ \emph{doesn't} produce $w_{\bj}$}]\right] \leq 2\rejrate \eps.
    \]
    By Markov's inequality we deduce that except for at most a
    $4\rejrate \eps$ fraction of coordinates $j \in [n]$ we have
    \[
        \Pr[\text{locally correcting $\Pi$ on  $X^{(j)}$ \emph{doesn't} produce $w_{j}$}] < 1/2.
    \]
    Combining this information with~\eqref{eqn:pcpp-correcting} we deduce that $w_j = w'_j$ except for at most a $4\rejrate\eps \leq \eps$ fraction of coordinates $j \in [n]$.  Since $w' \in \calC$ we conclude that $\dist(w,C) \leq \eps$, as desired.
\end{proof}
                                                    \index{PCPP|)}

You may feel that the doubly-exponential proof length $2^{2^n}$ in this theorem is quite bad, but bear in mind there are $2^{2^n}$ different properties~$\calC$.  Actually, giving a PCPP system for \emph{every} property is a bit overzealous since most properties are not interesting or natural.  A more reasonable goal would be to give efficient PCPP systems for all ``explicit'' properties.  A good way to formalize this is to consider properties decidable by polynomial-size circuits.
\iftex
Here we use the definition of general (De~Morgan) circuits from Exercise~\ref{ex:general-circuit}.
\fi
\ifblog
Here we consider a standard definition of general Boolean circuits
which allows for non-constant depth (unlike in Chapter~\ref{chap:circuits}):
\begin{definition}
    A \emph{(De Morgan) circuit}~$C$ over Boolean variables $x_1, \dots, x_n$ is a directed acyclic graph in which each node (``gate'') is labeled with either an~$x_i$ or with $\wedge$, $\vee$, or~$\neg$ (logical NOT).  Each~$x_i$ is used as label exactly once; the associated nodes are called ``input'' gates and must have in-degree~$0$.  Each~$\wedge$ and~$\vee$ node must have in-degree~$2$ and each~$\neg$ node must have in-degree~$1$.  Each node ``computes'' a function $\bn \to \bits$ of the inputs in the natural way (see Definition~\ref{def:constant-depth-circuit}).  Finally, one node of~$C$ is designated as the ``output'' gate, and~$C$ itself is said to compute the function computed by the output node.  We often identify~$C$ with the function it computes.

    We define the $\emph{size}$ of~$C$, denoted $\size(C)$, to be the number of nodes (which is always at least~$n$ and at most twice the number of arcs/wires).
\end{definition}
\fi
Given an $n$-variable circuit~$C$ we consider the set of strings which it ``accepts'' to be a property,
\begin{equation} \label{eqn:ppty-from-ckt}
    \calC = \{w \in \zo^n : C(w) = 1\}.
\end{equation}
For properties computed by modest-sized circuits~$C$ we may hope for PCPP systems with proof length much less than~$2^{2^n}$. We saw such a case in Example~\ref{eg:pcpp-linearity}.

                                                    \index{PCPP reduction|(}
Another advantage of considering ``explicit'' properties is that we can define a notion of \emph{constructing} a PCPP system, ``given'' a property.
A theorem of the form ``for each explicit property~$\calC$ there exists an efficient PCPP system\dots'' may not be useful, practically speaking, if its proof is nonconstructive.  We can formalize the issue as follows:
\begin{definition} \label{def:pcpp-reduction}
    A \emph{PCPP reduction} is an algorithm which takes as input a circuit~$C$ and outputs the \emph{description} of a PCPP system for the string property~$\calC$ decided by~$C$ as in~\eqref{eqn:ppty-from-ckt}, where $n$ is the number of inputs to~$C$.  If the output PCPP system always makes $\rquers$ queries, has proof length $\ell(n, \size(C))$ (for some function~$\ell$), and has rejection rate~$\rejrate > 0$, we say that the PCPP reduction has the same parameters.  Finally, the PCPP reduction should run in time $\poly(\size(C), \ell)$.
\end{definition}
(We haven't precisely specified what it means to output the description of a PCPP system; this will be explained more carefully in Section~\ref{sec:CSPs}.  In brief it means to list -- for each possible outcome of the tester's randomness -- which bits are queried and what predicate of them is used to decide acceptance.)

Looking back at the results on testing subclasses of dictatorship (Theorem~\ref{thm:test-any-dict-subclass}) and PCPPs for any property (Theorem~\ref{thm:pcpp-for-any-ppty}) we can see they have the desired sort of ``constructive'' proofs.  In Theorem~\ref{thm:test-any-dict-subclass} the local tester's description depends in a very simple way on the input~$1_S$.  As for Theorem~\ref{thm:pcpp-for-any-ppty}, it suffices to note that given an $n$-input circuit~$C$ we can write down its truth table (and hence the property it decides) in time $\poly(\size(C)) \cdot 2^n$, whereas the allowed running time is at least $\poly(\size(C), 2^{2^n})$.  Hence we may state:
\begin{theorem}                                     \label{thm:pcpp-builder0}
    There exists a $3$-query PCPP reduction with proof length $2^{2^n}$ (and rejection rate~$.001$).
\end{theorem}
In Exercise~\ref{ex:exponential-pcpp} you are asked to improve this result as follows:
\begin{theorem}                                     \label{thm:pcpp-builder1}
    There exists a $3$-query PCPP reduction with proof length $2^{\poly(\size(C))}$ (and positive rejection rate).
\end{theorem}
(The fact that we again have just~$3$ queries is explained by Exercise~\ref{ex:pcpps-have-3-queries}; there is a generic reduction from any constant number of queries down to~$3$.)

Indeed, there is a much more dramatic improvement:
\begin{named}{The PCPP Theorem}                                     \label{thm:pcpp-builder2}
    There exists a $3$-query PCPP reduction with proof length $\poly(\size(C))$ (and positive rejection rate).
\end{named}
This is (a slightly strengthened version of) the famous ``PCP~Theorem'' \cite{FGL+96,AS98,ALM+98}
                                                    \index{PCP Theorem}%
from the field of computational complexity, which is discussed later in this chapter.
Though the PCPP Theorem is far stronger than Theorem~\ref{thm:pcpp-builder0}, the latter is not unnecessary; it's actually an ingredient in Dinur's proof of the PCP~Theorem~\cite{Din07}, being applied only to circuits of ``constant'' size.  The current state of the art for PCPP length~\cite{Din07,BS08} is highly efficient:
\begin{theorem}                                     \label{thm:pcpp-builder3}
    There exists a $3$-query PCPP reduction with proof length $\size(C) \cdot \polylog(\size(C))$ (and positive rejection rate).
\end{theorem}
                                                    \index{PCPP reduction|)}

\section{CSPs and computational complexity}                                     \label{sec:CSPs}
                                                \index{CSP|(}%
                                                \index{constraint satisfaction problem|seeonly{CSP}}%
This section is about the computational complexity of constraint satisfaction problems (CSPs), a fertile area of application for analysis of Boolean functions.  To study it we need to introduce a fair bit of background material; in fact, this section will mainly consist of definitions.

In brief, a CSP is an algorithmic task in which a large number of ``variables'' must be assigned ``labels'' so as to satisfy given ``local constraints''.  We start by informally describing some examples:
\begin{examples} \label{eg:csps} \
    \begin{itemize}
        \item     In the ``Max-$3$-Sat'' problem,
                                                    \index{Max-$3$-Sat}%
                                                    \index{$3$-Sat|seeonly{Max-$3$-Sat}}%
        given is a CNF formula of width at most~$3$ over Boolean variables $x_1, \dots, x_n$.  The task is to find a setting of the inputs that satisfies (i.e., makes $\true$) as many clauses as possible.
        \item In the ``Max-Cut'' problem,
                                                    \index{Max-Cut}%
        given is an undirected graph $G = (V,E)$.  The task is to find a ``cut'' -- i.e., a partition of~$V$ into two parts -- so that as many edges as possible ``cross the cut''.
        \item In the ``Max-E$3$-Lin'' problem,
                                                    \index{Max-$3$-Lin}%
                                                    \index{$3$-Lin|seeonly{Max-$3$-Lin}}%
            given is a system of linear equations over~$\F_2$, each equation involving exactly~$3$ variables.  The system may in general be overdetermined; the task is to find a solution which satisfies as many equations as possible.
        \item In the ``Max-$3$-Coloring'' problem,
                                                    \index{Max-$3$-Coloring}%
            given is an undirected graph $G = (V,E)$.  The task is to color each vertex either red, green, or blue so as to make as many edges as possible bichromatic.
    \end{itemize}
\end{examples}
Let's rephrase the last two of these examples so that the descriptions have more in common.  In Max-E$3$-Lin we have a set of variables~$V$, to be assigned labels from the domain $\Omega = \F_2$.  Each constraint is of the form $v_1 + v_2 + v_3 = 0$ or $v_1 + v_2 + v_3 = 1$, where $v_1, v_2, v_3 \in V$.  In Max-$3$-Coloring we have a set of variables (vertices)~$V$ to be assigned labels from the domain $\Omega = \{\text{red}, \text{green}, \text{blue}\}$. Each constraint (edge) is a pair of variables, constrained to be labeled by unequal colors.

We now make  formal definitions which encompass all of the above examples:
\begin{definition}
    A constraint satisfaction problem (CSP) over \emph{domain} $\Omega$
                                                    \index{domain (CSP)}%
    is defined by a finite set of \emph{predicates}
                                                    \index{predicates (CSP)}%
    (``types of constraints'') $\predset$, with each $\pred \in \predset$ being of the form $\pred \co \Omega^r \to \{0,1\}$ for some \emph{arity}~$r$ (possibly different for different predicates).  We say that the \emph{arity}
                                                    \index{arity (CSP)}%
    of the CSP is the maximum arity of its predicates.
\end{definition}
Such a CSP is associated with an algorithmic task called ``Max-$\CSP(\predset)$'',
                                                \index{Max-CSP@Max-$\CSP(\predset)$|(}%
which we will define below.  First, though, let us see how the CSPs from Example~\ref{eg:csps} fit into the above definition.
\begin{itemize}
    \item Max-$3$-Sat: Domain $\Omega = \{\True, \False\}$; $\predset$ contains $14$ predicates: the $8$ logical OR functions on $3$ literals (variables/negated-variables), the $4$ logical OR functions on $2$ literals, and the $2$ logical OR functions on $1$ literal.
                                                    \index{Max-$3$-Sat}%
    \item Max-Cut: Domain $\Omega = \{-1,1\}$; $\predset = \{\neq\}$, where $\neq \co \bits^2 \to \{0,1\}$ is the ``not-equal'' predicate.
                                                    \index{Max-Cut}%
    \item Max-E$3$-Lin: Domain $\Omega = \F_2$; $\predset$ contains the two $3$-ary predicates \linebreak $(x_1,x_2,x_3) \mapsto x_1+x_2+x_3$ and $(x_1,x_2,x_3) \mapsto x_1+x_2+x_3 + 1$.
                                                    \index{Max-$3$-Lin}%
    \item Max-$3$-Coloring: Domain $\Omega = \{\text{red}, \text{green}, \text{blue}\}$; $\predset$ contains just the single not-equal predicate $\neq \co \Omega^2 \to \{0,1\}$.
                                                    \index{Max-$3$-Coloring}%
\end{itemize}
\begin{remark}                  \label{rem:csp-terms}
    Let us add a few words about traditional CSP terminology.  \emph{Boolean} CSPs refer to the case $|\Omega| = 2$.  If $\pred \co \bits^r \to \{0,1\}$ is a Boolean predicate we sometimes write ``Max-$\pred$'' to refer to the CSP where all constraints are of the form $\pred$ applied to \emph{literals};
                                                        \index{Max-$\pred$}%
    i.e., $\Predset = \{\pred(\pm v_1, \dots, \pm v_r)\}$.
    As an example, Max-E$3$-Lin could also be called Max-$\chi_{[3]}$.  The ``E$3$'' in the name Max-E$3$-Lin refers to the fact that all constraints involve ``E''xactly~$3$ variables.  Thus e.g.\ Max-$3$-Lin is the generalization in which $1$- and $2$-variable equations are allowed. Conversely, Max-E$3$-Sat is the special case of Max-$3$-Sat where each clause must be of width exactly~$3$ (a CSP which could also be called Max-$\OR_3$).
\end{remark}

To formally define the algorithmic task Max-$\CSP(\Predset)$, we begin by defining its input:
\begin{definition}                      \label{def:csp-instance}
    An \emph{instance} (or \emph{input}) $\instance$  of
                                                        \index{instance (CSP)}%
    Max-$\CSP(\predset)$ over variable set~$V$ is a list (multiset) of \emph{constraints}.  Each constraint $C \in \instance$ is a pair $C = (S,\pred)$, where $\pred \in \Predset$ and where the \emph{scope} $S = (v^1, \dots, v^r)$ is a tuple of \emph{distinct} variables from~$V$, with $r$ being the arity of $\pred$.  We always assume that each $v \in V$ participates in at least one constraint scope. The \emph{size} of an instance is the number of bits required to represent it; writing $n = |V|$ and treating $|\Omega|$, $|\Predset|$ and the arity of~$\Predset$ as constants, the size is between $n$ and $O(|\instance| \log n)$.
\end{definition}
\begin{remark}
    Let's look at how the small details of Definition~\ref{def:csp-instance} affect input graphs for Max-Cut.  Since an instance is a multiset of constraints, this means we allow graphs with parallel edges.  Since each scope must consist of distinct variables, this means we disallow graphs with self-loops. Finally, since each variable must participate in at least one constraint, this means input graphs must have no isolated vertices (though they may be disconnected).
\end{remark}

Given an assignment of labels for the variables, we are interested in the number of constraints that are ``satisfied''.  The reason we explicitly allow duplicate constraints in an instance is that we may want some constraints to be more important than others.  In fact it's more convenient to normalize by looking at the \emph{fraction} of satisfied constraints, rather than the number.  Equivalently, we can choose a constraint $\bC \sim \instance$ uniformly at random and look at the \emph{probability} that it is satisfied.  It will actually be quite useful to think of a CSP instance~$\instance$ as a probability distribution on constraints.  (Indeed, we could have more generally defined \emph{weighted CSPs} in which the constraints are given arbitrary nonnegative weights summing to~$1$; however, we don't want to worry about the issue of representing, say, irrational weights with finitely many bits.)
\begin{definition}
    An \emph{assignment} (or \emph{labeling}) for instance $\instance$
                                              \index{assignment (CSP)}%
    of Max-$\CSP(\Predset)$ is just a mapping $\asgn \co V \to \Omega$.   For constraint $C = (S,\pred) \in \instance$ we say that $\asgn$ \emph{satisfies} $C$ if $\pred(\asgn(S)) = 1$.  Here we use shorthand notation: if $S = (v^1, \dots, v^r)$ then $\asgn(S)$ denotes $(\asgn(v^1), \dots, \asgn(v^r))$.  The \emph{value} of $\asgn$, denoted
                                                \index{value (CSP)}%
    $\Val_\instance(\asgn)$, is the fraction of constraints in~$\instance$ that~$\asgn$ satisfies:
    \begin{equation} \label{eqn:csp-value}
        \Val_{\instance}(\asgn) = \Ex_{(\bS, \bpred) \sim \instance}[\bpred(\asgn(\bS))] \in [0,1].
    \end{equation}
    The \emph{optimum} value of $\instance$
                                                        \index{optimum value (CSP)}%
    is
    \[
        \Opt(\instance) = \max_{\asgn \co V \to \Omega} \{\Val_{\instance}(\asgn)\}.
    \]
    If $\Opt(\instance) = 1$, we say that $\instance$ is
                                                        \index{satisfiable}%
    \emph{satisfiable}.
\end{definition}

\begin{remark}                      \label{rem:variable-vs-asgn}
    In the literature on CSPs there is sometimes an unfortunate blurring between a variable and its assignment.  For example, a Max-E$3$-Lin instance may be written as
    \begin{align*}
        x_1 + x_2 + x_3 &= 0 \\
        x_1 + x_5 + x_6 &= 0 \\
        x_3 + x_4 + x_6 &= 1;
    \end{align*}
    then a particular assignment $x_1 = 0, x_2 = 1, x_3 = 0, x_4 = 1, x_5 = 1, x_6 = 1$ may be given.  Now there is confusion: Does $x_2$ represent the name of a variable or does it represent~$1$?  Because of this we prefer to display CSP instances with the name of the assignment~$\asgn$ present in the constraints.  That is, the above instance would be described as finding $\asgn \co \{x_1, \dots, x_6\} \to \F_2$ so as to satisfy as many as possible of the following:
    \begin{align*}
        \asgn(x_1) + \asgn(x_2) + \asgn(x_3) &= 0 \\
        \asgn(x_1) + \asgn(x_5) + \asgn(x_6) &= 0 \\
        \asgn(x_3) + \asgn(x_4) + \asgn(x_6) &= 1,
    \end{align*}
\end{remark}

Finally, we define the algorithmic task associated with a CSP:
\begin{definition}
    The algorithmic task Max-$\CSP(\predset)$ is defined as follows:
    The input is an instance $\instance$.  The goal is to output an assignment~$\asgn$ with as large a value as possible.
\end{definition}
                                                \index{Max-CSP@Max-$\CSP(\predset)$|)}%

\bigskip

                                                \index{CSP!equivalence with testing|(}%
Having defined CSPs, let us make a connection to the notion of a string testing algorithm from the previous section.  The connection is this: \emph{CSPs and string testing algorithms are the same object}.  Indeed, consider a CSP instance~$\instance$ over domain $\Omega$ with $n$ variables~$V$.   Fix an assignment $\asgn \co V \to \Omega$; we can also think of $\asgn$ as a string in $\Omega^n$ (under some ordering of~$V$).  Now think of a testing algorithm which chooses a constraint $(\bS, \bpred) \sim \instance$ at random, ``queries'' the string entry $\asgn(v)$ for each $v \in \bS$, and accepts if and only if the predicate $\bpred(\asgn(\bS))$ is satisfied.  This is indeed an $r$-query string testing algorithm, where $r$ is the arity of the CSP; the probability the tester accepts is precisely $\Val_{\instance}(\asgn)$.

Conversely, let $T$ be some  randomized testing algorithm for strings in $\Omega^n$. Assume for simplicity that $T$'s randomness comes from the uniform distribution over some sample space~$U$. Now suppose we enumerate all outcomes in~$U$, and for each we write the tuple of indices~$S$ that $T$ queries and the predicate $\pred \co \Omega^{|S|} \to \{0,1\}$ that~$T$ uses to make its subsequent accept/reject decision.  Then this list of scope/predicates pairs is precisely an instance of an $n$-variable CSP over $\Omega$.  The arity of the CSP is equal to the (maximum) number of queries that~$T$ makes and the predicates for the CSP are precisely those used by the tester in making its accept/reject decisions.  Again, the probability that $T$ accepts a string $F \in \Omega^n$ is equal to the value of~$\asgn$ as an assignment for the CSP.  (Our actual definition of string testers allowed any form of randomness, including, say, irrational probabilities; thus technically not every string tester can be viewed as a CSP.  However, it does little harm to ignore this technicality.)

In particular, this equivalence between string testers and CSPs lets us properly define ``outputting the description of a PCPP system'' as in Definition~\ref{def:pcpp-reduction} of PCPP reductions.
\begin{examples}
    The PCPP system for $\calO = \{w \in \F_2 : w_1 + \cdots + w_n = 1\}$ given in Example~\ref{eg:pcpp-linearity} can be thought of as an instance of the Max-$3$-Lin CSP over the $2n-1$ variables $\{w_1, \dots, w_n, \Pi_1, \dots, \Pi_{n-1}\}$.  The BLR linearity test for functions $\F_2^n \to \F_2$ can also be thought of as instance of Max-$3$-Lin over~$2^n$ variables (recall that function testers are string testers).  In this case we identify the variable set with $\F_2^n$; if $n = 2$ then the variables are named $(0,0)$, $(0,1)$, $(1,0)$, and $(1,1)$; and, if we write $F \co \F_2^2 \to \F_2$ for the assignment, the instance is

    {\tiny
    \begin{align*}
    F(0,0) + F(0,0) + F(0,0) &= 0 & F(0,1) + F(0,0) + F(0,1) &= 0 & F(1,0) + F(0,0) + F(1,0) &= 0 & F(1,1) + F(0,0) + F(1,1) &= 0  \\
    F(0,0) + F(0,1) + F(0,1) &= 0 & F(0,1) + F(0,1) + F(0,0) &= 0 & F(1,0) + F(0,1) + F(1,1) &= 0 & F(1,1) + F(0,1) + F(1,0) &= 0  \\
    F(0,0) + F(1,0) + F(1,0) &= 0 & F(0,1) + F(1,0) + F(1,1) &= 0 & F(1,0) + F(1,0) + F(0,0) &= 0 & F(1,1) + F(1,0) + F(0,1) &= 0  \\
    F(0,0) + F(1,1) + F(1,1) &= 0 & F(0,1) + F(1,1) + F(1,0) &= 0 & F(1,0) + F(1,1) + F(0,1) &= 0 & F(1,1) + F(1,1) + F(0,0) &= 0.
    \end{align*}
    }
    Cf.\ Remark~\ref{rem:variable-vs-asgn}; also, note the duplicate constraints.
\end{examples}
                                                \index{CSP!equivalence with testing|)}%

We end this section by discussing the computational complexity of finding high-value assignments for a given CSP -- equivalently, finding strings that make a given string tester accept with high probability.  Consider, for example, the task  of Max-Cut on $n$-vertex graphs.  Of course, given a Max-Cut instance one can always find the optimal solution in time roughly $2^n$, just by trying all possible cuts.  Unfortunately, this is not very efficient, even for slightly large values of~$n$.  In computational complexity theory, an algorithm is generally deemed ``efficient'' if it runs in time $\poly(n)$.  For some subfamilies of graphs there are $\poly(n)$-time algorithms for finding the maximum cut, e.g., bipartite graphs (Exercise~\ref{ex:bipartiteness-is-easy}) or planar graphs.  However, it seems very unlikely that there is a $\poly(n)$-time algorithm that is guaranteed to find an optimal Max-Cut assignment given any input graph.  This statement is formalized by a basic theorem from the field of computational complexity:
\begin{theorem}                                     \label{thm:max-cut-hard}
    The task of finding the maximum cut in a given input graph is ``$\NP$-hard''.
                                                    \index{NP-hard@$\NP$-hard}%
\end{theorem}
We will not formally define $\NP$-hardness in this book (though see Exercise~\ref{ex:cook} for some more explanation). Roughly speaking it means ``at least as hard as the Circuit-Sat problem'',
                                                    \index{Circuit-Sat}%
where ``Circuit-Sat'' is the following task: Given an $n$-variable Boolean circuit~$C$, decide whether or not $C$ is satisfiable (i.e., there exists $w \in \zo^n$ such that $C(w) = 1$).  It is widely believed that Circuit-Sat does not have a polynomial-time algorithm (this is the ``$\PTIME \neq \NP$'' conjecture).  In fact it is also believed that Circuit-Sat does not have a~$2^{o(n)}$-time algorithm.

For essentially all CSPs, including Max-E$3$-Sat, Max-E$3$-Lin, and Max-$3$-Coloring, finding an optimal solution is $\NP$-hard.  This motivates considering a relaxed goal:
\begin{definition}                          \label{def:approx-alg}
    Let $0 \leq \alpha \leq \beta \leq 1$.  We say that algorithm~$A$ is an \emph{$(\alpha,\beta)$-approximation algorithm} for Max-$\CSP(\predset)$ (pronounced ``$\alpha$ out of~$\beta$ approximation'')
                                            \index{alpha-beta-approximation@$(\alpha,\beta)$-approximation algorithm}%
                                            \index{approximation algorithm|seeonly{$(\alpha,\beta)$-approximation algorithm}}%
    if it has the following guarantee: on any instance with optimum value at least~$\beta$, algorithm~$A$ outputs an assignment of value at least~$\alpha$. In case~$A$ is a randomized algorithm, we only require that its output has value at least~$\alpha$ in expectation.
\end{definition}
\noindent A mnemonic here is that when the $\beta$est assignment has value~$\beta$, the $\alpha$lgorithm gets value~$\alpha$.
\begin{examples}                    \label{eg:approx-examples}
    Consider the following algorithm for Max-E$3$-Lin: Given an instance, output either the assignment $\asgn \equiv 0$ or the assignment $\asgn \equiv 1$, whichever has higher value.  Since either $0$ or $1$ occurs on at least half of the instance's ``right-hand sides'', the output assignment will always have value at least~$\half$.  Thus this is an efficient $(\half, \beta)$-approximation algorithm for any~$\beta$.  In the case $\beta = 1$ one can do better:  performing Gaussian elimination is an efficient $(1,1)$-approximation algorithm for Max-E$3$-Lin (or indeed Max-$r$-Lin for any~$r$).
                                            \index{Max-$3$-Lin}%

    As a far more sophisticated example, Goemans and Williamson~\cite{GW95} showed that there is an efficient (randomized) algorithm which $(.878 \beta, \beta)$-approximates Max-Cut for every~$\beta$.
                                            \index{Max-Cut}%
                                            \index{Goemans--Williamson Algorithm}%
\end{examples}
Not only is finding the optimal solution of a Max-E$3$-Sat instance $\NP$-hard, it's even $\NP$-hard on \emph{satisfiable} instances. In other words:
\begin{theorem}                                     \label{thm:cook-theorem}
    $(1,1)$-approximating Max-E$3$Sat is $\NP$-hard.  The same is true of Max-$3$-Coloring.
\end{theorem}
On the other hand, it's easy to $(1,1)$-approximate Max-$3$-Lin (Example~\ref{eg:approx-examples}) or Max-Cut (Exercise~\ref{ex:bipartiteness-is-easy}).  Nevertheless, the ``textbook'' $\NP$-hardness results for these problems imply the following:
\begin{theorem}
    $(\beta,\beta)$-approximating Max-E$3$-Lin is $\NP$-hard for any fixed $\beta \in (\frac12, 1)$.  The same is true of Max-Cut.
\end{theorem}

In some ways, saying that $(1,1)$-distinguishing Max-E$3$-Sat is $\NP$-hard is not necessarily that disheartening.  For example, if $(1-\delta,1)$-approximating Max-E$3$-Sat were possible in polynomial time for every $\delta > 0$, you might consider that ``good enough''.  Unfortunately, such a state of affairs is very likely ruled out:
\begin{theorem}                                     \label{thm:3sat-approx-hard}
    There exists a positive universal constant $\delta_0 > 0$ such that $(1-\delta_0, 1)$-approximating Max-E$3$-Sat is $\NP$-hard.
                                                    \index{PCP Theorem}%
\end{theorem}
In fact, Theorem~\ref{thm:3sat-approx-hard} is \emph{equivalent} to the ``PCP~Theorem'' mentioned in Section~\ref{sec:pcpps}.  It follows straightforwardly from the PCPP Theorem, as we now sketch:
\begin{proof}[Proof sketch]
    Let $\delta_0$ be the rejection rate in the PCPP~Theorem.  We want to show that $(1-\delta_0, 1)$-approximating Max-E$3$-Sat is at least as hard as the Circuit-Sat problem. Equivalently, we want to show that if there is an efficient algorithm~$A$ for $(1-\delta_0, 1)$-approximating Max-E$3$-Sat then there is an efficient  algorithm $B$ for Circuit-Sat.  So suppose $A$ exists and let $C$ be a Boolean circuit given as input to~$B$.  Algorithm~$B$ first applies to~$C$ the PCPP reduction given by the PCPP~Theorem.   The output is some arity-$3$ CSP instance $\instance$ over variables $w_1, \dots, w_n, \Pi_1, \dots, \Pi_\ell$, where $\ell \leq \poly(\size(C))$.  By Exercise~\ref{ex:pcpps-have-3-queries} we may assume that $\instance$ is an instance of Max-E$3$-Sat.  From the definition of a PCPP system, it is easy to check (Exercise~\ref{ex:check-pcp-thm}) the following:
    If $C$ is satisfiable then $\Opt(\instance) = 1$; and, if $C$ is not satisfiable then $\Opt(\instance) < 1 - \delta_0$.  Algorithm~$B$ now runs the supposed $(1-\delta_0,1)$-approximation algorithm~$A$ on~$\instance$ and outputs ``$C$ is satisfiable'' if and only if $A$ finds an assignment of value at least $1-\delta_0$.
\end{proof}

\section{Highlight: H{\aa}stad's hardness theorems}                 \label{sec:hastad}

In Theorem~\ref{thm:3sat-approx-hard} we saw that it is $\NP$-hard to $(1-\delta_0, 1)$-approximate Max-E$3$Sat for some positive but inexplicit constant~$\delta_0$.  You might wonder how large $\delta_0$ can be.  The natural limit here is~$\frac18$ because there is a very simple algorithm that satisfies a $\frac78$-fraction of the constraints in any Max-E$3$Sat instance:
\begin{proposition}                                     \label{prop:78-for-3sat}
    Consider the Max-E$3$-Sat algorithm that outputs a uniformly random assignment~$\asgn$.  This is a $(\frac78,\beta)$-approximation for any~$\beta$.
                                                \index{Max-$3$-Sat}%
\end{proposition}
\begin{proof}
    In instance $\calP$, each constraint is a logical OR of exactly~$3$ literals and will therefore be satisfied by~$\asgn$ with probability exactly $\frac78$.  Hence in expectation the algorithm will satisfy a $\frac78$-fraction of the constraints.
\end{proof}
\noindent (It's also easy to ``derandomize'' this algorithm, giving a deterministic guarantee of at least $\frac78$ of the constraints; see Exercise~\ref{ex:condit-expec}.)

This algorithm is of course completely brainless -- it doesn't even ``look at'' the instance it is trying to approximately solve.  But rather remarkably, it achieves the best possible approximation guarantee among all efficient algorithms (assuming $\PTIME \neq \NP$).  This is a consequence of the following 1997 theorem of \Hastad~\cite{Has01}, improving significantly on Theorem~\ref{thm:3sat-approx-hard}:
\begin{named}{\Hastad's 3-Sat Hardness}
    For any constant $\delta > 0$, it is $\NP$-hard to $(\frac78 + \delta, 1)$-approximate Max-E$3$-Sat.
                                                \index{Max-$3$-Sat!H{\aa}stad's hardness for}%
\end{named}
\Hastad gave similarly optimal hardness-of-approximation results for several other problems, including Max-E$3$-Lin:
\begin{named}{\Hastad's 3-Lin Hardness}
    For any constant $\delta > 0$, it is $\NP$-hard to $(\frac12 + \delta, 1 - \delta)$-approximate Max-E$3$-Lin.
                                                \index{Max-$3$-Lin!H{\aa}stad's hardness for}%
\end{named}
In this hardness theorem, both the ``$\alpha$'' and ``$\beta$'' parameters are optimal; as we saw in Example~\ref{eg:approx-examples} one can efficiently $(\half, \beta)$-approximate and also $(1,1)$-approximate Max-E$3$-Lin.

The goal of this section is to sketch the proof of the above theorems, mainly \Hastad's 3-Lin Hardness Theorem.  Let's begin by considering the 3-Sat hardness result.  If our goal is to increase the inexplicit constant $\delta_0$ in Theorem~\ref{thm:3sat-approx-hard}, it makes sense to look at how the constant arises. From the proof of Theorem~\ref{thm:3sat-approx-hard} we see that it's just the rejection rate in the PCPP~Theorem.  We didn't prove that theorem, but let's consider its length-$2^{2^n}$ analogue, Theorem~\ref{thm:pcpp-builder0}.  The key ingredient in the proof of Theorem~\ref{thm:pcpp-builder0} is the dictator test.  Indeed, if we strip away the few local correcting and consistency checks, we see that the dictator test component controls both the rejection rate \emph{and} the type of predicates output by the PCPP reduction.  This observation suggests that to get a strong hardness-of-approximation result for, say, Max-E$3$-Lin, we should seek a local tester for dictatorship which (a)~has a large rejection rate, and (b)~makes its accept/reject decision using $3$-variable linear equation predicates.

This approach (which of course needs to be integrated with efficient ``PCPP technology'') was suggested in a 1995 paper of Bellare, Goldreich, and Sudan~\cite{BGS95}.  Using it, they managed to prove $\NP$-hardness of $(1-\delta_0, 1)$-approximating Max-E$3$-Sat with the explicit constant $\delta_0 = .026$.  \Hastad's key conceptual contribution (originally from~\cite{Has96}) was showing that given known PCPP technology, it suffices to construct a certain kind of \emph{relaxed} dictator test.  Roughly speaking, dictators should still be accepted with probability~$1$ (or close to~$1$), but only functions which are ``very unlike'' dictators need to be rejected with substantial probability.  Since this is a weaker requirement than in the standard definition of a local tester, we can potentially achieve a much higher rejection rate, and hence a much stronger hardness-of-approximation result.

For these purposes, the most useful formalization of being ``very unlike a dictator'' turns out to be ``having no notable coordinates'' in the sense of Definition~\ref{def:small-influences}.
                            \index{epsilon delta-small stable influences@$(\eps,\delta)$-small stable influences}%
                            \index{notable coordinates}%
We make the following definition which is appropriate for Boolean CSPs.
\begin{definition}                                      \label{def:dvq}
    Let $\predset$ be a finite set of predicates over the domain $\Omega = \bits$.  Let $0 < \alpha < \beta \leq 1$ and let $\rejrate \co [0,1] \to [0,1]$ satisfy $\rejrate(\eps) \to 0$ as $\eps \to 0$.   Suppose that for each $n \in \N^+$ there is a local tester for functions $f \co \bits^n \to \bits$ with the following properties:
    \begin{itemize}
        \item If $f$ is a dictator then the test accepts with probability at least~$\beta$.
        \item If $f$ has no $(\eps,\eps)$-notable coordinates -- i.e., $\Inf^{(1-\eps)}_i[f] \leq \eps$ for all $i \in [n]$ -- then the test accepts with probability at most $\alpha + \rejrate(\eps)$.
        \item The tester's accept/reject decision uses predicates from $\predset$; i.e., the tester can be viewed as an instance of Max-$\CSP(\predset)$.
    \end{itemize}
    Then, abusing terminology, we call this family of testers an \emph{$(\alpha,\beta)$-\dvq using predicate set $\predset$}.
                                   \index{Dictator-vs.-@\dvq}%
\end{definition}
\begin{remark}                                      \label{rem:dvq-int}
    For very minor technical reasons, the above definition should actually be slightly amended. In this section we freely ignore the amendments, but for the sake of correctness we state them here.  One is a strengthening, one is a weakening.
    \begin{itemize}
        \item The second condition should be required even for functions $f \btI$; what this means is explained in Exercise~\ref{ex:dvq-with-bounded-range}.
        \item When the tester makes accept/reject decisions by applying $\pred \in \predset$ to query results $f(\bx^{(1)}), \dots, f(\bx^{(r)})$, it is \emph{allowed} that the query strings are not all distinct. (See Exercise~\ref{ex:fix-hastad-bug}.)
    \end{itemize}
\end{remark}
\begin{remark}
    It's essential in this definition that the ``error term'' $\rejrate(\eps) = o_\eps(1)$ be independent of~$n$.  On the other hand, we otherwise care very little about the rate at which it tends to~$0$; this is why we didn't mind using the same parameter~$\eps$ in the ``$(\eps,\eps)$-notable'' hypothesis.
\end{remark}

Just as the dictator test was the key component in our PCPP reduction (Theorem~\ref{thm:pcpp-builder0}), \dvqs are the key to obtaining strong hardness-of-approximation results.  The following result (essentially proved in Khot et~al.~\cite{KKMO07}) lets you obtain hardness results from \dvqs in a black-box way:
\begin{theorem}                                     \label{thm:ug-hardness-from-dict-tests}
    Fix a CSP over domain $\Omega = \bits$ with predicate set $\predset$.
                                    \index{Unique-Games}%
                                    \index{UG-hardness}%
                                   \index{Dictator-vs.-@\dvq!connection with hardness}%
    Suppose there exists an $(\alpha,\beta)$-\dvq using predicate set $\predset$.  Then for all $\delta > 0$, it is ``\UG-hard'' to $(\alpha + \delta, \beta - \delta)$-approximate Max-$\CSP(\predset)$.
\end{theorem}
In other words, the distinguishing parameters of a \dvq automatically translate to the distinguishing parameters of a hardness result (up to an arbitrarily small~$\delta$).

The advantage of Theorem~\ref{thm:ug-hardness-from-dict-tests} is that it reduces a problem about computational complexity to a purely Fourier-analytic problem, and a constructive one at that.  The theorem has two disadvantages, however.  The first is that instead of $\NP$-hardness -- the gold standard in complexity theory -- it merely gives ``\UG-hardness'', which roughly means ``at least as hard as the Unique-Games problem''.  We leave the definition of the Unique-Games problem to Exercise~\ref{ex:ug-def}, but suffice it to say it's not as universally believed to be hard as Circuit-Sat is.  The second disadvantage of Theorem~\ref{thm:ug-hardness-from-dict-tests} is that it only has $\beta - \delta$ rather than $\beta$.  This can be a little disappointing, especially when you are interested in hardness for satisfiable instances ($\beta = 1$), as in \Hastad's 3-Sat Hardness.  In his work, \hastad showed that both disadvantages can be erased provided you construct something similar to, but more complicated than, an $(\alpha, \beta)$-\dvq.  This is how the \Hastad 3-Sat and 3-Lin Hardness Theorems are proved.  Describing this extra complication is beyond the scope of this book; therefore we content ourselves with the following theorems:
\begin{theorem}                                     \label{thm:3sat-dvq}
    For any $0 < \delta < \frac18$, there exists a $(\frac78 + \delta,1)$-\dvq which uses logical OR functions on $3$ literals as its predicates.
\end{theorem}
\begin{theorem}                                     \label{thm:3lin-dvq}
    For any $0 < \delta < \frac12$, there exists a $(\frac12, 1-\delta)$-\dvq using $3$-variable $\F_2$-linear equations as its predicates.
\end{theorem}
Theorem~\ref{thm:3lin-dvq} will be proved below, while the proof of Theorem~\ref{thm:3sat-dvq} is left for Exercise~\ref{ex:3sat-dvq}. By applying Theorem~\ref{thm:ug-hardness-from-dict-tests} we immediately deduce the following weakened versions of \Hastad's Hardness Theorems:
\begin{corollary}                                       \label{cor:weak-hastad-3sat}
    For any $\delta > 0$, it is \UG-hard to $(\frac78 + \delta, 1-\delta)$-approximate Max-E$3$-Sat.
\end{corollary}
\begin{corollary}                                       \label{cor:weak-hastad-3lin}
    For any $\delta > 0$, it is \UG-hard to $(\frac12+\delta, 1-\delta)$-approximate Max-E$3$-Lin.
\end{corollary}
\begin{remark}
    For Max-E$3$-Lin, we don't mind the fact that Theorem~\ref{thm:ug-hardness-from-dict-tests} has $\beta-\delta$ instead of $\beta$ because our \dvq only accepts dictators with probability~$1-\delta$ anyway.  Note that the $1-\delta$ in Theorem~\ref{thm:3lin-dvq} cannot be improved to~$1$; see Exercise~\ref{ex:no-perf-comp-dvq}.)
\end{remark}
                                                \index{CSP|)}%

                                   \index{Dictator-vs.-@\dvq!for Max-E$3$-Lin|(}%
                                   \index{Max-$3$-Lin|see{\dvq for Max-E$3$-Lin}}%
To prove a result like Theorem~\ref{thm:3lin-dvq} there are two components: the design of the test, and its analysis.  We begin with the design.  Since we are looking for a test using $3$-variable linear equation predicates, the BLR~Test naturally suggests itself; indeed, all of its checks are of the form $f(x)+f(y)+f(z) = 0$.  It also accepts dictators with probability~$1$.  Unfortunately it's not true that it accepts functions with no notable coordinates with probability close to~$\half$.
There are two problems: the constant~$0$ function and ``large'' parity functions are both accepted with probability~$1$, despite having no notable coordinates.  The constant~$1$ function is easy to deal with: we can replace the BLR~Test by the ``Odd BLR Test''.
\begin{named}{Odd BLR Test}
    Given query access to $f \co \F_2^n \to \F_2$:
    \begin{itemize}
        \item Choose $\bx \sim \F_2^n$ and $\by \sim \F_2^n$ independently.
        \item Choose $\bb \sim \F_2$ uniformly at random and set $\bz = \bx + \by + (\bb, \bb, \dots, \bb) \in \F_2^n$.
        \item Accept if $f(\bx) + f(\by) + f(\bz) = \bb$.
    \end{itemize}
\end{named}
Note that this test uses both kinds of $3$-variable linear equations as its predicates.  For the test's analysis, we as usual switch to $\pm 1$ notation and think of testing $f(\bx)f(\by)f(\bz) = \bb$.  It is easy to show the following (see the proof of Theorem~\ref{thm:3lin-dvq}, or Exercise~\ref{ex:test-any-subspace} for a generalization):
\begin{proposition}                                     \label{prop:odd-blr-analysis}
    The Odd BLR Test accepts $f \btb$ with probability
    \[
        \half + \half \sum_{\substack{S \subseteq [n] \\ |S| \text{ odd}}} \wh{f}(S)^3  \leq \half + \half \max_{\substack{S \subseteq [n] \\ |S| \text{ odd}}} \{\wh{f}(S)\}.
    \]
\end{proposition}
This twist rules out the constant~$1$ function; it passes the Odd BLR Test with probability~$\half$.  It remains to deal with large parity functions.  \Hastad's innovation here was to add a small amount of \emph{noise} to the Odd~BLR~Test.  Specifically, given a small $\delta > 0$ we replace $\bz$ in the above test with $\bz' \sim N_{1-\delta}(\bz)$; i.e., we flip each of its bits with probability $\delta/2$.  If $f$ is a dictator, then there is only a~$\delta/2$ chance this will affect the test.  On the other hand, if $f$ is a parity of large cardinality, the cumulative effect of the noise will destroy its chance of passing the linearity test.  Note that parities of small odd cardinality will also pass the test with probability close to~$1$; however, we don't need to worry about them since they have notable coordinates.  We can now present \Hastad's \dvq for Max-E$3$-Lin.

\begin{proof}[Proof of Theorem~\ref{thm:3lin-dvq}]
    Given a parameter $0 < \delta < 1$, define the following test, which uses Max-E$3$-Lin predicates:
    \begin{named}{H{\aa}stad$_{\bdelta}$ Test}
        Given query access to $f \btb$:
        \begin{itemize}
            \item Choose $\bx, \by \sim \bn$ uniformly and independently.
            \item Choose bit $\bb \sim \bits$ uniformly and set $\bz = \bb \cdot (\bx \circ \by) \in \bn$ (where $\circ$ denotes entry-wise multiplication).
            \item Choose $\bz' \sim N_{1-\delta}(\bz)$.
            \item Accept if $f(\bx)f(\by)f(\bz') = \bb$.
        \end{itemize}
    \end{named}
    We will show that this is a $(\frac12, 1-\delta/2)$-\dvq.  First, let us analyze the test assuming $\bb = 1$.
    \begin{align*}
        \Pr[\text{H{\aa}stad$_{\delta}$ Test accepts $f$} \mid \bb = 1] &= \Ex[\half + \half f(\bx)f(\by)f(\bz')] \\
                                         &= \half + \half \Ex[f(\bx)\cdot f(\by) \cdot \T_{1-\delta}f(\bx\circ\by)]]\\
                                         &= \half + \half \Ex_{\bx}[f(\bx) \cdot (f \conv \T_{1-\delta}f)(\bx)] \\
                                         &= \half + \half \sum_{S \subseteq [n]} \wh{f}(S) \cdot \wh{f \conv \T_{1-\delta} f}(S)\\
                                         &= \half + \half \sum_{S \subseteq [n]} (1-\delta)^{|S|}\wh{f}(S)^3.
    \end{align*}
    On the other hand, when $\bb = -1$ we take the expectation of $\half - \half f(\bx)f(\by)f(\bz')$ and note that $\bz'$ is distributed as $N_{-(1-\delta)}(\bx \circ \by)$.  Thus
    \[
        \Pr[\text{H{\aa}stad$_{\delta}$ Test accepts $f$} \mid \bb = -1] = \half - \half \sum_{S \subseteq [n]} (-1)^{|S|}(1-\delta)^{|S|}\wh{f}(S)^3.
    \]
    Averaging the above two results we deduce
    \begin{equation} \label{eqn:hastad-test-acc}
        \Pr[\text{H{\aa}stad$_{\delta}$ Test accepts $f$}] = \half + \half \sum_{|S| \text{ odd}} (1-\delta)^{|S|}\wh{f}(S)^3.
    \end{equation}
    (Incidentally, by taking $\delta = 0$ here we obtain the proof of Proposition~\ref{prop:odd-blr-analysis}.)

    From~\eqref{eqn:hastad-test-acc} we see that if $f$ is a dictator, $f = \chi_S$ with $|S| = 1$, then it is accepted with probability~$1-\delta/2$.  (It's also easy to see this directly from the definition of the test.)  To complete the proof that we have a $(\half, 1-\delta/2)$-\dvq, we need to bound the probability that $f$ is accepted given that it has $(\eps,\eps)$-small stable influences.  More precisely, assuming
    \begin{equation} \label{eqn:hast-assump}
        \phantom{\quad \text{for all $i \in [n]$}} \Inf_i^{(1-\eps)}[f] = \sum_{S \ni i} (1-\eps)^{|S|-1} \wh{f}(S)^2 \leq \eps \quad \text{for all $i \in [n]$}
    \end{equation}
    we will show that
    \begin{equation} \label{eqn:hast-concl}
      \Pr[\text{H{\aa}stad$_{\delta}$ Test accepts $f$}] \leq \half + \half\sqrt{\eps}, \quad \text{provided } \eps \leq \delta.
    \end{equation}
    This is sufficient because we can take $\rejrate(\eps)$ in Definition~\ref{def:dvq} to be
    \[
        \rejrate(\eps) = \begin{cases} \half\sqrt{\eps} & \text{for $\eps \leq \delta$,} \\ \half & \text{for $\eps > \delta$.} \end{cases}
    \]
    Now to obtain~\eqref{eqn:hast-concl}, we continue from~\eqref{eqn:hastad-test-acc}:
    \begin{align*}
        \Pr[\text{H{\aa}stad$_{\delta}$ Test accepts $f$}] &\leq \half + \half \max_{|S| \text{ odd}}\{(1-\delta)^{|S|}\wh{f}(S)\} \cdot \sum_{|S| \text{ odd}} \wh{f}(S)^2 \\
        &\leq \half+ \half  \max_{|S| \text{ odd}}\{(1-\delta)^{|S|}\wh{f}(S)\} \\
        &\leq \half+ \half  \sqrt{\max_{|S| \text{ odd}}\{(1-\delta)^{2|S|}\wh{f}(S)^2\}} \\
        &\leq \half+ \half  \sqrt{\max_{|S| \text{ odd}}\{(1-\delta)^{|S|-1}\wh{f}(S)^2\}} \\
        &\leq \half+ \half  \sqrt{\max_{i \in [n]}\{\Inf_i^{(1-\delta)}[f]\}},
    \end{align*}
    where we used that $|S|$ odd implies $S$ nonempty.  And the above is indeed at most $\half + \half \sqrt{\eps}$ provided $\eps \leq \delta$, by~\eqref{eqn:hast-assump}.
\end{proof}
                                   \index{Dictator-vs.-@\dvq!for Max-E$3$-Lin|)}%

\section{Exercises and notes}   \label{sec:testing-notes}
\begin{exercises}
    \item \label{ex:local-test-to-standard} Suppose there is an $\rquers$-query local tester for property~$\calC$ with rejection rate~$\rejrate$. Show that there is a testing algorithm that, given inputs $0 < \eps, \delta \leq 1/2$, makes $O(\frac{\rquers \log(1/\delta)}{\rejrate \eps})$ (nonadaptive) queries to~$f$ and satisfies the following:
        \begin{itemize}
            \item If $f \in \calC$, then the tester accepts with probability~$1$.
            \item If $f$ is $\eps$-far from $\calC$, then the tester accepts with probability at most~$\delta$.
        \end{itemize}
    \item  \label{ex:match-test}  Let $\calM = \{(x, y) \in \zo^{2n}: x = y\}$, the property that a string's first half matches its second half.  Give a $2$-query local tester for $\calM$ with rejection rate~$1$.  (Hint: Locally test that $x \oplus y = (0, 0, \dots, 0)$.)
    \item Reduce the proof length in Example~\ref{eg:pcpp-linearity} to $n-2$.
    \item \label{ex:equality-string-test}  Verify the claim from Example~\ref{eg:string-local-testers} regarding the $2$-query tester for the property that a string has all its coordinates equal.  (Hint: Use $\pm 1$ notation.)
    \item \label{ex:odd-1s-untestable}  Let $\calO = \{w \in \F_2^n : \text{$w$ has an odd number of $1$'s}\}$. Let $T$ be any $(n-1)$-query string testing algorithm that accepts every $w \in \calO$ with probability~$1$.  Show that $T$ in fact accepts \emph{every} string~$v \in \F_2^n$ with probability~$1$ (even though $\dist(w,\calO) = \frac{1}{n} > 0$ for half of all strings~$w$).  Thus locally testing~$\calO$ requires~$n$ queries.
    \item \label{ex:no-2-query}  Let $T$ be a $2$-query testing algorithm for functions $\bn \to \bits$.   Suppose that $\calT$ accepts every dictator with probability~$1$.  Show that it also accepts $\Maj_{n'}$ with probability~$1$ for every odd $n' \leq n$.  This shows that there is no $2$-query local tester for dictatorship assuming $n > 2$.  (Hint: You'll need to enumerate all predicates on up to~$2$ bits.)
    \item  \label{ex:no-perf-comp-dvq}  For every $\alpha < 1$, show that there is no $(\alpha,1)$-\dvq using Max-E$3$-Lin predicates. (Hint: Consider large odd parities.)
    \item \label{ex:BGS-test}
        \begin{exercises}
            \item \label{ex:wedge-test} Consider the following $3$-query testing algorithm for $f \zotzo$.  Let $\bx, \by \sim \{0,1\}^n$ be independent and uniformly random, define $\bz \in \zo^n$ by $\bz_i = \bx_i \wedge \by_i$ for each $i \in [n]$, and accept if $f(\bx) \wedge f(\by) = f(\bz)$.  Let $p_k$ be the probability that this test accepts a parity function $\chi_S \zotzo$ with $|S| = k$. Show that $p_0 = p_1 = 1$ and that in general $p_k \leq \half + 2^{-|S|}$.  In fact, you might like to show that $p_k = \half + (\frac{3}{4} - \frac{1}{4}(-1)^k)2^{-k}$.  (Hint: It suffices to consider $k = n$ and then compute the correlation of $\chi_{\{1,\dots, n\}} \wedge \chi_{\{n+1, \dots, 2n\}}$ with the bent function $\IP_{2n}$.)
            \item Show how to obtain a $3$-query local tester for dictatorship by combining the following subtests: (i)~the Odd~BLR Test; (ii)~the test from part~\ref{ex:wedge-test}.
        \end{exercises}
    \item Obtain the largest explicit rejection rate in Theorem~\ref{thm:blr+nae2} that you can.  You might want to return to the Fourier expressions arising in  Theorem~\ref{thm:blr-test} and~\ref{thm:prob-condorcet}, as well as Exercise~\ref{ex:BLR}.  Can you improve your bound by doing the BLR and NAE~Tests with probabilities other than $1/2, 1/2$?
    \item \label{ex:distinguishing} \begin{exercises}
        \item  Say that $A$ is an \emph{$(\alpha,\beta)$-distinguishing} algorithm for Max-$\CSP(\predset)$ if it outputs `YES' on instances with value at least~$\beta$ and outputs `NO' on instances with value strictly less than~$\alpha$.
                        \index{alpha-beta-distinguishing@$(\alpha,\beta)$-distinguishing algorithm}%
            (On each instance with value in $[\alpha, \beta)$, algorithm~$A$ may have either output.)  Show that if there is an efficient $(\alpha,\beta)$-approximation algorithm for Max-$\CSP(\predset)$, then there is also an efficient $(\alpha,\beta)$-distinguishing algorithm for Max-$\CSP(\predset)$.
        \item Consider Max-$\CSP(\predset)$, where $\predset$ be a class of predicates that is closed under restrictions (to nonconstant functions); e.g., Max-$3$-Sat.  Show that if there is an efficient $(1,1)$-distinguishing algorithm, then there is also an efficient $(1,1)$-approximation algorithm. (Hint: Try out all labels for the first variable and use the distinguisher.)
        \end{exercises}
    \item \label{ex:3sat}
        \begin{exercises}
            \item Let $\phi$ be a CNF of size $s$ and width~$w \geq 3$ over variables $x_1, \dots, x_n$.  Show that there is an ``equivalent'' CNF $\phi'$ of size at most $(w-2)s$ and width~$3$ over the variables $x_1, \dots, x_n$ plus auxiliary variables $\Pi_1, \dots, \Pi_\ell$, with $\ell \leq (w-3)s$.  Here ``equivalent'' means that for every $x$ such that $\phi(x) = \True$ there exists $\Pi$ such that $\phi'(x,\Pi) = \True$; and, for every $x$ such that $\phi(x) = \False$ we have $\phi'(x,\Pi) = \false$ for all $\Pi$.
            \item \label{ex:e3sat} Extend the above so that every clause in $\phi'$ has width \emph{exactly} $3$ (the size may increase by $O(s)$).
        \end{exercises}
    \item \label{ex:pcpps-have-3-queries} Suppose there exists an $r$-query PCPP reduction $\calR_1$ with rejection rate~$\rejrate$.  Show that there exists a $3$-query PCPP reduction $\calR_2$ with rejection rate at least $\rejrate/(r2^r)$.  The proof length of $\calR_2$ should be  at most  $r2^r\cdot m$ plus the proof length of~$\calR_1$ (where $m$ is the description-size of $\calR_1$'s output) and the predicates output by the reduction should all be logical ORs applied to exactly three literals.  (Hint: Exercises~\ref{ex:triv-DNF},~\ref{ex:3sat}.)
    \item  \label{ex:cook}
        \begin{exercises}
            \item Give a polynomial-time algorithm $R$ that takes as input a general Boolean circuit~$C$ and outputs a width-$3$ CNF formula~$\phi$ with the following guarantee: $C$ is satisfiable if and only if $\phi$ is satisfiable.  (Hint: Introduce a variable for each gate in~$C$.)
            \item The previous exercise in fact formally justifies the following statement: ``$(1,1)$-distinguishing Max-$3$-Sat is $\NP$-hard''.
                                                    \index{NP-hard@$\NP$-hard}%
                                                    \index{Max-$3$-Sat}%
                (See Exercise~\ref{ex:distinguishing} for the definition of $(1,1)$-distinguishing.) Argue that, indeed, if $(1,1)$-distinguishing (or $(1,1)$-approximating) Max-$3$-Sat is in polynomial time, then so is Circuit-Sat.
            \item Prove Theorem~\ref{thm:cook-theorem}.  (Hint: Exercise~\ref{ex:3sat}\ref{ex:e3sat}.)
        \end{exercises}
    \item \label{ex:bipartiteness-is-easy} Describe an efficient $(1,1)$-approximation algorithm for Max-Cut.
    \item \label{ex:test-any-subspace} \begin{exercises}
            \item  Let $H$ be any subspace of $\F_2^n$ and let $\calH = \{\chi_\gamma \co \F_2^n \to \{-1,1\} \mid \gamma \in H^\perp\}$. Give a $3$-query local tester for $\calH$ with rejection rate~$1$.  (Hint: Similar to BLR, but with $\la \vphi_H \conv f, f \conv f \ra$.)
                                            \index{BLR (Blum--Luby--Rubinfeld) Test}
            \item Generalize to the case that $H$ is any affine subspace of $\F_2^n$.
           \end{exercises}
    \item \label{ex:pcpp-code}  Let $A$ be any affine subspace of $\F_2^n$.  Construct a $3$-query, length-$2^n$ PCPP system for~$A$ with rejection rate a positive universal constant.  (Hint: Given $w \in \F_2^n$, the tester should expect the proof $\Pi \in \bits^{2^n}$ to encode the truth table of~$\chi_w$.  Use Exercise~\ref{ex:test-any-subspace} and also a consistency check based on local correcting of $\Pi$ at $e_{\bi}$, where $\bi \in [n]$ is uniformly random.)
    \item \begin{exercises}
              \item Give a $3$-query, length-$O(n)$ PCPP system (with rejection rate a positive universal constant) for the class $\{w \in \F_2^n : \IP_{n}(w) = 1\}$,
                                                \index{inner product mod $2$ function}
                  where $\IP_{n}$ is the inner product mod $2$ function ($n$~even).
                \item Do the same for the complete quadratic function
                            \index{complete quadratic function}
                      $\CQ_n$ from Exercise~\ref{ex:compute-expansions}.  (Hint: Exercise~\ref{ex:general-circuit}.)
          \end{exercises}
    \item \label{ex:exponential-pcpp} In this exercise you will prove Theorem~\ref{thm:pcpp-builder1}.
        \begin{exercises}
            \item Let $D \in \F_2^{n \times n}$ be a nonzero matrix and suppose $\bx, \by \sim \F_2^n$ are uniformly random and independent.  Show that $\Pr[\by^\top D \bx \neq 0] \geq \frac14$.
            \item \label{ex:rank-one} Let $\gamma \in \F_2^n$ and $\Gamma \in \F_2^{n \times n}$. Suppose $\bx, \by \sim \F_2^n$ are uniformly random and independent.  Show that $\Pr[(\gamma^\top \bx)(\gamma^\top \by) = \Gamma \bullet (\bx \by^\top)]$ is $1$ if $\Gamma = \gamma \gamma^{\top}$ and is at most $\frac34$ otherwise.  Here we use the notation $B \bullet C = \sum_{i,j} B_{ij} C_{ij}$ for matrices $B, C \in \F_2^{n \times n}$.
            \item \label{ex:rank-one-cons} Suppose you are given query access to \emph{two} functions $\ell \co \F_2^n \to \F_2$ and $q \co \F_2^{n \times n} \to \F_2$.  Give a $4$-query testing algorithm with the following two properties (for some universal constant $\rejrate > 0$): (i)~if $\ell = \chi_\gamma$ and $q = \chi_{\gamma \gamma^\top}$ for some $\gamma \in \F_2^n$, the test accepts with probability~$1$; (ii)~for all $0 \leq \eps \leq 1$, if the test accepts with probability at least $1 - \gamma\cdot \eps$, then there exists some $\gamma \in \F_2^n$ such that $\ell$ is $\eps$-close to $\chi_\gamma$ and $q$ is $\eps$-close to~$\chi_{\gamma \gamma^\top}$.  (Hint: Apply the BLR~Test to $\ell$ and $q$, and use part~\ref{ex:rank-one} with local correcting on~$q$.)
            \item Let $L$ be a list of homogenous degree-$2$ polynomial equations over variables $w_1, \dots, w_n \in \F_2$. (Each equation is of the form $\sum_{i,j=1}^n c_{ij} w_i w_j = b$ for constants $b, c_{ij} \in \F_2$; we remark that $w_i^2 = w_i$.) Define the string property $\calL = \{w \in \F_2^n : w \text{ satisfies all equations in L}\}$. Give a $4$-query, length-$(2^n + 2^{n^2})$ PCPP system for $\calL$ (with rejection rate a positive universal constant).  (Hint: The tester should expect the truth table of $\chi_w$ and $\chi_{ww^\top}$.  You will need part~\ref{ex:rank-one-cons} as well as Exercise~\ref{ex:test-any-subspace} applied to~``$q$''.)
            \item Complete the proof of Theorem~\ref{thm:pcpp-builder1}. (Hints: given $w \in \zo^n$, the tester should expect a proof consisting of all gate values $\bar{w} \in \zo^{\size(C)}$ in $C$'s computation on~$w$, as well as truth tables of $\chi_{\bar{w}}$ and $\chi_{\bar{w}\bar{w}^\top}$.  Show that $\bar{w}$ being a valid computation of $C$ is encodable with a list of homogeneous degree-$2$ polynomial equations. Add a consistency check between~$w$ and~$\bar{w}$ using local correcting, and reduce the number of queries to~$3$ using Exercise~\ref{ex:pcpps-have-3-queries}.)
        \end{exercises}
    \item \label{ex:check-pcp-thm}  Verify the connection between $\Opt(\instance)$ and $C$'s satisfiability stated in the proof sketch of Theorem~\ref{thm:3sat-approx-hard}.  (Hint: Every string $w$ is $1$-far from the empty property.)
    \item \label{ex:rand-assignments} A \emph{randomized assignment}
                                                \index{randomized assignment}
            for an instance $\instance$ of a CSP over domain~$\Omega$ is a mapping $\bF$ that labels each variable in~$V$ with a \emph{probability distribution} over domain elements.  Given a constraint $(S,\pred)$ with $S = (v_1, \dots, v_r)$, we write $\pred(\bF(S)) \in [0,1]$ for the expected value of $\pred(\bF(v_1), \dots, \bF(v_r))$.  This is simply the probability that $\pred$ is satisfied when one actually draws from the domain-distributions assigned by~$\bF$.  Finally, we define the \emph{value} of $\bF$ to be $\val_\instance(\bF) = \Ex_{(\bS, \bpred) \sim \instance}[\bpred(\bF(\bS))]$.
          \begin{exercises}
               \item Suppose that $A$ is a deterministic algorithm that produces a randomized assignment of value~$\alpha$ on a given instance~$\instance$.  Show a simple modification to~$A$ that makes it a randomized algorithm that produces a (normal) assignment whose value is~$\alpha$ in expectation.  (Thus, in constructing approximation algorithms we may allow ourselves to output randomized assignments.)
               \item Let $A$ be the deterministic Max-E$3$-Sat algorithm that on every instance outputs the randomized assignment that assigns the uniform distribution on $\{0,1\}$ to each variable.  Show that this is a $(\frac78,\beta)$-approximation algorithm for any~$\beta$.  Show also that the same algorithm is a $(\frac12,\beta)$-approximation algorithm for Max-$3$-Lin.
               \item \label{ex:rand-asgn-bool} When the domain $\Omega$ is $\{-1,1\}$, we may model a randomized assignment as a function $f \co V \to [-1,1]$; here $f(v) = \mu$ is interpreted as the unique probability distribution on $\{-1,1\}$ which has mean~$\mu$.  Now given a constraint $(S,\pred)$ with $S = (v_1, \dots, v_r)$, show that the value of $f$ on this constraint is in fact $\pred(f(v_1), \dots, f(v_r))$, where we identify $\pred \co \bits^r \to \{0,1\}$ with its multilinear (Fourier) expansion.  (Hint: Exercise~\ref{ex:multilin-interp}.)
               \item Let $\predset$ be a collection of predicates over domain $\bits$, and define \linebreak $\nu = \min_{\pred \in \predset} \{\wh{\pred}(\emptyset)\}$.  Show that outputting the randomized assignment $f \equiv 0$ is an efficient $(\nu, \beta)$-approximation algorithm for Max-$\CSP(\predset)$.
          \end{exercises}
    \item  \label{ex:condit-expec} Let $\bF$ be a randomized assignment of value~$\alpha$ for CSP instance~$\instance$ (as in Exercise~\ref{ex:rand-assignments}).  Give an efficient deterministic algorithm that outputs a usual assignment~$F$ of value at least $\alpha$.  (Hint: Try all possible labelings for the first variable and compute the expected value that would be achieved if~$\bF$ were used for the remaining variables.  Pick the best label for the first variable and repeat.)
    \item \label{ex:dvq-with-bounded-range}  Given a local tester for functions $f \btb$, we can interpret it also as a tester for functions $f \btI$; simply view the tester as a CSP and view the acceptance probability as the value of $f$ when treated as a randomized assignment (as in  Exercise~\ref{ex:rand-assignments}\ref{ex:rand-asgn-bool}).  Equivalently, whenever the tester ``queries''~$f(x)$, imagine that what is returned is a random bit $\bb \in \bits$ whose mean is~$f(x)$.  This interpretation completes Definition~\ref{def:dvq} of \dvqs for functions $f \btI$ (see Remark~\ref{rem:dvq-int}).  Given this definition, verify that the H{\aa}stad$_\delta$~Test is indeed a $(\half, 1-\delta)$-\dvq.  (Hint: Show that~\eqref{eqn:hastad-test-acc} still holds for functions $f \btI$. There is only one subsequent inequality that uses that $f$'s range is $\bits$, and it still holds with range $[-1,1]$.)
    \item \label{ex:oddness-wlog}  Let $\predset$ be a finite set of predicates over domain $\Omega = \{-1,1\}$ that is closed under negating variables.
                                                    \index{folding}
        (An example is the scenario of Max-$\pred$ from Remark~\ref{rem:csp-terms}.)  In this exercise you will show that \dvqs using $\predset$ may assume $f \btI$ is odd without loss of generality.
        \begin{exercises}
            \item Let $T$ be an $(\alpha,\beta)$-\dvq using predicate set~$\predset$ that works under the assumption that $f \btI$ is odd.  Modify~$T$ as follows: Whenever it is about to query $f(x)$, with probability $\half$ let it use $f(x)$ and with probability $\half$ let it use $-f(-x)$.  Call the modified test~$T'$.  Show that the probability $T'$ accepts an arbitrary $f \btI$ is equal to the probability~$T$ accepts $f^\odd$ (recall Exercise~\ref{ex:odd-even}).
            \item Prove that $T'$ is an $(\alpha,\beta)$-\dvq using predicate set~$\predset$ for functions $f \btI$.
        \end{exercises}
    \item \label{ex:noisy-wlog}  This problem is similar to Exercise~\ref{ex:oddness-wlog} in that it shows you may assume that \dvqs are testing ``smoothed'' functions of the form $\T_{1-\delta} h$ for $h \btI$, so long as you are willing to lose~$O(\delta)$ in the probability that dictators are accepted.
        \begin{exercises}
            \item Let $U$ be an $(\alpha,\beta)$-\dvq using an arity-$r$ predicate set~$\predset$ (over domain $\bits$) which works under the assumption that the function $f \btI$ being tested is of the form $\T_{1-\delta} h$ for $h \btI$.  Modify~$U$ as follows: whenever it is about to query $f(x)$, let it draw $\by \sim N_{1-\delta}(x)$ and use $f(\by)$ instead.  Call the modified test~$U'$.  Show that the probability $U'$ accepts an arbitrary $h \btI$ is equal to the probability~$U$ accepts $\T_{1-\delta} h$.
            \item Prove that $U'$ is an $(\alpha,\beta - r\delta/2)$-\dvq using predicate set~$\predset$.
        \end{exercises}
    \item Give a slightly alternate proof of Theorem~\ref{thm:3lin-dvq} by using the original BLR~Test analysis and applying Exercises~\ref{ex:oddness-wlog},~\ref{ex:noisy-wlog}.
    \item Show that when using Theorem~\ref{thm:ug-hardness-from-dict-tests}, it suffices to have a ``Dictators-vs.-No-Influentials test'', meaning replacing $\Inf_i^{(1-\eps)}[f]$ in Definition~\ref{def:dvq} with just $\Inf_i[f]$.  (Hint: Exercise~\ref{ex:noisy-wlog}.)
    \item \label{ex:ug-def}  For $q \in \N^+$, \emph{Unique-Games($q$)} refers to the arity-$2$ CSP with domain $\Omega = [q]$ in which all $q!$ ``bijective'' predicates are allowed; here $\pred$ is ``bijective'' if there is a bijection $\pi \co [q] \to [q]$ such that $\pred(i,j) = 1$ iff $\pi(j) = i$.  Show that $(1,1)$-approximating Unique-Games($q$) can be done in polynomial time.
                                    \index{Unique-Games}%
        (The \emph{Unique Games Conjecture} of Khot~\cite{Kho02} states that for all $\delta > 0$ there exists $q \in \N^+$ such that $(\delta, 1-\delta)$-approximating Unique-Games($q$) is $\NP$-hard.)
    \item \label{ex:3-sat-to-3-lin} In this problem you will show that Corollary~\ref{cor:weak-hastad-3sat} actually follows directly from Corollary~\ref{cor:weak-hastad-3lin}.
        \begin{exercises}
            \item Consider the $\F_2$-linear equation $v_1 + v_2 + v_3 = 0$.  Exhibit a list of~$4$ clauses (i.e., logical ORs of literals) over the variables such that if the equation is satisfied, then so are all~$4$ clauses, but if the equation is not satisfied, then at most~$3$ of the clauses are.  Do the same for the equation $v_1 + v_2 + v_3 = 1$.
            \item Suppose that for every $\delta > 0$ there is an efficient algorithm for $(\frac78 + \delta, 1 - \delta)$-approximating Max-E$3$-Sat.  Give, for every $\delta > 0$, an efficient algorithm for $(\frac12 + \delta, 1 - \delta)$-approximating Max-E$3$-Lin.
            \item \label{ex:3-sat-to-3-lin-dvq} Alternatively, show how to transform any $(\alpha,\beta)$-\dvq using Max-E$3$-Lin predicates into a $(\frac34 + \frac14 \alpha, \beta)$-\dvq using Max-E$3$-Sat predicates.
        \end{exercises}
    \item \label{ex:3sat-dvq} In this exercise you will prove Theorem~\ref{thm:3sat-dvq}.
        \begin{exercises}
            \item Recall the predicate $\OXR$ from Exercise~\ref{ex:compute-expansions}.  Fix a small $0 < \delta < 1$.  The remainder of the exercise will be devoted to constructing a $(\frac34 + \delta/4,1)$-\dvq using Max-$\OXR$ predicates. Show how to convert this to a $(\frac78 + \delta/8,1)$-\dvq using Max-E$3$-Sat predicates. (Hint: Similar to Exercise~\ref{ex:3-sat-to-3-lin}\ref{ex:3-sat-to-3-lin-dvq}.)
                                                    \index{OXR function}%
            \item By Exercise~\ref{ex:oddness-wlog}, it suffices to construct a $(\frac34 + \delta/4,1)$-\dvq using the $\OXR$ predicate assuming $f \btI$ is odd.  \Hastad tests $\OXR(f(\bx), f(\by), f(\bz))$ where $\bx, \by, \bz \in \bn$ are chosen randomly as follows: For each $i \in [n]$ (independently), with probability $1-\delta$ choose $(\bx_i, \by_i, \bz_i)$ uniformly subject to $\bx_i\by_i\bz_i = -1$, and with probability $\delta$ choose $(\bx_i, \by_i, \bz_i)$ uniformly subject to $\by_i\bz_i = -1$.  Show that the probability this test accepts an odd $f \btI$ is
                \begin{equation} \label{eqn:oxr-bound}
                    \tfrac34 - \tfrac14 \Stab_{-\delta}[f] - \tfrac14\sum_{S \subseteq [n]} \wh{f}(S)^2 \E_{\bJ \subseteq_{1-\delta} S}[(-1)^{|\bJ|}\wh{f}(\bJ)],
                \end{equation}
                where $\bJ \subseteq_{1-\delta} S$ denotes that $\bJ$ is a $(1-\delta)$-random subset of~$S$ in the sense of Definition~\ref{def:delta-random-subset}.  In particular, show that dictators are accepted with probability~$1$.
            \item Upper-bound~\eqref{eqn:oxr-bound} by
                \[
                    \tfrac34 + \delta/4  + \tfrac14 \sqrt{(1-\delta)^t} + \tfrac14 \sum_{|S| \leq t} \wh{f}(S)^2 \E_{\bJ \subseteq_{1-\delta} S}[|\wh{f}(\bJ)|],
                \]
                or something stronger.  (Hint: Cauchy--Schwarz.)
            \item Complete the proof that this is a $(\frac34 + \delta/4, 1)$-\dvq, assuming $f$ is odd.
        \end{exercises}
    \item \label{ex:kkmo-reduction} In this exercise you will prove Theorem~\ref{thm:ug-hardness-from-dict-tests}.
                                    \index{UG-hardness}%
          Assume there exists an $(\alpha,\beta)$-\dvq $T$ using predicate set $\predset$ over domain $\bits$.  We define a certain efficient algorithm~$R$, which takes as input an instance $\calG$ of Unique-Games($q$) and outputs an instance $\instance$ of Max-$\CSP(\predset)$. For simplicity we refer to the variables~$V$ of the Unique-Games instance~$\calG$ as ``vertices'' and its constraints as ``edges''.  We also assume that when~$\calG$ is viewed as an undirected graph, it is regular.  (By a result of Khot--Regev~\cite{KR08} this assumption is without loss of generality for the purposes of the Unique Games Conjecture.)  The Max-$\CSP(\predset)$ instance $\instance$ output by algorithm~$R$ will have variable set $V \times \bits^{q}$, and we write assignments for it as collections of functions $(f_v)_{v \in V}$, where $f \co \bits^q \to \bits$.  The draw of a random of constraint for $\instance$ is defined as follows:
          \begin{itemize}
              \item Choose $\bu \in V$ uniformly at random.
              \item Draw a random constraint from the test~$T$; call it $\bpred(f(\bx^{(1)}), \dots, f(\bx^{(\br)}))$.
              \item Choose $\br$ random ``neighbors'' $\bv_1, \dots, \bv_{\br}$ of $\bu$ in $\calG$, independently and uniformly.  (By a neighbor of $\bu$, we mean a vertex $v$ such that either $(\bu,v)$ or $(v,\bu)$ is the scope of a constraint in~$\calG$.)  Since $\calG$'s constraints are bijective, we may assume that the associated scopes are $(\bu, \bv_1), \dots, (\bu, \bv_{\br})$ with bijections $\bpi_{1}, \dots, \bpi_{\br} \co [q] \to [q]$.
              \item Output the constraint $\bpred(f_{\bv_1}^{\bpi_1}(\bx^{(1)}), \dots, \bpred(f_{\bv_{\br}}^{\bpi_{\br}}(\bx^{(\br)}))$, where we use the permutation notation $f^\pi$ from Exercise~\ref{ex:golomb}.
          \end{itemize}
          \begin{exercises}
              \item Suppose $\Opt(\calG) \geq 1-\delta$.  Show that there is an assignment for $\instance$ with value at least $\beta - O(\delta)$ in which each $f_v$ is a dictator.  (You will use regularity of $\calG$ here.)  Thus $\Opt(\instance) \geq \beta - O(\delta)$.
              \item \label{ex:ug-nbr} Given an assignment $F = (f_v)_{v \in V}$ for $\instance$, introduce for each $u \in V$ the function $g_u \co \bits^q \to [-1,1]$ defined by $g(x) = \Ex_{\bv}[f_{\bv}^{\bpi}(x)]$, where $\bv$ is a random neighbor of $\bu$ in $\calG$ and $\bpi$ is the associated constraint's permutation. Show that $\val_{\instance}(F) = \E_{\bu \in V}[\val_{T}(g_{\bu})]$ (using the definition from Exercise~\ref{ex:dvq-with-bounded-range}).
              \item Fix an $\eps > 0$ and suppose that $\val_{\instance}(F) \geq s + 2\rejrate(\eps)$, where $\rejrate$ is the ``rejection rate'' associated with~$T$.  Show that for at least a $\rejrate(\eps)$-fraction of vertices $u \in V$, the set $\text{NbrNotable}_u = \{i \in [q] : \Inf_i^{(1-\eps)}[g_u] > \eps\}$ is nonempty.
              \item Show that for any $u \in V$, $i \in [q]$ we have $\E[\Inf^{(1-\eps)}_{\bpi^{-1}(i)}[f_{\bv}]] \geq \Inf_{i}^{(1-\eps)}[g_u]$, where $\bv$ is a random neighbor of $u$ and $\bpi$ is the associated constraint's permutation. (Hint: Exercise~\ref{ex:convex-functionals}.)
              \item For $v \in V$, define also the set $\text{Notable}_u = \{i \in [q] : \Inf_i^{(1-\eps)}[f_v] \geq \eps/2\}$.  Show that if $i \in \text{NbrNotable}_u$, then $\Pr_{\bv}[\bpi^{-1}(i) \in \text{Notable}_{\bv}] \geq \eps/2$, where $\bv$ and $\bpi$ are as in the previous part.
              \item Show that for every $u \in V$ we have $|\text{Notable}_u \cup \text{NbrNotable}_u| \leq O(1/\eps^2)$.  (Hint: Proposition~\ref{prop:few-stable-influences}.)
              \item Consider the following randomized assignment for $\calG$ (see Exericse~\ref{ex:rand-assignments}): for each $u \in V$, give it the uniform distribution on $\text{Notable}_u \cup \text{NbrNotable}_u$ (if this set is nonempty; otherwise, give it an arbitrary labeling).  Show that this randomized assignment has value $\Omega(\rejrate(\eps) \eps^5)$.
              \item Conclude Theorem~\ref{thm:ug-hardness-from-dict-tests}, where ``UG-hard'' means ``$\NP$-hard assuming the Unique Games Conjecture''.
          \end{exercises}
    \item \label{ex:fix-hastad-bug} Technically, Exercise~\ref{ex:kkmo-reduction} has a small bug: Since a \dvq using predicate set $\predset$ is allowed to use duplicate query strings in its predicates (see Remark~\ref{rem:dvq-int}), the reduction in the previous exercise does not necessarily output instances of Max-$\CSP(\predset)$ because our definition of CSPs requires that each scope consist of distinct variables.  In this exercise you will correct this bug. Let $M \in \N^+$ and suppose we modify the algorithm~$R$ from Exercise~\ref{ex:kkmo-reduction} to a new algorithm $R'$, producing an instance $\instance'$ with variable set $V \times [M] \times \bits^{q}$.  We now think of assignments to $\instance'$ as $M$-tuples of functions $f_v^1, \dots, f_v^M$, one tuple for each $v \in V$.  Further, thinking of $\instance$ as a function tester, we have $\instance'$ act as follows: Whenever $\instance$ is about to query $f_v(x)$, we have $\instance'$ instead query $f_v^{\bj}(x)$ for a uniformly random $\bj \in [M]$.
        \begin{exercises}
            \item Show that $\Opt(\instance) = \Opt(\instance')$.
            \item Show that if we delete all constraints in $\instance'$ for which the scope contains duplicates, then $\Opt(\instance')$ changes by at most $r^2/M$, where $r$ is the maximum arity of a constraint in~$\predset$.
            \item Show that the deleted version of $\instance'$ is a genuine instance of Max-$\CSP(\predset)$.  Since the constant $r^2/M$ can be arbitrarily small, this corrects the bug in Exercise~\ref{ex:kkmo-reduction}'s proof of Theorem~\ref{thm:ug-hardness-from-dict-tests}.
        \end{exercises}
\end{exercises}

\subsection*{Notes.}

The study of property testing was initiated by Rubinfeld and Sudan~\cite{RS96} and significantly expanded by Goldreich, Goldwasser, and Ron \cite{GGR98}; the stricter notion of local testability was introduced (in the context of error-correcting codes) by Friedl and Sudan~\cite{FS95}.  The first local tester for dictatorship was given by Bellare, Goldreich, and Sudan~\cite{BGS95,BGS98} (as in Exercise~\ref{ex:BGS-test}); it was later rediscovered by Parnas, Ron, and Samorodnitsky~\cite{PRS01,PRS02}.  The relevance of Arrow's Theorem to testing dictatorship was pointed out by Kalai~\cite{Kal02}.

The idea of assisting testers by providing proofs grew out of complexity-theoretic research on interactive proofs and PCPs; see the early work Erg{\"u}n, Kumar, and Rubinfeld~\cite{EKR99} and the references therein.  The specific definition of PCPPs was introduced independently by Ben-Sasson, Goldreich, Harsha, Sudan, and Vadhan~\cite{BGH+04} and by Dinur and Reingold~\cite{DR04} in 2004.  Both of these works obtained the PCPP~Theorem, relying on the fact that previous literature essentially already gave PCPP reductions of exponential (or greater) proof length: Ben-Sasson et~al.~\cite{BGH+04} observed that Theorem~\ref{thm:pcpp-builder1} can be obtained from Arora et.~al.~\cite{ALM+98} (their proof is Exercise~\ref{ex:exponential-pcpp}), while Dinur and Reingold~\cite{DR04} pointed out that the slightly easier Theorem~\ref{thm:pcpp-builder0} can be extracted from the work of Bellare, Goldreich, and Sudan~\cite{BGS98}. The proof we gave for Theorem~\ref{thm:pcpp-for-any-ppty} is inspired by the presentation in Dinur~\cite{Din07}.

The PCP~Theorem and its stronger forms (the PCPP~Theorem and Theorem~\ref{thm:pcpp-builder3}) have a somewhat remarkable consequence.  Suppose a researcher claims to prove a famous mathematical conjecture, say, ``$\PTIME \neq \NP$''.  To ensure maximum confidence in correctness, a journal might request the researcher submit a formalized proof, suitable for a mechanical proof-checking system.  If the submitted formalized proof~$w$ is a Boolean string of length~$n$, the proof-checker will be implementable by a circuit~$C$ of size~$O(n)$.  Notice that the string property~$\calC$ decided by~$C$ is nonempty if and only if there exists a (length-$n$) proof of $\PTIME \neq \NP$.
Suppose the journal applies Theorem~\ref{thm:pcpp-builder3} to~$C$ and requires the researcher submit the additional proof~$\Pi$ of length $n \cdot \polylog(n)$.  Now the journal can run a rather amazing testing algorithm, which reads just~$3$ bits of the submitted proof~$(w,\Pi)$. If the researcher's proof of $\PTIME \neq \NP$ is correct then the test will accept with probability~$1$.  On the other hand, if the test accepts with probability at least $1-\gamma$ (where $\gamma$ is the rejection rate in Theorem~\ref{thm:pcpp-builder3}), then~$w$ must be $1$-close to the set of strings accepted by~$C$.  This doesn't necessarily mean that~$w$ is a correct proof of $\PTIME \neq \NP$ -- but it does mean that $\calC$ is nonempty, and hence a correct proof of $\PTIME \neq \NP$ exists!  By querying a larger constant number of bits from~$(w,\Pi)$ as in Exercise~\ref{ex:local-test-to-standard}, say, $\lceil 30/\gamma\rceil$ bits, the journal can become $99.99$\% convinced that indeed $\PTIME \neq \NP$.

CSPs are very widely studied in computer science; it is impossible to survey the topic here.  In the case of Boolean CSPs various monographs~\cite{CKS01,KSTW01} contain useful background regarding complexity theory and approximation algorithms.  The notion of approximation algorithms and the derandomized $(\frac78,1)$-approximation algorithm for Max-E$3$-Sat (Proposition~\ref{prop:78-for-3sat}, Exercise~\ref{ex:condit-expec}) are due to Johnson~\cite{Joh74}.  Incidentally, there is also an efficient $(\frac78,1)$-approximation algorithm for Max-$3$-Sat~\cite{KZ97}, but both the algorithm and its analysis are extremely difficult, the latter requiring computer assistance~\cite{Zwi02}.

\Hastad's hardness theorems appeared in 2001~\cite{Has01}, building on earlier work~\cite{Has96,Has99}.  \Hastad~\cite{Has01} also proved $\NP$-hardness of $(\frac1p + \delta, 1-\delta)$-approximating Max-E$3$-Lin(mod~$p$) (for $p$ prime) and of $(\frac78, 1)$-approximating Max-$\CSP(\{\NAE_4\})$, both of which are optimal.  Using tools due to Trevisan et~al.~\cite{TSSW00}, \Hastad also showed $\NP$-hardness of $(\frac{11}{16} + \delta, \frac34)$-approximating Max-Cut, which is still the best known such result.  The best known inapproximability result for Unique-Games($q$) is
                                    \index{Unique-Games}%
$\NP$-hardness of $(\frac38 + q^{-\Theta(1)}, \half)$-approximation~\cite{OW12}.    Khot's influential Unique Games Conjecture dates from 2002~\cite{Kho02}; the peculiar name has its origins in a work of Feige and Lov{\'a}sz~\cite{FL92}. The generic Theorem~\ref{thm:ug-hardness-from-dict-tests}, giving \UG-hardness from \dvqs, is essentially from Khot et~al.~\cite{KKMO07}; the first explicit proof appearing in print may be due to Austrin~\cite{Aus08}.  (We remark that the terminology ``\dvq'' is not standard.) If one is willing to assume the Unique Games Conjecture, there is an almost-complete theory of optimal inapproximability due to Raghavendra~\cite{Rag09}. Many more inapproximability results, with and without the Unique Games Conjecture, are known; for some surveys, see those of Khot~\cite{Kho05,Kho10,Kho10a}.

As mentioned, Exercise~\ref{ex:BGS-test} is due to Bellare, Goldreich, and Sudan~\cite{BGS95} and to Parnas, Ron, and Samorodnitsky~\cite{PRS01}. The technique described in Exercise~\ref{ex:condit-expec} is known as the Method of Conditional Expectations.  The trick in Exercise~\ref{ex:oddness-wlog} is closely related to the notion of ``folding'' from the theory of PCPs.  The bug described in Exercise~\ref{ex:fix-hastad-bug} is rarely addressed in the literature; the trick used to overcome it appears in, e.g., Arora et~al.~\cite{ABH+05}. 

\chapter{Generalized domains}                                       \label{chap:generalized-domains}

                                                \index{product space domains|(}%
So far we have studied functions $f \co \{0,1\}^n \to \R$.  What about, say, $f \co \{0,1,2\}^n \to \R$?  In fact, very little of what we've done so far depends on the domain being~$\{0,1\}^n$; what it has mostly depended on is our viewing the domain as a \emph{product probability distribution}.  Indeed, much of analysis of Boolean functions carries over to the case of functions $f \co \Omega_1 \times \cdots \times \Omega_n \to \R$ where the domain has a product probability distribution $\pi_1 \otimes \cdots \otimes \pi_n$.
There are two main exceptions: the ``derivative'' operator $\D_i$ does not generalize to the case when $|\Omega_i| > 2$ (though the Laplacian operator $\Lap_i$ does), and the important notion of hypercontractivity (introduced in Chapter~\ref{chap:hypercontractivity}) depends strongly on the probability distributions~$\pi_i$.

In this chapter we focus on the case where all the $\Omega_i$'s are the same, as are the $\pi_i$'s. (This is just to save on notation; it will be clear that everything we do holds in the more general setting.)  Important classic cases include functions on the \emph{$p$-biased hypercube} (Section~\ref{sec:p-biased}) and functions on abelian groups (Section~\ref{sec:abelian}).  For the issue of generalizing the \emph{range} of functions -- e.g., studying functions $f \co \{0,1,2\}^n \to \{0,1,2\}$ -- see Exercise~\ref{ex:generalized-domain-range}.

\section{Fourier bases for product spaces}                                 \label{sec:product-spaces}

We will now begin to discuss functions on (finite) product probability spaces.
\begin{definition}                                 \label{def:general-inner-prod-space}
    Let $(\Omega, \pi)$ be a finite probability space with $|\Omega| \geq 2$ and assume $\pi$ has full support.
                                                    \index{L2@$L^2$}%
                                                    \index{product probability space}%
    For $n \in \N^+$ we write $L^2(\Omega^n, \pi^{\otimes n})$
                                                    \nomenclature[L2Omega]{$L^2(\Omega, \pi)$}{the inner product space of (square-integrable) functions $\Omega \to \R$ with inner product $\la f, g \ra = \Ex_{\bx \sim \pi}[f(\bx)g(\bx)]$}%
    for the (real) inner product space of functions $f \co \Omega^n \to \R$, with inner product
    \[
        \la f, g \ra = \Ex_{\bx \sim \pi^{\otimes n}}[f(\bx)g(\bx)].
    \]
    Here $\pi^{\otimes n}$ denotes the product probability distribution on $\Omega^n$.
                                            \nomenclature[pi x]{$\pi\xn$}{if $\pi$ is a probability distribution on $\Omega$, denotes the associated product probability distribution on $\Omega^n$}%
\end{definition}
\begin{example}  A simple example to keep in mind is $\Omega = \{a,b,c\}$ with $\pi(a) = \pi(b) = \pi(c) = 1/3$.   Here $a$, $b$, and $c$ are simply abstract set elements.
\end{example}
We can (and will) generalize to nondiscrete probability spaces, and to complex inner product spaces.  However, we will keep to the above definition for now.
\begin{notation}
    We will write $\unif$ for the uniform probability distribution on $\{-1,1\}$.
                                        \nomenclature[pi12]{$\unif$}{the uniform distribution on $\bits$}%
    Thus so far in this book we have been studying functions in $L^2(\bn, \unif^{\otimes n})$.  For simplicity, we will write this as $L^2(\bn)$.
                                        \nomenclature[L2-1]{$L^2(\bn)$}{denotes $L^2(\bn, \unif\xn)$}
\end{notation}
\begin{notation}
    Much of the notation we used for $L^2(\bn)$ extends naturally to the case of $L^2(\Omega^n, \pi\xn)$: e.g., $\|f\|_p = \Ex_{\bx \sim \pi\xn}[|f(\bx)|^p]^{1/p}$, or the restriction notation from Chapter~\ref{sec:restrictions}.
\end{notation}

As we described in Chapter~\ref{sec:basic-fourier-formulas}, the essence of Boolean Fourier analysis is in deriving combinatorial properties of a Boolean function $f \btR$ from its coefficients over a particular basis of~$L^2(\bn)$, the basis of parity functions.  We would like to achieve the same thing more generally for functions in $L^2(\Omega^n, \pi\xn)$.  We begin by considering vector space bases more generally.
\begin{definition}
    Let $|\Omega| = m$.  The \emph{indicator basis} (or \emph{standard basis}) for
                                        \index{indicator basis}%
    $L^2(\Omega, \pi)$ is just the set of $m$ indicator functions $(\indic{x})_{x \in \Omega}$,
    where
    \[
        \indic{x}(y) = \begin{cases} 1 & \text{if } y = x, \\ 0 & \text{if } y \neq x. \end{cases}
    \]
\end{definition}
\begin{fact}                                        \label{fact:l2-indic-basis}
    The indicator basis is indeed a basis for $L^2(\Omega,\pi)$ since the functions $(\indic{x})_{x \in \Omega}$ are nonzero, spanning, and orthogonal.  Hence $\dim(L^2(\Omega,\pi)) = m$.
\end{fact}

We will usually fix $\Omega$ and $\pi$ and then consider $L^2(\Omega^n, \pi\xn)$ for $n \in \N^+$.  Applying the above definition gives us an indicator basis $(\indic{x})_{x \in \Omega^n}$ for the $m^n$-dimensional space $L^2(\Omega^n, \pi\xn)$.  The representation of $f \in L^2(\Omega,\pi)$ in this basis is just $f = \sum_{x \in \Omega} f(x)\indic{x}$.  This is not very interesting; the coefficients are just the values of~$f$ so they don't tell us anything new about the function.  We would like a different basis that will generate useful ``Fourier formulas'' as in Chapter~\ref{sec:basic-fourier-formulas}.

For inspiration, let's look critically at the familiar case of $L^2(\bn)$. Here we used the basis of all parity functions, $\chi_S(x) = \prod_{i\in S} x_i$. It will be  helpful to think of the basis function $\chi_S \co \bn \to \R$ as follows: Identify~$S$ with its $0$-$1$ indicator vector and write
\begin{gather*}
    \chi_S(x) = \prod_{i=1}^n \phi_{S_i}(x_i), \qquad \text{where} \quad \phi_0 \equiv 1, \quad \phi_1 = \id.
\end{gather*}
(Here $\id$ is just the identity map $\id(b) = b$.) We will identify three properties of this basis which we'd like to generalize.

First, the parity basis is a \emph{product basis}.
                                                    \index{product basis}%
We can break down its ``product structure'' as follows: For each coordinate~$i \in [n]$ of the product domain $\bn$, the set $\{1, \id\}$ is a basis for the $2$-dimensional space $L^2(\bits, \unif)$.  We then get a basis for the $2^n$-dimensional product space $L^2(\bn)$ by taking all possible $n$-fold products.  More generally, suppose we are given an inner product space $L^2(\Omega, \pi)$ with $|\Omega| = m$.  Let $\phi_0, \dots, \phi_{m-1}$ be any  basis for this space.  Then the set of all products $\phi_{i_1} \phi_{i_2} \cdots \phi_{i_n}$ ($0 \leq i_j < m$) forms a basis for the space $L^2(\Omega^n, \pi\xn)$.

Second, it is convenient that the parity basis is \emph{orthonormal}.
                                                \index{orthonormal}%
We will later check that if a basis $\phi_0, \dots, \phi_{m-1}$ for $L^2(\Omega, \pi)$ is orthonormal, then so too is the associated product basis for $L^2(\Omega^n, \pi\xn)$. This relies on the fact that~$\pi\xn$ is the product distribution. For example, the parity basis for $L^2(\bn)$ is orthonormal because the basis $\{1, \id\}$ for $L^2(\bits, \unif)$ is orthonormal: $\E[1^2] = \Ex[\bx_i^2] = 1$, $\E[1 \cdot \bx_i] = 0$. Orthonormality is the property that makes Parseval's Theorem hold; in the general context, this means that if $f \in L^2(\Omega, \pi)$ has the representation $\sum_{i=0}^{m-1} c_i \phi_i$ then $\E[f^2] = \sum_{i=0}^{m-1} c_i^2$.

Finally, the parity basis contains the constant function~$1$. This fact leads to several of our pleasant Fourier formulas.  In particular, when you take an orthonormal basis $\phi_0, \dots, \phi_{m-1}$ for $L^2(\Omega, \pi)$ which has  $\phi_0 \equiv 1$, then $0 = \la \phi_0, \phi_i\ra = \Ex_{\bx \sim \pi} [\phi_i(\bx)]$ for all $i > 0$.   Hence if  $f \in L^2(\Omega,\pi)$ has the expansion $f= \sum_{i=0}^{m-1} c_i \phi_i$, then $\E[f] = c_0$ and $\Var[f] = \sum_{i > 0} c_i^2$.

\medskip

We encapsulate the second and third properties with a definition:
\begin{definition}
    A \emph{Fourier basis}
                                                        \index{Fourier basis}%
    for an inner product space $L^2(\Omega, \pi)$ is an orthonormal basis $\phi_0, \dots, \phi_{m-1}$ with $\phi_0 \equiv 1$.
\end{definition}
\begin{example}
    For each $n \in \N^+$, the $2^n$ parity functions $(\chi_S)_{S \subseteq [n]}$ form a Fourier basis for $L^2(\bn, \unif\xn)$.
\end{example}
\begin{remark}
    A Fourier basis for $L^2(\Omega, \pi)$ always exists because you can extend the  set $\{1\}$ to a basis and then perform the Gram--Schmidt process.  On the other hand, Fourier bases are not unique.  Even in the case of $L^2(\bits, \unif)$ there are two possibilities: the basis $\{1, \id\}$ and the basis $\{1, -\id\}$.
\end{remark}
\begin{example} \label{eg:3-ary-basis}
    In the case of $\Omega = \{a,b,c\}$ with $\pi(a) = \pi(b) = \pi(c) = 1/3$, one possible Fourier basis (see Exercise~\ref{ex:3-ary-basis}) is
    \[
        \phi_0 \equiv 1, \quad \begin{array}{l} \phi_1(a) = +\sqrt{2} \\ \phi_1(b) = -\sqrt{2}/2 \\  \phi_1(c) = -\sqrt{2}/2, \end{array} \quad \begin{array}{l} \phi_2(a) = 0 \\ \phi_2(b) = +\sqrt{6}/2, \\  \phi_2(c) = -\sqrt{6}/2. \end{array}
    \]
\end{example}

As mentioned, given a Fourier basis for $L^2(\Omega, \pi)$ you can construct a Fourier basis for any $L^2(\Omega^n, \pi\xn)$ by ``taking all $n$-fold products''.  To make this precise we need some notation.
\begin{definition}                          \label{def:multi-index}
    An $n$-dimensional \emph{multi-index}
                                                        \index{multi-index}%
    is a tuple $\alpha \in \N^n$.  We write
                                                \nomenclature[suppalpha]{$\supp(\alpha)$}{if $\alpha$ is a multi-index, denotes $\{i : \alpha_i \neq 0\}$}%
                                                \nomenclature[alpha1]{$\#\alpha$}{if $\alpha$ is a multi-index, denotes the number of nonzero components of~$\alpha$}%
                                                \nomenclature[alpha2]{$\lvert\alpha\rvert$}{if $\alpha$ is a multi-index, denotes $\sum_i \alpha_i$}%
    \[
        \supp(\alpha) = \{i : \alpha_i \neq 0\}, \quad\#\alpha = |\supp(\alpha)|, \quad |\alpha| = \sum_{i=1}^n \alpha_i.
    \]
                                                \nomenclature[N<m]{$\N_{< m}$}{$\{0, 1, \dots, m-1\}$}
    We may write $\alpha \in \N_{<m}^n$ when we want to emphasize that each $\alpha_i  \in \{0, 1, \dots, m-1\}$.
\end{definition}
\begin{definition}                                  \label{def:multi-index-phi}
    Given functions $\phi_0, \dots, \phi_{m-1} \in L^2(\Omega, \pi)$ and a multi-index $\alpha \in \N_{< m}^n$, we define $\phi_\alpha \in L^2(\Omega^n, \pi\xn)$
                                                \nomenclature[phialpha]{$\phi_\alpha$}{given functions $\phi_0, \dots, \phi_{m-1}$ and a multi-index $\alpha$, denotes $\prod_{i=1}^n \phi_{\alpha_i}$}%
    by
    \[
        \phi_\alpha(x) = \prod_{i=1}^n \phi_{\alpha_i}(x_i).
    \]
\end{definition}
Now we can show that products of Fourier bases are Fourier bases.
\begin{proposition}                                     \label{prop:fbasis-product}
    Let $\phi_0, \dots, \phi_{m-1}$ be a Fourier basis for $L^2(\Omega, \pi)$.  Then the collection $(\phi_\alpha)_{\alpha \in \N_{<m}^n}$ is a Fourier basis for $L^2(\Omega^n,\pi^{\otimes n})$ (with the understanding that $\alpha = (0, 0, \dots, 0)$  indexes the constant function~$1$).
\end{proposition}
\begin{proof}
    First we check orthonormality. For any multi-indices $\alpha, \beta \in \N_{<m}^n$ we have
    \begin{align*}
        \la \phi_\alpha, \phi_\beta \ra &= \Ex_{\bx \sim \pi\xn}[\phi_\alpha(\bx)\cdot \phi_\beta(\bx)] \\
        &= \Ex_{\bx \sim \pi\xn}\Bigl[\prod_{i = 1}^n \phi_{\alpha_i}(\bx_i) \cdot \prod_{i = 1}^n \phi_{\beta_i}(\bx_i)\Bigr] \\
        &= \prod_{i = 1}^n \Ex_{\bx_i \sim \pi}[\phi_{\alpha_i}(\bx_i) \cdot \phi_{\beta_i}(\bx_i)] \tag{since $\pi\xn$ is a product distribution} \\
        &= \prod_{i = 1}^n \bone_{\{\alpha_i = \beta_i\}} \tag{since $\{\phi_0, \dots, \phi_{m-1}\}$ is orthonormal} \\
        &= \bone_{\{\alpha = \beta\}}.
    \end{align*}
    This confirms that the collection $(\phi_\alpha)_{\alpha \in \N_{<m}^n}$ is orthonormal, and consequently  linearly independent.  It is therefore also a basis because it has cardinality~$m^n$, which we know is the dimension of $L^2(\Omega^n, \pi\xn)$ (see Fact~\ref{fact:l2-indic-basis}).
\end{proof}

Given a product Fourier basis as in Proposition~\ref{prop:fbasis-product}, we can express any $f \in L^2(\Omega^n, \pi\xn)$ as a linear combination of basis functions.  We will write $\wh{f}(\alpha)$ for the ``Fourier coefficient'' on $\phi_\alpha$ in this expression.
\begin{definition}                                  \label{def:general-fourier-coefficient}
    Having \emph{fixed} a Fourier basis $\phi_{0}, \dots, \phi_{m-1}$ for $L^2(\Omega, \pi)$,
                            \index{Fourier coefficient!product space domains}%
    every $f \in L^2(\Omega^n, \pi\xn)$ is uniquely expressible as
    \[
        f = \sum_{\alpha \in \N_{< m}^n} \wh{f}(\alpha) \phi_\alpha.
    \]
    This is the \emph{Fourier expansion} of $f$ with respect to the basis.
                                        \index{Fourier expansion!product space domains}%
    The real number $\wh{f}(\alpha)$ is called the \emph{Fourier coefficient of $f$ on $\alpha$} and it satisfies
    \[
        \wh{f}(\alpha) = \la f, \phi_\alpha \ra.
    \]
\end{definition}
\begin{example}  \label{eg:and2-3ary} Fix the Fourier basis as in Example~\ref{eg:3-ary-basis}.  Let $f \co \{a,b,c\}^2 \to \{0,1\}$ be the function which is~$1$ if and only if both inputs are~$c$.  Then you can check (Exercise~\ref{ex:and2-3ary}) that
\[
    f = \tfrac19 - \tfrac{\sqrt{2}}{18} \phi_{(1,0)} - \tfrac{\sqrt{6}}{18} \phi_{(2,0)}  - \tfrac{\sqrt{2}}{18} \phi_{(0,1)} - \tfrac{\sqrt{6}}{18} \phi_{(0,2)} + \tfrac{1}{18}\phi_{(1,1)} + \tfrac{\sqrt{12}}{36} \phi_{(2,1)}  + \tfrac{\sqrt{12}}{36} \phi_{(1,2)} + \tfrac{1}{6}\phi_{(2,2)}.
\]
\end{example}
The notation $\wh{f}(\alpha)$ may seem poorly chosen because it doesn't show the dependence on the basis.  However, the Fourier formulas we develop in the next section will have the property that \emph{they are the same for every product Fourier basis}.  We will show a basis-independent way of developing the formulas in Section~\ref{sec:orthogonal-decomposition}.

\section{Generalized Fourier formulas}                \label{sec:general-fourier-formulas}

In this section we will revisit a number of combinatorial/probabilistic notions and show that for functions $f \in L^2(\Omega^n, \pi\xn)$, these notions have familiar Fourier formulas that don't depend on the Fourier basis.

The orthonormality of Fourier bases gives us some formulas almost immediately:
\begin{proposition}                                     \label{prop:general-plancherel-etc}
    Let $f, g \in L^2(\Omega^n, \pi\xn)$.
                                       \index{Parseval's Theorem}%
                                       \index{Plancherel's Theorem}%
    Then for any fixed product Fourier basis, the following formulas hold:
    \begin{align*}
        \E[f] &= \wh{f}(0)\\
        \E[f^2] &= \sum_{\alpha \in \N_{< m}^n} \wh{f}(\alpha)^2\tag{Parseval} \\
        \Var[f] &= \sum_{\alpha \neq 0} \wh{f}(\alpha)^2 \\
        \la f, g \ra &= \sum_{\alpha \in \N_{<m}^n} \wh{f}(\alpha)\wh{g}(\alpha) \tag{Plancherel} \\
        \Cov[f,g] &= \sum_{\alpha \neq 0} \wh{f}(\alpha)\wh{g}(\alpha).
    \end{align*}
\end{proposition}
\begin{proof}
    We verify Plancherel's Theorem, from which the other identities follow (Exercise~\ref{ex:finish-general-plancherel}):
    \begin{align*}
          \la f, g \ra &= \Bigl\langle \sum_{\alpha \in \N_{< m}^n} \wh{f}(\alpha) \phi_\alpha, \sum_{\beta \in \N_{< m}^n} \wh{g}(\beta) \phi_\beta \Bigr\rangle \\
          &= \sum_{\alpha, \beta \in \N_{< m}^n} \wh{f}(\alpha) \wh{g}(\beta) \langle \phi_\alpha, \phi_\beta \rangle \\
          &= \sum_{\alpha \in \N_{< m}^n} \wh{f}(\alpha) \wh{g}(\alpha)
    \end{align*}
    by orthonormality of $(\phi_\alpha)_{\alpha \in \N_{<m}^n}$.
\end{proof}

We now give the key definition for developing basis-independent Fourier expansions.  In the case of $L^2(\bits)$ this definition appeared already in Exercise~\ref{ex:orthog1}.
\begin{definition}                      \label{def:projection-onto-coords}
    Let $J \subseteq [n]$ and write $\barJ = [n] \setminus J$.  Given $f \in L^2(\Omega^n, \pi\xn)$, the \emph{projection of $f$ on coordinates~$J$}
                                                            \index{projection onto coordinates}%
    is the function $f^{\subseteq J} \in L^2(\Omega^n, \pi\xn)$
                                \nomenclature[fJ1]{$f^{\subseteq J}$}{the function (depending only on the $J$ coordinates) defined by $f^{\subseteq J}(x) = \E_{\bx'_{\barJ}}[f(x_J, \bx'_{\barJ})]$; in particular, it's $\sum_{S \subseteq J}\wh{f}(S) \chi_S$ when $f \btR$}%
    defined by
    \[
        f^{\subseteq J}(x) = \Ex_{\bx' \sim \pi^{\otimes \barJ}}[f(x_J, \bx')],
    \]
    where $x_J \in \Omega^J$ denotes the values of $x$ in the $J$-coordinates.  In other words, $f^{\subseteq J}(x)$ is the expectation of~$f$ when the $\barJ$-coordinates of~$x$ are rerandomized.  Note that we take $f^{\subseteq J}$ to have $\Omega^n$ as its domain, even though it only depends on the coordinates in~$J$.

    Forming $f^{\subseteq J}$ is indeed the application of a projection linear operator to~$f$,
                                        \nomenclature[EJ]{$\uE_I$}{the expectation over coordinates $I$ operator}%
    namely the \emph{expectation over~$\barJ$ operator}, $\uE_{\barJ}$.
                                                \index{expectation operator}%
    We take this as the definition of the operator: $\uE_{\barJ} f = f^{\subseteq J}$.  When $\barJ = \{i\}$ is a singleton we write simply $\uE_i$.
\end{definition}
\begin{remark}
    This definition of $\uE_i$ is consistent with Definition~\ref{def:expectation-operator}.  You are asked to verify that $\uE_{\barJ}$ is indeed a projection, self-adjoint linear operator in Exercise~\ref{ex:expec-is-proj}.
\end{remark}
\begin{proposition}                                     \label{prop:projection-formula}
    Let $J \subseteq [n]$ and $f \in L^2(\Omega^n, \pi\xn)$.  Then for any  fixed product Fourier basis,
    \[
        f^{\subseteq J} = \sum_{\substack{\alpha \in \N_{< m}^n \\ \supp(\alpha) \subseteq J}} \wh{f}(\alpha)\,\phi_\alpha.
    \]
\end{proposition}
\begin{proof}
    Since $\uE_{\barJ}$ is a linear operator, it suffices to verify for all $\alpha$ that
    \[
        \phi_\alpha^{\subseteq J} = \begin{cases}
                                        \phi_\alpha & \text{if } \supp(\alpha) \subseteq J, \\
                                        0 & \text{otherwise}.
                                    \end{cases}
    \]
    If $\supp(\alpha) \subseteq J$, then $\phi_\alpha$ does not depend on the coordinates $\barJ$; hence indeed $\phi_\alpha^{\subseteq J}= \phi_\alpha$.  So suppose $\supp(\alpha) \not \subseteq J$.  Since $\phi_\alpha(x) = \bigl(\littleprod_{i\in J} \phi_{\alpha_i}(x_i)\bigr)\bigl(\littleprod_{i\in \barJ} \phi_{\alpha_i}(x_i)\bigr)$, we can write $\phi_\alpha = \phi_{\alpha_J} \cdot \phi_{\alpha_\barJ}$, where $\phi_{\alpha_J}$ depends only on the coordinates in~$J$, $\phi_{\alpha_{\barJ}}$ depends only on the coordinates in~$\barJ$, and $\E[\phi_{\alpha_{\barJ}}] = 0$ precisely because $\supp(\alpha) \not\subseteq J$.  Thus for every $x \in \Omega^n$,
    \[
        \phi_{\alpha}^{\subseteq J}(x) = \Ex_{\bx' \sim \pi^{\otimes \barJ}}[\phi_{\alpha_J}(x_J) \phi_{\alpha_{\barJ}}(\bx')] = \phi_{\alpha_J}(x_J)  \cdot \Ex_{\bx' \sim \pi^{\otimes \barJ}}[ \phi_{\alpha_{\barJ}}(\bx')] = 0
    \]
    as needed.
\end{proof}
\begin{corollary}                           \label{cor:f-depends}
    Let $f \in L^2(\Omega^n,\pi\xn)$ and fix a product Fourier basis.  If~$f$ depends only on the coordinates in $J \subseteq [n]$ then $\wh{f}(\alpha) = 0$ whenever $\supp(\alpha) \not \subseteq J$.
\end{corollary}
\begin{proof}
    This follows from Proposition~\ref{prop:projection-formula} because $f = f^{\subseteq J}$.
\end{proof}
\begin{corollary}                           \label{cor:general-expec-formula}
    Let $i \in [n]$ and $f \in L^2(\Omega^n, \pi\xn)$.  Then for any  fixed product Fourier basis,
    \[
        \uE_i f = \sum_{\alpha : \alpha_i = 0}  \wh{f}(\alpha)\,\phi_\alpha.
    \]
\end{corollary}

                                                    \index{influence!product space domains|(}
Let us now define influences for functions $f \in L^2(\Omega^n, \pi\xn)$.  In the case of $\Omega = \bits$, our definition of $\Inf_i[f]$ from  Chapter~\ref{sec:influences} was $\E[(\D_i f)^2]$.  However, the notion of a derivative operator does not make sense for more general domains~$\Omega$.  In fact, even in the case of $\Omega = \bits$ it isn't a basis-invariant notion: the choice of $\frac{f(x^{(i\mapsto 1)}) - f(x^{(i \mapsto -1)})}{2}$ rather than $\frac{f(x^{(i\mapsto -1)}) - f(x^{(i \mapsto 1)})}{2}$ is inherently arbitrary. Instead we can fall back on the \emph{Laplacian} operators, and take the identity $\Inf_i[f] = \la f, \Lap_i f \ra$ from
                                        \index{Laplacian operator!$i$th coordinate}%
Proposition~\ref{prop:dir-laplacian-facts} as a definition.
\begin{definition}
    Let $i \in [n]$ and $f \in L^2(\Omega^n, \pi\xn)$.  The \emph{$i$th coordinate Laplacian operator} $\Lap_i$ is the self-adjoint, projection linear operator defined by
    \[
        \Lap_i f = f - \uE_i f.
    \]
    The \emph{influence of coordinate $i$ on $f$} is defined to be
    \[
        \Inf_i[f] = \la f, \Lap_i f \ra = \la \Lap_i f, \Lap_i f \ra.
    \]
    The \emph{total influence}
                                                \index{total influence!product space domains}%
    of $f$ is defined to be $\Tinf[f] = \sum_{i=1}^n \Inf_i[f]$.
\end{definition}
\noindent You can think of $\Lap_i f$ as ``the part of $f$ which depends on the $i$th coordinate''.
\begin{proposition}                                     \label{prop:general-laplac-influence}
    Let $i \in [n]$ and $f \in L^2(\Omega^n, \pi\xn)$.  Then for any  fixed product Fourier basis,
    \begin{align*}
        \Lap_i f = \sum_{\alpha : \alpha_i \neq 0}  \wh{f}(\alpha)\,\phi_\alpha, \quad
        \Inf_i[f] = \sum_{\alpha : \alpha_i \neq 0}  \wh{f}(\alpha)^2, \quad
        \Tinf[f] = \sum_{\alpha}  \#\alpha \cdot \wh{f}(\alpha)^2,
    \end{align*}
\end{proposition}
\begin{proof}
    The first formula is immediate from Corollary~\ref{cor:general-expec-formula}, the second from Plancherel, and the third from summing over~$i$.
\end{proof}
Exercise~\ref{ex:dir-laplacian-facts2} asks you to verify the following formulas (cf.~Exercise~\ref{ex:inf-var}), which are often useful for computations:
\begin{proposition}                               \label{prop:dir-laplacian-facts2}
    Let $i \in [n]$ and $f \in L^2(\Omega^n, \pi\xn)$.  Then
    \[
        \Inf_i[f] = \Ex_{\bx \sim \pi\xn}[\Var_{\bx_i' \sim \pi}[f(\bx_1, \dots, \bx_{i-1}, \bx_i', \bx_{i+1}, \dots, \bx_n)]].
    \]
    If furthermore $f$'s range is $\bits$, then
    \[
        \Inf_i[f] = \E[|\Lap_i f|] = 2\Pr_{\substack{\bx \sim \pi\xn \\ \bx_i' \sim \pi}}[f(\bx) \neq f(\bx_1, \dots, \bx_{i-1}, \bx_i', \bx_{i+1}, \dots, \bx_n)].
    \]
\end{proposition}
\begin{example}  Let's continue Example~\ref{eg:and2-3ary}, in which $\{a,b,c\}$ has the uniform distribution and $f \co \{a,b,c\}^2 \to \{0,1\}$ is~$1$ if and only if both inputs are~$c$.  We compute $\Inf_1[f]$ two ways.  Using Proposition~\ref{prop:dir-laplacian-facts2} we have $\Var[f(\bx_1, a)] = \Var[f(\bx_1, b)] = 0$ and $\Var[f(\bx_1, c)] = \frac13 \cdot \frac23 = \frac29$ (because $f(\bx_1, c)$ is Bernoulli with parameter~$\frac13$); thus $\Inf_1[f] = \frac13 \cdot \frac29 = \frac{2}{27}$. Alternatively, using the formula from Proposition~\ref{prop:general-laplac-influence} as well as the Fourier expansion from Example~\ref{eg:and2-3ary}, we can compute
$\Inf_1[f] = (-\tfrac{\sqrt{2}}{18})^2 + (-\tfrac{\sqrt{6}}{18})^2 + (\tfrac{1}{18})^2 + (\tfrac{\sqrt{12}}{36})^2 +(\tfrac{\sqrt{12}}{36})^2 +(\tfrac16)^2 = \tfrac{2}{27}$.
\end{example}
                                                    \index{influence!product space domains|)}

\medskip

Next, we straightforwardly extend our definitions of the noise operator and noise stability to general product spaces.
\begin{definition}                                      \label{def:noise-general}
    Fix a finite product probability space $(\Omega^n, \pi\xn)$.
    For $\rho \in [0,1]$ and $x \in \Omega^n$ we write $\by \sim N_\rho(x)$ to denote that $\by \in \Omega^n$ is randomly chosen as follows: For each $i \in [n]$ independently,
    \[
        \by_i = \begin{cases} x_i & \text{with probability $\rho$,}\\
                            \text{drawn from $\pi$} & \text{with probability $1-\rho$.}
                 \end{cases}
    \]
    If $\bx \sim \pi\xn$ and $\by \sim N_\rho(\bx)$, we say that $(\bx, \by)$ is a \emph{$\rho$-correlated pair under~$\pi\xn$}. (This definition is symmetric in $\bx$ and $\by$.)
\end{definition}
\begin{definition}                                  \label{def:general-T-rho}
    For a fixed space $L^2(\Omega^n, \pi\xn)$ and $\rho \in [0,1]$, the \emph{noise operator with parameter $\rho$}
                                                \index{noise operator!product space domains}%
    is the linear operator $\T_\rho$ on functions $f \in L^2(\Omega^n,\pi\xn)$ defined by
    \[
        \T_\rho f(x) = \Ex_{\by \sim N_\rho(x)}[f(\by)].
    \]
    The \emph{noise stability of $f$ at~$\rho$}
                                            \index{noise stability!product space domains}%
    is
    \[
        \Stab_\rho[f] = \la f, \T_\rho f \ra = \Es{(\bx, \by) \text{ $\rho$-correlated}\\ \text{under } \pi\xn}[f(\bx) f(\by)].
    \]
\end{definition}
\begin{proposition}                                     \label{prop:general-trho}
    Let $\rho \in [0,1]$ and let $f \in L^2(\Omega^n, \pi\xn)$.  Then for any  fixed product Fourier basis,
    \[
        \T_\rho f = \sum_{\alpha \in \N_{<m}^n} \rho^{\#\alpha} \wh{f}(\alpha)\,\phi_\alpha, \qquad \Stab_\rho[f] = \sum_{\alpha \in \N_{<m}^n} \rho^{\#\alpha} \wh{f}(\alpha)^2.
    \]
\end{proposition}
\begin{proof}
    Let $\bJ$ denote a $\rho$-random subset of $[n]$; i.e., $\bJ$ is formed by including each $i \in [n]$ independently with probability~$\rho$.  Then by definition $T_\rho f (x) = \Ex_{\bJ} [f^{\subseteq \bJ}(x)]$, and so from Proposition~\ref{prop:projection-formula} we get
    \[
        T_\rho f (x) = \Ex_{\bJ} [f^{\subseteq \bJ}(x)] = \Ex_{\bJ} \Bigl[\sum_{\substack{\alpha \in \N_{<m}^n \\ \supp(\alpha) \subseteq \bJ}} \wh{f}(\alpha)\,\phi_\alpha(x)\Bigr]= \sum_{\alpha \in \N_{<m}^n} \rho^{\#\alpha} \wh{f}(\alpha)\,\phi_\alpha(x),
    \]
    since for a fixed $\alpha$, the probability of $\supp(\alpha) \subseteq \bJ$ is $\rho^{\#\alpha}$.  The formula for $\Stab_\rho[f]$ now follows from Plancherel.
\end{proof}
\begin{remark}                              \label{rem:general-Trho}
    The first formula in this proposition may be used to extend the definition of $\T_\rho f$ to values of $\rho$ outside $[0,1]$.
\end{remark}
We also define $\rho$-stable influences.  The factor of $\rho^{-1}$ in our definition is for consistency with the $L^2(\bn)$ case.
\begin{definition}                          \label{def:stable-influence-general}
    For $f \in L^2(\Omega^n, \pi\xn)$,  $\rho \in (0,1]$, and $i \in [n]$, the \emph{$\rho$-stable influence} of~$i$ on~$f$
                                        \index{stable influence!product space domains}%
    is
    \[
        \Inf_i^{(\rho)}[f] = \rho^{-1} \Stab_\rho[\Lap_i f] = \sum_{\alpha : \alpha_i \neq 0}  \rho^{\#\alpha-1} \wh{f}(\alpha)^2.
    \]
    We also define $\Tinf^{(\rho)}[f] = \sum_{i=1}^n \Inf_i^{(\rho)}[f]$.
\end{definition}
Just as in the case of $L^2(\bits^n)$ we can use stable influences to define the ``notable'' coordinates of a function, of which there is a bounded quantity.  A verbatim repetition of the proof of Proposition~\ref{prop:few-stable-influences} yields the following generalization:
\begin{proposition}                                     \label{prop:few-stable-influences-general}
    Suppose $f \in L^2(\Omega^n, \pi\xn)$ has $\Var[f] \leq 1$.  Given $0 < \delta < 1$, $0 < \eps \leq 1$, let $J = \{ i \in [n] : \Inf_i^{(1-\delta)}[f] \geq \eps\}$.  Then $|J| \leq \frac{1}{\delta \eps}$.
\end{proposition}

We end this section by discussing the ``degree'' of functions on general product spaces. For $f \in L^2(\bn)$ the Fourier expansion is a real polynomial; this yields an obvious definition for degree.  But for general $f \in L^2(\Omega^n, \pi\xn)$ the domain is just an abstract set so we need to look for a more intrinsic definition.  We take  our cue from Exercise~\ref{ex:degree}\ref{ex:degree-coord-free}:
                                                    \index{degree!product space domains}%
\begin{definition}                            \label{def:general-degree}
    Let $f \in L^2(\Omega^n, \pi\xn)$ be nonzero.  The \emph{degree} of~$f$, written $\deg(f)$, is the least $k \in \N$ such that $f$ is a sum of $k$-juntas (functions depending on at most~$k$ coordinates).
\end{definition}
\begin{proposition}                                     \label{prop:general-degree}
    Let $f \in L^2(\Omega^n, \pi\xn)$ be nonzero.  Then for any fixed product Fourier basis we have $\deg(f) = \max\{\#\alpha : \wh{f}(\alpha) \neq 0\}$.
\end{proposition}
\begin{proof}
    The inequality $\deg(f) \leq \max\{\#\alpha : \wh{f}(\alpha) \neq 0\}$ is immediate from the Fourier expansion:
    \[
        f = \sum_{\alpha : \wh{f}(\alpha) \neq 0} \wh{f}(\alpha)\,\phi_\alpha
    \]
    and each function $\wh{f}(\alpha)\,\phi_\alpha$ depends on at most~$\#\alpha$ coordinates.  For the reverse inequality, suppose $f = g_1 + \cdots + g_m$ where each $g_i$ depends on at most~$k$ coordinates.  By Corollary~\ref{cor:f-depends} each $g_i$ has its Fourier support on functions~$\phi_\alpha$ with $\#\alpha \leq k$.  But $\wh{f}(\alpha) = \wh{g_1}(\alpha) + \cdots + \wh{g_m}(\alpha)$,  so the same is true of~$f$.
\end{proof}

\section{Orthogonal decomposition}                              \label{sec:orthogonal-decomposition}

                                                \index{orthogonal decomposition|(}%
                                                \index{Efron--Stein decomposition|seeonly{orthogonal decomposition}}%
                                                \index{Hoeffding decomposition|seeonly{orthogonal decomposition}}%
                                                \index{ANOVA decomposition|seeonly{orthogonal decomposition}}%
In this section we describe a basis-free kind of ``Fourier expansion'' for functions on general product domains.  We will refer to it as the \emph{orthogonal decomposition} of $f \in L^2(\Omega^n, \pi\xn)$, though it goes by several other names in the literature: e.g., \emph{Hoeffding decomposition}, \emph{Efron--Stein decomposition}, or \emph{ANOVA decomposition}. The general idea is to express
\begin{equation} \label{eqn:pm1-orthog-decomp}
    f = \sum_{S \subseteq [n]} f^{=S}
\end{equation}
where each function $f^{=S} \in L^2(\Omega^n, \pi\xn)$ gives the ``contribution to $f$ coming from coordinates~$S$ (but not from any subset of~$S$)''.

To make this more precise, let's start with the familiar case of $f \btR$.  Here it is possible to define the functions $f^{=S} \btR$ simply by  $f^{=S} = \wh{f}(S)\,\chi_S$. (Later we will give an equivalent definition that doesn't involve the Fourier basis.) This definition satisfies~\eqref{eqn:pm1-orthog-decomp} as well as the following two properties:
\begin{enumerate}
    \item $f^{=S}$ depends only on the coordinates in~$S$.
    \item If $T \subsetneq S$ and $g$ is a function depending only on the coordinates in~$T$, then $\la f^{=S}, g \ra = 0$.
\end{enumerate}
These properties describe what we mean precisely when we say that $f^{=S}$ is the  ``contribution to $f$ coming from coordinates~$S$ (but not from any subset of~$S$)''. Furthermore, decomposition~\eqref{eqn:pm1-orthog-decomp} is \emph{orthogonal}, meaning $\la f^{=S}, f^{=T} \ra = 0$ whenever $S \neq T$.

To make this definition basis-free, recall the ``projection of~$f$ onto coordinates~$J$'', $f^{\subseteq J}$, from Exercise~\ref{ex:orthog1} and Definition~\ref{def:projection-onto-coords}.  You can think of $f^{\subseteq J}$ as the ``contribution to~$f$ coming from coordinates~$J$ (collectively)''.  It has a probabilistic definition not depending on any basis, and with the definition $f^{=S} = \wh{f}(S)\,\chi_S$ we have from Exercise~\ref{ex:orthog1} or Proposition~\ref{prop:projection-formula} that
\begin{equation} \label{eqn:orthog-invert-me}
    f^{\subseteq J} =  \sum_{S \subseteq J} f^{=S}.
\end{equation}
It is precisely by inverting~\eqref{eqn:orthog-invert-me} that we can give a basis-free definition of the functions~$f^{=S}$.

Let's do this inversion for a general $f \in L^2(\Omega^n, \pi\xn)$.  The projection functions $f^{\subseteq J} \in L^2(\Omega^n, \pi\xn)$ can be defined as in Definition~\ref{def:projection-onto-coords}.  If we want~\eqref{eqn:orthog-invert-me} to hold for $J = \emptyset$ then we should define
\[
    f^{=\emptyset} = f^{\subseteq \emptyset}
\]
(which is the constant function equal to~$\E[f]$).  Given this, if we want~\eqref{eqn:orthog-invert-me} to hold for singleton sets~$J = \{j\}$, then we need
\[
    f^{\subseteq \{j\}} =  f^{=\emptyset} + f^{=\{j\}} \quad\iff\quad f^{=\{j\}} = f^{\subseteq \{j\}} -  f^{\subseteq \emptyset}.
\]
In other words,
\[
    f^{= \{j\}}(x) = \Ex_{\bx \sim \pi\xn} [f \mid \bx_j = x_j] - \Ex_{\bx \sim \pi\xn} [f(\bx)].
\]
Notice this function only depends on the input value~$x_j$; it measures the change in expectation of~$f$ if you know the value~$x_j$.  Moving on to sets of cardinality~$2$, if we want~\eqref{eqn:orthog-invert-me} to hold for $J = \{i,j\}$, then we need
\begin{align*}
    f^{\subseteq \{i,j\}} &= f^{=\emptyset} + f^{=\{i\}} + f^{=\{j\}} + f^{=\{i,j\}} \\
     &= f^{\subseteq \emptyset} + (f^{\subseteq \{i\}} - f^{\subseteq \emptyset}) + (f^{\subseteq \{j\}} - f^{\subseteq \emptyset}) + f^{=\{i,j\}}
\end{align*}
and hence
\[
    f^{=\{i,j\}} = f^{\subseteq \{i,j\}} - f^{\subseteq \{i\}} - f^{\subseteq \{j\}} + f^{\subseteq \emptyset}.
\]
It's clear that we can continue this and define all the functions $f^{=S}$ by the principle of inclusion-exclusion.  To show this definition leads to an orthogonal decomposition we will need the following lemma:
\begin{lemma}                                       \label{lem:orthog-subsets}
    Let $f, g \in L^2(\Omega^n, \pi\xn)$. Assume that $f$ does not depend on any coordinate outside $I \subseteq [n]$, and~$g$ does not depend on any coordinate outside $J \subseteq [n]$.  Then
    $\la f, g \ra = \la f^{\subseteq I \cap J}, g^{\subseteq I \cap J} \ra$.
\end{lemma}
\begin{proof}
    We may assume without loss of generality that $I \cup J = [n]$. Given any $x \in \Omega^n$ we can break it into the parts $(x_{I \cap J}, x_{I \setminus J}, x_{J \setminus I})$.  We then have
    \begin{align*}
        \la f, g\ra = \Ex_{\bx_{I \cap J}, \bx_{I \setminus J}, \bx_{J \setminus I}}[f(\bx_{I \cap J}, \bx_{I \setminus J}) \cdot g(\bx_{I \cap J}, \bx_{J \setminus I})],
    \end{align*}
    where we have abused notation slightly by writing $f$ and $g$ as functions just of the coordinates on which they actually depend.  Since $\bx_{I \setminus J}$ and $\bx_{J \setminus I}$ are independent, the above equals
    \[
        \Ex_{\bx_{I \cap J}}\left[\Ex_{\bx_{I \setminus J}}[f(\bx_{I \cap J}, \bx_{I \setminus J})] \cdot \Ex_{\bx_{J \setminus I}}[g(\bx_{I \cap J}, \bx_{J \setminus I})] \right].
    \]
    But now $\Ex_{\bx_{I \setminus J}}[f(\bx_{I \cap J}, \bx_{I \setminus J})]$ is nothing more than $f^{\subseteq I \cap J}(\bx_{I \cap J})$, and similarly $\Ex_{\bx_{J \setminus I}}[g(\bx_{I \cap J}, \bx_{J \setminus I})] = g^{\subseteq I \cap J}(\bx_{I \cap J})$.  Thus the above equals
    \[
        \Ex_{\bx_{I \cap J}}[f^{\subseteq I \cap J}(\bx_{I \cap J}) \cdot g^{\subseteq I \cap J}(\bx_{I \cap J})] = \la f^{\subseteq I \cap J}, g^{\subseteq I \cap J} \ra. \qedhere
    \]
\end{proof}

We can now give the main theorem on orthogonal decomposition:
\begin{theorem}                                     \label{thm:orthogonal-decomposition}
    Let $f \in L^2(\Omega^n, \pi\xn)$.  Then $f$ has a unique decomposition as
    \[
        f = \sum_{S \subseteq [n]} f^{=S}
    \]
    where the functions $f^{=S} \in L^2(\Omega^n, \pi\xn)$ satisfy the following:
    \begin{enumerate}
        \item \label{item:first-orthog} $f^{=S}$ depends only on the coordinates in~$S$.
        \item \label{item:orthog-decomp-ortho} If $T \subsetneq S$ and $g \in L^2(\Omega^n, \pi\xn)$ depends only on the coordinates in~$T$, then $\la f^{=S}, g \ra = 0$.
    \end{enumerate}
    This decomposition has the following additional properties:
    \begin{enumerate} \setcounter{enumi}{2}
        \item \label{item:orthog-decomp-ortho2} Condition~\eqref{item:orthog-decomp-ortho} additionally holds whenever $S \not \subseteq T$.
        \item \label{item:orthog-decomp-ortho3} The decomposition is \emph{orthogonal}:  $\la f^{=S}, f^{=T} \ra = 0$ for $S \neq T$.
        \item \label{item:by-incl-excl} $\sum_{S \subseteq T} f^{=S} = f^{\subseteq T}$.
        \item \label{item:orthog-linear} For each $S \subseteq [n]$, the mapping $f \mapsto f^{=S}$ is a linear operator.
    \end{enumerate}
\end{theorem}
\begin{proof}
    We first show the existence of a decomposition satisfying~\eqref{item:first-orthog}--\eqref{item:orthog-linear}.  We then show uniqueness for decompositions satisfying~\eqref{item:first-orthog} and~\eqref{item:orthog-decomp-ortho}.  As suggested above, for each $S \subseteq [n]$ we define
    \[
        f^{=S} = \sum_{J \subseteq S} (-1)^{|S| - |J|} f^{\subseteq J},
    \]
    where the functions $f^{\subseteq J} \in L^2(\Omega^n, \pi\xn)$ are as in Definition~\ref{def:projection-onto-coords}.  Since each~$f^{\subseteq J}$ depends only on the coordinates in~$J$, condition~\eqref{item:first-orthog} certainly holds.  It is also immediate that condition~\eqref{item:by-incl-excl} holds by inclusion-exclusion; you are asked to prove this explicitly in Exercise~\ref{ex:orthog-incl-excl}.  Condition~\eqref{item:orthog-linear} also follows because each $f \mapsto f^{\subseteq J}$ is a linear operator, as discussed after Definition~\ref{def:projection-onto-coords}.

    We now verify~\eqref{item:orthog-decomp-ortho}.  Assume $T \subsetneq S$ and that $g \in L^2(\Omega^n, \pi\xn)$ only depends on the coordinates in~$T$.  We have
    \begin{equation} \label{eqn:pair-me}
        \la f^{=S}, g \ra = \sum_{J \subseteq S} (-1)^{|S| - |J|} \la f^{\subseteq J}, g \ra.
    \end{equation}
    Take any $i \in S \setminus T$ and pair up the summands in~\eqref{eqn:pair-me} as~$J'$, $J''$, where $J' \not \ni i$ and $J'' = J' \cup \{i\}$.  By Lemma~\ref{lem:orthog-subsets} we have
    \[
        \la f^{\subseteq J''}, g \ra = \la f^{\subseteq J'' \cap T}, g^{\subseteq T} \ra = \la f^{\subseteq J' \cap T}, g^{\subseteq T} \ra,
    \]
    the latter equality using $i \not \in T$.  But the signs $(-1)^{|S| - |J'|}$ and $(-1)^{|S| - |J''|}$ are opposite, so the summands in~\eqref{eqn:pair-me} cancel in pairs. This shows the sum is~$0$, confirming~\eqref{item:orthog-decomp-ortho}.

    We complete the existence proof by noting that $\eqref{item:orthog-decomp-ortho} \implies \eqref{item:orthog-decomp-ortho2} \implies \eqref{item:orthog-decomp-ortho3}$ (assuming~\eqref{item:first-orthog}). The first implication is because $\la f^{=S}, g \ra = \la f^{= S}, g^{\subseteq S \cap T} \ra$ when $g$ depends only on the coordinates in~$T$ (Lemma~\ref{lem:orthog-subsets}), and $S \cap T \subsetneq S$ when $S \not \subseteq T$.  The second implication is because $S \neq T$ implies either $S \not \subseteq T$ or $T \not \subseteq S$.

    It remains to prove the uniqueness statement.  Suppose~$f$ has two representations satisfying~\eqref{item:first-orthog} and~\eqref{item:orthog-decomp-ortho}.  By subtracting them we get a decomposition of the~$0$ function that satisfies~\eqref{item:first-orthog} and~\eqref{item:orthog-decomp-ortho}; our goal is to show that each function in this decomposition is the~$0$ function.  We can do this by showing that any decomposition satisfying~\eqref{item:first-orthog} and~\eqref{item:orthog-decomp-ortho} also satisfies ``Parseval's Theorem'': $\la f, f \ra = \sum_{S \subseteq [n]} \|f^{=S}\|_2^2$.  But this is an easy consequence of~\eqref{item:orthog-decomp-ortho3}, which we just noted is itself a consequence of~ \eqref{item:first-orthog} and~\eqref{item:orthog-decomp-ortho}.
\end{proof}

We can connect the orthogonal decomposition of~$f$ to its expansion under Fourier bases as follows:
\begin{proposition}                                     \label{prop:orthog-decomp-basis}
    Let $f \in L^2(\Omega^n, \pi\xn)$ have orthogonal decomposition $f = \sum_{S \subseteq [n]} f^{=S}$.  Fix any Fourier basis $\phi_0, \dots, \phi_{m-1}$ for $L^2(\Omega, \pi)$.  Then
    \begin{equation} \label{eqn:orthog-decomp-basis}
        f^{=S} = \sum_{\substack{\alpha \in \N^n_{< m} \\ \supp(\alpha) = S}} \wh{f}(\alpha)\,\phi_\alpha.
    \end{equation}
\end{proposition}
\begin{proof}
    This follows easily from the uniqueness part of Theorem~\ref{thm:orthogonal-decomposition}. If we take~\eqref{eqn:orthog-decomp-basis} as the definition of functions~$f^{=S}$, it is immediate that $\sum_S f^{=S} = f$ and that $f^{=S}$ depends only on the coordinates in~$S$.  Further, if~$g$ depends only on coordinates $T \subsetneq S$, then $f^{=S}$ and~$g$ have disjoint Fourier support by Corollary~\ref{cor:f-depends}; hence $\la f^{=S}, g \ra = 0$ by Plancherel (Proposition~\ref{prop:general-plancherel-etc}).
\end{proof}
\begin{example} \label{eg:orthog-decomp1}
    Let's compute the orthogonal decomposition of the function $f \co \{a,b,c\}^2 \to \{0,1\}$ from Example~\ref{eg:and2-3ary}.  Recall that in this example $\{a,b,c\}$ has the uniform distribution and $f(x_1,x_2) = 1$ if and only if $x_1 = x_2 = c$. First,
    \[
        f^{=\emptyset} = \E[f] = \tfrac{1}{9}.
    \]
    Next, for $i = 1,2$ we have that $f^{\subseteq \{i\}}(x)$ is~$\frac13$ if $x_i = c$ and $0$ otherwise; hence
    \[
        f^{= \{i\}}(x_1,x_2) = \begin{cases}
                                    +\frac{2}{9} & \text{if } x_i = c, \\
                                    -\frac{1}{9} & \text{else.}
                               \end{cases}
    \]
    Finally, it's easiest to compute $f^{=\{1,2\}}$ as $f - f^{=\emptyset} - f^{=\{1\}} - f^{=\{2\}}$; this yields
    \[
        f^{= \{1,2\}}(x_1,x_2) = \begin{cases}
                                    +\frac{4}{9} & \text{if } x_1 = x_2 = c, \\
                                    -\frac{2}{9} & \text{if exactly one of } x_1,\ x_2 \text{ is } c, \\
                                    +\frac{1}{9} & \text{if } x_1, x_2 \neq c.
                               \end{cases}
    \]
    You can check (Exercise~\ref{ex:orthog-decomp1}) that this is consistent with Proposition~\ref{prop:orthog-decomp-basis} and the Fourier expansion from Example~\ref{eg:and2-3ary}.
\end{example}
                                                \index{orthogonal decomposition|)}%

We can write all of the Fourier formulas from Section~\ref{sec:general-fourier-formulas} in terms of the orthogonal decomposition; e.g.,
\[
    \la f, g \ra = \sum_{S \subseteq [n]} \la f^{=S}, g^{=S} \ra, \quad \Inf_i[f] = \sum_{S \ni i} \|f^{=S}\|_2^2, \quad \T_\rho f = \sum_{S \subseteq [n]} \rho^{|S|} f^{=S}.
\]
These formulas can be proved either by using the connection from Proposition~\ref{prop:orthog-decomp-basis} or by reasoning directly from the defining Theorem~\ref{thm:orthogonal-decomposition}; see Exercise~\ref{ex:orthog-decomp-work}.  The orthogonal decomposition also gives us the natural way of stratifying~$f$ by degree; we end this section by generalizing some more definitions from Chapter~\ref{sec:basic-fourier-formulas}:
\begin{definition}
    For $f \in L^2(\Omega^n, \pi\xn)$ and $k \in \N$ we define the \emph{degree~$k$ part of~$f$} to be
                                                    \index{degree~$k$ part!general product space}%
                                                    \index{Fourier weight!general product space}%
    $f^{=k} = \sum_{|S| = k} f^{=S}$ and the \emph{weight of~$f$ at degree~$k$} to be $\W{k}[f] = \|f^{=k}\|_2^2$.  We also use notation like $f^{\leq k} = \sum_{|S| \leq k} f^{=S}$ and $\W{> k}[f] = \sum_{|S| > k} \|f^{=S}\|_2^2$.
\end{definition}

                                                \index{product space domains|)}%

\section{$p$-biased analysis}                                           \label{sec:p-biased}
                                            \index{p-biase Fourier analysis@$p$-biased Fourier analysis|seeonly{biased Fourier analysis}}%
                                            \index{biased Fourier analysis|(}%
Perhaps the most common generalized domain in analysis of Boolean functions is the case of the hypercube with ``biased'' bits.  In this setting we think of a random input in $\bn$ as having each bit independently equal to $-1$~($\true$) with probability~$p \in (0,1)$ and equal to $1$~($\false$) with probability~$q = 1-p$.  (We could also consider different parameters~$p_i$ for each coordinate; see Exercise~\ref{ex:general-p-biased}.)  In the notation of the chapter this means $L^2(\Omega^n, \pi_p\xn)$, where $\Omega = \{-1,1\}$ and $\pi_p$ is the distribution on~$\Omega$ defined by $\pi_p(-1) = p$, $\pi_p(1) = q$.
                                        \nomenclature[pip]{$\pi_p$}{the ``$p$-biased'' distribution on bits: $\pi_p(-1) = p$, $\pi_p(1) = 1-p$}%
This context is often referred to as \emph{$p$-biased Fourier analysis}, though it would be more consistent with our terminology if it were called ``$\mu$-biased'', where
\[
    \mu = \E_{\bx_i \sim \pi_p}[\bx_i] = q-p = 1-2p.
\]
One of the more interesting features of the setting is that we can fix a combinatorial Boolean function $f \btb$ and then consider its properties for various~$p$ between~$0$ and~$1$; we will discuss this further later in this section.  We will also sometimes use the abbreviated notation $\Pr_{\pi_p}[\cdot]$ in place of $\Pr_{\bx \sim \pi_p\xn}[\cdot]$, and similarly $\E_{\pi_p}[\cdot]$.
                                        \nomenclature[Prpip]{$\Pr_{\pi_p}[\cdot]$}{an abbreviation for $\Pr_{\bx \sim \pi_p\xn}[\cdot]$}%
                                        \nomenclature[Epip]{$\E_{\pi_p}[\cdot]$}{an abbreviation for $\E_{\bx \sim \pi_p\xn}[\cdot]$}%

The $p$-biased hypercube is one of the generalized domains where it can pay to look at an explicit Fourier basis.  In fact, since we have $|\Omega| = 2$ there is a \emph{unique} Fourier basis $\{\phi_0, \phi_1\}$ (up to negating~$\phi_1$).  For notational simplicity we'll write $\phi$ instead of $\phi_1$ and use ``set notation'' rather than multi-index notation:
\begin{definition}                                  \label{def:p-biased-phi}
    In the context of $p$-biased Fourier analysis we define the basis function $\phi \co \bits \to \R$ by
    \[
        \phi(x_i) = \frac{x_i - \mu}{\sigma},
    \]
    where
    \[
        \mu = \E_{\bx_i \sim \pi_p}[\bx_i] = q-p = 1- 2p, \quad \sigma = \stddev_{\bx_i \sim \pi_p}[\bx_i] = \sqrt{4pq} = 2\sqrt{p}\sqrt{1-p}.
    \]
    Note that $\sigma^2 = 1 - \mu^2$.  We also have the formula $\phi(1) = \sqrt{p/q}$, $\phi(-1) = -\sqrt{q/p}$.
\end{definition}
We will use the notation $\mu$ and $\sigma$ throughout this section. It's clear that $\{1, \phi\}$ is indeed a Fourier basis for $L^2(\bits, \pi_p)$ because $\E[\phi(\bx_i)] = 0$ and $\E[\phi(\bx_i)^2] = 1$ by design.
\begin{definition}
    In the context of $L^2(\bits^n, \pi_p\xn)$ we define the product Fourier basis functions $(\phi_S)_{S \subseteq [n]}$ by
    \[
        \phi_S(x) = \prod_{i \in S} \phi(x_i).
    \]
    Given $f \in L^2(\bits^n, \pi_p\xn)$ we write $\wh{f}(S)$ for the associated Fourier coefficient; i.e.,
    \[
        \wh{f}(S) = \Ex_{\bx \sim \pi_p\xn}[f(\bx)\,\phi_S(\bx)].
    \]
    Thus we have the biased Fourier expansion
    \[
        f(x) = \sum_{S \subseteq [n]} \wh{f}(S)\,\phi_S(x).
    \]
\end{definition}
Although the notation is very similar to that of the classic uniform-distribution Fourier analysis, we caution that in general,
\[
    \phi_S \phi_T \neq \phi_{S \symdiff T}.
\]
\begin{example}  Let $\chi_i \in L^2(\bits^n, \pi_p\xn)$ be the $i$th dictator function, $\chi_i(x) = x_i$, viewed under the $p$-biased distribution.  We have
    \[
        \phi(x_i) = \frac{x_i - \mu}{\sigma} \quad\implies\quad x_i = \mu + \sigma\phi(x_i),
    \]
    and the latter is evidently $f$'s (biased) Fourier expansion.
                                                \index{dictator!biased Fourier analysis}%
    That is,
    \[
        \wh{\chi_i}(\emptyset) = \mu, \quad \wh{\chi_i}(\{i\}) = \sigma, \quad \wh{\chi_i}(S) = 0 \text{ otherwise}.
    \]
\end{example}

This example lets us see a link between a function's ``usual'' Fourier expansion and its biased Fourier expansion.  (For more on this, see Exercise~\ref{ex:usual-vs-biased-expansion}.) Let's abuse notation a little by writing simply~$\phi_i$ instead of $\phi(x_i)$.  We have the formulas
\begin{equation} \label{eqn:p-biased-substitution}
    \phi_i = \frac{x_i - \mu}{\sigma} \quad\iff\quad x_i = \mu + \sigma \phi_i,
\end{equation}
and we can go from the usual Fourier expansion to the biased Fourier expansion simply by plugging in the latter.
\begin{example}
    Recall the ``selection function'' $\SEL \co \bits^3 \to \bits$ from Exercise~\ref{ex:compute-expansions}\ref{ex:SEL}; $\SEL(x_1, x_2, x_3)$ outputs~$x_2$ if $x_1 = -1$ and outputs~$x_3$ if $x_1 = 1$.  The usual Fourier expansion of $\SEL$ is
    \[
        \SEL(x_1, x_2, x_3) =
        \half x_2 + \half x_3 -\half x_1 x_2 + \half x_1 x_3.
    \]
    Using the substitution from~\eqref{eqn:p-biased-substitution} we get
    \begin{align}
        \SEL(x_1, x_2, x_3) &= \half (\mu + \sigma \phi_2) + \half (\mu + \sigma \phi_3) -\half (\mu + \sigma \phi_1) (\mu + \sigma \phi_2) + \half (\mu + \sigma \phi_1) (\mu + \sigma \phi_3) \nonumber\\
        &= \mu + (\half - \half \mu)\sigma \, \phi_{2} + (\half + \half \mu) \sigma \, \phi_{3} - \half \sigma^2 \, \phi_1 \phi_2 + \half \sigma^2 \, \phi_1 \phi_3. \label{eqn:sel-biased}
    \end{align}
    Thus if we write $\SEL\subp$ for the selection function thought of as an element of $L^2(\bits^3, \pi_p^{\otimes 3})$, we have
    \begin{gather*}
        \wh{\SEL\subp}(\emptyset) = \mu, \quad  \wh{\SEL\subp}(2) = (\half - \half \mu)\sigma, \quad  \wh{\SEL\subp}(3) = (\half + \half \mu)\sigma, \\
        \wh{\SEL\subp}(\{1,2\}) = -\half\sigma^2, \quad        \wh{\SEL\subp}(\{1,3\}) = \half\sigma^2, \quad \wh{\SEL\subp}(S) = 0 \text{ else}.
    \end{gather*}
    By the Fourier formulas of Section~\ref{sec:general-fourier-formulas} we can deduce, e.g., that $\E[\SEL\subp] = \mu$, $\Inf_1[\SEL\subp] = (-\half \sigma^2)^2 + (\half \sigma^2)^2 = \half \sigma^4$, etc.
\end{example}
Let's codify a piece of notation from this example:
\begin{notation}  \label{not:supmu}
    Let $f \btR$ and let $p \in (0,1)$.  We write $f\subp$ for the function when viewed as an element of $L^2(\bn, \pi_{p}\xn)$.
\end{notation}

\medskip
                                            \index{derivative operator!biased Fourier analysis|(}%
We now discuss derivative operators.  We would like to define an operator~$\D_i$ on $L^2(\bn, \pi_p\xn)$
that acts like differentiation on the biased Fourier expansion.  For example, referring to~\eqref{eqn:sel-biased} we would like to have
\[
    \D_3 \SEL\subp = (\half + \half \mu) \sigma +  \half \sigma^2 \, \phi_1.
\]
In general we are seeking $\frac{\partial}{\partial \phi_i}$ which, by basic calculus and the relationship~\eqref{eqn:p-biased-substitution}, satisfies
\[
    \frac{\partial}{\partial \phi_i} =     \frac{\partial x_i}{\partial \phi_i}  \cdot \frac{\partial}{\partial x_i} = \sigma \cdot  \frac{\partial}{\partial x_i}.
\]
Recognizing $\frac{\partial}{\partial x_i}$ as the ``usual'' $i$th derivative operator, we are led to the following:
\begin{definition}                          \label{def:biased-deriv}
    For $i \in [n]$, the \emph{$i$th (discrete) derivative} operator~$\D_i$ on \linebreak $L^2(\bn, \pi_p\xn)$ is defined by
    \[
        \D_i f(x) = \sigma \cdot   \frac{f(x^{(i\mapsto 1)}) - f(x^{(i \mapsto -1)})}{2}.
    \]
    Note that this defines a different operator for each value of~$p$.  We sometimes write the above definition as
    \[
        \D_{\phi_i} = \sigma \cdot \D_{x_i}.
    \]
    With respect to the biased Fourier expansion of $f \in L^2(\bits^n, \pi_p\xn)$ the operator $\D_i$ satisfies
    \begin{equation}                                        \label{eqn:p-biased-deriv}
        \D_i f = \sum_{S \ni i} \wh{f}(S)\,\phi_{S \setminus \{i\}}.
    \end{equation}
\end{definition}
                                            \index{derivative operator!biased Fourier analysis|)}%

Given this definition we can derive some additional formulas for influences,
                                                \index{influence!biased Fourier analysis}%
including a generalization of Proposition~\ref{prop:monotone-influences}:
\begin{proposition}                                     \label{prop:monotone-influences-biased}
    Suppose $f \in L^2(\bn, \pi_p\xn)$ is Boolean-valued (i.e., has range $\{-1,1\}$).  Then
    \[
        \Inf_i[f] = \sigma^2 \Pr_{\bx \sim \pi_p\xn}[f(\bx) \neq f(\bx^{\oplus i})]
    \]
    for each $i \in [n]$, and
    \[
        \Tinf[f] =  \sigma^2 \E_{\bx \sim \pi_p\xn}[\sens_f(\bx)].
    \]
    If furthermore~$f$ is monotone, then $\Inf_i[f] = \sigma \wh{f}(i)$.
\end{proposition}
\begin{proof}
    Using Definition~\ref{def:biased-deriv}'s notation we have
    \[
        \Inf_i[f] = \E_{\pi_p}[(\D_{\phi_i} f)^2] = \sigma^2 \E_{\pi_p}[(\D_{x_i} f)^2].
    \]
    Since $(\D_{x_i} f)^2$ is the $0$-$1$ indicator that~$i$ is pivotal for~$f$, the first formula follows.  The second formula follows by summing over~$i$.  Finally, when~$f$ is monotone we furthermore have that $(\D_{x_i} f)^2 = \D_{x_i} f$ and hence
    \[
        \Inf_i[f] = \sigma^2 \E_{\pi_p}[\D_{x_i} f] = \sigma \E_{\pi_p}[\D_{\phi_i} f] = \sigma \wh{f}(i),
    \]
    as claimed.
\end{proof}

\medskip

                                                \index{threshold phenomena|(}%
The remainder of this section is devoted to the topic of \emph{threshold phenomena} in Boolean functions.  Much of the motivation for this comes from theory of random graphs, which we now briefly introduce.
\begin{definition}
    Given an undirected graph $G$ on $v \geq 2$ vertices, we identify it with the string in $\{\True, \False\}^{\binom{v}{2}}$ which indicates which edges are present ($\True$) and which are absent~($\False$).  We write $\calG(v,p)$ for the distribution $\pi_{p}^{\otimes \binom{v}{2}}$; this is called the \emph{Erd\H{o}s--R\'{e}nyi random graph model}.
                                                    \index{random graph}%
                                                    \index{Erd\H{o}s--R\'{e}nyi random graph|seeonly{random graph}}%
                                                    \nomenclature[Gvp]{$\calG(v,p)$}{the Erd\H{o}s--R\'{e}nyi random graph distribution, $\pi_p^{\otimes \binom{v}{2}}$}%
    Note that if we permute the~$v$ vertices of a graph, this induces a permutation on the~$\binom{v}{2}$ edges.  \linebreak A ($v$-vertex) \emph{graph property} is a Boolean function $f \co \{\True, \False\}^{\binom{v}{2}} \to \{\True, \False\}$ that is invariant under all~$v!$ such permutations of its input; colloquially, this means that~$f$ ``does not depend on the names of the vertices''.
                                                    \index{graph property}%
\end{definition}
\noindent Graph properties are always transitive-symmetric functions in the sense of
                                        \index{transitive-symmetric function}%
Definition~\ref{def:transitive-symmetric}.
\begin{examples}
    The following are all $v$-vertex graph properties:
    \begin{align*}
        \text{Conn}(G) &= \True \text{ if $G$ is connected;} \\
        \text{3Col}(G) &= \True \text{ if $G$ is $3$-colorable;} \\
        \text{Clique}_{k}(G) &= \True \text{ if $G$ is contains a clique on at least $k$ vertices;} \\
        \maj_{n}(G) &= \True \textstyle \text{ (assuming $n = \binom{v}{2}$ is odd) if $G$ has at least $\binom{v}{2}/2$ edges;} \\
        \chi_{[n]}(G) &= \True \text{ if $G$ has an odd number of edges.} \\
    \end{align*}
    Note that each of these actually defines a family of Boolean functions, one for each value of~$v$; this is the typical situation in the study of graph properties.  An example of a function $f \co \{\True,\False\}^{\binom{v}{2}} \to \{\True, \False\}$ that is \emph{not} a graph property is the one defined by $f(G) = \True$ if vertex~\#$1$ has at least one neighbor; this~$f$ is not invariant under permuting the vertices.
\end{examples}
Graph properties which are \emph{monotone} are particularly nice to study;
                                        \index{graph property!monotone}%
                                        \index{monotone graph property|seeonly{graph property, monotone}}%
these are the ones for which adding edges can never make the property go from~$\True$ to~$\False$.  The properties $\text{Conn}$, $\text{Clique}_{k}$, and $\Maj_n$ defined above are all monotone, as is $\neg\text{3Col}$.  Now suppose we take a monotone graph property, say, $\text{Conn}$. A typical question in random graph theory would be, ``how many edges does a graph need to have before it is likely to be connected?'' Or more precisely, how does $\Pr_{\bG \sim \calG(v,p)}[\text{Conn}(\bG) = \True]$ vary as~$p$ increases from~$0$ to~$1$?

There's no need to ask this question just for graph properties.  Given any monotone Boolean function $f \co \{\True,\False\}^n \to \{\True, \False\}$ it is intuitively clear that when~$p$ increases from~$0$ to~$1$ this causes $\Pr_{\pi_p} [f(\bx) = \True]$ to increase from~$0$ to~$1$ (unless~$f$ is a constant function).  As illustration, we show a plot of $\Pr_{\pi_p} [f(\bx) = \True]$ versus~$p$ for the dictator function, $\AND_2$, and $\Maj_{101}$.

\myfig{.5}{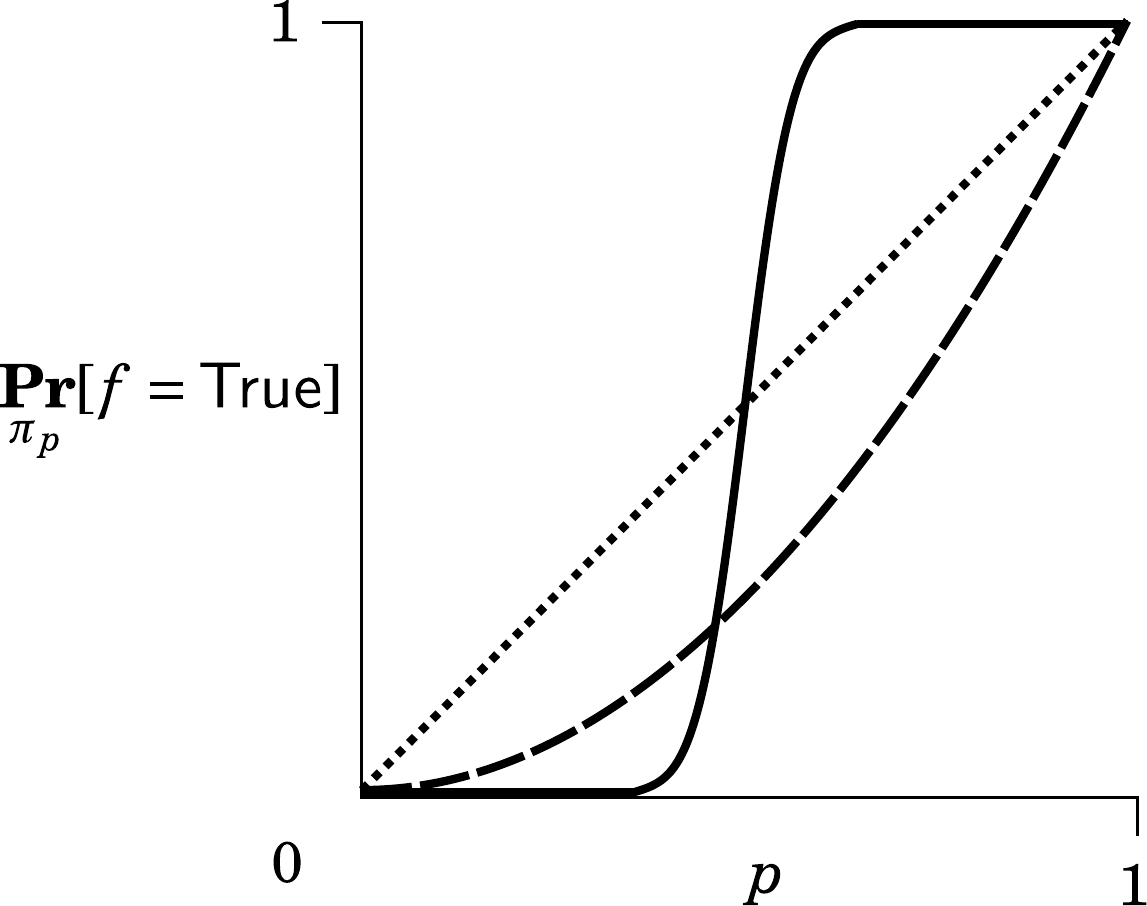}{Plot of $\Pr_{\pi_p} [f(\bx) = \True]$ versus $p$ for $f$ a dictator (dotted), $f = \AND_2$ (dashed), and $f = \Maj_{101}$ (solid)}{fig:threshold-examples}

The \emph{Margulis--Russo Formula} quantifies the rate at which $\Pr_{\pi_p} [f(\bx) = \True]$ increases with~$p$; specifically, it relates the slope of the curve at~$p$ to the total influence of~$f$ under $\pi_p\xn$.  To prove the formula we switch to~$\pm 1$ notation.
                                                    \index{Russo--Margulis Formula|seeonly{Margulis--Russo Formula}}%
                                                    \index{Margulis--Russo Formula}%
\begin{named}{Margulis--Russo Formula}
    Let $f \btR$.  Recalling Notation~\ref{not:supmu} and the relation $\mu = 1-2p$, we have
    \begin{equation} \label{eqn:marg-russo1}
        \frac{d}{d\mu}\E[f\subp] = \frac{1}{\sigma} \cdot \sum_{i=1}^n \wh{f\subp}(i).
    \end{equation}
    In particular, if $f \btb$ is monotone, then
    \begin{equation} \label{eqn:marg-russo2}
        \frac{d}{dp}\,\Pr_{\bx \sim \pi_p\xn}[f(\bx) = -1] = \frac{d}{d\mu}\E[f\subp]  = \frac{1}{\sigma^2} \cdot \Tinf[f\subp].
    \end{equation}
\end{named}
\begin{proof}
    Treating $f$ as a multilinear polynomial over $x_1, \dots, x_n$ we have
    \[
        \E[f\subp] = \T_\mu f(1, \dots, 1) = f(\mu, \dots, \mu)
    \]
    (this also follows from Exercise~\ref{ex:multilin-interp}).  By basic calculus,
    \[
        \frac{d}{d\mu}f(\mu, \dots, \mu) = \sum_{i=1}^n \D_{x_i} f(\mu, \dots, \mu).
    \]
    But
    \[
        \D_{x_i} f(\mu, \dots, \mu) = \E[\D_{x_i} f\subp] = \frac{1}{\sigma} \E[\D_{\phi_i} f\subp] = \frac{1}{\sigma} \wh{f\subp}(i),
    \]
    completing the proof of~\eqref{eqn:marg-russo1}.  As for~\eqref{eqn:marg-russo2}, the second equality follows immediately from Proposition~\ref{prop:monotone-influences-biased}. The first equality holds because $\mu = 1-2p$ and $\E[f] = 1-2\Pr[f = -1]$; the two factors of~$-2$ cancel.
\end{proof}
\begin{remark}
    If $f \co \{\True,\False\}^n \to \{\True, \False\}$ is a nonconstant monotone function, the Margulis--Russo Formula implies that $\Pr_{\pi_p} [f(\bx) = \True]$ is a strictly increasing function of~$p$, because $\Tinf[f^{(p)}]$ is always  positive.
\end{remark}

Looking again at Figure~\ref{fig:threshold-examples} we see that the plot for $\Maj_{101}$ looks very much like a step function, jumping from nearly~$0$ to nearly~$1$ around the critical value $p = 1/2$. For $\Maj_n$, this ``sharp threshold at~$p = 1/2$''
                                                \index{threshold, sharp}%
                                                \index{sharp threshold|seeonly{threshold, sharp}}%
becomes more and more pronounced as~$n$ increases.  This is clearly suggested by the Margulis--Russo Formula: the derivative of the curve at~$p = 1/2$ is equal to $\Tinf[\Maj_n]$ (the usual, uniform-distribution total influence), which has the very large value~$\Theta(\sqrt{n})$ (Theorem~\ref{thm:maj-maximizes-deg-1-sum}).  Such sharp thresholds exist for many Boolean functions; we give some examples:
\begin{examples} \label{eg:sharp-thresh}
    In Exercise~\ref{ex:condorcet-jury-theorem} you are asked to show that for every $\eps > 0$ there is a~$C$ such that
    \[
        \Pr_{\pi_{1/2 - C/\sqrt{n}}}[\Maj_n = \True] \leq \eps, \quad                 \Pr_{\pi_{1/2 + C/\sqrt{n}}}[\Maj_n = \True] \geq 1 - \eps.
    \]
    Regarding the Erd\H{o}s--R\'{e}nyi graph model, the following facts are known:
    \begin{align*}
        \Pr_{\bG \sim \calG(v,p)}[\text{Clique}_{\log v}(\bG) = \True] &\xrightarrow[v \to \infty]{}
            \begin{cases} 0 & \text{if } p < 1/4, \\
                          1 & \text{if } p > 1/4. \end{cases} \\
        \Pr_{\bG \sim \calG(v,p)}[\text{Conn}(\bG) = \True] &\xrightarrow[v \to \infty]{}
            \begin{cases} 0 & \text{if } p < \frac{\ln v}{v}(1-\tfrac{\log \log v}{\log v}), \\
                          1 & \text{if } p > \frac{\ln v}{v}(1+\tfrac{\log \log v}{\log v}). \end{cases} \\
    \end{align*}
\end{examples}
In the above examples you can see that the ``jump'' occurs at various values of~$p$. To investigate this phenomenon, we first single out the value for which $\Pr_{\pi_p}[f(\bx) = \True] = 1/2$:
\begin{definition}                  \label{def:crit-prob}
    Let $f \co \{\True,\False\}^n \to \{\True, \False\}$ be monotone and nonconstant.  The \emph{critical probability} for~$f$, denoted $p_c$, is the unique value in $(0,1)$ for which $\Pr_{\bx \sim \pi_p\xn}[f(\bx) = \True] = 1/2$. We also write $q_c = 1-p_c$, $\mu_c = q_c - p_c = 1-2p_c$, and $\sigma_c = \sqrt{4p_cq_c}$.
\end{definition}
\noindent In Exercise~\ref{ex:crit-defined} you are asked to verify that~$p_c$ is well defined.

Looking at the connectivity property from Example~\ref{eg:sharp-thresh} we see that not only does $\Pr_{\pi_p}[\text{Conn} = \True]$ jump from near~$0$ to near~$1$ in an  interval of the form $p_c \pm o(1)$, it actually makes the jump in an interval of the form $p_c(1 \pm o(1))$.  This latter phenomenon is (roughly speaking) what is meant by a ``sharp threshold''.  To investigate this further, suppose that $f$ is a (nonconstant) monotone function and $\Delta$ is the derivative of $\Pr_{\pi_p}[f(\bx) = \True]$ at $p = p_c$.  Intuitively, we would expect $\Pr_{\pi_p}[f(\bx) = \True]$ to jump from near~$0$ to near~$1$ in an interval of around~$p_c$ of width about~$1/\Delta$.  Thus a ``sharp threshold''
                                                \index{threshold, sharp}%
should roughly correspond to the case that $1/\Delta$ is small even compared to $\min(p_c, q_c)$.  The Margulis--Russo Formula says that $\Delta = \frac{1}{\sigma_c^2} \Tinf[f^{(p_c)}]$, and since $\min(p_c,q_c)$ is proportional to $4p_cq_c = \sigma_c^2$ it follows that $1/\Delta$ is ``small'' compared to $\min(p_c,q_c)$ if and only if~$\Tinf[f^{(p_c)}]$ is ``large''.  Thus we have a neat criterion:

\noindent \textbf{Sharp threshold principle:} \emph{Let $f \co \tf^n \to \tf$ be monotone.  Then, roughly speaking, $\Pr_{\pi_p}[f(\bx) = \True]$ has a ``sharp threshold'' if and only if~$f$ has ``large'' (``superconstant'') total influence under its critical probability distribution.}

Of course this should all be made a bit more precise; see Exercise~\ref{ex:precise-sharp-thresh} for details.  In light of this principle, we may try to prove that a given~$f$ has a sharp threshold by proving that $\Tinf[f^{(p_c)}]$ is not ``small''.  In turn, this strongly motivates the problem of ``characterizing'' Boolean-valued functions \linebreak $f \in L^2(\bn, \pi_p\xn)$ for which $\Tinf[f]$ is small.  Friedgut's Junta Theorem, mentioned at the end of Chapter~\ref{sec:low-degree} and proved in Chapter~\ref{sec:KKL}, tells us that in the uniform distribution case $p = 1/2$, the only way $\Tinf[f]$ can be small is if~$f$ is close to a junta.  In particular, any monotone graph property with $p_c = 1/2$ must have a very large derivative $\frac{d}{dp}\,\Pr_{\pi_p}[f = \True]$ at $p = p_c$: since the function is transitive-symmetric, all~$n$ coordinates are equally influential and it can't be close to a junta.  These results also hold so long as~$p$ is bounded away from~$0$ and~$1$; see Chapter~\ref{sec:hypercon-variants-apps}.  However, many interesting monotone graph properties have~$p_c$ very close to~$0$: e.g., connectivity, as we saw in Example~\ref{eg:sharp-thresh}.  Characterizing the functions $f \in L^2(\bn, \pi_p\xn)$ with small $\Tinf[f]$ when $p = o_n(1)$ is a trickier task; see the work of Friedgut, Bourgain, and Hatami described in Chapter~\ref{sec:friedgut-bourgain-hatami}.

\section{Abelian groups}                                            \label{sec:abelian}

            \nomenclature[C]{$\C$}{the complex numbers}%
            \nomenclature[Zm]{$\Z_m$}{the additive group of integers modulo~$m$}%

The previous section covered the case of $f \in L^2(\Omega^n, \pi\xn)$ with $|\Omega| = 2$; there, we saw it could be helpful to look at explicit Fourier bases.  When $|\Omega| \geq 3$ this is often \emph{not} helpful, especially if the only ``operation'' on the domain is equality. For example, if $f \co \{\textsf{Red}, \textsf{Green}, \textsf{Blue}\}^n \to \R$, then it's best to just work abstractly with the orthogonal decomposition.  However, if there is a notion of, say, ``addition'' in~$\Omega$, then there is a natural, canonical Fourier basis for $L^2(\Omega, \pi)$ when~$\pi$ is the uniform distribution.

More precisely, suppose the domain $\Omega$ is a finite abelian group~$G$, with operation~$+$ and identity~$0$.  We will consider the domain~$G$ under the uniform probability distribution~$\pi$; this is quite natural because~$\pi$ is \emph{translation-invariant}: $\pi(X) = \pi(t + X)$ for any $X \subseteq G$, $t \in G$. In this setting it is more convenient to allow functions with range the complex numbers; thus we come to the following definition:
\begin{definition}
    Let $G$ be a finite abelian group with operation~$+$ and identity~$0$.
    For $n \in \N^+$ we write $L^2(G^n)$
                                                    \nomenclature[L2Gn]{$L^2(G^n)$}{if $G$ is a finite abelian group, denotes the complex inner product space of functions $G^n \to \R$ with inner product $\la f, g \ra = \Ex_{\bx \sim G^n}[f(\bx)\overline{g(\bx)}]$}%
    for the complex inner product space of functions $f \co G^n \to \C$, with inner product
    \[
        \la f, g \ra = \Ex_{\bx \sim G^n}[f(\bx)\ol{g(\bx)}].
    \]
    Here and throughout this section $\bx \sim G^n$ denotes that~$\bx$ is drawn from the uniform distribution on~$G^n$.
\end{definition}
Everything we have done in this chapter for the real inner product space $L^2(\Omega^n, \pi\xn)$ generalizes easily to the case of a complex inner product; the main difference is that Plancherel's Theorem
                                                \index{Plancherel's Theorem!complex case}%
becomes
\[
    \la f, g \ra = \sum_{\alpha \in \N^n_{< m}} \wh{f}(\alpha) \ol{\wh{g}(\alpha)} = \sum_{S \subseteq [n]} \la f^{=S}, g^{=S} \ra.
\]
See Exercise~\ref{ex:complex-functions} for more.

                                                \index{character|(}%
A natural Fourier basis for $L^2(G)$ comes from a natural family of functions $G \to \C$, namely the \emph{characters}. These are defined to be the group homomorphisms from $G$ to $\C^\times$, where $\C^\times$ is the abelian group of nonzero complex numbers under multiplication.
\begin{definition}
    A \emph{character} of the (finite) group $G$ is a function $\chi \co G \to \C^\times$ which is a homomorphism; i.e., satisfies $\chi(x + y) = \chi(x) \chi(y)$.  Since $G$ is finite there is some $m \in \N^+$ such that $0 = x + x + \cdots + x$ ($m$~times) for each $x \in G$.  Thus $1 = \chi(0) = \chi(x)^m$, meaning the range of~$\chi$ is in fact contained in the $m$th roots of unity.  In particular, $|\chi(x)| = 1$ for all $x \in G$.
\end{definition}
We have the following easy facts:
\begin{fact}                                        \label{fact:more-characters}
    If $\chi$ and $\phi$ are characters of~$G$, then so are $\ol{\chi}$ and $\phi \cdot \chi$.
\end{fact}
\begin{proposition}                                        \label{prop:char-orthog}
    Let $\chi$ be a character of $G$.  Then either $\chi \equiv 1$ or $\E[\chi] = 0$.
\end{proposition}
\begin{proof}
    If $\chi \not \equiv 1$, pick some $y \in G$ such that $\chi(y) \neq 1$.  Since $\bx + y$ is uniformly distributed on~$G$ when $\bx \sim G$,
    \[
        \Ex_{\bx \sim G}[\chi(\bx)] = \Ex_{\bx \sim G}[\chi(\bx+y)] = \Ex_{\bx \sim G}[\chi(\bx)\chi(y)] = \chi(y)\Ex_{\bx \sim G}[\chi(\bx)].
    \]
    Since $\chi(y) \neq 1$ it follow that $\E[\chi(\bx)]$ must be~$0$.
\end{proof}
\begin{proposition}                                     \label{prop:char-orthon}
    The set of all characters of $G$ is orthonormal.  (As a consequence, $G$ has at most~$\dim(L^2(G)) = |G|$ characters.)
\end{proposition}
\begin{proof}
    First, if $\chi$ is a character, then $\la \chi, \chi \ra = \E[|\chi|^2] = 1$ because $|\chi| \equiv 1$.  Next, if~$\phi$ is another character distinct from~$\chi$ then  $\la \phi, \chi \ra = \E[\phi \cdot \ol{\chi}]$.  But $\phi \cdot \ol{\chi}$ is a character by  Fact~\ref{fact:more-characters}, and $\phi \cdot \ol{\chi} = \phi/\chi \not \equiv 1$ because $\phi$ and $\chi$ are distinct; here we used $\ol{\chi} = 1/\chi$ because $|\chi| \equiv 1$.  Thus $\la \phi, \chi \ra = 0$ by Proposition~\ref{prop:char-orthog}.
\end{proof}
As we will see next, $G$ in fact has exactly~$|G|$ characters.  It thus follows from Proposition~\ref{prop:char-orthon} that the set of all characters (which includes the constant~$1$ function) constitutes a Fourier basis for~$L^2(G)$.

To check that each finite abelian group~$G$ has $|G|$ distinct characters, we begin with the case of a cyclic group, $\Z_m$ for some~$m$.  In this case we know that every character's range will be contained in the $m$th roots of unity.
\begin{definition}                                     \label{def:Zm-chars}
    Fix an integer $m \geq 2$ and write $\omega$ for the $m$th root of unity $\exp(2\pi i/m)$.  For $0 \leq j < m$, we define $\chi_j \co \Z_m \to \C$ by $\chi_j(x) = \omega^{jx}$.  It is easy to see that these are distinct characters of~$\Z_m$.
\end{definition}
Thus the functions $\chi_0 \equiv 1, \chi_1, \dots, \chi_{m-1}$ form a Fourier basis for $L^2(\Z_m)$.
Furthermore, Proposition~\ref{prop:fbasis-product} tells us that we can get a Fourier basis for $L^2(\Z_m^n)$ by taking all products of these functions.
\begin{definition}                                     \label{def:Zmn-basis}
    Continuing Definition~\ref{def:Zm-chars}, let $n \in \N^+$.  For $\alpha \in \N_{< m}^n$ we define $\chi_\alpha \co \Z_m^n \to \C$ by
    \[
        \chi_{\alpha}(x) = \prod_{j =1}^n \chi_{\alpha_j}(x_j).
    \]
    These functions are easily seen to be (all of the) characters of the group~$\Z_m^n$, and they constitute a Fourier basis of~$L^2(\Z_m^n)$.
\end{definition}
Most generally, by the Fundamental Theorem of Finitely Generated Abelian Groups we know that any finite abelian~$G$ is a direct product of cyclic groups of prime-power order.  In Exercise~\ref{ex:general-group-duality} you are asked to check that you get all of the characters of~$G$ -- and hence a Fourier basis for~$L^2(G)$ -- by taking all products of the associated cyclic groups' characters.  In the remainder of the section we mostly stick to groups of the form~$\Z_m^n$ for simplicity.

Returning to the characters $\chi_0, \dots, \chi_{m-1}$ from Definition~\ref{def:Zm-chars}, it is easy to see (using $\omega^m = 1$) that they satisfy $\chi_{j} \cdot \chi_{j'} = \chi_{j + j' \pmod m}$ and also $1/\chi_{j} = \ol{\chi_j} = \chi_{-j \pmod m}$.  Thus the characters themselves form a group under multiplication, isomorphic to~$\Z_m$.  As in Chapter~\ref{sec:dec-trees}, we index them using the notation~$\wh{\Z_m}$.
                                                \index{dual group}%
More generally, indexing the Fourier basis/characters of~$L^2(\Z_m^n)$ by~$\wh{\Z_m^n}$ instead of multi-indices, we have:
\begin{fact}                                        \label{fact:Zmn-duality}
    The characters $(\chi_{\alpha})_{\alpha \in \wh{\Z_m^n}}$ of $\Z_m^n$ form a group under multiplication:
    \begin{itemize}
        \item $\chi_{\alpha} \cdot \chi_{\beta} = \chi_{\alpha + \beta}$,
        \item $1/\chi_{\alpha} = \ol{\chi_{\alpha}} = \chi_{-\alpha}$.
    \end{itemize}
\end{fact}
                                                \index{character|)}%

As mentioned, the salient feature of $L^2(G)$ distinguishing it from other spaces $L^2(\Omega, \pi)$ is that there is a notion of addition on the domain.  This means that \emph{convolution} plays a major role in its analysis.  We generalize the definition from the setting of~$\F_2^n$:
                                                \index{convolution}
\begin{definition}
    Let $f, g \in L^2(G)$.  Their \emph{convolution} is the function $f \conv g \in L^2(G)$
    defined by
    \begin{equation*}
        (f \conv g)(x) = \Ex_{\by \sim G}[f(\by) g(x - \by)] = \Ex_{\by \sim G}[f(x-\by)g(\by)].
    \end{equation*}
\end{definition}
Exercise~\ref{ex:conv-ac-general} asks you to check that convolution is associative and commutative, and that the following generalization of Theorem~\ref{thm:convolution-theorem} holds:
\begin{theorem}                                     \label{thm:convolution-theorem-general}
    Let $f, g \in L^2(G)$.  Then $\wh{f \conv g}(\alpha) = \wh{f}(\alpha) \wh{g}(\alpha)$.
\end{theorem}

We conclude this section by mentioning vector space domains.  When doing Fourier analysis over the group~$\Z_m^n$, it is natural for subgroups to arise.  Things are simplest when the only subgroups of~$\Z_m$ are the trivial ones, $\{0\}$ and~$\Z_m$; in this case, all subgroups will be isomorphic to $\Z_m^{n'}$ for some $n' \leq n$.  Of course, this simple situation occurs if and only if~$m$ is equal to some prime~$p$.  In that case, $\Z_p$  can be thought of as a field,
                                        \index{Fp@$\F_p$ (finite field)}%
                                        \nomenclature[Fp]{$\F_{p^\ell}$}{for $p$ prime and $\ell \in \N^+$, denotes the finite field of $p^\ell$ elements}%
$\Z_p^n$ as an $n$-dimensional vector space over this field, and its subgroups as subspaces.  We use the notation~$\F_p^n$ in this setting and write~$\wh{\F_p^n}$ to index the Fourier basis/characters; this generalizes the notation introduced for $p = 2$ in Chapter~\ref{sec:dec-trees}.  Indeed, all of the notions from Chapters~\ref{sec:dec-trees} and~\ref{sec:restrictions} regarding affine subspaces and restrictions thereto generalize easily to~$L^2(\F_p^n)$.

\section{Highlight: Randomized decision tree complexity}       \label{sec:monotone-dts}

                                                    \index{decision tree}%
A decision tree~$T$ for~$f \btb$ can be thought of as a deterministic algorithm which, given adaptive query access to the bits of an unknown string $x \in \bn$, outputs~$f(x)$. For example, to describe a natural decision tree for $f = \Maj_3$ in words: ``Query~$x_1$, then~$x_2$. If they are equal, output their value; otherwise, query and output~$x_3$.''  For a worst-case input (one where $x_1 \neq x_2$) this algorithm has a \emph{cost} of~$3$, meaning it makes~$3$ queries.  The cost of the worst-case input is the depth of the decision tree.

As is often the case with algorithms it can be advantageous to allow randomization.  For example, consider using the following randomized query algorithm for $\Maj_3$: ``Choose two distinct input coordinates at random and query them.  If they are equal, output their value; otherwise, query and output the third input coordinate.''  Now for \emph{every} input there is at least a~$1/3$ chance that the algorithm will finish after only~$2$ queries.  Indeed, if we define the cost of an input~$x$ to be the expected number of queries the algorithm makes on it, it is easy to see that the worst-case inputs for this algorithm have cost~$(1/3) \cdot 2 + (2/3) \cdot 3 = 8/3 < 3$.

Let's formalize the notion of a randomized decision tree:
                                       \index{decision tree!randomized}%
\begin{definition}
    Given $f \btR$, a \emph{(zero-error) randomized decision tree}~$\calT$ computing~$f$ is formally defined to be a probability distribution over (deterministic) decision trees that compute~$f$.  The \emph{cost} of~$\calT$ on input $x \in \bn$ is defined to be the expected number of queries $\bT$ makes on~$x$ when $\bT \sim \calT$. The cost of~$\calT$ itself is defined to be the maximum cost of any input.  Finally, the \emph{(zero-error) randomized decision tree complexity} of~$f$, denoted $\RDT(f)$,
                                        \nomenclature[RDT(f)]{$\RDT(f)$}{the zero-error randomized decision tree complexity of $f$}%
    is the minimum cost of a randomized decision tree computing~$f$.
\end{definition}

We can get further savings from randomization if we are willing to assume that the input~$x$ is chosen randomly.  For example, if $\bx \sim \{-1,1\}^3$ is uniformly random then any of the deterministic decision trees for $\Maj_3$ will make~$2$ queries with probability~$1/2$ and~$3$ queries with probability~$1/2$, for an overall expected $5/2 < 8/3 < 3$ queries.
\begin{definition}                                      \label{def:expected-depth}
    Let $\calT$ be a randomized decision tree.
                                       \index{decision tree!expected depth}%
    We define
    \begin{align}
        \delta_i(\calT) &= \Pr_{\substack{\bx \sim \bn,\\ \bT \sim \calT}}[\bT \text{ queries } \bx_i], \nonumber\\
        \Delta(\calT) &= \sum_{i=1}^n \delta_i(\calT) = \E_{\substack{\bx \sim \bn, \\ \bT \sim \calT}}[\# \text{ of coordinates queried by } \bT \text{ on } \bx]. \label{ex:expected-dt-cost}
    \end{align}
                                        \nomenclature[deltaiT]{$\delta_i^{(\pi)}(\calT)$}{the probability randomized decision tree $\calT$ queries coordinate $i$ when the input bits are chosen from the distribution~$\pi$}%
                                        \nomenclature[deltaf]{$\delta^{(\pi)}(f)$}{the revealment of $f$; i.e., $\min\{\max_i \delta_i^{(\pi)}(\calT) : \calT \text{ computes } f\}$}%
                                        \nomenclature[DeltaiT]{$\Delta^{(\pi)}(\calT)$}{the expected number of queries made by randomized decision tree $\calT$ when the input bits are chosen from the distribution~$\pi$}%
                                        \nomenclature[Delta(f)]{$\Delta^{(\pi)}(f)$}{the expected number of queries made by the best decision tree computing $f$ when the input bits are chosen from the distribution~$\pi$}%
                                        \index{revealment}%
    Given $f \btR$, we define $\Delta(f)$ to be the minimum of $\Delta(\calT)$ over all randomized decision trees~$\calT$ computing~$f$.

    We can also generalize these definitions for functions $f \in L^2(\Omega, \pi\xn)$.
                                                        \index{decision tree!product space domains}%
    A deterministic decision tree over domain~$\Omega$ is the natural generalization in which each internal query node has $|\Omega|$ outgoing edges, labeled by the elements of~$\Omega$.  We write $\delta_i^{(\pi)}(\calT), \Delta^{(\pi)}(\calT), \Delta^{(\pi)}(f)$ for the generalizations to trees over~$\Omega$; in the case of $L^2(\bn, \pi_p\xn)$ we use the superscript $(p)$ instead of $(\pi_p)$ for brevity.
\end{definition}
It follows immediately from the definitions that for any $f \in L^2(\Omega^n, \pi\xn)$,
\[
    \Delta^{(\pi)}(f) \leq \RDT(f) \leq \DT(f).
\]
\begin{remark}              \label{rem:Delta-det}
    In the definition of $\Delta^{(\pi)}(f)$ it is equivalent if we only allow deterministic decision trees; this is because in~\eqref{ex:expected-dt-cost} we can always choose the ``best'' deterministic~$T$ in the support of~$\calT$.
\end{remark}

\begin{example}     \label{eg:rec-maj-dt}
    It follows from our discussions that $\RDT(\Maj_3) \leq 8/3$ and $\Delta(\Maj_3) \leq 5/2$; indeed, it's not hard to show that both of these bounds are equalities.
    In Exercise~\ref{ex:rec-maj-dt} you are asked to generalize to the
                                            \index{recursive majority}%
    recursive majority of~$3$ function on $n = 3^d$ inputs; it satisfies $\DT(\Maj_3^{\otimes d}) = 3^d = n$, but
    \begin{align*}
        \RDT(\Maj_3^{\otimes d}) &\leq (8/3)^d = n^{\log_3(8/3)} \approx n^{.89},\\
        \Delta(\Maj_3^{\otimes d}) &\leq (5/2)^d = n^{\log_3(5/2)} \approx n^{.83}.
    \end{align*}
    Incidentally, these bounds are not asymptotically sharp; estimating $\RDT(\Maj_3^{\otimes d})$ in particular is a well-studied open problem.
\end{example}
\begin{example} \label{eg:or-rdt}
    In Exercise~\ref{ex:or-rdt} you are asked to show that for the logical OR function,
    $
        \Delta\subp(\OR_n) = \frac{1 - (1-p)^n}{p},
    $
    which is roughly~$2$ for $p = 1/2$ but is asymptotic to $n/(2 \ln 2)$ at the critical probability~$p_c$.
\end{example}

Example~\ref{eg:rec-maj-dt} illustrates a mildly surprising phenomenon: using randomness it's possible to evaluate certain unbiased $n$-bit functions~$f$ while reading only a~$1/n^{\Theta(1)}$ fraction of the input bits.  This is even more interesting when~$f$ is transitive-symmetric like $\Maj_3^{\otimes d}$. In that case it's not hard to show (Exercise~\ref{ex:trans-sym-delta}) that any randomized decision tree~$\calT$ computing~$f$ can be converted to one where $\Delta(\calT)$ remains the same but all $\delta_i(\calT)$ are equal to $\Delta(f)/n$.  Then~$f$ can be evaluated despite the fact that \emph{each} input bit is only queried with probability~$1/n^{\Theta(1)}$.
                                                    \index{revealment}%

In this section we explore the limits of this phenomenon.  In particular, a longstanding conjecture of Yao~\cite{Yao77} says that this is \emph{not} possible for monotone graph properties:
                                                    \index{Yao's Conjecture}%
\begin{named}{Yao's Conjecture} Let $f \btb$ be a nonconstant monotone $v$-vertex graph property, where $n = \binom{v}{2}$. Then $\RDT(f) \geq \Omega(n)$.
\end{named}
Toward this conjecture we will present a lower bound due to O'Donnell, Saks, Schramm, and Servedio~\cite{OSSS05}. (Two other incomparable bounds are discussed in the notes for this chapter.) It has the advantages that it works for the more general class of transitive-symmetric functions and that it even lower-bounds $\Delta^{(p_c)}(f)$:
                                                    \index{transitive-symmetric function!decision tree complexity}%
\begin{theorem}                                     \label{thm:osss-cor}
    Let $f \btb$ be a nonconstant monotone transitive-symmetric function with critical probability~$p_c$.  Then
    \[
        \Delta^{(p_c)}(f) \geq (n/\sigma_c)^{2/3}.
    \]
\end{theorem}

Theorem~\ref{thm:osss-cor} is essentially sharp in several interesting cases.  Whenever the critical probability~$p_c$ is $\Theta(1/n)$ or $1 - \Theta(1/n)$ then $\sigma_c = \Theta(1/\sqrt{n})$ and Theorem~\ref{thm:osss-cor} gives the strongest possible bound, $\Delta^{(p_c)}(f) \geq \Omega(n)$.  This occurs, e.g., for the~$\OR_n$ function (Example~\ref{eg:or-rdt}). Furthermore, Theorem~\ref{thm:osss-cor} can be tight up to a logarithmic factor when~$p_c = 1/2$ as the following theorem of Benjamini, Schramm, and Wilson shows:
\begin{theorem}                                     \label{thm:bsw}
    \cite{BSW05}. There exists an infinite family of monotone transitive-symmetric functions $f_n \btb$ with critical probability~$p_c = 1/2$ and $\Delta(f) \leq O(n^{2/3} \log n)$.
\end{theorem}

Theorem~\ref{thm:osss-cor} follows easily from two inequalities~\cite{OS06,OS07}, \cite{OSSS05}, which we now present:

                                                \index{OS Inequality}%
\begin{named}{OS Inequality}                       \label{thm:OS}
    Let $f \in L^2(\bn, \pi_p\xn)$.  Then
    $
        \sum_{i=1}^n \wh{f}(i) \leq \|f\|_2 \cdot \sqrt{\Delta\subp(f)}.
    $

    \noindent In particular, if $f$ has range $\bits$ and is monotone, then $\Tinf[f] \leq \sigma\sqrt{\Delta\subp(f)}$.
\end{named}
                                                \index{OSSS Inequality}%
\begin{named}{OSSS Inequality}  Let $f \in L^2(\Omega^n, \pi\xn)$ have range $\bits$ and let~$\calT$ be any randomized decision tree computing~$f$.  Then
\[
    \Var[f] \leq \sum_{i=1}^n \delta_i^{(\pi)}(\calT) \cdot \Inf_i[f].
\]
\end{named}
\begin{remark}                      \label{rem:every-dt-influential}
    An interesting corollary of the OSSS Inequality is that
    \[
        \MaxInf[f] \geq \Var[f]/\Delta^{(\pi)}(f) \geq \Var[f]/\DT(f) \geq \Var[f]/\deg(f)^3,
    \]
    the last inequality assuming $\Omega = \bits$.  See Exercise~\ref{ex:osss-variants}.
\end{remark}

These two inequalities can be thought of as strengthenings of basic Fourier inequalities which take into account the decision tree complexity of~$f$.  The OS~Inequality essentially generalizes the result that majority functions maximizes $\sum_{i=1}^n \wh{f}(i)$; i.e., Theorem~\ref{thm:maj-maximizes-deg-1-sum}. The OSSS~Inequality is a generalization of the Poincar\'{e} Inequality, discounting the influences of coordinates that are rarely read.

We will first derive the query complexity lower bound Theorem~\ref{thm:osss-cor} from the OS and OSSS~Inequalities. We will then prove the latter two inequalities.

\begin{proof}[Proof of Theorem~\ref{thm:osss-cor}]  We consider $f$ to be an element of $L^2(\bn, \pi_{p_c}\xn)$.  Let $\calT$ be a randomized decision tree achieving $\Delta^{(p_c)}(f)$.  In the OSSS~Inequality, we have $\Var[f] = 1$ since~$p_c$ is the critical probability and $\Inf_i[f] = \Tinf[f]/n$ for each $i \in [n]$ since~$f$ is transitive-symmetric.  Thus
\[
    1 \leq \sum_{i=1} \delta_i^{(p_c)}(\calT) \cdot \frac{\Tinf[f]}{n} \quad\implies\quad n \leq \Delta^{(p_c)}(f) \cdot \Tinf[f] \leq  \sigma \Delta^{(p_c)}(f)^{3/2},
\]
where we used the OS~Inequality. The theorem follows by rearranging.
\end{proof}

Now we prove the OS and OSSS~Inequalities, starting with the latter.  We will need a simple lemma that uses the decomposition $f = \uE_i f + \Lap_i f$.
\begin{lemma}                                       \label{lem:covariance-law}
    Let $f, g \in L^2(\Omega^n, \pi\xn)$ and let $j \in [n]$.  Given $\omega \in \Omega$, write $\restr{f}{}{\omega}$ for the restriction of~$f$ in which the $j$th coordinate is fixed to value~$\omega$, and similarly for~$g$.  Then \[
        \Cov[f,g] = \Ex_{\substack{\bomega, \bomega' \sim \pi \\ \text{\textnormal{independent}}}}[\Cov[\restr{f}{}{\bomega}, \restr{g}{}{\bomega'}]] + \la \Lap_j f, \Lap_j g \ra.
    \]
\end{lemma}
\begin{proof}
    Since the covariances and Laplacians are unchanged when constants are added, we may assume without loss of generality that $\E[f] = \E[g] = 0$.  Then  $\Cov[f,g] = \la f, g\ra$ and
    \begin{multline*}
        \Ex_{\bomega, \bomega'}[\Cov[\restr{f}{}{\bomega}, \restr{g}{}{\bomega'}]] = \Ex_{\bomega, \bomega'}[\la \restr{f}{}{\bomega},\restr{g}{}{\bomega'}\ra - \E[\restr{f}{}{\bomega}]\E[\restr{g}{}{\bomega'}]] \\=
        \Ex_{\bomega, \bomega'}[\la \restr{f}{}{\bomega},\restr{g}{}{\bomega'}\ra] - \E[f]\E[g]= \Ex_{\bomega, \bomega'}[\la \restr{f}{}{\bomega},\restr{g}{}{\bomega'}\ra] = \la \uE_j f, \uE_j g \ra.
    \end{multline*}
    Thus the stated equality reduces to the basic (Exercise~\ref{ex:E-L})
    identity
    \[
        \la f, g \ra = \la \uE_j f, \uE_j g \ra + \la \Lap_j f, \Lap_j g \ra. \qedhere
    \]
\end{proof}

\begin{proof}[Proof of the OSSS Inequality]
    More generally we show that if $g \btb$ is also an element of $L^2(\Omega^n, \pi\xn)$, then
    \begin{equation} \label{eqn:cov-osss}
        \Cov[f,g] \leq \sum_{i=1}^n \delta_i^{(\pi)}(\calT) \cdot \Inf_i[g].
    \end{equation}
    The result then follow by taking $g = f$.  We may also assume that $\calT = T$ is a single deterministic tree computing~$f$; this is because~\eqref{eqn:cov-osss} is linear in the quantities~$\delta_i^{(\pi)}(\calT)$.
    We prove~\eqref{eqn:cov-osss} by induction on the structure of~$T$.  If~$T$ is depth-$0$, then~$f$ must be a constant function; hence $\Cov[f,g] = 0$ and~\eqref{eqn:cov-osss} is trivial.  Otherwise, let~$j \in [n]$ be the coordinate queried at the root of~$T$. For each $\omega \in \Omega$, write~$T_\omega$ for the subtree of~$T$ given by the $\omega$-labeled child of the root. By applying Lemma~\ref{lem:covariance-law} and induction (noting that $T_\omega$ computes the restricted function $\restr{f}{}{\omega}$), we get
    \begin{align*}
        \Cov[f,g] &= \Ex_{\substack{\bomega, \bomega' \sim \pi \\ \text{\textnormal{independent}}}}[\Cov[\restr{f}{}{\bomega}, \restr{g}{}{\bomega'}]] + \la \Lap_j f, \Lap_j g \ra \\
        &\leq \Ex_{\bomega, \bomega' \sim \pi}\Bigl[\sum_{i\neq j} \delta_i^{(\pi)}(T_{\bomega}) \cdot \Inf_i[g_{\bomega'}]\Bigr]+ \la \Lap_j f, \Lap_j g \ra \\
        &= \sum_{i \neq j} \delta_i^{(\pi)}(T) \cdot \Inf_i[g] + \la f, \Lap_j g \ra \tag{\text{in part since $\E[\Lap_j g] = 0$}} \\
        &\leq \sum_{i \neq j} \delta_i^{(\pi)}(T) \cdot \Inf_i[g] + \E[|\Lap_j g|]   \tag{\text{since $|f| \leq 1$}} \\
        &= \sum_{i=1}^n \delta_i^{(\pi)}(T) \cdot \Inf_i[g],
    \end{align*}
    where the last step used $\delta_j^{(\pi)}(T) = 1$ and Proposition~\ref{prop:dir-laplacian-facts2}.  This completes the inductive proof of~\eqref{eqn:cov-osss}.
\end{proof}

Finally, we prove the OS~Inequality.  For this we require a definition.
\begin{definition}          \label{def:dt-process}
    Let $(\Omega, \pi)$ be a finite probability space and $T$ a deterministic decision tree over~$\Omega$.  The \emph{decision tree process}
                                                \index{decision tree process}%
    associated to~$T$ generates a random string~$\bx$ distributed according to~$\pi$ (and some additional random variables), as follows:
    \begin{enumerate}
        \item Start at the root node of~$T$; say it queries coordinate $j_1$.  Choose $\bx_{j_1} \sim \pi$ and follow the outgoing edge labeled by the outcome.
        \item Suppose the node of $T$ which is reached queries coordinate $\bj_2$.  Choose $\bx_{\bj_2} \sim \pi$ and follow the outgoing edge labeled by the outcome.
        \item Repeat until a leaf node is reached.  Then, define $\bJ = \{j_1, \bj_2, \bj_3, \dots\} \subseteq [n]$ to be the set of coordinates queried.
        \item Draw the as-yet-unqueried coordinates, denoted $\bx_{\ol{\bJ}}$,  from $\pi^{\otimes \ol{\bJ}}$.
    \end{enumerate}
    Despite the fact that the coordinates $\bx_i$ are drawn in a random, dependent order, it's not hard to see (Exercise~\ref{ex:dt-process}) that the final string $\bx = (\bx_{\bJ}, \bx_{\ol{\bJ}})$ is distributed according the product distribution $\pi^{\otimes n}$.
\end{definition}

\begin{proof}[Proof of the OS Inequality]
    We will prove the claim $\sum_{i=1}^n \wh{f}(i) \leq \|f\|_2 \cdot \sqrt{\Delta\subp(f)}$; the ``in particular'' statement follows immediately from Proposition~\ref{prop:monotone-influences-biased}.   Fix a deterministic decision tree~$T$ achieving $\Delta\subp(f)$ (see Remark~\ref{rem:Delta-det}) and let $\bx = (\bx_{\bJ}, \bx_{\ol{\bJ}})$ be drawn from the associated decision tree process.  Using the notation~$\phi$ from Definition~\ref{def:p-biased-phi} we have
    \[
        \sum_{i=1}^n \wh{f}(i) = \E_{\bJ, \bx_{\bJ}, \bx_{\ol{\bJ}}}[f(\bx) \littlesum_{i=1}^n \phi(\bx_i)] = \E_{\bJ, \bx_{\bJ}}[f(\bx_{\bJ}) \E_{\bx_{\ol{\bJ}}}[\littlesum_{i=1}^n \phi(\bx_i)]].
    \]
    Here we abused notation slightly by writing $f(\bx_{\bJ})$; in the decision tree process, $f$'s value is determined once~$\bx_{\bJ}$ is.  Since $\E[\phi(\bx_i)] = 0$ for each~$i \not \in \bJ$ we may continue:
    \begin{align*}
        \E_{\bJ, \bx_{\bJ}}[f(\bx_{\bJ}) \E_{\bx_{\ol{\bJ}}}[\littlesum_{i=1}^n \phi(\bx_i)]] &= \E_{\bJ, \bx_{\bJ}}[f(\bx_{\bJ}) \littlesum_{i=1}^n \bone_{\{i \in \bJ\}} \phi(\bx_i)] \\
        &\leq \sqrt{\E_{\bJ, \bx_{\bJ}}[f(\bx_{\bJ})^2]}\sqrt{\E_{\bJ, \bx_{\bJ}}\left[\Bigl(\littlesum_{i=1}^n \bone_{\{i \in \bJ\}} \phi(\bx_i)\Bigr)^2\right]},
    \end{align*}
    by Cauchy--Schwarz.  Now $\sqrt{\E_{\bJ, \bx_{\bJ}}[f(\bx_{\bJ})^2]}$ is simply $\|f\|_2$ since~$T$ computes~$f$. To complete the proof it suffices to show that
    \[
        \E_{\bJ, \bx_{\bJ}}\left[\Bigl(\littlesum_{i=1}^n \bone_{\{i \in \bJ\}} \phi(\bx_i)\Bigr)^2\right] = \Delta\subp(f).
    \]
    To see this, expand the square:
    \[
        \E_{\bJ, \bx_{\bJ}}\left[\Bigl(\littlesum_{i=1}^n \bone_{\{i \in \bJ\}} \phi(\bx_i)\Bigr)^2\right] = \sum_{i=1}^n \E_{\bJ, \bx_{\bJ}}[\bone_{\{i \in \bJ\}} \phi(\bx_i)^2] + \sum_{i \neq i'} \E_{\bJ, \bx_{\bJ}}[\bone_{\{i, i' \in \bJ\}}  \phi(\bx_i) \phi(\bx_{i'})].
    \]
    Conditioned on $i \in \bJ$ the quantity $\E[\phi(\bx_i)^2]$ is simply~$1$.  Thus
    \[
        \sum_{i=1}^n \E_{\bJ, \bx_{\bJ}}[\bone_{\{i \in \bJ\}} \phi(\bx_i)^2] = \sum_{i=1}^n \Pr[i \in \bJ] = \Delta\subp(f).
    \]

    It remains to show that $\E_{\bJ, \bx_{\bJ}}[\bone_{\{i, i' \in \bJ\}}  \phi(\bx_i) \phi(\bx_{i'})] = 0$ whenever $i \neq i'$. Suppose we condition on the event that $i, i' \in \bJ$ and we further condition on~$i$ being queried before $i'$ is queried.  Certainly this may affect the conditional distribution of~$\bx_i$, but the conditional distribution of~$\bx_{i'}$ remains~$\pi_p$; hence $\E[\phi(\bx_{i'})] = 0$ under this conditioning.  Of course the same argument holds when we condition on~$i'$ being queried before~$i$.  From this it follows that $\E_{\bJ, \bx_{\bJ}}[\bone_{\{i, i' \in \bJ\}}  \phi(\bx_i) \phi(\bx_{i'})]$ is indeed~$0$, completing the proof.
\end{proof}

\section{Exercises and notes}       \label{sec:domains-notes}
\begin{exercises}
    \item Explain how to generalize the definitions and results in Sections~\ref{sec:product-spaces} and~\ref{sec:general-fourier-formulas} to general finite product spaces $L^2(\Omega_1 \times \cdots \times \Omega_n, \pi_1 \times \cdots \times \pi_n)$.
    \item Verify that Definition~\ref{def:general-inner-prod-space} indeed defines a real inner product space.  (Where is the fact that $\pi$ has full support used?)
    \item Verify the formula for $\wh{f}(\alpha)$ in Definition~\ref{def:general-fourier-coefficient}.
    \item \label{ex:3-ary-basis} Verify that $\phi_0, \phi_1, \phi_2$ from Example~\ref{eg:3-ary-basis} indeed constitute a Fourier basis for $\Omega = \{a,b,c\}$ with the uniform distribution.
   \item \label{ex:and2-3ary} Verify the Fourier expansion in Example~\ref{eg:and2-3ary}.
    \item \label{ex:finish-general-plancherel} Complete the proof of Proposition~\ref{prop:general-plancherel-etc}.
    \item \label{ex:expec-is-proj} Prove that the expectation over~$I$ operator, $\uE_I$, is a linear operator on \linebreak $L^2(\Omega^n, \pi\xn)$ (i.e., $\uE_I (f+g) = \uE_I f + \uE_I g$), a projection (i.e., $\uE_I \circ \uE_I = \uE_I$), and self-adjoint (i.e., $\la f, \uE_I g \ra = \la \uE_I f, g \ra$).  Deduce that $\T_\rho$ is also self-adjoint.
    \item \label{ex:E-L} Show for any $f,g \in L^2(\Omega^n, \pi\xn)$ and $j \in [n]$ that $f = \uE_j f + \Lap_j f$ and that $\la f, g \ra = \la \uE_j f, \uE_j g \ra + \la \Lap_j f, \Lap_j g \ra$.
    \item \label{ex:dir-laplacian-facts2} Prove Proposition~\ref{prop:dir-laplacian-facts2}.  (Hint: Exercise~\ref{ex:Boolean-variance}.)
    \item \label{ex:laplacian-any-power-influence} Let $f \in L^2(\Omega^n, \pi\xn)$ have range $\{-1,1\}$.  Proposition~\ref{prop:dir-laplacian-facts2} tells us that $\|\Lap_i f\|_1 = \|\Lap_i f\|_2^2 = \Inf_i[f]$.
        \begin{exercises}
            \item \label{ex:weak-laplacian-any-power-influence} Show that $\|\Lap_i f\|_p^p \leq 2^p\Inf_i[f]$ for any $p \geq 1$.
            \item \label{ex:strong-laplacian-any-power-influence} In case $1 \leq p \leq 2$, show that in fact $\|\Lap_i f\|_p^p \leq \Inf_i[f]$.  (Hint: Use the general form of \Holder's inequality to bound $\|\Lap_i f\|_p$ in terms of $\|\Lap_i f\|_1$ and $\|\Lap_i f\|_2$.)
        \end{exercises}
    \item \label{ex:individual-T-gen} Generalize all of Exercise~\ref{ex:individual-T} to the setting of $L^2(\Omega^n, \pi\xn)$.  Caution: the two statements referring to $\rho \in [-1,1]$ should refer only to $\rho \in [0,1]$ in this more general setting.
    \item Assume $|\Omega| = m$ and let $\pi$ denote the uniform distribution on~$\Omega$.
        \begin{exercises}
            \item \label{ex:negrho-a} For $x \in \Omega^n$ and $\by \sim N_\rho(x)$, write a formula for $\Pr[\by_i = \omega]$ in terms of~$\rho$ (there are two cases depending on whether or not $x_i = \omega$).
            \item Verify that your formula defines a valid probability distribution on $\Omega$ even when $-\frac{1}{m-1} \leq \rho < 0$.  We may therefore extend the definition of $N_\rho$ to this case. (Cf.\ the second half of Definition~\ref{def:noise}.)
            \item Verify that for $\bx \sim \pi\xn$ and $\by \sim N_\rho(\bx)$, the distribution of $(\bx,\by)$ is symmetric in~$\bx$ and~$\by$.
            \item Show that when $\by \sim N_{-\frac{1}{m-1}}(x)$, each $\by_i$ is uniformly distributed on $\Omega \setminus \{x_i\}$.
            \item Verify that the formula for $\T_\rho$ from Proposition~\ref{prop:general-trho} continues to hold for $-\frac{1}{m-1} \leq \rho < 0$.  (Hint: Use the fact that it holds for $\rho \in [0,1]$ and that the formula in part~\ref{ex:negrho-a} is a polynomial in~$\rho$.)
        \end{exercises}
    \item Show that Definition~\ref{def:stable-influence-general} extends by continuity to
        \[
            \Inf_i^{(0)}[f] = \sum_{\substack{\#\alpha = 1 \\ \alpha_i \neq 0}} \wh{f}(\alpha)^2.
        \]
        Extend also Proposition~\ref{prop:few-stable-influences-general} to the case of $\delta = 1$.
    \item \label{ex:orthog-incl-excl} Prove explicitly that condition~\ref{item:by-incl-excl} holds in Theorem~\ref{thm:orthogonal-decomposition}.
    \item Prove that condition~\ref{item:orthog-linear} must hold in Theorem~\ref{thm:orthogonal-decomposition} directly from the uniqueness statement (i.e., without appealing to the explicit construction).
    \item Let $f \in L^2(\Omega^n, \pi\xn)$.  Prove directly from the defining Theorem~\ref{thm:orthogonal-decomposition} that $(f^{=S})^{\subseteq T}$ is equal to $f^{=S}$ if $S \subseteq T$ and is equal to~$0$ otherwise.
    \item \label{ex:martingale} Let $f \in L^2(\Omega^n, \pi\xn)$ and let $\bx \sim \pi\xn$. In this exercise you should think about how the (conditional) expectation of~$f$ changes as the random variables $\bx_1, \dots, \bx_n$ are revealed one at a time.
        \begin{exercises}
            \item Recalling that $f^{\subseteq [t]}(\bx)$ depends only on $\bx_1, \dots, \bx_t$, show that the sequence of random variables $(f^{\subseteq [t]}(\bx))_{t = 0 \dots n}$ is a martingale (where $f^{\subseteq [0]}$ denotes~$f^\emptyset$); i.e.,
                \[
                    \E[f^{\subseteq [t]}(\bx) \mid f^{\subseteq [0]}(\bx), \dots, f^{\subseteq [t-1]}(\bx)] = f^{\subseteq [t-1]}(\bx) \quad \forall t \in [n].
                \]
                (This is the \emph{Doob martingale} for~$f$.)
                                                        \index{martingale!Doob}%
            \item For each $t \in [n]$ define
                    \[
                        \mds{t}{f} = f^{\subseteq [t]} - f^{\subseteq [t-1]} = \sum_{\substack{S \subseteq [n] \\ \max(S) = t}} f^{=S}.
                    \]
                  Show that $\E[\mds{t}{f}(\bx) \mid f^{\subseteq [0]}(\bx), \dots, f^{\subseteq [t-1]}(\bx)] = 0$.  (Here $(\mds{t}{f})_{t = 1 \dots n}$ is the \emph{martingale difference sequence} for~$f$.)
                                                        \index{martingale difference sequence}%
        \end{exercises}
    \item \label{ex:orthog-decomp-work} For $f, g \in L^2(\Omega^n, \pi\xn)$, prove the following directly from Theorem~\ref{thm:orthogonal-decomposition}:
        \begin{gather*}
            \la f, g \ra = \sum_{S \subseteq [n]} \la f^{=S}, g^{=S} \ra\\
            \Inf_i[f] = \sum_{S \ni i} \|f^{=S}\|_2^2 \\
            \Tinf[f] =  \sum_{k=0}^n k \cdot \W{k}[f] \\
            \T_\rho (f^{=S}) = (\T_\rho f)^{=S} = \rho^k f^{=S} \\
            \Stab_\rho[f] = \sum_{k=0}^n \rho^k \cdot \W{k}[f].
        \end{gather*}
    \item \label{ex:f=Sbound} Let $f \in L^2(\Omega^n, \pi\xn)$ and let $S \subseteq [n]$.  Show that $\|f^{=S}\|_\infty \leq 2^{|S|} \|f\|_\infty$.
    \item \label{ex:orthog-decomp1} Explicitly verify that Proposition~\ref{prop:orthog-decomp-basis} holds for the function in Examples~\ref{eg:and2-3ary} and~\ref{eg:orthog-decomp1}.
    \item \label{ex:decomp-restriction} Let $f \in L^2(\Omega^n, \pi\xn)$ and let $i \in S \subseteq [n]$. Suppose we take $f^{=S}$ and restrict its $i$th coordinate to have value $\omega_i$, forming the subfunction $g = \restr{(f^{=S})}{}{\omega_i}$.  Show that $g = g^{=S \setminus \{i\}}$.  In particular, $\E[g] = 0$ assuming $|S| \geq 2$.
    \item \label{ex:hoeffding1} Let $f \in L^2(\Omega^n, \pi\xn)$ be a symmetric function.  Show that if $1 \leq |S| \leq |T| \leq n$, then $\frac{1}{|S|} \Var[f^{\subseteq S}] \leq \frac{1}{|T|} \Var[f^{\subseteq T}]$.
    \item \label{ex:condorcet-jury-theorem} Prove the sharp threshold statement about the majority function made in Example~\ref{eg:sharp-thresh}.  (Hint: Chernoff bound.)  In the social choice literature, this fact is known as the \emph{Condorcet Jury Theorem}.
    \item \label{ex:general-p-biased}  Let $p_1, \dots, p_n \in (0,1)$ and let $\pi = \pi_{p_1} \otimes \cdots \pi_{p_n}$ be the associated product distribution on~$\bn$.  Write $\mu_i = 1 - 2p_i$ and $\sigma_i = 2\sqrt{p_i}\sqrt{1-p_i}$.  Generalize Proposition~\ref{prop:monotone-influences-biased} to the setting of $L^2(\bn, \pi)$.
    \item \label{ex:usual-vs-biased-expansion} Let $f \btR$ and consider the general product distribution setting of Exercise~\ref{ex:general-p-biased}.
        \begin{exercises}
            \item For $S = \{i_1, \dots, i_k\} \subseteq [n]$, write $\D_{\phi_S}$ for $\D_{\phi_{i_1}} \circ \cdots \circ \D_{\phi_{i_k}}$ and similarly $\D_{x_S}$.  Show that $\D_{\phi_S} = \prod_{i \in S} \sigma_i \cdot \D_{x_S}$.
            \item Writing $f^{(\mu)}$ for the function $f$ viewed as an element of $L^2(\bn, \pi)$, show that
                \[
                    \wh{f\subp}(S) = \prod_{i \in S} \sigma_i \cdot \D_{x_S} f(\mu_1, \dots, \mu_n).
                \]
            \item Show that $\snorm{f\subp}_\infty \leq \prod_{i \in S} \sigma_i \cdot \|f\|_\infty$.
        \end{exercises}
    \item \label{ex:p-biased-influence-facts} \begin{exercises}
            \item \label{ex:biased-pivotal-influence}  Generalize Exercise~\ref{ex:one-way-inf} by showing that for
                                                        \index{pivotal}%
                $f \in L^2(\bn, \pi_p\xn)$ with range $\bits$,
                \[
                    \Pr_{\bx \sim \pi_p\xn}[i \text{ is $b$-pivotal for $f$ on } \bx] = \pi_p(b) \Inf_i[f]
                \]
                for $i \in [n]$ and $b \in \bits$.
            \item \label{ex:p-biased-dnf-width-influence} Generalize Proposition~\ref{prop:DNF-tinf}
                                        \index{total influence!DNF formulas}%
                  by showing that if $f \btb$ has $\DNFwidth(f) \leq w$, then $\Tinf[f\subp] \leq 4qw \leq 4w$, and if~$f$ has ${\CNFwidth(f) \leq w}$, then $\Tinf[f\subp] \leq 4pw \leq 4w$.
          \end{exercises}
    \item \label{ex:crit-defined} Fix any $\alpha \in (0,1)$.  Let $f \co \tf^n \to \tf$ be a nonconstant monotone function.  Show that there exists $p \in (0,1)$ such that $\Pr_{\pi_p}[f(\bx) = \True] = \alpha$.  (Hint: Intermediate Value Theorem.)
    \item \label{ex:precise-sharp-thresh}  Fix a small constant $0 < \eps < 1/2$.  Let $f \co \tf^n \to \tf$ be a nonconstant monotone function.  Let $p_0$ (respectively,~$p_c$,~$p_1$) be the unique value of~$p \in (0,1)$ such that $\Pr_{\pi_{p}}[f(\bx) = \True] = \eps$ (respectively,~$1/2$,~$1-\eps$).  (This is a valid definition by Exercise~\ref{ex:crit-defined}.) Define also $\sigma_c^2 = 4p_c(1-p_c)$. The \emph{threshold interval} for~$f$ is defined to be $[p_0, p_1]$, and $\delta = p_1 - p_0$ is the \emph{threshold width}.  Now let $(f_n)_{n \in \N}$ be a sequence of nonconstant monotone Boolean functions (usually ``naturally related'', with $f_n$'s input length an increasing function of~$n$).
                                                        \index{threshold, sharp}%
        Define the sequences $p_0(n)$, $p_c(n)$, $p_1(n)$, $\sigma_c^2(n)$, $\delta(n)$. We say that the family $(f_n)$ has a \emph{sharp threshold} if $\delta(n)/\sigma_c^2(n) \to 0$ as $n \to \infty$; otherwise, we say it has a \emph{coarse threshold}.  (Note: If $p_c(n) \leq 1/2$ for all~$n$, this is the same as saying that $\delta(n)/p_c(n) \to 0$.) Show that if~$(f_n)$ has a coarse threshold, then there exists~$C < \infty$, an infinite sequence $n_1 < n_2 < n_3 < \cdots$, and a sequence $(p(n_i))_{i \in \N}$ such that:
        \begin{itemize}
            \item $\eps < \Pr_{\pi_{p(n_i)}}[f_{n_i}(\bx) = \True] < 1-\eps$ for all~$i$;
            \item $\Tinf[f_{n_i}^{(p(n_i))}] \leq C$ for all~$i$.
        \end{itemize}
        (Hint: Margulis--Russo and the Mean Value Theorem.)
    \item \label{ex:somewhat-sharp-thresh} Let $f \btb$ be a nonconstant monotone function and let $F \co [0,1] \to [0,1]$ be the (strictly increasing) function defined by $F(p) = \Pr_{\pi_p}[f(\bx) = -1]$.  Let $p_c$ be the critical probability such that $F(p_c) = 1/2$.  Assume that $p_c \leq 1/2$. (This is without loss of generality since we can replace~$f$ by~$f^\booldual$.  We often think of $p_c \ll 1/2$.) The goal of this exercise is to show a weak kind of threshold result: roughly speaking, $F(p) = o(1)$ when $p = o(p_c)$ and $F(p) = 1-o(1)$ when $p = \omega(p_c)$.
        \begin{exercises}
                                                    \index{Margulis--Russo Formula}%
            \item Using the Margulis--Russo Formula and the Poincar\'e Inequality show that for all $0 < p < 1$,
            \[
                F'(p) \geq \frac{F(p)(1-F(p))}{p(1-p)}.
            \]
            \item   Show that for all $p \leq p_c$ we have $F'(p) \geq \frac{F(p)}{2p}$ and hence $\frac{d}{dp} \ln F(p) \geq \frac{1}{2p}$.
            \item \label{ex:somewhat-sharp-part} Deduce that for any $0 \leq p_0 \leq p_c$ we have $F(p_0) \leq \half \sqrt{p_0/p_c}$; i.e., $F(p_0) \leq \eps$ if $p_0 \leq (2\eps)^2 p_c$.
            \item Show that the factor $(2\eps)^2$ can be improved to $\Theta(\tau) \eps^{1+\tau}$ for any small constant $\tau > 0$.  (Hint: The quadratic dependence on~$\eps$ arose because we used $1-F(p) \geq 1/2$ for $p \leq p_c$; but from part~\ref{ex:somewhat-sharp-part} we have the improved bound $1-F(p) \geq 1-\tau$ once $p \leq (2\tau)^2 p_c$.)
            \item \label{ex:somewhat-sharp-part2} In the other direction, show that so long as $p_1 = \frac{1}{(2\eps)^2} p_c \leq 1/2$, we have $F(p_1) \geq 1-\eps$.  (Hint: Work with $\ln(1-F(p))$.)  In case $p_1 \leq 1/2$ does not hold, show that we at least have $F(1/2) \geq 1 - \sqrt{p_c/2}$.
            \item The bounds in part~\ref{ex:somewhat-sharp-part2} are not very interesting when~$p_c$ is close to~$1/2$.  Show that we also have $F(1-\delta) \geq 1 - \sqrt{\delta/2}$ (even when $p_c = 1/2$).
        \end{exercises}
    \item \label{ex:sharp-thresh-countereg} Consider the sequence of functions $f_n \co \tf^n \to \tf$ defined for odd~$n \geq 3$ as follows: $f_n(x_1, \dots, x_n) = \Maj_3(x_1, x_2, \Maj_{n-2}(x_3, \dots, x_n))$.
        \begin{exercises}
            \item Show that $f_n$ is monotone and has critical probability $p_c = 1/2$.
            \item Sketch a plot of $\Pr_{\pi_p}[f_n(\bx) = \True]$ versus $p$ (assuming~$n$ very large).
            \item Show that $\Tinf[f_n] = \Theta(\sqrt{n})$.
            \item Show that the sequence $f_n$ has a coarse threshold as defined in Exercise~\ref{ex:precise-sharp-thresh} (assuming~$\eps < 1/4$).
        \end{exercises}
    \item \label{ex:shapley}
            \begin{exercises}
                \item \label{ex:eqslices} Consider the following probability distributions on strings $\bx \in \F_2^n$:
                        \begin{enumerate}
                            \item First choose $\bk \sim \{0, 1, 2, \dots, n\}$ uniformly.  Then choose $\bx$ uniformly from the set of all strings of Hamming weight~$\bk$.
                            \item First choose a uniformly random ``path $\bpi$ from $(0, 0, \dots, 0)$ up to $(1, 1, \dots, 1)$''; i.e., let $\bpi$ be a uniformly random permutation from~$S_n$ and let $\bpi^{\leq i} \in \F_2^n$ denote the string whose $j$th coordinate is~$1$ if and only if $\pi(j) \leq i$.  Then choose $\bk \sim \{0, 1, 2, \dots, n\}$ uniformly and let $\bx$ be the ``$\bk$th string on the path'', namely~$\bpi^{\leq \bk}$.
                            \item First choose $\bp \sim [0,1]$.  Then choose $\bx \sim \pi_{\bp}\xn$.
                        \end{enumerate}
                      Show that these are in fact the same distribution.  (Hint: Imagine choosing~$n+1$ indistinguishable points uniformly from~$[0,1]$ and then randomly assigning them the labels ``$p$'', $1$, $2$, \dots, $n$.)
                                                    \index{Shapley value}%
                                                    \index{Shapley--Shubik index|seeonly{Shapley value}}%
                \item We denote by~$\eqslices^n$ the distribution on~$\F_2^{[n]}$ from part~\ref{ex:eqslices}; more generally, we use the notation $\eqslices^N$ for the distribution on~$\F_2^N$ where~$N$ is an abstract set of cardinality~$n$.  Given a nonempty $J \subseteq [n]$, show that if $\bx \sim \eqslices^n$ and $\bx_J \in \F_2^J$ denotes the restriction of~$\bx$ to coordinates~$J$, then $\bx_J$ has the distribution $\eqslices^J$.
                \item Let $f \co \F_2^n \to \R$ and fix $i \in [n]$.
                    The \emph{$i$th Shapley value} of~$f$ is defined to be
                    \[
                        \Shap_i[f] = \E_{\bx \sim \eqslices^n}[f(\bx^{(i \mapsto 1)}) - f(\bx^{(i \mapsto 0)})].
                    \]
                    Show that $\sum_{i=1}^n \Shap_i[f] = f(1, 1, \dots, 1) - f(0, 0, \dots, 0)$.
                \item Suppose $f \co \F_2^n \to \{0,1\}$ is monotone.  Show  $\Shap_i[f] = 4\int_0^1 \Inf_i[f\subp]\,dp$.
            \end{exercises}
    \item \label{ex:complex-functions} Explain how to generalize the definitions and results in Sections~\ref{sec:product-spaces},~\ref{sec:general-fourier-formulas} to the case of the \emph{complex} inner product space $L^2(\Omega^n, \pi\xn)$.  In particular, verify the following formulas from Proposition~\ref{prop:general-plancherel-etc}:
                                               \index{Parseval's Theorem!complex case}%
                                               \index{Plancherel's Theorem!complex case}%
        \begin{align*}
            \E[f] &= \wh{f}(0)\\
            \E[|f|^2] &= \E[\la f, f \ra] = \sum_{\alpha \in \N_{< m}^n} \la \wh{f}(\alpha), \wh{f}(\alpha) \ra =  \sum_{\alpha \in \N_{< m}^n} |\wh{f}(\alpha)|^2 \\
            \Var[f]  &= \la f - \E[f], f - \E[f] \ra = \sum_{\alpha \neq 0} |\wh{f}(\alpha)|^2 \\
            \la f, g \ra &= \sum_{\alpha \in \N_{<m}^n} \la \wh{f}(\alpha), \wh{g}(\alpha) \ra = \sum_{\alpha \in \N_{<m}^n} \wh{f}(\alpha)\ol{\wh{g}(\alpha)} \\
            \Cov[f,g] &= \la f - \E[f], g - \E[g] \ra = \sum_{\alpha \neq 0} \wh{f}(\alpha)\ol{\wh{g}(\alpha)}.
        \end{align*}
    \item \label{ex:generalized-domain-range}
        \begin{exercises}
            \item  \label{ex:generalized-domain-range-a} As in Exercise~\ref{ex:generalized-range}, explain how to generalize the definitions and results in Sections~\ref{sec:product-spaces},~\ref{sec:general-fourier-formulas} to the case of functions $f \co \Omega^n \to V$, where $V$ is a real inner product space with inner product $\la \cdot, \cdot \ra_V$.  Here the Fourier coefficients $\wh{f}(\alpha)$ will be elements of~$V$, and $\la f, g \ra$ is defined to be $\Ex_{\bx \sim \pi\xn}[\la f(\bx), g(\bx) \ra_V]$.  In particular, verify the formulas from Proposition~\ref{prop:general-plancherel-etc}, including Placherel: $\la f, g \ra = \sum_\alpha \la \wh{f}(\alpha), \wh{g}(\alpha) \ra_V$.
            \item \label{ex:simplex} For $\Sigma$ a finite set we write $\simplex_\Sigma$ for the set of all probability distributions over~$\Sigma$ (cf.~Exercise~\ref{ex:dvq-with-bounded-range}). Writing $|\Sigma| = m$, we also identify $\simplex_\Sigma$ with the standard convex simplex in $\R^{m}$, namely $\{\mu \in \R^{m} : \mu_1 + \cdots + \mu_m = 1, \mu_i \geq 0\ \forall i\}$ (where we assume some fixed ordering of~$\Sigma$).  Finally, we identify the~$m$ elements of~$\Sigma$ with the constant distributions in $\simplex_\Sigma$; equivalently, the vertices of the form $(0, \dots, 0, 1, 0, \dots, 0)$.  Given a function $f \co \Omega^n \to \Sigma$, often the most useful way to treat it analytically is to interpret it as a function $f \co \Omega^n \to \simplex_\Sigma \subset \R^m$ and then use the setting described in part~\ref{ex:generalized-domain-range-a}, with $V = \R^m$.  Using this idea, show that if $f \co \Omega^n \to \Sigma$ and $\pi$ is a distribution on $\Omega$, then
                \[
                    \Stab_\rho[f] = \Pr_{\bx \sim \pi\xn, \by \sim N_\rho(\bx)}[f(\bx) = f(\by)].
                \]
                (Here in $\Stab_\rho[f]$ we are interpreting $f$'s range as $\simplex_\Sigma \subset \R^m$, whereas in the expression $f(\bx) = f(\by)$ we are treating $f$'s range as the abstract set~$\Sigma$.)        \end{exercises}
    \item \label{ex:bow-ns-bound}  We say a function $f \in L^2(\Omega^n, \pi\xn)$ is a \emph{linear threshold function} if it is expressible as $f(x) = \sgn(\ell(x))$, where $\ell \co \Omega^n \to \R$ has degree at most~$1$ (in the sense of Definition~\ref{def:general-degree}).
        \begin{exercises}
            \item Given $\omega^{(+1)}, \omega^{(-1)} \in \Omega^n$ and $x \in \bn$, we introduce the notation $\omega^{(x)}$ for the string $(\omega_1^{(x_1)}, \dots, \omega_n^{(x_n)}) \in \Omega^n$.  Show that if $\bomega^{(+1)}, \bomega^{(-1)} \sim \pi\xn$ are drawn independently and $(\bx, \by) \sim \bn \times \bn$ is a $\rho$-correlated pair of binary strings, then $(\bomega^{(\bx)}, \bomega^{(\by)})$ is a $\rho$-correlated pair under $\pi\xn$.
            \item Let $f \in L^2(\Omega^n, \pi\xn)$ be a linear threshold function.  Given a pair \linebreak $\omega^{(+1)}, \omega^{(-1)} \in \Omega^n$, define $g_{\omega^{(+1)}, \omega^{(-1)}} \btb$ by $g_{\omega^{(+1)}, \omega^{(-1)}}(x) = f(\omega^{(x)})$.  Show that $g_{\omega^{(+1)}, \omega^{(-1)}}$ is a linear threshold function in the ``usual'' sense.
            \item Prove that Peres's Theorem (from Chapter~\ref{sec:peres}) applies to linear threshold functions in $L^2(\Omega^n, \pi\xn)$, with the same bounds.
        \end{exercises}
    \item \label{ex:general-group-duality}  Let $G$ be a finite abelian group.
                                                    \index{dual group}%
          We know by the Fundamental Theorem of Finitely Generated Abelian Groups that $G \cong \Z_{m_1} \times \cdots \Z_{m_n}$ where each $m_j$ is a prime power.
          \begin{exercises}
              \item Given $\alpha \in G$, define $\chi_\alpha \co G \to \C$ by
                  \[
                      \chi_\alpha(x) = \prod_{j = 1}^n \exp(2 \pi i \alpha_j x_j / m_j).
                  \]
                  Show $\chi_\alpha$ is a character of~$G$ and that the $\chi_\alpha$'s are distinct functions for distinct~$\alpha$'s.  Deduce that the set of all $\chi_\alpha$'s forms a Fourier basis for~$L^2(G)$.
              \item Show that this set of characters forms a group under multiplication and that this group is isomorphic to~$G$; i.e., generalize Fact~\ref{fact:Zmn-duality}. This is called the \emph{dual group} of~$G$ and it is written~$\wh{G}$.  We also identify the characters in~$\wh{G}$ with their indices~$\alpha$.
          \end{exercises}
    \item \label{ex:conv-ac-general} Verify that the convolution operation on $L^2(G)$ is associative and commutative, and that it satisfies $\wh{f \conv g}(\alpha) = \wh{f}(\alpha)\wh{g}(\alpha)$ for all $\alpha \in \wh{G}$.  (See Exercise~\ref{ex:general-group-duality} for the definition of~$\wh{G}$.)
    \item \label{ex:trans-sym-delta}
            \begin{exercises}
                                                \index{transitive-symmetric function}%
                                                \index{automorphism group}%
                \item \label{ex:trans-sym-delta-equal} Let $f \in L^2(\Omega^n, \pi\xn)$ be any transitive-symmetric function and let $\calT$ be a randomized decision tree computing~$f$.  Show that there exists a randomized decision tree $\calT'$ computing~$f$ with $\Delta^{(\pi)}(\calT') = \Delta^{(\pi)}(\calT)$ and such that $\delta_i^{(\pi)}(\calT')$ is the same for all $i \in [n]$.  (Hint: Randomize over the automorphism group $\Aut(f)$ and use Exercise~\ref{ex:transitive-regular}.)
                \item \label{ex:revealment-def} Given a randomized decision tree $\calT$, let $\delta^{(\pi)}(\calT) = \max_{i \in [n]}\{\delta_i^{(\pi)}(\calT)\}$.  Given $f \in L^2(\bn, \pi\xn)$, define $\delta^{(\pi)}(f)$ to be the minimum value of $\delta^{(\pi)}(\calT)$ over all $\calT$ which compute~$f$; this is called the \emph{revealment} of~$f$.
                                                        \index{revealment}%
                    Show that if $f$ is transitive-symmetric, then $\delta^{(\pi)}(f) = \frac{1}{n}\Delta^{(\pi)}(f)$.
            \end{exercises}
    \item \label{ex:rec-maj-dt}
                                                \index{recursive majority}%
          \begin{exercises}
                \item Show that $\DT(\Maj_3^{\otimes d}) = 3^d$, $\RDT(\Maj_3^{\otimes d}) \leq (8/3)^d$, and $\Delta(\Maj_3^{\otimes d}) \leq (5/2)^d$.
                \item Show that $\RDT(\Maj_3^{\otimes 2}) < (8/3)^2$.  How small can you make your upper bound?
          \end{exercises}
                                                \index{decision tree!randomized}%
    \item \label{ex:or-rdt}
        \begin{exercises}
            \item Show that for every deterministic decision tree~$T$ computing the logical OR function on~$n$ bits,
                \begin{multline*}
                    \qquad \qquad \Delta^{(p)}(T) = p \cdot 1 + (1-p)p \cdot 2 + (1-p)^2p \cdot 3 + \cdots \\
                    \cdots + (1-p)^{n-2}p \cdot (n-1) + (1-p)^{n-1} \cdot n = \frac{1 - (1-p)^n}{p}.
                \end{multline*}
                Deduce $\Delta^{(p)}(\OR_n) = \frac{1 - (1-p)^n}{p}$.
            \item Show that $\Delta^{(p_c)}(\OR_n) \sim n/(2 \ln 2)$ as $n \to \infty$, where $p_c$ denotes the critical probability for $\OR_n$.
        \end{exercises}
    \item \label{ex:and-or}  Let $\NAND \co \tf^2 \to \tf$ be the function that outputs $\true$ unless both its inputs are~$\true$.
        \begin{exercises}
            \item Show that for $d$ even, $\NAND^{\otimes d} = \tribes_{2,2}^{\otimes d/2}$.  (Thus the recursive NAND function is sometimes known as the AND-OR tree.)
            \item Show that $\DT(\NAND^{\otimes d}) = 2^d$.
            \item Show that $\RDT(\NAND) = 2$.
            \item For $b \in \tf$ and $\calT$ a randomized decision tree computing a function~$f$, let $\RDT_b(\calT)$ denote the maximum cost of $\calT$ among inputs~$x$ with $f(x) = b$. Show that there is a randomized decision tree~$\calT$ computing $\NAND$ with $\RDT_\false(\calT) = 3/2$.
            \item Show that $\RDT(\NAND^{\otimes 2}) \leq 3$.
            \item Show that there is a family of randomized decision trees $(\calT_d)_{d \in \N^+}$, with $\calT_d$ computing $\NAND^{\otimes d}$, satisfying the inequalities
                \begin{align*}
                    \RDT_{\false}(\calT_{d}) &\leq 2\RDT_{\true}(\calT_{d-1})\\
                    \RDT_{\true}(\calT_{d}) &\leq \RDT_{\false}(\calT_{d-1}) + (1/2)\RDT_{\true}(\calT_{d-1}).
                \end{align*}
            \item Deduce $\RDT(\NAND^{\otimes d}) \leq (\frac{1+\sqrt{33}}{4})^d \approx n^{.754}$, where $n = 2^d$.
        \end{exercises}
    \item   Let $\calC = \{\text{monotone } f \btb\mid \DT(f) \leq k\}$.  Show that $\calC$ is learnable from random examples with error~$\eps$ in time $n^{O(\sqrt{k}/\eps)}$.
                                    \index{decision tree!learning}%
        (Hint: OS~Inequality and Corollary~\ref{cor:learn-tinf}.)
    \item \label{ex:dt-process} Verify that the decision tree process described in Definition~\ref{def:dt-process} indeed generates strings distributed according to $\pi\xn$.  (Hint: Induction on the structure of the tree.)
    \item \label{ex:expected-depth-vs-size} Let $T$ be a deterministic decision tree of size~$s$.  Show that $\Delta(T) \leq \log s$.  (Hint: Let $\bP$ be a random root-to-leaf path chosen as in the decision tree process.  How can you bound the entropy of the random variable $\bP$?)
    \item \label{ex:osss-variants} Let $f \in L^2(\Omega^n, \pi\xn)$ be a nonconstant function with range $\bits$.
          \begin{exercises}
              \item Show that $\MaxInf[f] \geq \Var[f]/\Delta^{(\pi)}(f)$ (cf.~the KKL Theorem from Chapter~\ref{sec:tribes}).
              \item \label{ex:weak-AA} In case $\Omega = \bits$ show that $\MaxInf[f] \geq \Var[f] / \deg(f)^3$. (You should use the result of Midrij\=anis mentioned in the notes in Chapter~\ref{sec:spectral-structure-notes}.)
              \item Show that $\TInf[f] \geq \Var[f] / \delta^{(\pi)}(f)$, where
                                                \index{revealment}%
                    $\delta^{(\pi)}(f)$ is the revealment of~$f$, defined in Exercise~\ref{ex:trans-sym-delta}\ref{ex:revealment-def}.
          \end{exercises}
                                            \index{OSSS Inequality}%
    \item \label{ex:bsw-ex} Let $f \in L^2(\Omega^n, \pi\xn)$ have range $\bits$.
        \begin{exercises}
            \item Let $\calT$ be a randomized decision computing~$f$ and let $i \in [n]$.  Show that $\Inf_i[f] \leq \delta^{(\pi)}_i(\calT)$.  (Hint: The decision tree process.)
            \item Suppose $f$ is transitive-symmetric.  Show that $\Delta^{(\pi)}(f) \geq \sqrt{\Var[f] \cdot n}$. (Hint: Exercise~\ref{ex:trans-sym-delta}\ref{ex:revealment-def}.)  This result can be sharp up to an $O(\sqrt{\log n})$ factor even for an $f \btb$ with $\Var[f] = 1$; see~\cite{BSW05}.
        \end{exercises}
    \item \label{ex:jain-zhang} In this exercise you will give an alternate proof of the OSSS~Inequality that is sharp when $\Var[f] = 1$ and is weaker by only a factor of~$2$ when $\Var[f]$ is small. Let $f \in L^2(\Omega^n, \pi\xn)$ have range $\{-1,1\}$.  Given a randomized decision tree $\calT$ we write $\err(\calT) = \Pr_{\bx \sim \pi\xn}[\calT(\bx) \neq f(\bx)]$.
        \begin{exercises}
            \item Let $T$ be a depth-$k$ deterministic decision tree (not necessarily computing~$f$) whose root queries coordinate~$i$. Let $\calT$ be the distribution over deterministic trees of depth at most $k-1$ given by following a random outgoing edge from $T$'s root (according to~$\pi$). Show that $\err(\calT) \leq \err(T) + \half\Inf_i[f]$.
            \item  Let $\calT$ be a randomized decision tree of depth~$0$.  Show that $\err(\calT) \geq \min\{\Pr[f(\bx) = 1], \Pr[f(\bx) = -1]\}$.
            \item Prove by induction on depth that if $\calT$ is any randomized decision tree, then $\half \sum_{i=1}^n \delta_i^{(\pi)}(\calT) \cdot \Inf_i[f] \geq \min\{\Pr[f(\bx) = 1], \Pr[f(\bx) = -1]\} - \err(\calT)$.  Verify that this yields the OSSS Inequality when $\Var[f] = 1$ and in general yields the OSSS Inequality up to a factor of~$2$.
        \end{exercises}
    \item \label{ex:real-osss-countereg} Show that the OSSS Inequality fails for functions $f \btR$. (Hint: The simplest counterexample uses a decision tree with the shape in Figure~\ref{fig:osss-counterexample}.)
            \myfig{.5}{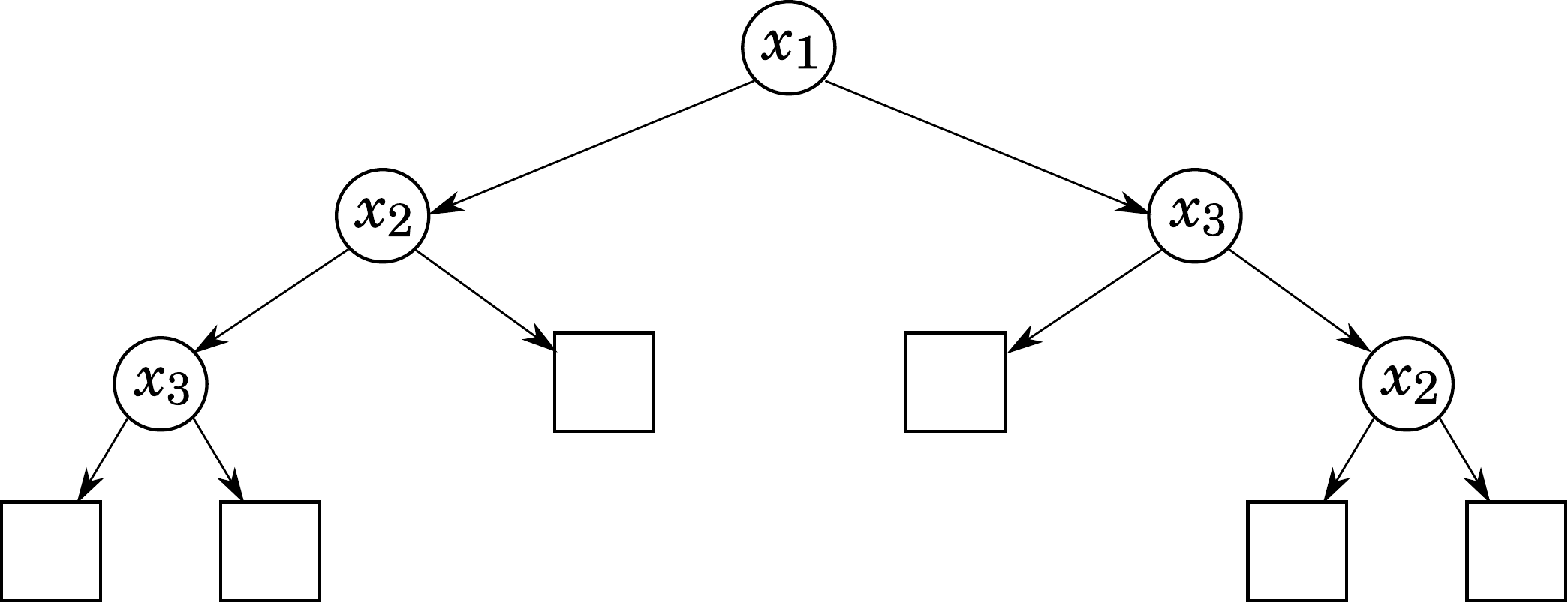}{The basis for a counterexample to the OSSS Inequality when $f \btR$}{fig:osss-counterexample}
        Can you make the ratio of the left-hand side to the right-hand side equal to $\frac{130 + 20\sqrt{3}}{157}$? Larger?
\end{exercises}

\subsection*{Notes.}

                                                \index{orthogonal decomposition}%
The origins of the orthogonal decomposition described in Section~\ref{sec:orthogonal-decomposition} date back to the work of Hoeffding~\cite{Hoe48} (see also von~Mises~\cite{vMis47}). Hoeffding's work introduced \emph{U-statistics}, i.e., functions~$f$ of independent random variables $\bX_1, \dots, \bX_n$ of the form $\avg_{1 \leq i_1 < \cdots < i_k \leq n} g(\bX_{i_1}, \dots, \bX_{i_k})$, where~$g \co \R^k \to \R$ is a symmetric function.  Such functions are themselves symmetric.  For these functions, Hoeffding introduced $f^{\subseteq S}$ (which, by symmetry, depends only on~$|S|$) and proved certain inequalities (e.g., those in Exercise~\ref{ex:hoeffding1}) relating $\Var[f]$ to the quantities $\|f^{\subseteq S}\|_2^2$, $\|f^{=S}\|_2^2$. Nonsymmetric functions~$f$ were considered only rarely in the subsequent three decades of statistics research. One notable exception comes in the work of H{\'a}jek~\cite{Haj68}, who effectively introduced~$f^{\leq 1}$, known as the \emph{H{\'a}jek projection} of~$f$.  Also, a work of Bourgain~\cite{Bou79} essentially describes the decomposition $f = \sum_{k} f^{=k}$. The first work that mentions the general orthogonal decomposition for not-necessarily-symmetric functions appears to be that of Efron and Stein~\cite{ES81} from the late 1970s. Efron and Stein's description is brief; the subsequent work of Karlin and Rinott~\cite{KR82} gives a more thorough development.  Efron and Stein's main result was a proof of the statement $\Var[f] \leq \Tinf[f]$ for symmetric~$f$; in the statistics literature this is known as the \emph{Efron--Stein Inequality}.
                                                \index{Efron--Stein Inequality|seeonly{Poincar\'{e} Inequality}}%
Steele~\cite{Ste86} extended this to the case of nonsymmetric~$f$ by a simple proof that used the Fourier basis approach to orthogonal decomposition.  This approach via Fourier bases originated in the work of Rubin and Vitale~\cite{RV80}; see also Takemura~\cite{Tak83} and Vitale~\cite{Vit84}. The terminology ``Fourier basis'' we use is not standard.

The $p$-biased hypercube distribution is strongly motivated by the Erd\H{o}s--R\'enyi~\cite{ER59} theory of random graphs (see e.g., Bollob{\'a}s and Riordan~\cite{BR08} for history) and by percolation theory (introduced in Broadbent and Hammersley~\cite{BH57}). Influences under the $p$-biased distribution -- and their connection to threshold phenomena -- were studied by Russo~\cite{Rus81,Rus82}.  The former work proved the Margulis--Russo formula independently of Margulis, who had proven it earlier~\cite{Mar74}.  Fourier analysis under the $p$-biased distribution seems to have been first introduced to the theoretical computer science literature by Furst, Jackson, and Smith~\cite{FJS91}, who extended the LMN learning algorithm for $\AC^0$ to this setting.  Talagrand~\cite{Tal93,Tal94} developed $p$-biased Fourier for the study of threshold phenomena, strengthening Margulis and Russo's work and proving the KKL~Theorem in the $p$-biased setting.  Similar results were obtained by Friedgut and Kalai~\cite{FK96b} using an earlier work of Bourgain, Kahn, Kalai, Linial, and Katznelson~\cite{BKK+92} that proved a version of the KKL~Theorem in the setting of general product spaces.  The statements about sharp thresholds for cliques and connectivity in Example~\ref{eg:sharp-thresh} are essentially due to Matula and to Erd\H{o}s--R\'{e}nyi, respectively; see, e.g., Bollob{\'a}s~\cite{Bol01}.  Weak threshold results similar to the ones in Exercise~\ref{ex:somewhat-sharp-thresh} were proved by Bollob\'{a}s and Thomason~\cite{BT87}, using the Kruskal--Katona Theorem rather than the Poincar\'e Inequality.

Fourier analysis on finite abelian groups -- and more generally, on locally compact abelian groups -- is an enormous subject upon which we have touched only briefly.  We cannot survey it here but refer instead to the standard textbook of Rudin~\cite{Rud62} and to the reader-friendly textbook of Terras~\cite{Ter99}, which focuses on finite groups.

One of the earliest works on randomized decision tree complexity is that of Saks and Wigderson~\cite{SW86}; they proved the contents of Exercise~\ref{ex:and-or}.  (We note that $\RDT(f)$ is usually denoted $R(f)$ in the literature, and $\DT(f)$ is usually denoted $D(f)$.) One basic lower bound in the area is that $\RDT(f) \geq \sqrt{\DT(f)}$ for any $f \btb$; in fact, this lower bound holds even for ``nondeterministic decision tree complexity'', as proved in~\cite{BI87,Tar89}.  Yao's Conjecture is also sometimes attributed to Richard Karp.  Regarding the recursive majority-of-$3$ function, Ravi Boppana was the first to point out that $\RDT(\Maj_3^{\otimes d}) = o(3^d)$ even though $\DT(\Maj_3^{\otimes d}) = 3^d$.  Saks and Wigderson noted the bound $\RDT(\Maj_3^{\otimes d}) \leq (8/3)^d$ and also that it is not optimal.  Following subsequent works~\cite{JKS03,She08a} the best known upper bound is $O(2.65^d)$~\cite{MNSX11} and the best known lower bound is $\Omega(2.55^d)$~\cite{Leo12}.

The proof of the OSSS Inequality we presented is essentially Lee's~\cite{Lee10}; the alternate proof from Exercise~\ref{ex:jain-zhang} is due to Jain and Zhang~\cite{JZ11}.  The Condorcet Jury Theorem (see Exercise~\ref{ex:condorcet-jury-theorem}) is from~\cite{dC85}. The Shapley value described in Exercise~\ref{ex:shapley} was introduced by the Nobelist Shapley~\cite{Sha53}; for more, see Roth~\cite{Rot88}.  Exercise~\ref{ex:bow-ns-bound} is from Blais, O'Donnell, and Wimmer~\cite{BOW10}. Exercises~\ref{ex:trans-sym-delta}\ref{ex:trans-sym-delta-equal} and~\ref{ex:bsw-ex} are from the work of Benjamini, Schramm, and Wilson~\cite{BSW05}; the term ``revealment'' was introduced by Schramm and Steif~\cite{SS10}. Exercise~\ref{ex:real-osss-countereg} is from~\cite{OSSS05}.  Related to this, it is extremely interesting to ask whether something like the result of Exercise~\ref{ex:osss-variants}\ref{ex:weak-AA} holds for functions $f \btI$.  It has been suggested that the answer is yes:
                                                    \index{Aaronson--Ambainis Conjecture}%
\begin{named}{Aaronson--Ambainis Conjecture} \cite{Aar08,AA11}  Let $f \btI$.  Then $\MaxInf[f] \geq \poly(\Var[f]/\deg(f))$.
\end{named}
\noindent If true, this conjecture would have significant consequences for the limitations of efficient quantum computation; see Aaronson and Ambainis~\cite{AA11}.  The best result in the direction of the conjecture, due to Dinur et~al.~\cite{DFKO07}, is the lower bound $\MaxInf[f] \geq \poly(\Var[f]/2^{\deg(f)})$.

\chapter{Basics of hypercontractivity}                                  \label{chap:hypercontractivity}
    In 1970, Bonami proved the following central result:
    \begin{named}{The Hypercontractivity Theorem}
                                                    \index{Hypercontractivity Theorem}%
        Let $f \btR$ and let ${1 \leq p \leq q \leq \infty}$.  Then $\|\T_\rho f\|_q \leq \|f\|_p$ for $0 \leq \rho \leq \sqrt{\tfrac{p-1}{q-1}}$.
    \end{named}
    As stated, this theorem may look somewhat opaque.  In this chapter we consider some special cases of it that are easier to understand, easier to prove, and that encompass almost all of the theorem's uses.  The proof of the full theorem is deferred to Chapter~\ref{chap:advanced-hypercon}.  The special cases in this chapter are the following:
                                                        \index{Bonami Lemma}%
                                                        \index{$(2,4)$-hypercontractivity|seeonly{Bonami Lemma}}%
    \begin{named}{Bonami Lemma}  Let $f \btR$ have degree $k$.  Then $\|f\|_4 \leq \sqrt{3}^k \|f\|_2$.
    \end{named}
    \noindent The fundamental idea of this statement is that if $\bx \sim \bn$ and $f \btR$ has low degree then the random variable~$f(\bx)$ is quite ``reasonable''; e.g., it is ``nicely'' distributed around its mean.  The Bonami Lemma has a very easy inductive proof and is already powerful enough to obtain many of the well-known applications of ``hypercontractivity'', including the KKL Theorem (proven at the end of this chapter) and the Invariance Principle.
                            \index{hypercontractivity!$(2,q)$- and $(p,2)-$}%
                            \index{$(2,q)$-hypercontractivity|seeonly{hypercontractivity, $(2,q)$- and $(p,2)$-}}%
    \begin{named}{$\mathbf{(2,q)}$-Hypercontractivity Theorem} Let $f \btR$ and let $2 \leq q \leq \infty$.  Then $\|\T_{1/\sqrt{q-1}} f\|_q \leq \|f\|_2$.  As a consequence, if $f$ has degree at most~$k$ then $\|f\|_q \leq \sqrt{q-1}^k \|f\|_2$.
    \end{named}
    \noindent This theorem quantifies the extent to which~$\T_\rho$ is a ``smoothing'' operator; equivalently, it gives even more control over the ``reasonableness'' of low-degree polynomials.  Its consequences include a generalization of the Level-$1$ Inequality (from Chapter~\ref{sec:weight-level-1}) to ``Level-$k$ Inequalities'', as well as a Chernoff-like tail bound for low-degree polynomials of random bits.

                            \index{p2-hypercontractivity@$(p,2)$-hypercontractivity|seeonly{hypercontractivity, $(2,q)$- and $(p,2)$-}}%
    \begin{named}{$\mathbf{(p,2)}$-Hypercontractivity Theorem} Let $f \btR$ and let $1 \leq p \leq 2$.  Then $\|\T_{\sqrt{p-1}} f\|_2 \leq \|f\|_p$.  Equivalently, $\Stab_\rho[f] \leq \|f\|_{1+\rho}^2$ for $0 \leq \rho \leq 1$.
    \end{named}
    \noindent This theorem is actually ``equivalent'' to the $(2,q)$-Hypercontractivity Theorem by virtue of \Holder's inequality.  When specialized to the case of $f \co \bn \to \{0,1\}$ it gives a precise quantification of the fact that the ``noisy hypercube graph'' is a ``small-set expander''.  Qualitatively, this means that if $A \subseteq \bn$ is ``small'', $\bx \sim A$, and $\by \sim N_\rho(x)$, then $\by$ is very unlikely to be in~$A$.

\newcommand{\rv}{\bX}
\section{Low-degree polynomials are reasonable}                 \label{sec:4th-moment}

As anyone who has worked in probability knows, a random variable can sometimes behave in rather ``unreasonable'' ways.  It may be never close to its expectation.  It might exceed its expectation almost always, or almost never.  It might have finite $1$st, $2$nd, and $3$rd moments, but an infinite $4$th moment.  All of this poor behavior can cause a lot of trouble -- wouldn't it be nice to have a class of ``reasonable'' random variables?

                                                \index{reasonable random variable}%
                                                \index{B-reasonable@$B$-reasonable|seeonly{reasonable random variable}}%
A very simple condition on a random variable that guarantees some good behavior is that its $4$th~moment is not too large compared to its $2$nd~moment.
\begin{definition}                              \label{def:reasonable}
    For a real number $B \geq 1$, we say that the real random variable $\bX$ is \emph{$B$-reasonable} if $\E[\bX^4] \leq B\E[\bX^2]^2$.  (Equivalently, if $\|\bX\|_4 \leq B^{1/4} \|\bX\|_2$.)
\end{definition}
The smaller $B$ is, the more ``reasonable'' $\bX$ is. This definition is scale-invariant (i.e., $c \bX$ is $B$-reasonable if and only if $\bX$ is, for $c \neq 0$) but not translation-invariant ($c + \bX$ and $\bX$ may not be equally reasonable).  The latter fact can sometimes be awkward, a point we'll address further in Section~\ref{sec:hypercontractivity}.  Indeed, we'll later encounter a few alternative conditions that also capture ``reasonableness''. For example, in Chapter~\ref{chap:invariance} we'll consider the analogous $3$rd~moment condition, $\E[|\bX|^3] \leq B \E[\bX^2]^{3/2}$.  Strictly speaking, the $4$th~moment condition is stronger: if $\bX$ is $B$-reasonable, then
\[
    \E[|\bX|^3] = \E[|\bX| \cdot \bX^2] \leq \sqrt{\E[\bX^2]}\sqrt{\E[\bX^4]} \leq \sqrt{B}\E[\bX^2]^{3/2};
\]
on the other hand, there exist random variables with finite $3$rd~moment and infinite $4$th~moment.  However, such unusual random variables almost never arise for us, and morally speaking the $4$th and $3$rd moment conditions are about equally good proxies for reasonableness.

\begin{examples}                        \label{eg:reasonable}
    If $\bx \sim \bits$ is uniformly random then $\bx$ is~$1$-reasonable.  If $\bg \sim \normal(0,1)$ is a standard Gaussian, then $\E[\bg^4] = 3$, so $\bg$ is $3$-reasonable.  If $\bu \sim [-1,1]$ is uniform, then you can calculate that it is $\frac95$-reasonable.  In all of these examples $B$ is a ``small'' constant, and we think of these random variables simply as ``reasonable''.  An example of an ``unreasonable'' random variable would be highly biased Bernoulli random variable; say, $\Pr[\by = 1] = 2^{-n}$, $\Pr[\by = 0] = 1-2^{-n}$, where~$n$ is large.  This~$\by$ is not $B$-reasonable unless $B \geq 2^n$.
\end{examples}

Let's give a few illustrations of why reasonable random variables are nice to work with.  First, they have slightly better tail bounds than what you would get out of the Chebyshev inequality:
\begin{proposition}                                     \label{prop:reasonable-conc}
    Let $\bX \not \equiv 0$ be $B$-reasonable.  Then $\Pr[|\bX| \geq t \|\bX\|_2] \leq B/t^4$ for all $t > 0$.
\end{proposition}
\begin{proof}
    This is immediate from Markov's inequality:
    \[
        \Pr[|\bX| \geq t\|\bX\|_2] = \Pr[\bX^4 \geq t^4\|\bX\|_2^4] \leq \frac{\E[\bX^4]}{t^4 \E[\bX^2]^2} \leq \frac{B}{t^4}. \qedhere
    \]
\end{proof}
More interestingly, they also satisfy \emph{anticoncentration} bounds;
                                                \index{anticoncentration}%
                                                \index{Paley--Zygmund inequality}%
                                                \index{second moment method|seeonly{Paley--Zygmund inequality}}%
e.g., you can \emph{upper}-bound the probability that they are near~$0$.
\begin{proposition}                                     \label{prop:reasonable-anticonc}
    Let $\bX \not \equiv 0$ be $B$-reasonable.  Then $\Pr[|\bX| > t \|\bX\|_2] \geq (1-t^2)^2/B$ for all $t \in [0,1]$.
\end{proposition}
\begin{proof}
    Applying the Paley--Zygmund inequality (also called the ``second moment method'') to~$\bX^2$, we get
    \[
        \Pr[|\bX| \geq t \|\bX\|_2] = \Pr[\bX^2 \geq t^2 \E[\bX^2]] \geq (1-t^2)^2\frac{\E[\bX^2]^2}{\E[\bX^4]} \geq \frac{(1-t^2)^2}{B}. \qedhere
    \]
\end{proof}
\noindent For a generalization of this proposition, see Exercise~\ref{ex:hc-anticonc}.

For a discrete random variable~$\rv$, a simple condition that guarantees reasonableness is that~$\rv$ takes on each of its values with nonnegligible probability:
\begin{proposition}                                     \label{prop:min-prob-implies-reasonable}
    Let $\rv$ be a discrete random variable with probability mass function~$\pi$.  Write
    \[
        \lambda = \min(\pi) = \min_{x \in \mathrm{range}(\rv)}\{\Pr[\rv = x]\}.
    \]
    Then $\rv$ is $(1/\lambda)$-reasonable.
\end{proposition}
\begin{proof}
    Let $M = \|\rv\|_\infty$.  Since $\Pr[|\rv| = M] \geq \lambda$ we get
    \[
        \E[\rv^2] \geq \lambda M^2 \quad\implies\quad M^2 \leq \E[\rv^2]/\lambda.
    \]
    On the other hand,
    \[
        \E[\rv^4] = \E[\rv^2 \cdot \rv^2] \leq M^2 \cdot \E[\rv^2],
    \]
    and thus $\E[\rv^4] \leq (1/\lambda)\E[\rv^2]^2$ as required.
\end{proof}
The converse to Proposition~\ref{prop:min-prob-implies-reasonable} is certainly not true. For example, if $\rv = \frac{1}{\sqrt{n}} \bx_1 + \cdots + \frac{1}{\sqrt{n}} \bx_n$ where $\bx \sim \bn$, then $\rv$ is very close to a standard Gaussian random variable (for~$n$ large) and is, unsurprisingly, $3$-reasonable.  On the other hand, the ``$\lambda$'' for this~$\rv$ is tiny, $2^{-n}$.

This discussion raises the issue of how you might try to construct an \emph{unreasonable} random variable out of independent uniform $\pm 1$ bits.  By Proposition~\ref{prop:min-prob-implies-reasonable}, at the very least you must use a lot of them. Furthermore, it also seems that they must be combined in a \emph{high-degree} way.  For example, to construct the unreasonable random variable~$\by$ from Example~\ref{eg:reasonable} requires degree~$n$: $\by = (1+\bx_1)(1+\bx_2)\cdots(1+\bx_n)/2^n$.

Indeed, the idea that high degree is required for unreasonableness is correct, as the following crucial result shows:
                                                    \index{Bonami Lemma}%
\begin{named}{The Bonami Lemma}  For each $k$, if $f \btR$ has degree at most~$k$ and $\bx_1, \dots, \bx_n$ are independent, uniformly random $\pm 1$ bits, then the random variable $f(\bx)$ is $9^k$-reasonable, i.e.,
\[
    \E[f^4] \leq 9^k \E[f^2]^2 \quad\iff\quad \|f\|_4 \leq \sqrt{3}^k \|f\|_2.
\]
\end{named}
In other words, \emph{low-degree polynomials of independent uniform $\pm 1$ bits are reasonable}.  As we will explain later, the Bonami Lemma is a special case of more general results in the theory of ``hypercontractivity''.  However, many key theorems using hypercontractivity -- e.g., the KKL Theorem, the Invariance Principle -- really need only the simple Bonami Lemma.  (We should also note that the name ``Bonami Lemma'' is not standard; however, the result was first proved by Bonami and it's often used as a lemma, so the name fits.  See the discussion in the notes in Section~\ref{sec:basic-hypercon-notes}.)

One pleasant thing about the Bonami Lemma is that once you decide to prove it by induction on~$n$, the proof practically writes itself.  The only ``non-automatic'' step is an application of Cauchy--Schwarz.

\begin{proof}[Proof of the Bonami Lemma]
    We assume $k \geq 1$ as otherwise~$f$ must be constant and the claim is trivial.  The proof is by induction on~$n$.  Again, if $n = 0$, then $f$ must be constant and the claim is trivial.  For $n \geq 1$ we can use the decomposition $f(x) = x_n \D_n f(x) + \uE_n f(x)$ (Proposition~\ref{prop:expectation-facts}), where $\deg(\D_n f) \leq k-1$, $\deg(\uE_n f) \leq k$, and the polynomials $\D_n f(x)$ and $\uE_n f(x)$ don't depend on~$x_n$.  For brevity we write $\boldf = f(\bx)$, $\bd = \D_nf(\bx)$, and $\be = \uE_n f(\bx)$.  Now
    \begin{align*}
        \E[\boldf^4] &= \E[(\bx_n \bd + \be)^4] \\
              &= \E[\bx_n^4 \bd^4] + 4 \E[\bx_n^3 \bd^3 \be] + 6 \E[\bx_n^2 \bd^2 \be^2] + 4 \E[\bx_n \bd \be^3] +  \E[\be^4] \\
              &= \E[\bx_n^4] \E[\bd^4] + 4 \E[\bx_n^3] \E[\bd^3 \be] + 6 \E[\bx_n^2] \E[\bd^2 \be^2] + 4 \E[\bx_n]\E[\bd \be^3] + \E[\be^4].
    \end{align*}
    In the last step we used the fact that $\bx_n$ is independent of $\bd$ and $\be$, since $\D_nf$ and $\uE_n f$ do not depend on $x_n$.  We now use $\E[\bx_n] = \E[\bx_n^3] = 0$ and $\E[\bx_n^2] = \E[\bx_n^4] = 1$ to deduce
    \begin{equation} \label{eqn:bonami-induct}
        \E[\boldf^4] = \E[\bd^4] + 6 \E[\bd^2 \be^2] + \E[\be^4].
    \end{equation}
    A similar (and simpler) sequence of steps shows that
    \begin{equation} \label{eqn:bonami-last}
        \E[\boldf^2] = \E[\bd^2] + \E[\be^2].
    \end{equation}
    To upper-bound~\eqref{eqn:bonami-induct}, recall that $\bd = \D_nf(\bx)$ where $\D_n f$ is a multilinear polynomial of degree at most $k-1$ depending on $n-1$ variables.  Thus we can apply the induction hypothesis to deduce $\E[\bd^4] \leq 9^{k-1} \E[\bd^2]^2$.  Similarly, $\E[\be^4] \leq 9^{k} \E[\be^2]^2$ since $\deg(\uE_nf) \leq k$.  To bound $\E[\bd^2 \be^2]$ we apply Cauchy--Schwarz, getting $\sqrt{\E[\bd^4]}\sqrt{\E[\be^4]}$ and letting us use induction again.  Thus we have
    \begin{align*}
        \E[\boldf^4] &\leq 9^{k-1} \E[\bd^2]^2 + 6 \sqrt{9^{k-1}\E[\bd^2]^2} \sqrt{9^{k}\E[\be^2]^2} + 9^{k} \E[\be^2]^2 \\
        &\leq 9^k \Bigl(\E[\bd^2]^2 + 2 \E[\bd^2]\E[\be^2] + \E[\be^2]^2\Bigr) = 9^k \Bigl(\E[\bd^2] + \E[\be^2]\Bigr)^2,
    \end{align*}
    where we used $9^{k-1} \E[\bd^2]^2 \leq 9^k \E[\bd^2]^2$.  In light of~\eqref{eqn:bonami-last}, this completes the proof.
\end{proof}
Some aspects of the sharpness of the Bonami Lemma are explored in Exercises~\ref{ex:deg-1-bonami}, \ref{ex:simple-bonami-lower-bound}, \ref{ex:super-bonami}, and~\ref{ex:number-of-matchings}.  Here we make one more observation.  At the end of the proof we used the wasteful-looking inequality $9^{k-1} \E[\bd^2]^2 \leq 9^k \E[\bd^2]^2$. Tracing back through the proof, it's easy to see that it would still be valid even if we just had $\E[\bx_i^4] \leq 9$ rather than $\E[\bx_i^4] = 1$.  For example, the Bonami Lemma holds not just if the $\bx_i$'s are random bits,  but if they are standard Gaussians, or are uniform on $[-1,1]$, or there are some of each.  We leave the following as Exercise~\ref{ex:bonami-lemma-cor}.
\begin{corollary}                                       \label{cor:bonami-lemma}
    Let $\bx_1, \dots, \bx_n$ be independent, not necessarily identically distributed, random variables satisfying $\E[\bx_i] = \E[\bx_i^3] = 0$.  (This holds if, e.g., each $-\bx_i$ has the same distribution as $\bx_i$.)  Assume also that each $\bx_i$ is $B$-reasonable.  Let $\boldf = F(\bx_1, \dots, \bx_n)$, where $F$ is a multilinear polynomial of degree at most~$k$.  Then $\boldf$ is $\max(B,9)^k$-reasonable.
\end{corollary}

As a first application of the Bonami Lemma, let us combine it with Proposition~\ref{prop:reasonable-anticonc} to show that a low-degree function is not too concentrated around its mean:
\begin{theorem}                                       \label{thm:low-deg-anticonc}
    Let $f \btR$ be a nonconstant function of degree at most~$k$; write $\mu = \E[f]$ and $\sigma = \sqrt{\Var[f]}$.  Then
    \[
        \Pr_{\bx \sim \bits}[|f(\bx) - \mu| > \half \sigma] \geq \tfrac{1}{16} 9^{1-k}.
    \]
\end{theorem}
\begin{proof}
    Let $g = \frac{1}{\sigma}(f - \mu)$, a function of degree at most~$k$ satisfying $\|g\|_2 = 1$.  By the Bonami Lemma, $g$~is $9^k$-reasonable.  The result now follows by applying Proposition~\ref{prop:reasonable-anticonc} to~$g$ with $t = \half$.
\end{proof}

Using this theorem, we can give a short proof of the FKN~Theorem
                                                    \index{FKN Theorem}%
from Chapter~\ref{sec:arrow}: If $f \btb$ has $\W{1}[f] = 1-\delta$ then $f$ is $O(\delta)$-close to $\pm \chi_i$ for some $i \in [n]$.
\begin{proof}[Proof of the FKN Theorem]
    Write $\ell = f^{=1}$, so $\E[\ell^2] = 1-\delta$ by assumption.  We may assume without loss of generality that $\delta \leq \frac{1}{1600}$.  The goal of the proof is to show that $\Var[\ell^2]$ is small; specifically we'll show that $\Var[\ell^2] \leq 6400\delta$.  This will complete the proof because (using Exercise~\ref{ex:fkn-helper} for the first equality below)
    \begin{align*}
        \tfrac12 \Var[\ell^2]
                 = \sum_{i \neq j} \wh{f}(i)^2\wh{f}(j)^2
                 &= \Bigl(\littlesum_{i=1}^n \wh{f}(i)^2\Bigr)^2 - \littlesum_{i=1}^n \wh{f}(i)^4 \\
                 &= (1-\delta)^2 - \littlesum_{i=1}^n \wh{f}(i)^4 \geq (1-2\delta) - \littlesum_{i=1}^n \wh{f}(i)^4
    \end{align*}
    and hence $\Var[\ell^2] \leq 6400\delta$ implies
    \[
        1 - 3202\delta \leq \littlesum_{i=1}^n \wh{f}(i)^4 \leq \max_i\{\wh{f}(i)^2\} \littlesum_{i=1}^n \wh{f}(i)^2
                 \leq \max_i\{\wh{f}(i)^2\} \leq \max_i\{|\wh{f}(i)|\},
    \]
    as required.

    To bound $\Var[\ell^2]$ we first apply Theorem~\ref{thm:low-deg-anticonc} to the degree-$2$ function $\ell^2$; this yields
    \[
        \Pr\Bigl[\bigl|\ell^2 - (1-\delta)\bigr| \geq \half \sqrt{\Var[\ell^2]}\Bigr] \geq \tfrac{1}{16} 9^{1-2} = \tfrac{1}{144}.
    \]
    Now suppose by way of contradiction that $\Var[\ell^2] > 6400\delta$; then the above implies
    \begin{equation} \label{eqn:fkn1}
        \tfrac{1}{144} \leq \Pr\Bigl[\bigl|\ell^2 - (1-\delta)\bigr| > 40\sqrt{\delta}\Bigr] \leq \Pr\Bigl[\bigl|\ell^2 - 1\bigr| > 39\sqrt{\delta}\Bigr].
    \end{equation}
    This says that $|\ell|$ is frequently far from~$1$.  Since $|f| = 1$ always, we can deduce that $|f - \ell|^2$ is frequently large. More precisely, a short calculation (Exercise~\ref{ex:fkn-calculation}) shows that $(f-\ell)^2 \geq 169\delta$ whenever $|\ell^2 - 1| > 39\sqrt{\delta}$.  But now~\eqref{eqn:fkn1} implies $\E[(f-\ell)^2] \geq \frac{1}{144} \cdot 169 \delta > \delta$, a contradiction since $\E[(f-\ell)^2] = 1- \W{1}[f] = \delta$ by assumption.
\end{proof}

\section{Small subsets of the hypercube are noise-sensitive}                \label{sec:sse-intro}

An immediate consequence of the Bonami Lemma is that for any $f \btR$ and $k \in \N$,
\begin{equation} \label{eqn:2-4-hypercon-k}
    \|\T_{1/\sqrt{3}} f^{=k}\|_4 = \tfrac{1}{\sqrt{3}^k} \|f^{=k}\|_4 \leq \|f^{=k}\|_2.
\end{equation}
This is a special case of the \emph{$(2,4)$-Hypercontractivity Theorem} (whose name will be explained shortly), which says that the assumption of degree-$k$ homogeneity is not necessary:
                                            \index{$(2,4)$-hypercontractivity}%
\begin{named}{$\mathbf{(2,4)}$-Hypercontractivity Theorem} Let $f \btR$.  Then
\[
    \|\T_{1/\sqrt{3}} f\|_4 \leq \|f\|_2.
\]
\end{named}
It almost looks as though you could prove this theorem simply by summing~\eqref{eqn:2-4-hypercon-k} over~$k$.  In fact that proof strategy can be made to work given a few extra tricks (see Exercise~\ref{ex:bonami-2-4-tricks}), but it's just as easy to repeat the induction technique used for the Bonami Lemma.
\begin{proof}
    We'll prove $\E[\T_{1/\sqrt{3}} f(\bx)^4] \leq \E[f(\bx)^2]^2$ using the same induction as in the Bonami Lemma.
    Retaining the notation $\bd$ and $\be$, and using the shorthand $\T = \T_{1/\sqrt{3}}$, we have
    \[
        \T \boldf = \bx_n \cdot \tfrac{1}{\sqrt{3}} \T \bd + \T \be.
    \]
    Similar computations to those in the Bonami Lemma proof yield
    \begin{align*}
        \E[(\T \boldf)^4] &= \bigl(\tfrac{1}{\sqrt{3}}\bigr)^4\E[(\T \bd)^4] + 6 \bigl(\tfrac{1}{\sqrt{3}}\bigr)^2 \E[(\T \bd)^2(\T \be)^2] + \E[(\T \be)^4] \\
                          &\leq \E[(\T \bd)^4] + 2 \E[(\T \bd)^2(\T \be)^2] + \E[(\T \be)^4] \\
                          &\leq \E[(\T \bd)^4] + 2 \sqrt{\E[(\T \bd)^4]}\sqrt{\E[(\T \be)^4]} + \E[(\T \be)^4] \\
                          &\leq \E[\bd^2]^2 + 2 \E[\bd^2]\E[\be^2] + \E[\be^2]^2 \\
                          &= \bigl(\E[\bd^2] + \E[\be^2]\bigr)^2 = \E[\boldf^2]^2,
    \end{align*}
    where the second inequality is Cauchy--Schwarz, the third is induction, and the final equality is a simple computation analogous to~\eqref{eqn:bonami-last}.
\end{proof}

The name ``hypercontractivity'' in this theorem describes the fact that not only is $\T_{1/\sqrt{3}}$ a ``contraction'' on~$L^2(\bits^n)$ -- meaning $\|\T_{1/\sqrt{3}} f\|_2 \leq \|f\|_2$ for all~$f$ (Exercise~\ref{ex:T-contracts}) -- it's even a contraction when viewed as an operator from $L^2(\bits^n)$ to $L^4(\bits^n)$.  You should think of hypercontractivity theorems as quantifying the extent to which $\T_\rho$ is a ``smoothing'', or ``reasonable-izing'' operator.

Unfortunately the quantity $\|\T_{1/\sqrt{3}}f\|_4$ in the $(2,4)$-Hypercontractivity Theorem does not have an obvious combinatorial meaning.  On the other hand, the quantity
\[
    \|\T_{1/\sqrt{3}}f\|_2 = \sqrt{\la \T_{1/\sqrt{3}}f, \T_{1/\sqrt{3}}f \ra} = \sqrt{\la f, \T_{1/\sqrt{3}} \T_{1/\sqrt{3}} f\ra} = \sqrt{\Stab_{1/3}[f]},
\]
does have a nice combinatorial meaning.  And we can make this quantity appear in the Hypercontractivity Theorem via a simple trick from analysis, just using the fact that $\T_{1/\sqrt{3}}$ is a self-adjoint operator.  We ``flip the norms across~$2$'' using \Holder's inequality:
                                                \index{Holder inequality@\Holder inequality}%
\begin{named}{$\mathbf{(4/3,2)}$-Hypercontractivity Theorem} Let $f \btR$.  Then
\[
\|\T_{1/\sqrt{3}} f\|_2 \leq \|f\|_{4/3};
\]
i.e.,
    \begin{equation} \label{eqn:4/3-2}
        \Stab_{1/3}[f] \leq \|f\|_{4/3}^2.
    \end{equation}
\end{named}
\begin{proof}
    Writing $\T = \T_{1/\sqrt{3}}$ for brevity we have
    \begin{equation}                    \label{eqn:hc-holder}
        \|\T f\|_2^2 = \la \T f, \T f \ra = \la f, \T \T f \ra \leq \|f\|_{4/3} \|\T \T f\|_4 \leq \|f\|_{4/3} \|\T f\|_2
    \end{equation}
    by \Holder's inequality and the $(2,4)$-Hypercontractivity Theorem.  Dividing through by $\|\T f\|_2$ (which we may assume is nonzero) completes the proof.
\end{proof}
In the inequality~\eqref{eqn:4/3-2} the left-hand side is a natural quantity.  The right-hand side is just~$1$ when $f \btb$, which is not very interesting. But if we instead look at $f \co \bn \to \{0,1\}$ we get something very interesting:
\begin{corollary}                                       \label{cor:sse-1/3}
    Let $A \subseteq \bn$ have volume $\alpha$; i.e., let $1_A \co \bn \to \zo$ satisfy $\E[1_A] = \alpha$.  Then
    \[
        \Stab_{1/3}[1_A] = \Pr_{\substack{\bx \sim \bn \\ \by \sim N_{1/3}(\bx)}}[\bx \in A, \by \in A] \leq \alpha^{3/2}.
    \]
    Equivalently (for $\alpha > 0$),
    \[
        \Pr_{\substack{\bx \sim A \\ \by \sim N_{1/3}(\bx)}}[\by \in A] \leq \alpha^{1/2}.
    \]
\end{corollary}
\begin{proof}
     This is immediate from inequality~\eqref{eqn:4/3-2}, since
     \[
        \|1_A\|_{4/3}^2 = \Bigr(\E_{\bx}[|1_A(\bx)|^{4/3}]^{3/4}\Bigr)^2 = \E_{\bx}[1_A(\bx)]^{3/2} = \alpha^{3/2}. \qedhere
     \]
\end{proof}

See Section~\ref{sec:hypercon-apps} for the generalization of this corollary to noise rates other than~$1/3$.

\begin{example}
    Assume $\alpha = 2^{-k}$, $k \in \N^+$, and $A$ is a subcube of codimension~$k$; e.g., $1_A \co \F_2^n \to \zo$ is the logical $\AND$ function on the first~$k$ coordinates.  For every $x \in A$, when we form $\by \sim N_{1/3}(x)$ we'll have $\by \in A$ if and only if the first~$k$ coordinates of $x$ do not change, which happens with probability $(2/3)^k = (2/3)^{\log(1/\alpha)} = \alpha^{\log(3/2)} \approx \alpha^{.585} \leq \alpha^{1/2}$.  In fact, the bound~$\alpha^{1/2}$ in Corollary~\ref{cor:sse-1/3} is essentially sharp when~$A$ is a Hamming ball; see Exercise~\ref{ex:gaussquad-est}.
\end{example}

We can phrase Corollary~\ref{cor:sse-1/3} in terms of the \emph{expansion} in a certain graph:
                    \index{noisy hypercube graph}%
                    \index{rho-stable hypercube graph@$\rho$-stable hypercube graph|seeonly{noisy hypercube graph}}%
\begin{definition}
    For $n \in \N^+$ and $\rho \in [-1,1]$, the $n$-dimensional \emph{$\rho$-stable hypercube graph} is the edge-weighted, complete directed graph on vertex set $\bn$ in which the weight on directed edge $(x,y) \in \bn \times \bn $ is equal to $\Pr[(\bx,\by) = (x,y)]$ when $(\bx,\by)$ is a $\rho$-correlated pair.  If $\rho = 1-2\delta$ for $\delta \in [0,1]$, we also call this the \emph{$\delta$-noisy hypercube graph}.  Here the weight on~$(x,y)$ is $\Pr[(\bx, \by) = (x,y)]$ where $\bx \sim \bn$ is uniform and $\by$ is formed from $\bx$ by negating each coordinate independently with probability~$\delta$.
\end{definition}
\begin{remark}
    The edge weights in this graph are nonnegative and sum to~$1$.  The graph is also ``regular'' in the sense that for each $x \in \bn$ the sum of all the edge weight leaving (or entering)~$x$ is $2^{-n}$.  You can also consider the graph to be undirected, since the weight on~$(x,y)$ is the same as the weight on~$(y,x)$; in this viewpoint, the weight on the undirected edge $(x,y)$ would be $2^{1-n}\delta^{\hamdist(x,y)}(1-\delta)^{n-\hamdist(x,y)}$.
    In fact, the graph is perhaps best thought of as the discrete-time Markov chain on state space $\bn$ in which a step from state $x \in \bn$ consists of moving to state $\by \sim N_\rho(x)$.  This is a reversible chain with the uniform stationary distribution. Each discrete step is equivalent to running the ``usual'' \emph{continuous-time} Markov chain on the hypercube for time $t = \ln(1/\rho)$ (assuming $\rho \in [0,1]$).
\end{remark}
With this definition in place, we can see Corollary~\ref{cor:sse-1/3} as saying that the $1/3$-stable (equivalently, $1/3$-noisy) hypercube graph is a ``small-set expander'':
                                                \index{expansion!small-set}%
                                                \index{small-set expansion|seeonly{expansion!small-set}}%
given any small $\alpha$-fraction of the vertices~$A$, almost all of the edge weight touching~$A$ is on its boundary. More precisely, if we choose a random vertex $\bx \in A$ and take a random edge out of~$\bx$ (with probability proportional to its edge weight), we end up outside~$A$ with probability at least $1 - \alpha^{1/2}$.  You can compare this with the discussion surrounding the Level-$1$ Inequality in Section~\ref{sec:weight-level-1}, which is the analogous statement for the $\rho$-stable hypercube graph ``in the limit $\rho \to 0^+$''.  The appropriate statement for general~$\rho$ is appears in Section~\ref{sec:hypercon-apps} as the ``Small-Set Expansion Theorem''.

\medskip

Corollary~\ref{cor:sse-1/3} would apply equally well if $1_A$ were replaced by a function $g \co \bn \to \{-1,0,1\}$, with $\alpha$ denoting $\Pr[g \neq 0] = \E[|g|] = \E[g^2]$.  This situation occurs naturally when $g = \D_i f$ for some Boolean-valued $f \btb$.  In this case $\Stab_{1/3}[g] = \Inf^{(1/3)}_i[f]$, the $1/3$-stable influence of~$i$ on~$f$. We conclude that for a Boolean-valued function, if the influence of~$i$ is small then its $1/3$-stable influence is much smaller:
                                        \index{stable influence}%
\begin{corollary}                                       \label{cor:1/3-stab-infl}
    Let $f \btb$.  Then $\Inf_i^{(1/3)}[f] \leq \Inf_i[f]^{3/2}$ for all~$i$.
\end{corollary}

We remark that the famous KKL~Theorem (stated in Chapter~\ref{sec:tribes}) more or less follows by summing the above inequality over $i \in [n]$; if you're impatient to see its proof you can skip directly to Section~\ref{sec:KKL} now.\\

Let's take one more look at the ``small-set expansion result'', Corollary~\ref{cor:sse-1/3}.  Since noise stability roughly measures how ``low'' a function's Fourier weight is, this corollary implies that a function $f \co \bn \to \{0,1\}$ with small mean~$\alpha$ cannot have much of its Fourier weight at low degree.  More precisely, for any $k \in \N$ we have
\begin{equation} \label{eqn:weak-level-k}
    \alpha^{3/2} \geq \Stab_{1/3}[f] \geq (1/3)^k \W{\leq k}[f] \quad\implies\quad \W{\leq k}[f] \leq 3^k \alpha^{3/2}.
\end{equation}
For $k = 1$ this gives $\W{\leq 1}[f] \leq 3\alpha^{3/2}$, which is nontrivial but not as strong as the Level-1 Inequality from Section~\ref{sec:weight-level-1}.  But~\eqref{eqn:weak-level-k} also gives us ``level-$k$ inequalities''
                                            \index{Level-$k$ Inequalities}%
for larger values of~$k$. For example,
\[
    \W{\leq .25 \log(1/\alpha)}[f] \leq \alpha^{-.25 \log 3 + 3/2} \leq \alpha^{1.1} \ll \alpha = \|f\|_2^2;
\]
i.e., almost all of $f$'s Fourier weight is above degree $.25 \log(1/\alpha)$.  We will give slightly improved versions of these level-$k$ inequalities in Section~\ref{sec:hypercon-apps}.

\section{$(2,q)$- and $(p,2)$-hypercontractivity for a single bit}          \label{sec:hypercontractivity}

Although you can get a lot of mileage out of studying the $4$-norm of random variables, it's also natural to consider other norms.  For example, we would get improved versions of our concentration and anticoncentration results, Propositions~\ref{prop:reasonable-conc} and~\ref{prop:reasonable-anticonc}, if we could bound the higher norms of a random variable in terms of its $2$-norm.  As we'll see, we can also get stronger ``level-$k$ inequalities''  by bounding the $(2+\eps)$-norm of a Boolean function for small $\eps > 0$.

We started with the $4$-norm due to the simplicity of the proofs of the Bonami Lemma and the $(2,4)$-Hypercontractivity Theorem.  To generalize these results to other norms it's a bit more elegant to work with the latter.  Partly this is because it's ``formally stronger'' (see Theorem~\ref{thm:general-bonami}).  But the main reason is that the hypercontractivity version alleviates the inelegant issue that being ``$B$-reasonable'' is not translation-invariant.  Thus instead of generalizing the condition that $\|\rho \bX\|_4 \leq \|\bX\|_2$ (``$\bX$ is $\rho^{-4}$-reasonable'') we'll  generalize the condition that $\|a + \rho b \bX\|_4 \leq \|a + b\bX\|_2$ (cf.\ the $n = 1$ case of the $(2,4)$-Hypercontractivity Theorem).

                                                    \index{hypercontractivity|(}%
\begin{definition}      \label{def:hypercon}
    Let $1 \leq p \leq q \leq \infty$ and let $0 \leq \rho < 1$. We say that a real random variable $\bX$ (with $\|\bX\|_q < \infty$) is
    \emph{$(p,q,\rho)$-hypercontractive} if
    \[
            \|a + \rho b \bX\|_q \leq \|a + b \bX\|_p \quad \text{for all constants } a, b \in \R.
    \]
\end{definition}
\begin{remark}              \label{rem:hypercon-basics}
    By homogeneity, it suffices to check the condition for $a = 1$, $b \in \R$ or for $a \in \R$, $b = 1$ (cf.~Exercise~\ref{ex:hc-basics0}\ref{ex:hc-homog}). It's also true (Exercise~\ref{ex:hc-basics2}) that if $\bX$ is $(p,q,\rho)$-hypercontractive then it is  $(p,q,\rho')$-hypercontractive for $\rho' < \rho$ as well.
\end{remark}
In Exercise~\ref{ex:hc-basics1} you will show that if~$\bX$ is hypercontractive then $\E[\bX]$ must be~$0$.  Thus hypercontractivity, like reasonableness, is not a translation-invariant notion. Nevertheless, the fact that the definition involves translation by an arbitrary~$a$ greatly facilitates proofs by induction.  For example, an elegant property we gain from the definition is the following (Exercise~\ref{ex:sum-hc-hc}):
                                                    \index{hypercontractivity!preserved by sums}%
\begin{proposition}                                     \label{prop:sum-hc-hc}
    Let $\bX$ and $\bY$ be independent $(p,q,\rho)$-hypercontractive random variables.  Then $\bX + \bY$ is also $(p,q,\rho)$-hypercontractive.
\end{proposition}

The $n = 1$ case of our $(2,4)$-Hypercontractivity Theorem precisely says that a single uniformly random $\pm 1$ bit~$\bx$ is $(2,4,1/\sqrt{3})$-hypercontractive; \linebreak the $(4/3,2)$-Hypercontractivity Theorem says that the bit $\bx$ is also $(4/3,2,1/\sqrt{3})$-hypercontractive.
                            \index{hypercontractivity!$(2,q)$- and $(p,2)$-|(}%
We'll spend the remainder of this section generalizing these facts to $(2,q,\rho)$- and $(p,2,\rho)$-hypercontractivity for other values of~$p$ and~$q$. We remark that in our study of hypercontractivity we'll focus mainly on the cases of $p = 2$ or $q = 2$.
The study of hypercontractivity  for $p, q \neq 2$ and for random variables other than uniform $\pm 1$ bits is deferred to Chapter~\ref{chap:advanced-hypercon}.
                                                    \index{hypercontractivity|)}%

We now consider hypercontractivity of a uniformly random $\pm 1$ bit~$\bx$. We know that $\bx$ is $(2,q,1/\sqrt{3})$-hypercontractive for $q = 4$; what about other values of~$q$?  Things are most pleasant when $q$ is an even integer because then you don't need to take the absolute value when computing $\|a + \rho b \bX\|_q$.  So let's try $q = 6$.
\begin{proposition}                                     \label{prop:2-6-hypercon}
    For $\bx$ a uniform $\pm 1$ bit, we have $\|a + \rho b \bx\|_6 \leq \|a + b \bx\|_2$ for all $a, b \in \R$ if (and only if) $\rho \leq 1/\sqrt{5}$.  That is, $\bx$ is $(2,6,1/\sqrt{5})$-hypercontractive.
\end{proposition}
\begin{proof}
    Raising the inequality to the $6$th power, we need to show
    \begin{equation} \label{eqn:2-6-hypercon1}
        \E[(a+\rho b \bx)^6] \leq \E[(a+b \bx)^2]^3.
    \end{equation}
    The result is trivial when $a = 0$; otherwise, we may assume $a = 1$ by homogeneity.  We expand both quantities inside expectations and use the fact that $\E[\bx^k]$ is $0$ when $k$ is odd and $1$ when~$k$ is even. Thus~\eqref{eqn:2-6-hypercon1} is equivalent to
    \begin{equation} \label{eqn:2-6-hypercon2}
        1+ 15 \rho^2 b^2 + 15 \rho^4 b^4 + \rho^6 b^6 \leq (1 + b^2)^3 = 1 + 3b^2 + 3b^4 + b^6.
    \end{equation}
    Comparing the two sides term-by-term we see that the coefficient on $b^2$ is the limiting factor:  in order for~\eqref{eqn:2-6-hypercon2} to hold for all $b \in \R$ it is sufficient that $15 \rho^2 \leq 3$; i.e., $\rho \leq 1/\sqrt{5}$.  By considering $b \to 0$ it's also easy to see that this condition is necessary.
\end{proof}
If you repeat this analysis for the case of $q = 8$ you'll find that again the limiting factor is the coefficient on $b^2$, and that $\bx$ is $(2,8,\rho)$-hypercontractive if (and only if) $\binom{8}{2} \rho^2 \leq \binom{4}{1}$; i.e., $\rho \leq 1/\sqrt{7}$.  In light of this it is natural to guess that the following is true:
\begin{theorem}                                     \label{thm:hypercon-bit}
    Let $\bx$ be a uniform $\pm 1$ bit and let $q \in (2, \infty]$.  Then $\|a + \rho b \bx \|_q \leq \|a+b \bx\|_2$ for all $a,b \in \R$ assuming $\rho \leq 1/\sqrt{q-1}$.

    Equivalent statements are that $\|a + (1/\sqrt{q-1}) b \bx \|_q^2 \leq a^2 + b^2$, that  $\bx$ is $(2,q,1/\sqrt{q-1})$-hypercontractive, and that $\|\T_{1/\sqrt{q-1}} f\|_q \leq \|f\|_2$ holds for any $f \co \bits \to \R$.
\end{theorem}
For $q$ an even integer it is not hard (see Exercise~\ref{ex:2-2s-hypercon}) to prove Theorem~\ref{thm:hypercon-bit} just as we did for $q = 6$.  Indeed, the proof works even under more general moment conditions on~$\bx$, as in Corollary~\ref{cor:bonami-lemma}.
Unfortunately, obtaining Theorem~\ref{thm:hypercon-bit} for all real $q > 2$ takes some more tricks.  A natural idea is to try forging ahead as in Proposition~\ref{prop:2-6-hypercon}, using the series expansions for $(1+\rho b x)^q$ and $(1+b^2)^{q/2}$ provided by the Generalized Binomial Theorem.  However, even when $|b| < 1$ (so that convergence is not an issue) there is a difficulty because the coefficients in the expansion of $(1+b^2)^{q/2}$ are sometimes negative.

Luckily, this issue of negative coefficients in the series expansion goes away if you try to prove the analogous $(p,2,\rho)$-hypercontractivity statement.  Thus the slick proof of Theorem~\ref{thm:hypercon-bit} proceeds by first proving that statement, then ``flipping the norms across~$2$''.
\begin{theorem}                                     \label{thm:hypercon-bit-2}
    Let $\bx$ be a uniform $\pm 1$ bit and let $1 \leq p < 2$.  Then $\|a + \rho b \bx \|_2 \leq \|a+b \bx\|_p$ for all $a,b \in \R$ assuming $0 \leq \rho \leq \sqrt{p-1}$.  That is, $\bx$ is $(p,2,\sqrt{p-1})$-hypercontractive.
\end{theorem}
\begin{proof}
    By Remark~\ref{rem:hypercon-basics} we may assume $a = 1$ and $\rho = \sqrt{p-1}$.  By Exercise~\ref{ex:nonneg-hc-trick} we may also assume without loss of generality that $1+bx \geq 0$ for $x \in \bits$; i.e., that $|b| \leq 1$. It then suffices to prove the result for all $|b| < 1$ because the $|b| = 1$ case follows by continuity.  Writing $b = \eps$ for the sake of intuition, we need to show
    \begin{align}
        \|1 + \sqrt{p-1}\cdot \eps \bx\|_2^p &\leq \|1+\eps\bx\|_p^p \nonumber\\
        \quad\iff\quad \E[(1+\sqrt{p-1}\cdot \eps \bx)^2]^{p/2} &\leq \E[(1+\eps \bx)^p]. \label{eqn:hypercon-proveme}
    \end{align}
    Here we were able to drop the absolute value on the right-hand side because $|\eps| < 1$.  The left-hand side of~\eqref{eqn:hypercon-proveme} is
    \begin{equation} \label{eqn:hypercon-in-light}
        (1+(p-1) \eps^2)^{p/2} \leq 1 + \tfrac{p(p-1)}{2} \eps^2,
    \end{equation}
    where we used the inequality $(1+t)^{\theta} \leq 1+\theta t$ for $t \geq 0$ and $0 \leq \theta \leq 1$ (easily proved by comparing derivatives in~$t$).  As for the right-hand side of~\eqref{eqn:hypercon-proveme}, since $|\eps \bx| < 1$ we may use the Generalized Binomial Theorem to show it equals
    \begin{align*}
        &\E\left[1 + p \eps\bx + \tfrac{p(p-1)}{2!} \eps^2\bx^2 + \tfrac{p(p-1)(p-2)}{3!} \eps^3\bx^3 + \tfrac{p(p-1)(p-2)(p-3)}{4!} \eps^4 \bx^4 + \cdots\right] \\
        =\ &1 + p \eps\E[\bx] + \tfrac{p(p-1)}{2!} \eps^2\E[\bx^2] + \tfrac{p(p-1)(p-2)}{3!} \eps^3\E[\bx^3] + \tfrac{p(p-1)(p-2)(p-3)}{4!} \eps^4\E[\bx^4] + \cdots \\
        =\ & 1 + \tfrac{p(p-1)}{2} \eps^2  + \tfrac{p(p-1)(p-2)(p-3)}{4!} \eps^4 + \tfrac{p(p-1)(p-2)(p-3)(p-4)(p-5)}{6!} \eps^6 + \cdots.
    \end{align*}
    In light of~\eqref{eqn:hypercon-in-light}, to verify~\eqref{eqn:hypercon-proveme} it suffices to note that each ``post-quadratic'' term above,
    \[
        \tfrac{p(p-1)(p-2)(p-3) \cdots (p-(2k-1))}{(2k)!}\eps^{2k},
    \]
    is nonnegative. This follows from $1 \leq p \leq 2$: the numerator has two positive factors and an even number of negative factors.
\end{proof}

To deduce Theorem~\ref{thm:hypercon-bit} from Theorem~\ref{thm:hypercon-bit-2} we again just need to flip the norms across~$2$ using the fact that $\T_\rho$ is self-adjoint. This is accomplished by taking $\Omega = \bits$, $\pi = \unif$, $q = 2$, $T = \T_{\sqrt{p-1}}$, and $C = 1$ in the following proposition (and noting that $1/\sqrt{p'-1} = \sqrt{p - 1}$):
                                                        \index{dual norm}%
\begin{proposition}                                     \label{prop:dual-norm}
    Let $T$ be a self-adjoint operator on $L^2(\Omega, \pi)$, let $1 \leq p,\,q \leq \infty$, and let $p',\,q'$ be their conjugate \Holder indices.  Assume $\|T f\|_q \leq C\|f\|_p$ for all~$f$.  Then $\|T g\|_{p'} \leq C\|g\|_{q'}$ for all~$g$.
\end{proposition}
\begin{proof}
    This follows from
    \[
        \|T g\|_{p'} = \sup_{\|f\|_p = 1} \la f, Tg \ra = \sup_{\|f\|_p = 1} \la Tf, g \ra \leq \sup_{\|f\|_p = 1} \|Tf\|_q \|g\|_{q'} \leq C\|g\|_{q'},
    \]
    where the first equality is the sharpness of \Holder's inequality, the second equality holds because~$T$ is self-adjoint, the subsequent inequality is \Holder's, and the final inequality uses the hypothesis $\|T f\|_{q} \leq C\|f\|_p$.
\end{proof}

At this point we have established that if $\bx$ is a uniform $\pm 1$ bit, then it is $(2,q,1/\sqrt{q-1})$-hypercontractive and $(p,2,\sqrt{p-1})$-hypercontractive.  In the next section we will give a very simple induction which transforms these facts into the full $(2,q)$- and $(p,2)$-Hypercontractivity Theorems stated at the beginning of the chapter.

\section{Two-function hypercontractivity and induction}                         \label{sec:hypercon-tensorize}

At this point we have established that if $f \co \bits \to \R$ then for any $p \leq 2 \leq q$,
\[
    \|\T_{\sqrt{p-1}} f\|_2 \leq \|f\|_p, \qquad     \|\T_{1/\sqrt{q-1}} f\|_q \leq \|f\|_2.
\]
We would like to extend these facts to the case of general $f \btR$; i.e., establish the $(p,2)$- and $(2,q)$-Hypercontractivity Theorems stated at the beginning of the chapter.  A natural approach is induction.

                            \index{tensorization|seeonly{hypercontractivity, induction}}%
                            \index{hypercontractivity!induction|(}
In analysis of Boolean functions, there are two methods for proving statements about $f \btR$ by induction on~$n$.
                                                \index{induction}%
One method, which might be called ``induction by derivatives'', uses the decomposition $f(x) = x_n \D_nf(x) + \uE_n f(x)$. We saw this approach in our inductive proof of the Bonami Lemma.  The other method, which might be called ``induction by restrictions'', goes via the subfunctions $f_{\pm 1}$ obtained by restricting the $n$th coordinate of~$f$ to $\pm 1$.  We saw this approach in our proof of the OSSS Inequality in Chapter~\ref{sec:monotone-dts}.  In both methods we reduce inductively from one function~$f$ to two functions: either $\D_n f$ and $\uE_n f$, or $f_{-1}$ and $f_{+1}$.  Because of this, when trying to prove a fact by induction on~$n$ it's often helpful to try proving a generalized fact about \emph{two} functions.  Our proof of the OSSS Inequality gives a good example of this technique.

                                                    \index{Hypercontractivity Theorem!Two-Function|(}%
So to facilitate induction, let's find a two-function version of the hypercontractivity statements we've proven so far.  Perhaps the most natural statement we've seen is the noise-stability rephrasing of the $(4/3,2)$-Hypercontractivity Theorem, namely $\Stab_{1/3}[f] \leq \|f\|_{4/3}^2$.  At least in the case $n = 1$, our work in the previous section (Theorem~\ref{thm:hypercon-bit-2}) generalizes this to $\Stab_{p-1}[f] \leq \|f\|_{p}^2$ for $1 \leq p \leq 2$. I.e.,
\[
    \Stab_\rho[f] = \Es{(\bx, \by) \\ \text{ $\rho$-correlated}}[f(\bx) f(\by)] \leq \|f\|_{1+\rho}^2
\]
for $0 \leq \rho \leq 1$.  Looking at this, you might naturally guess a (correct) generalization for two functions $f, g \btR$,  namely
\begin{equation}                                        \label{eqn:two-fn-warmup}
    \Es{(\bx, \by) \\ \text{ $\rho$-correlated}}[f(\bx) g(\by)] \leq \|f\|_{1+\rho} \|g\|_{1+\rho}.
\end{equation}
We have a nice interpretation of this inequality when $f, g \co \bn \to \{0,1\}$ are indicators of subsets $A, B \subseteq \bn$ as in Corollary~\ref{cor:sse-1/3}; it gives an upper bound on the probability of going from~$A$ to~$B$ in one step on the $\rho$-stable hypercube graph.  This bound is sharp when~$A$ and~$B$ have the same volume, but for~$A$ and~$B$ of different sizes you might imagine it's helpful to measure~$f$ and~$g$ by different norms in~\eqref{eqn:two-fn-warmup}. To see what we can expect, let's break up the $\rho$-correlation in~\eqref{eqn:two-fn-warmup} into two parts; say, write
\[
    \rho = \sqrt{rs}, \qquad 0 \leq r, s \leq 1,
\]
and use
\[
    \Es{(\bx, \by) \\ \text{ $\sqrt{rs}$-correlated}}[f(\bx) g(\by)] = \E[\T_{\sqrt{r}} f \cdot  \T_{\sqrt{s}} g].
\]
Then Cauchy--Schwarz implies
\begin{equation} \label{eqn:two-function-n1}
    \Es{(\bx, \by) \\ \text{ $\rho$-correlated}}[f(\bx) g(\by)] = \E[\T_{\sqrt{r}} f \cdot  \T_{\sqrt{s}} g] \leq \|\T_{\sqrt{r}} f\|_2  \|\T_{\sqrt{s}} g\|_2 \leq \|f\|_{1+r} \|g\|_{1+s},
\end{equation}
where the last step used $(p,2)$-hypercontractivity -- which we have so far only proven in the case $n = 1$ (Theorem~\ref{thm:hypercon-bit-2}).  The inequality~\eqref{eqn:two-function-n1}, restated below, is precisely the desired two-function version of the $(2,q)$- and $(p,2)$-Hypercontractive Theorems.
                                                \index{Hypercontractivity Theorem!Two-Function}%
\begin{named}{(Weak) Two-Function Hypercontractivity Theorem} Let $f, g \btR$, let $0 \leq r, s \leq 1$, and assume $0 \leq \rho \leq \sqrt{rs} \leq 1$.  Then
\[
    \Es{(\bx, \by) \\ \textnormal{ $\rho$-correlated}}[f(\bx) g(\by)] \leq \|f\|_{1+r} \|g\|_{1+s}.
\]
\end{named}
We call this the ``Weak'' Two-Function Hypercontractivity Theorem because the hypothesis $r, s \leq 1$ is not actually necessary; see Chapter~\ref{sec:full-hypercon-for-bits}.  As mentioned, we have so far established this theorem in the case $n = 1$.  However, the beauty of hypercontractivity in this form is that it extends to general~$n$ by an almost trivial induction.  The form of the induction is ``induction by restrictions''.  (It's also possible -- but a little trickier -- to extend the $(2,q)$-Hypercontractivity Theorem from $n = 1$ to general~$n$ via ``induction by derivatives''; see Exercise~\ref{ex:deriv-induct-2q-hc}.) For future use, we will write the induction in more general notation.
\begin{named}{Two-Function Hypercontractivity Induction Theorem}
    Let $0 \leq \rho \leq 1$ and assume that
    \[
         \Es{(\bx, \by) \\ \text{ $\rho$-correlated}}[f(\bx) g(\by)] \leq \|f\|_p \|g\|_q
    \]
    holds for every $f, g \in L^2(\Omega, \pi)$.  Then the inequality also holds for every $f, g \in L^2(\Omega^n, \pi\xn)$.
\end{named}
\begin{proof}
     The proof is by induction on~$n$, with the $n = 1$ case holding by assumption.  For $n > 1$, let $f, g \in L^2(\Omega^n, \pi\xn)$ and let $(\bx, \by)$ denote a $\rho$-correlated pair under~$\pi\xn$.  We'll use the notation $\bx = (\bx', \bx_n)$ where $\bx' = (\bx_1, \dots, \bx_{n-1})$, and similar notation for~$\by$.  Note that $(\bx', \by')$ and $(\bx_n, \by_n)$ are both $\rho$-correlated pairs (of length $n-1$ and $1$, respectively). We'll also write $f_{x_n} = \restr{f}{[n-1]}{x_n}$ for the restriction of~$f$ in which the last coordinate is fixed to value~$x_n$, and similarly for~$g$.  Now
    \[
        \Ex_{(\bx, \by)}[f(\bx) g(\by)] = \Ex_{(\bx_n, \by_n)} \Ex_{(\bx', \by')}[f_{\bx_n}(\bx') g_{\by_n}(\by')]
        \leq \Ex_{(\bx_n, \by_n)}[\|f_{\bx_n}\|_p \|g_{\by_n}\|_q]
    \]
    by induction.  If we write $F \in L^2(\Omega, \pi)$ for the function $x_n \mapsto \|f_{x_n}\|_p$ and similarly write $G(y_n) = \|g_{y_n}\|_q$, then we may continue the above as
    \[
        \Ex_{(\bx_n, \by_n)}[\|f_{\bx_n}\|_p \|g_{\by_n}\|_q] = \Ex_{(\bx_n, \by_n)}[F(\bx_n) G(\by_n)] \leq \|F\|_{p, \bx_n} \|G\|_{q, \by_n},
    \]
    where we used the base case of the induction.  Finally,
    \[
        \|F\|_{p, \bx_n} = \E_{\bx_n}[|F(\bx_n)|^p]^{1/p} = \E_{\bx_n}[\|f_{\bx_n}\|_p^p]^{1/p} =  \bigl(\E_{\bx_n} \Ex_{\bx'} |f_{\bx_n}(\bx')|^p]\bigr)^{1/p} = \|f\|_p
    \]
    by definition, and similarly for~$\|G\|_{q,\by_n}$.  Thus we have established $\E[f(\bx)g(\by)] \leq \|f\|_p\|g\|_q$, completing the induction.
\end{proof}
\begin{remark}
    More generally, if we assume the inequality holds over each of $(\Omega_1, \pi_1), \dots, (\Omega_n, \pi_n)$, then it also holds over $(\Omega_1 \times \cdots \times \Omega_n, \pi_1 \otimes \cdots \otimes \pi_n)$; the only change needed to the proof is notational.
\end{remark}
                            \index{hypercontractivity!induction|)}

At this point, we have fully established the Weak Two-Function Hypercontractivity Theorem.
                                                    \index{Hypercontractivity Theorem!Two-Function|)}%
By taking $g = f$ and $r = s = \rho$ in the theorem we obtain the full $(p,2)$-Hypercontractivity Theorem stated at the beginning of the chapter.  Finally, by applying Proposition~\ref{prop:dual-norm} we also obtain the $(2,q)$-Hypercontractivity Theorem for all $f \btR$.

                            \index{hypercontractivity!$(2,q)$- and $(p,2)$-|)}%

\section{Applications of hypercontractivity}                                    \label{sec:hypercon-apps}
With the $(2,q)$- and $(p,2)$-Hypercontractivity Theorems in hand, let's revisit some applications we saw in Sections~\ref{sec:4th-moment} and~\ref{sec:sse-intro}.  We begin by deducing a generalization of the Bonami Lemma:
\begin{theorem}                                       \label{thm:general-bonami}
    Let $f \btR$ have degree at most~$k$. Then $\|f\|_q \leq \sqrt{q-1}^k \|f\|_2$ for any $q \geq 2$.
\end{theorem}
\begin{proof}
    We have
    \[
        \|f\|_q^2 = \|\T_{1/\sqrt{q-1}} \T_{\sqrt{q-1}} f\|_q^2 \leq \|\T_{\sqrt{q-1}} f\|_2^2
    \]
    using the $(2,q)$-Hypercontractivity Theorem.  (Here we are  extending the definition of $\T_\rho$ to $\rho > 1$ via $\T_\rho f = \sum_j \rho^j f^{=j}$; see also Remark~\ref{rem:general-Trho}.)  The result now follows since
    \[
        \|\T_{\sqrt{q-1}} f\|_2^2  = \sum_{j = 0}^k (q-1)^j \W{j}[f] \leq (q-1)^k \sum_{j=0}^k \W{j}[f] = (q-1)^k \|f\|_2^2. \qedhere
    \]
\end{proof}
Using a trick similar to the one in our proof of the $(4/3,2)$-Hypercontractivity Theorem you can use this to deduce $\|f\|_2 \leq (1/\sqrt{p-1})^k \|f\|_p$ when $f$ has degree~$k$ for any $1 \leq p \leq 2$; see Exercise~\ref{ex:weak-p-2-bonami}.  However, a different trick yields a strictly better result, including a finite bound for $p = 1$:
\begin{theorem}                         \label{thm:p-2-bonami}
    Let $f \btR$ have degree at most~$k$. Then $\|f\|_2 \leq e^{k} \|f\|_1$.  More generally, for $1 \leq p \leq 2$ it holds that $\|f\|_2 \leq (e^{\frac2p - 1})^k\|f\|_p$.
\end{theorem}
\begin{proof}
    We prove the statement about the $1$-norm, leaving the case of general $1 \leq p \leq 2$ to Exercise~\ref{ex:p-2-bonami}.
    For $\eps > 0$, let $0 < \theta < 1$ be the solution of $\frac12 = \frac{\theta}{1} + \frac{1-\theta}{2+\eps}$ (namely, $\theta = \half\frac{\eps}{1+\eps}$).  Applying the general version of \Holder's inequality and then Theorem~\ref{thm:general-bonami}, we get
    \[
        \|f\|_2 \leq \|f\|_{2+\epsilon}^{1-\theta} \|f\|_1^{\theta}  \leq \sqrt{1+\epsilon}^{k(1-\theta)}\|f\|_{2}^{1-\theta} \|f\|_1^\theta.
    \]
    Dividing by $\|f\|_2^{1-\theta}$ (which we may assume is nonzero) and then raising the result to the power of $1/\theta$ yields
    \[
        \|f\|_2 \leq \left((1+\epsilon)^{\frac{1-\theta}{2\theta}}\right)^k \|f\|_1 =  \left((1+\epsilon)^{\frac{1}{\eps} +\frac12}\right)^k \|f\|_1.
    \]
    The result follows by taking the limit as $\eps \to 0$.
\end{proof}

In the linear case of $k = 1$, Theorems~\ref{thm:general-bonami} and~\ref{thm:p-2-bonami} taken together show that $c_p \|\sum_i a_i \bx_i\|_2 \leq \|\sum_i a_i \bx_i\|_p \leq C_p \|\sum_i a_i \bx_i\|_2$ for some constants $0 < c_p < C_p$ depending only on~$p \in [1,\infty)$.  This fact is known as Khintchine's Inequality.
                                                \index{Khintchine(--Kahane) Inequality}%

Theorem~\ref{thm:general-bonami} can be used to get a strong concentration bound for degree-$k$ Boolean functions.  Chernoff tells us that the probability a linear form $\sum a_i \bx_i$ exceeds~$t$ standard deviations decays like $\exp(-\Theta(t^2))$. The following theorem generalizes this to degree-$k$ forms, with decay $\exp(-\Theta(t^{2/k}))$:
\begin{theorem}                                     \label{thm:low-deg-conc}
    Let $f \btR$ have degree at most $k$.  Then for any $t \geq \sqrt{2e}^{k}$ we have
    \[
        \Pr_{\bx \sim \bn}[|f(\bx)| \geq t \|f\|_2] \leq \exp\left(-\tfrac{k}{2e} t^{2/k}\right).
    \]
\end{theorem}
\begin{proof}
    We may assume $\|f\|_2 = 1$ without loss of generality.  Let $q \geq 2$ be a parameter to be chosen later.  By Markov's inequality,
    \[
        \Pr[|f(\bx)| \geq t] = \Pr[|f(\bx)|^q \geq t^q] \leq \frac{\E[|f(\bx)|^q]}{t^q}.
    \]
    By Theorem~\ref{thm:general-bonami} we have
    \[
        \E[|f(\bx)|^q] \leq (\sqrt{q-1}^k)^q \|f\|_2^q = (q-1)^{(k/2)q} \leq q^{(k/2)q}.
    \]
    Thus $\Pr[|f(\bx)| \geq t] \leq (q^{k/2}/t)^q$.  It's not hard to see that the $q$ that minimizes this expression should be just slightly less than~$t^{2/k}$. Specifically, by choosing $q = t^{2/k}/e \geq 2$ we get
    \[
        \Pr[|f(\bx)| \geq t] \leq \exp(-(k/2)q) = \exp\left(-\tfrac{k}{2e} t^{2/k}\right)
    \]
    as claimed.
\end{proof}

We can use Theorem~\ref{thm:p-2-bonami} to get a ``one-sided'' analogue of Theorem~\ref{thm:low-deg-anticonc}, showing that a low-degree function exceeds its mean with noticeable probability:
\begin{theorem}                                     \label{thm:one-sided-low-deg-anticonc}
    Let $f \btR$ be a nonconstant function of degree at most~$k$.  Then
    \[
        \Pr_{\bx \sim \bn}\bigl[f(\bx) > \E[f]\bigr] \geq \tfrac14 e^{-2k}.
    \]
\end{theorem}
\begin{proof}
    We may assume $\E[f] = 0$ without loss of generality.  We then have
    \[
        \half \|f\|_1 = \half \left(\E[f \cdot \bone_{\{f(\bx) > 0\}}] -  \E[f \cdot (1-\bone_{\{f(\bx) >0\}})]\right)  = \E[f \cdot \bone_{\{f(\bx) > 0\}}];
    \]
    hence,
    \[
        \tfrac14 \|f\|_1^2 = \E[f \cdot \bone_{\{f(\bx) > 0\}}]^2 \leq \E[f^2]\cdot \E[\bone_{\{f(\bx) > 0\}}^2] \leq e^{2k}\|f\|_1^2 \cdot \Pr[f(\bx) > 0]
    \]
    using Cauchy--Schwarz and Theorem~\ref{thm:p-2-bonami}. The result follows.
\end{proof}

Next we turn to noise stability.  Using the $(p,2)$-Hypercontractivity Theorem we can immediately deduce the following generalization of Corollary~\ref{cor:sse-1/3}:
                                                \index{expansion!small-set}%
                                                \index{Small-Set Expansion Theorem}%
\begin{named}{Small-Set Expansion Theorem}
    Let $A \subseteq \bn$ have volume $\alpha$; i.e., let $1_A \co \bn \to \zo$ satisfy $\E[1_A] = \alpha$.  Then for any $0 \leq \rho \leq 1$,
    \[
        \Stab_{\rho}[1_A] = \Pr_{\substack{\bx \sim \bn \\ \by \sim N_{\rho}(\bx)}}[\bx \in A, \by \in A] \leq \alpha^{\frac{2}{1+\rho}}.
    \]
    Equivalently (for $\alpha > 0$),
    \[
        \Pr_{\substack{\bx \sim A \\ \by \sim N_{\rho}(\bx)}}[\by \in A] \leq \alpha^{\frac{1-\rho}{1+\rho}}.
    \]
\end{named}
In other words, the $\delta$-noisy hypercube is a small-set expander for any~$\delta > 0$: the probability that one step from a random $\bx \sim A$ stays inside~$A$ is at most $\alpha^{\delta/(1-\delta)}$.  It's also possible to derive a ``two-set'' generalization of this fact using the Two-Function Hypercontractivity Theorem; we defer the discussion to Chapter~\ref{sec:full-hypercon-for-bits} since the most general result requires the non-weak form of the theorem.  We can also obtain the generalization of Corollary~\ref{cor:1/3-stab-infl}:
                                         \index{stable influence}%
\begin{corollary}                                       \label{cor:stab-infl-bound}
    Let $f \btb$.  Then for any $0 \leq \rho \leq 1$ we have $\Inf_i^{(\rho)}[f] \leq \Inf_i[f]^{\frac{2}{1+\rho}}$ for all~$i$.
\end{corollary}

Finally, from the Small-Set Expansion Theorem we see that indicators of small-volume sets are not very noise-stable and hence can't have much of their Fourier weight at low levels.  Indeed, using hypercontractivity we can deduce the Level-1 Inequality from Chapter~\ref{sec:weight-level-1}
                                    \index{Level-1 Inequality}
and also generalize it to higher degrees.
\begin{named}{Level-$k$ Inequalities}  Let $f \co \bn \to \{0,1\}$ have mean $\E[f] = \alpha$ and let $k \in \N^+$ be at most $2\ln(1/\alpha)$.
                                    \index{Level-$k$ Inequalities}%
Then
\[
    \W{\leq k}[f]  \leq \left(\tfrac{2e}{k}\ln(1/\alpha)\right)^k \alpha^2.
\]
\end{named}
\begin{proof}
    By the Small-Set Expansion Theorem,
    \[
        \W{\leq k}[f] \leq \rho^{-k} \Stab_{\rho}[f] \leq \rho^{-k} \alpha^{2/(1+\rho)} \leq \rho^{-k} \alpha^{2(1-\rho)}
    \]
    for any $0 < \rho \leq 1$.  Basic calculus shows the right-hand side is minimized when $\rho = \frac{k}{2\ln(1/\alpha)} \leq 1$; substituting this into $\rho^{-k} \alpha^{2(1-\rho)}$ yields the claim.
\end{proof}
For the case $k = 1$, a slightly different argument gives the sharp Level-1 Inequality $\W{1}[f] \leq 2 \alpha^2 \ln(1/\alpha)$; see Exercise~\ref{ex:level-1-ineq}.

\section{Highlight: The Kahn--Kalai--Linial Theorem}            \label{sec:KKL}

Recalling the social choice setting of Chapter~\ref{sec:social-choice}, consider a $2$-candidate, $n$-voter election using a monotone voting rule $f \btb$.  We assume the impartial culture assumption (that the votes are independent and uniformly random), but with a twist: one of the candidates, say $b \in \bits$, is able to secretly bribe~$k$ voters, fixing their votes to~$b$.  (Since~$f$ is monotone, this is always the optimal way for the candidate to fix the bribed votes.)  How much can this influence the outcome of the election? This question was posed by Ben-Or and Linial in a 1985 work~\cite{BL85,BL90}; more precisely, they were interested in designing (unbiased) voting rules~$f$ that  minimize the effect of any bribed $k$-coalition.

Let's first consider $k = 1$.  If voter~$i$ is bribed to vote for candidate~$b$ (but all other votes remain uniformly random), this changes the bias of~$f$ by $b\wh{f}(i) = b \Inf_i[f]$.  Here we used the assumption that~$f$ is monotone (i.e., Proposition~\ref{prop:monotone-influences}).  This led Ben-Or and Linial to the question of which unbiased $f \btb$ has the least possible maximum influence:
\begin{definition}
    Let $f \btR$.  The \emph{maximum influence} of~$f$
                                                    \index{influence!maximum}%
    is
    \[
        \MaxInf[f] = \max \{\Inf_i[f] : i \in [n]\}.
    \]
\end{definition}
Ben-Or and Linial constructed the (nearly) unbiased $\tribes_n \btb$ function (from Chapter~\ref{sec:tribes})
                                                    \index{tribes function}%
and noted that $\MaxInf[\tribes_n] = O(\frac{\log n}{n})$.  They further conjectured \emph{every} unbiased function~$f$ has $\MaxInf[f] = \Omega(\frac{\log n}{n})$.  This conjecture was famously proved by Kahn, Kalai, and Linial~\cite{KKL88}:
                                                    \index{KKL Theorem|(}%
\begin{named}{Kahn--Kalai--Linial (KKL) Theorem}  For
any $f \btb$,
\[
    \MaxInf[f] \geq \Var[f] \cdot \Omega\Bigl(\frac{\log n}{n}\Bigr).
\]
\end{named}
Notice that the theorem says something sensible even for very biased functions~$f$, i.e., those with low variance.  The variance of $f$ is indeed the right ``scaling factor'' since
\[
    \frac{1}{n} \Var[f] \leq \MaxInf[f] \leq \Var[f]
\]
holds trivially, by the Poincar\'e Inequality and Exercise~\ref{ex:inf-at-most-var}.

Before proving the KKL Theorem, let's see an additional consequence for Ben-Or and Linial's problem.
\begin{proposition}                                     \label{prop:monotone-bribing}
    Let $f \btb$ be monotone and assume $\E[f] \geq -.99$.  Then there exists a subset $J \subseteq [n]$ with $|J| \leq O(n/\log n)$ that if ``bribed to vote~$1$'' causes the outcome to be~$1$ almost surely; i.e.,
    \begin{equation} \label{eqn:bribery-goal}
        \E[\restr{f}{\ol{J}}{(1, \dots, 1)}] \geq .99.
    \end{equation}
    Similarly, if $\E[f] \leq .99$ there exists $J \subseteq [n]$ with $|J| \leq O(n/\log n)$ such that $\E[\restr{f}{\ol{J}}{(-1, \dots, -1)}] \leq -.99$.
\end{proposition}
\begin{proof}
    By symmetry it suffices to prove the result regarding bribery by candidate~$+1$.  The candidate executes the following strategy: First, bribe the voter~$i_1$ with the largest influence on~$f_0 = f$; then bribe the voter~$i_2$ with the largest influence on $f_1 = f^{(i_1 \mapsto 1)}$; then bribe the voter~$i_3$ with the largest influence on~$f_2 = f^{(i_1, i_2 \mapsto 1)}$; etc.  For each $t \in \N$ we have
    \[
        \E[f_{t+1}] \geq \E[f_t] + \MaxInf[f_t].
    \]
    If after $t$ bribes the candidate has not yet achieved~\eqref{eqn:bribery-goal} we have $-.99 \leq \E[f_t] < .99$; thus $\Var[f_t] \geq \Omega(1)$ and the KKL~Theorem implies $\MaxInf[f_t] \geq \Omega(\frac{\log n}{n})$.  Thus the candidate will achieve a bias of at least~$.99$ after bribing at most $(.99 - (-.99))/\Omega(\frac{\log n}{n}) = O(n/\log n)$ voters.
\end{proof}
Thus in any monotone election scheme, there is always a candidate~$b \in \bits$ and a $o(1)$-fraction of the voters that~$b$ can bribe such that the election becomes $99$\%-biased in $b$'s favor.  And if the election scheme was not terribly biased to begin with, then \emph{both} candidates have this ability.  For a more precise version of this result, see Exercise~\ref{ex:precise-bribing}; for a nonmonotone version, see Exercise~\ref{ex:polarized-bribing}.  Note also that although the~$\tribes_n$ function is essentially optimal for standing up to a single bribed voter, it is quite bad at standing up to bribed coalitions: by bribing just a single tribe (DNF term) -- about $\log n$ voters -- the outcome can be completely forced to~$\True$. Nevertheless, Proposition~\ref{prop:monotone-bribing} is close to sharp: Ajtai and Linial~\cite{AL93} constructed an unbiased monotone function $f \btb$ such that bribing any set of at most $\eps n/\log^2 n$ voters changes the expectation by at most~$O(\eps)$.

\medskip

The remainder of this section is devoted to the proof of the KKL~Theorem and some variants.  As mentioned earlier, the proof quickly follows from summing Corollary~\ref{cor:1/3-stab-infl} over all coordinates; but let's give a more leisurely description.  We'll focus on the main case of interest: showing that $\MaxInf[f] \geq \Omega(\frac{\log n}{n})$ when $f$ is unbiased (i.e., $\Var[f] = 1$).  If $f$'s total influence is at least, say, $.1 \log n$, then even the \emph{average} influence is $\Omega(\frac{\log n}{n})$.  So we may as well assume $\Tinf[f] \leq .1 \log n$.

This leads us to the problem of characterizing (unbiased) functions with small total influence.  (This is the same issue that arose at the end of Chapter~\ref{sec:p-biased} when studying sharp thresholds.)  It's helpful to think about the case that the total influence is \emph{very} small -- say $\Tinf[f] \leq K$ where $K = 10$ or $K = 100$, though we eventually want to handle $K = .1 \log n$.  Let's think of $f$ as the indicator of a volume-$1/2$ set $A \subset \bn$, so $\frac{\Tinf[f]}{n}$ is the fraction of Hamming cube edges on the boundary of~$A$.  The edge-isoperimetric inequality (or Poincar\'e Inequality) tells us that $\Tinf[f] \geq 1$: at least a $\frac{1}{n}$ fraction of the cube's edges must be on~$A$'s boundary, with dictators and negated-dictators being the minimizers.
                                            \index{Poincar\'{e} Inequality}%
                                            \index{isoperimetric inequality!Hamming cube}%
Now what can we say if $\Tinf[f] \leq K$; i.e., $A$'s boundary has only $K$ times more edges than the minimum? Must~$f$ be ``somewhat similar'' to a dictator or negated-dictator?  Kahn, Kalai, and Linial showed that the answer is yes: $f$~must have a coordinate with influence at least~$2^{-O(K)}$.  This should be considered very large (and dictator-like), since a~priori all of the influences could have been equal to~$\frac{K}{n}$.
                                            \index{KKL Theorem!edge-isoperimetric version}%
\begin{named}{KKL Edge-Isoperimetric Theorem}  Let $f \btb$ be nonconstant and let $\wt{\Tinf}[f] = \Tinf[f]/\Var[f] \geq 1$ (which is just $\Tinf[f]$ if $f$ is unbiased).  Then
\[
    \MaxInf[f] \geq \left(\tfrac{9}{\wt{\Tinf}[f]^2}\right)\cdot 9^{-\wt{\Tinf}[f]}.
\]
\end{named}
This theorem is sharp for $\wt{\Tinf}[f] = 1$ (cf.~Exercises~\ref{ex:weight-1-dictator},~\ref{ex:erics-fkn}), and it's nontrivial (in the unbiased case) for $\Tinf[f]$ as large as $\Theta(\log n)$.  This last fact lets us complete the proof of the KKL~Theorem as originally stated:

\begin{proof}[Proof of the KKL Theorem from the Edge-Isoperimetric version]
    ~\linebreak We may assume $f$ is nonconstant.  If $\wt{\Tinf}[f] = \Tinf[f]/\Var[f] \geq .1 \log n$, then we are done: the total influence is at least $.1 \Var[f] \cdot \log n$ and hence $\MaxInf[f] \geq .1 \Var[f] \cdot \frac{\log n}{n}$.  Otherwise, the KKL Edge-Isoperimetric Theorem implies
    \[
        \MaxInf[f] \geq \Omega\left(\tfrac{1}{\log^2 n}\right) \cdot 9^{-.1 \log n} = \wt{\Omega}(n^{-.1 \log 9}) = \Omega(n^{-.317}) \gg  \Var[f] \cdot \Omega\left(\tfrac{\log n}{n}\right). \qedhere
    \]
\end{proof}
\noindent (You are asked  to be careful about the constant factors in Exercise~\ref{ex:careful-constant-kkl}.)\\

We now turn to proving the KKL Edge-Isoperimetric Theorem.  The high-level idea is to look at the contrapositive: supposing all of $f$'s influences are small, we want to show its total influence must be large.  The assumption here is that each derivative $\D_i f$ is a $\{-1,0,1\}$-valued function which is nonzero only on a ``small'' set. Hence ``small-set expansion''
                                                \index{expansion!small-set}%
implies that each derivative has ``unusually large'' noise sensitivity.  (We are really just repeating Corollary~\ref{cor:1/3-stab-infl} in words here.)  In turn this means that for each $i \in [n]$, the Fourier weight of~$f$ on coefficients containing~$i$ must be quite ``high up''. Since this holds for all~$i$ we deduce that \emph{all} of $f$'s Fourier weight must be quite ``high up'' -- hence $f$ must have ``large'' total influence.  We now make this story formal:

\begin{proof}[Proof of the KKL Edge-Isoperimetric Theorem]
    We treat only the case that $f$ is unbiased, leaving the general case to Exercise~\ref{ex:finish-kkl} (see also the version for product space domains in Chapter~\ref{sec:hypercon-variants-apps}).  The theorem is an immediate consequence of the following chain of inequalities:
    \[
    3 \cdot 3^{-\Tinf[f]} \ \stackrel{\text{(a)}}{\leq} \  3\Stab_{1/3}[f] \ \stackrel{\text{(b)}}{\leq} \  \Tinf^{(1/3)}[f]
     \ \stackrel{\text{(c)}}{\leq} \  \sum_{i=1}^n \Inf_i[f]^{3/2} \ \stackrel{\text{(d)}}{\leq} \  \MaxInf[f]^{1/2} \cdot \Tinf[f].
    \]
    The key inequality is~(c), which comes from summing Corollary~\ref{cor:1/3-stab-infl} over all coordinates~$i \in [n]$.  Inequality~(d) is immediate from $\Inf_i[f]^{3/2} \leq \MaxInf[f]^{1/2} \cdot \Inf_i[f]$.  Inequality~(b) is trivial from the Fourier formulas (recall Fact~\ref{fact:stable-tinf}):
    \[
        \Tinf^{(1/3)}[f] = \sum_{|S| \geq 1} |S| (1/3)^{|S|-1}\wh{f}(S)^2  \geq 3 \sum_{|S| \geq 1} (1/3)^{|S|}  \wh{f}(S)^2 = 3\Stab_{1/3}[f]
    \]
    (the last equality using $\wh{f}(\emptyset) = 0$).  Finally, inequality~(a) is quickly proved using the spectral sample: for $\bS \sim \specsamp{f}$ we have
    \begin{equation} \label{eqn:finish-kkl1}
        3\Stab_{1/3}[f] = 3 \sum_{S \subseteq [n]} (1/3)^{|S|} \wh{f}(S)^2 = 3 \Ex[3^{-|\bS|}] \geq 3 \cdot 3^{-\Ex[|\bS|] } = 3 \cdot 3^{-\Tinf[f]},
    \end{equation}
    the inequality following from convexity of $s \mapsto 3^{-s}$.
\end{proof}
                                                     \index{KKL Theorem|)}%

\medskip

We end this chapter by deriving an even stronger version of the KKL Edge-Isoperimetric Theorem, and deducing Friedgut's Junta Theorem (from the end of Chapter~\ref{sec:low-degree}) as a consequence.
                                                    \index{Friedgut's Junta Theorem|(}%
The KKL Edge-Isoperimetric Theorem tells us that if $f$ is unbiased and $\Tinf[f] \leq K$ then $f$ must look somewhat like a $1$-junta, in the sense of having a coordinate with influence at least~$2^{-O(K)}$.  Friedgut's Junta Theorem shows that in fact~$f$ must essentially be a $2^{O(K)}$-junta.
To obtain this conclusion, you really just have to sum Corollary~\ref{cor:1/3-stab-infl} only over the coordinates which have small influence on~$f$.  It's also possible to get even stronger conclusions if~$f$ is known to have particularly good low-degree Fourier concentration.
In aid of this, we'll start by proving the following somewhat technical-looking result:
\begin{theorem}                                     \label{thm:KKL-Friedgut-generalized}
    Let $f \btb$. Given $0 < \eps \leq 1$ and $k \geq 0$, define
    \[
        \tau = \frac{\eps^2}{\Tinf[f]^2}9^{-k}, \qquad J = \{j \in [n] : \Inf_j[f] \geq \tau\}, \qquad \text{so }|J| \leq (\Tinf[f]^3/\eps^2) 9^k.
    \]
    Then $f$'s Fourier spectrum is $\eps$-concentrated on
    \[
        \calF = \{S : S \subseteq J\} \cup \{S : |S| > k\}.
    \]
    In particular, suppose $f$'s Fourier spectrum is also $\eps$-concentrated on degree up to~$k$.  Then $f$'s Fourier spectrum is $2\eps$-concentrated on
    \[
        \calF' = \{S : S \subseteq J, |S| \leq k\},
    \]
    and $f$ is $\eps$-close to a $|J|$-junta $h \co \bits^J \to \bits$.
\end{theorem}
\begin{proof}
    Summing  Corollary~\ref{cor:1/3-stab-infl} just over $i \not \in J$ we obtain
    \[
        \sum_{i \not \in J} \Inf^{(1/3)}_i[f] \leq \sum_{i \not \in J} \Inf_i[f]^{3/2} \leq \max_{i \not \in J} \{\Inf_i[f]^{1/2}\} \cdot \sum_{i \not \in J} \Inf_i[f] \leq \tau^{1/2} \cdot \Tinf[f] \leq 3^{-k} \eps,
    \]
    where the last two inequalities used the definitions of $J$ and $\tau$, respectively.      On the other hand,
    \begin{multline*}
         \sum_{i \not \in J} \Inf^{(1/3)}_i[f] = \sum_{i \not \in J} \sum_{S \ni i} (1/3)^{|S|-1} \wh{f}(S)^2
         = \sum_{S} |S \cap \barJ| \cdot 3^{1-|S|} \wh{f}(S)^2 \\
          \geq \sum_{S\not \in \calF} |S \cap \barJ| \cdot 3^{1-|S|} \wh{f}(S)^2 \geq 3^{-k} \sum_{S\not \in \calF} \wh{f}(S)^2.
    \end{multline*}
    Here the last inequality used that  $S \not \in \calF$ implies $|S \cap \barJ| \geq 1$ and  $3^{1-|S|} \geq 3^{-k}$.  Combining these two deductions yields $\sum_{S \not \in \calF} \wh{f}(S)^2 \leq \eps$,
    as claimed.

    As for the second part of the theorem, when $f$'s Fourier spectrum is $2\eps$-concentrated on~$\calF'$ it follows from Proposition~\ref{prop:sign-conc} that $f$ is $2\eps$-close to the Boolean-valued $|J|$-junta $\sgn(f^{\subseteq J})$.  From     Exercise~\ref{ex:avrim-trick} we may deduce that~$f$ is in fact $\eps$-close to some $h \co \bits^J \to \bits$.
\end{proof}
\begin{remark}                                          \label{rem:KKL-Friedgut-generalized}
    As you are asked to show in Exercise~\ref{ex:KKL-Friedgut-generalized}, by using Corollary~\ref{cor:stab-infl-bound} in place of Corollary~\ref{cor:1/3-stab-infl}, we can achieve junta size $(\Tinf[f]^{2+\eta}/\eps^{1+\eta}) \cdot C(\eta)^k$ in Theorem~\ref{thm:KKL-Friedgut-generalized} for any $\eta> 0$, where $C(\eta) = (2/\eta+1)^2$.
\end{remark}
In Theorem~\ref{thm:KKL-Friedgut-generalized} we may always take $k = \Tinf[f]/\eps$, by the ``Markov argument'' Proposition~\ref{prop:tinf-concentration}.  Thus we obtain as a corollary:
\begin{named}{Friedgut's Junta Theorem}
        Let $f \btb$ and  let $0 < \eps \leq 1$.  Then $f$ is $\eps$-close to an $\exp(O(\TInf[f]/\eps))$-junta. Indeed, there is a set $J \subseteq [n]$ with $|J| \leq \exp(O(\Tinf[f]/\eps))$ such that $f$'s Fourier spectrum is $2\eps$-concentrated on $\{S \subseteq J : |S| \leq \Tinf[f]/\eps\}$.
\end{named}
                                                    \index{junta}%
                                                    \index{Friedgut's Junta Theorem|)}%
As mentioned, we can obtain stronger results for functions~$f$ that are $\eps$-concentrated up to degree much less than $\Tinf[f]/\eps$.  Width-$w$ DNFs, for example, are $\eps$-concentrated on degree up to~$O(w \log(1/\eps))$ (by Theorem~\ref{thm:dnf-low-degree}).  Thus:
                                                        \index{DNF!width}%
\begin{corollary}                                       \label{cor:dnf-width-junta}
    Any width-$w$ DNF is $\eps$-close to a $(1/\eps)^{O(w)}$-junta.
\end{corollary}
Uniformly noise-stable functions do even better. From Peres's Theorem
                                                        \index{Peres's Theorem}%
                                                        \index{uniformly noise-stable}%
                                                        \index{linear threshold function}%
we know that linear threshold functions are $\eps$-concentrated up to degree~$O(1/\eps^2)$.  Thus Theorem~\ref{thm:KKL-Friedgut-generalized} and Remark~\ref{rem:KKL-Friedgut-generalized} imply:
\begin{corollary}                                       \label{cor:ltf-junta}
    Let $f \btb$ be a linear threshold function and let $0 < \eps, \eta \leq 1/2$. Then $f$ is $\eps$-close to a junta on $\Tinf[f]^{2+\eta} \cdot (1/\eta)^{O(1/\eps^2)}$ coordinates.
\end{corollary}
\noindent Assuming $\eps$ is a small universal constant we can take $\eta = 1/\log(O(\Tinf[f]))$ and deduce that every LTF is $\eps$-close to a junta on $\Tinf[f]^2 \cdot \polylog(\Tinf[f])$ coordinates.  This is essentially best possible since $\Tinf[\Maj_n] = \Theta(\sqrt{n})$, but $\Maj_n$ is not even $.1$-close to any $o(n)$-junta.  By virtue of Theorem~\ref{thm:kane} on the uniform noise stability of PTFs, we can also get this conclusion for any constant-degree PTF.
                                                        \index{polynomial threshold function}%

One more interesting fact we may derive is that every Boolean function has a Fourier coefficient that is at least inverse-exponential in the square of its total influence:
\begin{corollary}                                       \label{cor:low-inf-large-coeff}
    Assume $f \btb$ satisfies $\Var[f] \geq 1/2$.  Then there exists $S \subseteq [n]$ with $0 < |S| \leq O(\Tinf[f])$ such that $\wh{f}(S)^2 \geq \exp(-O(\Tinf[f]^2))$.
\end{corollary}
\begin{proof}
    Taking $\eps = 1/8$ in Friedgut's Junta Theorem we get a set of coordinates $J$ with $|J| \leq \exp(O(\Tinf[f]))$ such that $f$ has Fourier weight at least $1-2\eps = 3/4$ on ${\calF = \{S \subseteq J : |S| \leq 8\TInf[f]\}}$.  Since $\wh{f}(\emptyset)^2 = 1 - \Var[f] \leq 1/2$ we conclude that~$f$ has Fourier weight at least $1/4$ on $\calF' = \calF \setminus \{\emptyset\}$.  But $|\calF'| \leq |J|^{8\Tinf[f]} = \exp(O(\Tinf[f]^2))$, so the result follows by the Pigeonhole Principle.   (Here we used that $(1/4)\exp(-O(\Tinf[f]^2)) = \exp(-O(\Tinf[f]^2))$ because $\Tinf[f] \geq \Var[f] \geq \frac12$.)
\end{proof}
\begin{remark}
    Of course, if $\Var[f] < 1/2$, then $f$ has a large empty Fourier coefficient: $\wh{f}(\emptyset)^2 \geq 1/2$.  For a more refined version of Corollary~\ref{cor:low-inf-large-coeff}, see Exercise~\ref{ex:low-inf-large-coeff}.
\end{remark}

It is an open question whether Corollary~\ref{cor:low-inf-large-coeff} can be improved to give a Fourier coefficient satisfying $\wh{f}(S)^2 \geq \exp(-O(\Tinf[f]))$ (but see Exercise~\ref{ex:min-EI-for-monotone}).

\section{Exercises and notes}                   \label{sec:basic-hypercon-exercises}
\begin{exercises}
    \item \label{ex:reasonable-not-transl} For every $1 < b < B$ show that there is a $b$-reasonable random variable~$\bX$ such that $1+\bX$ is not $B$-reasonable.
    \item \label{ex:deg-1-bonami} For $k = 1$, improve the $9$ in the Bonami Lemma to $3$.  More precisely, suppose $f \btR$ has degree at most~$1$ and that $\bx_1, \dots, \bx_n$ are independent $3$-reasonable random variables satisfying $\E[\bx_i] = \E[\bx_i^3] = 0$. (For example, the $\bx_i$'s may be uniform~$\pm 1$ bits.)  Show that $f(\bx)$ is also $3$-reasonable.  (Hint: By direct computation, or by running through the Bonami Lemma proof with $k = 1$ more carefully.)
    \item \label{ex:simple-bonami-lower-bound} Let $k$ be a positive multiple of~$3$ and let $n \geq 2k$ be an integer.  Define $f \btR$ by
        \[
            f(x) = \sum_{\substack{S \subseteq [n] \\ |S| = k}} x^S.
        \]
        \begin{exercises}
            \item Show that
            \[
                \E[f^4] \geq \frac{\binom{n}{k/3,\,k/3,\,k/3,\,k/3,\,k/3,\,k/3,\,n-2k}}{\binom{n}{k}^2} \E[f^2]^2,
            \]
            where the numerator of the fraction is a multinomial coefficient -- specifically, the number of ways of choosing six disjoint size-$k/3$ subsets of~$[n]$.  (Hint: Given such size-$k/3$ subsets, consider quadruples of size-$k$ subsets that hit each size-$k/3$ subset twice.)
            \item Using Stirling's Formula, show that
            \[
                \lim_{n \to \infty} \frac{\binom{n}{k/3,\,k/3,\,k/3,\,k/3,\,k/3,\,k/3,\,n-2k}}{\binom{n}{k}^2} = \Theta(k^{-2} 9^k).
            \]
            Deduce the following lower bound for the Bonami Lemma: $\|f\|_4 \geq \Omega(k^{-1/2}) \cdot \sqrt{3}^k \|f\|_2$.  (In fact, $\|f\|_4 = \Theta(k^{-1/4}) \cdot \sqrt{3}^k \|f\|_2$ and such an upper bound holds for all~$f$ homogeneous of degree~$k$; see Exercise and~\ref{ex:number-of-matchings}\ref{ex:janson-matchings}.)
        \end{exercises}
    \item \label{ex:bonami-lemma-cor} Prove Corollary~\ref{cor:bonami-lemma}.
    \item \label{ex:fkn-calculation} Let $0 \leq \delta \leq \frac{1}{1600}$ and let $f$, $\ell$ be real numbers satisfying $|\ell^2 - 1| > 39\sqrt{\delta}$ and $|f| = 1$.  Show that $|f - \ell|^2 \geq 169\delta$.  (This is a loose estimate; stronger ones are possible.)
    \item \label{ex:bonami-2-4-tricks}  Theorem~\ref{thm:general-bonami} shows that the $(2,4)$-Hypercontractivity Theorem implies the Bonami Lemma.  In this exercise you will show the reverse implication.
                                            \index{Bonami Lemma}%
                                            \index{$(2,4)$-hypercontractivity}%
        \begin{exercises}
            \item \label{ex:bonami-trick-a} Let $f \btR$.  For a fixed $\delta \in (0,1)$, use the Bonami Lemma to show that
                \[
                    \|\T_{(1-\delta)/\sqrt{3}} f\|_4 \leq \sum_{k=0}^\infty (1-\delta)^k \|f^{=k}\|_2 \leq \tfrac{1}{\delta} \|f\|_2.
                \]
            \item \label{ex:bonami-trick-b} For $g \btR$ and $d \in \N^+$, let $g^{\oplus d} \co \bits^{dn} \to \R$ be the function defined by $g^{\oplus d}(x^{(1)}, \dots, x^{(d)}) = g(x^{(1)})g(x^{(2)}) \cdots g(x^{(d)})$ (where each $x^{(i)} \in \bn$).  Show that $\|\T_\rho(g^{\oplus d})\|_p = \|\T_\rho g\|_p^d$ holds for every $p \in \R^+$ and $\rho \in [-1,1]$.  Note the special case $\rho = 1$.
            \item Deduce from parts~\ref{ex:bonami-trick-a} and~\ref{ex:bonami-trick-b} that in fact $\|\T_{(1-\delta)/\sqrt{3}} f\|_4 \leq \|f\|_2$. (Hint: Apply part~\ref{ex:bonami-trick-a} to $f^{\oplus d}$ for larger and larger~$d$.)
            \item Deduce that in fact $\|\T_{1/\sqrt{3}} f\|_4 \leq \|f\|_2$; i.e., the $(2,4)$-Hypercontractivity Theorem follows from the Bonami Lemma. (Hint: Take the limit as $\delta \to 0^+$.)
        \end{exercises}
    \item \label{ex:nonneg-hc-trick} Suppose we wish to show that $\|\T_\rho f\|_q \leq \|f\|_p$ for all $f \btR$.  Show that it suffices to show this for all nonnegative~$f$.  (Hint: Exercise~\ref{ex:T-and-abs}.)
    \item \label{ex:projection-bounded} Fix $k \in \N$.
                                            \index{projection!low-degree}%
        The goal of this exercise is to show that ``projection to degree~$k$ is a bounded operator in all $L^p$ norms, $p > 1$''.  Let $f \btR$.
        \begin{exercises}
            \item \label{ex:proj-bound-a} Let $q \geq 2$.  Show that $\|f^{\leq k}\|_q \leq \sqrt{q-1}^k \|f\|_q$. (Hint: Use Theorem~\ref{thm:general-bonami} to show the stronger statement $\|f^{\leq k}\|_q \leq \sqrt{q-1}^k \|f\|_2$.)
            \item Let $1 < q \leq 2$.  Show that $\|f^{\leq k}\|_q \leq (1/\sqrt{q-1})^k \|f\|_q$.  (Hint: Either give a similar direct proof using the $(p,2)$-Hypercontractivity Theorem, or explain how this follows from part~\ref{ex:proj-bound-a} using the dual norm Proposition~\ref{prop:dual-norm}.)
        \end{exercises}
    \item \label{ex:hc-basics0}  Let $\bX$ be $(p,q,\rho)$-hypercontractive.
        \begin{exercises}
            \item \label{ex:hc-homog} Show that $c\bX$ is $(p,q,\rho)$-hypercontractive for any $c \in \R$.
            \item Show that $\rho \leq \frac{\|\bX\|_p}{\|\bX\|_q}$.
        \end{exercises}
    \item \label{ex:hc-basics1} Let $\bX$ be $(p,q,\rho)$-hypercontractive.  (For simplicity you may want to assume~$\bX$ is a discrete random variable.)
        \begin{exercises}
            \item \label{ex:hc-mean0} Show that $\E[\bX]$ must be $0$.  (Hint: Taylor expand $\|1 + \rho \eps \bX\|_r$ to one term around $\eps = 0$; note that $\rho < 1$ by definition.)
            \item \label{ex:hc-max} Show that $\rho \leq \sqrt{\frac{p-1}{q-1}}$.  (Hint: Taylor expand $\|1 + \rho \eps \bX\|_r$ to two terms around $\eps = 0$.)
        \end{exercises}
    \item \label{ex:hc-basics2}
        \begin{exercises}
            \item Suppose $\E[\bX] = 0$.  Show that $\bX$ is $(q, q, 0)$-hypercontractive for all $q \geq 1$. (Hint: Use monotonicity of norms to reduce to the case $q = 1$.)
            \item Show further that $\bX$ is $(q, q, \rho)$-hypercontractive for all $0 \leq \rho < 1$.  (Hint: Write $(a + \rho \bX) = (1-\rho) a + \rho(a+\bX)$ and employ the triangle inequality for $\| \cdot \|_q$.)
            \item Show that if $\bX$ is $(p,q,\rho)$-hypercontractive, then it is also $(p, q, \rho')$-hypercontractive for all $0 \leq \rho' < \rho$.  (Hint: Use the previous exercise along with Exercise~\ref{ex:hc-basics1}\ref{ex:hc-mean0}.)
        \end{exercises}
    \item \label{ex:hc-anticonc} Let $\bX$ be a (nonconstant) $(2,4,\rho)$-hypercontractive random variable.  The goal of this exercise is to show the following anticoncentration
                                                    \index{anticoncentration}%
        result:    For all $\theta \in \R$ and $0 < t < 1$,
        \[
            \Pr[|\bX - \theta| > t \|\bX\|_2] \geq (1-t^2)^2\rho^4.
        \]
        \begin{exercises}
            \item Reduce to the case $\|\bX\|_2 = 1$.
            \item Letting $\bY = (\bX - \theta)^2$, show that $\E[\bY] = 1+\theta^2$ and $\E[\bY^2] \leq (\rho^{-2} + \theta^2)^2$.
            \item Using the Paley--Zygmund inequality, show that
                    \[
                        \Pr[|\bX - \theta| > t] \geq \left(\frac{\rho^2(1-t^2) + \rho^2 \theta^2}{1+\rho^2\theta^2}\right)^2.
                    \]
            \item Show that the right-hand side above is minimized for $\theta = 0$, thereby completing the proof.
        \end{exercises}
    \item \label{ex:noise-stab-for-large-range}  Let $m \in \N^+$ and let $f \co \bn \to [m]$ be ``unbiased'', meaning $\Pr[f(\bx) = i] = \frac{1}{m}$ for all $i \in [m]$.  Let $0 \leq \rho \leq 1$ and let $(\bx, \by)$ be a $\rho$-correlated pair.  Show that $\Pr[f(\bx) = f(\by)] \leq (1/m)^{(1-\rho)/(1+\rho)}$.  (More generally, you might show that this is an upper bound on $\Stab_\rho[f]$ for all $f \co \bn \to \simplex_m$ with $\E[f] = (\frac1m, \dots, \frac1m)$; see Exercise~\ref{ex:generalized-domain-range}.)
    \item \label{ex:weak-p-2-bonami}
        \begin{exercises}
            \item \label{ex:weak-p-2-a} Let $f \btR$ have $\deg(f) \leq k$.  Prove that $\|f\|_2 \leq (1/\sqrt{p-1})^k \|f\|_p$ for any $1 \leq p \leq 2$ using the \Holder inequality strategy from our proof of the $(4/3,2)$-Hypercontractivity Theorem, together with Theorem~\ref{thm:general-bonami}.
            \item Verify that $\exp(\frac{2}{p}-1) < 1/\sqrt{p-1}$ for all $1 \leq p < 2$; i.e., the trickier Theorem~\ref{thm:p-2-bonami} strictly improves on the bound from part~\ref{ex:weak-p-2-a}.
        \end{exercises}
    \item \label{ex:p-2-bonami}  Prove Theorem~\ref{thm:p-2-bonami} in full generality.  (Hint: Let $\theta$ be the solution of $\frac12 = \frac{\theta}{p} + \frac{1-\theta}{2+\eps}$.  You will need to show that        $\frac{1-\theta}{2\theta} = (\frac2p-1)\frac{1}{\eps} + (\frac1p - \frac12)$.)
    \item \label{ex:deriv-induct-2q-hc}  As mentioned, it's possible to deduce the $(2,q)$-Hypercontractivity Theorem from the $n = 1$ case using induction by derivatives.  From this one can also obtain the $(p,2)$-Hypercontractivity Theorem via Proposition~\ref{prop:dual-norm}.  Employing the notation $\bx = (\bx', \bx_n)$, $\T = \T_{1/\sqrt{q-1}}$, $\bd = \D_nf(\bx')$, and $\be = \uE_nf(\bx')$, fill in details and justifications for the following proof sketch:
        \begin{multline*}
             \|\T_{1/\sqrt{q-1}} f\|_q^2 = \E_{\bx'} \Bigl[\E_{\bx_n}\bigl[|\T \be + (1/\sqrt{q-1}) \bx_n \T \bd|^q\bigr]\Bigr]^{2/q} \leq \E_{\bx'} \bigl[((\T \be)^2 + (\T \bd)^2)^{q/2}\bigr]^{2/q} \\
             = \|(\T \be)^2 + (\T \bd)^2\|_{q/2} \leq \|(\T \be)^2\|_{q/2} + \|(\T \bd)^2\|_{q/2} = \|\T \be\|_{q}^2 + \|\T \bd\|_{q}^2 \leq \|\be\|_2^2 + \|\bd\|_2^2 = \|f\|_2^2.
        \end{multline*}
    \item \label{ex:hypercon-straddling-2}  Deduce the $p < 2 < q$ cases of the Hypercontractivity Theorem from the $(2,q)$- and $(p,2)$-Hypercontractivity Theorems.  (Hint: Use the semigroup property of~$\T_\rho$,
                                                \index{semigroup property}%
                                                \index{Hypercontractivity Theorem}%
        Exercise~\ref{ex:semigroup}.)
    \item \label{ex:level-1-ineq} Let $f \co \bn \to \{0,1\}$ have $\E[f] = \alpha$.
        \begin{exercises}
            \item Show that $\W{1}[f] \leq \frac{1}{\rho}(\alpha^{2/(1+\rho)} - \alpha^2)$ for any $0 < \rho \leq 1$.
                                                                \index{Level-1 Inequality}%
            \item Deduce the sharp Level-1 Inequality $\W{1}[f] \leq 2\alpha^2 \ln(1/\alpha)$.  (Hint: Take the limit $\rho \to 0^+$.)
        \end{exercises}
    \item \label{ex:level-k-uncert} For $f \co \bn \to \{0,1\}$ with $\E[f] = \alpha$, show that $\W{\leq k}[f] = o(\alpha)$ (as $\alpha \to 0$) provided $k \leq .373\ln(1/\alpha)$.    \item Show that the KKL~Theorem fails for functions $f \btI$, even under the assumption $\Var[f] \geq \Omega(1)$. (Hint: $f(x) = \trunc_{[-1,1]}(\frac{x_1 + \cdots + x_n}{\sqrt{n}})$.)
    \item \label{ex:learn-monotone} \begin{exercises}
            \item Show $\calC = \{f \btb \mid \Tinf[f] \leq O(\sqrt{\log n})\}$ is learnable from queries to any constant error $\eps > 0$ in time~$\poly(n)$.   (Hint: Theorem~\ref{thm:KKL-Friedgut-generalized}.)
            \item Show  $\calC = \{\text{monotone } f \btb \mid \Tinf[f] \leq O(\sqrt{\log n})\}$ is learnable from random examples to any constant error $\eps > 0$ in time~$\poly(n)$.
            \item Show that $\calC = \{\text{monotone } f \btb \mid \DTsize(f) \leq \poly(n)\}$ is learnable from random examples to any constant error $\eps > 0$ in time~$\poly(n)$. (Hint: the OS~Inequality and Exercise~\ref{ex:expected-depth-vs-size}.)
                                                \index{OS Inequality}%
                                                \index{decision tree!learning}%
                                                \index{monotone function!learning}%
          \end{exercises}
    \item \label{ex:o-w} Deduce the following generalization of the $(2,q)$-Hypercontractivity Theorem:  Let $f \btR$, $q \geq 2$, and assume $0 \leq \rho \leq 1$ satisfies $\rho^\lambda  \leq 1/\sqrt{q-1}$ for some $0 \leq \lambda \leq 1$.
                                                    \index{hypercontractivity|)}%
        Then
        \[
            \|\T_\rho f\|_q \leq \|\T_\rho f\|_2^{1-\lambda} \|f\|_2^\lambda.
        \]
        (Hint: Show $\|\T_\rho f\|_q^2 \leq \sum_S (\rho^{2|S|}\wh{f}(S)^2)^{1-\lambda} \cdot (\wh{f}(S)^2)^{\lambda}$ and use \Holder.)
    \item Let $f \btI$, let $0 \leq \eps \leq 1$, and assume $q \geq 2+2\eps$.  Show that
            \[
                \|\T_{1-\eps} f\|_q^q \leq \|\T_{\frac{1}{\sqrt{1+2\eps}}} f\|_q^q \leq (\|f\|_2^2)^{1+\eps}.
            \]
    \item \label{ex:gaussquad-est} Recall the Gaussian quadrant probability $\GaussQuad_\rho(\mu)$ defined in Exercise~\ref{ex:ball-stab} by $\GaussQuad_\rho(\mu) = \Pr[\bz_1 > t, \bz_2 > t]$, where $\bz_1, \bz_2$ are standard Gaussians with correlation $\E[\bz_1\bz_2] = \rho$ and~$t$ is defined by $\olPhi(t) = \mu$.
                                                \index{Gaussian quadrant probability}%
                                                \index{Small-Set Expansion Theorem}%
                                                \index{noisy hypercube graph}%
        The goal of this exercise is to show that for fixed $0 < \rho < 1$ we have the estimate
        \begin{equation} \label{eqn:gaussquad-goal}
            \GaussQuad_\rho(\mu) = \wt{\Theta}(\mu^{\frac{2}{1+\rho}})
        \end{equation}
        as $\mu \to 0$.  In light of Exercise~\ref{ex:ball-stab}, this will show that the Small-Set Expansion Theorem for the $\rho$-stable hypercube graph is essentially sharp due to the example of Hamming balls of volume~$\mu$.
        \begin{exercises}
            \item \label{ex:gaussquad-heuristic} First let's do an imprecise ``heuristic'' calculation.  We have $\Pr[\bz_1 > t] = \Pr[\bz_1 \geq t] = \mu$ by definition. Conditioned on a Gaussian being at least~$t$ it is unlikely to be much more than~$t$, so let's just pretend that $\bz_1 = t$.  Then the conditional distribution of $\bz_2$ is $\rho t + \sqrt{1-\rho^2} \by$, where $\by \sim \normal(0,1)$ is an independent Gaussian.  Using the fact that $\olPhi(u) \sim \phi(u)/u$ as $u \to \infty$, deduce that $\Pr[\bz_2 > t \mid \bz_1 = t] = \wt{\Theta}(\mu^{\frac{1-\rho}{1+\rho}})$ and ``hence''~\eqref{eqn:gaussquad-goal} holds.
            \item Let's now be rigorous.  Recall that we are treating $0 < \rho < 1$ as fixed and letting $\mu \to 0$ (hence $t \to \infty$).  Let $\phi_\rho(z_1,z_2)$ denote the joint pdf of $\bz_1, \bz_2$ so that
                  \[
                    \GaussQuad_\rho(\mu) = \int_{t}^\infty \int_t^\infty \phi_{\rho}(z_1, z_2)\, dz_1\,dz_2.
                  \]
                  Derive the following similar-looking integral:
                  \begin{equation} \label{eqn:gauss-quad-extra}
                      \int_{t}^\infty \int_t^\infty (z_2 - \rho z_1)(z_1 - \rho t) \phi_{\rho}(z_1, z_2)\, dz_1\,dz_2 = \frac{(1-\rho^2)^{3/2}}{2\pi} \exp\left(-\frac{2}{1+\rho} \frac{t^2}{2}\right)
                  \end{equation}
                  and show that the right-hand side is $\wt{\Theta}(\mu^{\frac{2}{1+\rho}})$.
            \item Show that
                \[
                    \Pr\left[\bz_1 > \tfrac{t-1}{\rho}\right] = \int_{\frac{t-1}{\rho}}^\infty \phi(z_1)\,dz_1 = \wt{\Theta}(\mu^{\frac{1}{\rho^2}}),
                \]
                and that this is asymptotically smaller than $\wt{\Theta}(\mu^{\frac{2}{1+\rho}})$.
            \item Deduce~\eqref{eqn:gaussquad-goal}.  (Hint: Try to arrange that the extraneous factors $(z_2 - \rho z_1)$, $(z_1 - \rho t)$ in~\eqref{eqn:gauss-quad-extra} are both at least~$1$.)
        \end{exercises}
    \item \label{ex:CInf} Let $f \btb$, let $J \subseteq [n]$, and write $\ol{J} = [n] \setminus J$.  Define the \emph{coalitional influence}
                                        \index{influence!coalitional}%
                                        \nomenclature[InfJf]{$\CInf_J[f]$}{the coalitional influence of $J \subseteq [n]$ on $f \btb$, namely $\Pr_{\bz \sim \bits^{\ol{J}}}[\restr{f}{J}{\bz} \text{ is not constant}]$}%
        of~$J$ on~$f$ to be
        \[
            \CInf_J[f] = \Pr_{\bz \sim \bits^{\ol{J}}}[\restr{f}{J}{\bz} \text{ is not constant}].
        \]
        Furthermore, for $b \in \{-1, +1\}$ define the \emph{coalitional influence toward~$b$} of~$J$ on~$f$ to be
                                            \nomenclature[InfJfb]{$\CInf^{\,b}_J[f]$}{for $b \in \bits$, equals $\Pr_{\bz \sim \bits^{\ol{J}}}[\restr{f}{J}{\bz} \not \equiv -b] - \Pr[f = b]$}%
        \begin{align*}
            \CInf_J^{\,b}[f] &= \Pr_{\bz \sim \bits^{\ol{J}}}[\restr{f}{J}{\bz} \text{ can be made $b$}] - \Pr[f = b] \\
        &= \Pr_{\bz \sim \bits^{\ol{J}}}[\restr{f}{J}{\bz} \not \equiv -b] - \Pr[f = b].
        \end{align*}
        For brevity, we'll sometimes write $\CInf_J^{\,\pm}[f]$ rather than $\CInf_J^{\,\pm 1}[f]$.
        \begin{exercises}
            \item Show that for coalitions of size~$1$ we have $\Inf_i[f] = \CInf_{\{i\}}[f] = 2\CInf_{\{i\}}^{\,\pm}[f]$.
            \item Show that $0 \leq \CInf_J^{\,\pm}[f] \leq 1$.
            \item Show that $\CInf_J[f] = \CInf_J^{\,+}[f] + \CInf_J^{\,-}[f]$.
            \item \label{ex:CInf-mono} Show that if $f$ is monotone, then
                    \[
                        \CInf^{\,b}_J[f] = \Pr[\restr{f}{\ol{J}}{(b, \dots, b)} = b] - \Pr[f = b].
                    \]
            \item Show that $\CInf_J[\chi_{[n]}] = 1$ for all~$J \neq \emptyset$.
            \item Supposing we write $t = |J|/\sqrt{n}$, show that $\CInf^{\,\pm}_J[\Maj_n] = \Phi(t) - \half \pm o(1)$ and hence $\CInf_J[\Maj_n] = 2\Phi(t) - 1 \pm o(1)$.  Thus $\CInf_J[\Maj_n] = o(1)$ if $|J| = o(\sqrt{n})$ and $\CInf_J[\Maj_n] = 1 - o(1)$ if $|J| = \omega(\sqrt{n})$.  (Hint: Central Limit Theorem.)
            \item Show that $\max \{\CInf_J^{\True}[\tribes_n] : |J| \leq \log n\} = 1/2 + \Theta(\frac{\log n}{n})$. On the other hand, show that $\max \{\CInf_J^{\False}[\tribes_n] : |J| \leq k\} \leq k \cdot O(\frac{\log n}{n})$. Deduce that for some positive constant~$c$ we have $\max \{\CInf_J[\tribes_n] : |J| \leq c n/\log n\} \leq .51$.  (Hint: Refer to Proposition~\ref{prop:tribes-facts}.)
        \end{exercises}
    \item \label{ex:friedgut-tight-on-tribes}  Show that the exponential dependence on $\Tinf[f]$ in Friedgut's Junta Theorem is necessary.  (Hint: Exercise~\ref{ex:tribes-not-a-junta}.)
    \item \label{ex:precise-bribing} Let $f \btb$ be a monotone function with $\Var[f] \geq \delta > 0$, and let $0 < \eps <   1/2$ be given.
        \begin{exercises}
            \item Improve Proposition~\ref{prop:monotone-bribing} as follows: Show that there exists $J \subseteq [n]$ with $|J| \leq O(\log \frac{1}{\eps \delta}) \cdot \frac{n}{\log n}$ such that $\E[\restr{f}{\ol{J}}{(1, \dots, 1)}] \geq 1-\eps$. (Hint: How many bribes are required to move~$f$'s mean outside the interval $[1-2\eta, 1-\eta]$?)
            \item \label{ex:prec-bribe-b} Show that there exists $J \subseteq [n]$ with $|J| \leq O(\log \frac{1}{\eps \delta}) \cdot \frac{n}{\log n}$ such that $\CInf_J[f] \geq 1 - \eps$.  (Hint: Use Exercise~\ref{ex:CInf}\ref{ex:CInf-mono} and take the union of two influential sets.)
        \end{exercises}
    \item \label{ex:polarized-bribing} Let $f \btb$.
        \begin{exercises}
            \item \label{ex:polar-bribe-a} Let $f^* \btb$ be the ``monotonization'' of~$f$
                                                        \index{polarization}%
                    as defined in Exercise~\ref{ex:polarization}.  Show that $\CInf_J^{\,b}[f^*] \leq \CInf_J^{\,b}[f]$ for all $J \subseteq [n]$ and $b \in \bits$, and hence also $\CInf_J[f^*] \leq \CInf_J[f]$.
            \item Let $\Var[f] \geq \delta > 0$ and let $0 < \eps <   1/2$ be given.  Show that there exists $J \subseteq [n]$ with $|J| \leq O(\log \frac{1}{\eps \delta}) \cdot \frac{n}{\log n}$ such that $\CInf_J[f] \geq 1 - \eps$.  (Hint: Combine part~\ref{ex:polar-bribe-a} with Exercise~\ref{ex:precise-bribing}\ref{ex:prec-bribe-b}.)
        \end{exercises}
    \item \label{ex:finish-kkl}
            Establish the general-variance case of the KKL Edge-Isoperimetric Theorem.  (Hint: You'll need to replace~\eqref{eqn:finish-kkl1} with
                \[
                    3 \sum_{|S| \geq 1} (1/3)^{|S|} \wh{f}(S)^2 \geq 3 \Var[f] \cdot 3^{-\Tinf[f]/\Var[f]}.
                \]
                Use the same convexity argument, but applied to the random variable~$\bS$ that takes on each outcome $\emptyset \neq S \subseteq [n]$ with probability $\wh{f}(S)^2/\Var[f]$.)
    \item \label{ex:careful-constant-kkl} The goal of this exercise is to attain the best known constant factor in the statement of the KKL~Theorem.
        \begin{exercises}
            \item By using Corollary~\ref{cor:stab-infl-bound} in place of Corollary~\ref{cor:1/3-stab-infl}, obtain the following generalization of the KKL Edge-Isoperimetric Theorem: For any (nonconstant) $f \btb$ and $0 < \delta < 1$,
                \[
                     \MaxInf[f] \geq \left(\tfrac{1+\delta}{1-\delta}\right)^{\frac{1}{\delta}} \left(\tfrac{1}{\wt{\Tinf}[f]}\right)^{\frac{1}{\delta}} \cdot \left(\tfrac{1-\delta}{1+\delta}\right)^{\frac{1}{\delta}\wt{\Tinf}[f]},
                \]
                where $\wt{\Tinf}[f]$ denotes $\Tinf[f]/\Var[f]$. (Hint: Write $\rho = \frac{1-\delta}{1+\delta}$.) Deduce that for any constant $C > e^2$ we have
                \[
                    \MaxInf[f] \geq \wt{\Omega}(C^{-\wt{\Tinf}[f]}).
                \]
            \item More carefully, show that by taking $\delta = \frac{1}{2\wt{\Tinf}[f]^{1/3}}$ we can achieve
                \[
                     \MaxInf[f] \geq \exp(-2\wt{\Tinf}[f]) \cdot e^2 \cdot \left(\tfrac{1}{\wt{\Tinf}[f]}\right)^{2\wt{\Tinf}[f]^{1/3}} \cdot \exp(-\tfrac14 \wt{\Tinf}[f]^{1/3}).
                \]
                (Hint: Establish $\left(\tfrac{1-\delta}{1+\delta}\right)^{\frac{1}{\delta}} \geq \exp(-2-\delta^2)$ for $0 < \delta \leq 1/2$.)
            \item By distinguishing whether or not $\wt{\Tinf}[f] \geq \half (\ln n - \sqrt{\log n})$, establish the following form of the KKL~Theorem:  For any $f \btb$,
                \[
                    \MaxInf[f] \geq \half \Var[f] \cdot \frac{\ln n}{n} (1-o_n(1)).
                \]
        \end{exercises}
    \item \label{ex:KKL-Friedgut-generalized}  Establish the claim in Remark~\ref{rem:KKL-Friedgut-generalized}.
    \item \label{ex:low-inf-large-coeff} Show that if $f \btb$ is nonconstant, then there exists $S \subseteq [n]$ with $0 < |S| \leq O(\Tinf[f]/\Var[f])$ such that $\wh{f}(S)^2 \geq \exp(-O(\Tinf[f]^2/\Var[f]^2))$.  (Hint: By mimicking Corollary~\ref{cor:low-inf-large-coeff}'s proof you should be able to establish the lower bound $\Omega(\Var[f]) \cdot \exp(-O(\Tinf[f]^2/\Var[f]^2))$.  To show that this quantity is also $\exp(-O(\Tinf[f]^2/\Var[f]^2))$, use Theorem~\ref{thm:edge-iso}.)
    \item \label{ex:min-EI-for-monotone}  Let $f \btb$ be a nonconstant \emph{monotone} function.  Improve on Corollary~\ref{cor:low-inf-large-coeff} by showing that there exists $S \neq \emptyset$ satisfying $\wh{f}(S)^2 \geq \exp(-O(\Tinf[f]/\Var[f]))$.  (Hint: You can even get $|S| \leq 1$; use the  KKL Edge-Isoperimetric Theorem and Proposition~\ref{prop:monotone-influences}.)
                                            \index{Fourier sparsity}%
    \item \label{ex:sparsity-hypercon}  Let $f \btR$.  Prove that $\|f\|_4 \leq \ssparsity{f}^{1/4} \|f\|_2$.
    \item \label{ex:multifunction-hypercon}  Let $q = 2r$ be a positive even integer, let $\rho = 1/\sqrt{q-1}$, and let $f_1, \dots, f_r \btR$.  Generalize the $(2,q)$-Hypercontractivity Theorem by showing that
        \[
            \E\left[\prod_{i=1}^r (\T_\rho f_i)^2\right] \leq \prod_{i=1}^r \E[f_i^2].
        \]
        (Hint: \Holder's inequality.)
    \item \label{ex:2-2s-hypercon} In this exercise you will give a simpler, stronger version of Theorem~\ref{thm:hypercon-bit} under the assumption that $q = 2r$ is a positive even integer.
        \begin{exercises}
            \item Using the idea of Proposition~\ref{prop:2-6-hypercon}, show that if $\bx$ is a uniformly random $\pm 1$ bit then $\bx$ is $(2,q,\rho)$-hypercontractive if and only if $\rho \leq 1/\sqrt{q-1}$.
            \item \label{ex:even-int-strong} Show the same statement for any random variable $\bx$ satisfying $\E[\bx^2] = 1$ and
                    \[
                        \E[\bx_i^{2j-1}] = 0, \quad \E[\bx_i^{2j}] \leq (2r-1)^j \frac{\binom{r}{j}}{\binom{2r}{2j}}\quad \text{for all integers $1 \leq j \leq r$.}
                    \]
            \item Show that none of the even moment conditions in part~\ref{ex:even-int-strong} can be relaxed.
        \end{exercises}
    \item \label{ex:super-bonami}  Let $q = 2r$ be a positive even integer and let $f \btR$ be homogeneous of degree~$k \geq 1$ (i.e., $f = f^{=k}$).  The goal of this problem is to improve slightly on the generalized Bonami Lemma, Theorem~\ref{thm:general-bonami}.
        \begin{exercises}
            \item Show that
                \begin{equation} \label{eqn:super-bonami-1}
                    \E[f^q] = \sum \wh{f}(S_1)  \cdots \wh{f}(S_q) \leq \sum |\wh{f}(S_1)|  \cdots |\wh{f}(S_q)|,
                \end{equation}
                where the sum is over all tuples $S_1, \dots, S_k$ satisfying $S_1 \symdiff  \cdots \symdiff S_q = \emptyset$.
            \item Let~$G$ denote the complete $q$-partite graph over vertex sets $V_1, \dots, V_q$, each of cardinality~$k$.  Let $\calM$ denote the set of all perfect matchings in~$G$.  Show that the right-hand side of~\eqref{eqn:super-bonami-1} is equal to
                \begin{equation} \label{eqn:super-bonami-2}
                    \frac{1}{(k!)^q}\sum_{M \in \calM}\ \sum_{\ell : M \to [n]} |\wh{f}(T_1(M,\ell))| \cdots  |\wh{f}(T_q(M,\ell))|,
                \end{equation}
                where $T_j(M,\ell)$ denotes $\bigcup \{\ell(e) : e \in M, e \cap V_j \neq \emptyset\}$.
            \item Show that \eqref{eqn:super-bonami-2} is equal to
                \begin{equation} \label{eqn:super-bonami-3}
                      \frac{1}{(rk)! \cdot (k!)^q}\sum_{\ol{M} \in \ol{\calM}}\ \sum_{i_1 = 1}^n \sum_{i_2 = 1}^n \cdots \sum_{i_{rk} = 1}^n |\wh{f}(U_1(\ol{M},i_1, \dots, i_{rk}))| \cdots  |\wh{f}(U_q(\ol{M},i_1, \dots, i_{rk}))|,
                \end{equation}
                where $\ol{\calM}$ is the set of \emph{ordered} perfect matchings of~$G$, and now \linebreak $U_j(\ol{M}, i_1, \dots, i_{rk})$ denotes $\bigcup \{i_t : \ol{M}(t) \cap V_j \neq \emptyset\}$.
             \item Show that for any $\ol{M} \in \ol{\calM}$ we have
                \begin{multline*}
                    \sum_{i_1 = 1}^n \sum_{i_2 = 1}^n \cdots \sum_{i_{rk} = 1}^n |\wh{f}(U_1(\ol{M},i_1, \dots, i_{rk}))| \cdots  |\wh{f}(U_q(\ol{M},i_1, \dots, i_{rk}))| \\ \leq \left(\sum_{j_1, \dots, j_k = 1}^n \wh{f}(\{j_1, \dots, j_k\})^2\right)^r
                \end{multline*}
                (Hint: Use Cauchy--Schwarz $rk$ times.)
             \item Deduce that
                    $
                        \|f\|_q^q \leq \frac{1}{(rk)! \cdot (k!)^q} \cdot |\ol{\calM}| \cdot (k!)^r \|f\|_2^{2r}
                    $
                   and hence
                   \[
                        \|f\|_q \leq \frac{|\calM|^{1/q}}{\sqrt{k!}} \|f\|_2.
                   \]
        \end{exercises}
    \item \label{ex:number-of-matchings}  The goal of this problem is to estimate $|\calM|$ from Exercise~\ref{ex:super-bonami} so as to give a concrete improvement on Theorem~\ref{thm:general-bonami}.
        \begin{exercises}
            \item \label{ex:kiener} Show that for $q = 4$, $k = 2$ we have $|\calM| = 60$.
            \item Show that $|\calM| \leq (qk-1)!!$.  (Hint: Show that $(qk-1)!!$ is the number of perfect matchings in the complete graph on~$qk$ vertices.)  Deduce $\|f\|_q \leq \sqrt{q}^k \|f\|_2$.
            \item \label{ex:bonami-matchings} Show that $|\ol{\calM}| \leq (\frac{2r-1}{r})^{rk} (rk)!^2$, and thereby deduce
                \[
                    \|f\|_q \leq C_{q,k} \cdot \sqrt{q-1}^k \|f\|_2,
                \]
                where $C_{q,k} = \left(\frac{(rk)!}{k!^r r^{rk}}\right)^{1/q}$.
              (Hint: Suppose that the first $t$ edges of the perfect matching have been chosen; show that there are $(\frac{2r-1}{r})(rk - t)^2$ choices for the next edge.  The worst case is if the vertices used up so far are spread equally among the~$q$ parts.)
            \item Give a simple proof that $C_{q,k} \leq 1$, thereby obtaining Theorem~\ref{thm:general-bonami}.
            \item Show that in fact $C_{q,k} = \Theta(1) \cdot k^{-1/4 + 1/(2q)}$.  (Hint: Stirling's Formula.)
            \item \label{ex:janson-matchings} Can you obtain the improved estimate
                  \[
                      \frac{|\calM|^{1/q}}{\sqrt{k!}} = \Theta_q(1) \cdot k^{-1/4} \cdot \sqrt{q-1}^k?
                  \]
                  (Hint: First exactly count -- then estimate -- the number of perfect matchings with exactly~$e_{ij}$ edges between parts~$i$ and~$j$.  Then sum your estimate over a range of the most likely values for~$e_{ij}$.)
        \end{exercises}
\end{exercises}

\subsection*{Notes.}                        \label{sec:basic-hypercon-notes}

                                                \index{hypercontractivity|(}
The history of the Hypercontractivity Theorem is complicated.  Its earliest roots are in the work of Paley~\cite{Pal32} from 1932; he showed that for $1 < p < \infty$ there are constants $0 < c_p < C_p < \infty$ such that $c_p \|\Sq f\|_p \leq \|f\|_p \leq C_p \|\Sq f\|_p$ holds for any $f \btR$.  Here $\Sq f = \sum_{t=1}^n \sqrt{\sum_{t = 1}^n (\mds{t}{f})^2}$ is the ``square function'' of~$f$, and $\mds{t}{f} = \sum_{S : \max(S) = t} \wh{f}(S)\,\chi_S$ is the
                                                \index{martingale difference sequence}%
martingale difference sequence for~$f$ defined in Exercise~\ref{ex:martingale}.   The main task in Paley's work is to prove the statement when~$p$ is an even integer; other values of~$p$ follow by the Riesz(--Thorin) interpolation theorem.  Using this result, Paley showed the following hypercontractivity result: If $f \btR$ is homogeneous of degree~$2$, then $c'_p \|f\|_2 \leq \|f\|_p \leq C'_p \|f\|_2$ for any $p \in \R^+$. Some extensions of Paley's work are in~\cite{Wat64}.

In 1968 Bonami~\cite{Bon68} stated the following variant of Theorem~\ref{thm:general-bonami}:  If $f \btR$ is homogeneous of degree~$k$, then for all $q \geq 2$, $\|f\|_q \leq c_k \sqrt{q} \|f\|_2$, where the constant $c_k$ may be taken to be~$1$ if~$q$ is an even integer.  She remarks that this theorem can be deduced from Paley's result but with a much worse (exponential) dependence on~$q$.  The proof she gives is combinatorial and actually only treats the case $k = 2$ and~$q$ an even integer; it is similar to Exercise~\ref{ex:super-bonami}.

Independently in 1969, Kiener~\cite{Kie69} published his Ph.D.\ thesis, which extended Paley's hypercontractivity result as follows: If $f \btR$ is homogeneous of degree~$k$, then $c_{p,k} \|f\|_2 \leq \|f\|_p \leq C_{p,k} \|f\|_2$ for any $p \in \R^+$.  The proof is an induction on~$k$, and again the bulk of the work is the case of even integer~$p$.  Kiener also gave a long combinatorial proof showing that if $f \btR$ is homogeneous of degree~$2$, then $\E[f^4] \leq 51 \E[f^2]^2$.  (Exercise~\ref{ex:number-of-matchings}\ref{ex:kiener} improves this~$51$ to~$15$.)

Also independently in 1969, Schreiber~\cite{Sch69} considered multilinear polynomials~$f$ over a general orthonormal sequence $\bx_1, \dots, \bx_n$ of centered real (or complex) random variables.  He showed that if~$f$ has degree at most~$k$, then for any even integer $q \geq 4$ it holds that $\|f\|_q \leq C \|f\|_2$, where $C$ depends only on~$k$,  $q$, and the $q$-norms of the $\bx_i$'s.  Again, the proof is very similar to Exercise~\ref{ex:super-bonami}; Schreiber does not estimate his analogue of $|\calM|$ but merely notes that it's finite.  Schreiber was interested mainly in the case that the $\bx_i$'s are Gaussian; indeed, his 1969 work~\cite{Sch69} is a generalization of his earlier work~\cite{Sch67} specific to the Gaussian case.

In 1970, Bonami published her Ph.D.\ thesis~\cite{Bon70}, which contains the full Hypercontractivity Theorem as stated at the beginning of the chapter.  Her proof follows the standard template seen in essentially all proofs of hypercontractivity: first an elementary proof for the case $n = 1$ and then an induction to extend to general~$n$. She also gives the sharper combinatorial result appearing in Exercises~\ref{ex:super-bonami} and~\ref{ex:number-of-matchings}\ref{ex:bonami-matchings}.  (The stronger bound from Exercise~\ref{ex:number-of-matchings}\ref{ex:janson-matchings} is due to Janson~\cite[Remark 5.20]{Jan97}.)  As in Corollary~\ref{cor:bonami-lemma}, Bonami notes that her combinatorial proof can be extended to a general sequence of symmetric orthonormal random variables, at the expense of including factors of $\|\bx_i\|_q$ into the bound.  She points out that this includes the Gaussian case independently studied by Schreiber.

Bonami's work was published in French, and it remained unknown to most English-language mathematicians for about a decade.  In the late 1960s and early 1970s, researchers in quantum field theory developed the theory of hypercontractivity for the Gaussian analogue of $\T_\rho$, namely, the Ornstein--Uhlenbeck operator~$\U_\rho$.  This is now recognized as essentially being a \emph{special case} of hypercontractivity for bits, in light of the fact that $\frac{\bx_1 + \cdots + \bx_n}{\sqrt{n}}$ tends to a Gaussian as $n \to \infty$ by the CLT (see Chapter~\ref{sec:gaussian}).  We summarize here some of the work in this setting.  In 1966 Nelson~\cite{Nel66} showed that $\|\U_{1/\sqrt{q-1}} f\|_q \leq C_q \|f\|_2$ for all $q \geq 2$. Glimm~\cite{Gli68} gave the alternative result that for each $q \geq 2$ there is a sufficiently small $\rho_q > 0$ such that  $\|\U_{\rho_q} f\|_q \leq \|f\|_2$.  Segal~\cite{Seg70} observed that hypercontractive results can be proved by induction on the dimension~$n$.  In 1973 Nelson~\cite{Nel73} gave the full Hypercontractivity Theorem in the Gaussian setting: $\|\U_{\sqrt{(p-1)/(q-1)}} f\|_q \leq \|f\|_p$ for all $1 \leq p < q \leq \infty$.  He also proved the combinatorial Exercise~\ref{ex:super-bonami}. The equivalence to the Two-Function
                                                    \index{Hypercontractivity Theorem!Two-Function}%
Hypercontractivity Theorem is from the work of Neveu~\cite{Nev76}.

In 1975 Gross~\cite{Gro75} introduced the notion of Log-Sobolev
                                                \index{Log-Sobolev Inequality}%
Inequalities (see Exercise~\ref{ex:log-sob}) and showed how to deduce hypercontractivity inequalities from them.  He established the Log-Sobolev Inequality for $1$-bit functions, used induction (citing Segal) to obtain it for $n$-bit functions, and then used the CLT to transfer results to the Gaussian setting.  (For some earlier results along these lines, see the works of Federbush and Gross~\cite{Fed69,Gro72}.) This gave a new proof of Nelson's result and also independently established Bonami's full Hypercontractivity Theorem.  Also in 1975, Beckner~\cite{Bec75} published his Ph.D.\ thesis, which proved a sharp form of the hypercontractive inequality for purely complex~$\rho$. (It is unfortunate that the influential paper of Kahn, Kalai, and Linial~\cite{KKL88} miscredited the Hypercontractivity Theorem to Beckner.)  The case of general complex~$\rho$ was subsequently treated by Weissler~\cite{Wei79}, with the sharp result being obtained by Epperson~\cite{Epp89}. Weissler~\cite{Wei80} also appears to have been the first to make the connection between this line of work and Bonami's thesis.

Independently of all this work, the $(q,2)$-Hypercontractivity Theorem was  reproved (without sharp constant) in the Banach spaces community by Rosenthal~\cite{Ros76} in 1975, using methods similar to those of Paley and Kiener.  For additional early references, see M{\"u}ller~\cite[Chapter~1]{Mul05}.

The \emph{term} ``hypercontractivity'' was introduced in a work of Simon and H{\o}egh-Krohn~\cite{SH72};  Definition~\ref{def:hypercon} of a hypercontractive random variable is due to Krakowiak and Szulga~\cite{KS88}.  The short inductive proof of the Bonami Lemma may have appeared first in Mossel, O'Donnell, and Oleszkiewicz~\cite{MOO05}.  Theorems~\ref{thm:p-2-bonami} and~\ref{thm:one-sided-low-deg-anticonc} appear in Janson~\cite{Jan97}. Theorem~\ref{thm:low-deg-conc} dates back to Pisier and Zinn and to Borell~\cite{PZ78,Bor79}.  As discussed further in the notes to Chapter~\ref{chap:advanced-hypercon}, the Small-Set Expansion Theorem originates in the work of Ahlswede and
G{\'a}cs~\cite{AG76}. The Level-$k$ Inequalities appear in several places but can probably be fairly credited to Kahn, Kalai, and Linial~\cite{KKL88}. The optimal constants for Khintchine's Inequality were established by Haagerup~\cite{Haa82}; see also Nazarov and Podkorytov~\cite{NP00}.  They always occur either when $\sum_i a_i \bx_i$ is just $\frac{1}{\sqrt{2}} \bx_1 + \frac{1}{\sqrt{2}} \bx_2$ or in the limiting Gaussian case of $a_i \equiv \frac{1}{\sqrt{n}}$, $n \to \infty$.
                                                \index{hypercontractivity|)}

Ben-Or and Linial's work~\cite{BL85,BL90} was motivated both by game theory and by the Byzantine Generals problem~\cite{LSP82} from distributed computing; the content of Exercise~\ref{ex:CInf} is theirs.  In turn it motivated the watershed paper by Kahn, Kalai, and Linial~\cite{KKL88}.  (See also the intermediate work of Chor and Ger{\'e}b-Graus~\cite{CGG87}.)
                                                \index{KKL Theorem}%
The ``KKL Edge-Isoperimetric Theorem'' (which is essentially a strengthening of the basic KKL~Theorem) was first explicitly proved by Talagrand~\cite{Tal94} (possibly independently of Kahn, Kalai, and Linial~\cite{KKL88}?); he also treated the $p$-biased case.  There is no known combinatorial proof of the KKL~Theorem (i.e., one which does not involve real-valued functions). However, several slightly different analytic proofs are known; see Falik and Samorodnitsky~\cite{FS07}, Rossignol~\cite{Ros06}, and O'Donnell and Wimmer~\cite{OW13}. The explicit lower bound on the ``KKL constant'' achieved in Exercise~\ref{ex:careful-constant-kkl} is the best known; it appeared first in Falik and Samorodnitsky~\cite{FS07}. It is still a factor of~$2$ away from the best known upper bound, achieved by the tribes function.

Friedgut's Junta Theorem dates from 1998~\cite{Fri98}.   The observation that its junta size can be improved for functions which have $\W{k}[f] \leq \eps$ for $k \ll \Tinf[f]/\eps$ was independently made by Li-Yang Tan in 2011; so was the consequence Corollary~\ref{cor:ltf-junta} and its extension to constant-degree PTFs.  A stronger result than Corollary~\ref{cor:ltf-junta} is known: Diakonikolas and Servedio~\cite{DS09} showed that every LTF is $\eps$-close to a $\Tinf[f]^2 \poly(1/\eps)$-junta.  As for Corollary~\ref{cor:dnf-width-junta}, it's incomparable with a result from Gopalan, Meka, and Reingold~\cite{GMR12}, which shows that every width-$w$ DNF is $\eps$-close to a $(w \log(1/\eps))^{O(w)}$-junta.

Exercise~\ref{ex:simple-bonami-lower-bound} was suggested to the author by Krzysztof Oleszkiewicz.  \linebreak Exercise~\ref{ex:hc-anticonc} is from Gopalan et~al.~\cite{GOWZ10}. Exercise~\ref{ex:learn-monotone} appears in O'Donnell and Servedio~\cite{OS07}; Exercise~\ref{ex:o-w} appears in O'Donnell and Wu \cite{OW09}.   The estimate in Exercise~\ref{ex:gaussquad-est} is from de~Klerk, Pasechnik, and Warners \cite{dKPW04} (see also works of Rinott and Rotar'~\cite{RR01} and Khot et~al.~\cite{KKMO07}).  Exercises~\ref{ex:precise-bribing} and~\ref{ex:polarized-bribing} are due to Kahn, Kalai, and Linial~\cite{KKL88}. Exercise~\ref{ex:sparsity-hypercon} was suggested to the author by John Wright.  Exercise~\ref{ex:2-2s-hypercon} appears in Kauers et~al.~\cite{KOTZ16}.

\chapter{Advanced hypercontractivity}         \label{chap:advanced-hypercon}

In this chapter we complete the proof of the Hypercontractivity Theorem for uniform~$\pm 1$ bits.  We then generalize the $(p,2)$ and $(2,q)$ statements to the setting of arbitrary product probability spaces, proving the following:
\begin{named}{The General Hypercontractivity Theorem}
            \index{Hypercontractivity Theorem!General}%
            \index{hypercontractivity}%
            \index{General Hypercontractivity Theorem|seeonly{Hypercontractivity Theorem, General}}
    Let $(\Omega_1, \pi_1)$, \dots, $(\Omega_n, \pi_n)$ be finite probability spaces, in each of which every outcome has probability at least~$\lambda$.  Let $f \in L^2(\Omega_1 \times \cdots \times \Omega_n, \pi_1 \otimes \cdots \otimes \pi_n)$.  Then for any $q > 2$ and $0 \leq \rho \leq \frac{1}{\sqrt{q-1}} \cdot \lambda^{1/2-1/q}$,
    \[
        \|\T_\rho f\|_q \leq \|f\|_2 \quad\text{and}\quad \|\T_\rho f\|_2 \leq \|f\|_{q'}.
    \]
    (And in fact, the upper bound on $\rho$ can be slightly relaxed to the value stated in Theorem~\ref{thm:discrete-hypercon}.)
\end{named}
We can thereby extend all the consequences of the basic Hypercontractivity Theorem for $f \btR$ to functions $f \in L^2(\Omega^n, \pi\xn)$, except with quantitatively worse parameters depending on~``$\lambda$''.  We also introduce the technique of randomization/symmetrization and show how it can sometimes eliminate this dependence on~$\lambda$. For example, it's used to prove Bourgain's Sharp Threshold Theorem, a characterization of Boolean-valued $f \in L^2(\Omega^n, \pi\xn)$ with low total influence that has no dependence at all on~$\pi$.

\section{The Hypercontractivity Theorem for uniformly random bits}        \label{sec:full-hypercon-for-bits}

In this section we'll prove the full Hypercontractivity Theorem for uniform~$\pm 1$ bits
                                                    \index{Hypercontractivity Theorem|(}%
stated at the beginning of Chapter~\ref{chap:hypercontractivity}:
\begin{named}{The Hypercontractivity Theorem}
    Let $f \btR$ and let ${1 \leq p \leq q \leq \infty}$.  Then $\|\T_\rho f\|_q \leq \|f\|_p$ for $0 \leq \rho \leq \sqrt{\tfrac{p-1}{q-1}}$.
\end{named}
Actually, when neither $p$ nor $q$ is~$2$,  the following equivalent form of theorem seems easier to interpret:
                                                \index{Hypercontractivity Theorem!Two-Function|(}%
\begin{named}{Two-Function Hypercontractivity Theorem} Let $f, g \btR$, let $r, s \geq 0$, and assume $0 \leq \rho \leq \sqrt{rs} \leq 1$.  Then
\[
    \Es{(\bx, \by) \\ \textnormal{ $\rho$-correlated}}[f(\bx) g(\by)] \leq \|f\|_{1+r} \|g\|_{1+s}.
\]
\end{named}
\noindent As a reminder, the only difference between this theorem and its ``weak'' form (proven in Chapter~\ref{sec:hypercon-tensorize}) is that we don't assume $r, s \leq 1$.  Below we will show that the two theorems \emph{are} equivalent, via \Holder's inequality.  Given the Two-Function Hypercontractivity Induction Theorem from Chapter~\ref{sec:hypercon-tensorize}, this implies that to prove the Hypercontractivity Theorem for general~$n$ we only need to prove it for~$n = 1$.  This is an elementary but technical inequality, which we defer to the end of the section.

Before carrying out these proofs, let's take some time to interpret the Two-Function Hypercontractivity Theorem.  One interpretation is simply as a generalization of \Holder's inequality.  Consider the case that the strings~$\bx$ and~$\by$ in the theorem are fully correlated; i.e., $\rho = 1$.  Then the theorem states that
\begin{equation}                                \label{eqn:its-holder}
    \E[f(\bx)g(\bx)] \leq \|f\|_{1+r} \|g\|_{1+1/r}
\end{equation}
because the condition $\sqrt{rs} = 1$ is equivalent to $s = 1/r$.  This statement is identical to \Holder's inequality, since $(1+r)' = 1+1/r$. \Holder's inequality is often used to ``break the correlation'' between two random variables; in the absence of any information about how~$f$ and~$g$ correlate then we can at least bound $\E[f(\bx)g(\bx)]$  by the product of certain norms of~$f$ and~$g$.  (If~$f$ and~$g$ have different ``sizes'', then \Holder lets us choose different norms for them; if~$f$ and~$g$ have roughly the same ``size'', then we can take $r = s = 1$ and get Cauchy--Schwarz.)  Now suppose we are considering $\E[f(\bx)g(\by)]$ for $\rho$-correlated $\bx,\by$ with $\rho < 1$.  In this case we might hope to improve~\eqref{eqn:its-holder} by using smaller norms on the right-hand side; in the extreme case of independent $\bx, \by$ (i.e., $\rho = 0$) we can use $\E[f(\bx)g(\by)] = \E[f] \E[g] \leq \|f\|_1 \|g\|_1$.
The Two-Function Hypercontractivity Theorem gives a precise interpolation between these two cases; the smaller the correlation~$\rho$ is, the smaller the norms we may take on the right-hand side.

In the case that $f$ and $g$ have range $\{0,1\}$, these ideas yield another interpretation of the Two-Function Hypercontractivity Theorem, namely a two-set generalization of the Small-Set Expansion Theorem:
                                                \index{expansion!small-set}%
                                                \index{Small-Set Expansion Theorem!generalized}%
\begin{named}{Generalized Small-Set Expansion Theorem}
    Let $0 \leq \rho \leq 1$.  Let $A,B \subseteq \bn$ have volumes $\exp(-\frac{a^2}{2}),\ \exp(-\frac{b^2}{2})$ and assume $0 \leq \rho a \leq b \leq a$.  Then
    \[
        \Pr_{\substack{(\bx, \by) \\ \textnormal{ $\rho$-correlated}}}[\bx \in A, \by \in B] \leq \exp\left(-\tfrac12 \tfrac{a^2 - 2\rho ab + b^2}{1-\rho^2}\right).
    \]
\end{named}
\begin{proof}
    Apply the Two-Function Hypercontractivity Theorem with $f = 1_A$, $g = 1_B$ and minimize the right-hand side by selecting $r = \rho \frac{b - \rho a}{a - \rho b}$, $s = \rho \frac{a-\rho b}{b-\rho a}$.
\end{proof}
\begin{remark}
     When~$a$ and~$b$ are not too close the optimal choice of~$s$ in the proof exceeds~$1$. Thus the Generalized Small-Set Expansion Theorem really needs the full (non-weak) Two-Function Hypercontractivity Theorem; equivalently, the full Hypercontractivity Theorem.
     Also note that the assumption $b \geq \rho a$ is needed to prevent~$r < 0$.
\end{remark}
\begin{remark}                                  \label{rem:gen-sse-sharpness}
    This theorem is essentially sharp in the case that $A$ and $B$ are concentric Hamming balls; see Exercise~\ref{ex:gen-sse-sharpness}.  In the case $b = a$ we recover the Small-Set Expansion Theorem.  In the case $b = \rho a$ we get only the trivial bound that $\Pr[\bx \in A, \by \in B] \leq \exp(-\frac{a^2}{2}) = \Pr[\bx \in A]$. However, not much better than this can be expected; in the concentric Hamming ball case it indeed holds that $\Pr[\bx \in A, \by \in B] \sim \Pr[\bx \in A]$ whenever $b < \rho a$.
\end{remark}
\begin{remark}          \label{rem:reverse-two-set}
    There is also a \emph{reverse} form of the Hypercontractivity Theorem and its Two-Function version; see Exercises~\ref{ex:reverse-hypercon1}--\ref{ex:reverse-hypercon3}. It directly implies the following:
    \begin{named}{Reverse Small-Set Expansion Theorem}
                        \index{Reverse Small-Set Expansion Theorem|seeonly{Small-Set Expansion Theorem, Reverse}}%
                        \index{Small-Set Expansion Theorem!Reverse}%
        Let $0 \leq \rho \leq 1$.  Let $A,B \subseteq \bn$ have volumes $\exp(-\frac{a^2}{2}),\ \exp(-\frac{b^2}{2})$, where $a, b \geq 0$. Then
        \[
            \Pr_{\substack{(\bx, \by) \\ \text{ $\rho$-correlated}}}[\bx \in A, \by \in B] \geq \exp\left(-\tfrac12 \tfrac{a^2 + 2\rho ab + b^2}{1-\rho^2}\right).
        \]
    \end{named}
\end{remark}

We now turn to the proofs.  We begin by showing that the Hypercontractivity Theorem and the Two-Function version are indeed equivalent.  This is a consequence of the following general fact (take $T = \T_\rho$, $p = 1+r$, $q = 1+1/s$):
\begin{proposition}                                     \label{prop:one-to-two-function-hc}
    Let $T$ be an operator on $L^2(\Omega, \pi)$ and let $1 \leq p,\,q \leq \infty$.  Then
    \begin{equation} \label{eqn:hc-one-to-two}
        \|Tf\|_q \leq \|f\|_p
    \end{equation}
    holds for all $f \in L^2(\Omega, \pi)$ if and only if
    \begin{equation} \label{eqn:hc-two-to-one}
            \la Tf, g\ra \leq \|f\|_p \|g\|_{q'}
    \end{equation}
    holds for all $f,g \in L^2(\Omega, \pi)$.
\end{proposition}
\begin{proof}
    For the ``only if'' statement, $\la Tf, g \ra \leq \|Tf\|_q \|g\|_{q'} \leq \|f\|_p \|g\|_{q'}$ by \Holder's inequality and~\eqref{eqn:hc-one-to-two}.  As for the ``if'' statement, by \Holder's inequality and~\eqref{eqn:hc-two-to-one} we have
    \[
        \|T f\|_{q} = \sup_{\|g\|_{q'} = 1} \la Tf, g \ra \leq \sup_{\|g\|_{q'} = 1} \|f\|_p \|g\|_{q'}= \|f\|_p. \qedhere
    \]
\end{proof}

Now suppose we prove the Hypercontractivity Theorem in the case $n = 1$.  By the above proposition we deduce the Two-Function version in the case $n = 1$.  Then the Two-Function Hypercontractivity Induction Theorem from Chapter~\ref{sec:hypercon-tensorize} yields the general-$n$ case of the Two-Function Hypercontractivity Theorem.  Finally, applying the above proposition again we get the general-$n$ case of the Hypercontractivity Theorem, thereby completing all needed proofs.  These observations all hold in the context of more general product spaces, so let's record the following for future use:
                            \index{hypercontractivity!induction}
\begin{named}{Hypercontractivity Induction Theorem}
    Let $0 \leq \rho \leq 1$, $1 \leq p, q \leq \infty$, and assume that $\|\T_\rho f\|_q \leq \|f\|_p$ holds for every $f \in L^2(\Omega_1, \pi_1), \dots, L^2(\Omega_n, \pi_n)$.  Then it also holds for every $f \in L^2(\Omega_1 \times \cdots \times \Omega_n, \pi_1 \otimes \cdots \otimes \pi_n)$.
\end{named}
                                                \index{Hypercontractivity Theorem!Two-Function|)}%
\begin{remark}
    In traditional proofs of the Hypercontractivity Theorem for $\pm 1$ bits, this theorem is proven directly; it's a slightly tricky induction by derivatives (see Exercise~\ref{ex:poly-hc-hc}).  For more general product spaces the same direct induction strategy also works but the notation becomes quite complicated.
\end{remark}

Our remaining task, therefore, is to prove the Hypercontractivity Theorem in the case $n = 1$; in other words, to show that a uniformly random $\pm 1$ bit is $(p,q,\sqrt{(p-1)/(q-1)})$-hypercontractive.  This fact is often called the ``Two-Point Inequality''
                                                    \index{Two-Point Inequality}%
because (for fixed $p$, $q$, and $\rho$) it's just an ``elementary'' inequality about two real variables.
\begin{named}{Two-Point Inequality}
    Let $1 \leq p \leq q \leq \infty$ and let $0 \leq \rho \leq \sqrt{(p-1)/(q-1)}$.  Then $\|\T_\rho f\|_q \leq \|f\|_p$ for any $f \co \bits \to \R$.  Equivalently (for $\rho \neq 1$), a uniformly random bit $\bx \sim \bits$ is $(p,q,\rho)$-hypercontractive; i.e., $\|a + \rho b \bx \|_q \leq \|a+b \bx\|_p$ for all $a,b \in \R$.
\end{named}
\begin{proof}
    As in Section~\ref{sec:hypercontractivity}, our main task will be to prove the inequality for $1 \leq p < q \leq 2$.  Having done this, the $2 \leq p < q \leq \infty$ cases follow from Proposition~\ref{prop:dual-norm}, the $p < 2 < q$ cases follow using the semigroup property of~$\T_\rho$ (Exercise~\ref{ex:hypercon-straddling-2}), and the $p = q$ cases follow from Exercise~\ref{ex:T-contracts} (or continuity).     The proof for $1 \leq p < q \leq 2$ will be very similar to that of Theorem~\ref{thm:hypercon-bit-2} (the $q = 2$ case). As in that proof we may reduce to the case that $\rho = \sqrt{(p-1)/(q-1)}$, $a = 1$, and $b = \eps$ satisfies $|\eps| < 1$. It then suffices to show
    \begin{align}
        \|1 + \rho \eps \bx \|_q^p &\leq \|1+\eps \bx\|_p^p  \nonumber\\
        \iff\quad \left(\half (1+\rho\eps)^q + \half (1-\rho\eps)^q\right)^{p/q} &\leq \half (1+\eps)^p + \half (1-\eps)^p \nonumber\\
        \iff\quad \left(1 + \sum_{k = 1}^\infty \tbinom{q}{2k} \rho^{2k} \eps^{2k}\right)^{p/q} &\leq 1 + \sum_{k = 1}^\infty \tbinom{p}{2k} \eps^{2k}. \label{eqn:hypercon2-proveme}
    \end{align}
    Again we used $|\eps| < 1$ to drop the absolute value signs and justify the Generalized Binomial Theorem. For each of the binomial coefficients on the left in~\eqref{eqn:hypercon2-proveme} we have
    \[
        \tbinom{q}{2k} = \tfrac{q(q-1)(q-2)(q-3)\cdots(q-(2k-2))(q-(2k-1))}{(2k)!} =         \tfrac{q(q-1)(2-q)(3-q)\cdots((2k-2)-q)((2k-1)-q)}{(2k)!} \geq 0.
    \]
    (Here we reversed an even number of signs, since $1 \leq q \leq 2$.  We will later do the same when expanding $\tbinom{p}{2k}$.) Thus we can again employ the inequality $(1+t)^\theta \leq 1+\theta t$ for $t \geq 0$ and $0 \leq \theta \leq 1$ to deduce that the left-hand side of~\eqref{eqn:hypercon2-proveme} is at most
    \[
        1 + \sum_{k = 1}^\infty \tfrac{p}{q}\tbinom{q}{2k} \rho^{2k} \eps^{2k} =  1 + \sum_{k = 1}^\infty \tfrac{p}{q}\left(\tfrac{p-1}{q-1}\right)^{k}\tbinom{q}{2k} \eps^{2k}.
    \]
    We can now complete the proof of~\eqref{eqn:hypercon2-proveme} by showing the following term-by-term inequality: for all $k \geq 1$,
    \begin{align*}
        \tfrac{p}{q}\left(\tfrac{p-1}{q-1}\right)^{k}\tbinom{q}{2k} &\leq \tbinom{p}{2k} \\
        \iff\quad   \tfrac{p}{q} \left(\tfrac{p-1}{q-1}\right)^{k} \tfrac{q(q-1)(2-q)\cdots((2k-1)-q)}{(2k)!}  &\leq \tfrac{p(p-1)(2-p)\cdots((2k-1)-p)}{(2k)!} \\
        \iff\quad   \tfrac{2-q}{\sqrt{q-1}} \cdot \tfrac{3-q}{\sqrt{q-1}} \cdots \tfrac{(2k-1) -q}{\sqrt{q-1}} &\leq \tfrac{2-p}{\sqrt{p-1}} \cdot \tfrac{3-p}{\sqrt{p-1}} \cdots \tfrac{(2k-1) -p}{\sqrt{p-1}}.
    \end{align*}
    And indeed this inequality holds factor-by-factor. This is because $p < q$ and $\frac{j-r}{\sqrt{r-1}}$ is a decreasing function of $r \geq 1$ for all $j \geq 2$, as is evident from $\frac{d}{dr} \frac{j-r}{\sqrt{r-1}} = -\frac{j-2 + r}{2(r-1)^{3/2}}$.
\end{proof}

\begin{remark}
    The upper-bound $\rho \leq \sqrt{(p-1)/(q-1)}$ in this theorem is best possible; see Exercise~\ref{ex:hc-basics1}\ref{ex:hc-max}.
\end{remark}
                                                    \index{Hypercontractivity Theorem|)}%

\section{Hypercontractivity of general random variables}                                \label{sec:general-hypercon}
                                                    \index{hypercontractivity|(}%
Let's now study hypercontractivity for general random variables. By the end of this section we will have proved the General Hypercontractivity Theorem stated at the beginning of the chapter.

Recall Definition~\ref{def:hypercon} which says that $\bX$ is $(p,q,\rho)$-hypercontractive if $\E[|\bX|^q] < \infty$ and
\[
    \|a + \rho b \bX\|_q \leq \|a + b \bX\|_p \quad \text{for all constants } a, b \in \R.
\]
(By homogeneity, it's sufficient to check this either with~$a$ fixed to~$1$ or with~$b$ fixed to~$1$.)
Let's also collect  some additional basic facts regarding the concept:
\begin{fact}                                        \label{fact:hc-basics}
    Suppose $\bX$ is $(p,q,\rho)$-hypercontractive ($1 \leq p \leq q \leq \infty$, $0 \leq \rho < 1$).  Then:
    \begin{enumerate}
        \item \label{fact:hc0} $\E[\bX] = 0$ (Exercise~\ref{ex:hc-basics1}).
        \item \label{fact:hc1} $c\bX$ is $(p,q,\rho)$-hypercontractive for any $c \in \R$ (Exercise~\ref{ex:hc-basics0}).
        \item $\bX$ is $(p,q,\rho')$-hypercontractive for any $0 \leq \rho' < \rho$ (Exercise~\ref{ex:hc-basics2}).
        \item \label{fact:hc-rho-bound} $\rho \leq  \sqrt{\frac{p-1}{q-1}}$ and $\rho \leq \frac{\|\bX\|_p}{\|\bX\|_q}$ (Exercises~\ref{ex:hc-basics1},~\ref{ex:hc-basics0}).
    \end{enumerate}
\end{fact}
\begin{proposition}                                     \label{prop:hc-holder}
    Let $\bX$ be $(2,q,\rho)$-hypercontractive.  Then $\bX$ is also $(q',2,\rho)$-hypercontractive, where $q'$ is the conjugate \Holder index of~$q$.
\end{proposition}
\begin{proof}
    The deduction is essentially the same as~\eqref{eqn:hc-holder} from Chapter~\ref{sec:sse-intro}. Since $\E[\bX] = 0$ (Fact~\ref{fact:hc-basics}\eqref{fact:hc0}) we have
    \[
        \|a + \rho b\bX\|_2^2 = \E[a^2 + 2\rho ab \bX + \rho^2 b^2 \bX^2] = \E[(a+b\bX)(a+\rho^2 b\bX)].
    \]
    By \Holder's inequality and then the $(2,q,\rho)$-hypercontractivity of~$\bX$ this is at most
    \[
        \|a+b\bX\|_{q'} \|a+\rho^2b\bX\|_q \leq   \|a+b\bX\|_{q'} \|a+\rho b\bX\|_2.
    \]
    Dividing through by $\|a+\rho b\bX\|_2$ (which can't be~$0$ unless $\bX \equiv 0$) gives $\|a+\rho b \bX\|_2 \leq \|a+b\bX\|_{q'}$ as needed.
\end{proof}
\begin{remark}
    The converse does not hold; see Exercise~\ref{ex:hc-holder}.
\end{remark}
\begin{remark}
    As mentioned in Proposition~\ref{prop:sum-hc-hc}, the sum of independent hypercontractive random variables is equally hypercontractive.  Furthermore, low-degree polynomials of independent hypercontractive random variables are ``reasonable''.  See Exercises~\ref{ex:sum-hc-hc} and~\ref{ex:poly-hc-hc}.
\end{remark}

Given $\bX$, $p$, and $q$, computing the largest~$\rho$ for which $\bX$ is $(p,q,\rho)$-hypercontractive
can often be quite a chore.  However, if you're not overly concerned about constant factors then things become much easier.  Let's focus on the most useful case, $p = 2$ and $q > 2$.  By Fact~\ref{fact:hc-basics}\eqref{fact:hc1} we may assume  $\|\bX\|_2 = 1$.  Then we can ask:
\begin{question} \label{ques:hc} Let $\E[\bX] = 0$, $\|\bX\|_2 = 1$, and assume $\|\bX\|_q < \infty$.  For what $\rho$ is $\bX$ $(2,q,\rho)$-hypercontractive?
\end{question}
In this section we'll answer the question by showing that $\rho =  \Theta_q(1/\|\bX\|_q)$ is sufficient.  By the second part of Fact~\ref{fact:hc-basics}\eqref{fact:hc-rho-bound}, $\rho \leq 1/\|\bX\|_q$ is also necessary.  So for a mean-zero random variable~$\bX$, the largest~$\rho$ for which $\bX$ is $(2,q,\rho)$-hypercontractive is always within a constant (depending only on~$q$) of $\frac{\|\bX\|_2}{\|\bX\|_q}$.

Let's arrive at this result in steps, introducing the useful techniques of \emph{symmetrization} and \emph{randomization} along the way.
                                         \index{randomization/symmetrization|(}%
When studying hypercontractivity of a random variable~$\bX$, things are much more convenient if $\bX$ is a \emph{symmetric} random variable, meaning $-\bX$ has the same distribution as~$\bX$.  One advantage of symmetric random variables~$\bX$ is that they have $\E[\bX^k] = 0$ for all odd $k \in \N$.
                                         \index{symmetric random variable}%
Using this it is easy to prove (Exercise~\ref{ex:reasonable-to-hc-symm}) the following fact, similar to Corollary~\ref{cor:bonami-lemma}. (The proof similar to that of Proposition~\ref{prop:2-6-hypercon}.)
\begin{proposition}                                     \label{prop:reasonable-to-hc-symm}
    Let $\bX$ be a symmetric random variable with $\|\bX\|_2 = 1$.  Assume that $\|\bX\|_4 = C$ (and hence $\bX$ is ``$C^4$-reasonable'').
                                             \index{reasonable random variable}%
    Then $\bX$ is $(2,4,\rho)$-hypercontractive if and only if $\rho \leq \min(\frac{1}{\sqrt{3}}, \frac{1}{C})$.
\end{proposition}

Given a symmetric random variable~$\bX$, the \emph{randomization} trick is to replace~$\bX$ by the identically distributed random variable $\br\bX$, where $\br \sim \bits$ is an independent uniformly random bit.  This trick sometimes lets you reduce a probabilistic statement about~$\bX$ to a related one about~$\br$.
\begin{theorem}                                     \label{thm:symm-mom-to-hc}
    Let $\bX$ be a symmetric random variable with $\|\bX\|_2 = 1$ and let $\|\bX\|_q = C$, where $q > 2$.  Then $\bX$ is $(2, q, \rho)$-hypercontractive for $\rho = \frac{1}{C\sqrt{q-1}}$.
\end{theorem}
\begin{proof}
    Let $\br \sim \bits$ be uniformly random and let  $\wt{\bX}$ denote $\bX/C$.  Then for any $a \in \R$,
    \begin{align*}
        \|a + \rho \bX\|_q^2 & = \|a + \rho \br \bX\|_q^2  \tag{by symmetry of $\bX$} \\
        &= \E_{\bX}\left[ \E_{\br}[|a+\rho \br \bX|^q]\right]^{2/q}\\
        &\leq \E_{\bX}\left[ \E_{\br}[|a+\tfrac{1}{C}\br \bX|^2]^{q/2}\right]^{2/q} \tag{$\br$ is $(2,q,\frac{1}{\sqrt{q-1}})$-hypercontractive} \\
        &= \E_{\bX}[(a^2 + \wt{\bX}^2)^{q/2}]^{2/q} \tag{Parseval}\\
        &= \|a^2 + \wt{\bX}^2\|_{q/2} \tag{norm with respect to $\bX$}\\
        &\leq a^2 + \|\wt{\bX}^2\|_{q/2} \tag{triangle inequality for $\|\cdot\|_{q/2}$}\\
        &= a^2 + \|\wt{\bX}\|_q^2 \\
        &= a^2 + 1 = a^2 + \E[\bX^2] = \|a + \bX\|_2^2,
    \end{align*}
    where the last step also used $\E[\bX] = 0$.
\end{proof}

Next, if $\bX$ is not symmetric then we can use a \emph{symmetrization} trick
to make it so.  One way to do this is to replace~$\bX$ with the symmetric random variable $\bX - \bX'$, where $\bX'$ is an independent copy of~$\bX$.  In general $\bX - \bX'$ has similar properties to~$\bX$.  In particular, if $\E[\bX] =0$ we can compare norms using the following one-sided bound:
\begin{lemma}                                       \label{lem:symmetrization1}
    Let $\bX$ be a random variable satisfying $\E[\bX] = 0$ and $\|\bX\|_q < \infty$, where $q \geq 1$.  Then for any $a \in \R$,
    \[
        \|a + \bX\|_q \leq \|a + \bX - \bX'\|_q,
    \]
    where $\bX'$ denotes an independent copy of~$\bX$.
\end{lemma}
\begin{proof}
    We have
    \[
        \|a + \bX\|_q^q = \E[|a + \bX|^q] = \E[|a + \bX - \E[\bX']|^q],
    \]
    where we used the fact that $\E[\bX' \mid \bX] \equiv 0$. But now
    \[
        \E[|a + \bX - \E[\bX']|^q] =  \E[|\E[a + \bX - \bX']|^q]  \leq \E[|a + \bX - \bX'|^q] = \|a + \bX - \bX'\|_q^q,
    \]
    where we used convexity of $t \mapsto |t|^q$.
\end{proof}
A combination of the randomization and symmetrization tricks is to replace an arbitrary random variable~$\bX$ by $\br \bX$, where $\br \sim \bits$ is an independent uniformly random bit.  This often lets you extend results about symmetric random variables to the case of general mean-zero random variables.  For example, the following hypercontractivity lemma lets us reduce to the case of a symmetric random variable while only ``spending'' a factor of~$\half$:
\begin{lemma}                                       \label{lem:symmetrization-randomization1}
    Let $\bX$ be a random variable satisfying $\E[\bX] = 0$ and $\|\bX\|_q < \infty$, where $q \geq 1$.  Then for any $a \in \R$,
    \[
        \|a + \half \bX\|_q \leq \|a + \br \bX\|_q,
    \]
    where $\br \sim \bits$ is an independent uniformly random bit.
\end{lemma}
\begin{proof}
    Letting $\bX'$ be an independent copy of $\bX$ we have
    \begin{align*}
    \|a + \half \bX\|_q &\leq \|a + \half \bX - \half \bX'\|_q \tag{Lemma~\ref{lem:symmetrization1} applied to $\half\bX$}\\
    &= \|a + \br(\half \bX - \half \bX')\|_q \tag{since $\half \bX - \half \bX'$ is symmetric}\\
    &= \|\half a + \half \br\bX + \half a - \half \br \bX'\|_q \\
    &\leq \|\half a + \half \br\bX\|_q + \|\half a - \half \br\bX'\|_q \tag{triangle inequality for $\|\cdot\|_q$}\\
    &= \|\half a + \half \br\bX\|_q + \|\half a + \half \br\bX' \|_q \tag{$-\br$ distributed as $\br$}\\
    &= \|a + \br\bX\|_q. \tag*{\qedhere}
    \end{align*}
\end{proof}

By employing these randomization/symmetrization techniques we obtain a $(2,q)$-hypercontractivity statement for all mean-zero random variables $\bX$ with $\frac{\|\bX\|_q}{\|\bX\|_2}$ bounded, giving a good answer to Question~\ref{ques:hc}:
\begin{theorem}                                     \label{thm:reasonable-to-hc}
    Let $\bX$ satisfy $\E[\bX] = 0$, $\|\bX\|_2 = 1$, $\|\bX\|_q = C$, where $q > 2$.  Then $\bX$ is $(2, q, \frac{1}{2} \rho)$-hypercontractive for $\rho = \frac{1}{C\sqrt{q-1}}$. (If $\bX$ is symmetric, then the factor of $\half$ may be omitted.)
\end{theorem}
\begin{proof}
    By Lemma~\ref{lem:symmetrization-randomization1} we have
    \[
        \|a + \half \rho \bX\|_q^2 \leq \|a + \rho \br \bX\|_q^2.
    \]
    Since $\br\bX$ is a symmetric random variable satisfying $\|\br\bX\|_2 = 1$, $\|\br\bX\|_q = C$, Theorem~\ref{thm:symm-mom-to-hc} implies
    \[
        \|a + \rho \br \bX\|_q^2 \leq \|a + \br \bX\|_2^2 = a^2 + 1 = \|a + \bX\|_2^2.
    \]
    This completes the proof.
\end{proof}
                                         \index{randomization/symmetrization|)}%

If $\bX$ is a discrete random variable then instead of computing $\frac{\|\bX\|_2}{\|\bX\|_q}$ it can sometimes be convenient to use a bound based on the minimum value of $\bX$'s probability mass function.  The following is a simple generalization of Proposition~\ref{prop:min-prob-implies-reasonable}, whose proof is left for Exercise~\ref{ex:min-prob-implies-q-norm-bound}:
\begin{proposition}                                     \label{prop:min-prob-implies-q-norm-bound}
    Let $\bX$ be a discrete random variable with probability mass function~$\pi$.  Write
    \[
        \lambda = \min(\pi) = \min_{x \in \mathrm{range}(\bX)}\{\Pr[\bX = x]\}.
    \]
    Then for any $q > 2$ we have $\|\bX\|_q \leq (1/\lambda)^{1/2 - 1/q} \cdot \|\bX\|_2$.

    As a consequence of Theorem~\ref{thm:reasonable-to-hc}, if in addition $\E[\bX] = 0$ then $\bX$ \linebreak is $(2, q, \half \rho)$-hypercontractive for $\rho = \frac{1}{\sqrt{q-1}} \cdot \lambda^{1/2 - 1/q}$, and $\bX$ is also $(q',2,\half \rho)$-hypercontractive by Proposition~\ref{prop:hc-holder}.  (If $\bX$ is symmetric then the factor of $\half$ may be omitted.)
\end{proposition}

For each $q > 2$, the value $\rho = \Theta_q(\lambda^{1/2 - 1/q})$ in Proposition~\ref{prop:min-prob-implies-q-norm-bound} has the optimal dependence on~$\lambda$, up to a constant.  In fact, a perfectly sharp version of Proposition~\ref{prop:min-prob-implies-q-norm-bound} is known.  The most important case is when $\bX$ is a $\lambda$-biased bit; more precisely, when $\bX = \phi(\bx_i)$ for $\bx_i \sim \pi_\lambda$ in the notation of Definition~\ref{def:p-biased-phi}. In that case, the below theorem (whose very technical proof is left to Exercises~\ref{ex:wolff}--\ref{ex:lo-wolff}) is due to Lata{\l}a and Oleszkiewicz~\cite{LO00}. The case of general discrete random variables is a reduction to the two-valued case due to Wolff~\cite{Wol07}.
                                            \index{hypercontractivity!biased bits}%
\begin{theorem}                                     \label{thm:discrete-hypercon}
    Let $\bX$ be a mean-zero discrete random variable and let $\lambda < 1/2$ be the least value of its probability mass function, as in Proposition~\ref{prop:min-prob-implies-q-norm-bound}.  Then for $q > 2$ it holds that $\bX$ is $(2,q, \rho)$-hypercontractive and $(q',2,\rho)$-hypercontractive for
    \begin{equation} \label{eqn:LO-bound}
         \rho =                  \sqrt{\frac{\exp(u/q) - \exp(-u/q)}{\exp(u/q')- \exp(-u/q')}} = \sqrt{\frac{\sinh(u/q)}{\sinh(u/q')}},  \text{ with $u$ defined by } \exp(-u) = \tfrac{\lambda}{1-\lambda}.
    \end{equation}
    This value of $\rho$ is optimal, even under the assumption that $\bX$ is two-valued.
\end{theorem}
\begin{remark} \label{rem:LO-bound}
    It's not hard to see that for $\lambda \to 1/2$ (hence $u \to 0$) we get $\rho \to \sqrt{\frac{1/q - (-1/q)}{1/q' - (-1/q')}} = \frac{1}{\sqrt{q-1}}$, consistent with the Two-Point Inequality from Section~\ref{sec:full-hypercon-for-bits}. Also, for $\lambda \to 0$ (hence $u \to \infty$) we get $\rho \sim \sqrt{\frac{\lambda^{-1/q}}{\lambda^{-1/q'}}} = \lambda^{1/2 - 1/q}$, showing that Proposition~\ref{prop:min-prob-implies-q-norm-bound} is sharp up to a $q$-dependent constant.  Exercise~\ref{ex:LO-bound} asks you to investigate the function defining~$\rho$ in~\eqref{eqn:LO-bound} more carefully.  In particular, you'll show that $\rho \geq \frac{1}{\sqrt{q-1}} \cdot \lambda^{1/2 - 1/q}$ holds for all $\lambda$.  Hence we can omit the factor of~$\half$ from the simpler bound in Proposition~\ref{prop:min-prob-implies-q-norm-bound} even for nonsymmetric random variables.
\end{remark}
\begin{corollary}                                       \label{cor:thm:discrete-hypercon}
    Let $(\Omega, \pi)$ be a finite probability space, $|\Omega| \geq 2$, in which every outcome has probability at least~$\lambda$.  Let $f \in L^2(\Omega, \pi)$.  Then for any $q > 2$ and $0 \leq \rho \leq \frac{1}{\sqrt{q-1}} \cdot \lambda^{1/2-1/q}$,
    \[
        \|\T_\rho f\|_q \leq \|f\|_2 \quad\text{and}\quad \|\T_\rho f\|_2 \leq \|f\|_{q'}.
    \]
\end{corollary}
\begin{proof}
    Recalling Chapter~\ref{sec:orthogonal-decomposition}, this follows from the decomposition $f(x) = f^{\emptyset} + f^{=\{1\}}$, under which $\T_\rho f = f^{\emptyset} + \rho f^{=\{1\}}$. Note that for $\bx \sim \pi$ the random variable $f^{=\{1\}}(\bx)$ has mean zero, and the least value of its probability mass function is at least~$\lambda$.
\end{proof}
                        \index{Hypercontractivity Theorem!General}%
\noindent The General Hypercontractivity Theorem stated at the beginning of the chapter now follows by applying the Hypercontractivity Induction Theorem from Section~\ref{sec:full-hypercon-for-bits}.
                                                    \index{hypercontractivity|)}%

\section{Applications of general hypercontractivity}                    \label{sec:hypercon-variants-apps}

In this section we will collect some applications of the General Hypercontractivity Theorem, including generalizations of the facts from Section~\ref{sec:hypercon-apps}.  We begin by bounding the $q$-norms of low-degree functions.  The proof is essentially the same as that of Theorem~\ref{thm:general-bonami}; see Exercise~\ref{ex:general-hypercon-induct}.
\begin{theorem}                                     \label{thm:general-hypercon-induct}
    In the setting of the General Hypercontractivity Theorem, if~$f$ has degree at most $k$, then
    \[
        \|f\|_q \leq (\sqrt{q-1} \cdot \lambda^{1/q-1/2})^k \|f\|_2.
    \]
\end{theorem}

Next we turn to an analogue of Theorem~\ref{thm:p-2-bonami}, getting a relationship between the $2$-norm and the $1$-norm for low-degree functions.  The proof (Exercise~\ref{ex:general-2-1-bound}) needs $(2,q,\rho)$-hypercontractivity with~$q$ tending to~$2$, so to get the most elegant statement requires appealing to the sharp bound from Theorem~\ref{thm:discrete-hypercon}:
\begin{theorem}                                     \label{thm:general-2-1-bound}
    In the setting of the General Hypercontractivity Theorem, if $f$ has degree at most~$k$, then
    \[
        \|f\|_2 \leq c(\lambda)^k \|f\|_1,  \quad \text{where } c(\lambda) = \sqrt{\tfrac{1-\lambda}{\lambda}}^{1/(1-2\lambda)}.
    \]
    We have $c(\lambda) \sim 1/\sqrt{\lambda}$ as  $\lambda \to 0$, $c(\lambda) \to e$ as $\lambda \to \half$, and in general, $c(\lambda) \leq e/\sqrt{2\lambda}$.
\end{theorem}
Just as in Chapter~\ref{sec:hypercon-apps} we obtain (Exercise~\ref{ex:general-one-sided-low-deg-anticonc}) the following as a corollary:
\begin{theorem}                                     \label{thm:general-one-sided-low-deg-anticonc}
    In the setting of the General Hypercontractivity Theorem, if $f$ is a nonconstant function of degree at most~$k$, then
    \[
        \Pr_{\bx \sim \pi\xn}\bigl[f(\bx) > \E[f]\bigr] \geq \tfrac14 (e^2/2\lambda)^{-k} \geq (15/\lambda)^{-k}.
    \]
\end{theorem}

Extending Theorem~\ref{thm:low-deg-conc}, the concentration bound for degree-$k$ functions, is straightforward (see Exercise~\ref{ex:general-low-deg-conc}).  We again get that the probability of exceeding~$t$ standard deviations decays like $\exp(-\Theta(t^{2/k}))$, though the constant in the $\Theta(\cdot)$ is linear in~$\lambda$:
\begin{theorem}                                     \label{thm:general-low-deg-conc}
    In the setting of the General Hypercontractivity Theorem, if $f$ has degree at most~$k$, then for any $t \geq \sqrt{2e/\lambda}^{k}$,
    \[
        \Pr_{\bx \sim \pi\xn}[|f(\bx)| \geq t \|f\|_2] \leq \lambda^k \exp\left(-\tfrac{k}{2e} \lambda t^{2/k}\right).
    \]
\end{theorem}

Next, we give a generalization of the Small-Set Expansion Theorem, the proof being left for Exercise~\ref{ex:sse-general}.
                    \index{Small-Set Expansion Theorem!product space domains}%
\begin{theorem}                                     \label{thm:sse-general}
    Let $(\Omega, \pi)$ be a finite probability space, $|\Omega| \geq 2$, in which every outcome has probability at least~$\lambda$.  Let $A \subseteq \Omega^n$ have ``volume''~$\alpha$; i.e., suppose $\Pr_{\bx \sim \pi\xn}[\bx \in A] = \alpha$.  Let $q \geq 2$. Then for any
    \[
        0 \leq \rho \leq \tfrac{1}{q-1} \cdot \lambda^{1 - 2/q}
    \]
    (or even $\rho$ as large as the square of the quantity in Theorem~\ref{thm:discrete-hypercon}) we have
    \[
        \Stab_{\rho}[1_A] = \Pr_{\substack{\bx \sim \pi\xn \\ \by \sim N_{\rho}(\bx)}}[\bx \in A, \by \in A] \leq \alpha^{2-2/q}.
    \]
\end{theorem}
\noindent Similarly, we can generalize Corollary~\ref{cor:stab-infl-bound}, bounding the stable influence of a coordinate by a power of the usual influence:
                                        \index{stable influence!product space domains}%
\begin{theorem}                         \label{thm:gen-stab-infl-bound}
    In the setting of Theorem~\ref{thm:sse-general}, if $f \co \Omega^n \to \{-1,1\}$, then
    \[
        \rho \Inf_i^{(\rho)}[f] \leq \Inf_i[f]^{2-2/q}.
    \]
    for all $i \in [n]$.  In particular, by selecting $q = 4$ we get
    \begin{equation} \label{eqn:gen-stab-infl-bound4}
        \sum_{S \ni i} (\sqrt{\lambda}/3)^{|S|} \|f^{=S}\|_2^2 \leq \Inf_i[f]^{3/2}.
    \end{equation}
\end{theorem}
\begin{proof}
    Applying the General Hypercontractivity Theorem to $\Lap_i f$ and squaring we get
    \[
        \|\T_{\sqrt{\rho}} \Lap_i f\|^2_2 \leq \|\Lap_i f\|_{q'}^2.
    \]
    By definition, the left-hand side is $\rho \Inf_i^{(\rho)}[f]$.  The right-hand side is \linebreak $(\|\Lap_i f\|_{q'}^{q'})^{2-2/q}$, and $\|\Lap_i f\|_{q'}^{q'} \leq \Inf_i[f]$ by Exercise~\ref{ex:laplacian-any-power-influence}\ref{ex:strong-laplacian-any-power-influence}.
\end{proof}

The KKL Edge-Isoperimetic Theorem in this setting now follows by an almost verbatim repetition of the proof from Chapter~\ref{sec:KKL}.
                                                \index{KKL Theorem!product space domains}
\begin{named}{KKL Isoperimetric Theorem for general product space domains}
    In the setting of the General Hypercontractivity Theorem, suppose $f$ has range $\{-1,1\}$ and is nonconstant. Let $\wt{\Tinf}[f]  = \Tinf[f]/\Var[f] \geq 1$.  Then
    \[
        \MaxInf[f] \geq \tfrac{1}{\wt{\Tinf}[f]^2}\cdot (9/\lambda)^{-\wt{\Tinf}[f]}.
    \]
    As a consequence, $\MaxInf[f] \geq \Omega(\frac{1}{\log(1/\lambda)}) \cdot \Var[f] \cdot \frac{\log n}{n}$.
\end{named}
\begin{proof}
    (Cf.~Exercise~\ref{ex:finish-kkl}.) The proof is essentially identical to the one in Chapter~\ref{sec:KKL}, but using~\eqref{eqn:gen-stab-infl-bound4} from Theorem~\ref{thm:gen-stab-infl-bound}. Summing this inequality over all~$i \in [n]$ yields
    \begin{equation} \label{eqn:gen-KKL}
        \sum_{S \subseteq [n]} |S| (\sqrt{\lambda}/3)^{|S|} \|f^{=S}\|_2^2 \leq \sum_{i=1}^n \Inf_i[f]^{3/2} \leq \MaxInf[f]^{1/2} \cdot \Tinf[f].
    \end{equation}
    On the left-hand side above we will drop the factor of $|S|$ for $|S| > 0$.  We also introduce a set-valued random variable $\bS$ defined by $\Pr[\bS = S] = \|f^{=S}\|_2^2/\Var[f]$ for $S \neq \emptyset$.  Note that $\E[|\bS|] = \wt{\Tinf}[f]$.  Thus
    \[
        \text{LHS}\eqref{eqn:gen-KKL} \geq \Var[f] \cdot \E_{\bS}[(\sqrt{\lambda}/3)^{|\bS|}] \geq \Var[f] \cdot (\sqrt{\lambda}/3)^{\E[|\bS|]} = \Var[f] \cdot (\sqrt{\lambda}/3)^{\wt{\Tinf}[f]},
    \]
    where we used that $s \mapsto (\sqrt{\lambda}/3)^s$ is convex.     The first statement of the theorem now follows after rearrangement. As for the second statement, there is some universal $c > 0$ such that
    \[
        \wt{\Tinf}[f] \leq c \cdot \tfrac{1}{\log(1/\lambda)} \cdot \log n \quad\implies\quad \tfrac{1}{\wt{\Tinf}[f]^2}\cdot (9/\lambda)^{-\wt{\Tinf}[f]} = O(1/\lambda)^{-\wt{\Tinf}[f]} \geq \frac{1}{\sqrt{n}},
    \]
    say, in which case our lower bound for $\MaxInf[f]$ is $\frac{1}{\sqrt{n}} \gg \frac{\log n}{n}$.  On the other hand,
    \[
        \wt{\Tinf}[f] \geq c \cdot \tfrac{1}{\log(1/\lambda)} \cdot \log n \quad\implies\quad \Tinf[f] \geq \Omega(\tfrac{1}{\log(1/\lambda)}) \cdot \Var[f] \cdot \log n,
    \]
    in which case even the average influence of~$f$ is $\Omega(\frac{1}{\log(1/\lambda)}) \cdot \Var[f] \cdot \frac{\log n}{n}$.
\end{proof}
Similarly, essentially no extra work is required to generalize Theorem~\ref{thm:KKL-Friedgut-generalized} and Friedgut's Junta Theorem to general product space domains; see Exercise~\ref{ex:friedgut-product-spaces}.  For example, we have:
                                                    \index{Friedgut's Junta Theorem!product space domains}%
\begin{named}{Friedgut's Junta Theorem for general product space domains}
    In the setting of the General Hypercontractivity Theorem, if $f$ has range $\{-1,1\}$ and $0 < \eps \leq 1$, then $f$ is $\eps$-close to a $(1/\lambda)^{O(\TInf[f]/\eps)}$-junta $h \co \Omega^n \to \bits$ (that is, \linebreak $\Pr_{\bx \sim \pi\xn}[f(\bx) \neq h(\bx)] \leq \eps$).
\end{named}

                                                \index{threshold, sharp|(}%
We conclude this section by establishing ``sharp thresholds'' -- in the sense of Chapter~\ref{sec:p-biased} -- for  monotone transitive-symmetric functions with critical probability in the range $[1/n^{o(1)}, 1-1/n^{o(1)}]$. Let $f \btb$ be a nonconstant monotone function and define the (strictly increasing) curve $F \co [0,1] \to [0,1]$ by $F(p) = \Pr_{\bx \sim \pi_p\xn}[f(\bx) = -1]$.  Recall that the critical probability~$p_c$ is defined to be the value such that $F(p_c) = 1/2$; equivalently, such that $\Var[f^{(p_c)}] = 1$.  Recall also the Margulis--Russo Formula,
                                                    \index{Margulis--Russo Formula}%
which says that
\[
    \frac{d}{dp} F(p) = \frac{1}{\sigma^2} \cdot \Tinf[f^{(p)}],
\]
where
\[
    \sigma^2 = \sigma^2(p) = \Var_{\pi_p}[\bx_i] = 4p(1-p) = \Theta(\min(p, 1-p)).
\]
\begin{remark}                              \label{rem:sigma-p}
    Since we will not be concerned with constant factors, it's helpful in the following discussion to mentally replace $\sigma^2$ with $\min(p, 1-p)$. In fact it's even more helpful to always assume $p \leq 1/2$ and replace $\sigma^2$ with~$p$.
\end{remark}

Now suppose~$f$ is a
                                        \index{transitive-symmetric function}%
                                        \index{graph property}%
transitive-symmetric function, e.g., a graph property.  This means that all of its influences are the same, i.e.,
\[
    \Inf_i[f^{(p)}] = \MaxInf[f^{(p)}] = \frac{1}{n} \Tinf[f^{(p)}]
\]
for all $i \in [n]$.  It thus follows from the  KKL Theorem for general product spaces that
\[
    \Tinf[f^{(p)}] \geq \Omega\bigl(\tfrac{1}{\log(1/\min(p,1-p))}\bigr) \cdot \Var[f^{(p)}] \cdot \log n;
\]
hence
\begin{equation}                            \label{eqn:every-mono-fcn-sharp-thresh}
    \frac{d}{dp} F(p) \geq \Var[f^{(p)}] \cdot \Omega\bigl(\tfrac{1}{\sigma^2 \ln(e/\sigma^2)}\bigr)  \cdot \log n.
\end{equation}
(As mentioned in Remark~\ref{rem:sigma-p}, assuming $p \leq 1/2$ you can read $\sigma^2 \ln(e/\sigma^2)$ as $p \log(1/p)$.)

If we take $p = p_c$ in inequality~\eqref{eqn:every-mono-fcn-sharp-thresh} we conclude that $F(p)$ has a large derivative at its critical probability: $F'(p_c) \geq \Omega(\tfrac{1}{p_c \log(1/p_c)}) \cdot \log n$, assuming $p_c \leq 1/2$.   In particular if $\log(1/p_c) \ll \log n$ -- that is, $p_c > 1/n^{o(1)}$ -- then $F'(p_c) = \omega(\tfrac{1}{p_c})$.  This suggests that~$f$ has a ``sharp threshold''; i.e., $F(p)$ jumps from near~$0$ to near~$1$ in an interval of the form $p_c(1\pm o(1))$.  However, largeness of $F'(p_c)$ is not quite enough to establish a sharp threshold (see Exercise~\ref{ex:sharp-thresh-countereg}); we need to have~$F'(p)$ large \emph{throughout} the range of~$p$ near~$p_c$ where $\Var[f^{(p)}]$ is large.  Happily, inequality~\eqref{eqn:every-mono-fcn-sharp-thresh} provides precisely this.
\begin{remark}
    Even if we are only concerned about monotone functions $f$ with $p_c = 1/2$, we still need the KKL Theorem for general product spaces to establish a sharp threshold.  Though $F'(1/2) \geq \Omega(\log n)$ can be derived using just the uniform-distribution KKL Theorem from Chapter~\ref{sec:KKL}, we also need to know that $F'(p) \geq \Omega(\log n)$ continues to hold for~$p = 1/2 \pm O(1/\log n)$.
\end{remark}
Making the above ideas precise, we can establish the following result of Friedgut and Kalai~\cite{FK96b} (cf.~Exercises~\ref{ex:precise-sharp-thresh}, \ref{ex:somewhat-sharp-thresh}):
\begin{theorem}                                     \label{thm:every-mono-fcn-sharp-thresh}
    Let $f \btb$ be a nonconstant, monotone, transitive-symmetric function and let $F \co [0,1] \to [0,1]$ be the strictly increasing function defined by $F(p) = \Pr_{\bx \sim \pi_p\xn}[f(\bx) = -1]$.  Let $p_c$ be the critical probability such that $F(p_c) = 1/2$ and assume without loss of generality that $p_c \leq 1/2$. Fix $0 < \eps < 1/4$ and let
    \[
        \eta = B \log(1/\eps) \cdot \frac{\log(1/p_c)}{\log n},
    \]
    where $B > 0$ is a certain universal constant.    Then assuming $\eta \leq 1/2$,
    \[
        F(p_c\cdot (1-\eta)) \leq \eps, \qquad F(p_c\cdot(1+\eta)) \geq 1 - \eps.
    \]
\end{theorem}
\begin{proof}
    Let $p$ be in the range $p_c\cdot (1\pm\eta)$. By the assumption $\eta \leq 1/2$ we also have $\tfrac12 p_c \leq p \leq \tfrac32 p_c \leq \tfrac34$.   It follows that the quantity $\sigma^2\ln(e/\sigma^2)$ in the KKL corollary~\eqref{eqn:every-mono-fcn-sharp-thresh} is within a universal constant factor of $p_c\log(1/p_c)$.  Thus for all $p$ in the range $p_c\cdot (1\pm\eta)$ we obtain
    \[
        F'(p) \geq \Var[f^{(p)}] \cdot \Omega\bigl(\tfrac{1}{p_c\log(1/p_c)}\bigr) \cdot \log n.
    \]
    Using $\Var[f^{(p)}] = 4 F(p)(1-F(p))$, the definition of~$\eta$, and a suitable choice of~$B$, this is equivalent to
    \begin{equation}            \label{eqn:fk-ineq}
            F'(p) \geq   \frac{2\ln(1/2\eps)}{\eta p_c} F(p)(1-F(p)).
    \end{equation}
    We now show that~\eqref{eqn:fk-ineq} implies that $F(p_c-\eta p_c) \leq \eps$ and leave the implication $F(p_c +\eta p_c) \geq 1-\eps$ to Exercise~\ref{ex:other-dir-every-mono-sharp}.  For $p \leq p_c$ we have $1-F(p) \geq 1/2$ and hence
    \[
        F'(p) \geq   \frac{\ln(1/2\eps)}{\eta p_c} F(p)  \quad\implies\quad
    \frac{d}{dp} \ln F(p) = \frac{F'(p)}{F(p)} \geq \frac{\ln(1/2\eps)}{\eta p_c}.
    \]
    It follows that
    \[
        \ln F(p_c - \eta p_c) \leq \ln F(p_c) - \ln(1/2\eps) = \ln(1/2) - \ln(1/2\eps) = \ln \eps;
    \]
    i.e., $F(p_c-\eta p_c) \leq \eps$ as claimed.
\end{proof}
This proof establishes that every monotone  transitive-symmetric function with critical probability at least~$1/n^{o(1)}$ (and at most $1 - 1/n^{o(1)}$) has a sharp threshold.  Unfortunately, the restriction on the critical probability can't be removed.  The simplest example illustrating this is the logical OR function $\OR_n \co \{\True, \False\}^{n} \to \{\True, \False\}$ (equivalently, the graph property of containing an edge), which has  critical probability~$p_c \sim \frac{\ln 2}{n}$.  Even though $\OR_n$ is transitive-symmetric, it has constant total influence at its critical probability, $\Tinf[\OR_n^{(p_c)}] \sim 2\ln2$. Indeed, $\OR_n$ doesn't have a sharp threshold; i.e., it's not true that $\Pr_{\pi_p}[\OR_n(\bx) = \True] = 1-o(1)$ for $p = p_c(1+o(1))$. For example, if $\bx$ is drawn from the $(2p_c)$-biased distribution we still just have $\Pr[\OR_n(\bx) = \True] \approx 3/4$.  On the other hand, most ``interesting'' monotone transitive-symmetric functions \emph{do} have a sharp threshold; in Section~\ref{sec:friedgut-bourgain-hatami} we'll derive a more sophisticated method for establishing this.
                                                \index{threshold, sharp|)}%

\section{More on randomization/symmetrization}   \label{sec:rand-symm}
In Section~\ref{sec:hypercon-variants-apps} we collected a number of consequences of the General Hypercontractivity Theorem for functions $f \in L^2(\Omega^n, \pi\xn)$.  All of these had a dependence on ``$\lambda$'', the least probability of an outcome under~$\pi$. This can sometimes be quite expensive; for example, the KKL Theorem and  its consequence Theorem~\ref{thm:every-mono-fcn-sharp-thresh} are trivialized when $\lambda = 1/n^{\Theta(1)}$.

                                         \index{randomization/symmetrization|(}%
However, as mentioned in Section~\ref{sec:general-hypercon}, when working with \emph{symmetric} random variables~$\bX$,  the ``randomization'' trick sometimes lets us reduce to the analysis of uniformly random $\pm 1$ bits (which have $\lambda = 1/2$).  Further, Lemma~\ref{lem:symmetrization-randomization1} suggests a way of ``symmetrizing'' general mean-zero random variables (at least if we don't mind applying $\T_{\frac12}$).  In this section we will develop the randomization/symmetrization technique more thoroughly and see an application: bounding the $L^p \to L^p$ norm of the ``low-degree projection'' operator.

Informally, applying the randomization/symmetrization technique to $f \in L^2(\Omega^n,\pi\xn)$ means introducing~$n$ independent uniformly random bits $\br = (\br_1, \dots, \br_n) \sim \bn$ and then ``multiplying the $i$th input to~$f$ by~$\br_i$''.  Of course $\Omega$ is just an abstract set so this doesn't quite make sense. What we really mean is ``multiplying $\Lap_i f$, the $i$th part of $f$'s Fourier expansion (orthogonal decomposition), by~$\br_i$''.  Let's see some examples:
\begin{example}                 \label{eg:rand-symm1}
    Let $f \btR$ be a usual Boolean function with Fourier expansion
    \[
        f(x) = \sum_{S \subseteq [n]} \wh{f}(S)\prod_{i\in S} x_i.
    \]
    Its randomization/symmetrization will be the
                                            \nomenclature[f~]{$\randomize{f}$}{the randomization/symmetrization of~$f$, defined by $\randomize{f}(r,x) = \sum_{S} \br^{S} f^{=S}(\bx)$}%
    function
    \[
        \randomize{f}(r,x) = \sum_{S \subseteq [n]} \wh{f}(S) \prod_{i \in S}r_ix_i = \sum_{S \subseteq [n]} \wh{f}(S)\,x^Sr^S.
    \]
    The key observation is that for random inputs~$\bx, \br \sim \bn$, the random variables $f(\bx)$ and $\randomize{f}(\br,\bx)$ are \emph{identically distributed}. This is simply because $\bx_i$ is a symmetric random variable, so it has the same distribution as $\br_i \bx_i$.
\end{example}
\begin{example}                 \label{eg:rand-symm2}
    Let's return to Examples~\ref{eg:3-ary-basis} and~\ref{eg:and2-3ary} from Chapter~\ref{sec:product-spaces}.  Here we had $\Omega = \{a,b,c\}$ with $\pi$ the uniform distribution, and we defined a certain Fourier basis $\{\phi_0 \equiv 1, \phi_1, \phi_2\}$.  A typical $f \co \Omega^3 \to \R$ here might look like
    \begin{align*}
        f(\bx_1,\bx_2,\bx_3) = \tfrac13 &- \tfrac14 \cdot\phi_1(\bx_1) + \tfrac32 \cdot\phi_{2}(\bx_1) +  \cdot\phi_1(\bx_2) + \tfrac12 \cdot\phi_{2}(\bx_2) - \tfrac23 \cdot\phi_2(\bx_3) \\
        &{}+\tfrac{1}{6}\cdot\phi_1(\bx_1)\cdot\phi_2(\bx_3) + \tfrac18 \cdot\phi_1(\bx_2)\cdot\phi_1(\bx_3)  \\
        &{}-\tfrac{1}{10} \cdot\phi_1(\bx_1)\cdot\phi_2(\bx_2)\cdot\phi_3(\bx_3) + \tfrac{1}{5} \cdot\phi_2(\bx_1)\cdot\phi_2(\bx_2)\cdot\phi_2(\bx_3).
    \end{align*}
    The randomization/symmetrization of this function would be the following function $\randomize{f} \in L^2(\bits^3 \times \Omega^3, \unif^{\otimes 3} \otimes \pi^{\otimes 3})$:
    \begin{align*}
        \randomize{f}(\br,\bx) = \tfrac13 &- \tfrac14 \phi_1(\bx_1)\cdot\br_1 + \tfrac32 \phi_{2}(\bx_1)\cdot\br_1 +  \phi_1(\bx_2)\cdot\br_2 + \tfrac12 \phi_{2}(\bx_2)\cdot\br_2 - \tfrac23 \phi_2(\bx_3)\cdot\br_3 \\
        &{}+\tfrac{1}{6}\phi_1(\bx_1)\cdot\phi_2(\bx_3)\cdot\br_1\br_3 + \tfrac18 \phi_1(\bx_2)\cdot\phi_1(\bx_3)\cdot\br_2\br_3  \\
        &{}-\tfrac{1}{10} \phi_1(\bx_1)\cdot\phi_2(\bx_2)\cdot\phi_3(\bx_3)\cdot\br_1\br_2\br_3 + \tfrac{1}{5} \phi_2(\bx_1)\cdot\phi_2(\bx_2)\cdot\phi_2(\bx_3)\cdot\br_1\br_2\br_3.
    \end{align*}
    There's no obvious way to compare the distributions of $f(\bx)$ and $\randomize{f}(\br,\bx)$.  However, looking carefully at Example~\ref{eg:3-ary-basis} we see that the basis function $\phi_2$ has the property that $\phi_2(\bx_i)$ is a symmetric real random variable when $\bx_i \sim \pi$.  In particular, $\br_i\cdot\phi_2(\bx_i)$ has the same distribution as $\phi_2(\bx_i)$.  Therefore if $g \in L^2(\Omega^n,\pi\xn)$ has the lucky property that its Fourier expansion happens to only use~$\phi_2$ and never uses~$\phi_1$, then we \emph{do} have that $g(\bx)$ and $\randomize{g}(\br,\bx)$ are identically distributed.
\end{example}

Let's give a formal definition of randomization/symmetrization.
\begin{definition}
    Let $f \in L^2(\Omega^n, \pi\xn)$.  The \emph{randomization/symmetrization} of~$f$ is the function $\randomize{f} \in L^2(\bits^n \times \Omega^n, \unif\xn \otimes \pi\xn)$ defined by
    \begin{equation}    \label{eqn:randsymm-def}
        \randomize{f}(r,x) = \sum_{S \subseteq [n]} r^{S} f^{=S}(x),
    \end{equation}
    where we recall the notation $r^S = \prod_{i \in S} r_i$.
\end{definition}
\begin{remark}
    Another way of defining $\randomize{f}$ is to stipulate that for each $x \in \Omega^n$, the function $\restr{\randomize{f}}{}{x} \btR$ is defined to be the Boolean function whose Fourier coefficient on~$S$ is $f^{=S}(x)$.  (This is more evident from~\eqref{eqn:randsymm-def} if you swap the positions of $r^{S}$ and $f^{=S}(x)$.)
\end{remark}
In light of this remark, the basic Parseval formula for Boolean functions implies that for all $x \in \Omega^n$,
\[
    \|\restr{\randomize{f}}{}{x}\|_{2,\br}^2 = \sum_{S \subseteq [n]} f^{=S}(x)^2.
\]
(The notation $\|\cdot\|_{2,\br}$ emphasizes that the norm is computed with respect to the random inputs~$\br$.)
If we take the expectation of the above over $\bx \sim \pi\xn$, the left-hand side becomes $\|\randomize{f}\|_{2,\br,\bx}^2$ and the right-hand side becomes $\|f\|_{2,\bx}^2$, by Parseval's formula for $L^2(\Omega^n,\pi\xn)$.  Thus:
\begin{proposition}                                     \label{prop:randomized-same-2-norm}
    Let $f \in L^2(\Omega^n,\pi\xn)$.  Then $\|\randomize{f}\|_2 = \|f\|_2$.
\end{proposition}
Thus randomization/symmetrization doesn't change $2$-norms.  What about $q$-norms for $q \neq 2$?  As discussed in Examples~\ref{eg:rand-symm1} and~\ref{eg:rand-symm2}, there may be cases where~$f$'s Fourier expansion is already symmetric; in such cases $\randomize{f}(\br,\bx)$ and $f(\bx)$ will have identical distributions, so their $q$-norms will be identical.  The essential feature of the randomization/symmetrization technique is that even for general~$f$ the $q$-norms don't change much -- if you are willing to apply $\T_\rho$ for some constant~$\rho$:
\begin{theorem}                                     \label{thm:rand-symm}
    For $f \in L^2(\Omega^n,\pi\xn)$ and $q > 1$,
    \begin{equation} \label{eqn:rand-symm-thm1}
        \|\T_{\frac12} f\|_q \leq \|\randomize{f}\|_q \leq \|\T_{c_q^{-1}} f\|_q.
    \end{equation}
    Equivalently,
\[
        \|\randomize{\T_{c_q} f}\|_q \leq \|f\|_q \leq \|\randomize{\T_2 f}\|_q.
\]
    Here $0 < c_q \leq 1$ is a constant depending only on~$q$; in particular, we may take $c_4 = c_{4/3} = \frac25$.
\end{theorem}
The two inequalities in~\eqref{eqn:rand-symm-thm1} are not too difficult to prove; for example, you might already correctly guess that the left-hand inequality follows from our first randomization/symmetrization Lemma~\ref{lem:symmetrization-randomization1} and an induction.  We'll give the proofs at the end of this section. But first, let's illustrate how you might use them by solving the following basic problem concerning low-degree projections:
                                        \index{projection!low-degree|(}%
                                        \index{low-degree projection|seeonly{projection, low-degree}}%
\begin{question} \label{ques:low-degree-proj-norm} Let $k \in \N$, let $1 < q < \infty$, and let $f \in L^2(\Omega^n, \pi\xn)$.  Can $\|f^{\leq k}\|_q$ be much larger than $\|f\|_q$?  To put the question in reverse, suppose $g \in L^2(\Omega^n, \pi\xn)$ has degree at most~$k$; is it possible to make the $q$-norm of~$g$ much smaller by adding terms of degree exceeding~$k$ to its Fourier expansion?
\end{question}
The question has a simple answer if $q = 2$: in this case we have $\|f^{\leq k}\|_2 \leq \|f\|_2$ always.  This follows from Paresval:
\begin{equation} \label{eqn:norm-proj2}
    \|f^{\leq k}\|_2^2 = \sum_{j = 0}^k \W{j}[f] \leq \sum_{j = 0}^n \W{j}[f] = \|f\|_2^2.
\end{equation}
When $q \neq 2$ things are not so simple, so let's first consider the most familiar setting of $\Omega = \bits$, $\pi = \unif$.  In this case we can relate the $q$-norm and the $2$-norm via the Hypercontractivity Theorem:
\begin{proposition}                                     \label{prop:projection-bound-uniform}
    Let $k \in \N$ and let $g \btR$.  Then for $q \geq 2$ we have  $\|g^{\leq k}\|_q \leq \sqrt{q-1}^k \|g\|_q$ and for $1 < q \leq 2$ we have  $\|g^{\leq k}\|_q \leq (1/\sqrt{q-1})^k \|g\|_q$.
\end{proposition}
This proposition is an easy consequence of the Hypercontractivity Theorem and already appeared as Exercise~\ref{ex:projection-bounded}.  The simplest case, $q = 4$, follows from the Bonami Lemma alone:
\begin{equation} \label{eqn:low-deg-proj-4-bounded}
    \|g^{\leq k}\|_4 \leq \sqrt{3}^k \|g^{\leq k}\|_2 \leq \sqrt{3}^k \|g\|_2 \leq \sqrt{3}^k \|g\|_4.
\end{equation}

Now let's consider functions $f \in L^2(\Omega^n, \pi\xn)$ on general product spaces; for simplicity, we'll continue to focus on the case $q = 4$.  One possibility is to repeat the above proof using the General Hypercontractivity Theorem (more specifically, Theorem~\ref{thm:general-hypercon-induct}). This would give us $\|f^{\leq k}\|_4 \leq \sqrt{3/\lambda}^k \|f\|_4$.  However, we will see that it's possible to get a bound completely independent of~$\lambda$ -- i.e., independent of $(\Omega, \pi)$ -- using randomization/symmetrization.

First, suppose we are in the lucky case described in Example~\ref{eg:rand-symm2} in which $f$'s Fourier spectrum only uses symmetric basis functions.  In this case $f^{\leq k}(\bx)$ and $\randomize{f^{\leq k}}(\br,\bx)$ have the same distribution for any~$k$, and we can leverage the $L^2(\bits)$ bound~\eqref{eqn:low-deg-proj-4-bounded} to get the same result for~$f$. First,
\[
    \|f^{\leq k}\|_4 = \|\randomize{f^{\leq k}}\|_4 = \left\|\ \|\restr{\randomize{f^{\leq k}}}{}{\bx}(\br)\|_{4, \br}\ \right\|_{4, \bx}.
\]
For each outcome $\bx = x$, the inner function $g(r) = \restr{\randomize{f^{\leq k}}}{}{x}(r)$ is a degree-$k$ function of $r \in \bn$.  Therefore we can apply~\eqref{eqn:low-deg-proj-4-bounded} with this~$g$ to deduce
\[
    \left\|\ \|\restr{\randomize{f^{\leq k}}}{}{\bx}(\br)\|_{4, \br}\ \right\|_{4, \bx} \leq
    \left\|\ \sqrt{3}^k \|\restr{\randomize{f}}{}{\bx}(\br)\|_{4, \br}\ \right\|_{4, \bx} = \sqrt{3}^k \|\randomize{f}\|_4 = \sqrt{3}^k \|f\|_4.
\]
Thus we see that we can deduce~\eqref{eqn:low-deg-proj-4-bounded} ``automatically'' for these luckily symmetric~$f$, with no dependence on~``$\lambda$''. We'll now show that we can get something similar for a completely general~$f$ using the randomization/symmetrization Theorem~\ref{thm:rand-symm}.  This will cause us to lose a factor of $(2\cdot\tfrac52)^k$, due to application of $\T_2$ and $\T_{\frac52}$; to prepare for this, we first extend the calculation in~\eqref{eqn:low-deg-proj-4-bounded} slightly.
\begin{lemma}                                       \label{lem:low-deg-T-bounded}
    Let $k \in \N$ and let $g \btR$.  Then for any $0 < \rho \leq 1$,
    \[
        \|g^{\leq k}\|_4 \leq (\sqrt{3}/\rho)^k \|\T_\rho g\|_4.
    \]
\end{lemma}
\begin{proof}
    We have
    \[
        \|g^{\leq k}\|_4 \leq \sqrt{3}^k \|g^{\leq k}\|_2 \leq (\sqrt{3}/\rho)^k \|\T_\rho g\|_2 \leq (\sqrt{3}/\rho)^k \|\T_\rho g\|_4.
    \]
    Here the first inequality is Bonami's Lemma and the second is because
    \[
        \|g^{\leq k}\|_2^2 = \sum_{j = 0}^k \W{j}[f] \leq (1/\rho^2)^{k} \sum_{j = 0}^k \rho^{2j} \W{j}[f] \leq (1/\rho^2)^{k} \sum_{j = 0}^n \rho^{2j} \W{j}[f]  =
        (1/\rho^2)^{k} \|\T_\rho g\|_2^2. \qedhere
    \]
\end{proof}
We can now give a good answer to Question~\ref{ques:low-degree-proj-norm}, showing that low-degree projection doesn't substantially increase any $q$-norm:
\begin{theorem}                                     \label{thm:projection-bound-general}
    Let $k \in \N$ and let $f \in L^2(\Omega^n, \pi\xn)$.  Then for $q > 1$ we have $\|f^{\leq k}\|_q \leq C_q^k \|f\|_q$.  Here~$C_q$ is a constant depending only on~$q$; in particular we may take $C_4, C_{4/3} = 5\sqrt{3} \leq 9$.
\end{theorem}
\begin{proof}
    We will give the proof for $q = 4$; the other cases are left for Exercise~\ref{ex:projection-bound-general}.  Using the randomization/symmetrization Theorem~\ref{thm:rand-symm},
    \[
        \|f^{\leq k}\|_4 \leq \|\randomize{\T_2 f^{\leq k}}\|_4  = \left\|\ \|\restr{\randomize{\T_2 f^{\leq k}}}{}{\bx}(\br)\|_{4, \br}\ \right\|_{4,\bx}.
    \]
    For a given outcome $\bx = x$, let's write $g = \restr{\randomize{\T_2 f}}{}{x} \btR$, so that we have $\|g^{\leq k}(\br)\|_4$ on the inside above.  For clarity, we remark that~$g$ is the Boolean function whose Fourier coefficient on~$S$ is $2^{|S|} f^{=S}(x)$.  We apply Lemma~\ref{lem:low-deg-T-bounded} to this~$g$, with $\rho = \frac15$. Note that $\T_\rho g$ is then the Boolean function whose Fourier coefficient on~$S$ is $(\tfrac25)^{|S|} f^{=S}(x)$;  i.e., it is $\restr{\randomize{\T_{\frac25} f}}{}{x}$. Thus we deduce
    \[
        \left\|\ \|\restr{\randomize{\T_2 f^{\leq k}}}{}{\bx}(\br)\|_{4, \br}\ \right\|_{4,\bx}
            \leq \left\|\ (5\sqrt{3})^k\|\restr{\randomize{\T_{\frac15} f}}{}{\bx}(\br)\|_{4, \br}\ \right\|_{4,\bx} = (5\sqrt{3})^k\|\randomize{\T_{\frac25} f}\|_4 \leq (5\sqrt{3})^k\|f\|_4,
    \]
    where the last step is the ``un-randomization/symmetrization'' inequality from Theorem~\ref{thm:rand-symm}.
\end{proof}
                                        \index{projection!low-degree|)}%

The remainder of this section is devoted to the proof of Theorem~\ref{thm:rand-symm}, which lets us compare norms of a function and its randomization/symmetrization.  It will help to view randomization/symmetrization from an operator perspective. To do this, we need to slightly extend our~$\T_\rho$ notation, allowing for ``different noise rates on different coordinates''.
                                        \index{noise operator!applied to individual coordinates}%
                                        \nomenclature[Trhoi]{$\T_\rho^i$}{the operator defined by $\T^i_\rho f(x) = \rho f + (1-\rho) \uE_i f$}%
                                        \nomenclature[Trhor]{$\T_{r}$}{for $r \in \R^n$, denotes the operator defined by $\T^1_{r_1} \T^2_{r_2} \cdots \T^n_{r_n}$}%
\begin{definition}              \label{def:general-T}
    For $i \in [n]$ and $\rho \in \R$, let $\T^i_\rho$ be the operator on $L^2(\Omega^n, \pi\xn)$ defined by
    \begin{equation}    \label{eqn:single-T}
        \T^i_\rho f = \rho f + (1-\rho) \uE_i f = \uE_i f + \rho \Lap_i f = \sum_{S \not \ni i} f^{=S} + \rho \sum_{S \ni i} f^{=S}.
    \end{equation}
    Furthermore, for $r = (r_1, \dots, r_n) \in \R^n$, let $\T_r$ be the operator on $L^2(\Omega^n, \pi\xn)$ defined by $\T_r = \T^1_{r_1} \T^2_{r_2} \cdots \T^n_{r_n}$.  From the third formula in~\eqref{eqn:single-T} we have
    \begin{equation} \label{eqn:single-T2}
        \T_r f = \sum_{S \subseteq [n]} r^S\,f^{=S},
    \end{equation}
    where we use the notation $r^S = \prod_{i \in S} r_i$.  In particular, $\T_{(\rho, \dots, \rho)}$ is the usual $\T_\rho$ operator.
    We remark that when $r \in [0,1]^n$ we have
    \[
        \T_r f(x) = \E_{\by_1 \sim N_{r_1}(x_1), \dots, \by_n \sim N_{r_n}(x_n)}[f(\by_1, \dots, \by_n)].
    \]
\end{definition}
These generalizations of the noise operator behave the way you would expect; you are referred to Exercise~\ref{ex:individual-T-gen} for some basic properties.  Now comparing~\eqref{eqn:single-T2} and~\eqref{eqn:randsymm-def} reveals the connection to randomization/symmetrization:
\begin{fact}                                        \label{fact:rand-symm-T-def}
    For $f \in L^2(\Omega^n, \pi\xn)$, $x \in \Omega^n$, and $r \in \bn$,
    \[
        \randomize{f}(r,x) = \T_r f(x).
    \]
\end{fact}
In other words, randomization/symmetrization of~$f$ means applying \linebreak $\T_{(\pm 1, \pm 1, \dots, \pm 1)}$ to~$f$ for a random choice of signs.  We use this viewpoint to prove Theorem~\ref{thm:rand-symm}, which we do in two steps:
\begin{theorem}                                     \label{thm:randomize-symmetrize}
    Let $f \in L^2(\Omega^n, \pi\xn)$.  Then for any $q \geq 1$,
    \begin{equation} \label{eqn:rand-symm}
        \|\T_{\frac12} f(\bx)\|_{q,\bx} \leq \|\T_{\br} f(\bx)\|_{q, \br, \bx}
    \end{equation}
    for $\bx \sim \pi\xn$, $\br \sim \bn$.  In other words, $\|\T_{\frac12} f\|_q \leq \|\randomize{f}\|_q$.
\end{theorem}
\begin{proof}
    In brief, the result follows from our first randomization/symmetrization result, Lemma~\ref{lem:symmetrization-randomization1}, and an induction.  To fill in the details, we begin by showing that if $h \in L^2(\Omega, \pi)$ is any one-input function and $\bomega \sim \pi$,  $\bb \sim \bits$, then
    \begin{equation} \label{eqn:symm-h}
        \|\T_{\frac12} h(\bomega)\|_{q, \bomega} \leq \|\T_{\bb} h(\bomega)\|_{q, \bb, \bomega}.
    \end{equation}
    This follows immediately from Lemma~\ref{lem:symmetrization-randomization1} because $h^{=\{1\}}(\bx)$ is a mean-zero random variable (cf.~the proof of Corollary~\ref{cor:thm:discrete-hypercon}).     Next, we show that for any $g \in L^2(\Omega^n, \pi\xn)$ and any $i \in [n]$,
    \begin{equation} \label{eqn:symm-g}
        \|\T^i_{\frac12} g(\bx)\|_{q,\bx} \leq \|\T^i_{\br_i} g(\bx)\|_{q, \br_i, \bx}.
    \end{equation}
    Assuming $i = 1$ for notational simplicity, and writing $x = (x_1, x')$ where $x' = (x_2, \dots, x_n)$, we have
    \[
        \|\T^i_{\frac12} g(\bx)\|_{q,\bx} = \left\|\ \|\T^i_{\frac12} g(\bx_1, \bx')\|_{q,\bx_1}\ \right\|_{q,\bx'} = \left\|\ \|(\T_{\frac12} \restr{g}{}{\bx'})(\bx_1)\|_{q,\bx_1}\ \right\|_{q,\bx'}.
    \]
    (You are asked to carefully justify the second equality here in Exercise~\ref{ex:justify-T-partial-restriction-commute}.) Now for each outcome of $\bx'$ we can apply~\eqref{eqn:symm-h} with $h = \restr{g}{}{\bx'}$ to deduce
    \[
        \left\|\ \|(\T_{\frac12} \restr{g}{}{\bx'})(\bx_1)\|_{q,\bx_1}\ \right\|_{q,\bx'} \leq
        \left\|\ \|(\T_{\br_1} \restr{g}{}{\bx'})(\bx_1)\|_{q,\bx_1, \br_1}\ \right\|_{q,\bx'}
        = \|\T^i_{\br_i} g(\bx)\|_{q, \br_i, \bx}.
    \]
    Finally, we illustrate the first step of the induction.  For distinct indices $i,j$,
    \[
        \|\T^i_{\frac12} \T^j_{\frac12} f(\bx)\|_{q,\bx} \leq \|\T^i_{\br_i} \T^j_{\frac12} f(\bx)\|_{q, \br_i, \bx}
    \]
    by applying~\eqref{eqn:symm-g} with $g = \T^j_{\frac12} f$.  Then
    \[
        \|\T^i_{\br_i} \T^j_{\frac12} f(\bx)\|_{q, \br_i, \bx} =
        \left\|\ \|\T^i_{\br_i} \T^j_{\frac12} f(\bx)\|_{q, \bx}\ \right\|_{q, \br_i} =
        \left\|\ \|\T^j_{\frac12} \T^i_{\br_i}  f(\bx)\|_{q, \bx}\ \right\|_{q, \br_i},
    \]
    where we used that $\T^i_{\rho_i}$ and $\T^j_{\rho_j}$ commute. Now for each outcome of~$\br_i$ we can apply~\eqref{eqn:symm-g} with $g = \T^i_{\br_i} f$ to get
    \[
        \left\|\ \|\T^j_{\frac12} \T^i_{\br_i}  f(\bx)\|_{q, \bx}\ \right\|_{q, \br_i} \leq
        \left\|\ \|\T^j_{\br_j} \T^i_{\br_i}  f(\bx)\|_{q, \br_j, \bx}\ \right\|_{q, \br_i} = \|\T^i_{\br_i} \T^j_{\br_j} f(\bx)\|_{q,\br_i,\br_j,\bx}.
    \]
    Thus we have shown
    \[
        \|\T^i_{\frac12} \T^j_{\frac12} f(\bx)\|_{q,\bx} \leq \|\T^i_{\br_i} \T^j_{\br_j} f(\bx)\|_{q,\br_i,\br_j,\bx}.
    \]
    Continuing the induction in the same way completes the proof.
\end{proof}

To prove the ``un-randomization/symmetrization'' inequality in Theorem \ref{thm:rand-symm}, we first establish an elementary lemma about mean-zero random variables:
\begin{lemma}                                       \label{lem:unsymmetrize}
    Let $q \geq 2$.  Then there is a small enough $0 < c_q \leq 1$ such that
    \[
        \|a - c_q \bX\|_q \leq \|a + \bX\|_q
    \]
    for any $a \in \R$ and any random variable $\bX$ satisfying $\E[\bX] = 0$ and $\|\bX\|_q < \infty$.   In particular we may take $c_4 = \frac25$.
\end{lemma}
\begin{proof}
    We will only prove the statement for $q = 4$; you are asked to establish the general case in Exercise~\ref{ex:general-unsymm-lemma}.  By homogeneity we may assume $a = 1$; then raising the inequality to the $4$th power we need to show
    \[
        \E[(1 - c \bX)^4] \leq \E[(1 + \bX)^4]
    \]
    for small enough~$c$.  Expanding both sides and using $\E[\bX] = 0$, this is equivalent to
    \begin{equation} \label{eqn:unsymm}
        \E[(1 - c^4) \bX^4 + (4+4c^3)\bX^3 + (6 - 6c^2)\bX^2] \geq 0.
    \end{equation}
    It suffices to find $c$ such that
    \begin{equation} \label{eqn:unsymm-ex}
        (1 - c^4) x^2 + (4+4c^3)x + (6 - 6c^2) \geq 0 \quad \forall x \in \R;
    \end{equation}
    then we can multiply \eqref{eqn:unsymm-ex} by $x^2$ and take expectations to obtain~\eqref{eqn:unsymm}.  This last problem is elementary, and Exercise~\ref{ex:c-root} asks you to find the largest~$c$ that works (the answer is $c \approx .435$).  To see that $c = \frac25$ suffices, we use the fact that $x \geq -\frac29 x^2 - \frac98$ for all~$x$ (because the difference of the left- and right-hand sides is $\frac{1}{72}(4x+9)^2$). Putting this into~\eqref{eqn:unsymm-ex}, it remains to ensure
    \[
        (\tfrac19-\tfrac89c^3-c^4)x^2 + (\tfrac32 -6c^2-\tfrac92c^3) \geq 0 \quad \forall x \in \R,
    \]
    and when $c = \frac25$ this is the trivially true statement $\frac{161}{5625}x^2+\frac{63}{250} \geq 0$.
\end{proof}
\begin{theorem}                                     \label{thm:unsymmetrize}
    Let $f \in L^2(\Omega^n,\pi\xn)$. Then for any $q > 1$,
    \[
        \|\T_{c_q \br} f(\bx)\|_{q,\br,\bx} \leq \| f(\bx)\|_{q, \bx}
    \]
    for $\bx \sim \pi\xn$, $\br \sim \bn$.  In other words, $\|\randomize{\T_{c_q} f}\|_q \leq \|f\|_q$.  Here $0 < c_q \leq 1$ is a constant depending only on~$q$; in particular we may take $c_4, c_{4/3} = \frac25$.
\end{theorem}
\begin{proof}
    In fact, we can show that for \emph{every} outcome $\br = r \in \bn$ we have
    \[
            \|\T_{c_q r} f(\bx)\|_{q,\bx} \leq \| f(\bx)\|_{q, \bx}
    \]
    for sufficiently small $c_q > 0$.     Note that on the left-hand side we have
    \[
        \|\T^1_{\pm c_q} \T^2_{\pm c_q} \cdots \T^n_{\pm c_q} f(\bx)\|_{q, \bx}.
    \]
    We know that $\T^i_\rho$ is a contraction in~$L^q$ for any $\rho \geq 0$ (Exercise~\ref{ex:individual-T-gen}).  Hence it suffices to show that $\T^i_{-c_q}$ is a contraction in~$L^q$, i.e., that
    \begin{equation}                \label{eqn:unsymm1}
        \|\T^i_{-c_q} g(\bx)\|_{q,\bx} \leq \|g(\bx)\|_{q,\bx}
    \end{equation}
    for all $g \in L^2(\Omega^n, \pi\xn)$.  Similar to the proof of Theorem~\ref{thm:randomize-symmetrize}, it suffices to show
    \begin{equation} \label{eqn:negative-contract}
        \|\T_{-c_q} h\|_q \leq \|h\|_q
    \end{equation}
    for all one-input functions $h \in L^2(\Omega, \pi)$, because then~\eqref{eqn:unsymm1} holds pointwise for all outcomes of $\bx_1, \dots, \bx_{i-1}, \bx_{i+1}, \dots, \bx_n$.  By Proposition~\ref{prop:dual-norm}, if we prove~\eqref{eqn:negative-contract} for some~$q$, then the same constant~$c_q$ works for the conjugate \Holder index~$q'$; thus we may restrict attention to $q \geq 2$.  Now the result follows from Lemma~\ref{lem:unsymmetrize} by taking $a = h^{=\emptyset}$ and $\bX = h^{=\{1\}}(\bx)$.
\end{proof}
                                         \index{randomization/symmetrization|)}%

\section{Highlight: General sharp threshold theorems}   \label{sec:friedgut-bourgain-hatami}

                                                \index{threshold, sharp}%
In Chapter~\ref{sec:p-biased} we described the problem of ``threshold phenomena'' for monotone functions $f \btb$. As~$p$ increases from~$0$ to~$1$, we are interested in whether $\Pr_{\bx \sim \pi_p\xn}[f(\bx) = -1]$ has a ``sharp threshold'', jumping quickly from near~$0$ to near~$1$ around the critical probability $p = p_c$.  The ``sharp threshold principle'' tells us that this occurs (roughly speaking) if and only if the total influence of~$f$ under its critical distribution, $\Tinf[f^{(p_c)}]$, is~$\omega(1)$.  (See Exercise~\ref{ex:precise-sharp-thresh} for more precise statements.)  This motivates finding a characterization of functions with small total influence.  Indeed, finding such a characterization is a perfectly natural question even for not-necessarily-monotone Boolean-valued functions $f \in L^2(\Omega^n,\pi\xn)$.
                                                \index{total influence!product space domains}%

For the usual uniform distribution on $\bn$, Friedgut's Junta Theorem from Chapter~\ref{sec:KKL} provides a very good characterization: $f \btb$ can only have $O(1)$ total influence if it's (close to) an $O(1)$-junta.
                                                    \index{Friedgut's Junta Theorem!product space domains}%
By the version of Friedgut's Junta Theorem for general product spaces (Section~\ref{sec:hypercon-variants-apps}), the same holds for Boolean-valued $f \in L^2(\bn, \pi_p\xn)$ so long as~$p$ is not too close to~$0$ or to~$1$.  However, for $p$ as small as $1/n^{\Theta(1)}$, the ``junta''-size promised by Friedgut's Junta Theorem may be larger than~$n$.  (Cf.~the breakdown of Friedgut and Kalai's sharp threshold result Theorem~\ref{thm:every-mono-fcn-sharp-thresh} for $p \leq 1/n^{\Theta(1)}$.) This is a shame, as many natural graph properties for which we'd like to show a sharp threshold -- e.g., (non-)$3$-colorability -- have $p = 1/n^{\Theta(1)}$. At a technical level, the reason for the breakdown for very small~$p$ is the dependence on the ``$\lambda$'' parameter in the General Hypercontractivity Theorem.  But there's a more fundamental reason for its failure, as suggested by the example at the end of Section~\ref{sec:hypercon-variants-apps}: Friedgut's Junta Theorem simply isn't true for such small~$p$.
\begin{examples}               \label{eg:biased-low-influence}
    Here are some examples of Friedgut's Junta Theorem failing for small~$p$:
    \begin{itemize}
        \item The logical OR function $\OR_n \btb$ has critical probability~$p_c \sim \frac{\ln 2}{n}$,
                                                    \index{OR function}%
            and its total influence at this probability is $\Tinf[\OR_n^{(p_c)}] \sim 2\ln2$, a small constant.  Yet it's easy to see that under the $p_c$-biased distribution, $\OR_n$ is not even, say, $.1$-close to any junta on~$o(n)$ coordinates.  (That is, for every $o(n)$-junta $h$, $\Pr_{\bx \sim \pi_{p_c}\xn}[f(\bx) \neq h(\bx)] > .1$.)
        \item Consider the function~$f \btb$ that is $\true$ ($-1$) if and only if there exists a ``run'' of three consecutive~$-1$'s in its input.  (We allow  runs to ``wrap around'', thus making~$f$ a transitive-symmetric function.) It's not hard to show that the critical probability for this~$f$ satisfies $p_c = \Theta(1/n^{1/3})$.  Furthermore, since~$f$ is a computable by a DNF of width~$3$, Exercise~\ref{ex:p-biased-influence-facts}\ref{ex:p-biased-dnf-width-influence} shows that $\Tinf[f^{(p_c)}] \leq 12$, a small constant.  But again, this~$f$ is not close to any $o(n)$-junta under the $p_c$-biased distribution.  A similar example is $\mathrm{Clique}_3 \co \tf^{\binom{v}{2}} \to \tf$, the graph property of containing a triangle.
    \end{itemize}
\end{examples}
We see from these examples that for $p$ very small, we can't hope to show that low-influence functions are close to juntas.  However, these counterexample functions still have low complexity in a weaker sense -- they are computable by narrow DNFs.  Indeed, Friedgut~\cite{Fri99} suggests this as a characterization:
                                                \index{Friedgut's Conjecture}%
\begin{named}{Friedgut's Conjecture}
    There is a function $w \co \R^+ \times  (0,1) \to \R^+$ such that the following holds: If  $f \co \tf^n \to \tf$ is a monotone function, $0 < p \leq 1/2$, and $\Tinf[f\subp] \leq K$, then $f$ is $\eps$-close     under $\pi_p\xn$ to a monotone DNF of width at most $w(K,\eps)$.
\end{named}
\noindent The assumption of monotonicity is essential in this conjecture; see Exercise~\ref{ex:parity-not-a-dnf}.

Short of proving his conjecture, Friedgut managed to show:
\begin{named}{Friedgut's Sharp Threshold Theorem}
    The above conjecture holds when $f$ is a graph property.
\end{named}
                                                \index{Friegut's Sharp Threshold Theorem}%
                                                \index{graph property!monotone}%
This gives a very good characterization of monotone graph properties with low total influence, one that works no matter how small~$p$ is.  Friedgut also extended his result to monotone hypergraph properties; this was sufficient for him to show that several interesting hypergraph (or hypergraph-like) properties have sharp thresholds -- for example, the property of a random
                                                \index{threshold, sharp}%
$3$-uniform hypergraph containing a perfect matching, or the property of a random width-$3$ DNF formula being a tautology.  (Interestingly, for neither of these properties do we know precisely where the critical probability~$p_c$ is; nevertheless, we know there is a sharp threshold around it.) Roughly speaking one needs to show that at the critical probability, these properties can't be well-approximated by narrow DNFs because they are almost surely not determined just by ``local'' information about the (hyper)graph.  This kind of deduction takes some effort in random graph theory and we won't discuss it further here beyond Exercise~\ref{ex:non-3-colorability}; for a survey, see Friedgut~\cite{Fri05}.

Friedgut's proof is rather long and it relies heavily on the function being a graph or hypergraph property.  Following Friedgut's work, Bourgain~\cite{Bou99} gave a shorter proof of an alternative characterization.  Bourgain's characterization is not as strong as Friedgut's for monotone graph properties; however, it has the advantage that it works for low-influence functions on \emph{any} product probability space. (In particular, there is no monotonicity assumption since the domain need not be~$\tf^n$.)  We first make a quick definition and then state Bourgain's theorem.
\begin{definition}
    Let $f \in L^2(\Omega^n, \pi\xn)$ be $\bits$-valued.  For $T \subseteq [n]$, $y \in \Omega^T$, and $\tau > 0$, we say that the restriction $y_T$ is a \emph{$\tau$-booster} if $f^{\subseteq T}(y) \geq \E[f] + \tau$.  (Recall that $f^{\subseteq T}(y) = \E[\restr{f}{\ol{T}}{y}]$.) In case $\tau < 0$ we say that $y_T$ is a $\tau$-booster if $f^{\subseteq T}(y) \leq \E[f] - |\tau|$.
\end{definition}
                                                \index{Bourgain's Sharp Threshold Theorem|(}%
\begin{named}{Bourgain's Sharp Threshold Theorem}
    Let $f \in L^2(\Omega^n, \pi\xn)$ be $\bits$-valued with $\Tinf[f] \leq K$.  Assume $\Var[f] \geq .01$.  Then there is some $\tau$ (either positive or negative) with $|\tau| \geq \exp(-O(K^2))$ such that
    \[
        \Pr_{\bx \sim \pi\xn}[\exists T \subseteq [n], |T| \leq O(K) \text{ such that $\bx_T$ is a } \tau\text{-booster}] \geq |\tau|.
    \]
\end{named}
\noindent (We emphasize that here and throughout, the constants hidden in the $O(\cdot)$ are absolute and do not depend on~$\Omega$ or~$\pi$.)

Thinking of $K$ as an absolute constant, the above theorem says that for a typical input string~$x$, there is a large chance that it contains a constant-sized substring that is an $\Omega(1)$-booster for~$f$.  In the particular case of monotone $f \in L^2(\tf^n, \pi_p\xn)$ with $p$ small, it's not hard to deduce (Exercise~\ref{ex:weak-bourgain}) that in fact there exists a $T$ with $|T| \leq O(K)$ such that restricting all coordinates in~$T$ to be $\true$ increases $\Pr_{\pi_p\xn}[f = \true]$ by $\exp(-O(K^2))$.  This is a qualitatively weaker conclusion than what you get from Friedgut's Sharp Threshold Theorem when~$f$ is a graph property with $\Tinf[f] \leq O(1)$ -- in that case, by taking~$T$ to be any of the width-$O(1)$ terms in the approximating DNF one can increase $\Pr_{\pi_p\xn}[f = \True]$ not just by~$\Omega(1)$ but up to almost~$1$.  Nevertheless, Bourgain's theorem apparently suffices to deduce any of the sharp thresholds results obtainable from Friedgut's theorem~\cite{Fri05}.  For a very high-level sketch of how Bourgain's theorem would apply in the case of $3$-colorability of random graphs, see Exercise~\ref{ex:non-3-colorability}.

The last part of this section will be devoted to proving Bourgain's Sharp Threshold Theorem.  Before doing this, we add a remark.  Hatami~\cite{Hat12} has significantly generalized Bourgain's work, establishing the following characterization of Boolean-valued functions with low total influence:
                                                \index{Hatami's Theorem}%
\begin{named}{Hatami's Theorem}
    Let $f \in L^2(\Omega^n, \pi\xn)$ be a $\bits$-valued function with $\Tinf[f] \leq K$.  Then for every $\eps > 0$, the function $f$ is $\eps$-close (under $\pi\xn$) to an $\exp(O(K^3/\eps^3))$-``pseudo-junta'' $h \co \Omega^n \to \bits$.
\end{named}
                                                \index{pseudo-junta}%
The term ``pseudo-junta'' is defined in Exercise~\ref{ex:pseudo-junta}.  A $K$-pseudo-junta~$h$ has the property that $\Tinf[h] \leq 4K$; thus Hatami's Theorem shows that having $O(1)$ total influence is essentially equivalent to being an $O(1)$-pseudo-junta.  A downside of the result, however, is that being a $K$-pseudo-junta is not a ``syntactic'' property; it depends on the probability distribution~$\pi\xn$.\\

Let's now turn to proving Bourgain's Sharp Threshold Theorem.  In fact, Bourgain proved the theorem as a corollary of the following main result:
\begin{theorem}                                     \label{thm:bourgain-sharp}
    Let $(\Omega, \pi)$ be a finite probability space
    and let $f \co \Omega^n \to \{-1,1\}$. Let $0 < \eps < 1/2$ and write $k = \Tinf[f]/\eps$.
    Then for each $x \in \Omega^n$ it's possible to define a set of ``notable coordinates'' $J_x \subseteq [n]$ satisfying
    $|J_x| \leq \exp(O(k))$ such that
    \[
        \Ex_{\bx \sim \pi\xn}\left[\sum_{S \not \in \calF_{\bx}} f^{=S}(\bx)^2\right] \leq 2\eps.
    \]
    Here
    $
        \calF_x = \{S : S \subseteq J_x, |S| \leq k\}
    $,
    a collection always satisfying
    $
        |\calF_x| \leq  \exp(O(k^2))
    $.
\end{theorem}
You may notice that this theorem looks extremely similar to Friedgut's Junta Theorem from Chapter~\ref{sec:KKL} (and the $\exp(-O(\Tinf[f]^2))$ quantity in Bourgain's Sharp Threshold Theorem looks similar to the Fourier coefficient lower bound in Corollary~\ref{cor:low-inf-large-coeff}).
                                                \index{Friedgut's Junta Theorem}%
Indeed, the only difference between Theorem~\ref{thm:bourgain-sharp} and Friedgut's Junta Theorem is that in the latter, the ``notable coordinates''~$J$ can be ``named in advance'' -- they're simply the coordinates~$j$ with $\Inf_j[f] = \sum_{S \ni j} \wh{f}(S)^2$ large.  By contrast, in Theorem~\ref{thm:bourgain-sharp} the notable coordinates depend on the input~$x$.  As we will see in the proof, they are precisely the coordinates~$j$ such that $\sum_{S \ni j} f^{=S}(x)^2$ is large.  Of course, in the setting of $f \btb$ we have $f^{=S}(x)^2 = \wh{f}(S)^2$ for all~$x$, so the two definitions coincide.  But in the general setting of $f \in L^2(\Omega^n, \pi\xn)$ it makes sense that we can't name the notable coordinates in advance and rather have to ``wait until~$x$ is chosen''. For example, for the $\OR_n$ function as in Example~\ref{eg:biased-low-influence}, there are no notable coordinates to be named in advance, but once~$x$ is chosen the few coordinates on which $x$ takes the value $\true$ (if any exist) will be the notable ones.

The proof of Theorem~\ref{thm:bourgain-sharp} mainly consists of adding the randomization/symmetrization
                                                \index{randomization/symmetrization}%
technique to the proof of Friedgut's Junta Theorem (more precisely, Theorem~\ref{thm:KKL-Friedgut-generalized}) to avoid dependence on the minimum probability of~$\pi$.
This randomization/symmetrization is applied to what are essentially  the key inequalities in that proof:
\[
    \|\T_{\frac{1}{\sqrt{3}}} \Lap_i f\|_2^2 \leq \|\Lap_i f\|_{4/3}^2 = \|\Lap_i f\|_{4/3}^{2/3} \cdot \|\Lap_i f\|_{4/3}^{4/3} \leq \|\Lap_i f\|_{4/3}^{2/3} \cdot \Inf_i[f].
\]
(The last inequality here is Exercise~\ref{ex:laplacian-any-power-influence}\ref{ex:strong-laplacian-any-power-influence}.)
The overall proof needs one more minor twist: since we work on a ``per-$x$'' basis and not in expectation, it's possible that the set of notable coordinates can be improbably large.  (Think again about the example of $\OR_n$; for $\bx \sim \pi_{1/n}\xn$ we expect only a constant number of coordinates of $\bx$ to be $\true$, but it's not always uniformly bounded.)  This is combated using the principle that low-degree functions are ``reasonable'' (together with randomization/symmetrization).

\begin{proof}[Proof of Theorem~\ref{thm:bourgain-sharp}]
    By the simple ``Markov argument'' (see Proposition~\ref{prop:tinf-concentration}) we have
    \[
        \Ex_{\bx \sim \pi\xn}\left[\sum_{|S| > k} f^{=S}(\bx)^2\right] = \sum_{|S| > k} \|f^{=S}\|_2^2 \leq \Tinf[f] / k = \eps.
    \]
    Thus it suffices to define the sets $J_x$ so that
    \begin{equation}                            \label{eqn:bourgain-final}
        \Ex_{\bx \sim \pi\xn}\left[\sum_{|S| \leq k,\ S \not \subseteq J_{\bx}} f^{=S}(\bx)^2\right] \leq \eps.
    \end{equation}
    We'll first define  ``notable coordinate'' sets $J'_x \subseteq [n]$ which almost do the trick:
    \[
        J'_x = \left\{ j \in [n] : \sum_{S \ni j}  f^{=S}(x)^2 \geq \tau\right\}, \quad \tau = c^{-k}.
    \]
    (where $c > 1$ is a universal constant).    Using this definition, the main effort of the proof will be to show
    \begin{equation}                \label{eqn:bourgain-main-goal}
        \Ex_{\bx \sim \pi\xn}\left[\sum_{|S| \leq k,\ S \not \subseteq J'_{\bx}} f^{=S}(\bx)^2\right] \leq \eps/2.
    \end{equation}
    This looks better than~\eqref{eqn:bourgain-final}; the only problem is that the sets $J'_x$ don't always satisfy $|J'_x| \leq \exp(O(k))$ as needed. However, ``in expectation'' $|J'_x|$ ought not be much larger than~$1/\tau = c^k$.  Thus we introduce the event
    \[
        \text{``$J'_{\bx}$ is too big''} \quad\iff\quad |J'_{\bx}| \geq C^k
    \]
    (where $C > c$ is another universal constant) and  define
    \[
        J_x = \begin{cases}
                  J'_x & \text{if $J'_x$ is not too big,} \\
                  \emptyset & \text{if $J'_x$ is too big.}
              \end{cases}
    \]
    The last part of the proof will be to show that
    \begin{equation}                            \label{eqn:bourgain-too-big}
        \Ex_{\bx \sim \pi\xn}\left[\bone[\text{$J'_{\bx}$ is too big}] \cdot \sum_{0 < |S| \leq k} f^{=S}(\bx)^2\right] \leq \eps/2.
    \end{equation}
    Together,~\eqref{eqn:bourgain-too-big} and~\eqref{eqn:bourgain-main-goal} establish~\eqref{eqn:bourgain-final}.  We will first prove~\eqref{eqn:bourgain-main-goal} and then prove~\eqref{eqn:bourgain-too-big}.  As a small aside, we'll see that for both inequalities we could obtain a bound much less than $\eps/2$ if desired.

    To prove~\eqref{eqn:bourgain-main-goal}, we mimic the proof of Theorem~\ref{thm:KKL-Friedgut-generalized} but add in randomization/symmetrization.  The key step is encapsulated in the following lemma.  Note that the lemma also holds with the more natural definition $g = \Lap_i f$; the additional $\T_{\frac25}$ is to facilitate future ``un-randomization/symmetrization''.
    \begin{lemma}                                       \label{lem:bourgain-thm-lemma}
        Fix $x \in \Omega^n$ and $i \not \in J'_x$.  Then writing $g = \T_{\frac25} \Lap_i f$ we have
        \[
            \|\T_{\frac{1}{\sqrt{3}}} \restr{\randomize{g}}{}{x}\|_2^2 \leq \tau^{1/3} \|\restr{\randomize{g}}{}{x}\|_{4/3}^{4/3}.
        \]
    \end{lemma}
    \begin{proof}
        Here $\randomize{g}$ is the randomization/symmetrization of~$g$, so $\restr{\randomize{g}}{}{x} = \restr{\randomize{g}}{}{x}(r)$ is a function on the uniform-distribution hypercube.  Applying the basic $(4/3,2)$-Hypercontractivity Theorem we have
        \[
            \|\T_{\frac{1}{\sqrt{3}}} \restr{\randomize{g}}{}{x}\|_2^2 \leq
            \|\restr{\randomize{g}}{}{x}\|_{4/3}^2 =
            (\|\restr{\randomize{g}}{}{x}\|_{4/3}^2)^{1/3} \cdot \|\restr{\randomize{g}}{}{x}\|_{4/3}^{4/3} \leq
            (\|\restr{\randomize{g}}{}{x}\|_{2}^2)^{1/3} \cdot \|\restr{\randomize{g}}{}{x}\|_{4/3}^{4/3}.
        \]
        But by the usual Parseval Theorem,
        \[
            \|\restr{\randomize{g}}{}{x}\|_{2}^2 = \sum_{S \subseteq [n]} g^{=S}(x)^2 = \sum_{S \ni i} (2/5)^{2|S|} f^{=S}(x)^2 \leq \sum_{S \ni i} f^{=S}(x)^2 \leq \tau,
        \]
        the last inequality due to the assumption that $i \not \in J'_x$.
    \end{proof}
    We now establish~\eqref{eqn:bourgain-main-goal}:
    \begin{align}
        \Ex_{\bx}\left[\sum_{|S| \leq k,\ S \not \subseteq J'_{\bx}} f^{=S}(\bx)^2\right]
        &\leq (5\sqrt{3}/2)^{2k} \cdot \Ex_{\bx}\left[\sum_{S \not \subseteq J'_{\bx}} (\T_{\frac{2}{5\sqrt{3}}} f^{=S})(\bx)^2\right] \nonumber\\
        &\leq 20^{k} \cdot \Ex_{\bx}\left[\sum_{i \not \in J'_{\bx}} \sum_{S \ni i} (\T_{\frac{2}{5\sqrt{3}}} f^{=S})(\bx)^2\right] \nonumber\\
        &= 20^k \cdot \Ex_{\bx}\left[\sum_{i \not \in J'_{\bx}} \|\T_{\frac{1}{\sqrt{3}}} \restr{\randomize{g^i}}{}{\bx}\|_2^2\right] \tag{for $g^i = \T_{\frac25} \Lap_i f$}\\
        &\leq 20^k \tau^{1/3} \cdot \Ex_{\bx}\left[\sum_{i \not \in J'_{\bx}} \| \restr{\randomize{g^i}}{}{\bx}\|_{4/3}^{4/3}\right] \tag{Lemma~\ref{lem:bourgain-thm-lemma}}\\
        &\leq 20^k \tau^{1/3} \cdot \sum_{i=1}^n \|\Lap_i f\|_{4/3}^{4/3} \tag{Theorem~\ref{thm:rand-symm}}\\
        &\leq 20^k \tau^{1/3} \cdot \sum_{i=1}^n \Inf_i[f] \tag{Exercise~\ref{ex:laplacian-any-power-influence}\ref{ex:strong-laplacian-any-power-influence}}\\
        &= 20^k \tau^{1/3} \cdot \Tinf[f] = (20c^{-1/3})^k k \eps \leq \eps/2, \nonumber
    \end{align}
    the last inequality because $(20 c^{-1/3})^{k}k \leq 1/2$ for all $k \geq 0$ once $c$ is a large enough constant.

    The last task in the proof is to establish~\eqref{eqn:bourgain-too-big}.  Using Cauchy--Schwarz,
    \begin{multline}
        \Ex_{\bx \sim \pi\xn}\left[\bone[\text{$J'_{\bx}$ is too big}] \cdot \sum_{0 < |S| \leq k} f^{=S}(\bx)^2\right] \\ \leq \sqrt{\E_{\bx}\left[\bone[\text{$J'_{\bx}$ is too big}]^2\right]}\sqrt{\E_{\bx}\left[\Bigl(\sum_{0 < |S| \leq k} f^{=S}(\bx)^2\Bigr)^2\right]}.
            \label{eqn:bourgain-too-big1}
    \end{multline}
    For the first factor on the right of~\eqref{eqn:bourgain-too-big1} we use Markov's inequality:
    \begin{multline}
        \E_{\bx}\left[\bone[\text{$J'_{\bx}$ is too big}]^2\right] = \Pr_{\bx}[\text{$J'_{\bx}$ is too big}] = \Pr_{\bx}[|J'_{\bx}| \geq C^k]\\
        \leq C^{-k} \E_{\bx}[|J'_{\bx}|]
        \leq C^{-k} \E_{\bx}\left[\Bigl(\sum_{i=1}^n \sum_{S\ni i} f^{=S}(\bx)^2\Bigr)/\tau\right]
        = C^{-k} c^k  \cdot \Tinf[f]. \label{eqn:bourgain-too-big2}
    \end{multline}
    As for the second factor on the right of~\eqref{eqn:bourgain-too-big1}, let's write $h = \T_{\frac25}(f - f^{=\emptyset})$.  (We are being slightly finicky about $f^{=\emptyset}$ just in case it's very large.) Then
    \begin{align}
        \E_{\bx}\left[\Bigl(\sum_{0 < |S| \leq k} f^{=S}(\bx)^2\Bigr)^2\right]
        &\leq (5/2)^{4k} \cdot \E_{\bx}\left[\left(\sum_{S \neq \emptyset} (\T_{\frac25} f^{=S})(\bx)^2\right)^2\right] \nonumber\\
        &= (5/2)^{4k} \cdot \E_{\bx}\left[\|\restr{\randomize{h}}{}{\bx}\|_2^4\right]\nonumber\\
        &\leq 40^k \cdot \E_{\bx}\left[\|\restr{\randomize{h}}{}{\bx}\|_4^4\right]\nonumber\\
        &\leq 40^k \cdot \|f - f^{=\emptyset}\|_4^4\tag{Theorem~\ref{thm:rand-symm}} \\
        &\leq 40^k \cdot 2^2 \E_{\bx}[(f - f^{=\emptyset})^2] \tag{since $|f - f^{=\emptyset}| \leq 2$ always}\\
        &= 4 \cdot 40^k \cdot \Var[f] \leq 4 \cdot 40^k \cdot \Tinf[f]. \label{eqn:bourgain-too-big3}
    \end{align}
    Substituting~\eqref{eqn:bourgain-too-big2} and~\eqref{eqn:bourgain-too-big3} into~\eqref{eqn:bourgain-too-big1} gives
    \begin{multline*}
        \Ex_{\bx \sim \pi\xn}\left[\bone[\text{$J'_{\bx}$ is too big}] \cdot \sum_{0 < |S| \leq k} f^{=S}(\bx)^2\right] \\ \leq \sqrt{C^{-k} c^k \cdot 4 \cdot 40^k} \cdot  \Tinf[f] =  2(\tfrac{40c}{C})^{k/2} k \eps \leq \eps/2,
    \end{multline*}
    the last inequality again holding for all $k \geq 0$ once $C$ is chosen large enough compared to~$c$.
\end{proof}

We end this chapter by deducing Bourgain's Sharp Threshold Theorem from  Theorem~\ref{thm:bourgain-sharp}.
\begin{proof}[Proof of Bourgain's Sharp Threshold Theorem]
    We take $\eps = .001$ in Theorem~\ref{thm:bourgain-sharp} and obtain the associated collections of subsets~$\calF_x$, where each $|\calF_x| \leq \exp(O(K^2))$ and each $S \in \calF_x$ satisfies $|S| \leq O(K)$.  Using the fact that $f^{=\emptyset}(x)^2 = 1 - \Var[f] \leq .99$ for each~$x$ we get
    \[
        \E_{\bx \sim \pi\xn} \left[\sum_{S \in \calF_\bx \setminus \{\emptyset\}} f^{=S}(\bx)^2\right] \geq 1 - 2\eps - .99 = .008.
    \]
    We always have $|\calF_x \setminus \{\emptyset\}| \leq \exp(O(K^2))$, and there's also no harm in assuming $|\calF_x \setminus \{\emptyset\}|  > 0$.  It follows that
    \[
        \E_{\bx \sim \pi\xn} \left[\max_{S \in \calF_\bx \setminus \{\emptyset\}} \{f^{=S}(\bx)^2\}\right] \geq \frac{.008}{\exp(O(K^2))} = \exp(-O(K^2)).
    \]
    Thus for each $x$ we can define a set $S_x$ with $0 < |S_x| \leq O(K)$ such that
    \begin{equation} \label{eqn:bourgain-calculation}
        \E_{\bx \sim \pi\xn} \left[f^{=S_\bx}(\bx)^2\right] \geq \exp(-O(K^2)).
    \end{equation}
    By Exercise~\ref{ex:f=Sbound} we have $|f^{=S_x}(x)| \leq 2^{|S_x|} \leq 2^{O(K)}$ and hence  $f^{=S_\bx}(\bx)^2 \leq \exp(O(K))$ always. It follows from~\eqref{eqn:bourgain-calculation} that we must have
    \[
        \Pr_{\bx \sim \pi\xn} \left[f^{=S_\bx}(\bx)^2 \geq \exp(-O(K^2))\right] \geq \exp(-O(K^2)).
    \]
    We will complete the proof by showing that whenever $f^{=S_\bx}(\bx)^2 \geq \exp(-O(K^2))$ occurs, there exists $T \subseteq S_{\bx}$ such that $\bx_T$ is a $\pm \exp(-O(K^2))$-booster for~$f$. So we either have a $+\exp(-O(K^2))$-booster with probability at least $\half \exp(-O(K^2))$, or a $-\exp(-O(K^2))$-booster with probability at least $\half \exp(-O(K^2))$; either way, the proof will be complete.

    Assume then that $f^{=S_x}(x)^2 \geq \exp(-O(K^2))$; equivalently,
    \[
        |f^{=S_x}(x)| \geq \exp(-O(K^2)).
    \]
    Let's now work with $g = f - \E[f]$.  Of course $g^{=T} = f^{=T}$ for all $T \neq \emptyset$; since $S_x \neq \emptyset$ the above inequality tells us that $|g^{=S_x}(x)| \geq \exp(-O(K^2))$.  Recall the formula
    \[
        g^{=S_x}(x) = \sum_{\emptyset \neq T \subseteq S_x} (-1)^{|S_x|-|T|}g^{\subseteq T}(x);
    \]
    we dropped the $T = \emptyset$ term since it's~$0$.  As there are only $2^{|S_x|}-1 = \exp(O(K))$ terms in the above sum, we deduce there must exist some~$T \subseteq S_x$ with $0 < |T| \leq O(K)$ such that
    \[
        |g^{\subseteq T}(x)| \geq \exp(-O(K^2))/\exp(O(K)) = \exp(-O(K^2)).
    \]
    But $g^{\subseteq T} = f^{\subseteq T} - \E[f]$, so the above gives us $|f^{\subseteq T}(x) - \E[f]| \geq \exp(-O(K^2))$.  This precisely says that $x_T$ is a $\pm \exp(-O(K^2))$-booster, as desired.
\end{proof}

For a relaxation of the assumption $\Var[f] \geq .01$ in this theorem, see Exercise~\ref{ex:refine-bourgain-thresh}.

                                                \index{Bourgain's Sharp Threshold Theorem|)}%

\section{Exercises and notes}
\begin{exercises}
    \item \label{ex:triangle-variant} Let $\bX$ be a random variable and let $1 \leq r \leq \infty$.  Recall that the triangle (Minkowski) inequality implies that for real-valued functions $f_1, f_2$,
        \[
            \|f_1(\bX) + f_2(\bX)\|_r \leq \|f_1(\bX)\|_r  + \|f_2(\bX)\|_r.
        \]
        More generally, if $w_1, \dots, w_m$ are nonnegative reals $f_1, \dots, f_m$ are real functions, then
        \[
            \|w_1 f_1(\bX) + \cdots + w_m f_m(\bX)\|_r \leq w_1 \|f_1(\bX)\|_r + \cdots + w_m \|f_m(\bX)\|_r.
        \]
        Still more generally, if $\bY$ is a random variable independent of~$\bX$ and $f(\bX,\bY)$ is a (measurable) real-valued function, then it holds that
        \[
            \bigl\| \E_{\bY}[f(\bX,\bY)] \bigr\|_{r, \bX} \leq \Ex_{\bY}[\|f(\bX,\bY)\|_{r,\bX}].
        \]
        Using this last fact, show that whenever $0 < p \leq q \leq \infty$,
        \[
            \left\|\ \| f(\bX,\bY)\|_{p,\bY}\ \right\|_{q, \bX} \leq \left\|\ \| f(\bX,\bY)\|_{q,\bX}\ \right\|_{p, \bY}.
        \]
        (Hint: Raise the inequality to the power of~$p$ and use $r = q/p$.)
    \item \label{ex:sum-hc-hc}  The goal of this exercise is to prove Proposition~\ref{prop:sum-hc-hc}: If $\bX$ and $\bY$ are independent $(p,q,\rho)$-hypercontractive random variables, then so is $\bX+\bY$.
                                                        \index{hypercontractivity!preserved by sums}%
        Let $a, b \in \R$.
        \begin{exercises}
            \item First obtain
                \[
                    \|a+\rho b(\bX+\bY)\|_{q, \bX,\bY} \leq  \left\|\ \|a+\rho b \bX + b\bY\|_{p, \bY}\ \right\|_{q,\bX}.
                \]
            \item \label{ex:sum-hc-hc-b} Next, upper-bound this by
            \[
                \left\|\ \|a+b\bY + \rho b \bX\|_{q, \bX}\ \right\|_{p,\bY}.
            \]
            (Hint: Exercise~\ref{ex:triangle-variant}.)
            \item \label{ex:sum-hc-hc-c} Finally, upper-bound this by
            \[
                \left\|\ \|a+b\bY + b \bX\|_{p, \bX}\ \right\|_{p,\bY} = \|a+b(\bX+\bY)\|_{p,\bX,\bY}.
            \]
        \end{exercises}
    \item \label{ex:poly-hc-hc}  Let $\bX_1, \dots, \bX_n$ be independent $(p,q,\rho)$-hypercontractive random variables.  Let $F(x) = \sum_{S \subseteq [n]} \wh{F}(S)x^S$ be an $n$-variate multilinear polynomial.  Define formally the multilinear polynomial $\T_\rho F(x) = \sum_{S \subseteq [n]} \rho^{|S|}\wh{F}(S)x^S$. The goal of this exercise is to show
        \begin{equation}                \label{eqn:poly-hc-hc}
            \|\T_\rho F(\bX_1, \dots, \bX_n)\|_q \leq \|F(\bX_1, \dots, \bX_n)\|_p.
        \end{equation}
                                    \index{hypercontractivity!induction}%
        Note that this result yields an alternative deduction of the Hypercontractivity Theorem for $\pm 1$ bits from the Two-Point Inequality.  A (notationally intense) generalization of this exercise can also be used as an alternative inductive strategy for deducing the General Hypercontractivity Theorem from Proposition~\ref{prop:min-prob-implies-q-norm-bound} or Theorem~\ref{thm:discrete-hypercon}.
        \begin{exercises}
            \item Why is Exercise~\ref{ex:sum-hc-hc} a special case of~\eqref{eqn:poly-hc-hc}?
            \item Begin the inductive proof of~\eqref{eqn:poly-hc-hc} by showing that the base case $n = 0$ is trivial.
            \item For the case of general~$n$, first establish
                  \[
                      \|\T_\rho F(\bX)\|_q \leq \left\| \ \|\T'_\rho E(\bX') + \bX_n \T'_\rho D(\bX') \|_{p, \bX_n} \ \right\|_{q, \bX'},
                  \]
                  where we are using the notation $x' = (x_1, \dots, x_{n-1})$, $F(x) = E(x') + x_n D(x')$, and $\T_\rho'$ for the operator acting formally on $(n-1)$-variate multilinear polynomials.
            \item Complete the inductive step, using steps similar to Exercises~\ref{ex:sum-hc-hc}\ref{ex:sum-hc-hc-b},\ref{ex:sum-hc-hc-c}. (Hint: For $X_n$ a real constant, why is $\T'_\rho E(\bX') + X_n \T'_\rho D(\bX') = \T'_\rho (E + X_n D)(\bX')$?)
        \end{exercises}
    \item \label{ex:hc-holder} This exercise is concerned with the possibility of a converse for Proposition~\ref{prop:hc-holder}.
        \begin{exercises}
            \item In our proof of the Two-Point Inequality we used Proposition~\ref{prop:dual-norm} to deduce that a uniform bit $\bx \sim \bits$ is $(p,q,\rho)$-hypercontractivity if it's $(q',p',\rho)$-hypercontractive.  Why can't we use Proposition~\ref{prop:dual-norm} to deduce this for a general random variable $\bX$?
            \item  For each $1 < p < 2$,  exhibit a random variable $\bX$ that is $(p, 2, \rho)$-hypercontractive (for some~$\rho$) but not $(2,p',\rho)$-hypercontractive.
        \end{exercises}
    \item \label{ex:gen-sse-sharpness}
        \begin{exercises}
            \item Regarding Remark~\ref{rem:gen-sse-sharpness}, heuristically justify (in the manner of Exercise~\ref{ex:gaussquad-est}\ref{ex:gaussquad-heuristic}) the following statement: If $A, B \subseteq \bn$ are concentric Hamming balls with volumes~$\exp(-\frac{a^2}{2})$ and~$\exp(-\frac{b^2}{2})$ and $\rho a \leq b \leq a$ (where $0 < \rho < 1$), then
            \[
                \Pr_{\substack{(\bx, \by) \\ \text{ $\rho$-correlated}}}[\bx \in A, \by \in B] \gtrapprox \exp\left(-\tfrac12 \tfrac{a^2 - 2\rho ab + b^2}{1-\rho^2}\right);
            \]
            and further, if $b < \rho a$, then $\Pr[\bx \in A, \by \in B] \sim \Pr[\bx \in A]$. Here you should treat~$\rho$ as fixed and $a, b \to \infty$.
                                \index{Small-Set Expansion Theorem!Reverse}%
            \item Similarly, heuristically justify that the Reverse Small-Set Expansion Theorem is essentially sharp by considering diametrically opposed Hamming balls.
        \end{exercises}
    \item \label{ex:reverse-hypercon1}  The goal of this exercise (and Exercise~\ref{ex:reverse-hypercon2}) is to prove the Reverse Hypercontractivity Theorem and its equivalent Two-Function version:
                            \index{Reverse Hypercontractivity Theorem|seeonly{Hypercontractivity Theorem, Reverse}}%
                            \index{Hypercontractivity Theory!Reverse}%
        \begin{named}{Reverse Hypercontractivity Theorem}
            Let $f \btR^{\geq 0}$ be a nonnegative function and let $-\infty \leq q < p \leq 1$.  Then $\|\T_\rho f\|_q \geq \|f\|_p$ for $0 \leq \rho \leq \sqrt{(1-p)/(1-q)}$.
        \end{named}
        \begin{named}{Reverse Two-Function Hypercontractivity Theorem}
            Let \linebreak $f, g \btR^{\geq 0}$ be nonnegative, let $r, s \leq 0$, and assume $0 \leq \rho \leq \sqrt{rs} \leq 1$.  Then
            \[
                \Es{(\bx, \by) \\ \text{ $\rho$-correlated}}[f(\bx) g(\by)] \geq \|f\|_{1+r} \|g\|_{1+s}.
            \]
        \end{named}
        \noindent Recall that for $-\infty < p < 0$ and for positive functions $f \in L^2(\Omega, \pi)$ the ``norm'' $\|f\|_p$ retains the definition $\E[f^p]^{-1/p}$. (The cases of $p = -\infty$, $p = 0$, and nonnegative functions are defined by appropriate limits; in particular $\|f\|_{-\infty}$ is the minimum of $f$'s values, $\|f\|_0$ is the geometric mean of $f$'s values, and $\|f\|_p$ is $0$ whenever $f$ is not everywhere positive.  We also define $p'$ by $\frac1p + \frac1{p'} = 1$, with $0' = 0$.)

        The Reverse Two-Function Hypercontractivity Theorem can be thought of as a generalization of the lesser known ``reverse \Holder inequality'' in the setting of $L^2(\bn, \unif\xn)$:
        \begin{named}{Reverse \Holder inequality}
            Let $f \in L^2(\Omega, \pi)$ be a positive function.  Then for any $p < 1$,
            \[
                \|f\|_p = \inf\ \{ \E[fg] : g > 0, \|g\|_{p'} = 1\}.
            \]
            In particular, for $r < 0$ and $f, g > 0$ we have $\E[fg] \geq \|f\|_{1+r}\|g\|_{1+1/r}$.
        \end{named}
        \begin{exercises}
            \item Show that to prove these two Reverse Hypercontractivity Theorems it suffices to consider the case of $f, g \btR^+$, i.e., strictly positive functions.
            \item Show that the Reverse Two-Function Hypercontractivity Theorem is equivalent (via the reverse \Holder inequality) to the Reverse Hypercontractivity Theorem.
            \item Reduce the Reverse Two-Function Hypercontractivity Theorem to the $n = 1$ case.  (Hint: Virtually identical to the Two-Function Hypercontractivity Induction.)  Further reduce to following:
                                                    \index{Two-Point Inequality!Reverse}%
                \begin{named}{Reverse Two-Point Inequality}
                    Let $-\infty \leq q < p \leq 1$ and let $0 \leq \rho \leq \sqrt{(1-p)/(1-q)}$.  Then $\|\T_\rho f\|_q \geq \|f\|_p$ for any $f \co \bits \to \R^+$.
                \end{named}
        \end{exercises}
    \item \label{ex:reverse-hypercon2}  The goal of this exercise is to prove the Reverse Two-Point Inequality.
        \begin{exercises}
            \item Similar to the non-reverse case, the main effort is proving the inequality assuming that $0 < q < p \leq 1$ and that $\rho = \sqrt{(1-p)/(1-q)}$.  Do this by mimicking the proof of the Two-Point Inequality. (Hint: You will need the inequality $(1+t)^\theta \geq 1 + \theta t$ for $\theta \geq 1$, and you will need to show that $\tfrac{j-r}{\sqrt{1-r}}$ is an increasing function of~$r$ on $[0,1)$ for all $j \geq 2$.)
            \item Extend to the case of $0 \leq \rho \leq \sqrt{(1-p)/(1-q)}$.  (Hint: Use the fact that for any $f \co \bn \to \R^{\geq 0}$ and $-\infty \leq p \leq q \leq \infty$ we have $\|f\|_p \leq \|f\|_q$.  You can prove this generalization of Exercise~\ref{ex:norm-monotonicity} by reducing to the case of negative $p$ and~$q$ to the case of positive $p$ and~$q$.)
            \item Establish the $q = -\infty$ case of the Reverse Two-Point Inequality.
            \item Show that the cases $-\infty < q < p < 0$ follow by ``duality''.  (Hint: Like Proposition~\ref{prop:dual-norm} but with the reverse \Holder inequality.)
            \item Show that the cases $q < 0 < p$ follow by the semigroup property of~$\T_\rho$.
            \item Finally, treat the cases of $p = 0$ or $q = 0$.
         \end{exercises}
    \item Give a simple proof of the $n = 1$ case of the Reverse Two-Function Hypercontractivity Theorem when $r = s = -1/2$.  (Hint: Replace $f$ and $g$ by $f^2$ and $g^2$; then you don't even need to assume $f$ and $g$ are nonnegative.)  Can you also give a simple proof when $r = s = -1 + 1/k$ for integers $k > 2$? \item \label{ex:reverse-hypercon3} By selecting $\text{``}r\text{''} = -\rho \frac{\rho a + b}{a+\rho b}$ and $\text{``}s\text{''} = -\rho \frac{a +\rho b}{\rho a+ b}$, prove
                                    \index{Small-Set Expansion Theorem!Reverse}%
    the Reverse Small-Set Expansion Theorem mentioned in Remark~\ref{rem:reverse-two-set}. (Hint: The negative norm of a $0$-$1$-indicator is~$0$, so be sure to verify no negative norms arise.)
    \item \label{ex:justify-T-partial-restriction-commute} Let $g \in L^2(\Omega^n, \pi\xn)$. Writing $x = (x_1, x')$, where $x' = (x_2, \dots, x_n)$, carefully justify the following identity of one-input functions:  $\restr{(\T^1_{\rho} g)}{}{x'} = \T_\rho (\restr{g}{}{x'})$.  (Hint: You may want to refer to Exercise~\ref{ex:decomp-restriction}.)
    \item \label{ex:reasonable-to-hc-symm}  Prove Proposition~\ref{prop:reasonable-to-hc-symm}.
    \item \label{ex:symm1}  Let
                                         \index{randomization/symmetrization}%
            $\bX$ be a random variable and let $\bY$ denote its symmetrization $\bX-\bX'$, where $\bX'$ is an independent copy of~$\bX$.  Show for any $t, \theta \in \R$ that $\Pr[|\bY| \geq t] \leq 2\Pr[|\bX - \theta| \geq t/2]$.
    \item \label{ex:general-unsymm-lemma} The goal of this exercise is to establish Lemma~\ref{lem:unsymmetrize}.
        \begin{exercises}
            \item Show that we may take $c_2 = 1$ (and that equality holds).  Henceforth assume $q > 2$.
            \item By following the idea of our $q = 4$ proof, reduce to showing that there exists $0 < c_q < 1$ such that
                \[
                    |1 - c_qx|^q + c_qqx - 1\leq |1+x|^q - qx - 1\quad \forall x \in \R.
                \]
            \item Further reduce to showing there exists $0 < c_q < 1$ such that
                \begin{equation} \label{eqn:unsymm-ex1}
                    \frac{|1 - c_qx|^q + c_qqx - 1}{x^2} \leq \frac{|1+x|^q - qx - 1}{x^2} \quad \forall x \in \R.
                \end{equation}
                Here you should also establish that both sides are continuous functions of $x \in \R$ once the value at $x = 0$ is defined appropriately.
            \item Show that there exists $M > 0$ such that for \emph{every} $0 < c_q < \half$, inequality~\eqref{eqn:unsymm-ex1} holds once $|x| \geq M$.  (Hint: Consider the limit of both sides as $|x| \to \infty$.)
            \item Argue that it suffices to show that
                \begin{equation} \label{eqn:unsymm-ex2}
                    \frac{|1+x|^q - qx -1}{x^2} \geq \eta
                \end{equation}
                for some universal positive constant $\eta > 0$.  (Hint: A uniform continuity argument for $(x,c_q) \in [-M,M] \times [0,\half]$.)
            \item Establish~\eqref{eqn:unsymm-ex2}.  (Hint: The best possible $\eta$ is~$1$, but to just achieve some positive~$\eta$, argue using Bernoulli's inequality that $\frac{|1+x|^q - qx -1}{x^2}$ is everywhere positive and then observe that it tends to~$\infty$ as $|x| \to \infty$.)
            \item Possibly using a different argument, what is the best asymptotic bound you can achieve for~$c_q$?  Is $c_q \geq \Omega(\frac{\log q}{q})$ possible?
        \end{exercises}
    \item \label{ex:c-root} Show that the largest $c$ for which inequality~\eqref{eqn:unsymm-ex} holds is the smaller real root of $c^4-2c^3-2c+1 = 0$, namely, $c \approx .435$.
    \item \label{ex:bourgain-hypercon}
                                             \index{randomization/symmetrization}%
            \begin{exercises}
                \item Show that $1 + 6c^2 x^2 + c^4 x^4 \leq 1 + 6x^2 + 4x^3 + x^4$ holds for all $x \in \R$ when $c = 1/2$.  (Can you also establish it for $c \approx .5269$?)
                \item Show that if $\bX$ is a random variable satisfying $\E[\bX] = 0$ and $\|\bX\|_4 < \infty$, then $\|a + \half \br \bX\|_4 \leq \|a + \bX\|_4$ for all $a \in \R$, where $\br \sim \bits$ is a uniformly random bit independent of~$\bX$.  (Cf.~Lemma~\ref{lem:symmetrization-randomization1}.)
                \item Establish the following improvement of Theorem~\ref{thm:unsymmetrize} in the case of $q = 4$: for all   $f \in L^2(\Omega^n, \pi\xn)$,
                    \[
                        \|\T_{\frac12 \br} f(\bx)\|_{4,\br,\bx} \leq \|f(\bx)\|_{4, \bx}
                    \]
                    (where $\bx \sim \pi\xn$, $\br \sim \bn$).
          \end{exercises}
    \item \label{ex:projection-bound-general} Complete the proof of Theorem~\ref{thm:projection-bound-general}.  (Hint: You'll need to rework Exercise~\ref{ex:projection-bounded} as in Lemma~\ref{lem:low-deg-T-bounded}.)
    \item \label{ex:min-prob-implies-q-norm-bound}  Prove Proposition~\ref{prop:min-prob-implies-q-norm-bound}.
                                                \index{hypercontractivity!general product probability spaces|(}%
    \item \label{ex:LO-bound} Recall from~\eqref{eqn:LO-bound} the function $\rho = \rho(\lambda)$ defined for $\lambda \in (0,1/2)$  (and fixed $q > 2$) by
            \[
                \rho = \rho(\lambda) = \sqrt{\frac{\exp(u/q) - \exp(-u/q)}{\exp(u/q')- \exp(-u/q')}} = \sqrt{\frac{\sinh(u/q)}{\sinh(u/q')}},
            \]
            where $u = u(\lambda)$ is defined by $\exp(-u) = \tfrac{\lambda}{1-\lambda}$.
        \begin{exercises}
            \item \label{ex:hypercon-const-increasing} Show that $\rho$ is an increasing function of $\lambda$.  (Hint: One route is to reduce to showing that $\rho^2$ is a decreasing function of $u \in (0,\infty)$, reduce to showing that $q\tanh(u/q)$ is an increasing function of $q \in (1, \infty)$, reduce to showing $\frac{\tanh r}{r}$ is a decreasing function of $r \in (0,\infty)$, and reduce to showing $\sinh(2r) \geq 2r$.)
            \item Verify the following statements from Remark~\ref{rem:LO-bound}:
                \begin{align*}
                    \text{for fixed $q$ and $\lambda \to 1/2$,} \quad &\rho \to \frac{1}{\sqrt{q-1}}; \\
                    \text{for fixed $q$ and $\lambda \to 0$,} \quad &\rho \sim \lambda^{1/2 - 1/q}.
                \end{align*}
                Also show:
                \[
                    \text{for fixed $\lambda$ and $q \to \infty$,} \quad \rho \sim \sqrt{\frac{u}{\sinh u}} \sqrt{\frac{1}{q}}, \\
                \]
                and $\sqrt{\frac{u}{\sinh u}} \sim \sqrt{2\lambda\ln(1/\lambda)}$ for $\lambda \to 0$.
            \item Show that $\rho \geq \frac{1}{\sqrt{q-1}} \lambda^{1/2-1/q}$ holds for all $\lambda$.
        \end{exercises}
    \item \label{ex:wolff}    Let $(\Omega, \pi)$ be a finite probability space, $|\Omega| \geq 2$, in which every outcome has probability at least~$\lambda$.  Let $1 < p < 2$ and $0 < \rho < 1$.  The goal of this exercise is to prove the result of Wolff~\cite{Wol07} that, subject to $\|\T_\rho f\|_2 = 1$, every $f \in L^2(\Omega,\pi)$ that minimizes $\|f\|_p$ takes on at most two values (and there is at least one minimizing~$f$).
        \begin{exercises}
            \item We consider the equivalent problem of minimizing $F(f) = \|f\|_p^p$ subject to $G(f) = \|\T_\rho f\|_2^2 = 1$.  Show that both $F(f)$ and $G(f)$ are $\calC^1$ functionals (identifying functions $f$ with points in $\R^\Omega$).
            \item Argue from continuity that the minimum value for $\|f\|_p^p$ subject to $\|\T_\rho f\|_2^2 = 1$ is attained.  Henceforth write $f_0$ to denote any minimizer; the goal is to show that $f_0$ takes on at most two values.
            \item Show that $f_0$ is either everywhere nonnegative or everywhere nonpositive.  (Hint: By homogeneity our problem is equivalent to maximizing $\|\T_\rho f\|_2$ subject to $\|f\|_p = 1$; now use Exercise~\ref{ex:T-and-abs}.) Replacing $f_0$ by $|f_0|$ if necessary, henceforth assume~$f_0$ is nonnegative.
            \item Show that $\grad F(f_0) = \pi \cdot p f_0^{p-1}$ and $\grad G(f_0) = \pi \cdot 2 \T_{\rho^2} f_0$.  Here $\pi \cdot g$ signifies the pointwise product of functions on~$\Omega$, with $\pi$ thought of as a function $\Omega \to \R^{\geq 0}$.  (Hint: For the latter, write $G(f) = \la \T_{\rho^2} f, f \ra$.)
            \item Use the method of Lagrange Multipliers to show that $c f_0^{p-1} = T_{\rho^2} f_0$ for some $c \in \R^+$.  (Hint: You'll need to note that $\grad G(f_0) \neq 0$.)
            \item Writing $\mu = \E[f_0]$, argue that each value $y = f(\omega)$ satisfies the equation
                  \begin{equation} \label{eqn:wolff-equation}
                    c y^{p-1} = \rho^2 y + (1-\rho^2) \mu.
                  \end{equation}
            \item Show that~\eqref{eqn:wolff-equation} has at most two solutions for $y \in \R^+$, thereby completing the proof that $f_0$ takes on at most two values.  (Hint: Strict concavity of $y^{p-1}$.)
            \item Suppose $q > 2$. By slightly modifying the above argument, show that subject to $\|g\|_2 = 1$, every $g \in L^2(\Omega,\pi)$ that maximizes $\|\T_{\rho} g\|_q$ takes on at most two values (and there is at least one maximizing~$g$).  (Hint: At some point you might want to make the substitution $g  = \T_\rho f$; note that~$g$ is two-valued if~$f$ is.)
        \end{exercises}
    \item \label{ex:lo} Fix $1 < p < 2$ and $0 < \lambda < 1/2$.  Let $\Omega = \{-1,1\}$ and $\pi = \pi_\lambda$, meaning $\pi(-1) = \lambda$, $\pi(1) = 1-\lambda$.  The goal of this exercise is to show the result of Lata{\l}a and Oleszkiewicz~\cite{LO00}: the largest value of $\rho$ for which $\|\T_\rho f\|_2 \leq \|f\|_p$ holds for all $f \in L^2(\Omega, \pi)$ is as given in Theorem~\ref{thm:discrete-hypercon}; i.e., it satisfies \begin{equation} \label{eqn:LO-bound2}
            \rho^2 = r^* = \frac{\exp(u/p') - \exp(-u/p')}{\exp(u/p)- \exp(-u/p)},
        \end{equation}
        where $u$ is defined by $\exp(-u) = \tfrac{\lambda}{1-\lambda}$.  (Here we are using $p = q'$ to facilitate the proof; we get the $(2,q)$-hypercontractivity statement by Proposition~\ref{prop:dual-norm}.)
        \begin{exercises}
            \item Let's introduce the notation $\alpha = \lambda^{1/p}$, $\beta = (1-\lambda)^{1/p}$.  Show that
            \[
                r^* = \frac{\alpha^p\beta^{2-p} - \alpha^{2-p}\beta^p}{\alpha^2-\beta^2}.
            \]
            \item Let  $f \in L^2(\Omega, \pi)$.  Write $\mu = \E[f]$ and $\delta = \D_1 f = \hat{f}(1)$.  Our goal will be to show
            \begin{equation} \label{eqn:lo-goal}
                 \mu^2 + \delta^2 r^* = \|T_{\sqrt{r^*}} f\|_2^2  \leq \|f\|_p^2.
            \end{equation}
            In the course of doing this, we'll also exhibit a nonconstant function~$f$ that makes the above inequality sharp.  Why does this establish that no larger value of~$\rho$ is possible?
            \item Show that without loss of generality we may assume
                \[
                    f(-1) = \frac{1+y}{\alpha}, \quad f(1) = \frac{1-y}{\beta}
                \]
                for some $-1 < y < 1$.   (Hint: First use Exercise~\ref{ex:T-and-abs} and a continuity argument to show that we may assume $f > 0$; then use homogeneity of~\eqref{eqn:lo-goal}.)
            \item The left-hand side of~\eqref{eqn:lo-goal} is now a quadratic function of~$y$.  Show that our $r^*$ is precisely such that
                \[
                    \text{LHS}\eqref{eqn:lo-goal} = Ay^2 + C
                \]
                for some constants $A, C$; i.e., $r^*$ makes the linear term in~$y$ drop out.  (Hint: Work exclusively with the $\alpha, \beta$ notation and recall from Definition~\ref{def:biased-deriv} that $\delta^2 = \lambda(1-\lambda)(f(1)-f(-1))^2 = \alpha^p \beta^p(f(1)-f(-1))^2$.)
            \item Compute that
                \begin{equation} \label{eqn:loA}
                    A = 2\frac{\beta^{p-1}-\alpha^{p-1}}{\beta-\alpha}.
                \end{equation}
                (Hint: You'll want to multiply the above expression by $\alpha^p + \beta^p = 1$.)
            \item Show that
                \[
                    \text{RHS}\eqref{eqn:lo-goal} = ((1+y)^p + (1-y)^p)^{2/p}.
                \]
                Why does it now suffice to show~\eqref{eqn:lo-goal} just for $0 \leq y < 1$?
            \item Let $y^* = \frac{\beta-\alpha}{\beta+\alpha} > 0$.  Show that if $y = -y^*$, then $f$ is a constant function and both sides of~\eqref{eqn:lo-goal} are equal to $\frac{4}{(\alpha+\beta)^2}$.
            \item \label{ex:lo-guts} Deduce that both sides of~\eqref{eqn:lo-goal} are equal to $\frac{4}{(\alpha+\beta)^2}$ for $y = y^*$.  Verify that after scaling, this yields the following nonconstant function for which~\eqref{eqn:lo-goal} is sharp: $f(x) = \exp(-xu/p)$.
            \item Write $y = \sqrt{z}$ for $0 \leq z < 1$.  By now we have reduced to showing
                \[
                    Az+C \leq ((1+\sqrt{z})^p + (1-\sqrt{z})^p)^{2/p},
                \]
                knowing that both sides are equal when $\sqrt{z} = y^*$.  Calling the expression on the right~$\phi(z)$, show that
                \[
                    \frac{d}{dz} \phi(z)\Bigr|_{\sqrt{z}=y^*} = A.
                \]
                (Hint: You'll need $\alpha^p+\beta^p = 1$, as well as the fact from part~\ref{ex:lo-guts} that $\phi(z) = \frac{4}{(\alpha+\beta)^2}$ when $\sqrt{z}=y^*$.) Deduce that we can complete the proof by showing that $\phi(z)$ is convex for $z \in [0,1)$.
            \item Show that $\phi$ is indeed convex on $[0,1)$ by showing that its derivative is a nondecreasing function of~$z$. (Hint: Use the Generalized Binomial Theorem as well as $1 < p < 2$ to show that $(1+\sqrt{z})^p + (1-\sqrt{z})^p$ is expressible as $\sum_{j=0}^\infty b_j z^j$ where each~$b_j$ is positive.)
        \end{exercises}
    \item \label{ex:lo-wolff}  Complete the proof of Theorem~\ref{thm:discrete-hypercon}.  (Hint: Besides Exercises~\ref{ex:wolff} and~\ref{ex:lo}, you'll also need Exercise~\ref{ex:LO-bound}\ref{ex:hypercon-const-increasing}.)
                                                    \index{hypercontractivity!general product probability spaces|)}%
    \item \label{ex:entropy}
        \begin{exercises}
            \item Let $\Phi \co [0, \infty) \to \R$ be defined by $\Phi(x) = x \ln x$, where we take $0\ln0 = 0$.  Verify that $\Phi$ is a smooth, strictly convex function.
            \item \label{ex:entropy-facts} Consider the following:
                                                    \index{entropy functional}%
                                                    \nomenclature[Entf]{$\Ent[f]$}{for a nonnegative function on a probability space, denotes $\E[f \ln f] - \E[f] \ln \E[f]$}%
                    \begin{definition}
                        Let $g \in L^2(\Omega, \pi)$ be a nonnegative function.  The \emph{entropy} of~$g$ is defined by
                        \[
                            \Ent[g] = \E_{\bx \sim \pi}[\Phi(g(\bx))] - \Phi\Bigl(\E_{\bx \sim \pi}[g(\bx)]\Bigr).
                        \]
                    \end{definition}
                Verify that $\Ent[g] \geq 0$ always, that $\Ent[g] = 0$ if and only if~$g$ is constant, and that $\Ent[c g] = c\Ent[g]$ for any constant $c \geq 0$.
            \item Suppose $\vphi$ is a probability density on $\bn$ (recall Definition~\ref{def:prob-density}).  Show that $\Ent[\vphi] = \dkl{\vphi}{\unif\xn}$, the Kullback--Leibler divergence of the uniform  distribution from~$\vphi$ (more precisely, the distribution with density~$\vphi$).
        \end{exercises}
                                                \index{Log-Sobolev Inequality|(}%
    \item \label{ex:log-sob} The goal of this exercise is to establish:
    \begin{named}{The Log-Sobolev Inequality}
        Let $f\btR$.  Then ${\half \Ent[f^2] \leq \Tinf[f]}$.
    \end{named}
        \begin{exercises}
            \item Writing $\rho = e^{-t}$, the $(p,2)$-Hypercontractivity Theorem tells us that
                    \[
                        \|\T_{e^{-t}} f\|_2^2 \leq \|f\|^2_{1+\exp(-2t)}
                    \]
                 for all $t \geq 0$.                    Denote the left- and right-hand sides as $\textrm{LHS}(t)$,  $\textrm{RHS}(t)$.  Verify that these are smooth functions of $t \in [0,\infty)$ and that $\textrm{LHS}(0) = \textrm{RHS}(0)$. Deduce that $\textrm{LHS}'(0) \leq \textrm{RHS}'(0)$.
            \item Compute $\textrm{LHS}'(0) = -2 \Tinf[f]$.  (Hint: Pass through the Fourier representation; cf.~Exercise~\ref{ex:laplacian-as-deriv}.)
            \item Compute $\textrm{RHS}'(0) = -\Ent[f^2]$, thereby deducing the Log-Sobolev Inequality.  (Hint: As an intermediate step, define $F(t) = \E[|f|^{1+\exp(-2t)}]$ and show that $\textrm{RHS}'(0) = F(0)\ln F(0) + F'(0)$.)
        \end{exercises}
    \item \label{ex:log-sob-vs-poincare}
        \begin{exercises}
            \item Let $f \btR$. Show that $\Ent[(1+\eps f)^2] \sim 2 \Var[f] \eps^2$ as $\eps \to 0$.
                                                \index{Poincar\'{e} Inequality}%
            \item Deduce the Poincar\'e Inequality for~$f$ from the Log-Sobolev Inequality.
        \end{exercises}
    \item \label{ex:edge-iso-from-log-sob}
                                                \index{isoperimetric inequality!Hamming cube}%
                                                \index{expansion!small-set}%
        \begin{exercises}
            \item Deduce from the Log-Sobolev Inequality that for $f \btb$ with $\alpha = \min\{\Pr[f = 1], \Pr[f=-1]\}$,
                \begin{equation} \label{eqn:weak-edge-iso}
                    2 \alpha \ln(1/\alpha) \leq \Tinf[f].
                \end{equation}
                This is off by a factor of $\ln 2$ from the optimal edge-isoperimetric inequality Theorem~\ref{thm:edge-iso}. (Hint: Apply the inequality to either $\half - \half f$ or $\half + \half f$.)
            \item Give a more streamlined direct derivation of~\eqref{eqn:weak-edge-iso} by differentiating the Small-Set Expansion Theorem.
        \end{exercises}
    \item This exercise gives a direct proof of the Log-Sobolev Inequality.
        \begin{exercises}
            \item The first step is to establish the $n = 1$ case.  Toward this, show that we may assume~$f \co \bits \to \R$ is nonnegative and has mean~$1$. (Hints: Exercise~\ref{ex:abs-decreases-influences}, Exercise~\ref{ex:entropy}\ref{ex:entropy-facts}.)
            \item Thus it remains to establish $\half \Ent[(1+b\bx)^2] \leq b^2$ for $b \in [-1,1]$.  Show that $g(b) = b^2 - \half \Ent[(1+b\bx)^2]$ is smooth on $[-1,1]$ and satisfies $g(0) = 0$, $g'(0) = 0$, and $g''(b) = \frac{2b^2}{1+b^2} + \ln\frac{1+b^2}{1-b^2} \geq 0$ for $b \in (-1,1)$.  Explain why this completes the proof of the $n = 1$ case of the Log-Sobolev Inequality.
            \item \label{ex:direct-log-sob-lemma} Show that for any two functions $f_+, f_- \btR$,
                    \[
                        \left(\tfrac{\sqrt{\E[f_+^2]} - \sqrt{\E[f_-^2]}}{2}\right)^2 \leq \E\left[\Bigl(\tfrac{f_+ - f_-}{2}\Bigr)^2\right].
                    \]
                    (Hint: The triangle inequality for $\| \cdot \|_2$.)
            \item Prove the Log-Sobolev Inequality via ``induction by restrictions'' (as described in Section~\ref{sec:hypercon-tensorize}).  (Hint: For the right-hand side, establish $\Inf[f] = \E[(\tfrac{f_+ - f_-}{2})^2] + \half \Tinf[f_+] + \half \Tinf[f_-]$.  For the left-hand side, apply induction, then the $n = 1$ base case, then part~\ref{ex:direct-log-sob-lemma}.)
        \end{exercises}
                                                \index{Log-Sobolev Inequality|)}%
    \item \label{ex:general-log-sob}
        \begin{exercises}
            \item By following   the strategy of Exercise~\ref{ex:log-sob}, establish the following:
                                                \index{Log-Sobolev Inequality!product space domains}%
        \begin{named}{Log-Sobolev Inequality for general  product space domains}
            Let $f \in L^2(\Omega^n, \pi\xn)$ and write $\lambda = \min(\pi)$, $\lambda' = 1-\lambda$,  $\exp(-u) = \frac{\lambda}{\lambda'}$.  Then $               \half \varrho \Ent[f^2] \leq \Tinf[f]$, where
            \[
                \varrho = \varrho(\lambda) = \frac{\tanh(u/2)}{u/2} = 2 \frac{\lambda' - \lambda}{\ln \lambda' - \ln \lambda}.
            \]
        \end{named}
            \item Show that $\varrho(\lambda) \sim 2/\ln(1/\lambda))$ as $\lambda \to 0$.
            \item Let $f \btb$ and treat $\bits^n$ as having the $p$-biased distribution $\pi_p\xn$.  Write $q = 1-p$. Show that if $\alpha = \min\{\Pr_{\pi_p}[f = 1], \Pr_{\pi_p}[f = -1]\}$, then
                \[
                    4\frac{q - p}{\ln q - \ln p} \alpha \ln(1/\alpha) \leq \Tinf[f\subp]
                \]
                and hence, for $p \to 0$,
                \begin{equation} \label{eqn:weak-p-biased-isoperim}
                    \alpha \log_p \alpha \leq (1+o_p(1))p \cdot \E_{\bx \sim \pi_p\xn}[\sens_f(\bx)].
                \end{equation}
                We remark that~\eqref{eqn:weak-p-biased-isoperim} is known to hold without the~$o_p(1)$ for all $p \leq 1/2$.
        \end{exercises}
    \item \label{ex:general-hypercon-induct} Prove Theorem~\ref{thm:general-hypercon-induct}.  (Hint: Recall Proposition~\ref{prop:general-trho}.)
    \item \label{ex:general-polynomial-bonami} Let $\bX_1, \dots, \bX_n$ be independent $(2,q,\rho)$-hypercontractive random variables and let $F(x) = \sum_{|S| \leq k} \wh{F}(S)\,x^S$ be an $n$-variate multilinear polynomial of degree at most~$k$.  Show that
        \[
            \|F(\bX_1, \dots, \bX_n)\|_q \leq (1/\rho)^k \|F(\bX_1, \dots, \bX_n)\|_2.
        \]
        (Hint: You'll need Exercise~\ref{ex:poly-hc-hc}.)
    \item \label{ex:simple-p-biased-hc-bound}  Let $0 < \lambda \leq 1/2$ and let $(\Omega, \pi)$ be a finite probability space in which some outcome $\omega_0 \in \Omega$ has $\pi(\omega_0) = \lambda$.  (For example, $\Omega = \bits$, $\pi = \pi_\lambda$.)  Define $f \in L^2(\Omega, \pi)$ by setting $f(\omega_0) = 1$, $f(\omega) = 0$ for $\omega \neq \omega_0$. For $q \geq 2$, compute $\|f\|_q/\|f\|_2$ and deduce (in light of the proof of Theorem~\ref{thm:general-hypercon-induct}) that Corollary~\ref{cor:thm:discrete-hypercon} cannot hold for $\rho > \lambda^{1/2 - 1/q}$.
    \item \label{ex:general-2-1-bound} Prove Theorem~\ref{thm:general-2-1-bound}.
    \item \label{ex:general-one-sided-low-deg-anticonc} Prove Theorem~\ref{thm:general-one-sided-low-deg-anticonc}.
    \item \label{ex:general-low-deg-conc} Prove Theorem~\ref{thm:general-low-deg-conc}.  (Hint: Immediately worsen~$q-1$ to $q$ so that finding the optimal choice of~$q$ is easier.)
    \item \label{ex:sse-general} Prove Theorem~\ref{thm:sse-general}.
    \item \label{ex:friedgut-product-spaces}  Prove Friedgut's Junta Theorem for general product spaces as stated in Section~\ref{sec:hypercon-variants-apps}.
    \item \label{ex:other-dir-every-mono-sharp} Show that \eqref{eqn:fk-ineq} implies $F(p_c + \eta p_c) \geq 1-\eps$ in the proof of Theorem~\ref{thm:every-mono-fcn-sharp-thresh}.  (Hint: Consider $\frac{d}{dp} \ln(1-F(p))$.)
    \item Justify the various calculations and observations in Example~\ref{eg:biased-low-influence}.
    \item \label{ex:parity-not-a-dnf}
          \begin{exercises}
              \item Let $p = \frac1n$ and let $f \in L^2(\bn, \pi_p\xn)$ be any Boolean-valued function.  Show that $\Tinf[f] \leq 4$.  (Hint: Proposition~\ref{prop:monotone-influences-biased}.)
              \item Let us specialize to the case $f = \chi_{[n]}$.  Show that $f$ is not $.1$-close to any width-$O(1)$ DNF (under the $\frac1n$-biased distribution, for $n$ sufficiently large).  This shows that the assumption of monotonicity can't be removed from Friedgut's Conjecture.  (Hint: Show that fixing any constant number of coordinates cannot change the bias of $\chi_{[n]}$ very much.)
          \end{exercises}
    \item \label{ex:pseudo-junta}     A function $h \co \Omega^n \to \Sigma$ is said to expressed as a \emph{pseudo-junta} if the following hold:
                                        \index{pseudo-junta}%
        There are ``juntas'' $f_1, \dots, f_m \co \Omega^n \to \tf$ with domains $J_1, \dots, J_m \subseteq [n]$ respectively.  Further, $g \co (\Omega \cup \{\ast\})^n \to \Sigma$, where $\ast$ is a new symbol not in~$\Omega$.  Finally, for each input $x \in \Omega^n$ we have $h(x) = g(y)$, where for $j \in [n]$,
        \[
            y_j = \begin{cases}
                    x_j & \text{if $j \in J_i$ for some $i$ with $f_i(x) = \true$,} \\
                    \ast & \text{else}.
                \end{cases}
        \]
        An alternative explanation is that on input~$x$, the junta $f_i$ decides whether the coordinates in its domain are ``notable''; then, $h(x)$~must be determined based only on the set of all notable coordinates.  Finally, if $\pi$ is a distribution on~$\Omega$, we say that the pseudo-junta has \emph{width-$k$ under~$\pi\xn$} if
        \[
            \E_{\bx \sim \pi\xn}[\#\{j : \by_j \neq \ast\}] \leq k;
        \]
        in other words, the expected number of notable coordinates is at most~$k$.  For $h \in L^2(\Omega^n, \pi\xn)$ we simply say that~$h$ is a \emph{$k$-pseudo-junta}.  Show that if such a $k$-pseudo-junta $h$ is $\bits$-valued, then $\Tinf[f] \leq 4k$. (Hint: Referring to the second statement in Proposition~\ref{prop:dir-laplacian-facts2}, consider the notable coordinates for both $\bx$ and $\bx' = (\bx_i, \dots, \bx_{i-1}, \bx'_i, \bx_{i+1}, \dots, \bx_n)$.)
    \item \label{ex:weak-bourgain} Establish the following further consequence of Bourgain's Sharp Threshold Theorem: Let $f \co \tf^n \to \tf$ be a  monotone function with $\Tinf[f\subp] \leq K$.  Assume $\Var[f] \geq .01$ and $0 < p \leq \exp(-cK^2)$, where $c$ is a large universal constant. Then there exists $T \subseteq [n]$ with $|T| \leq O(K)$ such that
        \[
            \Pr_{\bx \sim \pi_p\xn}[f(\bx) = \True \mid \bx_i = \True \text{ for all $i \in T$}] \geq \Pr_{\bx \sim \pi_p\xn}[f(\bx)= \True] + \exp(-O(K^2)).
        \]
        (Hint: Bourgain's Sharp Threshold Theorem yields a booster either toward~$\True$ or toward~$\False$.  In the former case you're easily done; to rule out the latter case, use the fact that $p|T| \ll \exp(-O(K^2))$.)
    \item  \label{ex:refine-bourgain-thresh}  Suppose that in Bourgain's Sharp Threshold Theorem we drop the assumption that $\Var[f] \geq .01$.  (Assume at least that $f$ is nonconstant.) Show that there is some $\tau$ with $|\tau| \geq \stddev[f] \cdot \exp(-O(\Tinf[f]^2/\Var[f]^2))$ such that
        \[
            \Pr_{\bx \sim \pi\xn}[\exists T \subseteq [n], |T| \leq O(\Tinf[f]/\Var[f]) \text{ such that $\bx_T$ is a } \tau\text{-booster}] \geq |\tau|.
        \]
        (Cf.~Exercise~\ref{ex:low-inf-large-coeff}.)
    \item \label{ex:non-3-colorability} In this exercise we give the beginnings of the idea of how Bourgain's Sharp Threshold Theorem can be used to show sharp thresholds for interesting monotone properties.  We will consider $\neg\text{3Col}$, the property of a random $v$-vertex graph $\bG \sim \calG(v,p)$ being
                                                    \index{threshold, sharp}%
                                                    \index{random graph}%
        non-$3$-colorable.
        \begin{exercises}
            \item Prove that the critical probability $p_c$ satisfies $p_c \leq O(1/v)$; i.e., establish that there is a universal constant~$C$ such that 
                \[
                    \Pr[\bG \sim \calG(v, C/v) \text{ is $3$-colorable}] = o_n(1).
                \]
                (Hint: Union-bound over all potential $3$-colorings.)
            \item Toward showing (non-)$3$-colorability has a sharp threshold, suppose the property had constant total influence at the critical probability.  Bourgain's Sharp Threshold Theorem would imply that there is a $\tau$ of constant magnitude such that for $\bG \sim \calG(v,p_c)$, there is a $|\tau|$ chance that $\bG$ contains a $\tau$-boosting induced subgraph~$\bG_T$.  There are two cases, depending on the sign of~$\tau$.  It's easy to rule out that the boost is in favor of $3$-colorability; the absence of a few edges shouldn't increase the probability of $3$-colorability by much (cf.~Exercise~\ref{ex:weak-bourgain}).  On the other hand, it might seem plausible that the \emph{presence} of a certain constant number of edges chould boost the probability of non-$3$-colorability by a lot.  For example, the presence of a $4$-clique immediately boosts the probability to~$1$.  However, the point is that \emph{at the critical probability} it is very unlikely that~$\bG$ contains a $4$-clique (or indeed, any ``local'' witness to non-$3$-colorability). Short of showing this, prove at least that the expected number of $4$-cliques in $\bG \sim \calG(v,p)$ is $o_v(1)$ unless $p = \Omega(v^{-2/3}) \gg p_c$.
        \end{exercises}
\end{exercises}

\subsection*{Notes.}                        \label{sec:adv-hypercon-notes}

As mentioned, the standard template introduced by Bonami~\cite{Bon70} for proving the Hypercontractivity Theorem for $\pm 1$ bits is to first prove the Two-Point Inequality, and then do the induction described in Exercise~\ref{ex:poly-hc-hc}.  Bonami's original proof of the Two-Point Inequality reduced to the $1 \leq p < q \leq 2$ case as we did, but then her calculus was a little more cumbersome.  We followed the proof of the Two-Point Inequality appearing in Janson~\cite{Jan97}.  Another approach to proving the Hypercontractivity Theorem is to derive it from the Log-Sobolev Inequality (Exercise~\ref{ex:log-sob}), as was done by Gross~\cite{Gro75}.

Our use of two-function hypercontractivity theorems to facilitate an inductive proof (and avoid the use of Exercise~\ref{ex:triangle-variant}) follows the communication/coding theory viewpoint of Ahlswede and G{\'a}cs~\cite{AG76}.  (We were also inspired by Mossel et~al.~\cite{MOR+06}, Barak et~al.~\cite{BBH+12}, and Kauers et~al.~\cite{KOTZ16}.) Ahlswede and G{\'a}cs established the close connection between hypercontractivity and small-set expansion in general product spaces, and independently obtained the sharp Hypercontractivity Theorem for $\pm 1$ bits, relying in part on a result of Witsenhausen~\cite{Wit75}.

Our statement of the Generalized Small-Set Expansion Theorem is
                                        \index{Small-Set Expansion Theorem!generalized}%
                                        \index{Small-Set Expansion Theorem!Reverse}%
modeled after the almost identical Reverse Small-Set Expansion Theorem, first proved by Mossel et~al.~\cite{MOR+06}.  The Reverse Hypercontractivity Inequality itself is due to Borell~\cite{Bor82}; the presentation in Exercises~\ref{ex:reverse-hypercon1}--\ref{ex:reverse-hypercon3} follows Mossel et~al.~\cite{MOR+06}. For more on reverse hypercontractivity, including the very surprising fact that the Reverse Hypercontractivity Inequality
                                    \index{Hypercontractivity Theorem!Reverse}%
holds with no change in constants for every product probability space, see Mossel, Oleszkiewicz, and Sen~\cite{MOS12b}.

As mentioned in Chapter~\ref{chap:hypercontractivity} the definition of a hypercontractive random variable is due to Krakowiak and Szulga~\cite{KS88}. Many of the basic facts from Section~\ref{sec:general-hypercon} (and also Exercise~\ref{ex:sum-hc-hc}) are from this work and the earlier work of Borell~\cite{Bor84};
                                    \index{hypercontractivity}%
see also various other works~\cite{KW92,Jan97,Szu98,MOO10}. As mentioned, the main part of Theorem~\ref{thm:discrete-hypercon} (the case of biased bits) is essentially from Lata{\l}a and Oleszkiewicz~\cite{LO00}; see also Oleszkiewicz~\cite{Ole03}.  Our Exercise~\ref{ex:lo} fleshes out (and slightly simplifies) their computations but introduces no new idea.  Earlier works~\cite{BKK+92,Tal94,FK96b,Fri98} had established forms of the General Hypercontractivity Theorem for $\lambda$-biased bits, giving as applications KKL-type theorems in this setting with the correct asymptotic dependence on~$\lambda$.  We should also mention that the sharp Log-Sobolev Inequality for product space domains (mentioned in Exercise~\ref{ex:general-log-sob}) was derived independently of Lata{\l}a and Oleszkiewicz's work by Higuchi and Yoshida~\cite{HY95} (without proof), by  Diaconis and Saloff-Coste~\cite{DS96} (with proof), and possibly also by Oscar Rothaus (see~\cite{BL98}).  Unlike in the case of uniform $\pm 1$ bits, it's not known how to derive Lata{\l}a and Oleszkiewicz's optimal biased hypercontractive inequality from the optimal biased Log-Sobolev Inequality.

Kahane~\cite{Kah68} has been credited with pioneering the randomization/ symmetrization trick for random variables.
                                                \index{randomization/symmetrization}%
The entirety of Section~\ref{sec:rand-symm} is due to Bourgain~\cite{Bou79}, though our presentation was significantly informed by the expertise of Krzysztof Oleszkiewicz (and our proof of Lemma~\ref{lem:unsymmetrize} is slightly different).  Like Bourgain, we don't give any explicit dependence for the constant~$C_q$ in Theorem~\ref{thm:projection-bound-general}; however, Kwapie\'{n}~\cite{Kwa10} has shown that one may take $C_{q'} = C_q = O(q/\log q)$ for $q \geq 2$.  Our proof of Bourgain's Theorem~\ref{thm:bourgain-sharp} follows the original~\cite{Bou99} extremely closely, though we also valued the easier-to-read version of Bal~\cite{Bal13}.

The biased edge-isoperimetric inequality~\eqref{eqn:weak-p-biased-isoperim} from Exercise~\ref{ex:general-log-sob} was proved by induction on~$n$, without the additional~$o_p(1)$ error, by Russo~\cite{Rus82} (and also independently by Kahn and Kalai~\cite{KK07}).  We remark that this work and the earlier~\cite{Rus81} already contain the germ of the idea that monotone functions with small influences have sharp thresholds.  Regarding the sharp threshold for $3$-colorability discussed in Exercise~\ref{ex:non-3-colorability}, Alon and Spencer \cite{AS08} contains a nice elementary proof of the fact that at the critical probability for $3$-colorability, every subgraph on $\eps v$ vertices is $3$-colorable, for some universal $\eps > 0$.  The existence of a sharp threshold for $k$-colorability was proven by Achlioptas and Friedgut~\cite{AF99}, with Achlioptas and Naor~\cite{AN05} essentially determining the location.

\chapter{Gaussian space and Invariance Principles}                         \label{chap:invariance}

The final destination of this chapter is a proof of the following theorem due to Mossel, O'Donnell, and Oleszkiewicz~\cite{MOO05a,MOO10}, first mentioned in Chapter~\ref{sec:majority}:
\begin{named}{Majority Is Stablest Theorem}
                                    \index{Majority Is Stablest Theorem}%
Fix $\rho \in (0,1)$.  Let $f \btI$ have $\E[f] = 0$.  Then, assuming $\MaxInf[f] \leq \eps$, or more generally that~$f$ has no $(\eps,\eps)$-notable coordinates,
\[
\Stab_\rho[f] \leq 1 - \tfrac{2}{\pi} \arccos \rho + o_\eps(1).
\]
\end{named}
\noindent This bound is tight; recalling Theorem~\ref{thm:maj-stab}, the bound $1 - \tfrac{2}{\pi} \arccos \rho$ is achieved by taking $f = \Maj_n$, the volume-$\half$ Hamming ball indicator,  for $n \to \infty$.  More generally, in Section~\ref{sec:MIST} we'll prove the General-Volume Majority Is Stablest Theorem,
                                    \index{Majority Is Stablest Theorem}%
which shows that for \emph{any} fixed volume, ``Hamming ball indicators have maximal noise stability among small-influence functions''.

There are two main ideas underlying this theorem.  The first is that ``functions on Gaussian space'' are a special case of small-influence Boolean functions. In other words, a Boolean function may always be a ``Gaussian function in disguise''.  This motivates \emph{analysis of Gaussian functions}, the topic introduced in Sections~\ref{sec:gaussian} and~\ref{sec:hermite}.  It also means that a prerequisite for proving the (General-Volume) Majority Is Stablest Theorem is proving its Gaussian special cases, namely, Borell's Isoperimetric Theorem
                                            \index{Borell's Isoperimetric Theorem}%
(Section~\ref{sec:borell}) and the Gaussian Isoperimetric Inequality (Section~\ref{sec:gaussian-surface-area}).  In many ways, working in the Gaussian setting is nicer because tools like rotational symmetry and differentiation are available.

The second idea is the converse to the first: In Section~\ref{sec:invariance} we prove the \emph{Invariance Principle}, a generalization of the Berry--Esseen Central Limit Theorem, which shows that any low-degree (or uniformly noise-stable) Boolean function with small influences is approximable by a Gaussian function.  In fact, the Invariance Principle roughly shows that given such a Boolean function, if you plug \emph{any} independent mean-$0$, variance-$1$ random variables into its Fourier expansion, the distribution doesn't change much.  In Section~\ref{sec:MIST} we use the Invariance Principle to prove the Majority Is Stablest Theorem by reducing to its Gaussian special case, Borell's Isoperimetric Theorem.

\section{Gaussian space and the Gaussian noise operator}              \label{sec:gaussian}
                                                    \index{analysis of Gaussian functions|(}%
We begin with a few definitions concerning Gaussian space.
                                                    \index{Gaussian space}%
\begin{notation}
    Throughout this chapter we write $\vphi$ for the pdf of a standard Gaussian random variable, $\vphi(z) = \frac{1}{\sqrt{2\pi}}\exp(-\half z^2)$.  We also write $\Phi$~for its cdf, and $\olPhi$ for the complementary cdf $\olPhi(t) = 1 - \Phi(t) = \Phi(-t)$.  We write $\bz \sim \normal(0,1)^n$ to denote that~$\bz = (\bz_1, \dots, \bz_n)$ is a random vector in~$\R^n$ whose components~$\bz_i$ are independent Gaussians.  Perhaps the most important property of this distribution is that it's rotationally symmetric; this follows because the pdf at~$z$ is $\frac{1}{(2\pi)^{n/2}}\exp(-\half(z_1^2 + \cdots + z_n^2))$, which depends only on the length~$\|z\|_2^2$ of~$z$.
\end{notation}
\begin{definition}
    For $n \in \N^+$ and $1 \leq p \leq \infty$ we write $L^p(\R^n,\gamma)$ for the space of Borel functions $f \co \R^n \to \R$ that have finite $p$th moment $\|f\|_p^p$ under the Gaussian measure (the ``$\gamma$'' stands for Gaussian).  Here for a function $f$ on Gaussian space we use the notation
    \[
        \|f\|_p = \E_{\bz \sim \normal(0,1)^n}[|f(\bz)|^p]^{1/p}.
    \]
                                                        \nomenclature[Lp]{$L^p(\R^n,\gamma)$}{the space of Borel functions $f \co \R^n \to \R$ satisfying $\E_{\bz \sim \normal(0,1)^n}[|f(\bz)|^p] < \infty$}%
    All functions $f \co \R^n \to \R$ and sets $A \subseteq \R^n$ are henceforth assumed to be Borel without further mention.
\end{definition}
\begin{notation}
    When it's clear from context that $f$ is a function on Gaussian space we'll use shorthand notation like $\E[f] = \Ex_{\bz \sim \normal(0,1)^n}[f(\bz)]$.  If $f = 1_A$ is the $0$-$1$ indicator of a subset $A \subseteq \R^n$ we'll also write
    \[
        \gvol(A) = \E[1_A] = \Pr_{\bz \sim \normal(0,1)^n}[\bz \in A]
    \]
    for the \emph{Gaussian volume} of~$A$.
                                                    \index{Gaussian volume}%
                                                    \nomenclature[volg]{$\gvol(A)$}{$\Pr_{\bz \sim \normal(0,1)^n}[\bz \in A]$, the Gaussian volume of~$A$}%
\end{notation}
\begin{notation}
    For $f, g \in L^2(\R^n,\gamma)$ we use the inner product notation $\la f, g \ra = \E[fg]$, under which $L^2(\R^n,\gamma)$ is a separable Hilbert space.
\end{notation}

If you're only interested in Boolean functions $f \btb$ you might wonder why it's necessary to study Gaussian space.  As discussed at the beginning of the chapter, the reason is that functions on Gaussian space are \emph{special cases} of Boolean functions.  Conversely, even if you're only interested in studying functions of Gaussian random variables, sometimes the easiest proof technique involves ``simulating'' the Gaussians using sums of random bits.  Let's discuss this in a little more detail.  Recall that the Central Limit Theorem
                                                \index{Central Limit Theorem}%
tells us that for $\bx \sim \bits^M$, the distribution of $\frac{1}{\sqrt{M}}(\bx_1 + \cdots + \bx_M)$ approaches that of a standard Gaussian as $M \to \infty$.  This is the sense in which a standard Gaussian random variable $\bz \sim \normal(0,1)$ can be ``simulated'' by random bits.  If we want~$d$ independent Gaussians we can simulate them by summing up $M$ independent $d$-dimensional vectors of random bits.
\begin{definition}
                                   \nomenclature[BitsToGaussians]{$\BitsToGaussians{M}{d}$}{on input the bit matrix $x \in \bits^{d \times M}$, has output $z \in \R^d$ equal to $\frac{1}{\sqrt{M}}$ times the column-wise sum of~$x$; if $d$ is omitted it's taken to be~$1$}
    The function $\BitsToGaussians{M}{} \co \bits^{M} \to \R$ is defined by
    \[
        \BitsToGaussians{M}{}(x) = \tfrac{1}{\sqrt{M}}(x_1 + \cdots + x_M).
    \]
    More generally, the function $\BitsToGaussians{M}{d} \co \bits^{dM} \to \R^d$ is defined on an input $x \in \bits^{d \times M}$, thought of as a matrix of column vectors $\vec{x}_1, \dots, \vec{x}_M \in \bits^d$, by
    \[
        \BitsToGaussians{M}{d}(x) = \tfrac{1}{\sqrt{M}}(\vec{x}_1 + \cdots + \vec{x}_M).
    \]
\end{definition}
                                         \index{Gaussian random variable!simulated by bits}%
Although $M$ needs to be large for this simulation to be accurate, many of the results we've developed in the analysis of Boolean functions $f \co \bits^M \to \R$ are independent of~$M$.  A further key point is that this simulation preserves polynomial degree: if $p(\bz_1, \dots, \bz_d)$ is a degree-$k$ polynomial applied to~$d$ independent standard Gaussians, the ``simulated version'' $p \circ\BitsToGaussians{M}{d} \co \bits^{dM} \to \R$ is a degree-$k$ Boolean function.  These facts allow us to transfer many results from the analysis of Boolean functions to the analysis of Gaussian functions.  On the other hand, it also means that to fully understand Boolean functions, we need to understand the ``special case'' of functions on Gaussian space: a Boolean function may essentially be a function on Gaussian space ``in disguise''.  For example, as we saw in Chapter~\ref{sec:maj-coefficients}, there is a sense in which the majority function $\Maj_n$ ``converges'' as $n \to \infty$; what it's converging to is the sign function on $1$-dimensional Gaussian space, $\sgn \in L^1(\R, \gamma)$.

We'll begin our study of Gaussian functions by developing the analogue of the most important operator on Boolean functions, namely the noise operator~$\T_\rho$.  Suppose we take a pair of $\rho$-correlated $M$-bit strings $(\bx, \bx')$ and use them to form approximate Gaussians,
\[
    \by = \BitsToGaussians{M}{}(\bx), \qquad \by' = \BitsToGaussians{M}{}(\bx').
\]
For each~$M$ it's easy to compute that $\E[\by] = \E[\by'] = 0$, $\Var[\by] = \Var[\by'] = 1$, and $\E[\by \by'] = \rho$.  As noted in Chapter~\ref{sec:majority}, a multidimensional version of the Central Limit Theorem (see, e.g., Exercises~\ref{ex:ltf-stab-error},~\ref{ex:multivar-be}) tells us that the joint distribution of $(\by,\by')$ converges to a pair of Gaussian random variables with the same properties.  We call these $\rho$-correlated Gaussians.
\begin{definition}
    For $-1 \leq \rho \leq 1$, we say that the random variables $(\bz, \bz')$ are \emph{$\rho$-correlated (standard) Gaussians}
                                            \index{correlated Gaussians}%
                                            \index{rho-correlated Gaussians@$\rho$-correlated Gaussians|seeonly{correlated Gaussians}}%
    if they are jointly Gaussian and satisfy $\E[\bz] = \E[\bz'] = 0$, $\Var[\bz] = \Var[\bz'] = 1$, and $\E[\bz \bz'] = \rho$.  In other words, if
    \[
        (\bz, \bz') \sim \normal\left(\begin{bmatrix} 0 \\ 0 \end{bmatrix}, \begin{bmatrix} 1 & \rho \\ \rho & 1 \end{bmatrix}\right).
    \]
    Note that the definition is symmetric in $\bz$, $\bz'$ and that each is individually distributed as $\normal(0,1)$.
\end{definition}
\begin{fact}                                \label{fact:corr-gaussians}
    An equivalent definition is to say that $\bz = \la \vec{u}, \vec{\bg}\ra$ and $\bz' = \la \vec{v}, \vec{\bg} \ra$, where $\vec{\bg} \sim \normal(0,1)^d$ and $\vec{u}, \vec{v} \in \R^d$ are any two unit vectors satisfying $\la \vec{u}, \vec{v} \ra = \rho$.  In particular we may choose $d = 2$, $\vec{u} = (1,0)$, and $\vec{v} = (\rho, \sqrt{1-\rho^2})$, thereby defining $\bz = \bg_1$ and $\bz' = \rho \bg_1 + \sqrt{1-\rho^2} \bg_2$.
\end{fact}
\begin{remark}                              \label{rem:sincos-corr}
    In Fact~\ref{fact:corr-gaussians} it's often convenient to write $\rho = \cos \theta$ for some $\theta \in \R$, in which case we may define the $\rho$-correlated Gaussians as $\bz = \la \vec{u}, \vec{\bg} \ra$ and $\bz' = \la \vec{v}, \vec{\bg}\ra$ for any unit vectors $\vec{u}, \vec{v}$ making an angle of~$\theta$; e.g., $\vec{u} = (1,0)$, $\vec{v} = (\cos\theta, \sin \theta)$.
\end{remark}
\begin{definition}
    For a fixed $z \in \R$ we say random variable $\bz'$ is a \emph{Gaussian $\rho$-correlated to~$z$}, written $\bz' \sim N_\rho(z)$, if $\bz'$ is distributed as $\rho z + \sqrt{1-\rho^2} \bg$ where $\bg \sim \normal(0,1)$.  By Fact~\ref{fact:corr-gaussians}, if we draw $\bz \sim \normal(0,1)$ and then form $\bz' \sim N_\rho(\bz)$, we obtain a $\rho$-correlated pair of Gaussians~$(\bz, \bz')$.
                                            \nomenclature[Nrhoz]{$N_\rho(z)$}{when $z \in \R^n$, denotes the probability distribution of $\rho z + \sqrt{1-\rho^2} \bg$ where $\bg \sim \normal(0,1)^n$}%
\end{definition}
\begin{definition}
    For $-1 \leq \rho \leq 1$ and $n \in \N^+$ we say that the $\R^n$-valued random variables $(\bz, \bz')$ are \emph{$\rho$-correlated $n$-dimensional Gaussian random vectors}
                                            \index{correlated Gaussians!vectors}%
    if each component pair $({\bz}_1, {\bz}'_1)$, \dots, $({\bz}_n, {\bz}'_n)$ is a $\rho$-correlated pair of Gaussians, and the~$n$ pairs are mutually independent.  We also naturally extend the definition of $\bz' \sim N_\rho(z)$ to the case of $z \in \R^n$; this means $\bz' = \rho z + \sqrt{1-\rho^2} \bg$ for $\bg \sim \normal(0,1)^n$.
\end{definition}
\begin{remark}                          \label{rem:rot-symm-cor-gauss}
     Thus, if $\bz \sim \normal(0,1)^n$ and then $\bz' \sim N_\rho(\bz)$ we obtain a $\rho$-correlated $n$-dimensional pair~$(\bz, \bz')$.  It  follows from this that $(\bz,\bz')$ has the same distribution as $(Q \bz, Q \bz')$ for any rotation $Q$ on $\R^n$.
\end{remark}

Now we can introduce the Gaussian analogue of the noise operator.
\begin{definition}
                                        \label{def:OU}%
                                        \index{Gaussian noise operator}%
                                        \index{noise operator!Gaussian|seeonly{Gaussian noise operator}}%
                                        \index{Ornstein--Uhlenbeck semigroup|seeonly{Gaussian noise operator}}%
                                        \index{Mehler transform|seeonly{Gaussian noise operator}}%
                                        \index{U$_\rho$|seeonly{Gaussian noise operator}}%
                                        \nomenclature[Urho]{$\U_\rho$}{the Gaussian noise operator: $\U_\rho f(z) = \Ex_{\bz' \sim N_\rho(z)}[f(\bz')]$}%
    For $\rho \in [-1,1]$, the \emph{Gaussian noise operator} $\U_\rho$ is the linear operator defined on the space of functions $f \in L^1(\R^n,\gamma)$ by
    \[
        \U_\rho f(z) = \E_{\bz' \sim N_\rho(z)}[f(\bz')] = \E_{\bg \sim \normal(0,1)^n}[f(\rho z + \sqrt{1-\rho^2} \bg)].
    \]
\end{definition}
\begin{fact}                            \label{fact:Urho-plug-in}
    (Exercise~\ref{ex:Urho-plug-in}.)  If $f \in L^1(\R^n,\gamma)$ is an $n$-variate multilinear polynomial, then $\U_\rho f(z) = f(\rho z)$.
\end{fact}
\begin{remark}
    Our terminology is nonstandard.  The Gaussian noise operators are usually collectively referred to as the \emph{Ornstein--Uhlenbeck semigroup} (or sometimes as the \emph{Mehler transforms}).  They are typically defined for $\rho = e^{-t} \in [0,1]$ (i.e., for $t \in [0,\infty]$) by
    \[
        \uP_t f(z) = \E_{\bg \sim \normal(0,1)^n}[f(e^{-t} z + \sqrt{1-e^{-2t}} \bg)] = \U_{e^{-t}} f(z).
    \]
    The term ``semigroup'' refers to the fact that the operators satisfy $\uP_{t_1} \uP_{t_2} = \uP_{t_1+t_2}$, i.e., $\U_{\rho_1} \U_{\rho_2} = \U_{\rho_1 \rho_2}$ (which holds for all $\rho_1, \rho_2 \in [-1,1]$; see Exercise~\ref{ex:semigroup-gaussian}).
                                                \index{semigroup property}%
\end{remark}
Before going further let's check that $\U_\rho$ is a bounded operator on all of $L^p(\R^n,\gamma)$ for $p \geq 1$; in fact, it's a contraction (cf.~Exercise~\ref{ex:T-contracts}):
\begin{proposition}                                     \label{prop:U-contracts}
    For each $\rho \in [-1,1]$ and $1 \leq p \leq \infty$ the operator $\U_\rho$ is a contraction on $L^p(\R^n,\gamma)$; i.e., $\|\U_\rho f\|_p \leq \|f\|_p$.
\end{proposition}
\begin{proof}
    The proof for $p = \infty$ is easy; otherwise, the result follows from Jensen's inequality, using that $t \mapsto |t|^p$ is convex:
    \begin{align*}
        \|\U_\rho f\|_p^p = \E_{\bz \sim \normal(0,1)^n}[|\U_\rho f(\bz)|^p] &= \E_{\bz \sim \normal(0,1)^n}\left[\left|\E_{\bz' \sim N_\rho(\bz)}[f(\bz')]\right|^p\right] \\
        &\leq \E_{\bz \sim \normal(0,1)^n}\left[\E_{\bz' \sim N_\rho(\bz)}[|f(\bz')|^p]\right] = \|f\|_p^p. \qedhere
    \end{align*}
\end{proof}
As in the Boolean case, you should think of the Gaussian noise operator as having a ``smoothing'' effect on functions.  As~$\rho$ goes from~$1$ down to~$0$, $\U_\rho f$ involves averaging~$f$'s values over larger and larger neighborhoods. In particular $\U_1$ is the identity operator, $\U_1f = f$, and $\U_0 f = \E[f]$, the constant function.  In Exercises~\ref{ex:Urho-smooths},~\ref{ex:Urho-strongly-cts} you are asked to verify the following facts, which say that for any~$f$, as $\rho \to 1^-$ we get a sequence of smooth~(i.e.,~$\calC^\infty$) functions $\U_\rho f$ that tend to~$f$.
\begin{proposition}                                     \label{prop:Urho-smooths}
    Let $f \in L^1(\R^n,\gamma)$ and let $-1 < \rho < 1$.  Then $\U_\rho f$ is a smooth function.
\end{proposition}
\begin{proposition}                                     \label{prop:Urho-strongly-cts}
    Let $f \in L^1(\R^n,\gamma)$.  As $\rho \to 1^-$ we have $\|\U_\rho f - f\|_1 \to 0$.
\end{proposition}
Having defined the Gaussian noise operator, we can also make the natural definition of Gaussian noise stability (for which we'll use the same notation as in the Boolean case):
\begin{definition}
    For $f \in L^2(\R^n,\gamma)$ and $\rho \in [-1,1]$, the \emph{Gaussian noise stability of $f$ at $\rho$} is defined to be
    \[
        \Stab_\rho[f] = \E_{\substack{(\bz, \bz') \text{ $n$-dimensional} \\ \text{$\rho$-correlated Gaussians}}}[f(\bz) f(\bz')] = \la f, \U_\rho f \ra = \la \U_\rho f, f \ra.
    \]
    (Here we used that $(\bz', \bz)$ has the same distribution as $(\bz, \bz')$ and hence $\U_\rho$ is self-adjoint.)
\end{definition}
\begin{example}                             \label{eg:sheppard}
    Let $f \co \R \to \{0,1\}$ be the $0$-$1$ indicator of the nonpositive halfline: $f = 1_{(-\infty, 0]}$.  Then
    \begin{equation}                    \label{eqn:01-sheppard}
        \Stab_\rho[f] = \E_{\substack{(\bz, \bz') \text{ $\rho$-correlated} \\ \text{standard Gaussians}}}[f(\bz) f(\bz')] = \Pr[\bz \leq 0, \bz' \leq 0] = \frac12 - \frac{1}{2}\frac{\arccos \rho}{\pi},
    \end{equation}
    with the last equality being \emph{Sheppard's Formula},
                                            \index{Sheppard's Formula}%
    which we stated in Section~\ref{sec:majority} and now prove.
\end{example}
\begin{proof}[Proof of Sheppard's Formula]
    Since $(-\bz, -\bz')$ has the same distribution as $(\bz, \bz')$, proving~\eqref{eqn:01-sheppard} is equivalent to proving
    \[
         \Pr[\bz \leq 0, \bz' \leq 0 \text{ or } \bz > 0, \bz' > 0] = 1  - \frac{\arccos \rho}{\pi}.
    \]
    The complement of the above event is the event that $f(\bz) \neq f(\bz')$ (up to measure~$0$); thus it's further equivalent to prove
    \begin{equation}                                    \label{eqn:elegant-sheppard}
       \Pr_{\substack{(\bz, \bz') \\\text{$\cos \theta$-correlated}}}[f(\bz) \neq f(\bz')] = \tfrac{\theta}{\pi}
    \end{equation}
    for all $\theta \in [0,\pi]$.  As in Remark~\ref{rem:sincos-corr}, this suggests defining $\bz = \la \vec{u}, \vec{\bg} \ra$, $\bz' = \la \vec{v}, \vec{\bg} \ra$, where $\vec{u}, \vec{v} \in \R^2$ is some fixed pair of unit vectors making an angle of~$\theta$, and $\vec{\bg}  \sim \normal(0,1)^2$.  Thus we want to show
    \[
       \Pr_{\vec{\bg} \sim \normal(0,1)^2}[\la \vec{u}, \vec{\bg} \ra \leq 0 \text{ \& }  \la \vec{v}, \vec{\bg} \ra > 0 \text{ or vice versa}] = \tfrac{\theta}{\pi}.
    \]
    But this last identity is easy: If we look at the diameter of the unit circle that is perpendicular to~$\vec{\bg}$, then the event above is equivalent (up to measure~$0$) to the event that this diameter ``splits''~$\vec{u}$ and $\vec{v}$.  By the rotational symmetry of~$\vec{\bg}$, the probability is evidently~$\theta$ (the angle between $\vec{u}, \vec{v}$) divided by~$\pi$ (the range of angles for the diameter).
\end{proof}

\begin{corollary}                               \label{cor:gstab-halfspace-origin}
    Let $H \subset \R^n$ be any halfspace (open or closed) with boundary hyperplane containing the origin.  Let $h = \pm 1_{H}$.  Then $\Stab_\rho[h] = 1 - \tfrac{2}{\pi} \arccos \rho$.
\end{corollary}
\begin{proof}
    We may assume~$H$ is open (since its boundary has measure~$0$).  By the rotational symmetry of correlated Gaussians (Remark~\ref{rem:rot-symm-cor-gauss}), we may rotate~$H$ to the form $H = \{z \in \R^n : z_1 > 0\}$.  Then it's clear that the noise stability of $h = \pm 1_H$ doesn't depend on~$n$, i.e., we may assume $n = 1$.  Thus $h = \sgn = 1-2f$, where $f = 1_{(-\infty, 0]}$ as in Example~\ref{eg:sheppard}.  Now if $(\bz, \bz')$ denote $\rho$-correlated standard Gaussians, it follows from~\eqref{eqn:01-sheppard} that
    \begin{align*}
        \Stab_\rho[h] = \E[h(\bz) h(\bz')] &= \E[(1-2f(\bz))(1-2f(\bz'))] \\ &= 1 - 4\E[f] + 4\Stab_\rho[f] = 1 - \tfrac{2}{\pi} \arccos \rho. \qedhere
    \end{align*}
\end{proof}
\begin{remark}                                      \label{rem:maj-stab}
    The quantity $\Stab_\rho[\sgn] = 1 - \tfrac{2}{\pi} \arccos \rho$ is also precisely the limiting noise stability of~$\maj_n$, as stated in Theorem~\ref{thm:maj-stab} and justified in Chapter~\ref{sec:majority}.
\end{remark}

\medskip

                                                    \index{Hypercontractivity Theorem!Gaussian|(}%
We've defined the key Gaussian noise operator $\U_\rho$ and seen (Proposition~\ref{prop:U-contracts}) that it's a contraction on all~$L^p(\R^n,\gamma)$.  Is it also hypercontractive?  In fact, we'll now show that the Hypercontractivity Theorem for uniform~$\pm 1$ bits holds identically in the Gaussian setting.  The proof is simply a reduction to the Boolean case, and it will use the following standard fact (see Janson~\cite[Theorem~2.6]{Jan97} or Teuwen~\cite[Section~1.3]{Teu12} for the proof in case of~$L^2$; to extend to other~$L^p$ you can use Exercise~\ref{ex:really-simple-functions}):

\begin{theorem}                                     \label{thm:polys-dense}
    For each $n \in \N^+$, the set of multivariate polynomials is dense in $L^p(\R^n,\gamma)$ for all $1 \leq p < \infty$.
\end{theorem}
\begin{named}{Gaussian Hypercontractivity Theorem}
Let $f, g \in L^1(\R^n,\gamma)$, let $r, s \geq 0$, and assume $0 \leq \rho \leq \sqrt{rs} \leq 1$.  Then
    \[
        \la f, \U_\rho g \ra = \la \U_\rho f, g \ra = \E_{\substack{(\bz, \bz') \textnormal{ $\rho$-correlated}\\ \textnormal{$n$-dimensional Gaussians}}}[f(\bz)g(\bz')] \leq \|f\|_{1+r}\|g\|_{1+s}.
    \]
\end{named}
\begin{proof}
    (We give a sketch; you are asked to fill in the details in Exercise~\ref{ex:gaussian-hypercon-theorem}.)  We may assume that $f \in L^{1+r}(\R^n,\gamma)$ and $g \in L^{1+s}(\R^n,\gamma)$.  We may also assume $f, g \in L^2(\R^n,\gamma)$ by a truncation and monotone convergence argument; thus the left-hand side is finite by Cauchy--Schwarz. Finally, we may assume that~$f$ and~$g$ are multivariate polynomials, using Theorem~\ref{thm:polys-dense}.   For fixed $M \in \N^+$ we consider  ``simulating'' $(\bz, \bz')$ using bits. More specifically, let $(\bx, \bx') \in \bits^{nM}  \times \bits^{nM}$ be a pair $\rho$-correlated random strings and define the joint $\R^n$-valued random variables $\by, \by'$ by
    \[
        \by = \BitsToGaussians{M}{n}(\bx), \qquad \by' = \BitsToGaussians{M}{n}(\bx').
    \]
    By a multidimensional Central Limit Theorem we have that
    \[
        \E[f(\by)g(\by')] \xrightarrow{M \to \infty} \E_{\substack{(\bz, \bz') \\ \text{$\rho$-correlated}}}[f(\bz)g(\bz')].
    \]
    (Since $f$ and $g$ are polynomials, we can even reduce to a Central Limit Theorem for bivariate monomials.)  We further have
    \[
        \E[|f(\by)|^{1+r}]^{1/(1+r)}  \xrightarrow{M \to \infty} \E_{\bz \sim \normal(0,1)^n}[|f(\bz)|^{1+r}]^{1/(1+r)}
    \]
    and similarly for~$g$.  (This can also be proven by the multidimensional Central Limit Theorem, or by the one-dimensional Central Limit Theorem together with some tricks.)  Thus it suffices to show
    \[
        \E[f(\by)g(\by')] \leq \E[|f(\by)|^{1+r}]^{1/(1+r)} \E[|g(\by')|^{1+s}]^{1/(1+s)}
    \]
    for any fixed~$M$.  But we can express $f(\by) = F(\bx)$ and $g(\by') = G(\bx')$ for some $F, G \co \bits^{nM} \to \R$ and so the above inequality holds by the Two-Function Hypercontractivity Theorem (for $\pm 1$ bits).
\end{proof}
An immediate corollary, using the proof of Proposition~\ref{prop:one-to-two-function-hc}, is the standard one-function form of hypercontractivity:
\begin{theorem}
    Let $1 \leq p \leq q \leq \infty$ and let $f \in L^p(\R^n,\gamma)$.  Then $\|\U_\rho f\|_q \leq \|f\|_p$ for $0 \leq \rho \leq \sqrt{\tfrac{p-1}{q-1}}$.
\end{theorem}
                                                    \index{Hypercontractivity Theorem!Gaussian|)}%

We conclude this section by discussing the Gaussian space analogue of the discrete Laplacian operator. Taking our cue from Exercise~\ref{ex:laplacian-as-deriv} we make the following definition:
\begin{definition}
                                                \index{Ornstein--Uhlenbeck operator}%
                                                \index{number operator|seeonly{Ornstein--Uhlenbeck operator}}%
    The \emph{Ornstein--Uhlenbeck operator}~$\Lap$ (also called the \emph{infinitesimal generator} of the Ornstein--Uhlenbeck semigroup, or the \emph{number operator}) is the linear operator acting on functions $f \in L^2(\R^n,\gamma)$ by
    \[
        \Lap f = \frac{d}{d\rho} \U_{\rho} f \Bigr|_{\rho = 1} = -\frac{d}{dt} \U_{e^{-t}} f \Bigr|_{t = 0}
    \]
    (provided $\Lap f$ exists in $L^2(\R^n, \gamma)$).     Notational warning: It is common to see this as the definition of $-\Lap$.
\end{definition}
\begin{remark}
We will not be completely careful about the domain of the operator~$\Lap$ in this section; for precise details, see Exercise~\ref{ex:gauss-Lap-domain}.
\end{remark}
\begin{proposition}                                     \label{prop:alternate-OU}
    Let $f \in L^2(\R^n,\gamma)$ be in the domain of $\Lap$, and further assume for simplicity that $f$~is $\calC^3$.  Then we have the formula
    \[
        \Lap f(x) = x \cdot \grad f(x) -\Delta f(x),
    \]
    where $\Delta$~denotes the usual Laplacian differential operator, $\cdot$~denotes the dot product, and $\grad$~denotes the gradient.
\end{proposition}
\begin{proof}
    We give the proof in the case $n = 1$, leaving the general case to Exercise~\ref{ex:alternate-OU}.  We have
    \begin{equation}                    \label{eqn:lap-formula}
        \Lap f(x) = -\lim_{t \to 0^+} \frac{\Ex_{\bz \sim \normal(0,1)}[f(e^{-t} x + \sqrt{1-e^{-2t}} \bz)] - f(x)}{t}.
    \end{equation}
    Applying Taylor's theorem to $f$ we have
    \[
        f(e^{-t} x + \sqrt{1-e^{-2t}} \bz) \approx f(e^{-t} x) + f'(e^{-t} x) \sqrt{1-e^{-2t}} \bz + \tfrac12 f''(e^{-t} x) (1-e^{-2t})\bz^2,
    \]
    where the $\approx$ denotes that the two quantities differ by at most $C (1-e^{-2t})^{3/2}|\bz|^3$ in absolute value, for some constant~$C$ depending on~$f$ and~$x$.  Substituting this into~\eqref{eqn:lap-formula} and using $\E[\bz] = 0$, $\E[\bz^2] = 1$, and that $\E[|\bz|^3]$ is an absolute constant, we get
    \[
        \Lap f(x) = -\lim_{t \to 0^+} \left(\frac{f(e^{-t} x) - f(x)}{t} + \frac{\tfrac12 f''(e^{-t} x) (1-e^{-2t})}{t}\right),
    \]
    using the fact that $\frac{(1-e^{-2t})^{3/2}}{t} \to 0$.  But this is easily seen to be $xf'(x) - f''(x)$, as claimed.
\end{proof}
An easy consequence of the semigroup property is the following:
\begin{proposition}                                     \label{prop:L-vs-U}
    The following equivalent identities hold:
    \begin{align*}
        \frac{d}{d\rho} \U_\rho f = \rho^{-1} \Lap \U_{\rho}& f = \rho^{-1} \U_{\rho} \Lap f, \\
        \frac{d}{dt} \U_{e^{-t}} f = -\Lap \U_{e^{-t}}& f = -\U_{e^{-t}} \Lap f.
    \end{align*}
\end{proposition}
\begin{proof}
    This follows from
    \begin{align*}
        \frac{d}{dt} \U_{e^{-t}} f(x) &= \lim_{\delta \to 0} \frac{\U_{e^{-t - \delta}} f(x) - \U_{e^{-t}} f(x) }{\delta} \\
        &= \lim_{\delta \to 0} \frac{\U_{e^{-\delta}} \U_{e^{-t}} f(x) - \U_{e^{-t}} f(x) }{\delta} = \lim_{\delta \to 0} \frac{\U_{e^{-t}} \U_{e^{-\delta}} f(x) - \U_{e^{-t}} f(x) }{\delta}. \qedhere
    \end{align*}
\end{proof}
We also have the following formula:
\begin{proposition}                                     \label{prop:dirichlet2}
    Let $f, g \in L^2(\R^n, \gamma)$ be in the domain of~$\Lap$, and further assume for simplicity that they are~$\calC^3$.  Then
    \begin{equation}                        \label{eqn:dirichlet2}
        \la f, \Lap g \ra = \la \Lap f, g \ra = \la \grad f, \grad g \ra.
    \end{equation}
\end{proposition}
\begin{proof}
    It suffices to prove the inequality on the right of~\eqref{eqn:dirichlet2}.  We again treat only the case of $n = 1$, leaving the general case to Exercise~\ref{ex:dirichlet2}.  Using Proposition~\ref{prop:alternate-OU},
    \begin{align*}
        \la \Lap f, g \ra &= \int_\R (x f'(x) - f''(x))g(x) \vphi(x)\,dx \\
                          &= \int_\R x f'(x) g(x)\vphi(x) \,dx  + \int_{\R} f'(x) (g \vphi)'(x)\,dx  \tag{integration by parts} \\
                          &= \int_\R x f'(x) g(x)\vphi(x) \,dx  + \int_{\R} f'(x) (g'(x) \vphi(x) + g(x) \vphi'(x))\,dx  \\
                          &= \int_{\R} f'(x) g'(x) \vphi(x)\,dx,
    \end{align*}
    using the fact that $\vphi'(x) = -x \vphi(x)$.
\end{proof}

Finally, by differentiating the Gaussian Hypercontractivity Inequality we obtain the Gaussian Log-Sobolev Inequality (see Exercise~\ref{ex:log-sob}; the proof is the same as in the Boolean case):
                                \index{Log-Sobolev Inequality!Gaussian}%
\begin{named}{Gaussian Log-Sobolev Inequality}
    Let $f \in L^2(\R^n, \gamma)$ be in the domain of~$\Lap$.  Then
    \[
        \half \Ent[f^2] \leq \E[\|\grad f\|^2].
    \]
\end{named}
It's tempting to use the notation $\Tinf[f]$ for $\E[\|\grad f\|^2]$; however, you have to be careful because this quantity is not equal to $\sum_{i=1}^n \E[\Var_{\bz_i}[f]]$ unless $f$ is a multilinear polynomial.  See Exercise~\ref{ex:gaussian-derivs-and-infls}.

\section{Hermite polynomials}                                   \label{sec:hermite}

                                        \index{Hermite polynomials|(}%
Having defined the basic operators of importance for functions on Gaussian space, it's useful to also develop the analogue of the Fourier expansion.  To do this we'll proceed as in Chapter~\ref{sec:product-spaces}, looking for a complete orthonormal ``Fourier basis'' for~$L^2(\R,\gamma)$, which we can extend to $L^2(\R^n,\gamma)$ by taking products.  It's natural to start with polynomials; by Theorem~\ref{thm:polys-dense} we know that the collection $(\phi_j)_{j \in \N}$, $\phi_j(z) = z^j$ is a complete basis for $L^2(\R,\gamma)$.
                                        \index{Fourier basis}%
To get an orthonormal (``Fourier'') basis we can simply perform the Gram--Schmidt process. Calling the resulting basis $(\he_j)_{j \in \N}$ (with ``$\he$'' standing for ``Hermite''), we get
\begin{equation}                \label{eqn:hermite-examples}
    \he_0(z) = 1, \quad \he_1(z) = z, \quad \he_2(z) = \frac{z^2-1}{\sqrt{2}}, \quad \he_3(z) = \frac{z^3 - 3z}{\sqrt{6}}, \quad \dots
\end{equation}
Here, e.g., we obtained $\he_3(z)$ in two steps. First, we made $\phi_3(z) = z^3$ orthogonal to $\he_0, \dots, \he_2$ as
\[
    z^3 - \la\bz^3, 1 \ra \cdot 1 - \la\bz^3, \bz \ra \cdot z - \la\bz^3, \tfrac{\bz^2 - 1}{\sqrt{2}} \ra \cdot \tfrac{z^2-1}{\sqrt{2}} = z^3 - 3z,
\]
where $\bz \sim \normal(0,1)$ and we used the fact that $\bz^3$ and $\bz^3 \cdot \tfrac{\bz^2 - 1}{\sqrt{2}}$ are odd functions and hence have Gaussian expectation~$0$.  Then we defined $\he_3(z) = \frac{z^3 - 3z}{\sqrt{6}}$ after determining that $\E[(\bz^3-3\bz)^2] = 6$.

Let's develop a more explicit definition of these Hermite polynomials.  The computations involved in the Gram--Schmidt process require knowledge of the moments of a Gaussian random variable $\bz \sim \normal(0,1)$.  It's most convenient to understand these moments through the moment generating function of~$\bz$, namely
\begin{equation}            \label{eqn:g-mgf}
    \E[\exp(t\bz)] = \tfrac{1}{\sqrt{2\pi}} \int_\R e^{tz} e^{-\frac12 z^2}\,dz = e^{\frac12 t^2} \tfrac{1}{\sqrt{2\pi}} \int_\R e^{-\frac12 (z-t)^2}\,dz = \exp(\half t^2).
\end{equation}
In light of our interest in the $\U_\rho$ operators, and the fact that orthonormality involves pairs of basis functions, we'll in fact study the moment generating function of a pair $(\bz, \bz')$ of $\rho$-correlated standard Gaussians. To compute it, assume $(\bz,\bz')$ are generated as in Fact~\ref{fact:corr-gaussians} with $\vec{u}, \vec{v}$ unit vectors in $\R^2$.  Then
\begin{align*}
    \E_{\substack{(\bz,\bz') \\ \text{$\rho$-correlated}}}[\exp(s\bz + t\bz')] &=  \E_{\substack{\bg_1, \bg_2 \sim \normal(0,1) \\ \text{independent}}}[\exp(s(u_1\bg_1 + u_2 \bg_2) + t(v_1\bg_1 + v_2 \bg_2))]\\
    &=  \E_{\bg_1 \sim \normal(0,1)}[\exp((su_1+tv_1)\bg_1)]\E_{\bg_2 \sim \normal(0,1)}[\exp((su_2+tv_2)\bg_2)]\\
    &=  \exp(\half (su_1+tv_1)^2)\exp(\half (su_2+tv_2)^2)\\
    &=  \exp(\half \|\vec{u}\|_2^2 s^2 + \la \vec{u},\vec{v} \ra st + \half \|\vec{v}\|_2^2 t^2) \\
    &=  \exp(\half (s^2 + 2\rho st + t^2)),
\end{align*}
where the third equality used~\eqref{eqn:g-mgf}. Dividing by $\exp(\half(s^2+t^2))$ it follows that
\begin{equation}                \label{eqn:hermite-dev}
    \E_{\substack{(\bz,\bz') \\ \text{$\rho$-correlated}}}[\exp(s\bz - \half s^2)\exp(t\bz' - \half t^2)] = \exp(\rho st) = \sum_{j=0}^\infty \frac{\rho^j}{j!} s^jt^j.
\end{equation}
Inside the expectation above we essentially have the expression $\exp(tz-\half t^2)$ appearing twice.  It's easy to see that if we take the power series in~$t$ for this expression, the coefficient on $t^j$ will be a polynomial in~$z$ with leading term~$\frac{1}{j!}z^j$.  Let's therefore write
\begin{equation}                                    \label{eqn:prob-herm}
    \exp(tz-\half t^2) = \sum_{j=0}^\infty \frac{1}{j!} H_j(z)t^j,
\end{equation}
where $H_j(z)$ is a monic polynomial of degree~$j$.  Now substituting this into~\eqref{eqn:hermite-dev} yields
\[
    \sum_{j,k=0}^\infty \frac{1}{j!k!}\E_{\substack{(\bz,\bz') \\ \text{$\rho$-correlated}}}[H_j(\bz)H_k(\bz')]s^jt^k = \sum_{j=0}^\infty \frac{\rho^j}{j!} s^jt^j.
\]
Equating coefficients, it follows that we must have
\[
    \E_{\substack{(\bz,\bz') \\ \text{$\rho$-correlated}}}[H_j(\bz)H_k(\bz')] = \begin{cases}
                                                                                    j!\rho^j & \text{if $j = k$,}\\
                                                                                    0 & \text{if $j \neq k$.}
                                                                                \end{cases}
\]
In particular (taking $\rho = 1$),
\begin{equation}                                \label{eqn:hermite-done2}
    \la H_j, H_k \ra = \begin{cases}
                                                                                    j! & \text{if $j = k$,}\\
                                                                                    0 & \text{if $j \neq k$};
                                                                                \end{cases}
\end{equation}
i.e., the polynomials $(H_j)_{j \in \N}$ are orthogonal.  Furthermore, since $H_j$ is monic and of degree~$j$, it follows that the $H_j$'s are precisely the polynomials that arise in the Gram--Schmidt orthogonalization of $\{1, z, z^2, \dots \}$.  We also see from~\eqref{eqn:hermite-done2} that the orthonormalized polynomials $(\he_j)_{j \in \N}$ are obtained by setting $\he_j = \frac{1}{\sqrt{j!}} H_j$.

Let's summarize and introduce the terminology for what we've deduced.
                                                \nomenclature[Hj]{$H_j$}{the $j$th probabilists' Hermite polynomial, defined by $\exp(tz-\frac12 t^2) = \sum_{j=0}^\infty \frac{1}{j"!} H_j(z)t^j$}% the quotation mark here and below are here intentionally, to fix some weird bug in \nomenclature
                                                \nomenclature[hea]{$\he_j$}{the $j$th (normalized) Hermite polynomial, $\he_j = \frac{1}{\sqrt{j"!}} H_j$}%
                                                \nomenclature[heb]{$\he_\alpha$}{for $\alpha \in \N^n$ a multi-index, the $n$-variate (normalized) Hermite polynomial $\he_\alpha(z) = \prod_{j=1}^n \he_{\alpha_j}(z_j)$}%
\begin{definition}                              \label{def:p-hermite}
    The \emph{probabilists' Hermite polynomials} $(H_j)_{j \in \N}$ are the univariate polynomials defined by the identity~\eqref{eqn:prob-herm}. An equivalent definition (Exercise~\ref{ex:alternate-hermite-formula-completethesquare}) is
    \begin{equation}                            \label{eqn:hermite-alternate}
        H_j(z) = \frac{(-1)^j}{\vphi(z)} \cdot \frac{d^j}{dz^j} \vphi(z).
    \end{equation}
    The \emph{normalized Hermite polynomials} $(\he_j)_{j \in \N}$ are defined by $\he_j = \frac{1}{\sqrt{j!}} H_j$; the first four are given explicitly in~\eqref{eqn:hermite-examples}.  For brevity we'll simply refer to the $\he_j$'s  as the ``Hermite polynomials'', though this is not standard terminology.
\end{definition}
\begin{proposition}                                     \label{prop:hermite-basis}
    The Hermite polynomials $(\he_j)_{j \in \N}$ form a complete orthonormal basis for $L^2(\R,\gamma)$. They are also a ``Fourier basis'', since $\he_0 = 1$.
\end{proposition}
\begin{proposition}                                     \label{prop:hermite-correlation}
    For any $\rho \in [-1,1]$ we have
    \[
        \E_{\substack{(\bz,\bz') \\ \text{$\rho$-correlated}}}[\he_j(\bz)\he_k(\bz')] = \la \he_j, \U_\rho \he_k \ra = \la \U_\rho \he_j, \he_k \ra =
                                                                                    \begin{cases}
                                                                                        \rho^j & \text{if $j = k$,}\\
                                                                                        0 & \text{if $j \neq k$}.
                                                                                    \end{cases}
    \]
\end{proposition}
                                                    \index{product basis}%
                                                        \index{Fourier basis}%
From this ``Fourier basis'' for $L^2(\R,\gamma)$ we can construct a ``Fourier basis'' for $L^2(\R^n,\gamma)$ just by taking products, as in Proposition~\ref{prop:fbasis-product}.
                                                        \index{Hermite polynomials!multivariate}%
\begin{definition}
    For a multi-index $\alpha \in \N^n$ we define the \emph{(normalized multivariate) Hermite polynomial} $\he_\alpha \co \R^n \to \R$ by
    \[
        \he_\alpha(z) = \prod_{j = 1}^n \he_{\alpha_j}(z_j).
    \]
    Note that the total degree of $\he_\alpha$ is $|\alpha| = \sum_j \alpha_j$.  We also identify a subset $S \subseteq [n]$ with its indicator $\alpha$ defined by $\alpha_j = 1_{j \in S}$; thus $\he_S(z)$ denotes $z^S = \prod_{j \in S} z_j$.
\end{definition}
\begin{proposition}                                     \label{prop:hermite-product}
    The Hermite polynomials $(\he_\alpha)_{\alpha \in \N^n}$ form a complete orthonormal (Fourier) basis for $L^2(\R^n,\gamma)$.  Further, for any $\rho \in [-1,1]$ we have
    \[
        \E_{\substack{(\bz,\bz') \\ \text{$\rho$-correlated}}}[\he_\alpha(\bz)\he_\beta(\bz')] = \la \he_\alpha, \U_\rho \he_\beta \ra = \la \U_\rho \he_\alpha, \he_\beta \ra =     \begin{cases}
                                                                                \rho^{|\alpha|} & \text{if $\alpha = \beta$,}\\
                                                                                0 & \text{if $\alpha \neq \beta$}.
                                                                            \end{cases}
    \]
\end{proposition}
                                        \index{Hermite polynomials|)}%

                                                        \index{Hermite expansion}%
We can now define the ``Hermite expansion'' of Gaussian functions.
\begin{definition}
    Every $f \in L^2(\R^n,\gamma)$ is uniquely expressible as
    \[
        f = \sum_{\alpha \in \N^n} \wh{f}(\alpha) \he_\alpha,
    \]
    where the real numbers $\wh{f}(\alpha)$ are called the \emph{Hermite coefficients of~$f$} and the convergence is in~$L^2(\R^n,\gamma)$; i.e.,
    \[
        \left\|f - \sum_{|\alpha| \leq k} \wh{f}(\alpha) \he_\alpha\right\|_2 \to 0 \quad \text{as $k \to \infty$}.
    \]
    This is called the \emph{Hermite expansion} of~$f$.
\end{definition}
\begin{remark}
    If $f \co \R^n \to \R$ is a multilinear polynomial, then it ``is its own Hermite expansion'':
    \[
        f(z) = \sum_{S \subseteq [n]} \wh{f}(S) z^S = \sum_{S \subseteq [n]} \wh{f}(S) \he_S(z) = \sum_{\alpha_1, \dots, \alpha_n \leq 1} \wh{f}(\alpha) \he_\alpha(z).
    \]
\end{remark}
\begin{proposition}                                     \label{prop:parseval}
                                   \index{Parseval's Theorem}%
                                   \index{Plancherel's Theorem}%
    The Hermite coefficients of $f \in L^2(\R^n,\gamma)$ satisfy the formula
    \[
        \wh{f}(\alpha) = \la f, \he_\alpha \ra,
    \]
    and for $f, g \in L^2(\R^n,\gamma)$ we have the Plancherel formula
    \[
        \la f, g \ra = \sum_{\alpha \in \N^n} \wh{f}(\alpha)\wh{g}(\alpha).
    \]
\end{proposition}
From this we may deduce:
\begin{proposition}                                     \label{prop:Urho-formula}
    For $f \in L^2(\R^n,\gamma)$, the function $\U_\rho f$ has Hermite expansion
    \[
        \U_\rho f = \sum_{\alpha \in \N^n} \rho^{|\alpha|} \wh{f}(\alpha) \he_\alpha
    \]
    and hence
    \[
        \Stab_\rho[f] = \sum_{\alpha \in \N^n} \rho^{|\alpha|} \wh{f}(\alpha)^2.
    \]
\end{proposition}
\begin{proof}
    Both statements follow from Proposition~\ref{prop:parseval}, with the first using
    \[
        \wh{\U_\rho f}(\alpha) = \la \U_\rho f, \he_\alpha \ra = \la \littlesum_{\beta} \U_\rho \wh{f}(\beta) \he_\beta, \he_\alpha \ra = \littlesum_\beta \wh{f}(\beta) \la \U_\rho \he_\beta, \he_\alpha \ra = \rho^{|\alpha|} \wh{f}(\alpha);
    \]
    we also used Proposition~\ref{prop:hermite-product} and the fact that $\U_\rho$ is a contraction in~$L^2(\R^n,\gamma)$.
\end{proof}
\begin{remark}
    When $f \co \R^n \to \R$ is a multilinear polynomial, this formula for $\U_\rho f$ agrees with the formula $f(\rho z)$ given in Fact~\ref{fact:Urho-plug-in}.
\end{remark}
\begin{remark}
    In a sense it's not very important to know the explicit formulas for the Hermite polynomials,~\eqref{eqn:hermite-examples},~\eqref{eqn:prob-herm}; it's usually enough just to know that the formula for $\U_\rho f$ from Proposition~\ref{prop:Urho-formula} holds.
\end{remark}
Finally, by differentiating the formula in Proposition~\ref{prop:Urho-formula} at $\rho = 1$ we deduce the following formula for the Ornstein--Uhlenbeck operator (explaining why it's sometimes called the number operator):
                                                    \index{Ornstein--Uhlenbeck operator}%
\begin{proposition}                                     \label{prop:ou-formula}
    For $f \in L^2(\R^n, \gamma)$ in the domain of $\Lap$ we have
    \[
        \Lap f = \sum_{\alpha \in \N^n} |\alpha| \wh{f}(\alpha) h_\alpha.
    \]
\end{proposition}
\noindent (Actually, Exercise~\ref{ex:gauss-Lap-domain} asks you to formally justify this and the fact that $f$~is in the domain of~$\Lap$ if and only if $\sum_{\alpha} |\alpha|^2 \wh{f}(\alpha)^2 < \infty$.) For additional facts about Hermite polynomials, see Exercises~\ref{ex:alternate-hermite-formula-completethesquare}--\ref{ex:krav-to-hermite}.

\section{Borell's Isoperimetric  Theorem}                                \label{sec:borell}

                                                \index{Borell's Isoperimetric Theorem|(}%
If we believe that the Majority Is Stablest Theorem should be true, then we also have to believe in its ``Gaussian special case''.  Let's see what this Gaussian special case is.  Suppose $f \co \R^n \to [-1,1]$ is a ``nice'' function (smooth, say, with all derivatives bounded) having $\E[f] = 0$. You're encouraged to think of~$f$ as (a smooth approximation to) the indicator~$\pm 1_A$ of some set $A \subseteq \R^n$ of Gaussian volume $\gvol(A) = \half$.  Now consider the Boolean function $g \co \bits^{nM} \to \{-1,1\}$ defined by
\[
    g = f \circ \BitsToGaussians{M}{n}.
\]
Using the multidimensional Central Limit Theorem, for any $\rho \in (0,1)$  we should have
\[
    \Stab_\rho[g] \xrightarrow{M \to \infty} \Stab_\rho[f],
\]
where on the left we have Boolean noise stability and on the right we have Gaussian noise stability. Using $\E[g] \to \E[f] = 0$, the Majority Is Stablest Theorem would tell us that
\[
    \Stab_\rho[g] \leq 1 - \tfrac{2}{\pi}\arccos \rho + o_\eps(1),
\]
where $\eps = \MaxInf[g]$.  But $\eps = \eps(M) \to 0$ as $M \to \infty$.  Thus we should simply have the Gaussian noise stability bound
\begin{equation}                                    \label{eqn:vol12-Borell}
    \Stab_\rho[f] \leq 1 - \tfrac{2}{\pi}\arccos \rho.
\end{equation}
(By a standard approximation argument this extends from ``nice'' $f \co \R^n \to [-1,1]$ with $\E[f] = 0$ to any measurable $f \co \R^n \to [-1,1]$ with $\E[f] = 0$.)  Note that the upper bound~\eqref{eqn:vol12-Borell} is achieved when~$f$ is the $\pm 1$-indicator of any halfspace through the origin; see Corollary~\ref{cor:gstab-halfspace-origin}.  (Note also that if $n = 1$ and $f = \sgn$, then the function $g$ is simply $\Maj_M$.)

The ``isoperimetric inequality''~\eqref{eqn:vol12-Borell} is indeed true, and is a special case of a theorem first proved by Borell~\cite{Bor85}.
                                    \index{Borell's Isoperimetric Theorem!volume-$\frac12$ case}%
\begin{named}{Borell's Isoperimetric Theorem (volume-${\frac{\mathbf 1}{\mathbf 2}}$ case)}
    Fix $\rho \in (0,1)$.  Then for any $f \in L^2(\R^n,\gamma)$ with range $[-1,1]$ and $\E[f] = 0$,
    \[
        \Stab_{\rho}[f] \leq 1 - \tfrac{2}{\pi}\arccos \rho,
    \]
    with equality if~$f$ is the $\pm 1$-indicator of any halfspace through the origin.
\end{named}
\begin{remark}                                  \label{rem:borell-set-to-function1}
    In Borell's Isoperimetric Theorem, nothing is lost by restricting attention to functions with range~$\{-1,1\}$, i.e., by considering only $f = \pm 1_A$ for $A \subseteq \R^n$.  This is because the case of range $[-1,1]$ follows straightforwardly from the case of range $\{-1,1\}$, essentially because $\sqrt{\Stab_\rho[f]} = \|\U_{\sqrt{\rho}} f\|_2$ is a convex functional of~$f$; see Exercise~\ref{ex:borell-function-from-set}.
\end{remark}
More generally, Borell showed that for any fixed volume~$\alpha \in [0,1]$, the maximum Gaussian noise stability of a set of volume~$\alpha$ is no greater than that of a halfspace of volume~$\alpha$.  We state here the more general theorem, using  range $\{0,1\}$ rather than range $\{-1,1\}$ for future notational convenience (and with Remark~\ref{rem:borell-set-to-function1} applying equally):
\begin{named}{Borell's Isoperimetric Theorem}
    Fix $\rho \in (0,1)$.  Then for any $f \in L^2(\R^n, \gamma)$ with range $[0,1]$ and $\E[f] = \alpha$,
    \[
        \Stab_{\rho}[f] \leq \GaussQuad_\rho(\alpha).
    \]
    Here $\GaussQuad_\rho(\alpha)$ is the \emph{Gaussian quadrant probability} function, discussed in Exercises~\ref{ex:ball-stab} and~\ref{ex:gaussquad}, and equal to $\Stab_\rho[1_H]$ for any (every) halfspace $H \subseteq \R^n$ having Gaussian volume $\gvol(H) = \alpha$.
\end{named}

We've seen that the volume-$\half$ case of Borell's Isoperimetric Theorem is a special case of the Majority Is Stablest Theorem, and similarly, the general version of Borell's theorem is a special case of the General-Volume Majority Is Stablest Theorem mentioned at the beginning of the chapter.  As a consequence, proving Borell's Isoperimetric Theorem is a \emph{prerequisite} for proving the General-Volume Majority Is Stablest Theorem.  In fact, our proof in Section~\ref{sec:MIST} of the latter will be a reduction to the former.

The proof of Borell's Isoperimetric Theorem itself is not too hard; one of five known proofs, the one due to Mossel and Neeman~\cite{MN12}, is outlined in Exercises~\ref{ex:mossel-neeman1}--\ref{ex:mossel-neeman4}. If our main goal is just to prove the basic Majority Is Stablest Theorem, then we only need the volume-$\half$ case of Borell's Isoperimetric Inequality.  Luckily, there's a very simple proof of this volume-$\half$ case for ``many'' values of~$\rho$, as we will now explain.

Let's first slightly rephrase the statement of Borell's Isoperimetric Theorem in the volume-$\half$ case.  By Remark~\ref{rem:borell-set-to-function1} we can restrict attention to sets; then the theorem asserts that among sets of Gaussian volume~$\half$, halfspaces through the origin have maximal noise stability, for each positive value of~$\rho$.  Equivalently, halfspaces through the origin have minimal noise \emph{sensitivity} under correlation $\cos \theta$, for $\theta \in (0,\frac{\pi}{2})$.  The formula for this minimal noise sensitivity was given as~\eqref{eqn:elegant-sheppard} in our proof of Sheppard's Formula.  Thus we have:
                                    \index{Borell's Isoperimetric Theorem!volume-$\frac12$ case}%
\begin{named}{Equivalent statement of the volume-${\frac{\mathbf 1}{\mathbf 2}}$ Borell Isoperimetric Theorem}
    Fix $\theta \in (0,\frac{\pi}{2})$.  Then for any $A \subset \R^n$ with $\gvol(A) = \half$,
    \[
       \Pr_{\substack{(\bz, \bz') \\\text{$\cos \theta$-correlated}}}[1_A(\bz) \neq 1_A(\bz')] \geq \tfrac{\theta}{\pi},
    \]
    with equality if~$A$ is any halfspace through the origin.
\end{named}

In the remainder of this section we'll show how to prove this formulation of the theorem whenever $\theta = \frac{\pi}{2\ell}$, where $\ell$ is a positive integer.  This gives the volume-$\half$ case of Borell's Isoperimetric Inequality for all $\rho$ of the form $\arccos\frac{\pi}{2\ell}$, $\ell \in \N^+$; in particular, for an infinite sequence of $\rho$'s tending to~$1$.  To prove the theorem for these values of~$\theta$, it's convenient to introduce notation for the following noise sensitivity variant:
\begin{definition}
    For $A \subseteq \R^n$ and $\delta \in \R$ (usually $\delta \in [0,\pi]$) we write $\RS_A(\delta)$ for the
                                            \index{rotation sensitivity}%
                                            \index{noise sensitivity!Gaussian|seeonly{rotation sensitivity}}%
                                            \nomenclature[RSA]{$\RS_A(\delta)$}{the rotation sensitivity of $A$ at $\delta$; i.e., $\Pr[1_A(\bz) \neq 1_A(\bz')]$ for a $\cos \delta$-correlated pair $(\bz,\bz')$}
    \emph{rotation sensitivity of $A$ at $\delta$},
    defined by
    \[
        \RS_A(\delta) = \Pr_{\substack{(\bz, \bz') \\ \cos \delta\text{-correlated}}}[1_A(\bz) \neq 1_A(\bz')].
    \]
\end{definition}
The key property of this definition is the following:
\begin{theorem}                                     \label{thm:rs-subadditive}
                                            \index{rotation sensitivity!subadditivity}%
    For any $A \subseteq \R^n$ the function $\RS_A(\delta)$ is subadditive; i.e.,
    \[
        \RS_A(\delta_1 + \cdots + \delta_\ell) \leq \RS_A(\delta_1) + \cdots + \RS_A(\delta_\ell).
    \]
    In particular, for any $\delta \in \R$ and $\ell \in \N^+$,
    \[
        \RS_A(\delta) \leq \ell \cdot \RS_A(\delta/\ell).
    \]
\end{theorem}
\begin{proof}
    Let $\bg, \bg' \sim \normal(0,1)^n$ be drawn independently and define $\bz(\theta) = (\cos\theta)\bg + (\sin\theta)\bg'$.  Geometrically, as $\theta$ goes from~$0$ to~$\frac{\pi}{2}$ the random vectors $\bz(\theta)$ trace from~$\bg$ to~$\bg'$ along the origin-centered ellipse passing through these two points.  The random vectors $\bz(\theta)$ are jointly normal, with each individually distributed as~$\normal(0,1)^n$.  Further, for each fixed $\theta, \theta' \in \R$ the pair $(\bz(\theta), \bz(\theta'))$ constitute $\rho$-correlated Gaussians with
    \[
        \rho = \cos\theta \cos\theta' + \sin\theta \sin\theta' = \cos(\theta' - \theta).
    \]
    Now consider the sequence $\theta_0, \dots, \theta_\ell$ defined by the partial sums of the $\delta_i$'s, i.e., $\theta_j = \sum_{i = 1}^j \delta_i$. We get that $\bz(\theta_0)$ and $\bz(\theta_\ell)$ are $\cos(\delta_1 + \cdots + \delta_\ell)$-correlated, and that $\bz(\theta_{j-1})$ and $\bz(\theta_{j})$ are $\cos\delta_j$-correlated for each $j \in [\ell]$.  Thus
    \begin{align}
        \RS_A(\delta_1 + \cdots + \delta_\ell) &= \Pr[1_A(\bz(\theta_0)) \neq 1_A(\bz(\theta_\ell))] \nonumber\\
        &\leq \sum_{j=1}^\ell \Pr[1_A(\bz(\theta_{j})) \neq 1_A(\bz(\theta_{j-1}))] = \sum_{j=1}^\ell \RS_A(\delta_j), \label{eqn:ellipse-union-bound}
    \end{align}
    where the inequality is the union bound.
\end{proof}

With this subadditivity result in hand, it's indeed easy to prove the equivalent statement of the volume-$\half$ Borell Isoperimetric Theorem for any~$\theta \in \{\frac{\pi}{4}, \frac{\pi}{6}, \frac{\pi}{8}, \frac{\pi}{10}, \dots\}$. As we'll see in Section~\ref{sec:MIST}, the case of $\theta = \frac{\pi}{4}$ can be used to give an excellent \UG-hardness result for the Max-Cut CSP.
                                                \index{Borell's Isoperimetric Theorem!volume-$\frac12$ case}%
\begin{corollary}                                   \label{cor:volume-half-borell}
    The equivalent statement of the volume-$\half$ Borell Isoperimetric Theorem holds whenever $\theta = \frac{\pi}{2\ell}$ for $\ell \in \N^+$.
\end{corollary}
\begin{proof}
    The exact statement we need to show is $\RS_A(\frac{\pi}{2\ell}) \geq \frac{1}{2\ell}$.  This follows by taking $\delta = \frac{\pi}{2}$ in Theorem~\ref{thm:rs-subadditive} because
    \[
        \RS_A(\tfrac{\pi}{2}) = \Pr_{\substack{(\bz, \bz') \\ 0\text{-correlated}}}[1_A(\bz) \neq 1_A(\bz')] = \tfrac12,
    \]
    using that $0$-correlated Gaussians are independent and that $\gvol(A) = \frac12$.
\end{proof}
\begin{remark}
    Although Sheppard's Formula already tells us that equality holds in this corollary when~$A$ is a halfspace through the origin, it's also not hard to derive this directly from the proof. The only inequality in the proof,~\eqref{eqn:ellipse-union-bound}, is an equality when~$A$ is a halfspace through the origin, because the elliptical arc can only cross such a halfspace~$0$ or~$1$ times.
\end{remark}
\begin{remark}                      \label{rem:odd-easy-borell}
    Suppose that $A \subseteq \R^n$ not only has volume~$\half$, it has the property that $x \in A$ if and only if $-x \not \in A$; in other words, the $\pm 1$-indicator of~$A$ is an odd function.  (In both statements, we allow a set of measure~$0$ to be ignored.)  An example set with this property is any halfspace through the origin.  Then $\RS_A(\pi) = 1$, and hence we can establish Corollary~\ref{cor:volume-half-borell} more generally for any $\theta \in \{\frac{\pi}{1}, \frac{\pi}{2}, \frac{\pi}{3}, \frac{\pi}{4}, \frac{\pi}{5}, \dots\}$ by taking $\delta = \pi$ in the proof.
\end{remark}
                                              \index{Borell's Isoperimetric Theorem|)}%

\section{Gaussian surface area and Bobkov's Inequality}             \label{sec:gaussian-surface-area}

                                            \index{Gaussian Isoperimetric Inequality|(}%
This section is devoted to studying the \emph{Gaussian Isoperimetric Inequality}.  This inequality is a special case of the Borell Isoperimetric Inequality (and hence also a special case of the General-Volume Majority Is Stablest Theorem); in particular, it's the special case arising from the limit $\rho \to 1^{-}$.

Restating Borell's theorem using rotation sensitivity we have that for any $A \subseteq \R^n$, if $H \subseteq \R^n$ is a halfspace with the same Gaussian volume as~$A$ then for all~$\eps$,
\[
    \RS_A(\eps) \geq \RS_H(\eps).
\]
Since $\RS_A(0) = \RS_H(0) = 0$, it follows that
\[
    \RS_A'(0^+) \geq \RS_H'(0^+).
\]
(Here we are considering the one-sided derivatives at~$0$, which can be shown to exist, though $\RS_A'(0^+)$ may equal~$+\infty$; see the notes at the end of this chapter.)  As will be explained shortly, $\RS_A'(0^+)$ is precisely $\sqrt{2/\pi} \cdot \gsa(A)$, where~$\gsa(A)$ denotes the ``Gaussian surface area''
                                        \index{Gaussian surface area|(}%
of~$A$.  Therefore the above inequality is equivalent to the following:
\begin{named}{Gaussian Isoperimetric Inequality}
    Let $A \subseteq \R^n$ have $\gvol(A) = \alpha$ and let $H \subseteq \R^n$ be any halfspace with $\gvol(H) = \alpha$.  Then $\gsa(A) \geq \gsa(H)$.
\end{named}
\begin{remark}                          \label{rem:gsa-halfspace}
    As shown in Proposition~\ref{prop:halfspace-gsa} below, the right-hand side in this inequality is equal to~$\Giso(\alpha)$, where~$\giso$ is the \emph{Gaussian isoperimetric function}, encountered earlier in Definition~\ref{def:giso} and
                                                    \index{Gaussian isoperimetric function}%
    defined by $\Giso = \vphi \circ \Phi^{-1}$.
\end{remark}

Let's now discuss the somewhat technical question of how to properly define $\gsa(A)$, the Gaussian surface area of a set~$A$.  Perhaps the most natural definition would be to equate it with the \emph{Gaussian Minkowski content} of the boundary~$\bdry A$ of~$A$,
                                                \index{Minkowski content|seeonly{Gaussian Minkowski content}}%
                                                \index{Gaussian Minkowski content|seeonly{Gaussian surface area}}%
                                                \nomenclature[gamma+]{$\gamma^+(\bdry A)$}{the Gaussian Minkowski content of~$\bdry A$}%
\begin{equation}                        \label{eqn:heuristic-gsa}
    \gamma^+(\bdry A) = \liminf_{\eps \to 0^+} \frac{\gvol(\{z : \dist(z, \bdry A) < \eps/2\})}{\eps}.
\end{equation}
(Relatedly, one might also consider the surface integral over $\bdry A$ of the Gaussian pdf~$\vphi$.)  Under the ``official'' definition of~$\gsa(A)$ we give below in Definition~\ref{def:unusual-gsa-def}, we'll indeed have $\gsa(A) = \gamma^+(\bdry A)$ whenever $A$ is sufficiently nice -- say, a disjoint union of closed, full-dimensional, convex sets.  However, the Minkowski content definition is not a good one in general because it's possible to have $\gamma^+(\bdry A_1) \neq \gamma^+(\bdry A_2)$ for some sets~$A_1$ and~$A_2$ that are equivalent up to measure~$0$. (For more information, see Exercise~\ref{ex:essential-boundary} and the notes at the end of this chapter.)

As mentioned above, one ``correct'' definition is~$\gsa(A) = \sqrt{\pi/2} \cdot \RS_A'(0^+)$.  This definition has the advantage of being insensitive to measure-$0$ changes to~$A$.  To connect this unusual-looking definition with Minkowski content, let's heuristically interpret $\RS_A'(0^+)$.  We start by thinking of it as $\frac{\RS_A(\eps)}{\eps}$ for ``infinitesimal~$\eps$''. Now $\RS_A(\eps)$ can be thought of as the probability that the line segment~$\bell$ joining two $\cos \eps$-correlated Gaussians crosses~$\bdry A$. Since $\sin \eps \approx \eps$, $\cos \eps \approx 1$ up to $O(\eps^2)$, we can think of these correlated Gaussians as $\bg$ and $\bg + \eps \bg'$ for independent $\bg, \bg' \sim \normal(0,1)^n$. When $\bg$ lands near~$\bdry A$, the length of~$\bell$ in the direction perpendicular to $\bdry A$ will, in expectation, be   $\eps \E[|\normal(0,1)|] = \sqrt{2/\pi} \eps$.  Thus $\RS_A(\eps)$ should essentially be~$\tfrac12\gvol(\{z : \dist(z, \bdry A) < \sqrt{2/\pi} \eps\})$ and we have heuristically justified
\begin{equation} \label{eqn:heur-gsa2}
    \sqrt{\pi/2} \cdot \RS_A'(0^+) =  \sqrt{\pi/2} \cdot \lim_{\eps \to 0^+} \frac{\RS_A(\eps)}{\eps}\ \mathop{=}^?\  \gamma^+(\bdry A).
\end{equation}

One more standard idea for the definition of~$\gsa(A)$ is~``$\E[\|\grad 1_A\|]$''.  This doesn't quite make sense since $1_A \in L^1(\R^n,\gamma)$ is not actually differentiable. However, we might consider replacing it with the limit of~$\E[\|\grad f_m\|]$ for a sequence~$(f_m)$ of smooth functions approximating~$1_A$.   To see why this notion should agree with the Gaussian Minkowski content $\gamma^+(\bdry A)$ for nice enough~$A$, let's suppose we have a smooth approximator~$f$ to~$1_A$ that agrees with~$1_A$ on ${\{z : \dist(z, \bdry A) \geq \eps/2\}}$ and is (essentially) a linear function on $\{z : \dist(z, \bdry A) < \eps/2\}$. Then $\|\grad f\|$ will be~$0$ on the former set and (essentially) constantly~$1/\eps$ on the latter (since it must climb from~$0$ to~$1$ over a distance of~$\eps$). Thus we indeed have
\[
    \E[\|\grad f\|] \approx \frac{\gvol(\{z : \dist(z, \bdry A) < \eps/2\})}{\eps} \approx \gamma^+(\bdry A),
\]
as desired.  We summarize the above technical discussion with the following definition/theorem, which is discussed further in the notes at the end of this chapter:
\begin{definition}          \label{def:unusual-gsa-def}
    For any $A \subseteq \R^n$, we define its \emph{Gaussian surface area} to be
    \[
        \gsa(A) = \sqrt{\pi/2} \cdot \RS_A'(0^+) \in [0, \infty].
    \]
    An equivalent definition is
    \[
        \gsa(A) = \inf\left\{\liminf_{m \to \infty} \E_{\bz \sim \normal(0,1)^n}[\|\grad f_m(\bz)\|]\right\},
    \]
    where the infimum is over all sequences $(f_m)_{m \in \N}$ of smooth $f_m \co \R^n \to [0,1]$ with first partial derivatives in $L^2(\R^n,\gamma)$ such that $\|f_m - 1_A\|_1 \to 0$.  Furthermore, this infimum is actually achieved by taking $f_m = \U_{\rho_m} f$ for any sequence $\rho_m \to 1^{-}$.  Finally, the equality $\gsa(A) = \gamma^+(\bdry A)$ with Gaussian Minkowski content holds if~$A$ is a disjoint union of closed, full-dimensional, convex sets.
\end{definition}

To get further acquainted with this definition, let's describe the Gaussian surface area of some basic sets.  We start with halfspaces, which as mentioned in Remark~\ref{rem:gsa-halfspace} have Gaussian surface area given by the Gaussian isoperimetric function.
\begin{proposition}                                     \label{prop:halfspace-gsa}
    Let $H \subseteq \R^n$ be any halfspace (open or closed) with $\gvol(H) = \alpha \in (0,1)$.  Then $\gsa(H) = \Giso(\alpha) = \vphi(\Phi^{-1}(\alpha))$.  In particular, if $\alpha = 1/2$ -- i.e., $H$'s boundary contains the origin -- then $\gsa(H) = \frac{1}{\sqrt{2\pi}}$.
\end{proposition}
\begin{proof}
    Just as in the proof of Corollary~\ref{cor:gstab-halfspace-origin}, by rotational symmetry we may assume~$H$ is a $1$-dimensional halfline, $H = (-\infty, t]$.  Since $\gvol(H) = \alpha$, we have $t = \Phi^{-1}(\alpha)$. Then $\gsa(H)$ is equal to
    \[
        \gamma^+(\bdry H) = \lim_{\eps \to 0^+} \frac{\gvol(\{z \in \R : \dist(z,\bdry H) < \tfrac{\eps}{2}\})}{\eps}
        = \lim_{\eps \to 0^+} \frac{\int_{t-\eps/2}^{t+\eps/2} \vphi(s)\,ds}{\eps} = \vphi(t) = \giso(\alpha). \qedhere
    \]
\end{proof}
Here are some more Gaussian surface area bounds:
\begin{examples}                        \label{eg:surface-areas}
    In Exercise~\ref{ex:gsa-of-intervals} you are asked to generalize the above computation and show that if $A \subseteq \R$ is the union of disjoint nondegenerate intervals $[t_1, t_2], [t_3, t_4], \dots, [t_{2m-1}, t_{2m}]$ then $\gsa(A) = \sum_{i=1}^{2m} \vphi(t_i)$.  Perhaps the next easiest example is when~$A \subseteq \R^n$ is an origin-centered ball; Ball~\cite{Bal93} gave an explicit formula for $\gsa(A)$ in terms of the dimension and radius, one which is always less than $\sqrt{\frac{2}{\pi}}$ (see Exercise~\ref{ex:ball-gsa}).
    This upper bound was extended to non-origin-centered balls in Klivans et~al.~\cite{KOS08}.  Ball also showed that every convex set $A \subseteq \R^n$ satisfies $\gsa(A) \leq O(n^{1/4})$; Nazarov \cite{Naz03} showed that this bound is tight up to the constant, using a construction highly reminiscent of Talagrand's Exercise~\ref{ex:talagrand-random-dnf}. As noted in Klivans et~al.~\cite{KOS08}, Nazarov's work  also immediately implies that an intersection of~$k$ halfspaces has Gaussian surface area at most $O(\sqrt{\log k})$ (tight for appropriately sized cubes in~$\R^k$), and that any cone in $\R^n$ with apex at the origin has Gaussian surface area at most~$1$.  Finally, by proving the ``Gaussian special case'' of the Gotsman--Linial Conjecture,
                                                \index{Gotsman--Linial Conjecture}%
    Kane~\cite{Kan11a} established that if $A \subseteq \R^n$ is a degree-$k$ ``polynomial threshold function'' -- i.e., $A = \{z : p(z) > 0\}$ for $p$ an $n$-variate degree-$k$ polynomial -- then $\gsa(A) \leq \frac{k}{\sqrt{2 \pi}}$. This is tight for every~$k$ (even when $n = 1$).
\end{examples}

Though we've shown that the Gaussian Isoperimetric Inequality follows from Borell's Isoperimetric Theorem, we now discuss some alternative proofs.  In the special case of sets of Gaussian volume~$\half$, we can again get a very simple proof using the subadditivity property of Gaussian rotation sensitivity, Theorem~\ref{thm:rs-subadditive}.
                                            \index{rotation sensitivity!subadditivity}%
That result easily yields the following kind of ``concavity property'' concerning Gaussian surface area:
\begin{theorem}                                     \label{thm:gsa-concavity}
    Let $A \subseteq \R^n$.  Then for any  $\delta > 0$,
    \[
        \sqrt{\pi/2} \cdot \frac{\RS_A(\delta)}{\delta} \leq \gsa(A).
    \]
\end{theorem}
\begin{proof}
    For $\delta > 0$ and $\eps = \delta/\ell$, $\ell \in \N^+$, Theorem~\ref{thm:rs-subadditive} is equivalent to
    \[
        \frac{\RS_A(\delta)}{\delta} \leq \frac{\RS_A(\eps)}{\eps}.
    \]
    Taking $\ell \to \infty$ hence $\eps \to 0^+$, the right-hand side becomes $\RS'_A(0^+) = \sqrt{2/\pi} \cdot \gsa(A)$.
\end{proof}
\noindent If we take $\delta = \pi/2$ in this theorem, the left-hand side becomes
\[
    \sqrt{2/\pi} \Pr_{\substack{\bz, \bz' \sim \normal(0,1)^n \\ \text{independent}}}[1_A(\bz) \neq 1_A(\bz')] = 2\sqrt{2/\pi} \cdot \gvol(A)(1-\gvol(A)).
\]
Thus we obtain a simple proof of the following result, which includes the Gaussian Isoperimetric Inequality in the volume-$\frac12$ case:
\begin{theorem}                                     \label{thm:weak-l1-poinc}
    Let $A \subseteq \R^n$.  Then
    \[
        2\sqrt{2/\pi} \cdot \gvol(A)(1-\gvol(A)) \leq \gsa(A).
    \]
    In particular, if $\gvol(A) = \frac12$, then we get the tight Gaussian Isoperimetric Inequality statement $\gsa(A) \geq \frac{1}{\sqrt{2\pi}} = \giso(\half)$.
\end{theorem}

As for the full Gaussian Isoperimetric Inequality, it's a pleasing fact that it can be derived by pure analysis of Boolean functions.  This was shown by Bobkov~\cite{Bob97}, who proved
                                            \index{Bobkov's Inequality|(}%
the following very interesting isoperimetric inequality about Boolean functions:
\begin{named}{Bobkov's Inequality}
    Let $f \co \bn \to [0,1]$.  Then
    \begin{equation}                                \label{eqn:bobkov}
        \giso(\E[f]) \leq \Ex_{\bx \sim \bn}\left[\|(\giso(f(\bx)), \grad f(\bx))\|\right].
    \end{equation}
    Here $\grad f$ is the discrete gradient (as in Definition~\ref{def:discrete-grad}) and $\| \cdot \|$ is the usual Euclidean norm (in $\R^{n+1}$). Thus to restate the inequality,
    \[
        \giso(\E[f]) \leq \Ex_{\bx \sim \bn}\left[\sqrt{\giso(f(\bx))^2 + \littlesum_{i=1}^n \D_if(\bx)^2}\right].
    \]
    In particular, suppose $f = 1_A$ is the $0$-$1$ indicator of a subset $A \subseteq \bn$.  Then since $\giso(0) = \giso(1) = 0$ we obtain $\giso(\E[1_A]) \leq \Ex[\|\grad 1_A\|]$.
\end{named}
\noindent As Bobkov noted, by the usual Central Limit Theorem argument one can straightforwardly obtain inequality~\eqref{eqn:bobkov} in the setting of functions $f \in L^2(\R^n, \gamma)$ with range~$[0,1]$, provided~$f$ is sufficiently smooth (for example, if~$f$ is in the domain of~$\Lap$; see Exercise~\ref{ex:gauss-Lap-domain}).   Then given $A \subseteq \R^n$, by taking a sequence of smooth approximations to~$1_A$ as in Definition~\ref{def:unusual-gsa-def}, the Gaussian Isoperimetric Inequality $\giso(\E[1_A]) \leq \gsa(A)$ is recovered.
                                                \index{Gaussian surface area|)}%
                                            \index{Gaussian Isoperimetric Inequality|)}%

Given $A \subseteq \bn$ we can write the quantity $\Ex[\|\grad 1_A\|]$ appearing in Bobkov's Inequality as
\begin{equation}                        \label{eqn:boolean-surface-area}
    \Ex[\|\grad 1_A\|] = \tfrac12 \cdot \E_{\bx \sim \bn}\bigl[\sqrt{\sens_A(\bx)}\bigr],
\end{equation}
using the fact that for $1_A \co \bn \to \{0,1\}$ we have
\[
    \D_i1_A(\bx)^2 = \tfrac14\cdot \bone[\text{coordinate $i$ is pivotal for $1_A$ on~$x$}].
\]
The quantity in~\eqref{eqn:boolean-surface-area} -- (half of) the expected square-root of the number of pivotal coordinates -- is an interesting possible notion of  ``Boolean surface area'' for sets $A \subseteq \bn$.  It was first essentially proposed by Talagrand~\cite{Tal93}.  By Cauchy--Schwarz it's upper-bounded by (half of) the square-root of our usual notion of boundary size, average sensitivity:
\begin{equation} \label{eqn:tal-surf-vs-surf}
    \E[\|\grad 1_A\|] \leq \sqrt{\E[\|\grad 1_A\|^2]} = \sqrt{\Tinf[1_A]}.
\end{equation}
(Note that $\Tinf[1_A]$ here is actually one quarter of the average sensitivity of~$A$, because we're using $0$-$1$ indicators as opposed to~$\pm 1$).  But the inequality in~\eqref{eqn:tal-surf-vs-surf} is often far from sharp. For example, while the majority function has average sensitivity~$\Theta(\sqrt{n})$, the expected square-root of its sensitivity is~$\Theta(1)$ because a~$\Theta(1/\sqrt{n})$-fraction of strings have sensitivity $\lceil n/2 \rceil$ and the remainder have sensitivity~$0$.
                                            \index{isoperimetric inequality!Hamming cube}%

Let's turn to the proof of Bobkov's Inequality. As you are asked to show in Exercise~\ref{ex:bobkov-induction}, the general-$n$ case of Bobkov's Inequality follows from the $n = 1$ case by a straightforward ``induction by restrictions''.  Thus just as in the proof of the Hypercontractivity Theorem, it suffices to prove the $n = 1$ ``two-point inequality'', an elementary inequality about two real numbers:
\begin{named}{Bobkov's Two-Point Inequality}
    Let $f \co \bits \to [0,1]$.  Then
    \[
        \giso(\E[f]) \leq \E[\|(\giso(f), \grad f)\|].
    \]
    Writing $f(x) = a + bx$, this is equivalent to saying that provided $a \pm b \in [0,1]$,
    \[
        \giso(a) \leq \half \|(\giso(a+b), b)\| + \half \|(\giso(a-b), b)\|.
    \]
\end{named}
\begin{remark}                                  \label{rem:giso-diffequ}
    The only property of $\giso$ used in proving this inequality is that it satisfies (Exercise~\ref{ex:giso-diffequ}) the differential equation $\calU \calU'' = -1$ on~$(0,1)$.
\end{remark}
Bobkov's proof of the two-point inequality was elementary but somewhat long and hard to motivate.  In contrast, Barthe and Maurey~\cite{BM00a} gave a fairly short proof of the inequality, but it used methods from stochastic calculus, namely It\^o's Formula.
                                                \index{It\^o's Formula}%
We present here an elementary discretization of the Barthe--Maurey proof.
\begin{proof}[Proof of Bobkov's Two-Point Inequality.]
    By symmetry and continuity we may assume $\delta \leq a-b < a+b \leq 1-\delta$ for some $\delta > 0$.  Let $\tau = \tau(\delta) > 0$ be a small quantity to be chosen later such that $b/\tau$ is an integer.  Let $\by_0, \by_1, \by_2, \dots$ be a random walk within $[a-b, a+b]$ that starts at $\by_0 = a$, takes independent equally likely steps of $\pm \tau$, and is absorbed at the endpoints $a \pm b$.  Finally, for $t \in \N$, define $\bz_t = \|(\giso(\by_t), \tau \sqrt{t})\|$.  The key claim for the proof is:
    \begin{claim}           \label{claim:bobkov}
         Assuming $\tau = \tau(\delta) > 0$ is small enough, $(\bz_t)_t$ is a submartingale with respect to~$(\by_t)_t$, i.e., $\E[\bz_{t+1} \mid \by_0, \dots, \by_t] = \E[\bz_{t+1} \mid \by_t] \geq \bz_t$.
    \end{claim}
    Let's complete the proof given the claim.  Let $\bT$ be the stopping time at which $\by_t$ first reaches~$a\pm b$.  By the Optional Stopping Theorem we have $\E[\bz_0] \leq \E[\bz_\bT]$; i.e.,
    \begin{equation}            \label{eqn:submartingale-deduction}
        \giso(a) \leq \E[\|(\giso(\bz_\bT), \tau\sqrt{\bT})\|].
    \end{equation}
    In the expectation above we can condition on whether the walk stopped at $a+b$ or~$a-b$.  By symmetry, both events occur with probability~$1/2$ and neither changes the conditional distribution of~$\bT$.  Thus we get
    \begin{align*}
        \giso(a) &\leq \half\E[\|(\giso(a+b), \tau \sqrt{\bT})\|] + \half\E[\|(\giso(a-b), \tau \sqrt{\bT})\|] \\
                 &\leq \half \|(\giso(a+b),  \sqrt{\E[\tau^2\bT]})\| + \half\|(\giso(a-b),  \sqrt{\E[\tau^2\bT]})\|,
    \end{align*}
    with the second inequality using concavity of $v \mapsto \sqrt{u^2 + v}$.  But it's a well-known fact (following immediately from Exercise~\ref{ex:walk-stopping}) that $\E[\bT] = (b/\tau)^2$.  Substituting this into the above completes the proof.

    It remains to verify Claim~\ref{claim:bobkov}.  Actually, although the claim is true as stated (see Exercise~\ref{ex:bobkov-submartingale}) it will be more natural to prove the following slightly weaker claim:
    \begin{equation}    \label{eqn:weaker-submartingale}
        \E[\bz_{t+1} \mid \by_t] \geq \bz_t - C_\delta \tau^3
    \end{equation}
    for some constant $C_\delta$ depending only on~$\delta$.  This is still enough to complete the proof: Applying the Optional Stopping Theorem to the submartingale~${(\bz_t + C_\delta \tau^3 t)_t}$ we get that~\eqref{eqn:submartingale-deduction} holds up to an additive~$C_\delta \tau^3 \E[\bT] = C_\delta b^2 \tau$.  Then continuing with the above we deduce Bobkov's Inequality up to~$C_\delta b^2 \tau$, and we can make~$\tau$ arbitrarily small.

    Even though we only need to prove~\eqref{eqn:weaker-submartingale}, let's begin a proof of the original Claim~\ref{claim:bobkov} anyway.  Fix $t \in \N^+$ and condition on $\by_t = y$.  If~$y$ is~$a \pm b$, then the walk is stopped and the claim is clear.  Otherwise, $\by_{t+1}$ is $y \pm \tau$ with equal probability, and we want to
    verify the following inequality (assuming $\tau > 0$ is sufficiently small as a function of~$\delta$, independent of~$y$):
    \begin{align}
        \|(\giso(y), \tau \sqrt{t})\|&\leq \half \|(\giso(y+\tau), \tau\sqrt{t+1})\| + \half \|(\giso(y-\tau), \tau\sqrt{t+1})\| \label{eqn:bobkov-verify}\\
        &=    \half \bigl\|\bigl(\sqrt{\giso(y+\tau)^2 + \tau^2}, \tau\sqrt{t}\bigr)\bigr\| + \half \bigl\|\bigl(\sqrt{\giso(y-\tau)^2 + \tau^2}, \tau\sqrt{t}\bigr)\bigr\|. \nonumber
    \end{align}
    By the triangle inequality, it's sufficient to show
    \[
        \giso(y) \leq \half \sqrt{\giso(y+\tau)^2 + \tau^2} + \half \sqrt{\giso(y-\tau)^2 + \tau^2},
    \]
    and this is actually necessary too, being the $t = 0$ case of~\eqref{eqn:bobkov-verify}.  (In fact, this is identical to Bobkov's Two-Point Inequality itself, except now we may assume~$\tau$ is sufficiently small.) Finally, since we actually only need the weakened submartingale statement~\eqref{eqn:weaker-submartingale}, we'll instead establish
    \begin{equation}        \label{eqn:bobkov-taylor}
        \giso(y) - C_\delta \tau^3 \leq \half \sqrt{\giso(y+\tau)^2 + \tau^2} + \half \sqrt{\giso(y-\tau)^2 + \tau^2}
    \end{equation}
    for some constant $C_\delta$ depending only on~$\delta$ and for every $\tau \leq \frac{\delta}{2}$.  We do this using  Taylor's theorem. Write $V_y(\tau)$ for the function of~$\tau$ on the right-hand side of~\eqref{eqn:bobkov-taylor}. For any $y \in [a-b,a+b]$ the function~$V_y$ is smooth on $[0,\frac{\delta}{2}]$ because~$\calU$ is a smooth, positive function on~$[\frac{\delta}{2}, 1 - \frac{\delta}{2}]$. Thus
    \[
        V_y(\tau) = V_y(0) + V_y'(0)\tau + \tfrac12 V_y''(0)\tau^2 + \tfrac16 V_y'''(\xi) \tau^3
    \]
    for some $\xi$ between~$0$ and~$\tau$.  The magnitude of $V_y'''(\xi)$ is indeed bounded by some~$C_\delta$ depending only on~$\delta$, using the fact that $\calU$ is smooth and positive on $[\frac{\delta}{2}, 1 - \frac{\delta}{2}]$.  But $V_y(0) = \giso(y)$, and it's straightforward to calculate that
    \[
        V_y'(0) = 0, \qquad V_y''(0) = \giso''(y) + 1/\giso(y) = 0,
    \]
    the last identity used the key property $\giso'' = -1/\giso$ mentioned in Remark~\ref{rem:giso-diffequ}.  Thus we conclude $V_y(\tau) \geq \giso(y) - C_\delta \tau^3$, verifying~\eqref{eqn:bobkov-taylor} and completing the proof.
\end{proof}

As a matter of fact, by a minor adjustment (Exercise~\ref{ex:gen-bobkov}) to this random walk argument we can establish the following generalization of Bobkov's Inequality:
\begin{theorem}                                     \label{thm:my-bobkov}
    Let $f \co \bn \to [0,1]$.  Then $\E[\|(\giso(\T_\rho f), \grad \T_\rho f)\|]$ is an increasing function of $\rho \in [0,1]$.  We recover Bobkov's Inequality by considering $\rho = 0, 1$.
\end{theorem}
                                            \index{Bobkov's Inequality|)}%
We end this section by remarking that De, Mossel, and Neeman~\cite{DMN13} have given a ``Bobkov-style'' Boolean inductive proof that yields both Borell's Isoperimetric Theorem and also the Majority Is Stablest Theorem (albeit with some aspects of the Invariance Principle-based proof appearing in the latter case); see Exercise~\ref{ex:DMN} and the notes at the end of this chapter.
                                                    \index{analysis of Gaussian functions|)}%

\section{The Berry--Esseen Theorem}                             \label{sec:berry-esseen}
                                            \index{Central Limit Theorem}%
                                            \index{Berry--Esseen Theorem|(}%
Now that we've built up some results concerning Gaussian space, we're motivated to try reducing problems involving Boolean functions to problems involving Gaussian functions.  The key tool for this is the Invariance Principle, discussed at the beginning of the chapter.  As a warmup, this section is devoted to proving (a form of) the Berry--Esseen Theorem.  As discussed in Chapter~\ref{sec:majority}, the Berry--Esseen Theorem is a quantitative form of the Central Limit Theorem for finite sums of independent random variables.  We restate it here:
\begin{named}{Berry--Esseen Theorem}
    Let $\bX_1, \dots, \bX_n$ be independent random variables with $\E[\bX_i] = 0$ and $\Var[\bX_i] = \sigma_i^2$, and assume $\sum_{i=1}^n \sigma_i^2 = 1$.  Let $\bS = \sum_{i=1}^n \bX_i$ and let $\bZ \sim \normal(0,1)$ be a standard Gaussian. Then for all $u \in \R$,
    \[
        |\Pr[\bS \leq u] - \Pr[\bZ \leq u]| \leq c \gamma,
    \]
    where
    \[
        \gamma = \sum_{i=1}^n \|\bX_i\|_3^3
    \]
    and $c$ is a universal constant.  (For definiteness, $c = .56$ is acceptable.)
\end{named}

In this traditional statement of Berry--Esseen, the error term $\gamma$ is a little opaque.  To say that~$\gamma$ is small is to simultaneously say two things: the random variables $\bX_i$ are all ``reasonable'' (as in Chapter~\ref{sec:4th-moment}); and, none is too dominant in terms of variance.  In Chapter~\ref{sec:4th-moment} we discussed several related notions of ``reasonableness'' for a random variable~$\bX$. It was convenient there to use the definition that $\|\bX\|_4^4$ is not much larger than $\|\bX\|_2^4$. For the Berry--Esseen Theorem it's more convenient (and slightly stronger) to use the analogous condition for the $3$rd moment.  (For the Invariance Principle it will be more convenient to use $(2,3,\rho)$- or $(2,4,\rho)$-hypercontractivity.)  The implication for Berry--Esseen is the following:
                                                \index{reasonable random variable}%
\begin{remark}                                  \label{rem:be-error}
    In the Berry--Esseen Theorem, if all of the $\bX_i$'s are ``reasonable'' in the sense that $\|\bX_i\|_3^3 \leq B \|\bX_i\|_2^3 = B \sigma_i^{3}$, then we can use the bound
    \begin{equation} \label{eqn:be-no-variance-dom}
        \gamma \leq B \cdot \max_i\{\sigma_i\},
    \end{equation}
    as this is a consequence of
    \[
        \gamma  =\sum_{i=1}^n \|\bX_i\|_3^3 \leq B \sum_{i=1}^n \sigma_i^{3} \leq B \cdot \max_i \{\sigma_i\} \cdot \sum_{i=1}^n \sigma_i^2  = B \cdot \max_i \{\sigma_i\}.
    \]
    (Cf.\ Remark~\ref{rem:simpler-BE}.)  Note that some ``reasonableness'' condition must hold if  $\bS = \sum_i \bX_i$ is to behave like a Gaussian. For example, if each $\bX_i$ is the ``unreasonable'' random variable which is $\pm \sqrt{n}$ with probability $\frac{1}{2n^2}$ each and~$0$ otherwise, then $\bS = 0$ except with probability at most $\frac{1}{n}$ -- quite unlike a Gaussian.  Further, even assuming reasonableness we still need a condition like~\eqref{eqn:be-no-variance-dom} ensuring that no~$\bX_i$ is too dominant (``influential'') in terms of variance.  For example, if $\bX_1 \sim \bits$ is a uniformly random bit and $\bX_2, \dots, \bX_n \equiv 0$, then $\bS \equiv \bX_1$, which is again quite unlike a Gaussian.
\end{remark}

There are several known ways to prove the Berry--Esseen Theorem; for example, using characteristic functions (i.e., ``real'' Fourier analysis), or Stein's Method.  We'll use the ``Replacement Method'' (also known as the Lindeberg Method, and similar to the ``Hybrid Method'' in theoretical cryptography).
                                                \index{Replacement Method}%
                                                \index{Lindeberg Method|seeonly{Replacement Method}}%
Although it doesn't always give the sharpest results, it's a very flexible technique which generalizes easily to higher-degree polynomials of random variables (as in the Invariance Principle) and random vectors.  The Replacement Method suggests itself as soon as the Berry--Esseen Theorem is written in a slightly different form: Instead of trying to show
\begin{equation} \label{eqn:replacement-no-suggest}
    \bX_1 + \bX_2 + \cdots + \bX_n \approx \bZ,
\end{equation}
where $\bZ \sim \normal(0,1)$, we'll instead try to show the equivalent statement
\begin{equation} \label{eqn:replacement-suggest}
    \bX_1 + \bX_2 + \cdots + \bX_n \approx \bZ_1 + \bZ_2 + \cdots + \bZ_n,
\end{equation}
where the $\bZ_i$'s are independent Gaussians with $\bZ_i \sim \normal(0,\sigma_i^2)$.  The statements~\eqref{eqn:replacement-no-suggest} and~\eqref{eqn:replacement-suggest} really are identical, since the sum of independent Gaussians is Gaussian, with the variances adding.  The Replacement Method proves~\eqref{eqn:replacement-suggest} by replacing the $\bX_i$'s with $\bZ_i$'s one by one.  Roughly speaking, we introduce the ``hybrid'' random variables
\newcommand{\Hyb}{\bH}
\[
    \Hyb_t = \bZ_1 + \cdots + \bZ_t + \bX_{t+1} + \cdots + \bX_n,
\]
show that $\Hyb_{t-1} \approx \Hyb_t$ for each $t \in [n]$, and then simply add up the~$n$ errors.

As a matter of fact, the Replacement Method doesn't really have anything to do with Gaussian random variables.  It actually seeks to show that
\[
    \bX_1 + \bX_2 + \cdots + \bX_n \approx \bY_1 + \bY_2 + \cdots + \bY_n
\]
whenever $\bX_1, \dots, \bX_n, \bY_1, \dots, \bY_n$ are independent random variables with ``matching first and second moments'', meaning $\E[\bX_i] = \E[\bY_i]$ and $\E[\bX_i^2] = \E[\bY_i^2]$ for each $i \in [n]$.  (The error will be proportional to $\sum_i (\|\bX_i\|^3 + \|\bY_i\|_3^3)$.) Another way of putting it (roughly speaking) is that the linear form $x_1 + \cdots + x_n$ is \emph{invariant} to what independent random variables you substitute in for $x_1, \dots, x_n$, so long as you always use the same first and second moments.  The fact that we can take the $\bY_i$'s to be Gaussians (with $\bY_i \sim \normal(\E[\bX_i], \Var[\bX_i])$) and then in the end use the fact that the sum of Gaussians is Gaussian to derive the simpler-looking
\[
    \bS = \sum_{i=1}^n \bX_i \approx \normal(\E[\bS], \Var[S])
\]
is just a pleasant bonus (and one that we'll no longer get once we look at \emph{nonlinear} polynomials of random variables in Section~\ref{sec:invariance}).  Indeed,  the remainder of this section will be devoted to showing that
\[
    \bS_X = \bX_1 + \cdots + \bX_n \quad \text{is ``close'' to} \quad \bS_Y = \bY_1 + \cdots + \bY_n
\]
whenever the $\bX_i$'s and $\bY_i$'s are independent, ``reasonable'' random variables with matching first and second moments.

To do this, we'll first have to discuss in more detail what it means for two random variables to be ``close''.
A traditional measure of closeness between two random variables $\bS_X$ and $\bS_Y$ is the ``cdf-distance'' used in the Berry--Esseen Theorem: $\Pr[\bS_X \leq u] \approx \Pr[\bS_Y \leq u]$ for every $u \in \R$.  But there are other natural measures of closeness too.  We might want to know that the absolute moments of $\bS_X$ and $\bS_Y$ are close; for example, that $\|\bS_X\|_1 \approx \|\bS_Y\|_1$. Or, we might like to know that $\bS_X$ and $\bS_Y$ stray from the interval $[-1,1]$ by about the same amount: $\E[\distint(\bS_X)] \approx \E[\distint(\bS_Y)]$.  Here we are using:
\begin{definition}
    For any interval $\emptyset \neq I \subsetneq \R$ the function $\dist_I \co \R \to \R^{\geq 0}$ is defined to measure the distance of a point from~$I$; i.e., $\dist_I(s) = \inf_{u \in I}\{|s-u|\}$.
\end{definition}
\newcommand{\test}{\psi}
All of the closeness measures just described can be put in a common framework: they are requiring $\E[\pred(\bS_X)] \approx \E[\pred(\bS_Y)]$ for various ``test functions'' (or ``distinguishers'') $\pred \co \R \to \R$.
                                                    \index{test functions}%

\myfig{1}{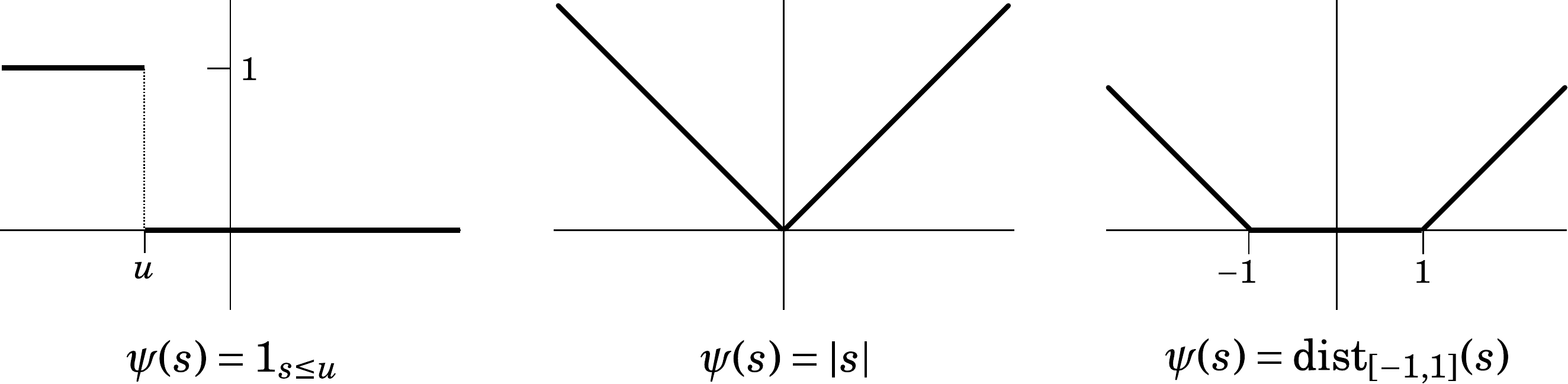}{The test functions $\pred$ used for judging $\Pr[\bS_X \leq u] \approx \Pr[\bS_Y \leq u]$, $\|\bS_X\|_1 \approx \|\bS_Y\|_1$, and $\E[\distint(\bS_X)] \approx \E[\distint(\bS_Y)]$, respectively}{fig:closeness-tests}

It would be nice to prove a version of the Berry--Esseen Theorem that showed closeness for all the test functions $\pred$ depicted in Figure~\ref{fig:closeness-tests}, and more.  What class of tests might we able to handle?  On one hand, we can't be \emph{too} ambitious.  For example, suppose each $\bX_i \sim \bits$, each $\bY_i \sim \normal(0,1)$, and $\psi(s) = 1_{s \in \Z}$.
Then $\E[\psi(\bS_X)] = 1$ because $\bS_X$ is supported on the integers, but $\E[\psi(\bS_Y)] = 0$ because $\bS_Y \sim \normal(0,n)$ is a continuous random variable.  On the other hand, there are some simple kinds of tests~$\psi$ for which we have exact equality. For example, if $\psi(s) = s$, then $\E[\psi(\bS_X)] = \E[\psi(\bS_Y)]$; this is by the assumption of matching first moments, $\E[\bX_i] = \E[\bY_i]$ for all~$i$.  Similarly, if $\psi(s) = s^2$, then
\begin{align}
    \E[\psi(\bS_X)] &= \E\Bigl[\bigl(\sum_i \bX_i\bigr)^2\Bigr] = \sum_{i} \E[\bX_i^2] + \sum_{i \neq j} \E[\bX_i\bX_j] \nonumber\\
    &= \sum_{i} \E[\bX_i^2] + \sum_{i \neq j} \E[\bX_i]\E[\bX_j]  \label{eqn:be-quadratic1}
\intertext{(using independence of the $\bX_i$'s); and also}
\label{eqn:be-quadratic2}
    \E[\psi(\bS_Y)] &= \sum_{i} \E[\bY_i^2] + \sum_{i \neq j} \E[\bY_i]\E[\bY_j].
\end{align}
The quantities~\eqref{eqn:be-quadratic1} and~\eqref{eqn:be-quadratic2} are equal because of the matching first and second moment conditions.

As a consequence of these observations we have $\E[\psi(\bS_X)] = \E[\psi(\bS_Y)]$ for any quadratic polynomial $\psi(s) = a + bs + cs^2$.  This suggests that to handle a general test~$\psi$ we try to approximate it by a quadratic polynomial up to some error; in other words, consider its $2$nd-order Taylor expansion.  For this to make sense the function~$\pred$ must have a continuous $3$rd derivative, and the error we incur will involve the magnitude of this derivative.  Indeed, we will now prove a variant of the Berry--Esseen Theorem for the class of $\calC^3$ test functions $\psi$ with $\psi'''$ uniformly bounded.  You might be concerned that this class doesn't contain \emph{any} of the interesting test functions depicted in Figure~\ref{fig:closeness-tests}.  But we'll be able to handle even those test functions with some loss in the parameters by using a simple ``hack'' -- approximating them by smooth functions, as suggested in Figure~\ref{fig:test-functions-hack}.
\myfig{.79}{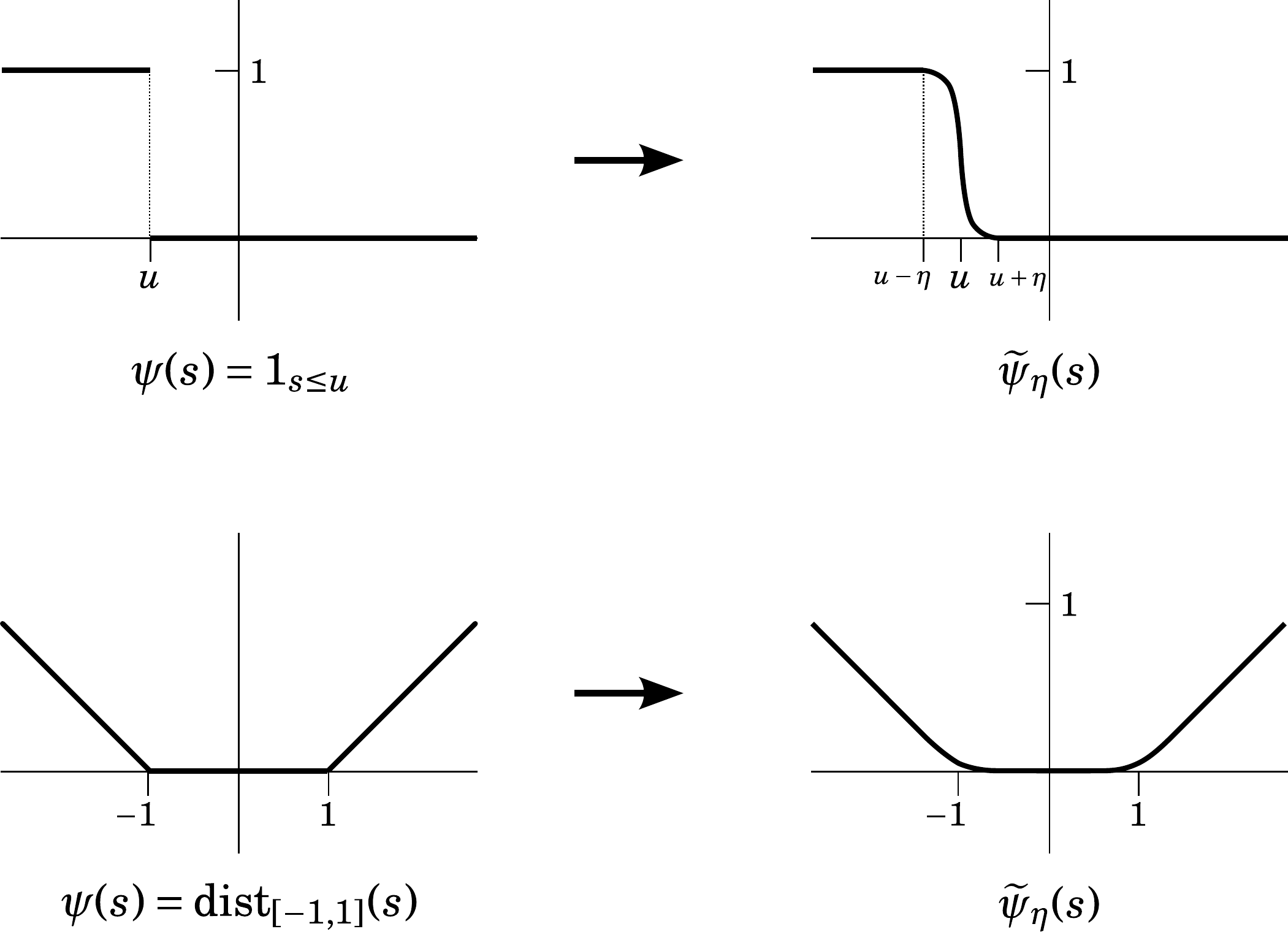}{The step function $\psi(s) = 1_{s \leq u}$ can be smoothed out on the interval $[u-\eta, u+\eta]$ so that the resulting function $\wt{\psi}_\eta$ satisfies $\|\wt{\psi}_\eta'''\|_\infty \leq O(1/\eta^3)$.  Similarly, we can smooth out $\psi(s) = \dist_{[-1,1]}(s)$ to a function $\wt{\psi}_\eta$ satisfying $\|\psi - \wt{\psi}\|_\infty \leq \eta$ and $\|\wt{\psi}_\eta'''\|_\infty \leq O(1/\eta^2)$.}{fig:test-functions-hack}

                                                \index{Invariance Principle!for sums of random variables}%
\begin{named}{Invariance Principle for Sums of Random Variables}
    Let $\bX_1, \dots, \bX_n$, $\bY_1, \dots, \bY_n$ be independent random variables with matching $1$st and $2$nd moments; i.e., $\E[\bX_i^k] = \E[\bY_i^k]$ for $i \in [n]$, $k \in \{1, 2\}$.  Write $\bS_X = \sum_i \bX_i$ and $\bS_Y = \sum_i \bY_i$. Then for any $\pred \co \R \to \R$ with continuous third derivative,
    \[
        \left| \E[\pred(\bS_X)] - \E[\pred(\bS_Y)] \right| \leq \tfrac16 \|\pred'''\|_\infty \cdot \gamma_{XY},
    \]
    where $\gamma_{XY} = \sum_i (\|\bX_i\|_3^3 + \|\bY_i\|_3^3)$.
\end{named}
\newcommand{\Hybm}{\bU}%
\begin{proof}
    The proof is by the Replacement Method.  For $0 \leq t \leq n$, define the ``hybrid'' random variable
    \[
        \Hyb_t = \bY_1 + \cdots + \bY_t + \bX_{t+1} + \cdots + \bX_n,
    \]
    so $\bS_X = \Hyb_0$ and $\bS_Y = \Hyb_n$.  Thus by the triangle inequality,
    \[
        \left| \E[\pred(\bS_X)] - \E[\pred(\bS_Y)] \right|
            \leq \sum_{t=1}^n \left| \E[\pred(\Hyb_{t-1})] - \E[\pred(\Hyb_t)] \right|.
    \]
    Given the definition of $\gamma_{XY}$, we can  complete the proof by showing that for each $t \in [n]$,
    \begin{align}
        \tfrac16 \|\pred'''\|_\infty \cdot (\E[|\bX_t|^3] + \E[|\bY_t|^3]) &\geq \left| \E[\pred(\Hyb_{t-1})] - \E[\pred(\Hyb_t)] \right| \nonumber\\
        &= \left| \E[\pred(\Hyb_{t-1}) - \pred(\Hyb_t)] \right| \nonumber\\
        &= \left| \E[\pred(\Hybm_t + \bX_t) -\pred(\Hybm_t + \bY_t)] \right|, \label{eqn:var-be-ineq}
    \end{align}
    where
    \[
        \Hybm_t = \bY_1 + \cdots + \bY_{t-1} + \bX_{t+1} + \cdots + \bX_n.
    \]
    Note that $\Hybm_t$ is independent of $\bX_t$ and $\bY_t$.     We are now comparing $\pred$'s values at $\Hybm_t + \bX_t$ and $\Hybm_t + \bY_t$, with the presumption that $\bX_t$ and $\bY_t$ are rather small compared to $\Hybm_t$.  This clearly suggests the use of Taylor's theorem: For all $u, \delta \in \R$,
    \[
        \pred(u + \delta) = \pred(u) + \pred'(u) \delta + \tfrac12 \pred''(u) \delta^2 + \tfrac16 \pred'''(u^*) \delta^3,
    \]
    for some $u^* = u^*(u,\delta)$ between $u$ and $u + \delta$.  Applying this pointwise with $u = \Hybm_t$, $\delta = \bX_t, \bY_t$ yields
    \begin{align*}
        \pred(\Hybm_t + \bX_t) &= \pred(\Hybm_t) +  \pred'(\Hybm_t) \bX_t + \tfrac12 \pred''(\Hybm_t) \bX_t^2 + \tfrac16 \pred'''(\Hybm_t^*) \bX_t^3\\
        \pred(\Hybm_t + \bY_t) &= \pred(\Hybm_t) +  \pred'(\Hybm_t) \bY_t + \tfrac12 \pred''(\Hybm_t) \bY_t^2 + \tfrac16 \pred'''(\Hybm_t^{**}) \bY_t^3
    \end{align*}
    for some random variables $\Hybm_t^*, \Hybm_t^{**}$.  Referring back to our goal of~\eqref{eqn:var-be-ineq}, what happens when we subtract these two identities and take expectations?  The $\pred(\Hybm_t)$ terms cancel.  The next difference is
    \[
        \E[\pred'(\Hybm_t) (\bX_t - \bY_t)] = \E[\pred'(\Hybm_t)]\cdot\E[\bX_t - \bY_t] = \E[\pred'(\Hybm_t)]\cdot 0 = 0,
    \]
    where the first equality used that $\Hybm_t$ is independent of $\bX_t$ and $\bY_t$, and the second equality used the matching $1$st moments of $\bX_t$ and $\bY_t$. An identical argument, using matching $2$nd moments, shows that the shows that the difference of the quadratic terms disappears in expectation.  Thus we're left only with the ``error term'':
    \begin{align*}
        \left| \E[\pred(\Hybm_t + \bX_t) -\pred(\Hybm_t + \bY_t)] \right| &=
        \tfrac16 \left| \E[\pred'''(\Hybm_t^*)\bX_t^3 -\pred'''(\Hybm_t^{**})\bY_t^3] \right| \\
        &\leq \tfrac16 \|\pred'''\|_\infty \cdot (\E[|\bX_t|^3] + \E[|\bY_t|^3]),
    \end{align*}
    where the last step used the triangle inequality.  This confirms~\eqref{eqn:var-be-ineq} and completes the proof.
\end{proof}

We can now give a Berry--Esseen-type corollary by taking the $\bY_i$'s to be Gaussians:
                                            \index{Berry--Esseen Theorem!Variant}%
\begin{named}{Variant Berry--Esseen Theorem}
    In the setting of the Berry--Esseen Theorem, for all $\calC^3$ functions $\pred \co \R \to \R$,
    \[
        \left| \E[\pred(\bS)] - \E[\pred(\bZ)] \right| \leq \tfrac16(1+2\sqrt{\tfrac{2}{\pi}})  \|\pred'''\|_\infty \cdot \gamma \leq .433 \|\pred'''\|_\infty \cdot \gamma.
    \]
\end{named}
\begin{proof}
    Applying the preceding  theorem with $\bY_i \sim \normal(0, \sigma_i^2)$ (and hence $\bS_Y \sim \normal(0, 1)$),
    it suffices to show that
    \begin{equation} \label{eqn:be-bound}
        \gamma_{XY} = \sum_{i=1}^n (\|\bX_i\|_3^3 + \|\bY_i\|_3^3) \leq  (1+2\sqrt{\tfrac{2}{\pi}}) \cdot \gamma = (1+2\sqrt{\tfrac{2}{\pi}}) \cdot \sum_{i=1}^n \|\bX_i\|_3^3.
    \end{equation}
    In particular, we just need to show that $\|\bY_i\|_3^3 \leq 2\sqrt{\tfrac{2}{\pi}}\|\bX_i\|_3^3$ for each~$i$.  This holds because Gaussians are extremely reasonable; by explicitly computing $3$rd absolute moments we indeed obtain
    \[
        \|\bY_i\|_3^3 = \sigma_i^3 \|\normal(0,1)\|_3^3 = 2\sqrt{\tfrac{2}{\pi}} \sigma_i^3 = 2\sqrt{\tfrac{2}{\pi}} \|\bX_i\|_2^3 \leq 2\sqrt{\tfrac{2}{\pi}} \|\bX_i\|_3^3. \qedhere
    \]
\end{proof}

This version of the Berry--Esseen Theorem is incomparable with the standard version.  Sometimes it can be stronger; for example, if for some reason we wanted to show $\E[\cos \bS] \approx \E[\cos \bZ]$
then the Variant Berry--Esseen Theorem gives this with error $.433\gamma$, whereas it can't be directly deduced from the standard Berry--Esseen at all.  On the other hand, as we'll see shortly, we can only obtain the standard Berry--Esseen conclusion from the Variant version with an error bound of $O(\gamma^{1/4})$ rather than $O(\gamma)$.

We end this section by describing the ``hacks'' which let us extend the Variant Berry--Esseen Theorem to cover certain non-$\calC^3$ tests~$\pred$.  As mentioned the idea is to smooth them out, or ``mollify'' them:
                                                    \index{mollification}%
                                                    \index{test functions!Lipschitz}%
\begin{proposition}                                     \label{prop:gaussian-mollify}
    Let $\psi \co \R \to \R$ be $c$-Lipschitz. Then for any $\eta > 0$ there exists $\wt{\psi}_\eta \co \R \to \R$ satisfying $\|\psi - \wt{\psi}_\eta\|_\infty \leq c \eta$ and $\|\wt{\psi}_\eta^{(k)}\|_\infty \leq C_k c/\eta^{k-1}$ for each $k \in \N^+$. Here $C_k$ is a constant depending only on~$k$, and $\wt{\psi}_\eta^{(k)}$ denotes the $k$th derivative of~$\wt{\psi}_\eta$.
\end{proposition}
\noindent The proof of this proposition is straightforward, taking $\wt{\psi}_\eta(s) = \displaystyle \E_{\bg \sim \normal(0,1)}[\psi(s+\eta \bg)]$; see Exercise~\ref{ex:gaussian-mollify}.

As $\eta \to 0$ this gives a better and better smooth approximation to $\psi$, but also a larger and larger value of $\|\wt{\psi}_\eta'''\|_\infty$.  Trading these off gives the following:
\begin{corollary}                                  \label{cor:lipschitz-be}
    In the setting of the Invariance Principle for Sums of Random Variables, if we merely have that $\pred \co \R \to \R$ is $c$-Lipschitz, then
    \[
        \left| \E[\pred(\bS_X)] - \E[\pred(\bS_Y)] \right| \leq O(c) \cdot \gamma_{XY}^{1/3}.
    \]
\end{corollary}
\begin{proof}
    Applying the Invariance Principle for Sums of Random Variables with the test $\wt{\pred}_\eta$ from Proposition~\ref{prop:gaussian-mollify} we get
    \[
        \left| \E[\wt{\pred}_\eta(\bS_X)] - \E[\wt{\pred}_\eta(\bS_Y)] \right| \leq O(c/\eta^2) \cdot \gamma_{XY}.
    \]
    But $\|\wt{\pred}_\eta - \pred\|_\infty \leq c\eta$ implies
    \[
        \left| \E[\wt{\pred}_\eta(\bS_X)] - \E[\pred(\bS_X)] \right| \leq \E[|\wt{\pred}_\eta(\bS_X) - \pred(\bS_X)|]  \leq c\eta
    \]
    and similarly for $\bS_Y$.  Thus we get
    \[
        \left| \E[\pred(\bS_X)] - \E[\pred(\bS_Y)] \right| \leq O(c) \cdot (\eta + \gamma_{XY}/ \eta^2)
    \]
    which yields the desired bound by taking $\eta = \gamma_{XY}^{1/3}$.
\end{proof}
\begin{remark}
    It's obvious that the dependence on~$c$ in this theorem should be linear in~$c$; in fact, since we can always divide $\pred$ by~$c$ it would have sufficed to prove the theorem assuming~$c = 1$.
\end{remark}

This corollary covers all Lipschitz tests, which suffices for the functions $\psi(s) = |s|$ and $\psi(s) = \dist_{[-1,1]}(s)$ from Figure~\ref{fig:closeness-tests}.  However, it still isn't enough for the test $\psi(s) = 1_{s \leq u}$ -- i.e., for establishing cdf-closeness as in the usual Berry--Esseen Theorem.  Of course, we can't hope for a smooth approximator $\wt{\psi}_\eta$ satisfying $|\wt{\psi}_\eta(s) - 1_{s \leq u}| \leq \eta$ for all~$s$ because of the discontinuity at~$u$.  However, as suggested in Figure~\ref{fig:test-functions-hack}, if we're willing to exclude $s \in [u - \eta, u+\eta]$ we can get an approximator with third derivative bound~$O(1/\eta^3)$, and thereby obtain (Exercises~\ref{ex:indicator-hack}, \ref{ex:be-levy}):
\begin{corollary}                                       \label{cor:be-levy}
    In the setting of the Invariance Principle for Sums of Random Variables, for all $u \in \R$ we have
    \[
        \Pr[\bS_Y \leq u - \eps] - \epsilon \leq \Pr[\bS_X \leq u] \leq \Pr[\bS_Y \leq u + \eps] + \eps
    \]
    for $\eps = O(\gamma_{XY}^{1/4})$; i.e., $\bS_X$ and $\bS_Y$ have  \emph{L\'evy distance}
                                                \index{L\'evy distance}%
    $\dlevy{\bS_X}{\bS_Y} \leq O(\gamma_{XY}^{1/4})$.
\end{corollary}
Finally, in the Berry--Esseen setting where $\bS_Y \sim \normal(0,1)$, we can appeal to the ``anticoncentration'' of
                                        \index{anticoncentration!Gaussians}%
Gaussians:
\[
    \Pr[\normal(0,1) \leq u + \eps] = \Pr[\normal(0,1) \leq u] + \Pr[u < \normal(0,1) \leq u+\eps] \leq \Pr[\normal(0,1) \leq u] + \tfrac{\eps}{\sqrt{2\pi}},
\]
and similarly for $\Pr[\normal(0,1) \leq u - \eps]$.  This lets us convert the L\'evy distance bound into a cdf-distance bound.  Recalling~\eqref{eqn:be-bound}, we immediately deduce the following weaker version of the classical Berry--Esseen Theorem:
\begin{corollary}                   \label{cor:weak-be}
    In the setting of the Berry--Esseen Theorem, for all $u \in \R$,
    \[
        \left| \Pr[\bS \leq u] - \Pr[\bZ \leq u \right| \leq O(\gamma^{1/4}),
    \]
    where the $O(\cdot)$ hides a universal constant.
\end{corollary}
\noindent Although the error bound here is weaker than necessary by a power of $1/4$, this weakness will be more than made up for by the ease with which the Replacement Method generalizes to other settings.  In the next section we'll see it applied to nonlinear polynomials of independent random variables.  Exercise~\ref{ex:multivar-be} outlines how to use it to give a Berry--Esseen theorem for sums of independent random vectors; as you'll see,
                                            \index{Berry--Esseen Theorem!multivariate}%
other than replacing Taylor's theorem with its multivariate form, hardly a symbol in the proof changes.
                                            \index{Berry--Esseen Theorem|)}

\section{The Invariance Principle}                                \label{sec:invariance}

                                                \index{Invariance Principles|(}
Let's summarize the Variant Berry--Esseen Theorem and proof from the preceding section, using slightly different notation.  (Specifically, we'll rewrite $\bX_i = a_i \bx_i$ where $\Var[\bx_i] = 1$, so $a_i = \pm \sigma_i$.)
                                            \index{Berry--Esseen Theorem!Variant}%
We showed that if $\bx_1, \dots, \bx_n, \by_1, \dots, \by_n$ are independent mean-$0$, variance-$1$ random variables, reasonable in the sense of having third absolute moment at most~$B$, and if $a_1, \dots, a_n$ are real constants assumed for normalization to satisfy $\sum_i a_i^2 = 1$, then
\begin{gather*}                                \label{eqn:be-generalizes}
    a_1 \bx_1 + \cdots + a_n \bx_n \approx a_1 \by_1 + \cdots + a_n \by_n, \\
     \text{with error bound proportional to } B \max\{|a_i|\}.
\end{gather*}
We think of this as saying that the linear form $a_1 x_1 + \cdots + a_n x_n$ is (roughly) \emph{invariant} to what independent mean-$0$, variance-$1$, reasonable random variables are substituted for the~$x_i$'s, so long as all  $|a_i|$'s are ``small'' (compared to the overall variance).  In this section we generalize this statement to degree-$k$ multilinear polynomial forms, $\sum_{|S| \leq k} a_S\,x^S$.  The appropriate generalization of the condition that ``all $|a_i|$'s are small'' is the condition that all ``influences'' $\sum_{S \ni i} a_S^2$ are small.  We refer to these nonlinear generalizations of Berry--Esseen as \emph{Invariance Principles}.

In this section we'll develop the most basic Invariance Principle, which involves replacing bits by Gaussians for a single Boolean function~$f$.  We'll show that this doesn't change the distribution of~$f$ much provided~$f$ has small influences and provided that~$f$ is of ``constant degree'' -- or at least, provided~$f$ is uniformly noise-stable
                                                \index{uniformly noise-stable}%
so that it's ``close to having constant degree''.  Invariance Principles in much more general settings are possible -- for example Exercises~\ref{ex:multifun-invariance} and~\ref{ex:general-invariance} describe variants which handle several functions applied to correlated inputs, and functions on general product spaces. Here we'll just focus on the simplest possible Invariance Principle,  which is already sufficient for the proof of the Majority Is Stablest Theorem in Section~\ref{sec:MIST}.
                                        \index{Majority Is Stablest Theorem}%

Let's begin with some notation.
\begin{definition}                                  \label{def:multilinear-poly}
    Let $F$ be a formal multilinear polynomial over the sequence of indeterminates $x = (x_1, \dots, x_n)$:
    \[
        F(x) = \sum_{S \subseteq [n]} \wh{F}(S) \prod_{i \in S} x_i,
    \]
    where the coefficients~$\wh{F}(S)$ are real numbers.  We introduce the notation
    \[
        \Var[F] = \sum_{S \neq \emptyset} \wh{F}(S)^2, \qquad \Inf_i[F] = \sum_{S \ni i} \wh{F}(S)^2.
    \]
\end{definition}
\begin{remark}                                        \label{rem:multlin-var-inf}
    To justify this notation, we remark that we'll always consider $F$ applied to a sequence $\bz = (\bz_1, \dots, \bz_n)$ independent random variables satisfying $\E[\bz_i] = 0$, $\E[\bz_i^2] = 1$.  Under these circumstances the collection of monomial random variables $\prod_{i \in S} \bz_i$ is orthonormal and so it's easy to see (cf.~Chapter~\ref{sec:general-fourier-formulas}) that
    \[
        \E[F(\bz)] = \wh{F}(\emptyset), \quad \E[F(\bz)^2] =  \sum_{S \subseteq [n]} \wh{F}(S)^2, \quad \Var[F(\bz)] = \Var[F] = \sum_{S \neq \emptyset} \wh{F}(S)^2.
    \]
    We also have $\E[\Var_{\bz_i}[F(\bz)]] = \Inf_i[F] = \sum_{S \ni i} \wh{F}(S)^2$, though we won't use this.
\end{remark}

As in the Berry--Esseen Theorem, to get good error bounds we'll need our random variables~$\bz_i$ to be ``reasonable''.  Sacrificing generality for simplicity in this section, we'll take the bounded $4$th-moment notion from Definition~\ref{def:reasonable} which will allow us to use the basic Bonami Lemma (more precisely, Corollary~\ref{cor:bonami-lemma}):
\begin{hypothesis}                                  \label{hyp:reasonable1}
    The random variable $\bz_i$ satisfies $\E[\bz_i] = 0$, $\E[\bz_i^2] = 1$, $\E[\bz_i^3] = 0$, and is ``$9$-reasonable'' in
                                                    \index{reasonable random variable}%
    the sense of Definition~\ref{def:reasonable}; i.e., $\E[\bz_i^4] \leq 9$.
\end{hypothesis}
\noindent The main examples we have in mind are that each~$\bz_i$ is either a uniform~$\pm 1$ random bit or a standard Gaussian.  (There are other possibilities, though; e.g., $\bz_i$ could be uniform on the interval $[-\sqrt{3}, \sqrt{3}]$.)

We can now prove the most basic Invariance Principle, for low-degree multilinear polynomials of random variables:
                                                \index{Invariance Principle!basic}%
\begin{named}{Basic Invariance Principle}
    Let $F$ be a formal $n$-variate multilinear polynomial of degree at most~$k \in \N$,
    \[
        F(x) = \sum_{S \subseteq [n], |S| \leq k} \wh{F}(S) \prod_{i \in S} x_i.
    \]
    Let $\bx = (\bx_1, \dots, \bx_n)$ and $\by = (\by_1, \dots, \by_n)$ be sequences of independent random variables, each satisfying Hypothesis~\ref{hyp:reasonable1}.  Assume $\pred \co \R \to \R$ is $\calC^4$ with $\|\pred''''\|_\infty \leq C$.  Then
    \begin{equation}                        \label{eqn:basic-inv-goal}
        \left|\E[\pred(F(\bx))] - \E[\pred(F(\by))]\right| \leq \tfrac{C}{12}\cdot 9^k \cdot \sum_{t=1}^n \Inf_t[F]^2.
    \end{equation}
\end{named}
\begin{remark}                                      \label{rem:basic-inv-3rd-mom}
    The proof will be very similar to the one we used for Berry--Esseen except that we'll take a $3$rd-order Taylor expansion rather than a $2$nd-order one (so that we can use the easy Bonami Lemma).  As you are asked to show in Exercise~\ref{ex:basic-inv-3rd-mom}, had we only required that $\test$ be $\calC^3$ and that the $\bx_i$'s and $\by_i$'s be $(2,3,\rho)$-hypercontractive with $2$nd moment equal to~$1$, then we could obtain
    \[
        \left|\E[\pred(F(\bx))] - \E[\pred(F(\by))]\right| \leq \tfrac{\|\pred'''\|_\infty}{3}\cdot (1/\rho)^{3k} \cdot \sum_{t=1}^n \Inf_t[F]^{3/2}.
    \]
\end{remark}
\begin{proof}
    The proof uses the Replacement Method.
                                                \index{Replacement Method}%
    For $0 \leq t \leq n$ we define
    \[
        \Hyb_t = F(\by_1, \dots, \by_{t}, \bx_{t+1}, \dots, \bx_{n}),
    \]
    so $F(\bx) = \Hyb_0$ and $F(\by) = \Hyb_n$.  We will show that
    \begin{equation}                                \label{eqn:basic-inv-step}
        \left|\E[\pred(\Hyb_{t-1}) - \pred(\Hyb_t)]\right| \leq \tfrac{C}{12}\cdot 9^k \cdot \Inf_t[F]^2;
    \end{equation}
    as in our proof of the Berry--Esseen Theorem, this will complete the proof after summing over~$t$ and using the triangle inequality. To analyze~\eqref{eqn:basic-inv-step} we  separate out the part of $F(x)$ that depends on~$x_t$; i.e., we write $F(x) = \uE_t F(x) + x_t \D_t F(x)$, where the formal polynomials $\uE_t F$ and $\D_t F$ are defined by
    \[
        \uE_tF(x) = \sum_{S \not \ni t} \wh{F}(S) \prod_{i \in S} x_i, \qquad \D_tF(x) = \sum_{S \ni t} \wh{F}(S) \prod_{i \in S \setminus \{t\}} x_i.
    \]
    Note that neither $\uE_t F$ nor $\D_t F$ depends on the indeterminate~$x_t$; thus we can define
\newcommand{\Invdel}{\bDelta}
    \begin{align*}
        \Hybm_t &= \uE_t F(\by_1, \dots, \by_{t-1}, \cdot, \bx_{t+1}, \dots, \bx_n), \\
        \Invdel_t &= \D_t F(\by_1, \dots, \by_{t-1}, \cdot, \bx_{t+1}, \dots, \bx_n),
    \end{align*}
    so that
    \[
        \Hyb_{t-1} = \Hybm_t + \Invdel_t\bx_t, \qquad         \Hyb_{t} = \Hybm_t + \Invdel_t\by_t.
    \]
    We now use a $3$rd-order Taylor expansion to bound~\eqref{eqn:basic-inv-step}:
    \begin{align*}
        \pred(\Hyb_{t-1}) &= \pred(\Hybm_t) + \pred'(\Hybm_t) \Invdel_t \bx_t + \half \pred''(\Hybm_t) \Invdel_t^2 \bx_t^2 + \tfrac16 \pred'''(\Hybm_t) \Invdel_t^3 \bx_t^3 + \tfrac{1}{24} \pred''''(\Hybm_t^*) \Invdel_t^4 \bx_t^4 \\
        \pred(\Hyb_{t}) &= \pred(\Hybm_t) + \pred'(\Hybm_t) \Invdel_t \by_t + \half \pred''(\Hybm_t) \Invdel_t^2 \by_t^2 + \tfrac16 \pred'''(\Hybm_t) \Invdel_t^3 \by_t^3 + \tfrac{1}{24} \pred''''(\Hybm_t^{**}) \Invdel_t^4 \by_t^4
    \end{align*}
    for some random variables $\Hybm_t^*$ and $\Hybm_t^{**}$.  As in the proof of the Berry--Esseen Theorem, when we subtract these and take the expectation there are significant simplifications.  The $0$th-order terms cancel.  As for the $1$st-order terms,
    \[
        \E[\pred'(\Hybm_t) \Invdel_t \bx_t - \pred'(\Hybm_t) \Invdel_t \by_t] =         \E[\pred'(\Hybm_t) \Invdel_t \cdot( \bx_t - \by_t)] = \E(\pred'(\Hybm_t) \Invdel_t] \cdot \E[\bx_t - \by_t] = 0.
    \]
    The second equality here crucially uses the fact that $\bx_t,\ \by_t$ are independent of $\Hybm_t,\ \Invdel_t$. The final equality only uses the fact that $\bx_t$ and~$\by_t$ have matching $1$st moments (and not the stronger assumption that both of these $1$st moments are~$0$).  The $2$nd- and $3$rd-order terms will similarly cancel, using the fact that $\bx_t$ and~$\by_t$ have matching $2$nd and $3$rd moments.  Finally, for the ``error'' term we'll just use $|\pred''''(\Hybm_t^*)|, |\pred''''(\Hybm_t^{**})| \leq C$ and the triangle inequality; we thus obtain
    \[
            \left|\E[\pred(\Hyb_{t-1}) - \pred(\Hyb_t)]\right| \leq \tfrac{C}{24} \cdot (\E[(\Invdel_t\bx_t)^4] + \E[(\Invdel_t\by_t)^4]).
    \]
    To complete the proof of~\eqref{eqn:basic-inv-step} we now just need to bound
    \[
        \E[(\Invdel_t\bx_t)^4],\ \E[(\Invdel_t\by_t)^4] \leq 9^k \cdot \Inf_t[F]^2,
    \]
    which we'll do using the Bonami Lemma.  We'll give the proof for $\E[(\Invdel_t\bx_t)^4]$, the case of $\E[(\Invdel_t\by_t)^4]$ being identical.  We have
    \[
        \Invdel_t \bx_t = \Lap_t F(\by_1, \dots, \by_{t-1}, \bx_t, \bx_{t+1}, \dots, \bx_n),
    \]
    where
    \[
        \Lap_t F(x) = x_t \D_t F(x) = \sum_{S \ni t} \wh{F}(S) \prod_{i \in S} x_i.
    \]
    Since $\Lap_t F$ has degree at most~$k$ we can apply the Bonami Lemma (more precisely, Corollary~\ref{cor:bonami-lemma}) to obtain
    \[
        \E[(\Invdel_t \bx_t)^4] \leq 9^k \E[\Lap_t F(\by_1, \dots, \by_{t-1}, \bx_t, \bx_{t+1}, \dots, \bx_n)^2]^2.
    \]
    But since $\by_1, \dots, \by_{t-1}, \bx_{t}, \dots, \bx_{n}$ are independent with mean~$0$ and $2$nd moment~$1$, we have (see Remark~\ref{rem:multlin-var-inf})
    \[
        \E[\Lap_t F(\by_1, \dots, \by_{t-1}, \bx_t, \bx_{t+1}, \dots, \bx_n)^2] = \sum_{S \subseteq [n]} \wh{\Lap_t F}(S)^2 = \sum_{S \ni t} \wh{F}(S)^2 = \Inf_t[F].
    \]
    Thus we indeed have $\E[(\Invdel_t \bx_t)^4] \leq 9^k \cdot \Inf_t[F]^2$, and the proof is complete.
\end{proof}

\begin{corollary}                                       \label{cor:simple-invariance}
    In the setting of the preceding theorem, if we furthermore have $\Var[F] \leq 1$ and $\Inf_t[F] \leq \eps$ for all $t \in [n]$, then
    \[
        \left|\E[\pred(F(\bx))] - \E[\pred(F(\by))]\right| \leq \tfrac{C}{12}\cdot k9^k \cdot \eps.
    \]
\end{corollary}
\begin{proof}
    We have $\sum_t \Inf_t[F]^2 \leq \eps \sum_t \Inf_t[F] \leq \eps \sum_{S} |S| \wh{F}(S)^2 \leq \eps k \Var[F]$.
\end{proof}
\begin{corollary}                                       \label{cor:simple-invariance-lipschitz}
    In the setting of the preceding corollary, if we merely have that $\pred \co \R \to \R$ is $c$-Lipschitz (rather than~$\calC^4$), then
    \[
        \left|\E[\pred(F(\bx))] - \E[\pred(F(\by))]\right| \leq O(c) \cdot 2^{k} \eps^{1/4}.
    \]
\end{corollary}
\begin{proof}
    Just as in the proof of Corollary~\ref{cor:lipschitz-be}, by using $\wt{\pred}_\eta$ from Proposition~\ref{prop:gaussian-mollify} (which has $\|\wt{\pred}_\eta''''\|_\infty \leq O(c/\eta^3)$) we obtain
    \[
        \left|\E[\pred(F(\bx))] - \E[\pred(F(\by))]\right| \leq O(c) \cdot (\eta + k 9^k \eps/\eta^3).
    \]
    The proof is completed by taking $\eta = \sqrt[4]{k 9^k \eps} \leq 2^k \eps^{1/4}$.
\end{proof}

Let's connect this last corollary back to the study of Boolean functions.  Suppose $f \btR$ has $\eps$-small influences (in the sense of Definition~\ref{def:small-influences}) and degree at most~$k$.  Letting $\bg = (\bg_1, \dots, \bg_n)$ be a sequence of independent standard Gaussians, Corollary~\ref{cor:simple-invariance-lipschitz} tells us that for any Lipschitz $\pred$ we have
\begin{equation} \label{eqn:cor-simple-inv}
    \left|\E_{\bx \sim \bn}[\pred(f(\bx))] - \E_{\bg \sim \normal(0,1)^n}[\pred(f(\bg))]\right| \leq O(2^{k} \eps^{1/4}).
\end{equation}
Here the expression ``$f(\bg)$'' is an abuse of notation indicating that the real numbers $\bg_1, \dots, \bg_n$ are substituted into $f$'s Fourier expansion (multilinear polynomial representation).

At first it may seem peculiar to substitute arbitrary real numbers into the Fourier expansion of a Boolean function.  Actually, if all the numbers being substituted are in the range $[-1,1]$ then there's a natural interpretation: as you were asked to show in Exercise~\ref{ex:multilin-interp}, if $\mu \in [-1,1]^n$, then $f(\mu) = \Ex[f(\by)]$ where $\by \sim \bn$ is drawn from the product distribution in which $\E[\by_i] = \mu_i$.  On the other hand, there doesn't seem to be any obvious meaning when real numbers outside the range~$[-1,1]$ are substituted into $f$'s Fourier expansion, as may certainly occur when we  consider~$f(\bg)$.

Nevertheless,~\eqref{eqn:cor-simple-inv} says that when~$f$ is a low-degree, small-influence function, the distribution of the random variable~$f(\bg)$ will be close to that of~$f(\bx)$.  Now suppose $f \btb$ is Boolean-valued and unbiased.  Then~\eqref{eqn:cor-simple-inv} might seem impossible; how could the continuous random variable~$f(\bg)$  essentially be~$-1$ with probability~$1/2$ and~$+1$ with probability~$1/2$?  The solution to this mystery is that there \emph{are no} low-degree, small-influence, unbiased Boolean-valued functions. This is a consequence of the OSSS~Inequality --
                                            \index{OSSS Inequality}%
more precisely, Exercise~\ref{ex:osss-variants}\ref{ex:weak-AA} -- which shows that in this setting we will always have $\eps \geq 1/k^3$ in~\eqref{eqn:cor-simple-inv}, rendering the bound very weak.  If the Aaronson--Ambainis Conjecture holds (see the notes in Chapter~\ref{sec:domains-notes}), a similar statement is true even for functions with range~$[-1,1]$.

The reason~\eqref{eqn:cor-simple-inv} is still useful is that we can apply it to small-influence, low-degree functions which are \emph{almost} $\{-1,1\}$-valued, or~$[-1,1]$-valued.  Such functions can arise from truncating a very noise-stable Boolean-valued function to a large but constant degree.  For example, we might profitably apply~\eqref{eqn:cor-simple-inv} to $f = \Maj_n^{\leq k}$ and then deduce some consequences for $\Maj_n(\bx)$ using the fact that $\E[(\Maj_n^{\leq k}(\bx) - \Maj_n(\bx))^2] = \W{> k}[\Maj_n] \leq O(1/\sqrt{k})$ (Corollary~\ref{cor:maj-asympt-asympt}).  Let's consider this sort of idea more generally:

\begin{corollary}                                       \label{cor:simple-invariance-lipschitz-trunc}
    Let $f \btR$ have $\Var[f] \leq 1$.  Let $k \geq 0$ and suppose $f^{\leq k}$ has $\eps$-small influences. Then for any $c$-Lipschitz $\pred \co \R \to \R$ we have
    \begin{equation}                                    \label{eqn:simple-inv-trunc1}
        \left|\E_{\bx \sim \bn}[\pred(f(\bx))] - \E_{\bg \sim \normal(0,1)^n}[\pred(f(\bg))]\right| \leq O(c) \cdot \bigl(2^{k} \eps^{1/4} + \|f^{> k}\|_2\bigr).
    \end{equation}
    In particular, suppose $h \btR$ has $\Var[h] \leq 1$ and no $(\eps,\delta)$-notable coordinates (we assume $\eps \leq 1$, $\delta \leq \frac{1}{20}$). Then
    \[
        \left|\E_{\bx \sim \bn}[\pred(\T_{1-\delta} h(\bx))] - \E_{\bg \sim \normal(0,1)^n}[\pred(\T_{1-\delta} h(\bg))]\right| \leq O(c) \cdot \eps^{\delta/3}.
    \]
\end{corollary}
\begin{proof}
    For the first statement we simply decompose $f = f^{\leq k} + f^{> k}$.  Then the left-hand side of~\eqref{eqn:simple-inv-trunc1} can be written as
    \begin{multline*}
        \left|\E[\pred(f^{\leq k}(\bx) + f^{> k}(\bx))] - \E[\pred(f^{\leq k}(\bg) + f^{> k}(\bg))]\right| \\
          \leq         \left|\E[\pred(f^{\leq k}(\bx))] - \E[\pred(f^{\leq k}(\bg))]\right| + c\E[|f^{>k}(\bx)|] + c\E[|f^{>k}(\bg)|],
    \end{multline*}
    using the fact that $\pred$ is $c$-Lipschitz.  The first quantity is at most $O(c) \cdot 2^{k} \eps^{1/4}$, by Corollary~\ref{cor:simple-invariance-lipschitz} (even if~$k$ is not an integer).  As for the other two quantities, Cauchy--Schwarz implies
    \[
        \E[|f^{>k}(\bx)|] \leq \sqrt{\E[f^{>k}(\bx)^2]} = \sqrt{\sum_{|S| > k} \wh{f}(S)^2} = \|f^{> k}\|_2,
    \]
    and the same bound also holds for $\E[|f^{>k}(\bg)|]$; this uses the fact that $\E[f^{>k}(\bg)^2] = \sum_{|S| > k} \wh{f}(S)^2$ just as in Remark~\ref{rem:multlin-var-inf}.  This completes the proof of~\eqref{eqn:simple-inv-trunc1}.

    As for the second statement of the corollary, let $f = \T_{1-\delta} h$.  The assumptions on $h$ imply that $\Var[f] \leq 1$ and that $f^{\leq k}$ has $\eps$-small influences for any~$k$; the latter is true because
    \[
        \Inf_i[f^{\leq k}] = \sum_{|S| \leq k, S \ni i} (1-\delta)^{2|S|} \wh{h}(S)^2 \leq \sum_{S \ni i} (1-\delta)^{|S| - 1} \wh{h}(S)^2 = \Inf_i^{(1-\delta)}[h] \leq \eps
    \]
    since $h$ has no $(\eps, \delta)$-notable coordinate.  Furthermore,
    \[
        \|f^{ > k}\|_2^2 = \sum_{|S| > k} (1-\delta)^{2|S|} \wh{h}(S)^2 \leq (1-\delta)^{2k} \Var[h] \leq  (1-\delta)^{2k} \leq \exp(-2k\delta)
    \]
    for any $k \geq 1$; i.e., $\|f^{ > k}\|_2 \leq \exp(-k\delta)$.  So applying the first part of the corollary gives
    \begin{equation}                \label{eqn:weird-balance}
        \left|\E[\pred(f(\bx))] - \E[\pred(f(\bg))]\right| \leq O(c) \cdot \bigl(2^{k} \eps^{1/4} + \exp(-k\delta)\bigr)
    \end{equation}
    for any~$k \geq 0$.  Choosing $k = \frac13 \ln(1/\eps)$, the right-hand side of~\eqref{eqn:weird-balance} becomes
    \[
        O(c) \cdot \left(\eps^{-(1/3)\ln 2} \eps^{1/4} + \eps^{\delta/3}\right) \leq O(c) \cdot \eps^{\delta/3},
    \]
    where the inequality uses the assumption $\delta \leq \frac{1}{20}$ (numerically, $\frac14 - \frac13 \ln 2 \approx \frac{1}{53}$). This completes the proof of the second statement of the corollary.
\end{proof}

Finally, if we think of the Basic Invariance Principle as the nonlinear analogue of our Variant Berry--Esseen Theorem, it's natural to ask for the nonlinear analogue of the Berry--Esseen Theorem itself, i.e., a statement showing cdf-closeness of~$F(\bx)$ and~$F(\bg)$.  It's straightforward to obtain a L\'{e}vy distance bound just as in the degree-$1$ case, Corollary~\ref{cor:be-levy}; Exercise~\ref{ex:basic-inv-levy} asks you to show the following:
\begin{corollary}                                       \label{cor:basic-inv-levy}
    In the setting of Corollary~\ref{cor:simple-invariance} we have the L\'evy distance bound     $\dlevy{F(\bx)}{F(\by)} \leq O(2^k\eps^{1/5})$.
                                                \index{L\'evy distance}%
    In the setting of Remark~\ref{rem:basic-inv-3rd-mom}  we have the  bound     $\dlevy{F(\bx)}{F(\by)} \leq (1/\rho)^{O(k)} \eps^{1/8}$.
\end{corollary}
Suppose we now want actual cdf-closeness in the case that $\by \sim \normal(0,1)^n$.  In the degree-$1$ (Berry--Esseen) case  we used the fact that degree-$1$ polynomials of independent Gaussians have good anticoncentration.  The analogous statement for higher-degree polynomials of Gaussians is not so easy to prove; however, Carbery and Wright~\cite[Theorem~$8$]{CW01} have obtained the following essentially optimal result:
                                                    \index{anticoncentration!polynomials of Gaussians|seeonly{Carbery--Wright Theorem}}%
                                                    \index{Carbery--Wright Theorem}%
\begin{named}{Carbery--Wright Theorem}
    Let $p \co \R^n \to \R$ be a polynomial (not necessarily multilinear) of degree at most~$k$, let $\bg \sim \normal(0,1)^n$, and assume $\E[p(\bg)^2] = 1$.  Then for all $\eps > 0$,
    \[
        \Pr[|p(\bg)| \leq \eps] \leq O(k \eps^{1/k}),
    \]
    where the $O(\cdot)$ hides a universal constant.
\end{named}

Using this theorem it's not hard (see Exercise~\ref{ex:simple-invariance-for-cdf-close}) to obtain:
\begin{theorem}                                   \label{thm:simple-invariance-for-cdf-close}
    Let $f \btR$ be of degree at most~$k$, with $\eps$-small influences and $\Var[f] = 1$.  Then for all $u \in \R$, \[
        \left| \Pr[f(\bx) \leq u] - \Pr[f(\bg) \leq u] \right| \leq O(k) \cdot \eps^{1/(4k+1)},
    \]
    where the $O(\cdot)$ hides a universal constant.
\end{theorem}
                                                    \index{Invariance Principles|)}

\section{Highlight: Majority Is Stablest Theorem}                                 \label{sec:MIST}

                                        \index{Max-Cut|(}%
                                        \index{Goemans--Williamson Algorithm|(}%
                                        \index{Majority Is Stablest Theorem}%
\newcommand{\cgw}{c_{\textnormal{GW}}}
The Majority Is Stablest Theorem (to be proved at the end of this section) was originally conjectured in~2004~\cite{KKMO04,KKMO07}.  The motivation came from studying the approximability of the Max-Cut~CSP.  Recall that Max-Cut is perhaps the simplest possible constraint satisfaction problem: the domain of the variables is~$\Omega = \{-1,1\}$ and the only constraint allowed is the binary non-equality predicate, $\neq \co \bits^2 \to \{0,1\}$.  As we mentioned briefly in Chapter~\ref{sec:CSPs}, Goemans and Williamson~\cite{GW95} gave a very sophisticated efficient algorithm using ``semidefinite programming'' which $(\cgw \beta, \beta)$-approximates Max-Cut for every~$\beta$, where $\cgw \approx .8786$ is a certain trigonometric constant.

Turning to hardness of approximation, we know from Theorem~\ref{thm:ug-hardness-from-dict-tests} (developed in~\cite{KKMO04}) that to prove \UG-hardness of $(\alpha+\delta, \beta-\delta)$-approximating Max-Cut, it suffices
                                    \index{UG-hardness}%
                                   \index{Dictator-vs.-@\dvq!connection with hardness}%
to construct an $(\alpha,\beta)$-\dvq which uses the predicate~$\neq$.  As we'll see in this section, the quality of the most natural such test can be easily inferred from the Majority Is Stablest Theorem.  Assuming that theorem (as Khot et~al.~\cite{KKMO04} did), we get a surprising conclusion: It's \UG-hard to approximate the Max-Cut CSP any better than the Goemans--Williamson Algorithm does.  In other words, the peculiar approximation guarantee of Goemans and Williamson on the very simple Max-Cut problem is optimal (assuming the Unique Games Conjecture).
                                    \index{Unique-Games}%

Let's demystify this somewhat, starting with a description of the Goemans--Williamson Algorithm.  Let $G = (V,E)$ be an $n$-vertex input graph for the algorithm; we'll write $(\bv, \bw) \sim E$ to denote that $(\bv, \bw)$ is a uniformly random edge (i.e., $\neq$-constraint) in the graph. The first step of the Goemans--Williamson Algorithm is to solve following optimization problem:
                                                    \index{semidefinite programming}%
                                                    \index{SDP|seeonly{semidefinite programming}}%
\begin{equation} \label{eqn:gw-sdp} \tag{SDP}
\begin{aligned}
    \text{maximize}& \quad  \E_{(\bv,\bw) \sim E}\left[\half - \half \la \vec{U}(\bv), \vec{U}(\bw) \ra\right] \\
    \text{subject to}& \quad \vec{U} \co V \to S^{n-1}.
\end{aligned}
\end{equation}
Here $S^{n-1}$ denotes the set of all unit vectors in~$\R^n$.  Somewhat surprisingly, since this optimization problem is a ``semidefinite program'' it can be solved in polynomial time using the Ellipsoid Algorithm.  (Technically, it can only be solved up to any desired additive tolerance~$\eps > 0$, but we'll ignore this point.)  Let's write $\SDPOpt(G)$ for the optimum value of~\eqref{eqn:gw-sdp}, and~$\Opt(G)$ for the optimum Max-Cut value for~$G$.  We claim that~\eqref{eqn:gw-sdp} is  a \emph{relaxation} of the Max-Cut CSP on input~$G$, and therefore
\[
    \SDPOpt(G) \geq \Opt(G).
\]
To see this, simply note that if $F^* \co V \to \{-1,1\}$ is an optimal assignment (``cut'') for~$G$ then we can define $\vec{U}(v) = (F^*(v), 0, \dots, 0) \in S^{n-1}$ for each $v \in V$ and achieve the optimal cut value $\Val_G(F^*)$ in~\eqref{eqn:gw-sdp}.

The second step of the Goemans--Williamson Algorithm might look familiar from Fact~\ref{fact:corr-gaussians} and Remark~\ref{rem:sincos-corr}.  Let $\vec{U}^* \co V \to S^{n-1}$ be the optimal solution for~\eqref{eqn:gw-sdp}, achieving~$\SDPOpt(G)$; abusing notation we'll write $\vec{U}^*(v) = \vec{v}$.  The algorithm now chooses $\vec{\bg} \sim \normal(0,1)^n$ at random and outputs the assignment~(cut) $\bF \co V \to \{-1,1\}$ defined by $\bF(v) = \sgn(\la \vec{v}, \vec{\bg} \ra)$.  Let's analyze the (expected) quality of this assignment.  The probability the algorithm's assignment~$\bF$ cuts a particular edge $(v,w) \in E$ is
\[
    \Pr_{\vec{\bg} \sim \normal(0,1)^n}[\sgn(\la \vec{v}, \vec{\bg} \ra) \neq  \sgn(\la \vec{w}, \vec{\bg} \ra)].
\]
This is precisely the probability that $\sgn(\bz) \neq \sgn(\bz')$ when $(\bz, \bz')$ is a pair of $\la \vec{v}, \vec{w}\ra$-correlated $1$-dimensional Gaussians.  Writing $\angle(\vec{v},\vec{w}) \in [0,\pi]$ for the angle between the unit vectors $\vec{v}, \vec{w}$, we conclude from Sheppard's Formula (see~\eqref{eqn:elegant-sheppard}) that
\[
    \Pr_{\vec{\bg}}[\bF \text{ cuts edge $(v,w)$}] = \frac{\angle(\vec{v}, \vec{w})}{\pi}.
\]
By linearity of expectation we can compute the expected value of the algorithm's assignment~$\bF$:
\begin{align}
    \E_{\vec{\bg}}[\val_G(\bF)] &= \Ex_{(\bv,\bw) \sim E}\left[\angle(\vec{\bv}, \vec{\bw})/\pi\right].  \label{eqn:gw-gets}
\intertext{On the other hand, by definition we have}
    \SDPOpt(G) &= \Ex_{(\bv,\bw) \sim E}\left[\half-\half\cos \angle(\vec{\bv}, \vec{\bw})\right]. \label{eqn:gw-sdpopt}
\end{align}
It remains to compare~\eqref{eqn:gw-gets} and~\eqref{eqn:gw-sdpopt}.  Define
\begin{equation}                    \label{eqn:cgw-def}
    \cgw = \min_{\theta \in [0,\pi]} \left\{\frac{\theta/\pi}{\half-\half \cos \theta}\right\} \approx .8786.
\end{equation}
Then from~\eqref{eqn:gw-gets} and~\eqref{eqn:gw-sdpopt} we immediately get
\[
    \E_{\vec{\bg}}[\val_G(\bF)]  \geq \cgw \cdot \SDPOpt(G) \geq \cgw \cdot \Opt(G);
\]
i.e., in expectation the Goemans--Williamson Algorithm delivers a cut of value at least $\cgw$ times the Max-Cut. In other words, it's a $(\cgw \beta, \beta)$-approximation algorithm, as claimed.
By being a little bit more careful about this analysis (Exercise~\ref{ex:careful-gw}) you can show following additional result:
\begin{theorem}                                     \label{thm:gw-behavior}
    \cite{GW95}. Let $\theta \in [\theta^*, \pi]$, where $\theta^* \approx .74\pi$ is the minimizing~$\theta$ in~\eqref{eqn:cgw-def} (also definable as the positive solution of $\tan(\theta/2) = \theta$).  Then on any graph~$G$ with $\SDPOpt(G) \geq \half - \half \cos \theta$, the Goemans--Williamson Algorithm produces a cut of (expected) value at least~$\theta/\pi$.  In particular, the algorithm is a $(\theta/\pi, \half - \half \cos \theta)$-approximation algorithm for Max-Cut.
\end{theorem}
\begin{example}                             \label{eg:5-cycle}
    Consider the Max-Cut problem on the $5$-vertex cycle graph~$\Z_5$.  The best bipartition of this graph cuts~$4$ out of the~$5$ edges; hence $\Opt(\Z_5) = \frac45$.  Exercise~\ref{ex:5-cycle} asks you to show that taking \[
        \vec{U}(v) = (\cos \tfrac{4\pi v}{5}, \sin \tfrac{4\pi v}{5}), \quad v \in \Z_5,
    \]
    in the semidefinite program~\eqref{eqn:gw-sdp} establishes that $\SDPOpt(\Z_5) \geq \half - \half \cos \frac{4\pi}{5}$. (These are actually unit vectors in~$\R^2$ rather than in~$\R^5$ as~\eqref{eqn:gw-sdp} requires, but we can pad out the last three coordinates with zeroes.) This example shows that the Goemans--Williamson analysis in Theorem~\ref{thm:gw-behavior} lower-bounding $\Opt(G)$ in terms of $\SDPOpt(G)$ cannot be improved (at least when $\SDPOpt(G) = \frac45$).  This is termed an optimal \emph{integrality gap}.  In fact, Theorem~\ref{thm:gw-behavior} also implies that $\SDPOpt(\Z_5)$ must equal $\half - \half \cos \frac{4\pi}{5}$, for if it were greater, the theorem would falsely imply that $\Opt(\Z_5) > \frac45$.  Note that the Goemans--Williamson Algorithm actually finds  the maximum cut  when run on the cycle graph~$\Z_5$.  For a related example, see Exercise~\ref{ex:4-cycle}.
\end{example}
                                        \index{Goemans--Williamson Algorithm|)}%

Now we explain the result of Khot et~al.~\cite{KKMO04}, that the Majority Is Stablest Theorem implies it's \UG-hard to approximate Max-Cut better than the Goemans--Williamson Algorithm does:
\begin{theorem}                                     \label{thm:kkmo}
    \cite{KKMO04}. Let $\theta \in (\frac{\pi}{2}, \pi)$.  Then for any $\delta > 0$ it's \UG-hard to $(\theta/\pi + \delta, \half - \half \cos\theta)$-approximate Max-Cut.
\end{theorem}
\begin{proof}
    It follows from Theorem~\ref{thm:ug-hardness-from-dict-tests} that we just need to construct a $(\theta/\pi, \half - \half \cos\theta)$-\dvq using the predicate~$\neq$.  (See Exercise~\ref{ex:kkmo-technical} for an extremely minor technical point.)  It's very natural to try the following, with $\beta = \half - \half \cos \theta \in (\frac12, 1)$:
                                                \index{Noise Sensitivity Test}%
                                                \index{Dictator-vs.-@\dvq}%
                                                \index{noise sensitivity}%
    \begin{named}{$\bbeta$-Noise Sensitivity Test}
        Given query access to $f \btb$:
        \begin{itemize}
            \item Choose $\bx \sim \bn$ and form $\bx'$ by reversing each bit of $\bx$ independently with probability~$\beta = \half - \half \cos\theta$.  In other words let $(\bx, \bx')$ be a pair of $\cos\theta$-correlated strings.  (Note that $\cos\theta < 0$.)
            \item Query $f$ at $\bx$, $\bx'$.
            \item Accept if $f(\bx) \neq f(\bx')$.
        \end{itemize}
    \end{named}
    \noindent By design,
    \begin{equation}                        \label{eqn:kkmo-acceptance}
        \Pr[\text{the test accepts } f] = \NS_\beta[f] = \half - \half \Stab_{\cos \theta}[f].
    \end{equation}
    (We might also express this as ``$\RS_f(\theta)$''.)
    In particular, if $f$ is a dictator, it's accepted with probability exactly~$\beta = \half - \half \cos \theta$.  To complete the proof that this is a $(\theta/\pi, \half - \half \cos\theta)$-\dvq, let's suppose $f \btI$ has no $(\eps,\eps)$-notable coordinates and show that~\eqref{eqn:kkmo-acceptance} is at most $\theta/\pi + o_\eps(1)$.  (Regarding $f$ having range $[-1,1]$, recall Remark~\ref{rem:dvq-int}.)

    At first it might look like we can immediately apply the Majority Is Stablest Theorem; however, the theorem's inequality goes the ``wrong way'' and the correlation parameter $\rho = \cos \theta$ is negative.  These two difficulties actually cancel each other out.  Note that
    \begin{align}
        \Pr[\text{the test accepts } f]  &= \half - \half \Stab_{\cos \theta}[f] \nonumber\\
        &= \half - \half \sum_{k=0}^n (\cos \theta)^{k} \W{k}[f] \nonumber\\
                                          &\leq \half + \half \sum_{\text{$k$ odd}} (-\cos \theta)^{k} \W{k}[f] \tag{since $\cos \theta < 0$} \nonumber\\
                                          &= \half + \half \Stab_{-\cos\theta}[f^\odd],    \label{eqn:finish-kkmo}
    \end{align}
    where $f^\odd \btI$ is the odd part of~$f$ (see Exercise~\ref{ex:odd-even}) defined by
    \[
        f^\odd(x) = \half(f(x) - f(-x)) = \sum_{|S|\text{ odd}} \wh{f}(S)\,x^S.
    \]
    Now we're really in a position to apply the Majority Is Stablest Theorem to~$f^\odd$, because $-\cos\theta \in (0,1)$, $\E[f^\odd] = 0$,  and $f^\odd$ has no $(\eps,\eps)$-notable coordinates (since it's formed from~$f$ by just dropping some terms in the Fourier expansion).  Using $-\cos\theta = \cos(\pi - \theta)$, the result is that
    \[
        \Stab_{-\cos\theta}[f^\odd] \leq 1 - \tfrac{2}{\pi} \arccos(\cos(\pi - \theta)) + o_\eps(1)
        = 2\theta/\pi - 1 + o_\eps(1).
    \]
     Putting this into~\eqref{eqn:finish-kkmo} yields
    \[
        \Pr[\text{the test accepts } f] \leq \half + \half (2\theta/\pi - 1 + o_\eps(1)) = \theta/\pi + o_\eps(1),
    \]
    as needed.
\end{proof}
\begin{remark}              \label{rem:ow08}
    There's actually still a mismatch between the algorithmic guarantee of Theorem~\ref{thm:gw-behavior} and the \UG-hardness result Theorem~\ref{thm:kkmo}, concerning the case of $\theta \in (\frac{\pi}{2}, \theta^*)$.  In fact, for these values of $\theta$ -- i.e., $\half \leq \beta \lessapprox .8446$ -- \emph{neither} result is sharp; see O'Donnell and Wu~\cite{OW08}.
\end{remark}

\begin{remark}                              \label{rem:easy-almost-gw}
    If we want to prove \UG-hardness of $(\theta'/\pi + \delta, \half - \half \cos\theta')$-approximating Max-Cut, we don't need the full version of Borell's Isoperimetric Theorem; we only need the volume-$\half$ case with parameter $\theta = \pi - \theta'$.
                                                    \index{Borell's Isoperimetric Theorem!volume-$\frac12$ case}%
    Corollary~\ref{cor:volume-half-borell} gave a simple proof of this result for $\theta = \frac{\pi}{4}$, hence $\theta' = \frac{3}{4} \pi$.  This yields \UG-hardness of $(\tfrac34 + \delta, \half + \frac{1}{2\sqrt{2}})$-approximating Max-Cut.  The ratio between $\alpha$ and $\beta$ here is approximately $.8787$, very close to the Goemans--Williamson constant $\cgw \approx .8786$.
\end{remark}
                                            \index{Max-Cut|)}%

Finally, we will prove the General-Volume Majority Is Stablest Theorem, by using the Invariance Principle to reduce it to Borell's Isoperimetric Theorem.
                                        \index{Majority Is Stablest Theorem|(}%
\begin{named}{General-Volume Majority Is Stablest Theorem}
    Let $f \co \bits^n \to [0,1]$.  Suppose that $\MaxInf[f] \leq \eps$, or more generally, that $f$ has no $(\eps, \frac{1}{\log(1/\eps)})$-notable coordinates.  Then for any $0 \leq \rho < 1$,
    \begin{equation}                                \label{eqn:mist}
        \Stab_\rho[f] \leq \GaussQuad_\rho(\E[f]) + O\bigl(\tfrac{\log \log (1/\eps)}{\log(1/\eps)}\bigr) \cdot \tfrac{1}{1-\rho}.
    \end{equation}
    (Here the $O(\cdot)$ bound has no dependence on~$\rho$.)
\end{named}
\begin{proof}
    The proof involves using the Basic Invariance Principle
                                                        \index{Invariance Principle!basic}%
    twice (in the form of Corollary~\ref{cor:simple-invariance-lipschitz-trunc}).  To facilitate this we introduce $f' = \T_{1-\delta} f$, where (with foresight) we choose
    \[
        \delta = 3 \tfrac{\log \log(1/\eps)}{\log(1/\eps)} \geq \tfrac{1}{\log(1/\eps)}.
    \]
    (We may assume $\eps$ is sufficiently small so that $0 < \delta \leq \frac{1}{20}$.)    Note that $\E[f'] = \E[f]$ and that
    \[
        \Stab_\rho[f'] = \sum_{S \subseteq [n]} \rho^{|S|} (1-\delta)^{2|S|} \wh{f}(S)^2 = \Stab_{\rho(1-\delta)^2}[f].
    \]
    But
    \begin{equation}                            \label{eqn:mist-tradeoff}
        \bigl|\Stab_{\rho(1-\delta)^2}[f] - \Stab_{\rho}[f]\bigr| \leq (\rho - \rho(1-\delta)^2) \cdot \tfrac{1}{1-\rho} \cdot \Var[f] \leq 2\delta \cdot \tfrac{1}{1-\rho}
    \end{equation}
    by Exercise~\ref{ex:stab-lipschitz}, and with our choice of~$\delta$ this can be absorbed into the error of~\eqref{eqn:mist}.  Thus it suffices to prove~\eqref{eqn:mist} with~$f'$ in place of~$f$.

\newcommand{\Sqr}{\textnormal{Sq}}
    Let $\Sqr \co \R \to \R$ be the continuous function which agrees with $t \mapsto t^2$ for $t \in [0,1]$ and is constant outside~$[0,1]$.  Note that $\Sqr$ is $2$-Lipschitz.  We will apply the second part of Corollary~\ref{cor:simple-invariance-lipschitz-trunc}  with ``$h$'' set to $\T_{\sqrt{\rho}} f$ (and thus $\T_{1-\delta} h = \T_{\sqrt{\rho}} f'$).  This is valid since the variance and $(1-\delta)$-stable influences of~$h$ are only smaller than those of~$f$.  Thus
    \begin{equation}                            \label{eqn:mist-halfway}
        \left|\E_{\bx \sim \bn}[\Sqr(\T_{\sqrt{\rho}}f'(\bx))] - \E_{\bg \sim \normal(0,1)^n}[\Sqr(\T_{\sqrt{\rho}} f'(\bg))]\right| \leq O(\eps^{\delta/3}) = O\bigl(\tfrac{1}{\log(1/\eps)}\bigr),
    \end{equation}
    using our choice of~$\delta$.  (In fact, it's trading off this error with~\eqref{eqn:mist-tradeoff} that led to our choice of~$\delta$.)
    Now $\T_{\sqrt{\rho}} f'(\bx) = \T_{(1-\delta)\sqrt{\rho}} f(\bx)$ is always bounded in~$[0,1]$, so
    \[
        \Sqr(\T_{\sqrt{\rho}} f'(\bx)) = (\T_{\sqrt{\rho}} f'(\bx))^2 \quad \implies \quad \E_{\bx \sim \bn}[\Sqr(\T_{\sqrt{\rho}}f'(\bx))] = \Stab_\rho[f'].
    \]
    Furthermore, $\T_{\sqrt{\rho}} f'(\bg)$ is the same as $\U_{\sqrt{\rho}} f'(\bg)$ because $f'$ is a multilinear polynomial. (Both are equal to $f'(\rho \bg)$; see Fact~\ref{fact:Urho-plug-in}.)  Thus in light of~\eqref{eqn:mist-halfway}, to complete the proof of~\eqref{eqn:mist} it suffices to show
    \begin{equation}                            \label{eqn:mist-suffices}
        \E_{\bg \sim \normal(0,1)^n}[\Sqr(\U_{\sqrt{\rho}} f'(\bg))] \leq \GaussQuad_\rho(\E[f']) + O\bigl(\tfrac{1}{\log(1/\eps)}\bigr).
    \end{equation}

    Define the function $F \co \R^n \to [0,1]$  by
    \[
        F(g) = \trunc_{[0,1]}(f'(g)) = \begin{cases}
                                            0 & \text{if $f'(g) < 0$,} \\
                                            f'(g) & \text{if $f'(g) \in [0,1]$,}\\
                                            1 & \text{if $f'(g) >1$.}
                                       \end{cases}
    \]
    We will establish the following two inequalities, which together imply~\eqref{eqn:mist-suffices}:
    \begin{gather}
         \left|\E_{\bg \sim \normal(0,1)^n}[\Sqr(\U_{\sqrt{\rho}} f'(\bg))] - \E_{\bg \sim \normal(0,1)^n}[\Sqr(\U_{\sqrt{\rho}} F(\bg))]\right|  \leq O\bigl(\tfrac{1}{\log(1/\eps)}\bigr),   \label{eqn:mist1} \\
         \E_{\bg \sim \normal(0,1)^n}[\Sqr(\U_{\sqrt{\rho}} F(\bg))] \leq \GaussQuad_\rho(\E[f']) + O\bigl(\tfrac{1}{\log(1/\eps)}\bigr). \label{eqn:mist2}
    \end{gather}
    Both of these inequalities will in turn follow from
    \begin{equation}                        \label{eqn:f'-to-F}
        \E_{\bg \sim \normal(0,1)^n}[|f'(\bg) - F(\bg)|] = \E_{\bg \sim \normal(0,1)^n}[\dist_{[0,1]}(f'(\bg))] \leq O\bigl(\tfrac{1}{\log(1/\eps)}\bigr).
    \end{equation}
    Let's show how~\eqref{eqn:mist1} and~\eqref{eqn:mist2} follow from~\eqref{eqn:f'-to-F}, leaving the proof of~\eqref{eqn:f'-to-F} to the end.  For~\eqref{eqn:mist1},
    \begin{align*}
        \left|\E[\Sqr(\U_{\sqrt{\rho}} f'(\bg))] - \E[\Sqr(\U_{\sqrt{\rho}} F(\bg))] \right|
           &\leq 2 \E[|\U_{\sqrt{\rho}} f'(\bg) - \U_{\sqrt{\rho}} F(\bg)|]\\
           &\leq 2 \E[|f'(\bg) - F(\bg)|] \leq O\bigl(\tfrac{1}{\log(1/\eps)}\bigr),
    \end{align*}
    where the first inequality used that $\Sqr$ is $2$-Lipschitz, the second inequality used the fact that $\U_{\sqrt{\rho}}$ is a contraction on $L^1(\R^n,\gamma)$, and the third inequality was~\eqref{eqn:f'-to-F}.
    As for~\eqref{eqn:mist2}, $\U_{\sqrt{\rho}} F$ is bounded in~$[0,1]$ since~$F$ is. Thus
    \[
        \E[\Sqr(\U_{\sqrt{\rho}} F(\bg))] = \E[(\U_{\sqrt{\rho}} F(\bg))^2] = \Stab_\rho[F] \leq \GaussQuad_\rho(\E[F(\bg)]),
    \]
    where we used Borell's Isoperimetric Theorem.
                                            \index{Borell's Isoperimetric Theorem}%
    But $|\E[F(\bg)] - \E[f'(\bg)]| \leq O\bigl(\tfrac{1}{\log(1/\eps)}\bigr)$ by~\eqref{eqn:f'-to-F}, and $\GaussQuad_\rho$ is easily shown to be $2$-Lipschitz (Exercise~\ref{ex:gaussquad}\ref{ex:GaussQuad-Lipschitz}).  This establishes~\eqref{eqn:mist2}.

    It therefore remains to show~\eqref{eqn:f'-to-F}, which we do by applying the Invariance Principle one more time.  Taking $\psi$ to be the $1$-Lipschitz function $\dist_{[0,1]}$ in Corollary~\ref{cor:simple-invariance-lipschitz-trunc} we deduce
    \begin{align*}
        \left|\E_{\bg \sim \normal(0,1)^n}[\dist_{[0,1]}(f'(\bg))] - \E_{\bx \sim \bn}[\dist_{[0,1]}(f'(\bx))]\right| \leq O(\eps^{\delta/3}) = O\bigl(\tfrac{1}{\log(1/\eps)}\bigr).
    \end{align*}
    But $\E[\dist_{[0,1]}f'(\bx)] = 0$ since $f'(\bx) = \T_{1-\delta} f(\bx) \in [0,1]$ always.  This establishes~\eqref{eqn:f'-to-F} and completes the proof.
\end{proof}
                                        \index{Majority Is Stablest Theorem|)}%

                                        \index{Arrow's Theorem}%
                                        \index{Condorcet Paradox}%
We conclude with one more application of the Majority Is Stablest Theorem.  Recall Kalai's version of Arrow's Theorem from Chapter~\ref{sec:arrow}, i.e., Theorem~\ref{thm:prob-condorcet}.  It states that in a $3$-candidate Condorcet election using the voting rule $f \btb$, the probability of having a Condorcet winner -- often called a \emph{rational outcome} -- is precisely $\tfrac34 - \tfrac34\Stab_{-1/3}[f]$.  As we saw in the proof of Theorem~\ref{thm:kkmo} near~\eqref{eqn:finish-kkmo}, this is in turn at most $\tfrac34 + \tfrac34\Stab_{1/3}[f^\odd]$, with equality if~$f$ is already odd.  It follows from the Majority Is Stablest Theorem  that among all voting rules with $\eps$-small influences (a condition all reasonable voting rules should satisfy), majority rule is the ``most rational''.  Thus we see that the principle of representative democracy can be derived using analysis of Boolean functions.

\section{Exercises and notes}                                   \label{sec:invariance-exercises}
\begin{exercises}
    \item \label{ex:really-simple-functions} Let $\calA$ be the set of all functions $f \co \R^n \to \R$ which are finite linear combinations of indicator functions of boxes. Prove that $\calA$ is dense in $L^1(\R^n, \gamma)$.
    \item \label{ex:gaussian-hypercon-theorem}  Fill in proof details for the Gaussian Hypercontractivity Theorem.

    \item \label{ex:Urho-plug-in} Prove Fact~\ref{fact:Urho-plug-in}. (Cf.~Exercise~\ref{ex:Trho-plug-in}.)
    \item \label{ex:semigroup-gaussian}
                                                    \index{semigroup property}%
        Show that $\U_{\rho_1} \U_{\rho_2} = \U_{\rho_1 \rho_2}$ for all $\rho_1, \rho_2 \in [-1,1]$.  (Cf.~Exercise~\ref{ex:semigroup}.)
    \item \label{ex:Urho-smooths}  Prove Proposition~\ref{prop:Urho-smooths}.  (Hint: For $\rho \neq 0$, write $g(z) = \U_\rho f(z)$ and show that $g(z/\rho)$ is a smooth function using the relationship between convolution and derivatives.)
    \item \label{ex:Urho-strongly-cts}
        \begin{exercises}
            \item Prove Proposition~\ref{prop:Urho-strongly-cts}. (Hint: First prove it for bounded continuous~$f$; then make an approximation and use Proposition~\ref{prop:U-contracts}.)
            \item Deduce more generally that for $f \in L^1(\R^n,\gamma)$ the map $\rho \mapsto \U_\rho f$ is ``strongly continuous'' on~$[0,1]$, meaning that for any $\rho \in [0,1]$ we have $\|\U_{\rho'} f - \U_\rho f\|_1 \to 0$ as $\rho' \to \rho$.  (Hint: Use Exercise~\ref{ex:semigroup-gaussian}.)
        \end{exercises}
    \item \label{ex:alternate-OU} Complete the proof of Proposition~\ref{prop:alternate-OU} by establishing the case of general~$n$.
    \item \label{ex:dirichlet2} Complete the proof of Proposition~\ref{prop:dirichlet2} by establishing the case of general~$n$.
                                                \index{Hermite polynomials|(}%
    \item \label{ex:alternate-hermite-formula-completethesquare}
        \begin{exercises}
            \item Establish the alternative formula~\eqref{eqn:hermite-alternate} for the probabilists' Hermite polynomials~$H_j(z)$ given in Definition~\ref{def:p-hermite}; equivalently, establish the formula
                \[
                    H_j(z) = (-1)^j \exp(\tfrac12 z^2) \cdot \left(\frac{d}{dz}\right)^j \exp(-\tfrac12 z^2).
                \]
                (Hint: Complete the square on the left-hand side of~\eqref{eqn:prob-herm}; then differentiate~$j$ times with respect to~$t$ and evaluate at~$0$.)
            \item \label{ex:hermite-deriv0} Establish the recursion
                    \[
                        H_j(z) = (z - \tfrac{d}{dz}) H_{j-1}(z) \quad\iff\quad \he_j(z) = \tfrac{1}{\sqrt{j}} \cdot (z - \tfrac{d}{dz}) h_{j-1}(z)
                    \]
                    for $j \in \N^+$, and hence the formula $H_j(z) = (z - \tfrac{d}{dz})^j 1$.
            \item Show that $\he_j(z)$ is an odd function of~$z$ if $j$ is odd and an even function of~$z$ if $j$ is even.
        \end{exercises}
    \item \label{ex:more-hermite-deriv}
        \begin{exercises}
            \item Establish the derivative formula for Hermite polynomials:
                    \[
                        H_j'(z) = j \cdot H_{j-1}(z) \quad\iff\quad \he_j'(z) = \sqrt{j} \cdot \he_{j-1}(z).
                    \]
            \item By combining this with the other formula for $H_j'(z)$ implicit in Exercise~\ref{ex:alternate-hermite-formula-completethesquare}\ref{ex:hermite-deriv0}, deduce the recursion
                    \[
                        H_{j+1}(z) = zH_j(z) - jH_{j-1}(z).
                    \]
            \item Show that $H_j(z)$ satisfies the second-order differential equation
                  \[
                        j H_j(z) = z H'_j(z) - H_j''(z).
                  \]
                  (It's equivalent to say that $\he_j(z)$ satisfies it.)                   Observe that this is consistent with Propositions~\ref{prop:alternate-OU} and~\ref{prop:ou-formula} and says that~$H_j$ (equivalently, $\he_j$) is an eigenfunction of the Ornstein--Uhlenbeck operator~$\Lap$, with eigenvalue~$j$.
        \end{exercises}
    \item Prove that
          \[
              H_j(x+y) = \sum_{i=0}^j \binom{j}{i} x^{j-i} H_i(y),
          \]
          and, relatedly, that for $p+q = 1$ we have
          \[
                h_k(\sqrt{p}x + \sqrt{q}y) = \sum_{i+j=k} \sqrt{\tbinom{k}{i,j}p^iq^j}\,h_i(x)h_j(y).
          \]
    \item \begin{exercises}
            \item By equating both sides of~\eqref{eqn:prob-herm} with
                  \[
                    \E_{\bg \sim \normal(0,1)}[\exp(t(z + i \bg))]
                  \]
                  (where $i = \sqrt{-1}$),  show that
                  \[
                      H_j(z) = \E_{\bg \sim \normal(0,1)}[(z + i \bg)^j].
                  \]
            \item  Establish the explicit formulas
                    \begin{align*}
                        H_j(z) &= \sum_{k=0}^{\lfloor j/2 \rfloor} (-1)^{k} \binom{j}{2k} \E_{\bg \sim \normal(0,1)}[\bg^{2k}] z^{j-2k} \\
                            &= j! \cdot \left(\frac{z^j}{0!! \cdot j!} - \frac{z^{j-2}}{2!! \cdot (j-2)!} + \frac{z^{j-4}}{4!!\cdot (j-4)!} - \frac{z^{j-6}}{6!! \cdot (j-6)!} + \cdots \right).
                    \end{align*}
          \end{exercises}
    \item  \label{ex:gaussian-derivs-and-infls}
        \begin{exercises}
            \item Establish the formula
                    \[
                        \E[\|\grad f\|^2] = \sum_{\alpha \in \N^n} |\alpha| \wh{f}(\alpha)^2
                    \]
                  for all $f \in L^2(\R^n, \gamma)$ (or at least for all $n$-variate polynomials~$f$).
            \item For $f \in L^2(\R^n, \gamma)$, establish the formula
                  \[
                    \sum_{i=1}^n \E[\Var_{\bz_i}[f]] = \sum_{\alpha \in \N^n} (\# \alpha)\wh{f}(\alpha)^2.
                  \]
          \end{exercises}
    \item \label{ex:krav-to-hermite} Show that for all $j \in \N$ and all $z \in \R$
                                                        \index{Kravchuk polynomials}%
          we have
          \[
              \tbinom{n}{j}^{-1/2} \cdot \krav^{(n)}_j\left(\frac{n}{2} - z \frac{\sqrt{n}}{2}\right) \xrightarrow{n \to \infty} \he_j(z),
          \]
          where $\krav^{(n)}_j$ is the Kravchuk polynomial of degree~$j$ from Exercise~\ref{ex:kravchuk} (with its dependence on $n$ indicated in the superscript).
                                                \index{Hermite polynomials|)}%
    \item \label{ex:essential-boundary}  Recall the definition~\eqref{eqn:heuristic-gsa} of the Gaussian Minkowski content of the boundary~$\bdry A$ of a set $A \subseteq \R^n$.  Sometimes the following very similar definition is also proposed for the Gaussian surface
                                        \index{Gaussian surface area}%
        area of~$A$:
        \[
            M(A) = \liminf_{\eps \to 0^+} \frac{\gvol(\{z : \dist(z, A) < \eps\}) - \gvol(A)}{\eps}.
        \]
        Consider the following subsets of~$\R$:
            \[
            A_1 = \emptyset, \quad A_2 = \{0\}, \quad A_3 = (-\infty, 0), \quad A_4 = (-\infty, 0], \quad A_5 = \R \setminus \{0\}, \quad A_6 = \R.
            \]
        \begin{exercises}
            \item Show that
                \begin{align*}
                    \gamma^+(A_1) &= 0 & M(A_1) &= 0 & \gsa(A_1) = 0 \\
                    \gamma^+(A_2) &= \tfrac{1}{\sqrt{2\pi}} & M(A_2) &= \sqrt{\tfrac{2}{\pi}} & \gsa(A_2) = 0 \\
                    \gamma^+(A_3) &= \tfrac{1}{\sqrt{2\pi}} & M(A_3) &= \tfrac{1}{\sqrt{2\pi}} & \gsa(A_3) = \tfrac{1}{\sqrt{2\pi}} \\
                    \gamma^+(A_4) &= \tfrac{1}{\sqrt{2\pi}} & M(A_4) &= \tfrac{1}{\sqrt{2\pi}} & \gsa(A_4) = \tfrac{1}{\sqrt{2\pi}} \\
                    \gamma^+(A_5) &= \tfrac{1}{\sqrt{2\pi}} & M(A_5) &= 0 & \gsa(A_5) = 0 \\
                    \gamma^+(A_6) &= 0 & M(A_6) &= 0 & \gsa(A_6) = 0. \\
                \end{align*}%
            \item For $A \subseteq \R^n$, the \emph{essential boundary} (or \emph{measure-theoretic boundary}) of~$A$ is defined to be
                \[
                    \partial_* A = \left\{ x \in \R^n : \lim_{\delta \to 0^+} \frac{\gvol(A \cap B_\delta(x))}{\gvol(B_\delta(x))} \neq 0, 1\right\},
                \]
                where $B_\delta(x)$ denotes the ball of radius~$\delta$ centered at~$x$.  In other words, $\partial_* A$ is the set of points where the ``local density of~$A$'' is strictly between~$0$ and~$1$.
                \label{ex:essential-boundary-content}
                Show that if we replace $\bdry A$ with $\bdry_* A$ in the definition~\eqref{eqn:heuristic-gsa} of the Gaussian Minkowski content of the boundary of~$A$, then
                we have the identity $\gamma^+(\bdry_* A_i) = \gsa(A_i)$ for all $1 \leq i \leq 6$.  Remark: In fact, the equality $\gamma^+(\bdry_* A) = \gsa(A)$ is known to hold for every set~$A$ such that $\bdry_* A$ is ``rectifiable''.
        \end{exercises}
    \item \label{ex:gsa-of-intervals} Justify the formula for the Gaussian surface area of unions of intervals stated in Example~\ref{eg:surface-areas}.
    \item \label{ex:ball-gsa}
        \begin{exercises}
            \item Let $B_r \subset \R^n$ denote the ball of radius~$r > 0$ centered at the origin.  Show that
                \begin{equation} \label{eqn:balls-gsa}
                    \gsa(B_r) = \frac{n}{2^{n/2} (n/2)!} r^{n-1} e^{-r^2/2}.
                \end{equation}
            \item Show that~\eqref{eqn:balls-gsa} is maximized when $r = \sqrt{n-1}$. (In case $n = 1$, this should be interpreted as $r \to 0^+$.)
            \item Let $S(n)$ denote this maximizing value, i.e., the value of~\eqref{eqn:balls-gsa} with $r = \sqrt{n-1}$.  Show that $S(n)$ decreases from $\sqrt{\frac{2}{\pi}}$ to a limit of $\frac{1}{\sqrt{\pi}}$ as~$n$ increases from~$1$ to~$\infty$.
        \end{exercises}
    \item \label{ex:gauss-Lap-domain}
        \begin{exercises}
            \item For $f \in L^2(\R^n, \gamma)$, show that $\Lap f$ is defined, i.e.,
                    \[
                        \lim_{t \to 0} \frac{f - \U_{e^{-t}} f}{t}
                    \]
                    exists in~$L^2(\R^n,\gamma)$, if and only if $\sum_{\alpha \in \N^n} |\alpha|^2 \wh{f}(\alpha)^2 < \infty$. (Hint: Proposition~\ref{prop:Urho-formula}.)
            \item Formally justify Proposition~\ref{prop:ou-formula}.
            \item Let $f \in L^2(\R^n, \gamma)$.  Show that $\U_\rho f$ is in the domain of~$\Lap$ for any $\rho \in (-1,1)$.
        \end{exercises}
        \noindent Remark: It can be shown that the $\calC^3$ hypothesis in Propositions~\ref{prop:alternate-OU} and~\ref{prop:dirichlet2} is not necessary (provided the derivatives are interpreted in the distributional sense); see, e.g., Bogachev~\cite[Chapter~1]{Bog98} for more details.
    \item \label{ex:gaussquad} This exercise is concerned with (a generalization of) the function appearing in Borell's Isoperimetric Theorem.
                                            \index{Borell's Isoperimetric Theorem}
        \begin{definition}
            For $\rho \in [-1,1]$ we define the \emph{Gaussian quadrant probability} function $\GaussQuad_\rho \co [0,1]^2 \to [0,1]$ by
                                                    \index{Gaussian quadrant probability}%
                                                    \nomenclature[Lambdarhomu]{$\GaussQuad_\rho(\alpha,\beta)$}{$\Pr[\bz_1 \leq t, \bz_2 \leq t']$, where $\bz_1, \bz_2$ are standard Gaussians with correlation $\E[\bz_1\bz_2] = \rho$, and $t = \Phi^{-1}(\alpha)$, $t' = \Phi^{-1}(\beta)$}%
            \[
                \GaussQuad_\rho(\alpha,\beta) = \Pr_{\substack{(\bz, \bz') \text{ $\rho$-correlated} \\ \text{standard Gaussians}}}[\bz \leq t, \bz' \leq t'],
            \]
            where $t$ and $t'$ are defined by $\Phi(t) = \alpha$, $\Phi(t') = \beta$.  This is a slight reparametrization of the bivariate Gaussian cdf.  We also use the shorthand notation
            \[
                \GaussQuad_\rho(\alpha) = \GaussQuad_\rho(\alpha, \alpha),
            \]
            which we encountered in Borell's Isoperimetric Theorem (and also in Exercises~\ref{ex:ball-stab} and~\ref{ex:gaussquad-est}, with a different, but equivalent, definition).
        \end{definition}
        \begin{exercises}
            \item Confirm the statement from Borell's Isoperimetric Theorem, that for every halfspace $H \subseteq \R^n$ with $\gvol(H) = \alpha$ we have $\Stab_\rho[1_H] = \GaussQuad_\rho(\alpha)$.
            \item Verify the following formulas:
                \begin{align*}
                    \GaussQuad_\rho(\alpha,\beta) &= \GaussQuad_\rho(\beta,\alpha), \\
                    \GaussQuad_0(\alpha,\beta) &= \alpha \beta, \\
                    \GaussQuad_1(\alpha, \beta) &= \min(\alpha, \beta), \\
                    \GaussQuad_{-1}(\alpha,\beta) &= \max(\alpha + \beta - 1, 0), \\
                    \GaussQuad_\rho(\alpha, 0) &= \GaussQuad_\rho(0, \alpha) = 0, \\
                    \GaussQuad_\rho(\alpha, 1) &= \GaussQuad_\rho(1, \alpha) = \alpha, \\
                    \GaussQuad_{-\rho}(\alpha, \beta) &= \alpha - \GaussQuad_{\rho}(\alpha, 1-\beta) = \beta - \GaussQuad_{\rho}(1-\alpha, \beta), \\
                    \GaussQuad_\rho(\half, \half) &= \half - \half\tfrac{\arccos \rho}{\pi}.
                \end{align*}
            \item Prove that $\GaussQuad_\rho(\alpha, \beta) \gtrless \alpha \beta$ according as $\rho \gtrless 0$, for all $0 < \alpha, \beta < 1$.
            \item \label{ex:gaussquad-deriv1} Establish
                  \[
                      \frac{d}{d\alpha} \GaussQuad_\rho(\alpha, \beta) = \Phi\left(\frac{t' - \rho t}{\sqrt{1-\rho^2}}\right), \quad \frac{d}{d\beta} \GaussQuad_\rho(\alpha, \beta) = \Phi\left(\frac{t - \rho t'}{\sqrt{1-\rho^2}}\right),
                  \]
                  where $t = \Phi^{-1}(\alpha)$, $t' = \Phi^{-1}(\beta)$ as usual.
            \item \label{ex:GaussQuad-Lipschitz} Show that
                  \[
                    |\GaussQuad_\rho(\alpha,\beta) - \GaussQuad_\rho(\alpha', \beta')| \leq |\alpha - \alpha'| + |\beta - \beta'|,
                  \]
                  and hence $\GaussQuad_\rho(\alpha)$ is a $2$-Lipschitz function of~$\alpha$.
        \end{exercises}
    \item \label{ex:bobkov-induction}  Show that the general-$n$ case of Bobkov's Inequality follows by induction from the $n = 1$ case.
                                                    \index{Bobkov's Inequality}%
    \item Let $f \btb$ and let $\alpha = \min\{\Pr[f = 1], \Pr[f=-1]\}$.  Deduce $\Tinf[f] \geq 4 \giso(\alpha)^2$ from Bobkov's Inequality.  Show that this recovers the edge-isoperimetric inequality for the Boolean cube (Theorem~\ref{thm:edge-iso}) up to a constant factor.  (Hint: For the latter problem, use Proposition~\ref{prop:giso-asympt}.)
    \item \label{ex:walk-stopping} Let $d_1, d_2 \in \N$. Suppose we take a simple random walk on~$\Z$, starting from the origin and moving by $\pm 1$ at each step with equal probability.  Show that the expected time it takes to first reach either~$-d_1$ or $+d_2$ is $d_1d_2$.
    \item \label{ex:bobkov-submartingale}  Prove Claim~\ref{claim:bobkov}.  (Hint: For the function $V_y(\tau)$ appearing in the proof of Bobkov's Two-Point Inequality, you'll want to establish that $V_y'''(0) = 0$ and that $V_y''''(0) = \frac{2+10\giso'(y)^2}{\giso(y)^3} > 0$.)
    \item \label{ex:gen-bobkov} Prove Theorem~\ref{thm:my-bobkov}.  (Hint: Have the random walk start at $\by_0 = a \pm \rho b$ with equal probability, and define $\bz_t = \|(\giso(\by_t), \rho b, \tau \sqrt{t})\|$.  You'll need the full generality of Exercise~\ref{ex:walk-stopping}.)
                                                        \index{Borell's Isoperimetric Theorem|(}%
    \item \label{ex:borell-function-from-set} Justify Remark~\ref{rem:borell-set-to-function1} (in the general-volume context)
        by showing that Borell's Isoperimetric Theorem for all functions in $K = \{f \co \R^n \to [0,1] \mid \E[f] = \alpha\}$ can be deduced from the case of functions in $\bdry K = \{ f \co \R^n \to \{0,1\} \mid \E[f] = \alpha\}$.  (Hint: As stated in the remark, the intuition is that $\sqrt{\Stab_\rho[f]}$
        is a norm and that~$K$ is a convex set whose extreme points are~$\bdry K$. To make this precise, you may want to use Exercise~\ref{ex:really-simple-functions}.)
    \item \label{ex:mossel-neeman1}  The goal of this exercise and Exercises~\ref{ex:mossel-neeman2}--\ref{ex:mossel-neeman4} is to give the proof of Borell's Isoperimetric Theorem due to Mossel and Neeman~\cite{MN12}.  In fact, their proof gives the following natural ``two-set'' generalization of the theorem (Borell's original work~\cite{Bor85} proved something even more general):
        \begin{named}{Two-Set Borell Isoperimetric Theorem}
            Fix $\rho \in (0,1)$ and $\alpha, \beta \in [0,1]$.  Then for any $A, B \subseteq \R^n$ with $\gvol(A) = \alpha$, $\gvol(B) = \beta$,
            \begin{equation}                \label{eqn:two-set-borell1}
                \Pr_{\substack{(\bz, \bz') \textnormal{ $\rho$-correlated} \\ \textnormal{$n$-dimensional Gaussians}}}[\bz \in A, \bz' \in B] \leq \GaussQuad_\rho(\alpha,\beta).
            \end{equation}
        \end{named}
        By definition of $\GaussQuad_\rho(\alpha,\beta)$, equality holds if~$A$ and~$B$ are parallel halfspaces.  Taking $\beta = \alpha$ and $B = A$ in this theorem gives Borell's Isoperimetric Theorem as stated in Section~\ref{sec:borell} (in the case of range $\{0,1\}$, at least, which is equivalent by Exercise~\ref{ex:borell-function-from-set}).  It's quite natural to guess that parallel halfspaces should maximize the ``joint Gaussian noise stability'' quantity on the left of~\eqref{eqn:two-set-borell1}, especially in light of Remark~\ref{rem:gen-sse-sharpness} from Chapter~\ref{sec:full-hypercon-for-bits} concerning the analogous
                                                    \index{Small-Set Expansion Theorem!generalized}%
        Generalized Small-Set Expansion Theorem.  Just as our proof of the Small-Set Expansion Theorem passed through the Two-Function Hypercontracitivity Theorem to facilitate induction, so too does the Mossel--Neeman proof
                                                    \index{Hypercontractivity Theorem!Two-Function|(}%
        pass through the following ``two-function version'' of Borell's Isoperimetric Theorem:
        \begin{named}{Two-Function Borell Isoperimetric Theorem}
            Fix $\rho \in (0,1)$ and let $f, g \in L^2(\R^n, \gamma)$ have range $[0,1]$.  Then
            \[
                \E_{\substack{(\bz, \bz') \textnormal{ $\rho$-correlated} \\ \textnormal{$n$-dimensional Gaussians}}}[\GaussQuad_\rho(f(\bz),g(\bz'))] \leq \GaussQuad_\rho\left(\E[f], \E[g]\right).
            \]
        \end{named}
        \begin{exercises}
            \item Show that the Two-Function Borell Isoperimetric Theorem implies the Two-Set Borell Isoperimetric Theorem and the Borell Isoperimetric Theorem (for functions with range~$[0,1]$).  (Hint: You may want to use facts from Exercise~\ref{ex:gaussquad}.)
            \item Show conversely that the Two-Function Borell Isoperimetric Theorem (in dimension~$n$) is implied by the Two-Set Borell Isoperimetric Theorem (in dimension~$n+1$).  (Hint: Given $f \co \R^n \to [0,1]$, define $A \subseteq \R^{n+1}$ by $(z,t) \in A \iff f(z) \geq \Phi(t)$.)
            \item Let $\ell_1,\ell_2 \co \R^n \to \R$ be defined by $\ell_i(z) = \la a, z \ra + b_i$ for some $a \in \R^n$, $b_1, b_2 \in \R$.  Show that equality occurs in the Two-Function Borell Isoperimetric Theorem if $f(z) = 1_{\ell_1(z) \geq 0}$, $g(z) = 1_{\ell_2(z) \geq 0}$ or if  $f(z) = \Phi(\ell_1(z))$, $g(z) = \Phi(\ell_2(z))$.
        \end{exercises}
    \item \label{ex:mossel-neeman2}  Show that the inequality in the Two-Function Borell Isoperimetric Theorem ``tensorizes'' in the sense that if it holds for $n = 1$, then it holds for all~$n$.  Your proof should not use any property of the function $\GaussQuad_\rho$, nor any property of the $\rho$-correlated $n$-dimensional Gaussian distribution besides the fact that it's a product distribution. (Hint: Induction by restrictions as in the proof of the Two-Function Hypercontractivity Induction Theorem from  Chapter~\ref{sec:hypercon-tensorize}.)
    \item \label{ex:mossel-neeman3} Let $I_1, I_2 \subseteq \R$ be open intervals and let $\mathcal{F} \co I_1 \times I_2 \to \R$ be $\calC^2$.  For $\rho \in \R$, define the matrix
        \[
            H_\rho\mathcal{F} = (H\mathcal{F})  \circ \begin{bmatrix} 1 & \rho \\ \rho & 1 \end{bmatrix},
        \]
        where $H \mathcal{F}$ denotes the Hessian of~$\mathcal{F}$ and~$\circ$ is the entrywise (Hadamard) product.  We say that $\mathcal{F}$ is \emph{$\rho$-concave} (this terminology introduced by Ledoux~\cite{Led13}) if $H_\rho \mathcal{F}$ is everywhere negative semidefinite.  Note that the $\rho = 1$ case corresponds to the usual notion of concavity, and the $\rho = 0$ case corresponds to concavity separately along the two coordinates.  The goal of this exercise is to show that the Gaussian quadrant probability $\GaussQuad_\rho$ function is $\rho$-concave for
                                        \index{Gaussian quadrant probability}%
        all $\rho \in (0,1)$.
        \begin{exercises}
            \item Extending Exercise~\ref{ex:gaussquad}\ref{ex:gaussquad-deriv1}, show that for any $\rho \in (-1,1)$,
                    \begin{align*}
                              \frac{d^2}{d\alpha^2} \GaussQuad_\rho(\alpha, \beta) &= -\frac{\rho}{\sqrt{1-\rho^2}} \cdot \frac{1}{\phi(t)} \cdot \phi\left(\frac{t' - \rho t}{\sqrt{1-\rho^2}}\right),
                    \end{align*}
                    and deduce a similar formula for $\frac{d^2}{d\beta^2} \GaussQuad_\rho(\alpha, \beta)$.
            \item Show that
                    \begin{align*}
                              \frac{d^2}{d\alpha\,d\beta} \GaussQuad_\rho(\alpha, \beta) &= \frac{1}{\sqrt{1-\rho^2}} \cdot \frac{1}{\phi(t')} \cdot \phi\left(\frac{t' - \rho t}{\sqrt{1-\rho^2}}\right),
                    \end{align*}
                    and deduce a similar (in fact, equal) formula for $\frac{d^2}{d\beta\,d\alpha} \GaussQuad_\rho(\alpha, \beta)$.
            \item Show that $\det (H_\rho \GaussQuad_\rho) = 0$ on all of $(0,1)^2$.
            \item Show that if $\rho \in (0,1)$, then $\frac{d^2}{d\alpha^2} \GaussQuad_\rho,\ \frac{d^2}{d\beta^2} \GaussQuad_\rho < 0$ on $(0,1)^2$. Deduce that $\GaussQuad_\rho$ is $\rho$-concave.
        \end{exercises}
    \item \label{ex:mossel-neeman4}  This exercise is devoted to Mossel and Neeman's proof~\cite{MN12} of the Two-Function Borell Isoperimetric Theorem in the case $n = 1$.  For another approach, see Exercise~\ref{ex:DMN}. By Exercise~\ref{ex:mossel-neeman2}, this is sufficient to establish the case of general~$n$.  (Actually, the proof in this exercise works essentially verbatim in the general~$n$ case, but we stick to $n = 1$ for simplicity.)
        \begin{exercises}
            \item  More generally, we intend to prove that for $f, g \co \R \to [0,1]$,
                \[
                    \lambda(\sigma) = \E_{\substack{(\bz, \bz') \text{ $\rho$-correlated} \\ \text{standard Gaussians}}}[\GaussQuad_\rho(\U_\sigma f(\bz),\U_\sigma g(\bz'))]
                \]
                is a nonincreasing function of $0 < \sigma < 1$ (cf.~Theorem~\ref{thm:my-bobkov}).  Obtain the desired conclusion by taking $\sigma \to 0^+, 1^-$.  (Hint: You'll need Exercises~\ref{ex:Urho-strongly-cts} and~\ref{ex:gaussquad}\ref{ex:GaussQuad-Lipschitz}.)
            \item Write $f_\sigma = \U_\sigma f$, $g_\sigma = \U_\sigma g$ for brevity, and write $\partial_i \GaussQuad_\rho$ ($i = 1,2$) for the partial derivatives of $\GaussQuad_\rho$.  Also let $\bh_1, \bh_2$ denote independent standard Gaussians.  Use the Chain Rule and Proposition~\ref{prop:L-vs-U} to establish
                \begin{align}
                    \sigma \lambda'(\sigma) = &\E[(\partial_1 \GaussQuad_\rho)(f_\sigma(\bh_1), g_\sigma (\rho \bh_1 + \sqrt{1-\rho^2} \bh_2)) \cdot \Lap f_\sigma (\bh_1)] \label{eqn:first-mn-deriv}\\
                    {}+ &\E[(\partial_2 \GaussQuad_\rho)(f_\sigma(\rho \bh_2 + \sqrt{1-\rho^2} \bh_1), g_\sigma (\bh_2)) \cdot \Lap g_\sigma (\bh_2)]. \label{eqn:second-mn-deriv}
                \end{align}
            \item Use Proposition~\ref{prop:dirichlet2} to show that the first expectation~\eqref{eqn:first-mn-deriv} equals
                \[
                    \E[(\partial_{11} \GaussQuad_\rho f)(f_\sigma, g_\sigma) \cdot (f_\sigma')^2 + \rho \cdot (\partial_{21} \GaussQuad_\rho f)(f_\sigma, g_\sigma) \cdot f_\sigma' \cdot g_\sigma'],
                \]
                where $f_\sigma, f_\sigma'$ are evaluated at $\bh_1$ and $g_\sigma, g_\sigma'$ are evaluated at $\rho \bh_1 + \sqrt{1-\rho^2}\bh_2$.  Give a similar formula for~\eqref{eqn:second-mn-deriv}.
            \item Deduce that
                \[
                    \sigma \lambda'(\sigma) = \E_{\substack{(\bz, \bz') \text{ $\rho$-correlated} \\ \text{standard Gaussians}}}\left[\begin{bmatrix} f_\sigma'(\bz) & g_\sigma'(\bz') \end{bmatrix} \cdot (H_\rho \GaussQuad_\rho)(f_\sigma(\bz), g_\sigma(\bz')) \cdot \begin{bmatrix} f_\sigma'(\bz) \\ g_\sigma'(\bz') \end{bmatrix}\right],
                \]
                where $H_\rho$ is as in Exercise~\ref{ex:mossel-neeman3}, and that indeed~$\lambda$ is a nonincreasing function.
        \end{exercises}
    \item \label{ex:DMN}
        \begin{exercises}
            \item Suppose the Two-Function Borell Isoperimetric Theorem were to hold for $1$-bit functions, i.e., for $f, g \co \bits \to [0,1]$.  Then the easy induction of Exercise~\ref{ex:mossel-neeman2} would extend the result to $n$-bit functions $f, g \co \bn \to [0,1]$; in turn, this would yield the  Two-Function Borell Isoperimetric Theorem for $1$-dimensional Gaussian functions (i.e., Exercise~\ref{ex:mossel-neeman4}), by the usual Central Limit Theorem argument.  Show, however, that dictator functions provide a counterexample to a potential ``$1$-bit Two-Function Borell Isoperimetric Theorem''.
            \item Nevertheless, the idea can be salvaged by proving a weakened version of the inequality for $1$-bit functions that has an ``error term'' that is a \emph{superlinear} function of $f$ and $g$'s ``influences''.  Fix $\rho \in (0,1)$ and some small $\eps > 0$.  Let $f, g \co \bits \to [\epsilon, 1-\eps]$.  Show that
                \[
                    \Ex_{\substack{(\bx, \bx') \\ \text{$\rho$-correlated}}}[\GaussQuad_\rho(f(\bx), g(\bx'))] \leq \GaussQuad_\rho(\E[f], \E[g]) + C_{\rho, \eps}\cdot (\E[|\D_1 f|^3] + \E[|\D_1 g|^3]),
                \]
                where $C_{\rho,\eps}$ is a constant depending only on~$\rho$ and~$\eps$.  (Hint: Perform a $2$nd-order Taylor expansion of~$\GaussQuad_\rho$ around~$(\E[f], \E[g])$; in expectation, the quadratic term should be
                \[
                    \begin{bmatrix} \D_1 f & \D_1 g \end{bmatrix} \cdot (H_\rho \GaussQuad_\rho)(\E[f],\E[g]) \cdot \begin{bmatrix} \D_1 f \\ \D_1 g \end{bmatrix}.
                \]
                As in Exercise~\ref{ex:mossel-neeman4}, show this quantity is nonpositive.)
            \item Extend the previous result by induction to obtain the following theorem of De, Mossel, and Neeman~\cite{DMN13}:
                \begin{theorem}                             \label{thm:DMN}
                    For each $\rho \in (0,1)$ and $\eps > 0$, there exists a constant~$C_{\rho, \eps}$ such that the following holds: If $f, g \co \bn \to [\eps, 1-\eps]$, then
                    \[
                        \Ex_{\substack{(\bx, \bx') \\ \text{$\rho$-correlated}}}[\GaussQuad_\rho(f(\bx), g(\bx'))] \leq \GaussQuad_\rho(\E[f], \E[g]) + C_{\rho, \eps}\cdot (\Delta_n[f] + \Delta_n[g]).
                    \]
                    Here we using the following inductive notation: $\Delta_1[f] = \E[|f - \E[f]|^3]$, and
                    \[
                        \Delta_n[f] = \E_{\bx_n \sim \bits}\left[\Delta_{n-1}[\restr{f}{}{\bx_n}]\right] + \Delta_1[f^{\subseteq \{n\}}].
                    \]
                \end{theorem}
            \item Prove by induction that $\Delta_n[f] \leq 8\sum_{i=1}^n \|\D_i f\|_3^3$.
            \item Suppose that $f,g \in L^2(\R, \gamma)$ have range $[\eps, 1-\eps]$ and are $c$-Lipschitz.  Show that for any $M \in \N^+$, the Two-Function Borell Isoperimetric Theorem holds for $f, g$ with an additional additive error of $O(M^{-1/2})$, where the constant in the $O(\cdot)$ depends only on $\rho$, $\epsilon$, and~$c$.  (Hint: Use $\BitsToGaussians{M}{}$.)
            \item By an approximation argument, deduce the Two-Function Borell Isoperimetric Theorem for general $f, g \in L^2(\R, \gamma)$ with range $[0,1]$; i.e., prove Exercise~\ref{ex:mossel-neeman4}.
        \end{exercises}
                                                 \index{Borell's Isoperimetric Theorem|)}%
    \item \label{ex:Tal1000}  Fix $0 < \rho < 1$ and suppose $f \in L^1(\R,\gamma)$ is nonnegative and satisfies $\E[f] = 1$.  Note that $\E[\U_\rho f] = 1$ as well.  The goal of this problem is to show that $\U_\rho f$ satisfies an improved Markov inequality: $\Pr[\U_\rho f > t] = O(\frac{1}{t \sqrt{\ln t}}) = o(\frac{1}{t})$ as $t \to \infty$.  This gives a quantitative sense in which $\U_\rho$ is a ``smoothing operator'': $\U_\rho f$ can never look too much like like a step function (the tight example for Markov's inequality).
        \begin{exercises}
            \item For simplicity, let's first assume $\rho = 1/\sqrt{2}$.  Given~$t > \sqrt{2}$, select $h > 0$ such that $\vphi(h) = 1/(\sqrt{\pi} t)$.  Show that $h \sim \sqrt{2\ln t}$.
            \item Let $H = \{z : \U_\rho f(z) > t\}$.  Show that if $H \subseteq (-\infty, -h] \cup [h, \infty)$, then we have $\Pr[\U_\rho f > t] \lesssim \frac{\sqrt{2/\pi}}{t \sqrt{\ln t}}$, as desired.  (Hint: You'll need $\olPhi(u) < \vphi(u)/u$.
            \item Otherwise, we wish to get a contradiction.  First, show that there exists $y \in (-h,h)$ and $\delta_0 > 0$ such that $\U_\rho f(z) > t$ for all $t \in (y-\delta_0, y+\delta_0)$.  (Hint: You'll need that $\U_\rho f$ is continuous; see Exercise~\ref{ex:Urho-smooths}.)
            \item For $0 < \delta < \delta_0$, define $g \in L^1(\R,\gamma)$ by $g(z) = \frac{1}{2\delta} 1_{(y-\delta, y+\delta)}$.  Show that $0 \leq \U_\rho g \leq \frac{1}{\sqrt{\pi}}$ pointwise.  (Hint: Why is $\U_\rho g(z)$ maximized at $\sqrt{2}y$?)
            \item Show that $\frac{1}{\sqrt{\pi}} \geq \la f, \U_\rho g \ra > t \E[g]$.
            \item Derive a contradiction by taking $\delta \to 0$, thereby showing that indeed $\Pr[\U_\rho f > t] \lesssim \frac{\sqrt{2/\pi}}{t \sqrt{\ln t}}$.
            \item Show that this result is tight by constructing an appropriate~$f$.
            \item Generalize the above to show that for any fixed $0 < \rho < 1$ we have $\Pr[\U_\rho f > t] \lesssim \frac{1}{\sqrt{\pi(1-\rho^2)}}\frac{1}{t \sqrt{\ln t}}$.
        \end{exercises}
    \item \label{ex:5-cycle} As described in Example~\ref{eg:5-cycle}, show that $\SDPOpt(\Z_5) \geq  \half - \half \cos \frac{4\pi}{5} =\frac58 + \frac{\sqrt{5}}{8}$.
                                        \index{Goemans--Williamson Algorithm|(}%
    \item \label{ex:careful-gw}  Prove Theorem~\ref{thm:gw-behavior}.
    \item \label{ex:gw-for-2lin} Consider the generalization of the Max-Cut CSP in which the variable set is~$V$, the domain is~$\bits$, and each constraint is an equality of two literals, i.e., it's of the form $b F(v) = b' F(v')$ for for some $v, v' \in V$ and $b, b' \in \bits$.  This CSP is traditionally called Max-E$2$-Lin.
                                                            \index{Max-$2$-Lin}%
        Given an instance~$\instance$, write $(\bv, \bv', \bb, \bb') \sim \instance$ to denote a uniformly chosen constraint.  The natural SDP relaxation (which can also be solved efficiently) is the following:
        \[
            \begin{aligned}
                \text{maximize}& \quad  \E_{(\bv,\bv',\bb, \bb') \sim \instance}\left[\half + \half \la \bb\vec{U}(\bv), \bb'\vec{U}(\bv') \ra\right]\\
                \text{subject to}& \quad \vec{U} \co V \to S^{n-1}.
            \end{aligned}
        \]
        Show that the Goemans--Williamson algorithm, when using this SDP, is a $(\cgw \beta, \beta)$-approximation algorithm for Max-E$2$Lin, and that it also has the same refined guarantee as in Theorem~\ref{thm:gw-behavior}.
    \item \label{ex:4-cycle} This exercise builds on Exercise~\ref{ex:gw-for-2lin}.  Consider the following instance~$\instance$ of Max-E$2$-Lin: The variable set is~$\Z_4$ and the constraints are
        \[
            F(0) = F(1), \quad F(1) = F(2), \quad F(2) = F(3), \quad F(3) = -F(0).
        \]
        \begin{exercises}
            \item Show that $\Opt(\instance) = \frac34$.
            \item Show that $\SDPOpt(\instance) \geq \frac{1}{2} + \frac{1}{2\sqrt{2}}$.  (Hint: Very similar to Exercise~\ref{ex:5-cycle}; you can use four unit vectors at $45^\circ$ angles in~$\R^2$.)
            \item Deduce that $\SDPOpt(\instance) = \frac{1}{2} + \frac{1}{2\sqrt{2}}$ and that this is an optimal SDP integrality gap for Max-E$2$Lin.  (Cf.~Remark~\ref{rem:easy-almost-gw}.)
        \end{exercises}
                                        \index{Goemans--Williamson Algorithm|)}%
    \item \label{ex:kkmo-technical}  In our proof of Theorem~\ref{thm:kkmo} it's stated that showing the $\beta$-Noise Sensitivity Test is a $(\theta/\pi, \half - \half \cos \theta)$-\dvq implies the desired \UG-hardness of $(\theta/\pi +\delta, \half - \half \cos \theta)$-approximating Max-Cut (for any constant~$\delta > 0$). There are two minor technical problems with this: First, the test can only actually be implemented when~$\beta$ is a rational number.  Second, even ignoring this, Theorem~\ref{thm:ug-hardness-from-dict-tests} only directly yields hardness of $(\theta/\pi +\delta, \half - \half \cos \theta - \delta)$-approximation.  Show how to overcome both technicalities.  (Hint: Continuity.)

    \item Use Corollary~\ref{cor:lipschitz-be} (and~\eqref{eqn:be-bound}) to show that in the setting of the Berry--Esseen Theorem, $|\|\bS\|_1 -\sqrt{2/\pi}| \leq O(\gamma^{1/3})$.  (Cf.~Exercise~\ref{ex:l1-be}.)
                                            \index{mollification|(}%
    \item \label{ex:gaussian-mollify} The goal of this exercise is to prove Proposition~\ref{prop:gaussian-mollify}.
        \begin{exercises}
            \item Reduce to the case $c = 1$.
            \item Reduce to the case $\eta = 1$. (Hint: Dilate the input by a factor of~$\eta$.)
            \item Assuming henceforth that $c = \eta = 1$, we define $\wt{\psi}(s) = \E[\psi(s+\bg)]$ for $\bg \sim \normal(0,1)$ as suggested; i.e., $\wt{\psi} = \psi \ast \vphi$, where $\vphi$ is the Gaussian pdf.  Show that indeed $\|\wt{\psi} - \psi\|_\infty \leq \sqrt{2/\pi} \leq 1$.
            \item To complete the proof we need to show that for all $s \in \R$ and $k \in \N^+$ we have $|\wt{\psi}^{(k)}(s)| \leq C_k$. Explain why, in proving this, we may assume $\psi(s) = 0$.  (Hint: This requires $k \geq 1$.)
            \item Assuming $\psi(s) = 0$, show $|\wt{\psi}^{(k)}(s)| = |\psi \ast \vphi^{(k)}(s)| \leq C_k$. (Hint: Show that $\vphi^{(k)}(s) = p(s) \vphi(s)$ for some polynomial~$p(s)$ and use the fact that Gaussians have finite absolute moments.)
        \end{exercises}
    \item \label{ex:gaussian-mollify-d} Establish the following multidimensional generalization of Proposition~\ref{prop:gaussian-mollify}:
            \begin{proposition}         \label{prop:gaussian-mollify-d}
                Let $\psi \co \R^d \to \R$ be $c$-Lipschitz. Then for any $\eta > 0$ there exists $\wt{\psi}_\eta \co \R^d \to \R$ satisfying $\|\psi - \wt{\psi}_\eta\|_\infty \leq c \sqrt{d} \eta$ and $\|\partial^\beta \wt{\psi}_\eta\|_\infty \leq C_{|\beta|} c \sqrt{d}/\eta^{|\beta|-1}$ for each multi-index $\beta \in \N^{d}$ with $|\beta| = \sum_i \beta_i \geq 1$, where $C_k$ is a constant depending only on~$k$.
            \end{proposition}
    \item \label{ex:mollifier}  In Exercise~\ref{ex:gaussian-mollify} we ``mollified'' a function $\psi$ by convolving it with the (smooth) pdf of a Gaussian random variable.  It's sometimes helpful to instead use a random variable with bounded support (but still with a smooth pdf on all of~$\R$).  Here we construct such a random variable.          Define $b \co \R \to \R$ by
          \[
              b(x) = \begin{cases}
                         \exp\left(-\tfrac{1}{1-x^2}\right) & \text{if $-1 < x < 1$,} \\
                         0 & \text{else.}
                     \end{cases}
          \]
        \begin{exercises}
            \item Verify that $b(x) \geq 0$ for all $x$ and that $b(-x) = b(x)$.
            \item Prove the following statement by induction on $k \in \N$:  On $(-1,1)$, the $k$th derivative of~$b$ at~$x$ is of the form $p(x)(1-x^2)^{-2k}\cdot b(x)$, where $p(x)$ is a polynomial.
            \item Deduce that $b$ is a smooth ($\calC^\infty$) function on~$\R$.
            \item Verify that $C = \int_{-1}^{1} b(x)\,dx$ satisfies $0 < C < \infty$ and that we can therefore define a real random variable $\by$, symmetric and supported on $(-1,1)$, with the smooth pdf $\wt{b}(y) = b(y)/C$.  Show also that for $k \in \N$, the numbers $c_k = \|\wt{b}^{(k)}\|_\infty$ are finite and positive, where $\wt{b}^{(k)}$ denotes the $k$th derivative of $\wt{b}$.
            \item Give an alternate proof of Exercise~\ref{ex:gaussian-mollify} using~$\by$ in place of~$\bg$.
        \end{exercises}
    \item  \label{ex:indicator-hack}
        Fix $u \in \R$, $\test(s) = 1_{s \leq u}$, and $0 < \eta < 1/2$.
        \begin{exercises}
            \item Suppose we approximate $\test$ by a smooth function $\wt{\pred}_\eta$ as in Exercise~\ref{ex:gaussian-mollify}, i.e., we define $\wt{\pred}_\eta(s) = \E[\pred(s + \eta \bg)]$ for $\bg \sim \normal(0,1)$.  Show that $\wt{\pred}_\eta$ satisfies the following properties:
                \begin{itemize}
                    \item $\wt{\pred}_\eta$ is a decreasing function with $\wt{\pred}_\eta(s) < \pred(s)$ for $s < u$ and $\wt{\pred}_\eta(s) > \pred(s)$ for $s > u$.
                    \item $|\wt{\pred}_\eta(s) - \pred(s)| \leq \eta$ provided $|s - u| \geq O(\eta \sqrt{\log(1/\eta)})$.
                    \item $\|\wt{\pred}_\eta^{(k)}\|_\infty \leq C_k/\eta^k$ for each $k \in \N$, where $C_k$ depends only on~$k$.
                \end{itemize}
            \item Suppose we instead approximate $\test$ by the function $\wt{\pred}_\eta(s) = \E[\pred(s + \eta \by)]$, where $\by$ is the random variable from Exercise~\ref{ex:mollifier}.  Show that~$\wt{\pred}_\eta$ satisfies the following slightly nicer properties:
                \begin{itemize}
                    \item $\wt{\pred}_\eta$ is a nonincreasing function which agrees with $\test$ on $(\infty, u-\eta]$ and on $[u+\eta, \infty)$.
                    \item $\wt{\pred}_\eta$ is smooth and satisfies $\|\wt{\pred}_\eta^{(k)}\|_\infty \leq C_k/\eta^{k}$ for each $k \in \N$, where~$C_k$ depends only on~$k$.
                \end{itemize}
        \end{exercises}
    \item \label{ex:be-levy}  Prove Corollary~\ref{cor:be-levy} by first proving
            \[
                \Pr[\bS_Y \leq u-2\eta] - O(\eta^{-3})\gamma_{XY} \leq \Pr[\bS_X \leq u] \leq \Pr[\bS_Y \leq u+2\eta] + O(\eta^{-3})\gamma_{XY}.
            \]
            (Hint: Obtain $\Pr[\bS_X \leq u-\eta] \leq \E[\wt{\psi}_\eta(\bS_X)] \approx \E[\wt{\psi}_\eta(\bS_Y)] \leq \Pr[\bS_Y \leq u+\eta]$ using properties from Exercise~\ref{ex:indicator-hack}.  Then replace $u$ with $u+2\eta$ and also interchange $\bS_X$ and $\bS_Y$.)
    \item \label{ex:power-hack}
        \begin{exercises}
            \item Fix $q \in \N$.  Establish the existence of a smooth function~$f_q \co \R \to \R$ that is~$0$ on $(-\infty, -\half]$ and that agrees with some polynomial of degree exactly~$q$ on $[\half, \infty)$.  (Hint: Induction on~$q$; the base case $q = 0$ is essentially Exercise~\ref{ex:indicator-hack}, and the induction step can be achieved by integration.)
            \item Deduce that for any prescribed sequence $a_0, a_1, a_2, \dots$ that is eventually constantly~$0$, there is a smooth function $g \co \R \to \R$ that is~$0$ on $(-\infty, -\half]$ and has $g^{(k)}(\half) = a_k$ for all~$k \in \N$.
            \item Fix a univariate polynomial~$p \co \R \to \R$.  Show that there is a smooth function $\wt{\psi} \co \R \to \R$ that agrees with~$p$ on $[-1,1]$ and is identically~$0$ on $(-\infty, -2] \cup [2, \infty)$.
        \end{exercises}
                                                    \index{mollification|)}%

    \item \label{ex:basic-inv-levy} Establish Corollary~\ref{cor:basic-inv-levy}.
    \item \label{ex:simple-invariance-for-cdf-close} Prove Theorem~\ref{thm:simple-invariance-for-cdf-close}.
                                                    \index{Invariance Principles|(}
    \item \label{ex:multivar-be} \begin{exercises}
        \item By following our proof of the $d = 1$ case and using the multivariate Taylor theorem, establish the following:
                                            \index{Berry--Esseen Theorem!multivariate|seeonly{Invariance Principle for sums of random vectors}}%
                                            \index{Invariance Principle!for sums of random vectors}%
            \begin{named}{Invariance Principle for Sums of Random Vectors}
                Let $\vec{\bX}_1, \dots, \vec{\bX}_n$, $\vec{\bY}_1, \dots, \vec{\bY}_n$ be independent $\R^d$-valued random variables with matching means and covariance matrices; i.e., $\E[\vec{\bX}_t] = \E[\vec{\bY}_t]$ and $\Cov[\vec{\bX}_t] = \Cov[\vec{\bY}_t]$ for all $t \in [n]$.  (Note that the~$d$ individual components of a particular $\vec{\bX}_t$ or $\vec{\bY}_t$ are not required to be independent.) Write $\vec{\bS}_X = \sum_{t=1}^n \vec{\bX}_t$ and $\vec{\bS}_Y = \sum_{t=1}^n \vec{\bY}_t$. Then for any $\calC^3$ function $\psi \co \R^d \to \R$ satisfying $\|\partial^\beta \psi\|_\infty \leq C$ for all $|\beta| = 3$,
                \[
                    \left| \E[\pred(\vec{\bS}_X)] - \E[\pred(\vec{\bS}_Y)] \right| \leq C \gamma_{\vec{X}\vec{Y}},
                \]
                where
                \[
                    \gamma_{\vec{X}\vec{Y}} = \sum_{\substack{\beta \in \N^d\\|\beta| = 3}} \frac{1}{\beta!} \sum_{t=1}^n \bigl(\E[|\vec{\bX}_t^\beta|] + \E[|\vec{\bY}_t^\beta|]\bigr).
                \]
            \end{named}
        \item \label{ex:multivar-be-holder-trick} Show that $\gamma_{\vec{X}\vec{Y}}$ satisfies
            \[
                \gamma_{\vec{X}\vec{Y}} \leq \frac{d^2}{6}\sum_{t = 1}^n \sum_{i=1}^d \bigl(\E[|\vec{\bX}_t^{3e_i}|] + \E[|\vec{\bY}_t^{3e_i}|] \bigr).
            \]
            Here $\vec{\bX}_t^{3e_i}$ denotes the cube of the $i$th component of vector $\vec{\bX}_t$, and similarly for~$\vec{\bY}_t$.  (Hint: $abc \leq \frac13(a^3+b^3+c^3)$ for $a,b,c\geq 0$.)
        \item Deduce multivariate analogues of the Variant Berry--Esseen Theorem, Remark~\ref{rem:be-error}, and Corollary~\ref{cor:lipschitz-be} (using Proposition~\ref{prop:gaussian-mollify-d}).
        \end{exercises}
    \item \label{ex:basic-inv-3rd-mom}  Justify Remark~\ref{rem:basic-inv-3rd-mom}.  (Hint: You'll need Exercise~\ref{ex:general-polynomial-bonami}.)
    \item \label{ex:multifun-invariance}
        \begin{exercises}
        \item Prove the following:
                                                    \index{Invariance Principle!multifunction}%
            \begin{named}{Multifunction Invariance Principle}
                Let $F^{(1)}$, \dots, $F^{(d)}$ be formal $n$-variate multilinear polynomials each of degree at most~$k \in \N$.  Let $\vec{\bx}_1, \dots, \vec{\bx}_n$ and $\vec{\by}_1, \dots, \vec{\by}_n$ be independent $\R^d$-valued random variables such that $\E[\vec{\bx}_t] = \E[\vec{\by}_t] = 0$ and $M_t = \Cov[\vec{\bx}_t] = \Cov[\vec{\by}_t]$ for each $t \in [n]$.  Assume each $M_t$ has all its diagonal entries equal to~$1$ (i.e., each of the~$d$ components of $\vec{\bx}_t$ has variance~$1$, and similarly for~$\vec{\by}_t$).  Further assume each component random variable $\vec{\bx}_t^{(j)}$ and $\vec{\by}_t^{(j)}$ is $(2,3,\rho)$-hypercontractive ($t \in [n]$, $j \in [d]$). Then for any $\calC^3$ function $\psi \co \R^d \to \R$ satisfying $\|\partial^\beta \psi\|_\infty \leq C$ for all $|\beta| = 3$,
                \[
                    \left|\E[\pred(\vec{F}(\vec{\bx}))] - \E[\pred(\vec{F}(\vec{\by}))]\right| \leq \tfrac{Cd^2}{3} \cdot (1/\rho)^{3k} \cdot \sum_{t=1}^n \sum_{j=1}^d \Inf_t[F^{(j)}]^{3/2}.
                \]
                Here we are using the following notation: If $\vec{\bz} = (\vec{\bz}_1, \dots, \vec{\bz}_n)$ is a sequence of $\R^d$-valued random variables, $\vec{F}(\vec{\bz})$ denotes the vector in $\R^d$ whose $j$th component is $F^{(j)}(\vec{\bz}_1^{(j)}, \dots, \vec{\bz}_n^{(j)})$.
            \end{named}
            (Hint: Combine the proofs of the Basic Invariance Principle and the Invariance Principle for Sums of Random Vectors, Exercise~\ref{ex:multivar-be}.  The only challenging part should be notation.)
        \item Show that if we further have $\Var[F^{(j)}] \leq 1$ and $\Inf_t[F^{(j)}] \leq \eps$ for all $j \in [d]$, $t \in [n]$, then
                \[
                    \left|\E[\pred(\vec{F}(\vec{\bx}))] - \E[\pred(\vec{F}(\vec{\by}))]\right| \leq \tfrac{Cd^3}{3} \cdot k(1/\rho)^{3k} \cdot \eps^{1/2}.
                \]
        \end{exercises}
    \item \label{ex:general-invariance}
        \begin{exercises}
            \item Prove the following:
                                                    \index{Invariance Principle!general product spaces}%
            \begin{named}{Invariance Principle in general product spaces}
                Let $(\Omega, \pi)$ be a finite probability space, $|\Omega| = m \geq 2$, in which every outcome has probability at least~$\lambda$. Suppose $f \in L^2(\Omega^n, \pi\xn)$ has degree at most~$k$; thus, fixing some Fourier basis $\phi_0, \dots, \phi_{m-1}$ for $L^2(\Omega, \pi)$, we have
                \[
                    f = \sum_{\substack{\alpha \in \N^n_{<m} \\ \#\alpha \leq k}} \wh{f}(\alpha)\phi_\alpha.
                \]
                Introduce indeterminates $x = (x_{i,j})_{i \in [n], j \in [m-1]}$ and let~$F$ be the formal $(m-1)n$-variate polynomial of degree at most~$k$ defined by
                \[
                    F(x) = \sum_{\#\alpha \leq k} \wh{f}(\alpha) \prod_{i \in \supp(\alpha)} x_{i, \alpha_i}.
                \]
                Then for any $\psi \co \R \to \R$ that is $\calC^3$ and satisfies $\|\psi'''\|_\infty \leq C$ we have
                \[
                    \quad \qquad \left|\E_{\bx \sim \bits^{(m-1)n}}[\pred(F(\bx))] - \E_{\bomega \sim \pi\xn}[\pred(f(\bomega))]\right| \leq \tfrac{C}{3} \cdot (2\sqrt{2/\lambda})^k \cdot \sum_{i=1}^n \Inf_i[f]^{3/2}.
                \]
            \end{named}
            (Hint: For $0 \leq t \leq n$, define $h_t \in L^2(\Omega^t \times \bits^{(m-1)(n-t)}, \pi^{\otimes t} \otimes \pi_{1/2}^{\otimes (m-1)(n-t)})$ via
            \[
                \qquad h_t(\bomega_1, \dots, \bomega_t, \bx_{t+1,1}, \dots, \bx_{n,m-1}) = \sum_{\#\alpha \leq k} \wh{f}(\alpha) \prod_{\substack{i \in \supp(\alpha) \\ i \leq t}} \phi_{\alpha_i}(\bomega_i) \prod_{\substack{i \in \supp(\alpha) \\ i > t}} \bx_{i,\alpha_i}.
            \]
            Express
            \[
                h_{t} = \uE_t h_{t} + \Lap_t h_{t} = \uE_t h_t + \sum_{j = 1}^m D_j \cdot \phi_j(\bomega_t)
            \]
            where
            \[
                D_j = \sum_{\alpha : \alpha_t = j} \wh{f}(\alpha) \prod_{\substack{i \in \supp(\alpha) \\ i < t}} \phi_{\alpha_i}(\bomega_i) \prod_{\substack{i \in \supp(\alpha) \\ i > t}} \bx_{i,\alpha_i},
            \]
            and note that $h_{t-1} = \uE_t h_t + \sum_{j = 1}^m D_j \cdot \bx_{t,j}$.)
            \item \label{ex:gen-inv-gaussian} In the setting of the previous theorem, show also that
                \[
                    \left|\E_{\bg \sim \normal(0,1)^{(m-1)n}}[\pred(F(\bg))] - \E_{\bomega \sim \pi\xn}[\pred(f(\bomega))]\right| \leq \tfrac{2C}{3} \cdot (2\sqrt{2/\lambda})^k \cdot \sum_{i=1}^n \Inf_i[f]^{3/2}.
                \]
                (Hint: Apply the Basic Invariance Principle in the form of Exercise~\ref{ex:basic-inv-3rd-mom}.  How can you bound the $(m-1)n$ influences of~$F$ in terms of the~$n$ influences of~$f$?)
        \end{exercises}
                                                    \index{Invariance Principles|)}%
    \item \label{ex:general-MIST}  Prove the following version of the General-Volume Majority Is Stablest Theorem in the setting of general product spaces:
                            \index{Majority Is Stablest Theorem!general product spaces}%
        \begin{theorem}                     \label{thm:MIST-general}
            Let $(\Omega, \pi)$ be a finite probability space in which each outcome has probability at least~$\lambda$.  Let $f \in L^2(\Omega^n, \pi\xn)$ have range~$[0,1]$. Suppose that $f$ has no $(\eps, \frac{1}{\log(1/\eps)})$-notable coordinates.  Then for any ${0 \leq \rho < 1}$,
            \[
                \Stab_\rho[f] \leq \GaussQuad_\rho(\E[f]) + O\bigl(\tfrac{\log \log (1/\eps)}{\log(1/\eps)}\bigr) \cdot \tfrac{\log(1/\lambda)}{1-\rho}.
            \]
        \end{theorem}
        (Hint: Naturally, you'll need Exercise~\ref{ex:general-invariance}\ref{ex:gen-inv-gaussian}.)
\end{exercises}

\subsection*{Notes.}                        \label{sec:invariance-notes}
                                                    \index{Gaussian space}%
The subject of Gaussian space is too enormous to be surveyed here; some recommended texts include Janson~\cite{Jan97} and Bogachev~\cite{Bog98}, the latter having an extremely thorough bibliography.  The Ornstein--Uhlenbeck semigroup dates back to the work of Uhlenbeck and Ornstein~\cite{UO30} whose motivation was to refine Einstein's theory of Brownian motion~\cite{Ein05} to take into account the inertia of the particle.
                                        \index{Gaussian noise operator}%
The relationship between the action of~$\U_\rho$ on functions and on Hermite expansions (i.e., Proposition~\ref{prop:hermite-correlation}) dates back even further, to Mehler~\cite{Meh66}.
                                        \index{Hermite polynomials}%
Hermite polynomials were first defined by Laplace~\cite{Lap11}, and then studied by Chebyshev~\cite{Che60} and Hermite~\cite{Her64}.  See Lebedev~\cite[Chapter~4.15]{Leb72} for a proof of the pointwise convergence of a piecewise-$\calC^1$ function's Hermite expansion.

As mentioned in Chapter~\ref{sec:basic-hypercon-notes}, the Gaussian Hypercontractivity Theorem is originally due to Nelson~\cite{Nel66} and now has many known proofs.
                                                \index{Log-Sobolev Inequality!Gaussian}%
                                                \index{Hypercontractivity Theorem!Gaussian}%
The idea behind the proof we presented -- first proving the Boolean hypercontractivity result and then deducing the Gaussian case by the Central Limit Theorem -- is due to Gross~\cite{Gro75} (see also Trotter~\cite{Tro58}). Gross actually used the idea to prove his Gaussian Log-Sobolev Inequality, and thereby deduced the Gaussian Hypercontractivity Theorem.  Direct proofs of the Gaussian Hypercontractivity Theorem have been given by Neveu~\cite{Nev76} (using stochastic calculus), Brascamp and Lieb~\cite{BL76a} (using rearrangement), and Ledoux~\cite{Led13} (using a variation on Exercises~\ref{ex:mossel-neeman1}--\ref{ex:mossel-neeman4}); direct proofs of the Gaussian Log-Sobolev Inequality have been given by Adams and Clarke \cite{AC79}, by Bakry and \'{E}mery~\cite{BE85a}, and by Ledoux~\cite{Led92}, the latter two using semigroup techniques.  Bakry's survey~\cite{Bak94} on these topics is also recommended.

                                            \index{Gaussian Isoperimetric Inequality|(}%
The Gaussian Isoperimetric Inequality was first proved independently by Borell~\cite{Bor75} and by Sudakov and Tsirel'son~\cite{ST78}.  Both works derived the result by taking the isoperimetric inequality on the sphere (due to L\'{e}vy~\cite{Lev22} and Schmidt~\cite{Sch48}, see also Figiel, Lindenstrauss, and Milman~\cite{FLM77}) and then taking ``Poincar\'e's limit'' -- i.e., viewing Gaussian space as a projection of the sphere of radius~$\sqrt{n}$ in~$n$ dimensions, with $n \to \infty$ (see L{\'e}vy~\cite{Lev22}, McKean~\cite{McK73}, and Diaconis and Freedman~\cite{DF87}).  Ehrhard~\cite{Ehr83} gave a different proof using a symmetrization argument intrinsic to Gaussian space. This may be compared to the alternate proof of the spherical isoperimetric inequality~\cite{Ben84} based on the ``two-point symmetrization'' of Baernstein and Taylor~\cite{BT76} (analogous to Riesz rearrangement in Euclidean space and to the polarization operation from Exercise~\ref{ex:polarization}).

                                        \index{Gaussian surface area}%
To carefully define Gaussian surface area for a broad class of sets requires venturing into the study of geometric measure theory and functions of bounded variation.  For a clear and comprehensive development in the Euclidean setting (including the remark in Exercise~\ref{ex:essential-boundary}\ref{ex:essential-boundary-content}), see the book by Ambrosio, Fusco, and Pallara~\cite{AFP00}.  There's not much difference between the Euclidean and finite-dimensional Gaussian settings; research on Gaussian perimeter tends to focus on the trickier infinite-dimensional case.  For a thorough development of surface area in this latter setting (which of course includes finite-dimensional Gaussian space as a special case) see the work of Ambrosio, Miranda, Maniglia, and Pallara~\cite{AMMP10}; in particular, Theorem~4.1 in that work gives several additional equivalent definitions for~$\gsa$ besides those in Definition~\ref{def:unusual-gsa-def}.  Regarding the fact that~$\RS_A'(0^+)$ is an equivalent definition, the Euclidean analogue of this statement was proven in Miranda et~al.~\cite{MPPP07} and the statement itself follows similarly~\cite{Mir13} using Ambrosio et~al.~\cite{AFR13}.  (Our heuristic justification of~\eqref{eqn:heur-gsa2} is similar to the one given by Kane~\cite{Kan11a}.)  Additional related results can be found in Hino~\cite{Hin10} (which includes the remark about convex sets at the end of Definition~\ref{def:unusual-gsa-def}), Ambrosio and Figalli~\cite{AF11}, Miranda et~al.~\cite{MNP12}, and Ambrosio et~al.~\cite{AFR13}.

The inequality of Theorem~\ref{thm:gsa-concavity} is explicit in Ledoux~\cite{Led94} (see also the excellent survey~\cite{Led96}); he used it to deduce the Gaussian Isoperimetric Inequality.  He also noted that it's essentially deducible from an earlier inequality of Pisier and Maurey~\cite[Theorem~2.2]{Pis86}.
                                            \index{rotation sensitivity}%
Theorem~\ref{thm:rs-subadditive}, which expresses the subadditivity of rotation sensitivity, can be viewed as a discretization of the Pisier--Maurey inequality. This theorem appeared in work of Kindler and O'Donnell~\cite{KO12}, which also made the observations about the volume-$\half$ case of Borell's Isoperimetric Theorem at the end of Section~\ref{sec:borell} and in Remark~\ref{rem:easy-almost-gw}.

                                            \index{Bobkov's Inequality}%
Bobkov's Inequality~\cite{Bob97} in the special case of Gaussian space had already been implicitly established by Ehrhard~\cite{Ehr84}; the striking novelty of Bobkov's work (partially inspired by Talagrand~\cite{Tal93}) was his reduction to the two-point Boolean inequality.  The proof of this inequality which we presented is, as mentioned a discretization of the stochastic calculus proof of Barthe and Maurey~\cite{BM00a}. (In turn, they were extending the stochastic calculus proof of Bobkov's Inequality in the Gaussian setting due to Capitaine, Hsu, and Ledoux~\cite{CHL97}.) The idea that it's enough to show that Claim~\ref{claim:bobkov} is ``nearly true'' by computing two derivatives -- as opposed to showing it's exactly true by computing four derivatives -- was communicated to the author by Yuval Peres.  Following Bobkov's paper, Bakry and Ledoux~\cite{BL96b} established Theorem~\ref{thm:my-bobkov} in very general infinite-dimensional settings including Gaussian space; Ledoux~\cite{Led98} further pointed out that the Gaussian version of Bobkov's Inequality has a very short and direct \linebreak semigroup-based proof.  See also Bobkov and G{\"o}tze~\cite{BG99} and Tillich and Z{\'e}mor~\cite{TZ00} for results similar to Bobkov's Inequality in other discrete settings.
                                            \index{Gaussian Isoperimetric Inequality|)}%

                                            \index{Borell's Isoperimetric Theorem}%
Borell's Isoperimetric Theorem is from Borell~\cite{Bor85}.  Borell's proof used ``Ehrhard symmetrization'' and actually gave much stronger results -- e.g., that if $f, g \in L^2(\R^n, \gamma)$ are nonnegative and $q \geq 1$, then $\la (\U_\rho f)^q, g\ra$ can only increase under simultaneous Ehrhard symmetrization of~$f$ and~$g$.  There are at least four other known proofs of the basic Borell Isoperimetric Theorem. Beckner~\cite{Bec92} observed that the analogous isoperimetric theorem on the sphere follows from two-point symmetrization; this yields the Gaussian result via Poincar\'e's limit (for details, see Carlen and Loss~\cite{CL90}).  (This proof is perhaps the conceptually simplest one, though carrying out all the technical details is a chore.) Mossel and Neeman~\cite{MN12} gave the proof based on semigroup methods outlined in Exercises~\ref{ex:mossel-neeman1}--\ref{ex:mossel-neeman4}, and later together with De~\cite{DMN12} gave a ``Bobkov-style'' Boolean proof (see Exercise~\ref{ex:DMN}). Finally, Eldan~\cite{Eld13} gave a proof using stochastic calculus.

                                            \index{Berry--Esseen Theorem}%
As mentioned in Section~\ref{sec:berry-esseen}  there are several known ways to prove the Berry--Esseen Theorem.  Aside from the original method (characteristic functions), there is also Stein's Method~\cite{Ste72,Ste86a}; see also, e.g.,~\cite{Bol84,BH84,CGS11}.  The Replacement Method approach we presented originates in the work of Lindeberg~\cite{Lin22}.  The mollification techniques used (e.g., those in Exercise~\ref{ex:mollifier}) are standard.
                                            \index{Invariance Principle}%
The Invariance Principle as presented in Section~\ref{sec:invariance} is from Mossel, O'Donnell, and Oleszkiewicz~\cite{MOO10}. Further extensions (e.g., Exercise~\ref{ex:multifun-invariance}) appear in the work of Mossel~\cite{Mos10}.  In fact the Invariance Principle dates back to the 1971 work of Rotar'~\cite{Rot73,Rot74}; therein he essentially proved the Invariance Principle for degree-$2$ multilinear polynomials (even employing the term ``influence'' as we do for the quantity in Definition~\ref{def:multilinear-poly}).  Earlier work on extending the Central Limit Theorem to higher-degree polynomials had focused on obtaining sufficient conditions for polynomials (especially quadratics) to have a Gaussian limit distribution; this is the subject of \emph{U-statistics}. Rotar' emphasized the idea of invariance and of allowing any (quadratic) polynomial with low influences.  Rotar' also credited Girko~\cite{Gir73} with related results in the case of positive definite quadratic forms.  In 1975, Rotar'~\cite{Rot75} generalized his results to handle multilinear polynomials of any constant degree, and also random vectors (as in Exercise~\ref{ex:multifun-invariance}).  (Rotar' also gave further refinements in 1979~\cite{Rot79}.)

The difference between the results of Rotar'~\cite{Rot75} and the results of Mossel~et~al. \cite{MOO10} comes in the treatment of the error bounds.  It's somewhat difficult to extract simple-to-state error bounds from Rotar'~\cite{Rot75}, as the error there is presented as a sum over $i \in [n]$ of expressions $\E[F(\bx) \bone_{|F(\bx)| > u_i}]$, where~$u_i$ involves $\Inf_i[F]$.  (Partly this is so as to generalize the statement of the Lindeberg CLT.)  Nevertheless, the work of Rotar' implies a L\'{e}vy distance bound as in Corollary~\ref{cor:basic-inv-levy}, with some inexplicit function~$o_\eps(1)$ in place of $(1/\rho)^{O(k)} \eps^{1/8}$.  By contrast, the work of Mossel et~al.~\cite{MOO10} shows that a straightforward combination of the Replacement Method and hypercontractivity yields good, explicit error bounds.  Regarding the Carbery--Wright Theorem~\cite{CW01},
                                                    \index{Carbery--Wright Theorem}%
an alternative exposition appears in Nazarov, Sodin, and Vol'berg~\cite{NSV02}.

Regarding the Majority Is Stablest Theorem (conjectured in Khot, Kindler, Mossel, and O'Donnell~\cite{KKMO04} and proved originally in Mossel, O'Donnell, and Oleszkiewicz~\cite{MOO05a}), it can be added that additional motivation for the conjecture came from Kalai~\cite{Kal02}. The fact that~\eqref{eqn:gw-sdp} is an efficiently computable relaxation for the Max-Cut problem dates back to the 1990 work of Delorme and Poljak~\cite{DP93}; however, they were unable to give an analysis relating its value to the optimum cut value.  In fact, they conjectured that the case of the $5$-cycle from Example~\ref{eg:5-cycle} had the worst ratio of $\Opt(G)$ to~$\SDPOpt(G)$.  Goemans and Williamson~\cite{GW94} were the first to give a sharp analysis of the SDP (Theorem~\ref{thm:gw-behavior}), at least for $\theta \geq \theta^*$.  Feige and Schechtman~\cite{FS02} showed an optimal integrality gap for the SDP for all values $\theta \geq \theta^*$ (in particular, showing an integrality gap ratio of~$\cgw$); interestingly, their construction essentially involved proving Borell's Isoperimetric Inequality (though they did it on the sphere rather than in Gaussian space).  Both before and after the Khot et~al.~\cite{KKMO04} UG-hardness result for Max-Cut there was a long line of work~\cite{Kar99,Zwi99,AS00,ASZ02,CW04,KV05,FL06,KO06} devoted to improving the known approximation algorithms and UG-hardness results, in particular for $\theta < \theta^*$.  This culminated in the results from O'Donnell and Wu~\cite{OW08} (mentioned in Remark~\ref{rem:ow08}), which showed explicit matching $(\alpha,\beta)$-approximation algorithms, integrality gaps, and UG-hardness results for all $\frac12 < \beta < 1$.  The fact that the best integrality gaps matched the best UG-hardness results proved not to be a coincidence; in contemporaneous work, Raghavendra~\cite{Rag08} showed that for \emph{any} CSP, \emph{any} SDP integrality gap could be turned into a matching \dvq.  This implies the existence of matching efficient $(\alpha,\beta)$-approximation algorithms and UG-hardness results for every CSP and every~$\beta$.  See Raghavendra's thesis~\cite{Rag09} for full details of his earlier publication~\cite{Rag08} (including some Invariance Principle extensions building further on Mossel~\cite{Mos10}); see also Austrin's work~\cite{Aus07,Aus10} for precursors to the Raghavendra theory.

Exercise~\ref{ex:Tal1000} concerns a problem introduced by Talagrand~\cite{Tal89}. Talagrand offers a \$1,000 prize~\cite{Tal06a} for a solution to the following  Boolean version of the problem: Show that for any fixed $0 < \rho < 1$ and for $f \co \bits^n \to \R^{\geq 0}$ with $\E[f] = 1$ it holds that $\Pr[\T_\rho f > t] = o(1/t)$ as $t \to \infty$.  (The rate of decay may depend on~$\rho$ but not, of course, on~$n$; in fact, a bound of the form $O(\frac{1}{t \sqrt{\log t}})$ is expected.)  The result outlined in Exercise~\ref{ex:Tal1000} (obtained together with James Lee) is for the very special case of $1$-dimensional Gaussian space; Ball, Barthe, Bednorz, Oleszkiewicz, and Wolff~\cite{BBB+13} obtained the same result and also showed a bound of $O(\frac{\log \log t}{t \sqrt{\log t}})$ for $d$-dimensional Gaussian space (but with the constant in the $O(\cdot)$ depending on~$d$).

The Multifunction Invariance Principle (Exercise~\ref{ex:multifun-invariance} and its special case Exercise~\ref{ex:multivar-be}) are from Mossel~\cite{Mos10}; the version for general product spaces (Exercise~\ref{ex:general-invariance}) is from Mossel, O'Donnell, and Oleszkiewicz~\cite{MOO10}.

                                        \index{analysis of Boolean functions|)}% 

\chapter*{Some tips}

\vspace{-.7in}
\begin{itemize}
    \item You might try using analysis of Boolean functions whenever you're faced with a problems involving Boolean strings in which both the uniform probability distribution and the Hamming graph structure play a role.  More generally, the tools may still apply when studying functions on (or subsets of)  product probability spaces.
    \item If you're mainly interested in unbiased functions, or subsets of volume~$\frac12$, use the representation $f \btb$.  If you're mainly interested in subsets of small volume, use the representation $f \co \bn \to \{0,1\}$.
    \item As for the domain, if you're interested in the operation of adding two strings (modulo~$2$), use~$\F_2^n$.  Otherwise use~$\bn$.
    \item If you have a conjecture about Boolean functions:
            \begin{itemize}
                \item Test it on dictators, majority, parity, tribes (and maybe recursive majority of~$3$).  If it's true for these functions, it's probably true.
                \item Try to prove it by induction on~$n$.
                \item Try to prove it in the special case of functions on Gaussian space.
            \end{itemize}
    \item Try not to prove any bound on Boolean functions $f \btb$ that involves the parameter~$n$.
    \item Analytically, the only multivariate polynomials we really know how to control are degree-$1$ polynomials. Try to reduce to this case if you can.
    \item Hypercontractivity is useful in two ways: (i)~It lets you show that low-degree functions of independent random variables behave ``reasonably''. (ii)~It implies that the noisy hypercube graph is a small-set expander.
    \item Almost any result about functions on the hypercube extends to the case of the $p$-biased cube, and more generally, to the case of functions on products of discrete probability spaces in which every outcome has probability at least~$p$ -- possibly with a dependence on~$p$, though.
    \item Every Boolean function consists of a junta part and Gaussian part.
\end{itemize}

\backmatter
\bibliographystyle{alpha}
\bibliography{odonnell-bib}

\printindex

\end{document}